\pdfoutput=1
\PassOptionsToPackage{dvipsnames}{xcolor}

\documentclass[10pt,a4paper,twoside,openright,titlepage,fleqn,%
               headinclude,footinclude,BCOR5mm,%
               numbers=noenddot,cleardoublepage=empty]{scrreprt}

\usepackage{graphicx} \usepackage[fleqn]{amsmath} \usepackage{amssymb,amsthm} \usepackage[utf8]{inputenc} \usepackage[square]{natbib} \usepackage{subcaption} \usepackage{chngpage} \usepackage{calc} \usepackage{ccicons} \usepackage{xspace} \usepackage{booktabs} \usepackage{multirow} \usepackage[english]{babel}
\usepackage{listings} \usepackage{scrhack} 
\usepackage[printonlyused,smaller,withpage]{acronym} \usepackage{siunitx} \usepackage{physics} \usepackage[mathscr]{euscript} \usepackage{lscape} \usepackage{mathtools} 
\usepackage{longtable} \usepackage{array} \usepackage{xtab} \usepackage{colortab} \usepackage{multicol} \usepackage{mdframed} \usepackage{tikz} \usepackage{tikz-cd} \usepackage{tabularx} \usepackage{alltt} \usepackage{float} \usepackage{lipsum}
\usepackage[ruled,vlined,algochapter,linesnumbered]{algorithm2e}

\usepackage[floatperchapter,dottedtoc,eulerchapternumbers,pdfspacing,linedheaders]{classicthesis}
\usepackage{arsclassica}

\usepackage{pdfpages}

\usepackage{imakeidx}
\makeindex[columns=2, title=Alphabetical Index, intoc]

\usepackage{hyperref} \usepackage[capitalise]{cleveref}  

\theoremstyle{plain}
\newtheorem{definition}{Definition}[section]

\theoremstyle{plain}
\newtheorem{theorem}{Theorem}[section]

\theoremstyle{plain}
\newtheorem{proposition}{Proposition}[section]

\theoremstyle{plain}
\newtheorem{remark}{Remark}[section]

\theoremstyle{plain}
\newtheorem{example}{Example}[section]

\theoremstyle{plain}
\newtheorem{lemma}{Lemma}[section]

\theoremstyle{plain}
\newtheorem{corollary}{Corollary}[section]

\theoremstyle{plain}
\newtheorem{disclaimer}{Disclaimer}[section]

\newcommand{\mycomment}[1]{}

\makeatletter
\newcommand\footnoteref[1]{\protected@xdef\@thefnmark{\ref{#1}}\@footnotemark}
\makeatother

\newcommand{\myName}{Juan Sebastián Flórez Jiménez}
\newcommand{\myTitle}{The Non-Commutative Brillouin Torus\\a Non-Commutative Geometry perspective}

\newcommand{\myGroup}{Grupo de Geometría y Lógica}
\def\myGroupLogo{figs/logo_unal.png}
\newcommand{\myDepartment}{Departamento de Matemáticas}
\newcommand{\myFaculty}{Facultad de Ciencias}
\newcommand{\myUni}{Universidad Nacional de Colombia}

\newcommand{\myGraduationYear}{2024}
\newcommand{\myGraduationMonth}{October}

\newcommand{\myProf}{Ph. D. Leonardo Arturo Cano}
\newcommand{\mySupervisor}{Ph. D. Andres Reyes Lega}

\hypersetup{pdfauthor={\myName}}
\hypersetup{pdfkeywords={}}
\hypersetup{pdfsubject={}}
\hypersetup{pdftitle={\myTitle}}

\begin{document}
\pagenumbering{roman}
\pagestyle{plain}

\newcolumntype{L}[1]{>{\raggedright\let\newline\\\arraybackslash}m{#1}}

\newcolumntype{C}[1]{>{\centering\let\newline\\\arraybackslash}m{#1}}

\newcolumntype{R}[1]{>{\raggedleft\let\newline\\\arraybackslash}m{#1}}


\includepdf{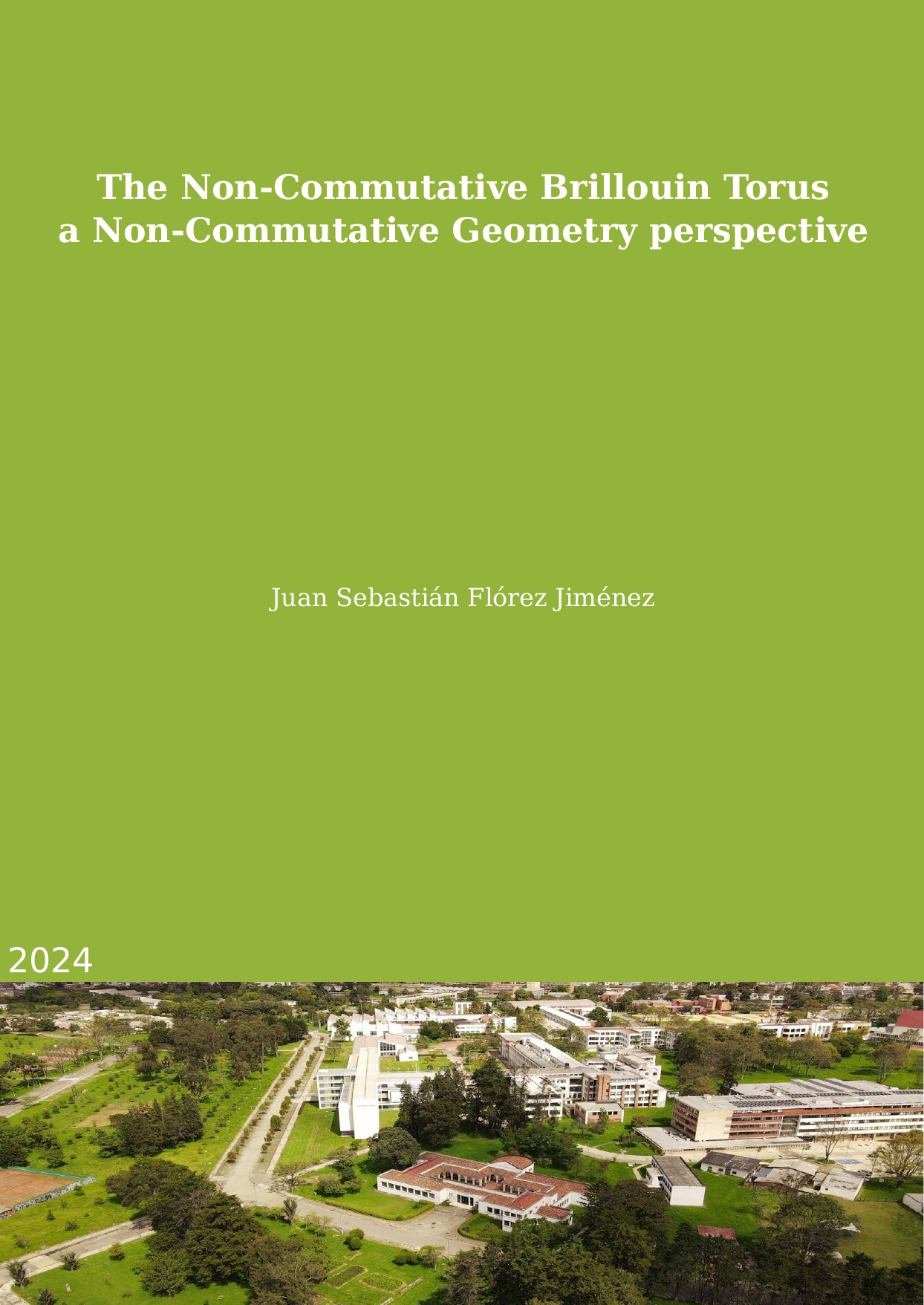}


\thispagestyle{empty}

\begin{titlepage}
  \begin{addmargin}[-1cm]{-3cm}
    \begin{center}
        \large  
        \hfill
        \vfill

        \begingroup
            \color{Maroon}\spacedallcaps{\myTitle}
        \endgroup
        
        \vspace{5cm}
        
        Tesis presentada como requisito para optar por el título de\\
        \vspace{2cm}
        Master en Ciencias - Matemáticas\\
        \vspace{4cm}
        por\\
        \bigskip
        \myName\\
        \bigskip
        \myGraduationMonth\xspace\myGraduationYear\\

        \vfill

    \end{center}  
  \end{addmargin}       
\end{titlepage}















    


          
\clearpage

\thispagestyle{empty}

\hfill
\vfill

\noindent\myName: \\ 
\textit{The Non-Commutative Brillouin Torus \\ a Non-Commutative Geometry perspective} \\ (\myGraduationYear)\\
\ccby\xspace This work is licensed under a Creative Commons Attribution 4.0 International License. To view a copy of this license, visit\\
\url{http://creativecommons.org/licenses/by/4.0/}.

\vspace{3em}


\vspace{3em}

\noindent{} This thesis has been developed in the:\\

\begin{tabular}{ll}
\parbox{0.15\textwidth}{\includegraphics[width=\linewidth]{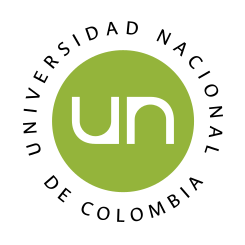}}
&
\parbox{0.7\textwidth}
{
  \myGroup\\
  \myDepartment\\
  \myFaculty\\
  \myUni\\
}       
\end{tabular}



\vspace{3em}

\noindent
\begin{tabular}{ll}
Main Advisor:  &  \myProf \\
Secondary Advisor:              &  \mySupervisor \\
\end{tabular}

\clearpage
\chapter*{Abstract}

Non-commutative geometry has found a fruitful field of examples in quantum mechanics and has provided rigorous mathematical tools to extract topological and geometrical information from those systems. This thesis is an incursion into some of the tools used by Jean Bellissard and collaborators for the analysis of homogeneous materials, having this in mind, our first step is to look into the objects used in such analysis, those are a pair of topological algebras devised to extend the realm of Fourier analysis over $\mathbb{T}^d$ into the study of tight-binding models for homogeneous materials. These algebras come as a generalization of $C(\mathbb{T}^d)$ and $C^{\infty}(\mathbb{T}^d)$, and are considered as a non-commutative smooth manifold in the field of Non-Commutative Geometry, we will refer to them as the Non-Commutative Brillouin Torus.

We explore how various techniques from the Fourier analysis over $\mathbb{T}^d$ can be translated into the Non-Commutative Brillouin Torus e.g. Fourier coefficients and Fejér summation, and those techniques allow us to capture the topological and smooth structure of such non-commutative smooth manifold. In this context, the topological structure is captured by a C* algebra while the smooth structure is captured by a particular type of Fréchet algebra, which is called a smooth sub-algebra. These results turn out to be the building blocks for the construction of topological invariants of Hamiltonians through the continuous cyclic cohomology of the smooth sub-algebra.

Another tool to study topological invariants of Hamiltonians is the K theory of C* algebras and smooth sub-algebras. Smooth sub-algebras and C* algebras share a similar functional calculus, which implies that smooth sub-algebras contain enough information to study the K theory of their C* algebras. This fact plays a crucial role in the identification of topological invariants of Hamiltonians. When the conditions are right (temperature close to $0$ and low electron density), it is one of the underlying causes of the quantization of the transversal conductivity in homogenous materials. 

This is yet another thesis about Non-Nommutative Geometry and topological insulators, however, we believe that it will be a useful asset for newcomers to the field. 
\chapter*{Agradecimientos}

Agradezco a Leonardo y Andrés por su guía y tutoría. Al trabajar en esta tesis hemos transitado un camino lleno de misterios y secretos escondidos, que nos ha llevado a aprender y explorar más allá de lo que originalmente pensabamos. Este camino nos deja con más preguntas que respuestas, y es así que nos maravillamos ante la belleza de nuestra pasión compartida, las matemáticas.

Agradezco a mi familia por apoyarme todos estos años, este es un logro tanto mio como de ellos, y así mismo quiero que sepan que en cada proyecto que yo emprenda en el futuro, habrá un pedacito de ellos allí presente. Agradezco a mis amigos, quienes me han apoyado de muchisimas maneras, constantemente recuerdo sus concejos, los cuales viven en mí como faros guiando mi trasegar. Valoro de sobremanera cada palabra, almuerzo y chiste compartidos, y deseo que estos momentos se multipliquen en el futuro.

Durante esta maestría he crecido tanto profesionalemnte como personalmente, le agradezco a cada persona que hizo parte de ese camino, en especial a las personas con quien compartí los salones de clase y apasionadas charlas sobre ciencia. Cada una de estas personas me ha ayudado a crecer y su apoyo ha sido fundamental en la concepción de este documento.

Mientras estuve matriculado en esta maestría no solamente estudíe matematícas, tuve la fortuna de explorar varios entornos tanto profesionales como personales, y cada uno de estas experiencias me han ayudado a conocerme mejor y explorar mis pasiones. Le agradezco a cada una de estas personas quienes dedicaron parte de su tiempo para compartir su vida conmigo.
\pagestyle{scrheadings} 
\clearpage
\setcounter{tocdepth}{2}

\tableofcontents

\cleardoublepage

\pagenumbering{arabic}
\chapter{Introduction}
\label{chap:introduction}

This thesis was conceived as we tried to work our way through the book \textit{Bulk and Boundary Invariants for Complex Topological Insulators} (\citep{prodan_bulk_2016}), as such, it is the result of our limitations, our curiosity, and our passion for sharing what we learned through this journey. In the aforementioned book the authors tackle the issue of defining topological invariants for physical systems with and without a boundary, in particular, they work with Hamiltonians over infinite lattices and such Hamiltonians correspond to tight binding models describing the dynamics of electrons in crystals.

The topological invariants defined in \citep{prodan_bulk_2016} are resilient against noise and take into account the effect of an external constant magnetic field, these are features that lay outside the realm of the Bloch theorem and require the introduction of new tools (\citep[Sections 1.5, 1.6, 1.7]{prodan_bulk_2016}). The tools introduced by Prodan et. al. belong to the field of Non-commutative geometry, and one of its objectives is the generalization of the even and odd Chern numbers \citep[Section 2.2.1 and Section 2.3.1]{prodan_bulk_2016}. In this setting, the even and odd Chern numbers are topological invariants associated with equivalence classes of Hamiltonians, and the equivalence relation corresponds to homotopic equivalence of Hamiltonians satisfying the bulk gap hypothesis (\cref{definition:bulk_gap_hypothesis}), the interested reader can refer to \cref{sec:motivation_from_physics} for more information on this equivalence relation.

The specific physical conditions under which the even and odd Chern numbers appear are irrelevant to the present document, instead, we are interested in the particular mathematical form these invariants take. Let $H$ be a self-adjoint bounded operator over the Hilbert space $L^2(\mathbb{Z}^d) \otimes \mathbb{C}^N$, also, denote by $S^y$ for $y \in \mathbb{Z}^d$ the bounded linear operator acting as $S^y \ket{x} = \ket{x+y}$ where $\ket{x}$ is the function from $\mathbb{Z}^d$ into $\mathbb{C}$ that takes the value $1$ for $x$ and the value $0$ for any other values. The prototypical Hamiltonian used in physics looks like $H = \sum_{y \in \mathcal{R}} S^y \otimes W_y$ with $W_y \in M_N(\mathbb{C})$, and the Fourier transform over $\mathbb{Z}^d$ provides a unique continuous function from $\mathbb{T}^d$ into $M_N(\mathbb{C})$ associated to $H$, this function takes the form,
$$ k \mapsto H_k, \; H_k = \sum_{y \in \mathcal{R}} e^{i\braket{y}{k}} W_y, \; H_k : \mathbb{C}^N  \to \mathbb{C}^N.$$ 

The map $k \mapsto H_k$ is smooth with respect to the norm topology over the Hilbert space $\mathbb{C}^N$, and the following formulae define topological invariants for Hamiltonians that satisfy the Bulk gap hypothesis, see \citep[equations 2.3 and 2.36]{prodan_bulk_2016}:
\begin{itemize}
    \item If $d$ is even, the \textbf{even Chern numbers} take the form,
    $$
    \mathrm{Ch}_d\left(P_{\text{Fermi}}\right)=\frac{(2 \pi \mathrm{i})^{\frac{d}{2}}}{\left(\frac{d}{2}\right) !} \sum_{\rho \in \mathcal{S}_d}(-1)^\rho \int_{\mathbb{T}^d} \frac{d k}{(2 \pi)^d} \operatorname{tr}\left(P_{\text{Fermi}}(k) \prod_{j=1}^d \frac{\partial P_{\text{Fermi}}(k)}{\partial k_{\rho(j)}}\right).
    $$
    \item If $d$ is odd, the \textbf{odd Chern numbers} take the form
    $$
    \mathrm{Ch}_d\left(U_{\text{Fermi}}\right)=\frac{i (\pi \mathrm{i})^{\frac{d-1}{2}}}{d!!} \sum_{\rho \in \mathcal{S}_d}(-1)^\rho \int_{\mathbb{T}^d} \frac{d k}{(2 \pi)^d} \operatorname{tr}\left( \prod_{l = 1}^{d} U_{\text{Fermi}}^*(k) \frac{\partial U_{\text{Fermi}}}{\partial k_{\rho(l)}}(k) \right).
    $$
\end{itemize}
In the previous formulae $P_{\text{Fermi}}$ and $U_{\text{Fermi}}$ are a projection and a unitary associated to the self-adjoint operator $H$ whose existence is guarantied by the presence of gaps in the spectrum of $H$, this technical detail is of no relevance to the main corpus of the document, however, the interested reader can refer to \citep[Section 2.4.2]{prodan_bulk_2016}, and also in \cref{sec:motivation_from_physics} we give a little more information. The range of $\mathrm{Ch}_d$ is $\mathbb{Z}$ (\citep[Corollary 5.7.2]{prodan_bulk_2016}), therefore, these are used to classify the homotopically invariant Hamiltonians satisfying the Bulk gap hypothesis. In the aforementioned presentation of the even and odd Chern numbers for Hamiltonians, the topological space $\mathbb{T}^d$ is called the \textbf{Brillouin torus}. The Chern numbers are an example of topological invariants. 

To generalize the even and odd Chern numbers into topological invariants that take into account the effect of an external constant magnetic field and are resilient against noise, Prodan et. al. look at $\mathrm{Ch}_d$ through the lens of continuous multilinear functionals over the Fréchet algebra $C^{\infty}(\mathbb{T}^d, M_N(\mathbb{C}))$, and come up with a new Fréchet algebra $\mathcal{B}$ and continuous multilinear functionals over $\mathcal{B}$. The Fréchet algebra $\mathcal{B}$ is dense inside a C* algebra $B$, the C* algebra $B$ is a C* algebra known as a twisted crossed product (\cref{definition:twsited_crossed_product}) and is a generalization of the C* algebra $C(\mathbb{T}^d, M_N(\mathbb{C}))$ (\cref{sec:trivial_twisting_actions}), additionally, $\mathcal{B}$ is a smooth sub algebra (\cref{def:smooth_sub_algebra}) of $B$. The elements of $B$ are families of operators over $L^2(\mathbb{Z}^d, \mathbb{C}^N)$, such that, the structure of those families encodes the noise of the physical systems and the presence of a constant external magnetic field. For more information on how the physical information is encoded in those families of Hamiltonians refer to \cref{sec:motivation_from_physics}. The literature (\citep[Chapter 3]{prodan_computational_2017})) refers to the pair $(B, \mathcal{B})$ as the \textbf{Non Commutative Brillouin Torus} (\cref{definition:non_commutative_brillouin_torus}).

The generalization of $\mathrm{Ch}_d$ proposed by Prodan et. al. takes the form of the paring between the continuous cyclic cohomology of $\mathcal{B}$ and the K theory of $B$, you can refer to \cref{chapter:cyclic_cohomology} for a brief introduction to cyclic cohomology and to \cref{chap:K_theory} for a brief introduction to K theory for C* algebras. The pairing between the continuous cyclic cohomology of $\mathcal{B}$ and the K theory of $B$ (\cref{proposition:pairing_topological_k_0_even_ciclyc_cohomology_C_star_algebras}, \cref{proposition:pairing_topological_k_1_odd_ciclyc_cohomology_C_star_algebras}) provides two abelian group homomorphism,
$$ \Psi_{0, \xi} : K_0(B) \to \mathbb{C}, \; \Psi_{1, \xi'} : K_1(B) \to \mathbb{C} $$
where $\xi, \; \xi'$ are elements of the continuous cyclic cohomology of $\mathcal{B}$. In \cref{chap:K_theory} we will see that $K_0(B)$ has information of the homotopically equivalent projections on matrix algebras over $B$ and $K_1(B)$ has information of homotopically equivalent unitaries on matrix algebras over $B$, so, the maps $ \Psi_{0, \xi}$ and $\Psi_{1, \xi'}$ provide a way of checking whether two projections or two unitaries are not homotopically equivalent, which in the physics context translates to finding out whether two families of Hamiltonians belong to different topological phases. In \citep[Proposition 4.2.4]{prodan_bulk_2016} it is proven that $K_0(B) \simeq \mathbb{Z}^{2^{d-1}}$ and $K_1(B) \simeq \mathbb{Z}^{2^{d-1}}$, under this context, for applications in physics it is relevant to identify those cyclic cocycles $\xi, \xi'$ such that the range of $\Psi_{0, \xi}, \; \Psi_{1, \xi'}$ is a discrete subgroup of $\mathbb{C}$ (\cref{sec:top_inva_over_the_non_commutative_brillouim_torus}).

Since the pairing between the continuous cyclic cohomology of $\mathcal{B}$ and the K theory of $B$ plays such an important role in physics, the purpose of the present document is to study a particular technical feature of that pairing, this detail is the following,

\begin{mdframed}
\begin{center}
\textbf{Thesis objective}    
\end{center}
We construct an unital C* algebra $B$ and a smooth sub algebra of $B$ denoted by $\mathcal{B}$, we refer to the pair $(B, \mathcal{B})$ as the Non Commutative Brillouin torus (\cref{definition:non_commutative_brillouin_torus}). Since the K theory groups are defined primarily for the C* algebra $B$, we extend the definition of $K_0(B)$ and $K_1(B)$ into the smooth sub algebra $\mathcal{B}$ to come up with two groups $K_0(\mathcal{B})$ and $K_1(\mathcal{B})$, then, we proof that $K_0(B)$ is isomorphic to $K_0(\mathcal{B})$ and $K_1(B)$ is isomorphic to $K_1(\mathcal{B})$ (\cref{remark:k_groups_and_the_Non_cOmmutative_brillouin_torus}). 
\end{mdframed}

This technical detail is key to the aforementioned pairing because the continuous cyclic cohomology of the smooth sub algebra $\mathcal{B}$ can not be directly related to the K-theory groups $K_0(B)$ and $K_1(B)$ of the C* algebra $B$, instead, it can be related to the K-theory groups of the smooth sub algebra $\mathcal{B}$, $K_0(\mathcal{B})$ (\cref{proposition:pairing_topological_k_0_even_ciclyc_cohomology}) and $K_1(\mathcal{B})$ (\cref{proposition:pairing_topological_k_1_odd_ciclyc_cohomology}). This happens because the continuous cyclic cohomology of $\mathcal{B}$ is constructed using continuous multilinear functionals over $\mathcal{B}$ (\cref{sec:cyclic cohomology definition}), and there are continuous multilinear functionals over $\mathcal{B}$ which cannot be extended to continuous multilinear functionals over $B$, for example, in the case of $B = C(\mathbb{T}^d)$ and $\mathcal{B} = C^{\infty}(\mathbb{T}^d)$ there are $d$ derivations over $\mathcal{B}$ which are continuous maps over $\mathcal{B}$ but cannot be extended to continuous maps over $B$ (\cref{sec:when_is_cyclic_cohomology_interesting}); the continuous cyclic cohomology of $B$ is tightly related to those $d$ derivations (\citep[Remark 5.1.3]{prodan_bulk_2016}).

To tackle the aforementioned technical detail we will take the following path:
\begin{enumerate}
    \item In \cref{chap:fourier_analysis} we will define the \textbf{Non-commutative Brillouin Torus}, that is, we will show how the pair of algebras $(B, \mathcal{B})$ is constructed. The C* algebra is a twisted crossed product which uses the group $\mathbb{Z}^d$ and is defined as the enveloping C* algebra of a Banach *-algebra (\cref{definition:twsited_crossed_product}), each element $b$ of $B$ is uniquely determined by a set of elements called its Fourier coefficients (\cref{definition:fourier_coefficients}), and $\mathcal{B}$ is a sub *-algebra of $B$ whose elements are characterised by the decay of their Fourier coefficients (\cref{proposition:characterization_smooth_sub_algebra_twsited_crossed_product}). The *-algebra $\mathcal{B}$ is a Fréchet algebra with a topology stronger than the topology of $B$ and invariant under the holomorphic functional calculus of $B$, thus, it becomes a smooth sub algebra of $B$ (\cref{corollary:smooth_elements_are_invariant_under_holomoprhic_and_c_infty_calculus}).
    \item In \cref{chap:K_theory} we briefly introduce the K-theory groups of $B$ and $\mathcal{B}$, and we proof the isomorphisms $K_0(B) \simeq K_0(\mathcal{B})$ and $K_1(B) \simeq K_1(\mathcal{B})$. The most important tool for such demonstration are \textbf{the holomorphic functional calculus for C* algebras} (\cref{theorem:holomorphic_functional_calculus_banach_algebras}) and \textbf{the holomorphic functional calculus for smooth sub algebras} (\cref{theorem:holomorphic_funct_calculus_smooth_sub_algebras}).  
\end{enumerate}

The approach taken in \citep{prodan_bulk_2016} is coined as a \textbf{non-commutative geometry approach} because instead of looking for a smooth manifold that replaces $\mathbb{T}^d$, they look for a Fréchet algebra that replaces $C^{\infty}(\mathbb{T}^d)$, which aligns with arguably the most important theme in Non-commutative geometry (\citep[Introduction]{connes_noncommutative_2014})
\begin{center}
    Replace topological/geometrical objects with algebraic/analytic objects.
\end{center}
Similarly, Prodan et. al. look for a C* algebra that replaces $C(\mathbb{T}^d)$. This approach is motivated by various important results from twentieth-century mathematicians, for example, the Gelfand representation theorem (\cref{theorem:gelfand_representation_theorem}) motivates the usage of non-commutative C* algebras as a formal dual of topological spaces in non-commutative geometry. 

In \cref{chap:non_commutative_geometry_and_topological_invariants_for_hamiltonians} we describe at a high level how the results exposed in this document fit within the formulation of topological invariants for effective models describe the dynamics of electrons on a crystal. \cref{chap:non_commutative_geometry_and_topological_invariants_for_hamiltonians} is potentially interesting for a reader with a physics background and uses various results from the appendices e.g. continuous cyclic cohomology (\cref{chapter:cyclic_cohomology}) and traces over twisted crossed products (\cref{section:fourier_analysis_weight_traces_and_states_apendix}).

\section{Comments on the structure of the document}
\label{sec:comments_on_the_structure_of_the_document}

This document consists of two parts, the main body of the document and the appendices, the purpose of the appendices is to provide a friendly introduction to the mathematical background of the main body of the document, in particular, the appendices are designed to be helpful for newcomers into the field of Non-Commutative Geometry, so, feel free to skip them and consult them when you consider it necessary.

\chapter{The Non Commutative Brillouin Torus}
\label{chap:fourier_analysis}

\textbf{The Non-Commutative Brillouin Torus} arises as a generalization of the $n$ dimensional torus, and we will explore how various concepts translate between the $n$ dimensional torus and \textbf{the Non-Commutative Brillouin Torus}, concepts like Fourier coefficients (\cref{definition:fourier_coefficients}), derivations (\cref{lemma:derivations_over_smooth_sub_algebra_twisted_crossed_product}) and the description as a universal C* algebra (\cref{lemma:twisted_crossed_products_with_Z}), for a comprehensive list of such concepts you can refer to \cref{sec:translating_concepts_from_T_n_to_non_comm_brillo_torus}. At a high level, the \textbf{The Non-Commutative Brillouin Torus} consists of two topological algebras, these objects are, a C* algebra with unit (\cref{def:C_star_algebra}) $ \boldsymbol{B}$ and a smooth sub algebra (\cref{def:smooth_sub_algebra}) of $ \boldsymbol{B}$ which is denoted by $ \boldsymbol{\mathcal{B}}$. 

In the spirit of Non-Commutative Geometry, the C* algebra $ \boldsymbol{B}$ contains the topological information of \textbf{The Non-Commutative Brillouin Torus} while the topological algebra $ \boldsymbol{\mathcal{B}}$ contains the differential geometry information of the \textbf{The Non-Commutative Brillouin Torus}. The C* algebra $ \boldsymbol{B}$ is a generalization of the C* algebras $ \boldsymbol{C(\mathbb{T}^d)}$, while the topological algebra $ \boldsymbol{\mathcal{B}}$ is a generalization of the Fréchet m-convex algebra $C^{\infty}(\mathbb{T}^d)$, in fact, $ \boldsymbol{\mathcal{B}}$ will turn out to be also a Fréchet m-convex algebra.

To come up with the definition of the \textbf{The Non-Commutative Brillouin Torus} we will take the following path:
\begin{enumerate}
    \item In \cref{sec:intro_twisted_crossed_products} we will start by looking into a particular family of C* algebras called the twisted crossed products, turns out that the C* algebra $ \boldsymbol{B}$ is a twisted crossed product. The twisted crossed products are C* algebras that are constructed using four ingredients: 
    \begin{itemize}
        \item a \textbf{C* algebra} $A$,
        \item a \textbf{locally compact group} $G$,
        \item an \textbf{action} $\alpha$ of $G$ over $A$,
        \item a \textbf{normalized cocyle} $\zeta$ over $G$,
    \end{itemize}
    Given these four ingredients is possible to define a C* algebra, and it is denoted by $ \boldsymbol{A \rtimes_{\alpha,\zeta} G}$.
    \item In \cref{sec:twisted_crossed_products_and_discrete_groups} we will look into a couple of results related to how these crossed products simplify when the group $G$ is an abelian discrete countable group.
    \item In \cref{section:twisted_crossed_products_with_Z} we will use the conceptual framework layout so far to look into twisted crossed product where $\boldsymbol{G = \mathbb{Z}^d}$, and we will provide a description of $ \boldsymbol{A \rtimes_{\alpha,\zeta} \mathbb{Z}^d}$ as a universal C* algebra, moreover, this description will come a generalization of the characterization of the algebra of continuous functions over the $d$ dimensional torus $\boldsymbol{C(\mathbb{T}^d)}$ as the C* algebra generated by $d$ commuting unitaries.
    \item In \cref{sec:Fourier_analysis_twistted_crossed_product} we will look at how the concept of Fourier coefficients from $C(\mathbb{T}^d)$ generalizes to the elements of the twisted crossed product $ \boldsymbol{A \rtimes_{\alpha,\zeta} \mathbb{Z}^d}$.
    \item In \cref{sec:derivations_twistted_crossed_product} we will use the concept of Fourier coefficients of element of $ \boldsymbol{A \rtimes_{\alpha,\zeta} \mathbb{Z}^d}$ to come up with a dense subalgebra of $ \boldsymbol{A \rtimes_{\alpha,\zeta} \mathbb{Z}^d}$, we will refer to this subalgebra as $ \boldsymbol{\mathcal{A}_{\alpha, \zeta}}$. The algebra $ \boldsymbol{\mathcal{A}_{\alpha, \zeta}}$ will be a smooth sub algebra of $ \boldsymbol{A \rtimes_{\alpha,\zeta} \mathbb{Z}^d}$ (\cref{def:smooth_sub_algebra}) and will have among other the following properties: 
    \begin{itemize}
        \item $ \boldsymbol{A \rtimes_{\alpha,\zeta} \mathbb{Z}^d}$ will be a Fréchet m-convex algebra (\cref{proposition:smooth_elements_as_D_infinity_Freche_subalgebra})
        \item $ \boldsymbol{A \rtimes_{\alpha,\zeta} \mathbb{Z}^d}$ will have $d$ continuous derivations (\cref{lemma:algebra_of_smooth_elements_is_Frechet_space})
        \item $\boldsymbol{\mathcal{A}_{\alpha,\zeta}} $ is invariant under the holomorphic functional calculus over $\boldsymbol{A \rtimes_{\alpha,\zeta} \mathbb{Z}^d}$ (\cref{corollary:smooth_elements_are_invariant_under_holomoprhic_and_c_infty_calculus})
    \end{itemize}
    \item In \cref{sec:intro_twsited_transformation_group_C_star_algebras} we will look into the case where $\boldsymbol{A = C(\Omega)}$ with $\Omega$ a compact space, and the action of $G$ on $C(\Omega)$ comes from a continuous action of $G$ on the compact topological space $\Omega$. These type of twisted crossed products are called twisted transformation group C* algebras.
    \item In \cref{sec:topological_algebras_for_disordered_crystals} we will construct a pair of topological algebras $(B, \mathcal{B})$ used for the study of disordered crystal under constant magnetic fields, these topological algebras are an example of a pair $ \boldsymbol{A \rtimes_{\alpha,\zeta} \mathbb{Z}^d, \mathcal{A}_{\alpha, \zeta}}$ and are referred in the literature as \textbf{The Non-Commutative Brillouin Torus}.
\end{enumerate}

In the path towards the definition of the \textbf{The Non-Commutative Brillouin Torus} we will use various results that lay the foundations of Non-Commutative Geometry, for example, the Gelfand-Naimark theorem (\cref{theorem:faithfull_universal_representation_c_star_algebras}) justifies the study of C* algebras through the study of operators over Hilbert spaces, and we will define the Fourier coefficients of an element in $ \boldsymbol{A \rtimes_{\alpha,\zeta} \mathbb{Z}^d}$ by looking at $ \boldsymbol{A \rtimes_{\alpha,\zeta} \mathbb{Z}^d}$ as an algebra of operators over the Hilbert space $L^2(\mathbb{Z}^d,H)$ (\cref{definition:fourier_coefficients}). Also, the representation of the C* algebra $ \boldsymbol{A \rtimes_{\alpha,\zeta} \mathbb{Z}^d}$ over $L^2(\mathbb{Z}^d,H)$ will be a generalization of the representation of the C* algebra $C(\mathbb{T})$ as multiplication operators over the Hilbert space $L^2(\mathbb{T})$ (\cref{remark:left_regular_representation_and_Foureir_coefficients}).

Most of the results exposed in the present chapter lay within the filed of Fourier analysis on C* algebras, this has been a prolific field where various of the methods from classic harmonic analysis have found a home, like summation techniques and representation theory \citep{bedos_fourier_2016, bedos_twisted_2009, bedos_fourier_2015}. Twisted crossed products and harmonic analysis have also found a generalization into smooth sub-algebras of C* algebras and Von Neumann algebras \citep{schulz-baldes_harmonic_2022}.

\section{Twisted crossed products}
\label{sec:twisted_crossed_products}

We will dive in twisted crossed product because those are key to the definition of Non-commutative Brillouin torus (\cref{sec:topological_algebras_for_disordered_crystals}).

\subsection{Introduction to Twisted crossed products}
\label{sec:intro_twisted_crossed_products}

A twisted crossed product is going to be a C* algebra which we will assign to a separable twisted dynamical system (\cref{definition:twsited_crossed_product}), and a separable twisted dynamical system will be defined by four ingredients (\cref{def:separable_twisted_dynamical_system})
\begin{itemize}
    \item \textbf{A separable C* algebra:} $A$ (\cref{def:separable_c_star_algebra}).
    \item \textbf{A second countable locally compact abelian group:} $G$.
    \item \textbf{An action of $G$ over $A$}: $\alpha: G \to \text{Aut}(A)$.
    \item \textbf{A normalized 2 cocycle over $G$}: $\zeta :  G \times G \to \mathbb{T}$.
\end{itemize}
Our definition of twisted crossed product is going to be a simplification of the general definition of a separable twisted crossed product which you can find in \citep{packer_twisted_1989} and \citep{busby_representations_1970}, those references work with normalized 2 cocycles that take values on $M(A)$ instead of $\mathbb{C}$, where $M(A)$ is the multiplier C* algebra of $A$ (\cref{remark:multiplier_algebras}). Before we delve into the definitions, we would like to make some statements about the groups $G$ and C* algebras $A$ which we will work with,

\begin{remark}[Simplifying our analysis]\label{remark:simplifying_our_analysis}
We will work with 
\begin{itemize}
    \item $G$ is a second countable locally compact Hausdorff abelian group,
    \item $A$ is a separable C* algebra, 
\end{itemize}
these two statements have some interesting consequences which are relevant for the upcoming discussions:
\begin{itemize}
    \item From \cref{theorem:faithfull_universal_representation_c_star_algebras} we know that $A$ has a faithful representation on a separable Hilbert space $H_A$.
    \item If $\mu$ is the Haar measure over $G$ (\cref{definition:Haar_measure}), then, \cref{section:algebra_continuous_functions_locally_comp_space} tell us that $\mu$ is both $\sigma$-compact and $\mu$-countably generated.
    \item Given that $A$ is separable, then, the statement in the previous item along with \cref{remark:what_happend_if_X_is_separable} guaranties that the three definitions of measurability for functions $f : G \to A$ taking values in Banach algebras coincide.
    \item  From \cref{prop:separability_Lp_spaces} we get that $L^p(G;A)$ ($L^p(G;H_A)$) are separable Banach spaces for any $1 \leq p < \infty$.
\end{itemize}
In the upcoming discussions we will refer to $\mu$ as the Haar measure over $G$, additionally, if $G$ is compact we will assume that the Haar measure is normalized, that is, $\mu(G) = 1$, or equivalently, the Haar measure on $\hat{G}$ is the counting measure, where $\hat{G}$ is the dual group of $G$ (\cref{definition:dual_group}).
\end{remark}

To define the twisted crossed products we need to establish some relations between $G$ and $A$, so, let's introduce some notation. Normalized 2 cocycles are going to be one of the sources of non-commutativity of the twisted crossed products,

\begin{definition}[normalized 2 cocycle \index{normalized 2 cocycle} (Definition 2.1 \citep{bedos_twisted_2009})]\label{definition:normalized_2_cocycle}
Let $G$ be an abelian locally compact group and $e$ the identify of $G$, then, a (normalized) 2-cocycle on $G$ with values in $\mathbb{T}$ is a map $\zeta: G \times G \rightarrow \mathbb{T}$ such that
$$
\begin{gathered}
\zeta(g, h) \zeta(g + h, k)=\zeta(h, k) \zeta(g, h + k) \quad(g, h, k \in G) \\
\zeta(g, e)=\zeta(e, g)=1 \quad(g \in G) .
\end{gathered}
$$
The set of all normalized 2-cocycles will be denoted by $Z^2(G, \mathbb{T})$.
\end{definition}

\begin{definition}[Twisting action\index{Twisting action} (Definition 2.1 \citep{packer_twisted_1989})]\label{definition:twisting_action}
Let $A$ be a separable C* algebra and $G$ a second countable locally compact Hausdorff abelian group, a \textbf{twisted action of} $\boldsymbol{G}$ \textbf{on} $\boldsymbol{A}$ is a pair $(\alpha, \zeta)$ of maps $\alpha: G \to \text{Aut}(A)$, $\zeta: G \times G \to \mathbb{T}$ satisfying
\begin{itemize}
    \item $\zeta$ is a normalized 2-cocycle that is measurable.
    \item For each $a \in A$ the map $s \mapsto \alpha(s)(a)$ is measurable.
    \item $\alpha(id_G)=Id_A$, with $Id_A: A \to A, \; Id(a) = a$ for all $a \in A$.
    \item $\alpha(s) \circ \alpha(t)=\alpha(s + t)$ for $s, t \in G$.
\end{itemize}
The term $\text{Aut}(A)$\index{$\text{Aut}(A)$} denotes the set of all C* automorphisms of the C* algebra $A$.
\end{definition}

\begin{definition}[Separable twisted dynamical system\index{Separable twisted dynamical system}]\label{def:separable_twisted_dynamical_system}
Let $A$ be a separable C* algebra and $G$ a second countable locally compact Hausdorff abelian group, let $(\alpha, \zeta)$ be a \textbf{twisted action of} $\boldsymbol{G}$ \textbf{on} $\boldsymbol{A}$ (\cref{definition:twisting_action}), then, we shall refer to the quadruple $(A,G,\alpha,\zeta)$ as a \textbf{separable twisted dynamical system}. 
\end{definition}

From the discussion in \cref{remark:simplifying_our_analysis} we know that the maps $\zeta$ and $s \mapsto \alpha(s)(a)$ are $\mu$-strongly measurable, thus, \cref{def:separable_twisted_dynamical_system} coincides with the definition of a separable twisted dynamical system provided in \citep[Definition 2.1]{busby_representations_1970} and we can rely on the results given in \citep{busby_representations_1970} to state the following key result,

\begin{theorem}[Twisted $L^1(G;A)$ (Theorem 2.2 \citep{busby_representations_1970})]\label{theorem:twisted_L1_G_A_is_banach_algebra}

Let $(A,G,\alpha,\zeta)$ be a separable twisted dynamical system, then, $L^1(G;A)$ provided with the norm $f \mapsto \int_{G} \| f (g) \| d \mu(g)$ becomes a Banach *-algebra under the following operations:
\begin{itemize}
    \item \textbf{Addition:} $(f + g)(s) := f(s) + g(s)$ for $f,g \in L^1(G;A), \; s \in G$.
    \item \textbf{Multiplication:} Let  $f,g \in L^1(G;A), \; x \in G$,
    $$ (f \cdot g)(x) := \int_G f(y) \alpha(y)(g(x - y)) \zeta(y, x-y) d \mu(y).$$
    \item \textbf{Involution:} Let  $f \in L^1(G;A), \; x \in G$,
    $$ f^*(x) := (\zeta(x, -x))^* [\alpha(x)(f(-x)^*)].$$
\end{itemize}
This Banach *-algebra is denoted by $L^1(G,A; \alpha,\zeta)$\index{$L^1(G,A; \alpha,\zeta)$}. 
\end{theorem}

\begin{definition}[Banach *-algebra  $L^1(G,A; \alpha,\zeta)$]\label{def:banahc_algebra_L_1_A_G_alpha_zeta}
Let $(A,G,\alpha,\zeta)$ be a separable twisted dynamical system, then, the elements of $L^1(G;A)$ provided with the operations described in \cref{theorem:twisted_L1_G_A_is_banach_algebra} and the usual norm of $L^1(G,A)$ becomes a Banach *-algebra which will be denoted by $L^1(G,A; \alpha,\zeta)$.
\end{definition}

\cref{def:banahc_algebra_L_1_A_G_alpha_zeta} gives us a generalization of the convolution in the Banach algebra $L^1(G)$ which is described in \cref{proposition:useful_inequalities_of_Bochner_L_p_spaces}, since we are adding both the action of $G$ over $A$ and the normalized 2 cocycle $\zeta$ into the multiplication, also, it provides a generalization to the involution given in \cref{example:L1_Bohner_spaces_and_convolution}.

\begin{lemma}[Properties of $L^1(G,A; \alpha,\zeta)$]\label{lemma:properties_of_L_1_A_G_alpha_zeta}
We have that $L^1(G,A; \alpha,\zeta)$ is a separable Banach *-algebra. .

\end{lemma}
\begin{proof}
 $L^1(G,A; \alpha,\zeta)$ has the same norm as $L^1(G,A)$, and $L^1(G,A)$ is separable by \cref{remark:simplifying_our_analysis}, therefore, $L^1(G,A; \alpha,\zeta)$ is a separable Banach *-algebra.
\end{proof}

It is possible to establish an equivalence relation between twisted actions, as follows

\begin{definition}[Equivalence relation of twisted actions (Definition 2.4 \citep{busby_representations_1970})]\label{def:equivalence_between_twsited_actions}
Let $(\alpha_0, \zeta_0)$ and $(\alpha_1, \zeta_1)$ be two twisted actions of $G$ on $A$, then we say that $(\alpha_0, \zeta_0)$ is equivalent to $(\alpha_1, \zeta_1)$ if there exists a measurable function $\rho: G \to \mathbb{T}$ such that
$$ \alpha_0 = \alpha_1, \; \zeta_0(x,y) = \frac{\rho(x) \rho(y)}{ \rho(x+y)} \zeta_1(x,y). $$
\end{definition}

\begin{theorem}[cf. Theorem 2.7 \citep{busby_representations_1970} (Equivalence relation between twisted actions)]\label{theorme:equivalence_relations_twisted_actions}
 Let $(\alpha_0, \zeta_0)$ and $(\alpha_1, \zeta_1)$ be two twisted actions of $G$ on $A$ that are equivalent (\cref{def:equivalence_between_twsited_actions}), then, the Banach *-algebras $L^1(G,A;\alpha_0,\zeta_0)$ $L^1(G,A;\alpha_1, \zeta_1)$ are isomorphic.
\end{theorem}

There are many equivalent ways of defining the twisted crossed product associated to the twisted dynamical system $(A,G, \alpha, \zeta)$, we take arguably the simplest one and then piggyback on results from \citep{packer_twisted_1989} to give a better characterization of the twisted crossed products. If $(A,G,\alpha,\zeta)$ is a separable twisted dynamical system, then, $L^1(G,A;\alpha,\zeta)$ is a Banach *-algebra, therefore, any *-representation of $L^1(G,A;\alpha,\zeta)$ is norm decreasing (\cref{proposition:automatic_continuity_banach_star_algebras}), and this implies that for any $f \in L^1(G,A;\alpha,\zeta)$ the value
$$ \| f \|_{\text{sup}}  = \sup_{\pi} \| \pi(f) \|$$
is finite, where $\pi$ runs over all representations of $L^1(G,A;\alpha,\zeta)$ over a separable Hilbert space. The map $f \mapsto \| f \|_{\text{sup}}$ is a seminorm over $L^1(G,A;\alpha,\zeta)$, also, the seminorm $\| \cdot \|_{\text{sup}}$ satisfies the C* property, that is, $\| f f^* \| = \|f\|^2$. Under this setting, define by $N_{\text{sup}}$ the sub *-algebra of $L^1(G,A;\alpha,\zeta)$ consisting of elements where the seminorm $\| \cdot \|_{\text{sup}}$ becomes zero, then, $L^1(G,A;\alpha,\zeta) / N_{\text{sup}}$ becomes a normed *-algebra where the norm satisfies the C* property, and we refer to norm over $L^1(G,A;\alpha,\zeta) / N_{\text{sup}}$ as $\| \cdot \|_{\text{sup}}$ also. The completition of $L^1(G,A;\alpha,\zeta) / N_{\text{sup}}$ with respect to $\| \cdot \|_{\text{sup}}$ is a C* algebra and is called the enveloping C* algebra of $L^1(G,A;\alpha,\zeta)$, this C* algebra is denoted by $C^*(L^1(G,A;\alpha,\zeta))$.

The C* algebra $C^*(L^1(G,A;\alpha,\zeta))$ is an example of an universal C* algebra, an universal C* algebra is a C* algebra that is defined through a set of elements and relations between those elements (\cref{sec:Universal_C_star_algebras}). The most common example of universal C* algebras are the enveloping C* algebras, this term is used when the set of elements are the elements of a *-algebra, and the relations between those elements are the definition of addition, multiplication and involution over the aforementioned *-algebra, much like $C^*(L^1(G,A;\alpha,\zeta))$ is defined by the elements and algebraic structure of $L^1(G,A;\alpha,\zeta)$. If $G$ is discrete and countable we will see that $C^*(L^1(G,A;\alpha,\zeta))$ is isomorphic to the enveloping C* algebra of $L^1_c(G,A;\alpha,\zeta)$ (\cref{proposition:twisted_crossed_product_restricts_to_discrete_conunatlbe_groups}), where the elements of $L^1_c(G,A;\alpha,\zeta)$ are functions with compact support and the addition, multiplication and involution are the same from $L^1(G,A;\alpha,\zeta)$ (\cref{lemma:sub_algebra_functions_with_support_inside_compact_sets}). 

The enveloping C* algebras have an universal property (\cref{proposition:factoring_representations_with_enveloping_C_star_algebras}), as expected from an object called universal, in the case of $C^*(L^1(G,A;\alpha,\zeta))$ this property states that, given any representation $\pi$ of $L^1(G,A;\alpha,\zeta)$, there is a C* homomorphism $\tilde{\pi}$ with domain $C^*(L^1(G,A;\alpha,\zeta))$ which coincides with $\pi$ when restricted to $L^1(G,A;\alpha,\zeta)$. According to \citep[Remark 2.6]{packer_twisted_1989}, the C* algebra $C^*(L^1(G,A;\alpha,\zeta))$ is the twisted crossed product associated to the separable twisted dynamical system $(A,G,\alpha,\zeta)$, therefore, we introduce the following definition, 

\begin{definition}[Twisted crossed product\index{Twisted crossed product} ]\label{definition:twsited_crossed_product}
Let $(A,G,\alpha,\zeta)$ be a separable twisted dynamical system, the enveloping C* algebra of $L^1(G,A;\alpha,\zeta)$ is called the twisted crossed product of $(A,G,\alpha,\zeta)$ and is denoted by $A \rtimes_{\alpha, \zeta}G$. Using the notation for universal C* algebras (\cref{sec:Universal_C_star_algebras}), we have that
$$ A \rtimes_{\alpha, \zeta} G:= C^*(L^1(G,A;\alpha,\zeta)),.$$
\end{definition}

\begin{lemma}[$A \rtimes_{\alpha, \zeta}G$ is separable]\label{lemma:twsited_crossed_product_is_separable}
Let $(A,G,\alpha,\zeta)$ be a separable twisted dynamical system and $A \rtimes_{\alpha, \zeta}G$ its associated twisted crossed product (\cref{definition:twsited_crossed_product}), then, $A \rtimes_{\alpha, \zeta}G$ is a separable C* algebra.
\end{lemma}
\begin{proof}
Let $\pi: L^1(G,A;\alpha,\zeta) \to C^*(L^1(G,A;\alpha,\zeta))$ be the *-homomorphism that comes from the definition of the enveloping C* algebra (\cref{sec:Universal_C_star_algebras}), since the representations of a Banach *-algebra are norm decreasing (\cref{proposition:automatic_continuity_banach_star_algebras}) we have that $\pi$ is norm decreasing, additionally, $\pi(L^1(G,A;\alpha,\zeta))$ is dense in $ C^*(L^1(G,A;\alpha,\zeta))$ by definition (\cref{definition:twsited_crossed_product}). Given that under our assumptions $L^1(G,A;\alpha,\zeta)$ is separable (\cref{lemma:properties_of_L_1_A_G_alpha_zeta}), the previous statements guarantee that $A \rtimes_{\alpha, \zeta}G$ is also separable.
\end{proof}


\subsection{Twisted crossed products and discrete groups}
\label{sec:twisted_crossed_products_and_discrete_groups}

In the present section we want to look into the case $G$ discrete and countable, and we want to describe $A \rtimes_{\alpha, \zeta}G$ as the enveloping C* algebra of the *-algebra $L^1_c(G,A;\alpha,\zeta)$. We want to have such description because it is simpler to operate with the elements of $L^1_c(G,A;\alpha,\zeta)$ than with the elements of $L^1(G,A;\alpha,\zeta)$, and this will give us a simpler definition of $A \rtimes_{\alpha, \zeta}G$ (\cref{lemma:descriptpon_twisted_crossed_product_discrete_group_simplified_relations}).

\begin{definition}\label{def:algebra_of_funcitons_with_supports_essentially_contained_in_compact_set}
Denote by $L^1(G,A;\alpha,\zeta)_c$\index{$L^1(G,A;\alpha,\zeta)_c$} the dense subset of $L^1(G,A;\alpha,\zeta)$ consisting of functions in $L^1(G,A;\alpha, \zeta)$ whose support is essentially contained inside a compact set of $G$.
\end{definition}

\begin{lemma}[Sub algebra of functions with support inside compact sets]\label{lemma:sub_algebra_functions_with_support_inside_compact_sets}
$L^1(G,A;\alpha,\zeta)_c$ is a dense sub algebra of $L^1(G,A;\alpha, \zeta)$ closed under involution.
\end{lemma}
\begin{proof}
Take $f,g \in L^1(G,A;\alpha, \zeta)$ such that $\text{sup}(f) \subset K_1,  \; \text{sup}(g) \subset K_2,$ with $K_i$ compact sets, let us denote
$$K_1 + K_2 = \{ g \in G : \exists g_1 \in K_1, g_2 \in K_2, \text{ st } g = g_1 + g_2 \}$$
then, if $y \notin K_1$ then $f(y) \alpha(y)(g(x - y)) \zeta(y, x-y) x = 0$. So, suppose that $y \in K_1$, then, if $x-y \notin K_2$ we also have that $f(y) \alpha(y)(g(x - y)) \zeta(y, x-y) x = 0$, thus, when $x \notin K_2 + K_1$ we have that $f(y) \alpha(y)(g(x - y)) \zeta(y, x-y) x = 0$, meaning that $f \cdot g$ has support inside $K_1 + K_2$. Since $K_1 + K_2$ is also compact (\citep[Proposition 4.3]{de_chiffre_haar_2011}), we have that the support of $f \cdot g$ is also contained in a compact set. The support of $f+g$ is contained in $K_1 \cup K_2$, which is also compact, and according to the involution described in \cref{theorem:twisted_L1_G_A_is_banach_algebra} the support of $f^*$ is contained in $-K_1$ which is also compact. The previous statements imply that $L^1(G,A;\alpha,\zeta)_c$ is a sub algebra of $L^1(G,A;\alpha, \zeta)$ closed under involution.

Given that the continuous functions with compact support from $G$ into $A$ are dense in $L^1(G,A;\alpha, \zeta)$ (\cref{lemma:L_p_bochner_spaces_for_locally_compact_hausdroff_spaces}), and any continuous function with compact support is Bochner integrable (\cref{example:Bochner_integral_continuos_funcion}), then $L^1(G,A;\alpha,\zeta)_c$ is dense inside $L^1(G,A;\alpha, \zeta)$.
\end{proof}

\begin{lemma}\label{proposition:twisted_crossed_product_restricts_to_discrete_conunatlbe_groups}
Let $(G,A,\alpha,\zeta)$ be a separable twisted dynamical system (\cref{def:separable_twisted_dynamical_system}) and $G$ a countable and discrete group, then, $A \rtimes_{\alpha, \zeta} G$ is the C* algebra generated by the set $\mathcal{G} = L^1(G,A;\alpha,\zeta)_c$ subject to the algebraic relations of $L^1(G,A;\alpha,\zeta)_c$, so
$$ A \rtimes_{\alpha,\zeta} G \simeq C^*(L^1(G,A;\alpha,\zeta)_c).$$
\end{lemma}
\begin{proof}
Before diving into the proof we introduce some notation, take $f \in L^1(G,A;\alpha,\zeta)$, then, we will identify $f$ with the formal sum $\sum_{s \in G} f(s) u_s$, such that, if $f, g \in L^1(G,A;\alpha,\zeta)$ then
\begin{itemize}
    \item $f^* \mapsto \sum_{ s \in G } \zeta(s, -s)^*  \alpha(s)(f(-s)^*) u_s $,
    \item $fg \mapsto \sum_{ k \in G } \left( \sum_{m \in G} \zeta(m, k-m) f(m) \alpha(m)(g(k-m)) \right) u_k$,
\end{itemize}
following the operations described in \cref{theorem:twisted_L1_G_A_is_banach_algebra}. This construction translates into $L^1(G,A;\alpha,\zeta)_c$, in which case the sum $\sum_{s \in G} f(s) u_s$ has a finite amount of terms.

Under this setup, the map $a \mapsto a u_0$ is a *-algebra homomorphism from $A$ into $L^1(G,A;\alpha,\zeta)_c$, and for every representation $\pi$ of $L^1(G,A;\alpha,\zeta)_c$ there is a corresponding representation of $A$ given by $a \mapsto \pi (a u_0)$. Since C* homomorphisms are norm decreasing (\cref{proposition:automatic_continuity_C_star_algebras}) we must have that $\|\pi (a u_0 ) \| \leq \| a\|$ for all $a \in A$. Take $\pi$ a representation of $L^1(G,A;\alpha,\zeta)_c$, then, from the C* property of the norm in C* algebras (\cref{def:C_star_algebra}) we get that
$$ \| \pi(a u_s) \pi(a u_s)^* \| = \| \pi(a u_s) \|^2, $$
thus, with the aid of the multiplication and involution on $L^1(G,A;\alpha,\zeta)_c$ we can check that
$$ (a u_s)^* = \zeta(-s,s)^* \alpha(-s)(a^*)u_{-s}, \;  (a u_s) (a u_s)^* = \zeta(-s,s)^* a a^* u_0, $$
and we get
$$ \| \pi(a u_s) \pi(a u_s)^* \| = \| \pi(a a^* u_0) \|^2 \leq \| a a^* \|^2 = \| a \|^2,  $$
which ultimately implies that $\| a u_s \| \leq \| a \|$ for any $s \in G$.

Given that the norm on $L^1(G,A;\alpha,\zeta)_c$ takes the following form
$$ \| \sum_{s \in G} f(s) u_s \|_1 = \sum_{s \in G} \| f(s) \|, $$
the bound $\| \pi(a u_s) \| \leq \|a\|$ tell us that for any representation $\pi$ of $L^1(G,A;\alpha,\zeta)_c$ we have $\| \pi(f)\| \leq \|f \|_1$ i.e. the representations of $L^1(G,A;\alpha,\zeta)_c$ are automatically continuous with respect to the norm of $L^1(G,A;\alpha,\zeta)$. Since the norm of all the representations of $L^1(G,A;\alpha,\zeta)_c$ is bounded, the enveloping C* algebra of $L^1(G,A;\alpha,\zeta)_c$ exists.

Any representation $\pi$ of $L^1(G,A;\alpha,\zeta)$ restricts to a representation of $L^1(G,A;\alpha,\zeta)_c$, hence, for any $f \in L^1(G,A;\alpha,\zeta)_c$,
$$ \sup_{\pi_1} \| \pi(f) \| \leq \sup_{\pi_0} \| \pi(f) \|,$$
where $\pi_0$ runs over all representations of $L^1(G,A;\alpha,\zeta)$ and $\pi_1$ runs over all representations of $L^1(G,A;\alpha,\zeta)_c$.

Take $\pi$ a representation of $L^1(G,A;\alpha,\zeta)_c$ over a separable Hilbert space $H$, then, given that $\| \pi(f) \| \leq \|f\|$ for any $f \in L^1(G,A;\alpha,\zeta)_c$ and $L^1(G,A;\alpha,\zeta)_c$ is dense in $L^1(G,A;\alpha,\zeta)$ (\cref{lemma:sub_algebra_functions_with_support_inside_compact_sets}), the following defines a representation of $L^1(G,A;\alpha,\zeta)$,
$$\pi_1 : L^1(G,A;\alpha,\zeta) \to B(H), \; \pi_1(f) = \lim_{n \to \infty} \pi(f_n)$$
with $\{f_n\}_{n \in \mathbb{N}} $ a sequence of elements of $L^1(G,A;\alpha,\zeta)$ converging to $f$. This implies that, for any $f \in L^1(G,A;\alpha,\zeta)_c$,
$$ \sup_{\pi_0} \| \pi(f) \| \leq \sup_{\pi_1} \| \pi(f) \|,$$
where $\pi_0$ runs over all representations of $L^1(G,A;\alpha,\zeta)$ and $\pi_1$ runs over all representations of $L^1(G,A;\alpha,\zeta)_c$.

Take $f \in L^1(G,A;\alpha,\zeta)_c$, then, its norm as an element of $C^*(L^1(G,A;\alpha,\zeta))$ is
$$ \sup_{\pi_0} \| \pi(f) \| , \; \pi_0 : L^1(G,A;\alpha,\zeta) \to B(H),$$
and its norm as an element of $C^*(L^1(G,A;\alpha,\zeta)_c)$ is
$$ \sup_{\pi_1} \| \pi(f) \| , \; \pi_1 : L^1(G,A;\alpha,\zeta)_c \to B(H),$$
therefore, the aforementioned inequalities imply that $f$ has the same norm as an element of $C^*(L^1(G,A;\alpha,\zeta)_c)$ and $C^*(L^1(G,A;\alpha,\zeta))$. 

Since $L^1(G,A;\alpha,\zeta)_c$ is dense in $L^1(G,A;\alpha,\zeta)$ and the representations of $L^1(G,A;\alpha,\zeta)$ are norm decreasing, $L^1(G,A;\alpha,\zeta)_c$ is also dense in the C* algebra $C^*(L^1(G,A;\alpha,\zeta))$. Under the current setup, following an argument similar to the one explained in \cref{lemma:extending_star_homomorphisms_into_C_star_homomorphisms} is possible to construct two C* homomorphisms 
$$\phi_0: C^*(L^1(G,A;\alpha,\zeta)) \to C^*(L^1(G,A;\alpha,\zeta)_c)$$
and 
$$\phi_1: C^*(L^1(G,A;\alpha,\zeta)_c) \to C^*(L^1(G,A;\alpha,\zeta)),$$
which restrict to isometries over $L^1(G,A;\alpha,\zeta)_c$, therefore, they must be isometries over $C^*(L^1(G,A;\alpha,\zeta))$, moreover, $\phi_0$ and $\phi_1$ are isomorphisms of C* algebras due to the automatic continuity of C* homomorphisms (\cref{proposition:automatic_continuity_C_star_algebras}). 
\end{proof}

Following the notation in the proof of \cref{proposition:twisted_crossed_product_restricts_to_discrete_conunatlbe_groups}, if $G$ is countable and discrete, denote by $au_s$ with the function from $G$ into $A$ that assigns $a$ to $s$ and $0$ to the other elements, using this notation, any member of $L^1(G,A;\alpha,\zeta)_c$ is a finite sum of objects like $au_s$. Following the definition of multiplication, addition and involution over $L^1(G,A;\alpha,\zeta)_c$ exposed in \cref{theorem:twisted_L1_G_A_is_banach_algebra}, we have the following,
\begin{itemize}
    \item \textbf{addition:} $a_0 u_s + a_1 u_s = (a_0 + a_1)u_s $ where $a_0 + a_1$ is an element of $A$.
    \item \textbf{multiplication:} $(a_0 u_s)( a_1 u_t) = (\zeta(s,t) a_0 \alpha(s)(a_1)) u_{s+t} $ where $\zeta(s,t) a_0 \alpha(s)(a_1)$ is an element of $A$.
    \item \textbf{Involution:} $(au_s)^* = \zeta(-s,s)^*\alpha(-s)(a^*) u_{-s}$ where $\zeta(-s,s)^*\alpha(-s)(a^*)$ is an element of $A$.
\end{itemize}
Henceforth, the *-algebra $L^1(G,A;\alpha,\zeta)_c$ is the universal *-algebra generated by the set $\{ a u_s :\; a \in A , s \in G\}$ and following the aforementioned relations.

\begin{lemma}\label{lemma:twisted_crossed_product_discrete_group_and_generators}
Let $(G,A,\alpha,\zeta)$ be a separable twisted dynamical system (\cref{def:separable_twisted_dynamical_system}) and $G$ a countable and discrete group, then, $ A \rtimes_{\alpha,\zeta} G$ is isomorphic to the C* algebra generated by the set 
$$\mathcal{R} =  \{ a u_s :\; a \in A , s \in G\},$$
subject to the following relations:
\begin{itemize}
    \item \textbf{addition:} $a_0 u_s + a_1 u_s = (a_0 + a_1)u_s $ where $a_0 + a_1$ is an element of $A$.
    \item \textbf{multiplication:} $(a_0 u_s)( a_1 u_t) = (\zeta(s,t) a_0 \alpha(s)(a_1)) u_{s+t} $ where $\zeta(s,t) a_0 \alpha(s)(a_1)$ is an element of $A$.
    \item \textbf{Involution:} $(au_s)^* = \zeta(-s,s)^*\alpha(-s)(a^*) u_{-s}$ where $\zeta(-s,s)^*\alpha(-s)(a^*)$ is an element of $A$.
\end{itemize}
\end{lemma}

If $G$ is discrete and countable, the relations between the elements $au_s$ established in \cref{lemma:twisted_crossed_product_discrete_group_and_generators} can be rewritten in a simpler form if look at the elements $au_s$ as a formal multiplication of two elements, the element $a$ and the element $u_s$. In this case, the relations established \cref{lemma:twisted_crossed_product_discrete_group_and_generators} are equivalent to the following relations
$$ u_s u_t = \zeta(s,t) u_{s+t}, \; \alpha(s)(a) u_s = u_s a, \; u_s^* = \zeta(-s,s)^*u_{-s}  $$
plus the algebraic relations between the objects of $A$, therefore, we have an alternative description of $A \rtimes_{\alpha, \zeta} G$

\begin{lemma}\label{lemma:descriptpon_twisted_crossed_product_discrete_group_simplified_relations}
Let $(G,A,\alpha,\zeta)$ be a separable twisted dynamical system (\cref{def:separable_twisted_dynamical_system}) and $G$ a countable and discrete group, then, $A \rtimes_{\alpha, \zeta} G$ is the C* algebra generated by the set $\mathcal{G} = \{ a u_s :\; a \in A , s \in G\}$ subject to the relations
$$ u_s u_t = \zeta(s,t) u_{s+t}, \; \alpha(s)(a) u_s = u_s a, \; u_s^* = \zeta(-s,s)^*u_{-s}  $$
plus the algebraic relations between the objects of $A$.
\end{lemma}

\subsection{Norm of the twisted crossed products}
\label{label:norm_twited_crossed_products}

According \citep[Proposition 11.2.5, item 3]{dales_introduction_2003} and \citep[Definition 9.2.3]{dales_introduction_2003}, a locally compact abelian group $G$ is said to be amenable\index{amenable group} if the enveloping C* algebra of $L^1(G)$ has a faithful representation over the Hilbert space $L^2(G)$ and that faithful representation takes the following form,
$$ \pi: C^*(L^1(G)) \to B(L^2(G)), \pi(f)(h)(s) = \int_G f(x) h(s-x) d \mu(x),$$
where $f \in L^1(G)$ and $h \in L^2(G)$. Turns out that all locally compact abelian groups are amenable (\citep[Proposition 11.1.3]{dales_introduction_2003}), this is an important fact for our analysis.

Let $\pi$ be a faithful representation of $A$ over the Hilbert space $H$, then, if is $G$ amenable, according to \citep[Theorem 3.11]{packer_twisted_1989} there is a faithful representation of $A \rtimes_{\alpha,\zeta} G$ over the Hilbert space $L^2(G,H)$, this faithful presentation is called the regular representation induced by $\pi$ and corresponds to the integrated form of a covariant representation of $(G,A,\alpha,\zeta)$ (\cref{def:integral_form_covariant_representation}). In this section, we will introduce the concepts necessary to understand the faithful representation of $A \rtimes_{\alpha,\zeta} G$ when $G$ is discrete, countable, and abelian, this faithful representation will come in handy to analyze the case $G=\mathbb{Z}^d$ (\cref{sec:Fourier_analysis_twistted_crossed_product}). The faithful representation of the C* algebra $A \rtimes_{\alpha,\zeta} \mathbb{Z}^d$ over $L^2(\mathbb{Z}^d, H)$ will allow us to define the Fourier coefficients of the elements of $A \rtimes_{\alpha,\zeta} \mathbb{Z}^d$ (\cref{sec:Fourier_analysis_twistted_crossed_product}).

\begin{definition}[Covariant representation of a twisted dynamical system\index{Covariant representation of a twisted dynamical system} (Definition 2.3 \citep{packer_twisted_1989})]\label{definition:representation_twsited_dynamical_system}
Let $(A,G,\alpha,\zeta)$ be a separable twisted dynamical system, then a pair $(\pi,\pi_U)$ is a \textbf{covariant representation} if it satisfies the following conditions,
\begin{enumerate}
    \item $\pi$ is a non-degenerate representation of $A$ on a separable Hilbert space $H$.
    \item $\pi_U: G \to U(H)_s$ is a measurable map.
    \item \textbf{Projective representation of $G$ on $H$ }(\citep[Definition 2.2]{bedos_twisted_2009}):  $\pi_U(s) \pi_U(t) = \zeta(s,t) \pi_U(s+t)$ for $s,t \in G$.
    \item $\pi(\alpha(s)(a)) = \pi_U(s) \pi(a)( \pi_U(s) )^*$ for $a \in A, \; s \in G$.
\end{enumerate}
Under this setup we have that $\pi_U(s)^* = \zeta(s,-s)^* \pi_U(-s)$ for $s \in G$.
\end{definition}

The measurability of the map $\pi_U$ is equivalent to saying that the maps $s \mapsto U(s)(\xi)$ and $s \mapsto \langle U(s)(\xi), \eta \rangle$ are measurable by \cref{proposition:measurability_in_unitary_group}, moreover, since $H$ is separable then the maps $s \mapsto U(s)(\xi)$ and $s \mapsto \langle U(s)(\xi), \eta \rangle$ are $\mu$-strongly measurable by the statements in \cref{remark:simplifying_our_analysis}.

According to \citep[Theorem 3.3]{busby_representations_1970}, given a covariant representation $(\pi,\pi_U)$ of $(A,G,\alpha,\zeta)$, is possible to come up with a representation of $L^1(G,A;\alpha,\zeta)$, such that, to each element $f \in L^1(G,A;\alpha,\zeta)$ corresponds a bounded operator over $H$ which is denoted by 
$$ \int_G \pi(f(s)) \pi_U(s) d \mu (s). $$
This operator is the unique bounded operator over $H$ such that, for any $\eta,\psi \in H$ we have that
$$ \langle  \int_G \pi(f(s)) \pi_U(s) d \mu (s) (\eta), \psi \rangle = \int_G \langle \pi(f(s)) \pi_U(s) (\eta), \psi \rangle d \mu (s). $$
In the case of $G$ a discrete, countable and abelian group, the existence of the operator aforementioned operator is guarantied by \cref{proposition:functions_compact_support_representations_L_1}. In the case of $G$ an abelian locally compact group (not necessarily discrete), the existence of the operator $ \int_G \pi(f(s)) \pi_U(s) d \mu (s) $ is guarantied by \cref{proposition:creating_operators_on_L_2_from_composition_of_measurealbe_functions}, additionally, this operator is defined using the weak topology of $H$ because its existence is guarantied by the one-to-one correspondence between sesquilinear forms and bounded linear operators over Hilbert spaces (\cref{proposition:correspondence_bounded_operators_and_sesquilinear_forms}), as explained in the proof of \cref{proposition:creating_operators_on_L_2_from_composition_of_measurealbe_functions}. In \citep[page 515]{busby_representations_1970} the operator $ \int_G \pi(f(s)) \pi_U(s) d \mu (s)$ is defined as the weak limit of sums of the form $\sum_{i=1}^{n}\pi(f(x_i))\pi_U (x_i) \mu(S_i) $ where $S_i$ is a measurable set with finite measure, which is equivalent to the definition using sesquilinear due to the Riesz representation theorem (\citep[Theorem 2.53]{allan_introduction_2011}).

The function $s \mapsto \|\pi(f(s)) \pi_U(s) \|$ is integrable as explained in the proof of \cref{proposition:creating_operators_on_L_2_from_composition_of_measurealbe_functions}, that is, $\int_{G} \|\pi(f(s)) \pi_U(s) \| d\mu(s) \leq \infty$, therefore, we may ask whether the operator $\int_G \pi(f(s)) \pi_U(s) d \mu (s)$ can be defined by the integral of the Banach algebra valued function $s \mapsto \pi(f(s)) \pi_U(s)$ by using a generalization of the Lebesgue integral, this type of integral is called the Bochner integral  (\cref{sec:Bochner_integral}). 
If $G$ is discrete and countable the operator $\int_G \pi(f(s)) \pi_U(s) d \mu (s)$ takes the form of a Bochner integral as explained in \cref{proposition:functions_compact_support_representations_L_1}, however, if $G$ is not a discrete and countable group some technicalities details may interfere in the definition of the integral of the function $s \mapsto \pi(f(s)) \pi_U(s)$, you can refer to the introduction of \cref{sec:oper_valu_func} for an example of such technical details.

\begin{definition}[cf. Theorem 3.3 \citep{busby_representations_1970} (Integral form of a covariant representation)]\label{def:integral_form_covariant_representation}
Let $(A,G,\alpha,\zeta)$ be a separable twisted dynamical system, let $(\pi,\pi_U)$ be a covariant representation of $(A,G,\alpha,\zeta)$ over a separable Hilbert space $H$, then, the representation $\Pi$ of $L^1(G,A;\alpha,\zeta)$ given by
$$\Pi(f) := \int_G \pi(f(s)) \pi_U(s) d \mu (s)$$
is called the integrated form of $(\pi,\pi_U)$\index{integrated form of $(\pi,\pi_U)$}.
\end{definition}

We say that a Banach algebra $B$ has a left-bounded approximate identity if there is a bounded sequence of elements $\{ c_i \}_{i \in \mathbb{N}}$ such that, for any $b \in B$ we have that $\lim_{n \to \infty} c_i b = b$. $B$ is said to have a right bounded approximate identity if there a bounded sequence $\{ c_i \}_{i \in \mathbb{N}}$ such that, for any $b \in B$ we have that $\lim_{n \to \infty} b c_i = b$. $B$ is said to have a two-sided approximate identity if there is a bounded sequence $\{ c_i \}_{i \in \mathbb{N}}$ that is both a left and right bounded approximate identity. According to \citep[Theorem 3.3]{busby_representations_1970}, if $(A,G,\alpha,\zeta)$ is a separable twisted dynamical system such that $L^1(G,A;\alpha,\zeta)$ has a norm bounded left approximate identity, then, there is a one-to-one correspondence between the covariant representations of $(A,G,\alpha,\zeta)$ and the representations of $L^1(G,A;\alpha,\zeta)$, additionally, if $(\pi,\pi_U)$ is covariant representation then its corresponding *-representation is given by the integrated form of  $(\pi,\pi_U)$. Since \citep[Proposition in appendix]{packer_twisted_1990} states that $L^1(G,A;\alpha,\zeta)$ has a two-sided norm bounded approximate identity, we can use \citep[Theorem 3.3]{busby_representations_1970} to provide the following characterization of the the representations of $L^1(G,A;\alpha,\zeta)$,

\begin{theorem}[Characterization of twisted crossed products]\label{theorem:characterization_of_twisted_crossed_products}
Let $(A,G,\alpha,\zeta)$ be a separable twisted dynamical system, then,
\begin{enumerate}
    \item There is a one-to-one correspondence between covariant representations of $(A,G,\alpha,\zeta)$ and *-representations of $L^1(G,A;\alpha,\zeta)$, such that, if $(\pi,\pi_U)$ is covariant representation then its corresponding *-representation is given by the integrated form of  $(\pi,\pi_U)$.
    \item  For every covariant representation $(\pi,\pi_U)$ of $L^1(G,A;\alpha,\zeta)$ into a separable Hilbert space $H$, or equivalently, for every representation $\Pi$ of $L^1(G,A;\alpha,\zeta)$ into a separable Hilbert space $H$, there is a C* algebra homomorphism 
    $$\phi: A \rtimes_{\alpha, \zeta} G \to B(H),$$
    such that $\phi(f) = \Pi(f)$ for $f \in L^1(G,A;\alpha,\zeta)$. 
\end{enumerate}
\end{theorem}
\begin{proof}
\begin{enumerate}
    \item It is a consequence of \citep[Theorem 3.3]{busby_representations_1970} as established in the preamble of this theorem.
    \item This is a consequence of the universal property of enveloping C* algebras stated in \cref{proposition:factoring_representations_with_enveloping_C_star_algebras}.
\end{enumerate}
\end{proof}

So far we have defined the twisted crossed product, but we have not checked that is not the trivial C* algebra $0$, for that, we need to provide a non-zero representation of $L^1(G,A;\alpha,\zeta)$, or equivalently (\cref{theorem:characterization_of_twisted_crossed_products}) a non zero covariant representation of $(A,G,\alpha,\zeta)$.  

\begin{proposition}\label{proposition:regular_representation}

Let $(A, G, \alpha, \zeta)$ be a separable twisted dynamical system and $\pi$ a faithful representation of the C* algebra $A$ on a separable Hilbert space $H$, then, the following pair of maps define a covariant representation of $(A, G, \alpha, \zeta)$ over $L^2(G,H)$:
$$
\begin{gathered}
\tilde{\pi} : A \to B(L^2(G, H)), \;  (\tilde{\pi}(a) (\xi))(t)=\pi\left(\alpha(t)(a)\right)(\xi(t)), \\
R_{\zeta} : G \to U(L^2(G, H)), \; \left(R_{\zeta}(s) (\xi)\right)(t)=\zeta(t, s)(\xi(t + s)) ,
\end{gathered}
$$
with $t,s \in G$, $a\in A$ and $\xi \in L^2(G, H)$.

\end{proposition}
\begin{proof}
We need to check that both $\tilde{\pi}(a)$ and $R_{\zeta}(s)$ are bounded linear operators over $L^2(G;H)$. First, lets check first that $\tilde{\pi}(a) (\xi), R_{\zeta}(s) (\xi) \in L^2(G;H)$:
\begin{itemize}
    \item $\tilde{\pi}(a)(\xi) \in L^2(G;H):$ By definition $s \mapsto \alpha(s)(a)$ is measurable for any $a \in A$ (\cref{def:separable_twisted_dynamical_system}), since $\pi$ is continuous, then $s \mapsto \pi(\alpha(s)(a))$ is also measurable. By \cref{prop:oper_valu_new_strong_measu_funcs_in_separable_case} we have that $s \mapsto \pi\left(\alpha(s)(a)\right)(\xi(s))$ is $\mu$-strongly measurable. $\alpha(s)$ is an *-automorphism and has inverse given by $\alpha(-s)$, since C* homomorphisms are norm decreasing (\cref{proposition:automatic_continuity_C_star_algebras}) we have that $\| \alpha(s)(a)\| = \|a \|$. Therefore,  

    $$ \int_G \|\pi\left(\alpha(s)(a)\right)(\xi(s))\|^2 d \mu(s) \leq  \int_G \| a\|^2 \|\xi(s)\|^2 d \mu(s) \leq \|a \|^2 \| \xi \|^2, $$
    consequently $\tilde{\pi}(a)(\xi) \in L^2(G;H)$.
    \item $R_{\zeta}(s) (\xi) \in L^2(G;H):$ By definition $(t,s) \mapsto \zeta(t,s)$ is measurable, thus $t \mapsto \zeta(t,s)$ is measurable for any $s \in G$. A similar argument as the one in the previous item tells us that $t \mapsto \zeta(t, s)(\xi(t + s))$ is $\mu$-strongly measurable, moreover, since the Haar measure is left invariant (\cref{definition:Haar_measure}) we have that
    $$ \int_G \|\zeta(t, s)(\xi(t + s))\|^2 d \mu(t) = \int_G \|\zeta(t-s, s)(\xi(t))\|^2 d \mu (t) = \| \xi\|^2, $$\
    consequently $R_{\zeta}(s) (\xi) \in L^2(G;H)$. Also, 
    $$ \int_G \langle R_{\zeta}(s) (\xi)(t), R_{\zeta}(s) (\eta)(t) \rangle d \mu(t) = \int_G \zeta(t, s) \zeta(t, s)^* \langle \xi(t+s), \eta(t+s) \rangle d \mu (t) = \langle \xi,\eta\rangle,  $$
    thus, $R_{\zeta}(s) \in U(L^2(G;H))$.
\end{itemize}
It is easy to check that $\tilde{\pi}$ is a *-representation of $A$. Also, $R_{\zeta}(s) = m(s) \circ T(-s) $, with $T$ the left regular representation and $m(s)(\xi)(t) = \zeta(t,s) \xi(t)$, so, we know that $ s \mapsto T(-s)(\xi)$ is continuous for any $\xi \in H$, hence, it is a measurable map. Since $\zeta: G \times G \to \mathbb{C}$ is measurable we can check that that $m(s) \in B(L^2(G;H))$ and $s \mapsto m(s)$ is a measurable map, thus, by \cref{prop:oper_valu_new_strong_measu_funcs_in_separable_case} we have that $s \mapsto m(s)(T(-s)(\xi))$ is a measurable map.

From the properties of the normalized 2 cocycles, we can check that
$$ R_{\zeta}(s) R_{\zeta}(t) (\xi) = R_{\zeta}(s)(l \mapsto \zeta(l, t)(\xi(l + t))) = (l \mapsto \zeta(l,s) \zeta(l+s, t)(\xi(l + t + s)) ),$$
so, using the properties of normalized 2 cocycles (\cref{definition:normalized_2_cocycle}) we get that
$$ R_{\zeta}(s) R_{\zeta}(t) (\xi) = \zeta(s,t) R_{\zeta}(s+t)(\xi). $$

Notice that, $ R_{\zeta}(s)^* (\xi)(t) = \zeta(t-s,s)^* \xi(t -s) $,
therefore,
$$R_{\zeta}(s) \tilde{\pi}(a) R_{\zeta}(s)^* (\xi)(t) = \pi(\alpha(s+t)(a))(\xi)(t),  $$
consequently,
$\tilde{\pi}(\alpha(s)(a)) = R_{\zeta}(s) \tilde{\pi}(a)( R_{\zeta}(s) )^*$ for $a \in A, \; s \in G$.
Thus, the pair $(\tilde{\pi},R_{\zeta})$ is a convariant representation of the separable twisted dynamical system $(A, G, \alpha, \zeta)$.
\end{proof}

The covariant representation presented in \cref{proposition:regular_representation} plays an important role in this document, therefore we introduce the following definition

\begin{definition}[Right regular representation\index{Right regular representation} (Definition 3.10 \citep{packer_twisted_1989})]\label{def:right_regular_representation_twisted_crossed_product}
Let $(A, G, \alpha, \zeta)$ be a separable twisted dynamical and $\pi:A \to B(H)$ a faithful representation with $H$ a separable Hilbert space, then, the covariant representation $(\tilde{\pi}, R_{\zeta})$ is called the right regular representation of $(A, G, \alpha, \zeta)$ associated to $\pi$. In the previous statement the representations $(\tilde{\pi}, R_{\zeta})$ take the following form, $\xi \in L^2(G, H)$
$$
\begin{gathered}
\tilde{\pi} : A \to B(L^2(G, H)), \;  (\tilde{\pi}(a) (\xi))(t)=\pi\left(\alpha(t)(a)\right)(\xi(t)), \\
R_{\zeta} : G \to U(L^2(G, H)), \; \left(R_{\zeta}(s) (\xi)\right)(t)=\zeta(t, s)(\xi(t + s)) ,
\end{gathered}
$$
with $t,s \in G$, $a\in A$ and $\xi \in L^2(G, H)$. The integrated form of $(\tilde{\pi}, R_{\zeta})$ (\cref{theorem:characterization_of_twisted_crossed_products}) is denoted by $\tilde{\Pi}$, that is, 
$$\tilde{\Pi}(f) :=  \int_G \tilde{\pi}(f(s)) R_{\zeta}(s) d \mu(s). $$
\end{definition}

According to \citep[Remark 3.12]{packer_twisted_1989} the norm of the operator $\int_G \tilde{\pi}(f(s)) R_{\zeta}(s) d \mu(s)$ provided in \cref{def:right_regular_representation_twisted_crossed_product} does not depend on the representation $\pi$ as long as it is a faithful representation, therefore, we have the following definition,

\begin{definition}[Reduced twisted crossed product\index{Reduced twisted crossed product}]\label{definition:reduced_crossed_product}
Let $(A, G, \alpha, \zeta)$ be a separable twisted dynamical and $\pi:A \to B(H)$ a faithful representation, denote by  $A \rtimes_{\zeta,\alpha,r} G$ the smallest sub C* algebra of $B(L^2(G,H))$ that contains $\tilde{\Pi}( L^1(G,A;\alpha,\zeta))$. $A \rtimes_{\zeta,\alpha,r} G$\index{$A \rtimes_{\zeta,\alpha,r} G$} is called the reduced twisted crossed product of $(A, G, \alpha, \zeta)$.
\end{definition}

\begin{lemma}\label{lemma:map_from_twisted_crossed_product_to_reduced_version}
There is a C* homomorphism
$$ \phi: A \rtimes_{\alpha, \zeta} G \to A \rtimes_{\zeta,\alpha,r} G ,$$
such that $\phi(f) = \int_G \tilde{\pi}(f(s)) R_{\zeta}(s) d \mu(s)$ for $f \in L^1(G,A;\alpha,\zeta)$.
\end{lemma}
\begin{proof}
Given that $A \rtimes_{\alpha,\zeta} G$ is the enveloping C* algebra of $L^1(G,A;\alpha,\zeta)$, the universal property of enveloping C* algebras (\cref{proposition:factoring_representations_with_enveloping_C_star_algebras}) states that given the *homomorphism of *-algebras 
$$ \tilde{\Pi}: L^1(G,A;\alpha,\zeta) \to B(L^2(G,H)) $$
there is a C* homomorphism 
$$ \phi: A \rtimes_{\alpha,\zeta} G \to B(L^2(G,H)) $$
such that $\phi|_{L^1(G,A;\alpha,\zeta)} = \tilde{\Pi}$. This is the desired homomorphism of C* algebras.
\end{proof}

\citep[Theorem 3.11]{packer_twisted_1989} states that, if $G$ is an amenable group then the C* homomorphism in \cref{lemma:map_from_twisted_crossed_product_to_reduced_version} becomes an isomorphism, or equivalently, that the twisted crossed product $A \rtimes_{\alpha, \zeta} G$ has a faithful representation over $L^2(G,H)$ and that faithful representation takes the form of $ \tilde{\Pi}$ when is restricted to $L^1(G,A;\alpha,\zeta)$. Given that all locally compact abelian groups are amenable (\citep[Proposition 11.1.3]{dales_introduction_2003}) and $G$ is assumed to be abelian in the setup of this document (\cref{def:separable_twisted_dynamical_system}), we have the following,

\begin{theorem}[cf. Theorem 3.11 \citep{packer_twisted_1989} (Norm of twisted crossed products of separable twisted dynamical systems)]\label{theorem:norm_of_twisted_crossed_products}
Let $(A, G, \alpha, \zeta)$ be a separable twisted dynamical, assume that $\pi : A \to B(H)$ is a faithful representation of $A$ on a separable Hilbert space, then, the twisted crossed product $A \rtimes_{\alpha, \zeta} G$ is isomorphic to the reduced twisted crossed product $A \rtimes_{\zeta,\alpha,r} G$ and the isomorphism is the given by the map described in \cref{lemma:map_from_twisted_crossed_product_to_reduced_version}.
\end{theorem}

In \cref{def:right_regular_representation_twisted_crossed_product} we provided a covariant representation of $(A,G, \alpha,\zeta)$, now, we proceed to provide another covariant representation that arises as a generalization of the left regular representation of $G$ on $L^2(G)$ (\cref{example:left_regular_representation_of_G}). In \cref{proposition:regular_representation} the map $R_{\zeta}: G \to U(H)$ has the letter $R$ in its name because it is a generalization of the right action of $G$ on $L^2(G)$, thus, we can look for a generalization of the left action of $G$ on $L^2(G)$ (\cref{example:left_regular_representation_of_G}).

\begin{definition}[Left regular representation\index{Left regular representation} of separable twisted dynamical systems]\label{def:left_regular_representation}
Let $(A,G,\alpha,\zeta)$ be a separable twisted dynamical system, let $\pi: A \to B(H)$ be a faithful representation of $A$ in a separable Hilbert space, set
$$
\begin{gathered}
(\overline{\pi}(a) (\xi))(t)=\pi\left(\alpha(-t)(a)\right)(\xi(t)), \\
\left(L_{\zeta}(s) (\xi)\right)(t)=\zeta(s, t-s)\xi(t - s),
\end{gathered}
$$
then $(\overline{\pi},L_{\zeta})$ is a covariant representation of $(A,G, \alpha,\zeta)$. The integrated form of $(\overline{\pi},L_{\zeta})$ is denoted by $\overline{\Pi}$.
\end{definition}

To show that $(\overline{\pi},L_{\zeta})$ is a covariant representation of you can follow an argument similar to the one in \cref{proposition:regular_representation}. Some literature refer $(\overline{\pi},L_{\zeta})$ as the regular covariant representation associated to $\pi$ (\citep[equations 3,4]{bedos_fourier_2015}), and is usually preferred over the right representation ($\tilde{\pi}, R_{\zeta}$) e.g. \citep{bedos_fourier_2015}, \citep{bedos_fourier_2016}. Moreover, in the physics literature the left regular representation is usually preferred, as exemplified by \citep[Section 3.1.2]{prodan_bulk_2016}. 



\begin{remark}[Reduced twisted crossed products and the left regular representation]\label{remark:reduced_crossed_products_and_left_regular_representation}
In the context of using amenable groups for twisted crossed products as we are currently doing, some authors (\citep{bedos_fourier_2016,bedos_fourier_2015,bedos_discrete_2011}) use the sub C* algebra of $B(L^2(G,H))$ where $\overline{\Pi}(L^1(G,A;\alpha,\zeta))$ is dense as the reduced twisted crossed product of $(A,G,\alpha,\zeta)$ instead of the sub C* algebra where $\tilde{\Pi}(L^1(G,A;\alpha,\zeta))$ is dense (\cref{definition:reduced_crossed_product}). Moreover, if we denote by $C_{\overline{\Pi}}$ the sub C* algebra of $B(L^2(G,H))$ where $\overline{\Pi}(L^1(G,A;\alpha,\zeta))$ is dense, the same authors mention that there is an isomorphism of C* algebras  $$ \overline{\phi}: A \rtimes_{\alpha, \zeta} G \to A \rtimes_{\alpha, \zeta, r} G ,$$
such that $\overline{\phi}(f) = \int_G \overline{\pi}(f(s)) L_{\zeta}(s) d \mu(s)$ for $f \in L^1(G,A;\alpha,\zeta)$, which in the present setup implies that $C_{\overline{\Pi}}$ is isomorphic to $A \rtimes_{\alpha,\zeta,r}G$ due to \cref{theorem:norm_of_twisted_crossed_products}. The isomorphism $\overline{\phi}$ plays an important role in the application of twisted crossed products to the study of physical systems as exposed in \cref{sec:motivation_from_physics}), and we will use it in \cref{sec:Fourier_analysis_twistted_crossed_product} to show how the definition of Fourier coefficient of elements of $A \rtimes_{\alpha,\zeta} \mathbb{Z}^d$ are a generalization of the Fourier coefficients of elements of $C(\mathbb{T}^d)$.
\end{remark}

In \cref{proposition:twisted_crossed_product_restricts_to_discrete_conunatlbe_groups} we checked that $A \rtimes_{\alpha,\zeta} G$ can be expressed in terms of generators when $G$ is discrete an countable, this is not the only nice thing we get when working with discrete and countable groups, in particular, the integrated form of the left regular representation of $L^1(G,A;\alpha,\zeta)$ (\cref{def:right_regular_representation_twisted_crossed_product}) takes the form of a Bochner integral, let us look on why this happens. 

\begin{proposition}[Functions with compact supports and representations of $L^1(G,A;\alpha,\zeta)$]\label{proposition:functions_compact_support_representations_L_1}
Let $(G,A,\alpha,\zeta)$ be a separable twisted dynamical system (\cref{def:separable_twisted_dynamical_system}) and $G$ a countable and discrete group, let $(\pi,\pi_U)$ be a covariant representation of $(G,A,\alpha,\zeta)$ over a separable Hilbert space $H$, then, for any $f \in L^1(G,A;\alpha,\zeta)$ the series
$$\sum_{s \in G} \pi(f(s)) \pi_U(s) $$
converges in the norm topology over $B(H)$ and is equal to the operator
$$\int_G \pi(f(s)) \pi_U(s) d \mu (s).$$
\end{proposition}
\begin{proof}
Let $f \in L^1(G,A;\alpha,\zeta)$, given that representations of C* algebras are norm decreasing (\cref{proposition:automatic_continuity_C_star_algebras}) we have that
$$ \| \pi(f(s)) \pi_U(s) \| \leq \| \pi(f(s)) \| \| \pi_U(s) \| \leq \| f(s) \|,  $$
therefore, 
$$ \| \sum_{s \in F} \pi(f(s)) \pi_U(s) \| \leq  \sum_{s \in F} \| f(s) \|  $$
for any $F$ a finite set of $G$. Given that $G$ has the discrete topology, the function $s \mapsto \pi(f(s)) \pi_U(s)$ can be approximated pointwise by finite sums of indicator functions over sets with finite measure, or equivalently, it is $\mu$-strongly measurable (\cref{def:mu_strong_measu}), moreover, the function $s \mapsto \| \pi(f(s)) \pi_U(s) \|$ belongs to $L^1(G)$, so, according to \cref{prop:bochn_integra_condition} the function $s \mapsto \pi(f(s)) \pi_U(s) $ is Bochner integrable, which implies that the series $\sum_{s \in G} \pi(f(s)) \pi_U(s)$ converges in the norm topology of $B(H)$. 

Since the series $\sum_{s \in G} \pi(f(s)) \pi_U(s)$ converges in the norm topology of $B(H)$, if $\psi \in H$ we have that 
$$ \left( \sum_{s \in G} \pi(f(s)) \pi_U(s) \right)(\psi) = \sum_{s \in G} \left( \pi(f(s)) \pi_U(s) (\psi) \right).  $$
Take $\psi, \eta \in H$, given that the inner product over $H$ is a continuous map of the form $H \times H \to \mathbb{C}$, we have that
$$ \langle \sum_{s \in G} \left( \pi(f(s)) \pi_U(s) (\psi) \right) , \eta \rangle = \sum_{s \in G} \langle \pi(f(s)) \pi_U(s) (\psi) , \eta \rangle,  $$
therefore, due to the one-to-one correspondence between bounded sesquilinear forms and bounded linear operators over Hilbert spaces (\cref{proposition:correspondence_bounded_operators_and_sesquilinear_forms}), the operator $\sum_{s \in G} \pi(f(s)) \pi_U(s)$ is the only bounded linear operator $T$ over $H$ such that
$$ \langle T(\psi), \eta \rangle = \int_{G} \langle \pi(f(s)) \pi_U(s) (\psi), \eta \rangle d\mu(s). $$
\end{proof}

An alternative argument for the previous proposition goes along the lines of \cref{proposition:weak_L1_and_Bochner_integrable_functions}, and could be applied for any Bochner intetrable function regardless of whether the group $G$ is discrete or not.

\subsection{Twisted crossed products with $\mathbb{Z}^{d}$}
\label{section:twisted_crossed_products_with_Z}

The C* algebras that represent physical systems will come from twisted crossed products with $G=\mathbb{Z}^d$ (\cref{sec:motivation_from_physics}), those twisted crossed products use normalized 2 cocycles that look like $$ \sigma_{\Theta} : \mathbb{Z}^d \times \mathbb{Z}^d \to \mathbb{T}, \sigma_{\Theta}(x,y) = \exp(i x^t \Theta y), $$
where $\Theta$ is a lower triangular matrix with zeros in the diagonal and entries on $[0,2 \pi)$. These normalized 2 cocycles are quite special, as the following statements assert,

\begin{definition}[Projectively equivalent normalized 2 cocycles\index{normalized 2 cocycle!Projectively equivalent} (Section 2 \citep{backhouse_projective_1970_part_1})]\label{definition:projective_equivalent_cocycles}

Let $\sigma_0$ and $\sigma_1$ be two normalized 2 cocycles on $\mathbb{Z}^d$, then they are called projectively equivalent if \citep[equation 2.9]{backhouse_projective_1970_part_1} there is a function $\nu: \mathbb{Z}^d \to \mathbb{T}$ such that
$$ \sigma_0(x,y) = \sigma_1(x,y) \frac{\nu(x+y)}{\nu(x) \nu(y)}. $$

\end{definition}

Notice that the equivalence relation in \cref{definition:projective_equivalent_cocycles} is the same equivalence relation for normalized 2 cocycles that are part of twisted actions (\cref{def:equivalence_between_twsited_actions}), hence, two projectively equivalent normalized 2 cocycles produce the same $L^1$ *-algebra. There is a special way of characterizing the equivalence classes of projectively equivalent normalized 2 cocycles, this can be done through skew-symmetric matrices or lower triangular matrices with zeros in the diagonal,

\begin{proposition}[Characterization of normalized 2 cocycles over $\mathbb{Z}^d$ (Theorem 3.2 \citep{backhouse_projective_1970_part_2})]\label{proposition:characterization_of_+cocycles_over_integers}
For every normalized 2 cocycle $\sigma$ on $\mathbb{Z}^d$, there is a normalized 2 cocycle $\sigma_{\Theta}$ to which it is projectively equivalent and takes the following form,
$$ \sigma_{\Theta} : \mathbb{Z}^d \times \mathbb{Z}^d \to \mathbb{T}, \sigma_{\Theta}(x,y) = \exp(i x^t \Theta y), $$
where $\Theta$ is a lower triangular matrix with zeros in the diagonal and entries on $[0,2 \pi)$ (skew-symmetric matrix with entries in $[- \pi, \pi)$). Moreover, if $\Theta_0 \neq \Theta_1$ then the normalized 2 cocycles $\sigma_{\Theta_0}$ and $\sigma_{\Theta_1}$ are not equivalent.

Also, if $\Theta$ is lower triangular with zeros in the diagonal, and 
$$\hat{\Theta}_{i,j} = \Theta_{i,j} \text{ if } i \leq j \text{ else } \hat{\Theta}_{i,j} = -\Theta_{j,i},$$
then, the cocyle generated by $\Theta$ is equivalent to the cocyle generated by $\hat{\Theta}$. 
\end{proposition}

The set of inequivalent normalized 2 cocycles described in \cref{proposition:characterization_of_+cocycles_over_integers} generate the non-commutative torus \citep[Section 12.2]{gracia-bondia_elements_2001}. 

From \cref{lemma:descriptpon_twisted_crossed_product_discrete_group_simplified_relations} we know that $A \rtimes_{\alpha, \zeta} \mathbb{Z}^d$ is isomorphic to the C* algebra generated by the set $$\mathcal{R} =  \{ a u_s :\; a \in A , s \in \mathbb{Z}^d\},$$
subject to the following relations
$$ u_s u_t = \zeta(s,t) u_{s+t}, \; \alpha(s)(a) u_s = u_s a, \; u_s^* = \zeta(-s,s)^*u_{-s},  $$
and the algebraic relations of $A$. 

Recall that $\mathbb{Z}^d$ is a group generated by $d$ generators, each one associated with one of the dimensions of $\mathbb{Z}^d$, that is, $\mathbb{Z}^d = \langle e_1, \dotsc, e_d \rangle$, with 
$$ e_j \in \mathbb{Z}^d, \; e_j(l) = 1 \text{ iff } l = j, \text{ and } 1 \leq j \leq d. $$ 
So, take $\zeta(x,y) = \exp(i x^t \Theta y)$ with $\Theta$ a lower triangular matrix entries in $[0, 2\pi)$ and zeros in the diagonal, as in \cref{proposition:characterization_of_+cocycles_over_integers}, then, $ \zeta(e_j, e_l) = \exp(i \Theta_{j,l})$, which implies,
\begin{itemize}
    \item Given that  $\Theta$ has zeros in the diagonal, if $j =l$ then $ \zeta(e_j, e_l) =1$, this implies that $u_{e_j} u_{e_j} = u_{2 e_{j}}$, also, $u_{e_j} u_{- e_j} = u_{-e_j} u_{e_j} = u_0$.
    \item Since $\Theta$ has no diagonal entries, we have that $u_{e_j}^* = u_{-e_j}$.
    \item Since $\Theta$ is lower triangular and has zeros in the diagonal, if $j \leq l$ then $u_{e_j} u_{e_l} = u_{e_j + e_l}$, thus, $u_{ne_j} u_{me_l} = u_{ne_j + me_l}$ for $n,m \in \mathbb{Z}$.
    \item If $j > l$ then $u_{e_j} u_{e_l} = \exp(i \Theta_{j,l}) u_{e_j + e_l}$, thus, $u_{ne_j} u_{me_l} = \exp(i nm\Theta_{j,l}) u_{ne_j + me_l}$ for $n,m \in \mathbb{Z}$.
    \item If $j \geq l$ then $u_{e_j} u_{e_l} = \exp(i \Theta_{j,l}) u_{e_l} u_{e_j}$, and if $j < l$ then $u_{e_j} u_{e_l} = \exp(-i \Theta_{l,j}) u_{e_l} u_{e_l}$.
\end{itemize}
If $s > 0$ denote $u_{e_j}^s = u_{e_j} \cdot \dotsc \cdot u_{e_j}$ ($s$ times), if $s=0$ denote $u_{e_j}^s = u_0$ and if $s < 0$ denote $u_{e_j}^s = u_{e_j}^* \cdot \dotsc \cdot u_{e_j}^*$ ($s$ times). For any $s = \sum_{1 \leq i \leq d} s_i e_i \in\mathbb{Z}^d$ denote,
$$ u^s = u_1^{s_1} \cdots u_d^{s_d}, $$
then, after some algebraic manipulations we get the following, 
\begin{itemize}
    \item If $s \in \mathbb{Z}^d$, then, $ u_s^* = u_d^{-s_d} \cdots u_1^{-s_1}$, moreover, $ u_s^* = \exp(i s^t \Theta s) u^{-s}$.
    \item If $s,l \in \mathbb{Z}^d$, then, $u^s u^l = \exp(i s^t \Theta l) u^{s+l}$.
    \item If $s,l \in \mathbb{Z}^d$, then, $u^s u^l = \exp(i s^t \hat{\Theta} l) u^{l} u^{s}$, where 
    $$\hat{\Theta}_{i,j} = \Theta_{i,j} \text{ if } i \leq j \text{ else } \hat{\Theta}_{i,j} = -\Theta_{j,i}.$$
\end{itemize}

The aforementioned relations motivate the following definition, whose name is justified by \cref{lemma:computation_of_Fourier_coefficients},

\begin{definition}[Algebra of generalized trigonometric polynomials\index{Algebra of generalized trigonometric polynomials}]\label{def:algebra_generalized_trigonometric_polynomials}
Let $(\mathbb{Z}^d,A,\alpha,\zeta)$ be a separable twisted dynamical system with $\zeta(x,y) = \exp(i x^t \Theta y)$, where $\Theta$ is a lower triangular matrix with zeros in the diagonal and entries in $[0,2\pi)$. Let $\{ e_j \}_{1 \leq j \leq d}$ be an ordered basis of $\mathbb{Z}^d$, also, let
$$\hat{\Theta}_{i,j} = \Theta_{i,j} \text{ if } i \leq j \text{ else } \hat{\Theta}_{i,j} = -\Theta_{j,i},$$
then, denote by $\mathcal{P}(\mathbb{Z}^d,A, \alpha,\Theta)$ the *-algebra generated by the set of elements 
$$\{ a u_j \; | \; a \in A, \; 1 \leq j \leq d \}, $$
subject to the following commutation relations,
\begin{itemize}
    \item $ \alpha(e_j)(a) u_j = u_j a$ and $ \alpha(-e_j)(a) u_j^* = u_j^* a$.
    \item $u_j u_l = \exp \left( i\hat{\Theta}_{j,l} \right) u_l u_j$.
    \item $u_j^* u_l = \exp \left( -i\hat{\Theta}_{j,l} \right) u_l u_j^*$.
    \item The algebraic relations of elements inside $A$.
\end{itemize}
The *algebra $\mathcal{P}(\mathbb{Z}^d,A, \alpha,\Theta)$\index{$\mathcal{P}(\mathbb{Z}^d,A, \alpha,\Theta)$} is called the algebra of \textbf{generalized trigonometric polynomials}. 
\end{definition}

The computations previous to \cref{def:algebra_generalized_trigonometric_polynomials} tell us that there us a canonical isomorphism of *-algebras,
$$ \phi : \mathcal{P}(\mathbb{Z}^d,A, \alpha,\Theta) \to L^1(\mathbb{Z}^d,A;\alpha,\zeta)_c ,$$
such that, $\phi(au_j)$ is the function from $\mathbb{Z}^d$ into $A$ that assigns $a$ to $e_j$ and $0$ elsewhere, and $\phi(au_j^*)$ is the function from $\mathbb{Z}^d$ into $A$ that assigns $a$ to $-e_j$ and $0$ elsewhere. This fact is really important because we are pursuing an algebraic characterization of the C* algebra $A \rtimes_{\alpha,\Theta} \mathbb{Z}^d$ with out resorting to the norm of the Banach algebra $L^1(\mathbb{Z}^d,A;\alpha,\zeta)$.

\begin{lemma}\label{lemma:algebra_generalized_trigo_polynomials_and_funcitons_with_compact_supports}
Let $(\mathbb{Z}^d,A,\alpha,\zeta)$ be a separable twisted dynamical system with $\zeta(x,y) = \exp(i x^t \Theta y)$, where $\Theta$ is a lower triangular matrix with zeros in the diagonal and entries in $[0,2\pi)$. The *-algebra $\mathcal{P}(\mathbb{Z}^d,A, \alpha,\Theta)$ is isomorphic to the *-algebra $L^1(\mathbb{Z}^d,A;\alpha,\zeta)_c$, the isomorphism maps $a u_j$ into the function that assigns $a $ to $e_j$ and $0$ else where, ans maps maps $a u_j^*$ into the function that assigns $a $ to $-e_j$ and $0$ else where.
\end{lemma}

Given the particular form that $L^1(G,A;\alpha,\zeta)_c$ takes when $G=\mathbb{Z}^d$, the following lemma is a restatement of the fact we have already explored in \cref{proposition:twisted_crossed_product_restricts_to_discrete_conunatlbe_groups}, and we provided it for easy reference inside the document,

\begin{lemma}[Twisted crossed products with $\mathbb{Z}^d$]\label{lemma:twisted_crossed_products_with_Z}
Let $\Theta$ be a lower triangular matrix with entries in $[0, 2 \pi)$, then, $\zeta(x,y) = \exp(i x^t \Theta y)$ is a normalized 2 cocycle over $\mathbb{Z}^d$. Under this setup, the twisted crossed product $A \rtimes_{\alpha, \zeta} \mathbb{Z}^d$ is denoted by $A \rtimes_{\alpha,\Theta} \mathbb{Z}^d$ and is isomorphic to the enveloping C* algebra of the *-algebra $\mathcal{P}(\mathbb{Z}^d,A, \alpha,\Theta)$, i.e. 
$$A \rtimes_{\alpha,\zeta} \mathbb{Z}^d \simeq C^*(\mathcal{P}(\mathbb{Z}^d,A, \alpha,\Theta)).$$ \index{$A \rtimes_{\alpha,\zeta} \mathbb{Z}^d$}
\end{lemma}
\begin{proof}
Given that the *-algebra $\mathcal{P}(\mathbb{Z}^d,A, \alpha,\Theta)$ is isomorphic to $L^1(\mathbb{Z}^d,A;\alpha,\zeta)_c$, then, this lemma is a consequence of \cref{proposition:twisted_crossed_product_restricts_to_discrete_conunatlbe_groups}.
\end{proof}

The description of  $A \rtimes_{\alpha, \zeta} \mathbb{Z}^d$ given in \cref{lemma:twisted_crossed_products_with_Z} is frequently used in the literature, for example, that is the description of the twisted crossed product given by Prodan et.al. in \citep[Definition 3.1.1]{prodan_bulk_2016}, such that setting $\Theta$ as lower triangular or skew-symmetric corresponds to using the Landau gauge for the left regular representations of $u_s$ (\citep[Section 3.1.2]{prodan_bulk_2016}) while using $\Theta$ as a skew-symmetric matrix corresponds to using the symmetric gauge for the left regular representations of $u_s$ (\citep[equation 3.13]{prodan_bulk_2016})

Motivated by \cref{proposition:characterization_of_+cocycles_over_integers}, from now on when we work with $G= \mathbb{Z}^d$ we will use the notation $A \rtimes_{\alpha, \Theta} \mathbb{Z}^d$\index{$A \rtimes_{\alpha, \Theta} \mathbb{Z}^d$} to refer to the twisted crossed product where $\zeta(x,y) = \exp(i x^t \Theta y)$.

\subsubsection{Trivial twisting actions}
\label{sec:trivial_twisting_actions}

 \cref{lemma:twisted_crossed_products_with_Z} tell us that $A \rtimes_{\alpha,\Theta} \mathbb{Z}^d$ is the C* algebra generated by the set 
$$\mathcal{R} =  \{ a u_j :\; a \in A , 1 \leq j \leq d \},$$
subject to the following relations
\begin{itemize}
    \item $ \alpha(e_j)(a) u_j = u_j a$ and $ \alpha(-e_j)(a) u_j^* = u_j^* a$.
    \item $u_j u_l = \exp \left( i\hat{\Theta}_{j,l} \right) u_l u_j$.
    \item $u_j^* u_l = \exp \left( -i\hat{\Theta}_{j,l} \right) u_l u_j^*$.
    \item The algebraic relations of elements inside $A$.
\end{itemize}
If $\alpha: \mathbb{Z}^d \to \text{Aut}(A)$ be the trivial action, that is, $\alpha(s)(a) = a$ for all $s \in \mathbb{Z}^d$, $a \in A$, and $\Theta$ is the lower triangular matrix that generates the trivial normalized 2 cocycle over $\mathbb{Z}^d$, that is, $\Theta_{i,j} = 0$ such that $\zeta(x,y) = 1$ for all $x,y \in \mathbb{Z}^d$, then, the *-algebra generated by $\mathcal{R}$ under the aforementioned commutation relations is isomorphic to the *-algebra
$$ A \odot \mathcal{J},  $$
where, $\mathcal{J}$ the *-algebra generated by $u_1, \dotsc, u_d$ subject to the commutation relations $u_i u_i^* = u_i^*u_i = 1, u_i u_j = u_j u_i$, and $A \odot \mathcal{J}$ is the algebraic tensor product of the *-algebras $A$ and $ \mathcal{J}$.

In the literature the C* algebra $C(\mathbb{T}^d)$ is usually presented as the enveloping C* algebra of $\mathcal{J}$ (\cref{section:n_torus}), also, it is a basic result from the theory of C* algebras that for any $C^*$ algebra $A$ we have that $C^*(A) \simeq A$ (\cref{sec:Universal_C_star_algebras}). So, in the context of a trivial action and a trivial normalized 2 cocycle, we get that $A \rtimes_{\alpha,\Theta} \mathbb{Z}^d$ is the tensor product of the C* algebras $C(\mathbb{T}^d)$ and $A$ (\cref{lemma:tensor_products_and_generated_C_star_algebras}). Additionally, given that $C(\mathbb{T}^d) \otimes A \simeq C(\mathbb{T}^d,A)$ (\cref{proposition:isomorphism_tensor_product_of_commutative_C_star_algebras}) we end up having that
$$ A \rtimes_{\alpha,\Theta} \mathbb{Z}^d \simeq C(\mathbb{T}^d,A), $$

Under the aforementioned setup, we can see that the twisted crossed products with $\mathbb{Z}^d$ are a generalization of the C* algebra $C(\mathbb{T}^d,A)$, where we impose non-trivial commutation relations between the generators of $C(\mathbb{T}^d)$ and the generators of $A$. 

\subsection{Fourier analysis}
\label{sec:Fourier_analysis_twistted_crossed_product}

\cref{theorem:norm_of_twisted_crossed_products} states that, if $\pi: A \to B(H)$ is a faithful representation of $A$ on a separable Hilbert space, then, 
$$ \tilde{\Pi}: L^1(G,A;\alpha,\Theta) \to B(L^2(\mathbb{Z}^d;H)) $$
extends to a faithful representation of C* algebra $A \rtimes_{\alpha,\Theta} \mathbb{Z}^d$, and we will denote that faithful representation as $\tilde{\Pi}$ also, that is,
$$ \tilde{\Pi}: A \rtimes_{\alpha,\Theta} \mathbb{Z}^d \to B(L^2(\mathbb{Z}^d;H)). $$
This faithful representation of $A \rtimes_{\alpha,\Theta} \mathbb{Z}^d$ will be the main tool to define the Fourier coefficients. 

Recall that the direct sum of Hilbert spaces $\sum_{s \in \mathbb{Z}^d}H$ is the Hilbert space whose elements are maps $f:\mathbb{Z}^d \to H$ such that $\sum_{s \in \mathbb{Z}^d} \| f(s) \|^2 < \infty$, and the inner product is defined as (\cref{defintion:direct_sum_hilbert_spaces})
$$\langle f,g \rangle = \sum_{s \in \mathbb{Z}^d} \langle f(s), g(s) \rangle, \; f,g \in \sum_{s \in \mathbb{Z}^d}H.$$
The elements of $\sum_{s \in \mathbb{Z}^d}H$ are denoted by $\sum_{s \in \mathbb{Z}^d}h_s$.

The representation $\tilde{\Pi}$ allows to look at each element of $ A \rtimes_{\alpha,\Theta} \mathbb{Z}^d$ as a bounded linear operator over the Hilbert space $L^2(\mathbb{Z}^d;H)$. The Hilbert space $L^2(\mathbb{Z}^d;H)$ is isomorphic to the tensor product of Hilbert spaces $L^2(\mathbb{Z}^d) \otimes H$ (\cref{proposition:tensor_product_and_L_2_functs}), the Hilbert space $L^2(\mathbb{Z}^d)$ has a canonical orthonormal basis given by $\{ e_s \}_{s \in \mathbb{Z}^d}$ where $e_s$ is the function from $\mathbb{Z}^d$ into $\mathbb{C}$ that takes the value $1$ in $s$ and $0$ elsewhere, under this setup the Hilbert space $L^2(\mathbb{Z}^d) \otimes H$ is isomorphic to the direct sum of Hilbert spaces $\sum_{\mathbb{Z}^d}H$ where the isomorphism from $L^2(\mathbb{Z}^d,H)$ to $L^2(\mathbb{Z}^d) \otimes H$ maps $(h_s)_{s \in \mathbb{Z}^d}$ to $\sum_{s \in \mathbb{Z}^d} e_s \otimes h_s$ (\cref{proposition:properties_tensor_product_Hilbert_spaces}). 

Given a bounded linear map $T$ over the Hilbert space $\sum_{\mathbb{Z}^d}H$, it is possible to think of $T$ as an infinite matrix with entries in $B(H)$ which we denote by $[T_{ij}]_{i,j \in \mathbb{Z}^d}$ (\cref{sec:tensor_product_of_bounded_opeartors_os_infinite_matrices}). Assume that $H_j := H$ for all $j \in \mathbb{Z}^d$, given $s \in \mathbb{Z}^d$ we introduce the bounded linear operators $U_s,\; V_s$,
$$ U_s : H \to \sum_{j \in \mathbb{Z}^d}H_j, \; U_s(x)=  \sum_{j \in \mathbb{Z}^d} x_j, \; x_j = x \text{ if } j = s \text{ else } x_j = 0,$$
and 
$$ V_s : \sum_{j \in \mathbb{Z}^d} H_{j} \to H, \;  V_s (\sum_{j \in \mathbb{Z}^d} x_j) = x_s.$$
Under this setup, we have that $T_{i,j} = V_{i}TU_{j}$, and the operator $T$ is uniquely determined by the set of elements of the form $T_{i,j}$ as follows (\cref{sec:tensor_product_of_bounded_opeartors_os_infinite_matrices}),
$$ T(x) = \sum_{i \in \mathbb{Z}^d} y_i, \; \text{with } y_i = \sum_{j \in \mathbb{Z}^d} T_{i,j}(x_{j}) .$$
Additionally, according to \cref{proposition:infinite_matrices_over_tensor_products} we have that $\| T \| \geq \| T_{i,j} \|$ for any $i,j \in \mathbb{Z}^d$ , and $\| T \| \leq \sum_{s \in \mathbb{Z}^d} \| T_{i,j } \|$. For an in depth discussion on how to analyze bounded linear operators over Hilbert spacer as infinite matrices refer to \citep[Section 2.6, Matrix representations]{kadison_fundamentals_1983_V1}.

The previous discussion implies that each element $p$ of $A \rtimes_{\alpha,\Theta} \mathbb{Z}^d$ is uniquely determined by a set of bounded operators over $H$, and the way of computing those operators is to check how $\tilde{\Pi}(p)$ acts over a vector of the form $e_i \otimes h$, with $i \in \mathbb{Z}^d$ and $h \in H$, or equivalently, how $\tilde{\Pi}(p)$ acts on a function of the form $f : \mathbb{Z}^d \to H$ such that $f(j) = h $ if $i =j$ and zero else where. If we look at how $\tilde{\Pi}(p)$ acts on the vector $e_i \otimes h$, with $i \in \mathbb{Z}^d$ and $h \in H$ we will be able to find the elements of the column $j$ of the infinite matrix representation of $\tilde{\Pi}(p)$, that is, we will be able to find all $T_{i,j}$ for $i \in \mathbb{Z}^d$. The previous argument is not only valid for $\tilde{\Pi}(p)$, in fact this procedure can be used for any operator over $L^2(\mathbb{Z}^d,H)$.

Let $(R_{\zeta}, \tilde{\pi})$ be the right regular covariant representation of $(\mathbb{Z}^d,A,\alpha,\zeta)$, take $e_j \otimes h$ with $h \in H$, also, take $a \in A$, then, by \cref{def:right_regular_representation_twisted_crossed_product} we have that,
$$R_{\zeta}(s)(e_j \otimes h) = \zeta(j-s, s) e_{j-s} \otimes h, \; \tilde{\pi}(a)(e_j \otimes h) = e_j \otimes \pi(\alpha(j)(a))(h), $$
therefore, the infinite matrices of $R_{\zeta},\; \tilde{\pi}$ take the form,
$$R_{\zeta}(s)(h) = [\zeta(m-s,s) \delta(l,m-s) I_H ]_{l,m \in \mathbb{Z}^d}, \; \tilde{\pi}(a)(h) = [\pi(\alpha(l)(a)) \delta(l,m)]_{l,m \in \mathbb{Z}^d}, $$
where $I_H$ is the identity operator over $H$. From the action of $R_{\zeta}(s),\; \tilde{\pi}(a)$ over $e_j \otimes h$ we get,
$$ \tilde{\pi}(a) R_{\zeta}(s) (e_j \otimes h) = \zeta(j-s, s) e_{j-s} \otimes \pi(\alpha(j-s)(a))(h), $$
therefore, the infinite matrix of $\tilde{\pi}(a) R_{\zeta}(s)$ takes the form,
$$\tilde{\pi}(a) R_{\zeta}(s)(h) = [\zeta(m-s,s) \delta(l,m-s) \pi(\alpha(l)(a)) ]_{l,m \in \mathbb{Z}^d}.$$

Take $f \in L^1(\mathbb{Z}^d,A,\alpha,\zeta)$, then, from \cref{proposition:functions_compact_support_representations_L_1} we know that the representation $\tilde{\Pi}$ evaluated over $f$ takes the form of the convergent sum of bounded operators,
$$ \tilde{\Pi}(f) = \sum_{s \in \mathbb{Z}^d} \tilde{\pi}(f(s)) R_{\zeta}(s). $$
We have that
$$ \tilde{\Pi}(f) (e_j \otimes h) = \sum_{s \in \mathbb{Z}^d} \tilde{\pi}(f(s)) R_{\zeta}(s) (e_j \otimes h) =  \sum_{s \in \mathbb{Z}^d} \zeta(j-s, s) e_{j-s} \otimes \pi(\alpha(j-s)(f(s)))(h), $$
therefore, the infinite matrix associated to $\tilde{\Pi}(f)$ takes the form,
$$ [\zeta(l,m-l) \pi(\alpha(l)(f(m-l)))]_{l,m \in \mathbb{Z}^d}. $$

\begin{lemma}\label{lemma:lower_bound_on_C_start_norm_in_twisted_crossed_product_with_Z}
Let $(\mathbb{Z}^d,A,\alpha,\zeta)$ be a separable twisted dynamical system with $\zeta(x,y) = \exp(i x^t \Theta y)$, take $f,g \in L^1(\mathbb{Z}^d,A,\alpha,\zeta)$, then, 
\begin{enumerate}
    \item $\| f \| \geq \| f(s) \|$ for any $s \in \mathbb{Z}^d$, where the norm of $f$ is computed as an element of $A \rtimes_{\alpha,\Theta} \mathbb{Z}^d$ and the norm of $f(s)$ is computed as an element of $A$.
    \item $\| f- g \| = 0$ iff $f(s) = g(s)$, where the norm of $f-g$ is computed as an element of $A \rtimes_{\alpha,\Theta} \mathbb{Z}^d$.
\end{enumerate}
\end{lemma}
\begin{proof}
\begin{enumerate}
    \item \cref{theorem:norm_of_twisted_crossed_products} tells us that there is a faithful representation of $A \rtimes_{\alpha,\Theta} \mathbb{Z}^d$ that takes the form of $\tilde{\Pi}$ when restricted to $L^1(\mathbb{Z}^d,A,\alpha,\zeta)$. By \cref{proposition:infinite_matrices_over_tensor_products} we have that $\| f \| \geq \|\zeta(l,m-l) \pi(\alpha(l)(f(m-l)))\| $ for any $l,m \in \mathbb{Z}^d$, also, due to the automatic continuity of C* homomorphisms (\cref{proposition:automatic_continuity_C_star_algebras}) we know that the automorphisms of C* algebras are isometries, so, since $\alpha(l)$ is an automorphisms of the C* algebra $A$, we have that
    $$ \| \zeta(l,m-l) \pi(\alpha(l)(f(m-l))) = \| f(m-l)\|. $$
    The previous statements implies that $\| f \| \geq \|f(m-l)\| $ for any $l,m \in \mathbb{Z}^d$.
    \item This is a consequence of the previous item.
\end{enumerate}
\end{proof}

\begin{lemma}\label{lemma:bound_on_the_norm_of_right_covariant_representation}
Let $(\mathbb{Z}^d,A, \alpha, \zeta)$ be a separable twisted dynamical system, then, $\| a \| = \| \tilde{\pi}(a) R_{\zeta}(s) \|$, where the norm of $a$ is its norm as an element of $A$.
\end{lemma}
\begin{proof}
We have that $\| \tilde{\pi}(a) R_{\zeta}(s) \| \leq \| \tilde{\pi}(a)\| \| R_{\zeta}(s) \|$, since $R_{\zeta}(s)$ is an unitary operator we have that $\| \tilde{\pi}(a) R_{\zeta}(s) \| \leq \| \tilde{\pi}(a) \|$. Given that C* homomorphisms are norm decreasing (\cref{proposition:automatic_continuity_C_star_algebras}), we have that $\| \tilde{\pi}(a) \| \leq a$, therefore, $\| \tilde{\pi}(a) R_{\zeta}(s) \| \leq \| a \|$.

From \cref{proposition:infinite_matrices_over_tensor_products} we know that, given an operator $T$ over $L^2(\mathbb{Z}^d,H)$ with an associated infinite matrix $[T_{i,j}]_{i,j \in \mathbb{Z}^d}$, we have that $\| T \| \geq \| T_{i,j}\|$ for any $i,j \in \mathbb{Z}^d$, therefore, in the context of $T = \tilde{\pi}(a) R_{\zeta}(s) $ the aforementioned statement implies that $\| \tilde{\pi}(a) R_{\zeta}(s)  \| \geq \| \pi(\alpha(l)(a)) \|$ for any $l \in \mathbb{Z}^d$. Since $\alpha(j)$ is an automorphism of the C* algebra $A$, it must be an isometry, which implies that $\| \tilde{\pi}(a) R_{\zeta}(s)  \| \geq \| a \|$.

Both inequalities $\| \tilde{\pi}(a) R_{\zeta}(s)  \| \geq \| a \|$ and $\| \tilde{\pi}(a) R_{\zeta}(s)  \| \leq \| a \|$ imply the desired result.
\end{proof}

\begin{lemma}\label{lemma:convergence_of_fourier_coefficients_on_polynomials}
Let $(\mathbb{Z}^d,A, \alpha, \zeta)$ be a separable twisted dynamical system with $\zeta(x,y) = \exp(i x^t \Theta y)$, take $\{ p_n \}_{n \in \mathbb{N}}$ a Cauchy sequence inside $A \rtimes_{\alpha,\Theta} \mathbb{Z}^d$ such that $p_n \in \mathcal{P}(\mathbb{Z}^d,A, \alpha,\Theta)$, then,
\begin{enumerate}
    \item Denote by $p$ the limit of $\{ p_n \}_{n \in \mathbb{N}}$, let $p_n = \sum_{s \in \mathbb{Z}^d} a_{n,s} u^s$, then, for any $s \in \mathbb{Z}^d$ the sequence $\{a_{n,s}\}_{n \in \mathbb{N}}$ is a Cauchy sequence inside $A$. 
    \item If $\{ q_n \}_{n \in \mathbb{N}}$ is another Cauchy sequence converging to $p$, let $q_n = \sum_{s \in \mathbb{Z}^d} b_{n,s} u^s$, then, for any $s \in \mathbb{Z}^d$, $\lim_{n \to \infty} a_{n,s} = \lim_{n \to \infty} b_{n,s}$. 
    \item Denote by $a_s$ the limit of the Cauchy sequence $\{a_{n,s}\}_{n \in \mathbb{N}}$, let $\tilde{\Pi}(p) = \lim_{n \to \infty} \tilde{\Pi}(p_n)$, then, $\tilde{\Pi}(p)$ has an associated infinite matrix\index{infinite matrix} given by 
    $$  [\zeta(l,m-l) \pi(\alpha(l)(a_{m-l}))]_{l,m \in \mathbb{Z}^d}.$$ 
\end{enumerate}
\end{lemma}
\begin{proof}
\begin{enumerate}
    \item This is a consequence of \cref{lemma:lower_bound_on_C_start_norm_in_twisted_crossed_product_with_Z} if we identify each element of $\mathcal{P}(\mathbb{Z}^d,A, \alpha,\Theta)$ with its corresponding element in $L^1(\mathbb{Z}^d,A,\alpha,\zeta)_c$ as in \cref{lemma:algebra_generalized_trigo_polynomials_and_funcitons_with_compact_supports}. 
    \item Denote by $a_s = \lim_{n \to \infty} a_{n,s}$ and $b_s = \lim_{n \to \infty} b_{n,s}$. We know that $p_n \to p$ and $q_n \to p$, therefore, $p_n - q_n \to 0$, since $p_n -q_n = \sum_{s \in \mathbb{Z}^d} (a_{n,s} - b_{n,s})u^s$ using \cref{lemma:lower_bound_on_C_start_norm_in_twisted_crossed_product_with_Z} we get that $a_{n,s} - b_{n,s} \to 0$, therefore, $\lim_{n \to \infty} a_{n,s} = \lim_{n \to \infty} b_{n,s}$. 
    \item Let us call by $\tilde{\Pi}(p)$ the element of $B(L^2(\mathbb{Z}^d,H))$ which is the limit of the elements $\{ \tilde{\Pi}(p_n) \}_{n \in \mathbb{N}}$, the element $\tilde{\Pi}(p)$ is the image of $p$ under the faithful representation of $A \rtimes_{\alpha,\Theta} \mathbb{Z}^d$ mentioned in (\cref{theorem:norm_of_twisted_crossed_products}). We have that
    $$ \tilde{\Pi}(p)(e_j \otimes h) = \lim_{n \to \infty} \tilde{\Pi}(p_n)(e_j \otimes h),  $$
    therefore,
    $$ \tilde{\Pi}(p)(e_j \otimes h) = \lim_{n \to \infty} \sum_{s \in \mathbb{Z}^d} \zeta(j-s, s) e_{j-s} \otimes \pi(\alpha(j-s)(a_{n,s}))(h).  $$
    Since $\{e_j \}_{j \in \mathbb{Z}^d}$ is a orthonormal basis of $\mathbb{Z}^d$ and both $\pi, \; \alpha(j-s)$ are continuous maps, we must have that
    $$ \tilde{\Pi}(p)(e_j \otimes h) =  \sum_{s \in \mathbb{Z}^d} \zeta(j-s, s) e_{j-s} \otimes \pi(\alpha(j-s)( \lim_{n \to \infty} a_{n,s}))(h), $$
    consequently,
    $$ \tilde{\Pi}(p)(e_j \otimes h) =  \sum_{s \in \mathbb{Z}^d} \zeta(j-s, s) e_{j-s} \otimes \pi(\alpha(j-s)( a_{s}))(h). $$
    The previous equation give us the desired information to provide the infinite matrix form of $\tilde{\Pi}(p)$.
\end{enumerate}
\end{proof}


\begin{definition}[Fourier coefficients\index{Fourier coefficient}]\label{definition:fourier_coefficients}
Let $(\mathbb{Z}^d,A, \alpha, \zeta)$ be a separable twisted dynamical system with $\zeta(x,y) = \exp(i x^t \Theta y)$, take $p \in A \rtimes_{\alpha,\Theta} \mathbb{Z}^d$ and $\{ p_n \}_{n \in \mathbb{N}}$ a sequence of elements of $\mathcal{P}(\mathbb{Z}^d,A, \alpha,\Theta)$ such that $p_n \to p$ in $A \rtimes_{\alpha,\Theta} \mathbb{Z}^d$, denote $p_n = \sum_{s \in \mathbb{Z}^d} a_{n,s} u^s$, then, given $s \in \mathbb{Z}^s$ we call $\lim_{n \to \infty} a_{n,s}$ the $s$ Fourier Coefficient of $p$ and we use the notation $\Phi_s(p)$\index{$\Phi_s(p)$ ($s$ Fourier coefficient of $p$)} to refer to that element.
\end{definition}

As a consequence of \cref{lemma:convergence_of_fourier_coefficients_on_polynomials} the $s$ Fourier coefficient of $p$ is in fact independent of the sequence $\{ p_n \}_{n \in \mathbb{N}}$ inside $\mathcal{P}(\mathbb{Z}^d,A, \alpha,\Theta)$ converging to $p$. Using \cref{lemma:convergence_of_fourier_coefficients_on_polynomials} we get the following,

\begin{lemma}\label{lemma:Fourier_coefficients_and_right_regular_representation}
Let $(\mathbb{Z}^d,A, \alpha, \zeta)$ be a separable twisted dynamical system with $\zeta(x,y) = \exp(i x^t \Theta y)$, take $p \in A \rtimes_{\alpha,\Theta} \mathbb{Z}^d$, then, the infinite matrix of $\tilde{\Pi}(p)$ takes the form,
$$[\zeta(l,m-l) \pi(\alpha(l)( \Phi_{m-l}(p)))]_{l,m \in \mathbb{Z}^d}.$$
\end{lemma}

The previous lemma implies that,

\begin{lemma}\label{lemma:bound_on_the_fourier_coefficients}
Let $(\mathbb{Z}^d,A, \alpha, \zeta)$ be a separable twisted dynamical system with $\zeta(x,y) = \exp(i x^t \Theta y)$, take $p \in A \rtimes_{\alpha,\Theta} \mathbb{Z}^d$, then, for any $s \in \mathbb{Z}^d$ we have that $\| \Phi_s(p) \| \leq \|p \|$.
\end{lemma}

The description of the Fourier coefficients of an element of $A \rtimes_{\alpha,\Theta} \mathbb{Z}^d$ provides a generalization of the Riemann-Lebesgue lemma\index{Riemann-Lebesgue lemma} (\cref{proposition:Fourier_transform_properties})

\begin{lemma}\label{lemma:generalization_riemann_lebesgue_lemma}
Let $(\mathbb{Z}^d,A, \alpha, \zeta)$ be a separable twisted dynamical system with $\zeta(x,y) = \exp(i x^t \Theta y)$, take $p \in A \rtimes_{\alpha,\Theta} \mathbb{Z}^d$, then $\lim_{|s| \to \infty} \Phi_s(p) =0$, where $|s| = \min \{ |s_i| \}$ with $s = (s_1, \cdots, s_d)$.
\end{lemma}
\begin{proof}
Let $\epsilon > 0$, then, let $q \in \mathcal{P}(\mathbb{Z}^d,A, \alpha,\Theta)$ such that $\| p-q \| \leq \epsilon$, then, using \cref{lemma:bound_on_the_fourier_coefficients} we know that $\| \Phi_s(p-q) \| \leq \epsilon$. Let $N$ be a non zero natural number such that $q_{s} = 0$ if $s \notin [-N,N]^d$, then, this implies that if $|s| > N $ then $\| \Phi_s(p) \| \leq \epsilon$.
\end{proof}

\begin{remark}[Left regular representation and Fourier coefficients]\label{remark:left_regular_representation_and_Foureir_coefficients}
Let $(\mathbb{Z}^d,A, \alpha, \zeta)$ be a separable twisted dynamical system with $\zeta(x,y) = \exp(i x^t \Theta y)$, let $(\overline{\pi},L_{\zeta})$ be the left regular representation of that dynamical system (\cref{def:left_regular_representation}), then, 
$$L_{\zeta}(s)(e_j \otimes h) = \zeta(s, j) e_{j+s} \otimes h, \; \overline{\pi}(a)(h) = e_j \otimes \pi(\alpha(-j)(a))(h), $$
that is,
$$L_{\zeta}(s)(h) = [\zeta(s,m) \delta(l-s,m)I_H ]_{l,m \in \mathbb{Z}^d}, \; \overline{\pi}(a)(h) = [\pi(\alpha(-l)(a)) \delta(l,m)]_{l,m \in \mathbb{Z}^d}, $$
therefore, we have that
$$ \overline{\pi}(a) L_{\zeta}(s) = [ \zeta(s,m) \pi(\alpha(-l)(a)) \delta(l-s,m) ]_{l,m \in \mathbb{Z}^d}.$$
$\overline{\pi}(a) L_{\zeta}(s)$ is a diagonal operator that arises as a generalization of the operator $\pi(a) \pi(e_{s})$ that is mentioned in \cref{section:Fourier_tranform_C_star_valued_functions_over_the_torus}. The operator $L_{\zeta}(s)$ is referred to as the dual magnetic translations under the Landau gauge (\citep[equation 2.7]{prodan_bulk_2016}), and provides the representation of the twisted crossed product $A \rtimes_{\alpha,\Theta}\mathbb{Z^d}$ on $L^2(\mathbb{Z}^d,H)$ under the Landau gauge (\citep[Section 3.1.2]{prodan_bulk_2016}).

Also, the matrix representation of $\overline{\Pi}(p)$ under the left regular representations becomes,
$$ \overline{\Pi}(p)_{l,m} = \zeta(l-m, m) \pi(\alpha(-l)(\Phi_{l-m}(p))), \; l,m \in \mathbb{Z}^d,$$
since the left regular representation of $ A \rtimes_{\alpha,\Theta} \mathbb{Z}^d$ is also a faithful representation (\cref{remark:reduced_crossed_products_and_left_regular_representation}), each $p$ an element of $ A \rtimes_{\alpha,\Theta} \mathbb{Z}^d$ is uniquely determined by the entries of the infinite matrix form of $ \overline{\Pi}(p)$. 

If $d=1$ and $\alpha, \Theta$ are trivial, then $A \rtimes_{\alpha,\zeta} \mathbb{Z}^d$ falls back to the C* algebra $C(\mathbb{T},A)$ (\cref{sec:trivial_twisting_actions}), and for any $f \in C(\mathbb{T},A)$ the representation $\overline{\Pi}$ falls back into the multiplication operator over $L^2(\mathbb{T},H)$ where $H$ is a separable Hilbert space where $A$ has a faithful representation, this is the case because the infinite matrix representation of $f$ as the multiplication operator has the following form over $L^2(\mathbb{Z},H)$ (\cref{section:Fourier_tranform_C_star_valued_functions_over_the_torus}),
$$[(\mathcal{F} \otimes Id_H)(f)(l-m) ]_{l,m \in \mathbb{Z}}.$$
\end{remark}

\begin{remark}[Canonical inclusion of $A$ into $A \rtimes_{\alpha,\Theta}\mathbb{Z}^d$]\label{remark:canonical_inclusion_of_twisted_crossed_products}
Let $i : A \to A \rtimes_{\alpha,\Theta}\mathbb{Z}^d$ given by $i(a) = au^0$, then from \cref{lemma:bound_on_the_norm_of_right_covariant_representation} we know that $\| i(a) \| = \|a\|$, and the commutation relations between $u^0$ and $a$, that is, $u^0 a u^0 = a u^0$, tell us that it is a *-homomorphims, thus, we have that $i$ is an injective C* homomorphism i.e. $A$ is a sub C* algebra of $A \rtimes_{\alpha,\Theta}\mathbb{Z}^d$. Also, under the left regular representation for every $a \in A$ we have that $\overline{\Pi}(i(a)) = [ \delta(l,m) \pi(\alpha(-l)(a))]_{l,m \in \mathbb{Z}^d} $,
which can be seen as an generalization of the operator $\pi(a) \otimes Id_{l^2(\mathbb{Z}^d)} = [\delta(l,m) \pi(a)]_{l,m \in \mathbb{Z}^d}$ that is mentioned in \cref{sec:tensor_product_of_bounded_opeartors_os_infinite_matrices}. 
\end{remark}

\subsubsection{Fourier coefficients as integrals}
\label{sec:fourier_coefficients_as_integrals}

Given an element of $L^1(\mathbb{T}^d)$, which we call $f$, the Fourier transform provides a function $\mathcal{F}(f)$ over $\mathbb{Z}^d$ given by (\cref{example:Fourier_transform_and_dual_groups}),
$$ \mathcal{F}(f)(s)) = \int_{\mathbb{T}^d} f(\lambda) \lambda^{-s} d \mu(\lambda), $$
having said this, our next step is to look on how $\Phi_s(p)$ can be computed as a Bochner integral We provide a subtle generalization of the computation of $\Phi_s(p)$ for crossed products presented in \citep[Section VIII.2]{davidson_c-algebras_1996}. Recall that the characters of the group $\mathbb{T}^d$ are given by (\cref{example:Fourier_transform_and_dual_groups})
$$ \gamma_{s}(\lambda) =  \left( \prod_{1 \leq j \leq d} \lambda_j^{s_j}  \right), \; s:= (s_1, \dots, s_d) \in \mathbb{Z}^d, \; \lambda := (\lambda_1, \cdots, \lambda_d) \in \mathbb{T}^d.$$
We will now construct an action\index{action of a group} of $\mathbb{T}^d$ over $A \rtimes_{\alpha,\Theta}\mathbb{Z}^d$ that uses all the characters of $\mathbb{T}^d$\index{characters of $\mathbb{T}^d$}, and that will be key to compute the Fourier coefficients of elements of $A \rtimes_{\alpha,\Theta}\mathbb{Z}^d$ as integrals. This approach is a generalization of the algebraic approach to Fourier coefficients of elements of $C(\mathbb{T},A)$ exposed in the literature (\cref{section:Fourier_tranform_C_star_valued_functions_over_the_torus}).  

\begin{lemma}[Continuous action of $\mathbb{T}^d$ over $A \rtimes_{\alpha,\Theta}\mathbb{Z}^d$]\label{lemma:action_of_torus_over_twisted_crossed_product}
There is a group homomorphism
$$ \tau: \mathbb{T}^d \to \text{Aut}( A \rtimes_{\alpha,\Theta}\mathbb{Z}^d),$$
such that, if $p \in A \rtimes_{\alpha,\Theta}\mathbb{Z}^d$, $s = \sum_{1 \leq j \leq d} e_j s_j \in \mathbb{Z}^d$ and $\lambda = (\lambda_1, \cdots, \lambda_d)$ then the Fourier coefficients (\cref{definition:fourier_coefficients}) of $\tau(\lambda)(p)$ are related with the Fourier coefficients of $p$ by the formula
$$ \Phi_s(\tau(\lambda)(p)) = \gamma_{s}(\lambda) \Phi_s(p). $$
Additionally, for any $p \in A \rtimes_{\alpha,\Theta}\mathbb{Z}^d$ the following map is continuous (strong continuity),
$$ \lambda \mapsto \tau(\lambda)(p).$$
\end{lemma}
\begin{proof}
Let $\lambda = (\lambda_1, \dotsc, \lambda_d) \in \mathbb{T}^d$, then, denote by $\tilde{\tau}_{\lambda}$ the *-algebra homormophism from $\mathcal{P}(\mathbb{Z}^d,A, \alpha,\Theta)$ into $\mathcal{P}(\mathbb{Z}^d,A, \alpha,\Theta)$ that takes the following form on the generators of $\mathcal{P}(\mathbb{Z}^d,A, \alpha,\Theta)$,
$$ \tilde{\tau}_{\lambda}(au_{j}) := \lambda_j a u_j.$$
Given $\sum_{s \in \mathbb{Z}^d} p(s)u_1^{s^1} \cdot \dotsc \cdot u_d^{s^d} \in \mathcal{P}(\mathbb{Z}^d,A, \alpha,\Theta)$, then
$$ \tilde{\tau}_{\lambda}(\sum_{s \in \mathbb{Z}^d} p(s)u_1^{s^1} \cdot \dotsc \cdot u_d^{s^d}) = \sum_{s \in \mathbb{Z}^d} \left( \prod_{1 \leq j \leq d} \lambda_j^{s_j}  \right) p(s) u_1^{s^1} \cdot \dotsc \cdot u_d^{s^d}, $$
moreover, the explicit description of $\tilde{\tau}_{\lambda}$ tells us that $\tau_{\lambda}$ is bijective. Also,we have that $(\hat{\tau}_{\lambda_0} \circ \hat{\tau}_{\lambda_1})(q) = (\hat{\tau}_{\lambda_0 + \lambda_1}) (q)$ for all $q \in \mathcal{P}(\mathbb{Z}^d,A, \alpha,\Theta)$.

Since, $A \rtimes_{\alpha,\Theta}\mathbb{Z}^d$ is the enveloping C* algebra of $\mathcal{P}(\mathbb{Z}^d,A, \alpha,\Theta)$ (\cref{lemma:twisted_crossed_products_with_Z}), then, the universal property of enveloping C* algebras (\cref{proposition:factoring_representations_with_enveloping_C_star_algebras}) implies that $\tilde{\tau}_{\lambda}$ extends to a C* algebra homomorphism 
$$\tau_{\lambda} : A \rtimes_{\alpha,\Theta}\mathbb{Z}^d \to A \rtimes_{\alpha,\Theta}\mathbb{Z}^d,$$
such that,
$$ \tau_{\lambda}(au_{j}) = \lambda_jau_{j}, \; \tau_{\lambda}(a u_j^*) = \lambda_j^* au_{j}^*.$$
Since $\tau_{\lambda} (\mathcal{P}(\mathbb{Z}^d,A, \alpha,\Theta)) = \mathcal{P}(\mathbb{Z}^d,A, \alpha,\Theta)$ the homomorphism $\tau_{\lambda}$ is surjective, moreover, it is bijection because for any $q \in \mathcal{P}(\mathbb{Z}^d,A, \alpha,\Theta)$ we have that $(\tau_{\lambda} \circ \tau_{-\lambda})(q) =q$. Since $\tau_\lambda$ is an automorphism at the level of $\mathcal{P}(\mathbb{Z}^d,A, \alpha,\Theta)$, then, it is also an automorphism at the level of the enveloping C* algebra of $\mathcal{P}(\mathbb{Z}^d,A, \alpha,\Theta)$, that is, $\tau_\lambda$ is an automorphism of $ A \rtimes_{\alpha,\Theta}\mathbb{Z}^d$.

Take, $p \in A \rtimes_{\alpha,\Theta}\mathbb{Z}^d$ and $q \in \mathcal{P}(\mathbb{Z}^d,A, \alpha,\Theta)$ such that $\| p - q \| \leq \epsilon/3$, then 
$$ \| \tau_{\lambda_0}(p) - \tau_{\lambda_1}(p) \| \leq \| \tau_{\lambda_0}(p) - \tau_{\lambda_0}(q) \| + \| \tau_{\lambda_0}(q) - \tau_{\lambda_1}(q) \| + \| \tau_{\lambda_1}(p) - \tau_{\lambda_1}(q) \|, $$
since $\tau_{\lambda}$ is an isometry we have that
$$\| \tau_{\lambda_0}(p) - \tau_{\lambda_1}(p) \| \leq \frac{2 \epsilon}{3} + \| \tau_{\lambda_0}(q) - \tau_{\lambda_1}(q) \|.$$
Consequently, in order to show that the map $\lambda \mapsto \tau(\lambda)(p)$ is continuous we just need to show that it is continuous when $p \in \mathcal{P}(\mathbb{Z}^d,A, \alpha,\Theta)$, which comes from the following observation of its value on the generators of $\mathcal{P}(\mathbb{Z}^d,A, \alpha,\Theta)$,
$$ \| \tau_{\lambda_0}(au_j) - \tau_{\lambda_1}(a u_j) \| = \| (\lambda_0 - \lambda_1)a u_j \| \leq | \lambda_0 - \lambda_1 |\| a \|.  $$
The previous statements tells us that the strongly continuous group homomorphism that we look for is $\tau_{\lambda}$, i.e. $\tau(\lambda) := \tau_{\lambda}$.

Now, take $p \in A \rtimes_{\alpha,\Theta}\mathbb{Z}^d$ and $\{ p_n \}_{n \in \mathbb{N}} \subset \mathcal{P}(\mathbb{Z}^d,A, \alpha,\Theta)$ such that $p_n \to p$, since $\tau(\lambda)$ is a C* homomorphism it is continuous, we have that
$$  \tau(\lambda)(p_n )\to \tau(\lambda)(p). $$
Given that the maps an element of $A \rtimes_{\alpha,\Theta}\mathbb{Z}^d$ into one of its Fourier coefficient is continuous (\cref{lemma:bound_on_the_fourier_coefficients}), we have that,
$$ \Phi_s(\tau(\lambda)(p)) = \lim_{n \to \infty}  \Phi_s(\tau(\lambda)(p_n)) =  \left( \prod_{1 \leq j \leq d} \lambda_j^{s_j}  \right) \lim_{n \to \infty} \Phi_s(p_n).$$
Since the definition of Fourier coefficients (\cref{definition:fourier_coefficients}) guaranties that 
$$\lim_{n \to \infty} \Phi_s(p_n) = \Phi_s(p),$$
we have that,
$$  \Phi_s(\tau(\lambda)(p)) =  \gamma_{s}(\lambda) \Phi_s(p).$$
\end{proof}

At the level of the description of $ A \rtimes_{\alpha,\Theta}\mathbb{Z}^d$ as an universal C* algebra $\tau$ can be intuitively thought of as replacing the generator $u_j$ by $\lambda_j u_j$. Now, we look at how the group homomorphism $\tau$ can be used to compute the zero coefficient of elements from $A \rtimes_{\alpha,\Theta}\mathbb{Z}^d$ as an intetral, 
 
\begin{proposition}[Computation of $\Phi_0$ using a Bochner integral\index{Bochner integral}]\label{proposition:continuous_expectation_of_twisted_crossed_products} 
The map 
$$ p \mapsto \Phi_0(p)$$
is a Banach space morphism from $A \rtimes_{\alpha,\Theta}\mathbb{Z}^d$ into $A$ and satisfies the following relation
$$ \Phi_0 (p) u^0 =  \int_{\mathbb{T}^d} \tau(\lambda)(p) d \mu(\lambda).$$
\end{proposition}
\begin{proof}
Take $p \in  A \rtimes_{\alpha,\Theta}\mathbb{Z}^d$, since the map $\lambda \mapsto \tau(\lambda)(p)$ is continuous and $\mathbb{T}^d$ is compact space, then by \cref{example:Bochner_integral_continuos_funcion} we know that it is Bochner integrable, thus, we can define
$$ \Psi(p):= \int_{\mathbb{T}^d} \tau(\lambda)(p) d \mu(\lambda) \in A \rtimes_{\alpha,\Theta}\mathbb{Z}^d.$$
By definition, $p \mapsto \Psi(p)$ is a vector space homomorphism, also, since we are working with normalized Haar measures (\cref{remark:simplifying_our_analysis}), by \cref{prop:bochn_integra_condition} we have that,
$$ \| \Psi(p) \| \leq \int_{\mathbb{T}^d} \|\tau(\lambda)(p) \| d \mu(\lambda) \leq  \| p \|,$$
thus, $\Phi$ is a contraction (continuous map).

In \cref{lemma:action_of_torus_over_twisted_crossed_product} we mentioned that $\tau(\lambda)(a u^0) = au^0$, this means that $i(A)$ is a fixed point of the map $\tau(\lambda)$, where $i$ is the canonical inclusion of $A$ on $A \rtimes_{\alpha,\Theta}\mathbb{Z}^d$ mentioned in \cref{remark:canonical_inclusion_of_twisted_crossed_products}, consequently, $\Psi(i(A)) = i(A)$. Take $s = \sum_{1 \leq j \leq d} e_j s_j \in \mathbb{Z}^d$, then 
$$ \Psi(a u^s) = \int_{\mathbb{T}^d} \tau(\lambda)(a u^s) d \mu(\lambda) = \int_{\mathbb{T}^d} \left( \prod_{1 \leq j \leq d} \lambda_j^{s_j}  \right) a u^s d \mu(\lambda). $$
Since the map $\lambda \mapsto (\lambda_1^{s_1}, \dotsc ,\lambda_d^{s_d})$ is a character of the group $\mathbb{T}^d$ (\cref{example:Fourier_transform_and_dual_groups}) we have that it is continuous (\cref{definition:dual_group}), thus, by \cref{remark:bochner_integral_and_multiplication} we have that
$$ \Psi(a u^s) = \left( \int_{\mathbb{T}^d}  \prod_{1 \leq j \leq d} \lambda_j^{s_j} d \mu(\lambda)  \right) a u^s. $$
Since the set $\{\lambda  \mapsto \prod_{1 \leq j \leq d} \lambda_j^{s_j} \}_{s \in \mathbb{Z}^d} $ is an orthonormal basis of $L^2(\mathbb{T}^d)$ (\cref{section:Fourier_analysis_and_ortonormal_basis}), we have that $\int_{\mathbb{T}^d}  \prod_{1 \leq j \leq d} \lambda_j^{s_j} d \mu(\lambda) = \delta(s,0)$, and
$$ \Psi(a u^s) = a u^s \delta(s,0). $$

Take $p \in \mathcal{P}(\mathbb{Z}^d,A, \alpha,\Theta)$ given by $p = \sum_{s \in \mathbb{Z}^d} p(s) u^s$ (finite sum), then, the previous discussion tells us that
$$ \Psi(p) = p(0)u^0.$$
Since the computation of the Fourier coefficients is a continuous map (\cref{lemma:bound_on_the_fourier_coefficients}), the fact that $\Psi$ is continuous tell us that for any $ p \in A \rtimes_{\alpha,\Theta}\mathbb{Z}^d$ with Fourier coefficients given by $\{ \Phi_s(p) \}_{s \in \mathbb{Z}^d}$ (\cref{definition:fourier_coefficients}), the map $\Psi$ takes the following form
$$ \Psi(p) = \Phi_0(p) u^0. $$
Since $i(A)$ is isomorphic to $A$ (\cref{remark:canonical_inclusion_of_twisted_crossed_products}), $\Phi_0(p)$ is the unique element of $A$ such that $\Psi(p) = \Phi_0(p) u^0$.
\end{proof} 

In the following discussion, we will show how the $s$ Fourier coefficient of elements of $A \rtimes_{\alpha,\zeta} \mathbb{Z}^d$ can be computed using an integral, and this integral resembles the one used in \cref{proposition:continuous_expectation_of_twisted_crossed_products} to compute $\Phi_0$. The map $\lambda \mapsto  \gamma_{s}(\lambda)$ is continuous for any $s \in \mathbb{Z}^d$ because is a character of the group $\mathbb{T}^d$, therefore, the map
$$ \lambda \mapsto  \gamma_{-s}(\lambda) \tau(\lambda)(p), \;  p \in A \rtimes_{\alpha,\Theta}\mathbb{Z}^d$$
is Bochner integrable by  \cref{example:Bochner_integral_continuos_funcion}.

For any $p \in A \rtimes_{\alpha,\zeta} \mathbb{Z}^d$ define
$$\Psi_{s}(p) := \int_{\mathbb{T}^d} \gamma_{-s}(\lambda) \tau(\lambda)(p) d \mu(\lambda), $$
then 
$$ \| \Psi_s(p) \| \leq \int_{\mathbb{T}^d} \| \gamma_{-s}(\lambda) \tau(\lambda)(p) \| d \mu(\lambda) \leq \| p \|.$$
The map $p \mapsto \Psi_s(p)$ is a linear map, since the Bochner integral is linear and $\tau(\lambda)$ is linear for every $\lambda \in \mathbb{T}^d$, therefore, the map $p \mapsto \Psi_s(p)$ is a continuous (contraction) Banach space map. 

Notice that, if $s, s' \in \mathbb{Z}^d$ then
$$ \Phi_{s'} (\gamma_{-s}(\lambda) \tau(\lambda)(p)) = \gamma_{-s}(\lambda) \gamma_{s'}(\lambda) \Phi_{s'}(p) = \gamma_{(s'-s)}(\lambda) \Phi_{s'}(p),$$
thus, if we take $q = \sum_{s \in \mathbb{Z}^d} q(s) u^s \in \mathcal{P}(\mathbb{Z}^d,A, \alpha,\Theta)$ (finite sum), we can use an argument similar to \cref{proposition:continuous_expectation_of_twisted_crossed_products} to proof that
$$ \Psi_s(q) = q(s)u^s, $$
and consequently, 
$$ q = \sum_{s \in \mathbb{Z}^d} \Psi_s(q). $$ 
Take $\{ p_n \}_{n \in \mathbb{N}} \subset \mathcal{P}(\mathbb{Z}^d,A, \alpha,\Theta)$ such that $p_n \to p$, then 
$$ \Psi_s(p_n) \to \Psi_s(p), \text{ and } \Psi_s(p_n) = p_n(s) u^s,$$
since $p_n(s)u^s \to \Phi_s(p)u^s$ (\cref{lemma:convergence_of_fourier_coefficients_on_polynomials}) and $\Psi_s$ is continuous, we have that 
$$\Psi_s(p) = \Phi_s(p) u^s.$$

The previous discussion give us the following lemma,

\begin{lemma}[Computation of the Fourier coefficients]\label{lemma:computation_of_Fourier_coefficients}
Given $p \in A \rtimes_{\alpha,\Theta}\mathbb{Z}^d$ with Fourier coefficients $\{ \Phi_s(p) \}_{s \in \mathbb{Z}^d}$ (\cref{definition:fourier_coefficients}), then, $\Phi_s(p)$ is the unique element of $A$ such that
$$ \int_{\mathbb{T}^d} \gamma_{-s}(\lambda) \tau(\lambda)(p) d \mu(\lambda) = \Phi_s(p) u^s. $$
Therefore, $\Phi_s(p)$ is the unique element of $A$ such that $\Psi_s(p) = \Phi_s(p) u^s$, and for any $p \in \mathcal{P}(\mathbb{Z}^d,A, \alpha,\Theta)$ we have that
$$ p = \sum_{s \in \mathbb{Z}^d} \Psi_s (p) = \sum_{s \in \mathbb{Z}^d} \Phi_s (p) u^s .$$

Additionally, for any $s \in \mathbb{Z}^d$, the maps
$$ \Psi_s : A \rtimes_{\alpha,\Theta}\mathbb{Z}^d \to A \rtimes_{\alpha,\Theta}\mathbb{Z}^d, \; \Phi_s : A \rtimes_{\alpha,\Theta}\mathbb{Z}^d  \to A$$
are continuous (contractions) Banach space maps, that is,
\begin{itemize}
    \item \textbf{Linearity:} take $p,q \in A \rtimes_{\alpha,\Theta}\mathbb{Z}^d $ and $\nu \in \mathbb{C}$, then
    \begin{itemize}
        \item $\Psi_s(\nu p+ q) = \nu \Psi_s(p) + \Psi_s(q)$
        \item $\Phi_s(\nu p+ q) = \nu \Phi_s(p) + \Phi_s(q)$
    \end{itemize}
    \item \textbf{Contraction:} take $p \in A \rtimes_{\alpha,\Theta}\mathbb{Z}^d $, then, $\| \Phi_s(p) \|= \| \Psi_s (p) \|$ (\cref{definition:fourier_coefficients}) and 
    $$ \| \Phi_s(p) \| \leq \| p \|.  $$
\end{itemize}
\end{lemma}

Notice that this form of computing the Fourier coefficients of $p \in A \rtimes_{\alpha,\Theta}\mathbb{Z}^d$ is similar in form to the computation of the Fourier coefficients of complex-valued functions over $\mathbb{T}^d$ (\cref{definition:Fourier_transform}), where we had an integral of a function $f(\lambda)$ times a character $ \gamma_{-s}(\lambda)$. In this case, we are still using characters inside the integral ($ \gamma_{-s}(\lambda)$), but we have replaced the function $f$ by the result of applying the map $\tau(\lambda)$ to an element of the C* algebra $A \rtimes_{\alpha,\Theta}\mathbb{Z}^d$. If $\alpha$ and $\Theta$ are trivial the computation of the Fourier coefficients given in \cref{lemma:computation_of_Fourier_coefficients} falls back into the computation of the Fourier coefficients on $C(\mathbb{T}^d,A)$ (\cref{section:Fourier_tranform_C_star_valued_functions_over_the_torus}), since $u^s$ would correspond to the function $ \lambda \mapsto \exp(i s \cdot \lambda) $ with $s, \lambda \in \mathbb{Z}^d$ (\cref{sec:trivial_twisting_actions}). 

Now, we can look into a generalization of the Fejér summation of functions over $\mathbb{T}^d$ (\cref{section:convergence_of_the_Fourier_series}), which will give a convergent sequence of elements in $\mathcal{P}(\mathbb{Z}^d,A, \alpha,\Theta)$ approximating elements in $A \rtimes_{\alpha,\Theta}\mathbb{Z}^d$,  

\begin{proposition}[Fejér summation approximation]\label{proposition:Fejer_sums_approximation}

Let $p \in A \rtimes_{\alpha,\Theta}\mathbb{Z}^d$ and $n \in \mathbb{N}$, let
$$V_n = \{(s_1, \dotsc, s_d) \in \mathbb{Z}^d : \; |s_j| \leq n \}.$$
Define
$$ p^{(n)}:= \sum_{s \in V_n} \left( \prod_{1 \leq j \leq d} \left( 1 - \frac{|s_j|}{n+1} \right) \right) \Phi_s(p) u^s, $$
then
$p^{(n)} \to p $ in $A \rtimes_{\alpha,\Theta}\mathbb{Z}^d$.
\end{proposition}
\begin{proof}
Note that
$$ p^{(n)} = \sum_{s \in V_n} \left( \prod_{1 \leq j \leq d} \left( 1 - \frac{|s_j|}{n+1} \right) \right) \Psi_s(p)$$
$$ = \int_{\mathbb{T}^d}  \sum_{s \in V_n} \left( \prod_{1 \leq j \leq d} \left( 1 - \frac{|s_j|}{n+1} \right) \right)  \left( \prod_{1 \leq j \leq d} \lambda_j^{s_j}  \right) \tau(\lambda)(p) d \mu(\lambda) .$$
The map
$$ K_n(\lambda) := \sum_{s \in V_n} \left( \prod_{1 \leq j \leq d} \left( 1 - \frac{|s_j|}{n+1} \right) \right)  \left( \prod_{1 \leq j \leq d} \lambda_j^{s_j}  \right) $$
is called the Fej\'er kernel in $d$ dimensions (\cref{section:convergence_of_the_Fourier_series}), so, we can express $p^{(n)}$ as,
$$ p^{(n)} = \int_{\mathbb{T}^d} K_n(-\lambda) \tau(\lambda)(p) d \mu(\lambda).$$
Since $K_n$ is a positive normalized kernel then 
$$ \|p^{(n)}\| \leq \int_{\mathbb{T}^d} \| \tau(\lambda)(p) \| |K_n(-\lambda) | d \mu(\lambda) = \| p \|, $$
thus the map $p \mapsto p^{(n)}$ is a continuous (contraction) vector space homomorphism.

For any $q \in \mathcal{P}(\mathbb{Z}^d,A, \alpha,\Theta)$ we have that
$$ \lim_{n \to \infty} q^{(n)} = q, $$
so, given $\epsilon>0$, take $q \in \mathcal{P}(\mathbb{Z}^d,A, \alpha,\Theta)$ such that $\| q -p \| \leq \epsilon$ and $M$ such that if $n \geq M$ then $ \| q^{(n)} - q \| \leq \epsilon$, under this set up
$$ \| p - p^{(n)} \| \leq \| q-p \| + \| q - q^{(n)} \| + \| (p-q)^{(n)} \| \leq 3 \epsilon, $$
consequently $p^{(n)} \to p$ as desired.
\end{proof}

\begin{remark}[Equivalent actions]\label{remark:equivalent_actions_over_twsited_crossed_product}
In \cref{lemma:action_of_torus_over_twisted_crossed_product} we have defined a continuous action of $\mathbb{T}^d$ over $A \rtimes_{\alpha,\Theta}\mathbb{Z}^d$, and the action was designed to provide a generalization for the procedure of computing the Fourier coefficients of a continuous function over $\mathbb{T}^d$. This is not the only action that could be defined, for example, some authors prefer to work with the action that results from setting $au_{j,\lambda} := \lambda_j^* a u_j, \;au_{j,\lambda}^* := \lambda_j a u_j^*$ (\citep[Section 3.3.1]{prodan_bulk_2016}), which generates a continuous action
$$ \tau': \mathbb{T}^d \to \text{Aut}( A \rtimes_{\alpha,\Theta}\mathbb{Z}^d),$$
such that 
$$ \Phi_s( \tau'(\lambda)(p)) =   \gamma_{-s}(\lambda) \Phi_s(p). $$
We have chosen to follow the action
$$ \Phi_s( \tau'(\lambda)(p)) =   \gamma_{s}(\lambda) \Phi_s(p) $$
because it fits better with the description of Fourier coefficients given in \citep[Section VIII.2]{davidson_c-algebras_1996} for crossed products with C* algebras with unit. 
\end{remark}

In the following definition, we use the notation from \cref{proposition:Fejer_sums_approximation}

\begin{definition}[Generalized Fourier series and generalized Fejér summation]\label{definition:generalized_fourier_sequence_and_fejer_summation}
Given $p \in A \rtimes_{\alpha,\Theta}\mathbb{Z}^d$, we refer to the sequence of elements
$$ \{ p^{(n)} \}_{n \in \mathbb{N}}, \; p^{(n)}:= \sum_{s \in V_n} \left( \prod_{1 \leq j \leq d} \left( 1 - \frac{|s_j|}{n+1} \right) \right) \Phi_s(p) u^s, $$\index{$p^{(n)}$ (Generalized Fejér summation of $p$)}
as \textbf{Generalized Fejér summation of $p$}.\index{Generalized Fejér summation} 
In a similar fashion, we refer to the sequence of elements
$$ \{ S^{(n)}(p) \}_{n \in \mathbb{N}}, \; S^{(n)}(p):= \sum_{s \in V_n} \Phi_s(p) u^s, $$\index{$S^{(n)}(p)$ (Generalized Fourier series of $p$)}
as the \textbf{Generalized Fourier series of $p$}.\index{Generalized Fourier series} 
\end{definition}

The generalized Fejér summation of an element of  $A \rtimes_{\alpha,\Theta}\mathbb{Z}^d$ provides a generalization of computing the convolution of an element in $C(\mathbb{T}^d)$ with the Fejér kernel (\cref{proposition:Fejer_sums_approximation}). Recall that the convolution of an element $f \in C(\mathbb{T}^d)$ with the Fejér kernel gives a sequence of trigonometric polynomials that converge uniformly to $f$ (\cref{section:convergence_of_the_Fourier_series}). Additionally, the Fej\'er summation can be used to define weights, traces and states over the C* algebra $A \rtimes_{\alpha,\Theta}\mathbb{Z}^d$ (\cref{section:fourier_analysis_weight_traces_and_states_apendix}).

\begin{lemma}[Fourier coefficients, multiplication and involution]\label{lemma:Fourier_coefficients_multiplication_and_involution}
Take $p,q \in A \rtimes_{\alpha,\Theta} \mathbb{Z}^d$, then,
\begin{enumerate}
    \item for $s,x \in \mathbb{Z}^d$ denote,
    $$  K_{x,s} =  \left( \prod_{1 \leq j \leq d} \left( 1 - \frac{|x_j|}{n+1} \right) \left( 1 - \frac{|(s-x)_j|}{n+1} \right) \right), $$
    then, 
    $$
    \Phi_s (pq) = \lim_{n \to \infty} \Phi_s (p^{(n)}q^{(n)}) = \lim_{n \to \infty } \sum_{s \in V_n} \left( \sum_{ x \in V_n} \mathrm{1}_{s-x \in V_n} K_{x,s} e^{i x^t \Theta (s-x)} \Phi_x (p) \alpha(x) (\Phi_{s - x}(q)) \right). $$
    \item $$ \Phi_s(p^*) = \lim_{n \to \infty} \Phi_s ((p^{(n)})^*) = e^{i s^t \Theta s } \alpha(s)(\Phi_{-s}(p)^*).$$
\end{enumerate}
\end{lemma}
\begin{proof}
Take $p,q \in \mathcal{P}(\mathbb{Z}^d,A, \alpha,\Theta) $ given by the finite sums
$$ p = \sum_{s \in \mathbb{Z}^d} \Phi_s (p) u^s, \; q = \sum_{s \in \mathbb{Z}^d} \Phi_s (q) u^s,$$
then, from the definition of the involution and multiplication on twisted dynamical systems (\cref{theorem:twisted_L1_G_A_is_banach_algebra}), we have that
$$ pq = \sum_{s \in\mathbb{Z}^d} \left( \sum_{x \in \mathbb{Z}^d} e^{i x^t \Theta (s-x)} \Phi_x (p) \alpha(x) (\Phi_{s - x}(q)) \right) u^s $$
and
$$ p^* = \sum_{s \in \mathbb{Z}^d} e^{ s^t \Theta s } \alpha(s)(\Phi_{-s}(p)^*) u^s.$$
In particular, if we use the facts that
\begin{itemize}
    \item $p^{(n)} \to p$ for any $p \in A \rtimes_{\alpha,\Theta}\mathbb{Z}^d$,
    \item $p^{(n)} q^{(n)} \to pq$ (multiplication is a continuous operation),
    \item $p \mapsto p^*$ is a continuous map,
    \item $p \mapsto \Phi_s (p)$ is a continuous map for any $s \in \mathbb{Z}^d$,
\end{itemize}
we have that, $p^{(n)} q^{(n)} \to pq$, so, denote
$$  K_{x,s} =  \left( \prod_{1 \leq j \leq d} \left( 1 - \frac{|x_j|}{n+1} \right) \left( 1 - \frac{|(s-x)_j|}{n+1} \right) \right), $$
then, 
$$
\sum_{s \in V_n} \left( \sum_{ x \in V_n} \mathrm{1}_{s-x \in V_n} K_{x,s} e^{i x^t \Theta (s-x)} \Phi_x (p) \alpha(x) (\Phi_{s - x}(q)) \right) u^s \xrightarrow[n \to \infty]{} pq,
$$
therefore, 
$$
\Phi_s (pq) = \lim_{n \to \infty} \Phi_s (p^{(n)}q^{(n)}) = \lim_{n \to \infty } \sum_{s \in V_n} \left( \sum_{ x \in V_n} \mathrm{1}_{s-x \in V_n} K_{x,s} e^{i x^t \Theta (s-x)} \Phi_x (p) \alpha(x) (\Phi_{s - x}(q)) \right). $$

Additionally,
$$ \Phi_s(p^*) = \lim_{n \to \infty} \Phi_s ((p^{(n)})^*) = e^{i s^t \Theta s } \alpha(s)(\Phi_{-s}(p)^*).$$

\end{proof}

Another way of checking how the Fourier coefficient behaves under involution is to take the definition of the Fourier coefficients from \cref{lemma:computation_of_Fourier_coefficients} as a Bochner integral, and use the dominated convergence theorem for the Bochner integral (\cref{proposition:properties_of_Bochner_integral}) along with the continuity of the involution, to interchange the involution and the integration, such that
$$ (\Phi_s(p) u^s)^* = (\int_{\mathbb{T}^d} \gamma_{-s}(\lambda) \tau(\lambda)(p) d \mu(\lambda))^* = \int_{\mathbb{T}^d} (\gamma_{-s}(\lambda) \tau(\lambda)(p))^* d \mu(\lambda) = \Phi_{-s}(p^*) u^{-s}. $$
Then, we just need to follow the description of the operations described in \cref{theorem:twisted_L1_G_A_is_banach_algebra} to get the relation between the Fourier coefficients of $p$ and the Fourier coefficients of $p^*$ described in \cref{lemma:Fourier_coefficients_multiplication_and_involution}.

Also, we can look for a generalization of the fact that $u^0$ is the unit of the C* algebra $C(\mathbb{T}^d)$.

\begin{lemma}[Multiplicative identity on $A \rtimes_{\alpha,\Theta}\mathbb{Z}^d$]\label{lemma:multiplicative_identity_of_twisted_crossed_product}
$A \rtimes_{\alpha,\Theta}\mathbb{Z}^d$ has a unit iff $A$ has a unit, and if $A$ has a unit then 
$$ 1_{A \rtimes_{\alpha,\Theta}\mathbb{Z}^d} = 1_A u^0:= u^0.  $$
\end{lemma}
\begin{proof}
Assume that $p \in A \rtimes_{\alpha,\Theta}\mathbb{Z}^d$ is a two sided identity for the multiplication on $A \rtimes_{\alpha,\Theta}\mathbb{Z}^d$, then, for any $a u^0$ we must have that
$$ (a u^0 ) p = p (a u^0) = a u^0.  $$
Following \cref{lemma:Fourier_coefficients_multiplication_and_involution} let's check how that looks on the Fourier coefficients,
\begin{itemize}
    \item $\Phi_s( (a u^0 ) p) = a \Phi_s (p),$
    \item $\Phi_s( p (a u^0 ) ) = \Phi_s (p) \alpha(s)(a),$
\end{itemize}
thus, $$ a \Phi_0 (p) = \Phi_0 (p) a $$ for all $a \in A$. This implies that $\Phi_0 (p)$ is a multiplicative identity on $A$, so, since the multiplicative identity of an algebra is unique we must have that $A$ has a unit and $1_A = \Phi_0 (p)$.

If $A$ has a multiplicative identity, we can use the fact that $\alpha(s)$ is a C* homomorphism to show that $\alpha(s)(1_A)= 1_A$, which implies that $1_A u^0$ is a multiplicative identity on $A \rtimes_{\alpha,\Theta}\mathbb{Z}^d$, and again, the uniqueness of the multiplicative identity tell us that 
$$ 1_{A \rtimes_{\alpha,\Theta}\mathbb{Z}^d} = 1_A u^0.  $$
\end{proof}

The previous lemma is a generalization of the fact that $u^0$ is the unit of the C* algebra $C(\mathbb{T}^d)$, that is, the

\subsection{Derivations and Fréchet algebras}\index{Fréchet algebra}
\label{sec:derivations_twistted_crossed_product}

As we claim in the introduction, a key element for the study of the bulk-boundary correspondence are pairs of topological algebras $(M, \mathcal{M})$, where $M$ is a C* algebra and $\mathcal{M}$ is a dense Fréchet sub *algebra of $M$ (\cref{sec:motivation_from_physics}). We will use the action of $\mathbb{T}^d$ over the twisted crossed product $A \rtimes_{\alpha,\Theta}\mathbb{Z}^d$ (\cref{lemma:action_of_torus_over_twisted_crossed_product}) to provide a Fréchet sub *-algebra along with derivations, this algebra comes as a generalization on how $C^{\infty}(\mathbb{T})$ arises as a sub *-algebra of $C(\mathbb{T})$ (\cref{section:Fourier_transform_and_Frechet_algebras}).

The torus $\mathbb{T}^d$ is not only a topological space, but it is also a Lie group, that is, it has a smooth structure that behaves well with respect to the group operations. We take advantage of this structure to define a dense sub *-algebra of $A \rtimes_{\alpha,\Theta}\mathbb{Z}^d$ that will turn out to be a Fréchet algebra, and more importantly, it will be a smooth sub algebra of $A \rtimes_{\alpha,\Theta}\mathbb{Z}^d$ (\cref{sec:smooth_subalgebras}).

We say that the map $\lambda \mapsto \tau(\lambda)(p)$ is smooth if for any $\lambda \in \mathbb{T}^d$ there exists a chart $(U_{\lambda}, \phi_{\lambda})$ that belongs to a smooth atlas of $\mathbb{T}^d$ such that,
\begin{itemize}
    \item $U_{\lambda}$ is a open neighbourhood of $\lambda$.
    \item $\tau(\cdot)(p) \circ \phi^{-1}_{\lambda} : \phi^{-1}_{\lambda}(U_{\lambda}) \to  A \rtimes_{\alpha,\Theta}\mathbb{Z}^d$ is smooth. 
\end{itemize}

Given that $\mathbb{T}^d$ is a Lie group, if $U$ is an open neighborhood of the origin, then, $Ug:= \{ v = g : v \in U\}$ is an open neighborhood of $g \in \mathbb{T}^d$, consequently, we only need to find a chart for the origin and translate it to find a chart centered in any other point. We use the canonical chart of the $d$ dimensional torus, which comes from the $d$ times cartesian product of the canonical atlas of the $1$ dimensional torus:
\begin{align*}
\phi^{-1}_{1_{\mathbb{T}}} : (-\pi, \pi ) \to \mathbb{T} &, \; \phi^{-1}_{1_{\mathbb{T}}}(\theta) = e^{i \theta}, \; U_{1_{\mathbb{T}}} := \phi^{-1}_{1_{\mathbb{T}}}( (-\pi, \pi ) )   \\
\phi^{-1}_{-1_{\mathbb{T}}} : (0, 2 \pi  ) \to \mathbb{T}  &, \;  \phi^{-1}_{-1_{\mathbb{T}}}(\theta) = e^{i \theta}, \; U_{-1_{\mathbb{T}}} := \phi^{-1}_{-1_{\mathbb{T}}}( (0, 2\pi ) ) \\
\end{align*} 
Using these charts you can compute the $2^d$ canonical charts of the $d$ dimensional torus, in particular, if we want them centered at $\lambda=(\lambda_1, \cdots, \lambda_d) \in \mathbb{T}^d$ and $-\lambda$, the chart that would be important for us is
$$ \phi^{-1}_{(\lambda_1, \cdots, \lambda_d)} : (\theta_1 - \pi, \theta_1 + \pi) \times \cdots \times (\theta_d - \pi, \theta_d + \pi)   \to \mathbb{T}^d, \text{ with } e^{i \theta_j} = \lambda_j, \text{ and }$$
$$ \phi^{-1}_{\lambda}(x_1, \cdots, x_d) := (e^{i x_1}, \cdots, e^{i x_d}), \; U_{\lambda}:= \phi^{-1}_{\lambda}((\theta_1 - \pi, \theta_1 + \pi) \times \cdots \times (\theta_d - \pi, \theta_d + \pi)).$$
Therefore, $\tau(\cdot)(p) \circ \phi^{-1}_{\lambda} $ is smooth if the map
$$ (x_1, \cdots, x_d) \mapsto \tau((e^{ix_1}, \cdots, e^{i x_d})) (p), \text{ with } x_j \in (\theta_j - \pi, \theta_j + \pi)$$
is smooth, which means that, given an arbitrary ordered $k$-tuple $(x_{j_1}, \cdots, x_{j_k})$ of elements from $\{ x_l\}_{\leq d}$, the function 
$$ (y_1, \cdots, y_d) \mapsto  \frac{ \partial^k \tau((e^{ix_1}, \cdots, e^{i x_d})) (p)}{ \partial x_{j_1} \cdots \partial x_{j_k}} (y_1, \cdots, y_d) $$
exist and is continuous.

\begin{definition}[Smooth sub algebra of $A \rtimes_{\alpha,\Theta}\mathbb{Z}^d$]\label{definition:smooth_sub_algebra_of_twisted_crossed_product_with_Z}

Let $(A,\mathbb{Z}^d,\alpha,\Theta)$ be a separable twisted dynamical system and $A \rtimes_{\alpha,\Theta}\mathbb{Z}^d$ its associated twisted crossed product C* algebra, then, denote
$$ \mathcal{A}_{\alpha, \Theta} = \{ p \in A \rtimes_{\alpha,\Theta}\mathbb{Z}^d | \lambda \mapsto \tau(\lambda)(p) \text{ is smooth} \}.$$
\index{$\mathcal{A}_{\alpha, \Theta}$}
\end{definition}

\begin{remark}[Technicalities on smoothness for C* valued functions]\label{remark:technialities_smoothnes_C_star_valued_functions}
The function $\tau(\cdot)(p) \circ \phi^{-1}_{\lambda} $ takes values in a C* algebra, instead of the real or complex scalar fields, thus, you may wonder if the results from traditional differential calculus apply to this setting,in this case, the definition of partial derivative is the usual one as exposed in \cref{remark:leibniz_rule_fn_values_in_topological_algebra}. In \cref{definition:smooth_sub_algebra_of_twisted_crossed_product_with_Z} we have taken a computational/analytical approach towards smoothness of functions from $\mathbb{T}^d$ into $A \rtimes_{\alpha,\Theta}\mathbb{T}^d$ instead of a geometrical approach, since we are not looking into general directional derivatives with respect to the local $\mathbb{R}^d$ structure of $\mathbb{T}^d$ but rather partial derivatives with respect to the canonical charts of $\mathbb{T}^d$. 
\end{remark}

Let's look into the particular form of $ \mathcal{A}_{\alpha, \Theta}$. Since computing the $s$ Fourier coefficient is a continuous map (\cref{lemma:computation_of_Fourier_coefficients}), that is, $p \mapsto \Psi_s (p)$ is continuous, then

\begin{flalign*}
& \Psi_s \left(\frac{\partial\tau((e^{iy_1}, \cdots, e^{i y_d})) (p) }{ \partial y_j} (x_1, \cdots, x_d) \right) \\
& = \Psi_s\left( \lim_{\delta x \to 0} \frac{\tau((e^{ix_1}, \cdots, e^{i (x_j + \delta x)} ,\cdots, e^{i x_d})) (p) - \tau((e^{ix_1}, \cdots, e^{ix_j} ,\cdots, e^{i x_d})) (p) }{ \delta x} \right) \\
& = \lim_{\delta x \to 0} \Psi_s\left( \frac{\tau((e^{ix_1}, \cdots, e^{i (x_j + \delta x)} ,\cdots, e^{i x_d})) (p) - \tau((e^{ix_1}, \cdots, e^{ix_j} ,\cdots, e^{i x_d})) (p) }{ \delta x} \right) \\
& = \lim_{\delta x \to 0} \frac{ \Psi_s ( \tau((e^{ix_1}, \cdots, e^{i (x_j + \delta x)} ,\cdots, e^{i x_d})) (p)) - \Psi_s ( \tau((e^{ix_1}, \cdots, e^{ix_j} ,\cdots, e^{i x_d})) (p) ) }{ \delta x}  \\
& = \lim_{\delta x \to 0} \frac{ e^{i s_1 x_1}\cdots e^{i s_i (x_j + \delta x)} \cdots e^{i s_d x_d} \Psi_s (p) - e^{i s_1 x_1} \cdots e^{i s_i x_j} \cdots e^{i s_d x_d} \Psi_s (p)}{ \delta x}  \\
& = \lim_{\delta x \to 0} \frac{ e^{i s_i (x_j + \delta x)} - e^{ix_j} }{ \delta x} e^{i s_1 x_1}\cdots e^{i s_{i-1} x_{i-1} } e^{i s_{i+1} x_{i+1}} \cdots e^{i s_d x_d} \Psi_s (p) \\
& = i s_j e^{i s_1 x_1}\cdots e^{i s_j x_{j}} \cdots e^{i s_d x_d} \Psi_s (p) \\
& = i s_j \Psi_s\left(\tau((e^{ix_1}, \cdots, e^{ix_j} ,\cdots, e^{i x_d})) (p) \right).
\end{flalign*}

Therefore, computing the $j$-th partial derivative amounts for multiplying the $s$ Fourier coefficient by $s_j$, generalizing how differentiation looks over the Fourier coefficients of elements in $C^{\infty}(\mathbb{T})$ (\cref{section:Fourier_transform_and_Frechet_algebras}). 

\begin{remark}[Commutativity of the partial derivatives]\label{remark:commutativity_of_partial_derivatives}
Given $p \in \mathcal{A}_{\alpha, \Theta}$, from the previous computation we get that the action of a partial derivation with respect to $x_j$ corresponds to multiplying the $s$ Fourier coefficients by $s_j$, so, given that multiplication is commutative over $\mathbb{C}$, we have that 
$$  \frac{ \partial^2 \tau((e^{ix_1}, \cdots, e^{i x_d})) (p)}{ \partial x_{j} \partial x_{k}} (y_1, \cdots, y_d) = \frac{ \partial^2 \tau((e^{ix_1}, \cdots, e^{i x_d})) (p)}{ \partial x_{k} \partial x_{j}} (y_1, \cdots, y_d) .$$
In  consequence, we do not need to take into account the order of the partial derivatives when deriving $\tau$, and for simplicity, we will use the notation
$$
\frac{\partial\tau((e^{ix_1}, \cdots, e^{i x_d})) (p) }{ \partial^{l_1} x_1 \cdots \partial^{l_d} x_d } (y_1, \cdots, y_d) := \frac{ \partial^k \tau((e^{ix_1}, \cdots, e^{i x_d})) (p)}{ \partial x_{j_1} \cdots \partial x_{j_k}} (y_1, \cdots, y_d).
$$
\end{remark}

Since any partial derivative of $\tau(\cdot)(p) \circ \phi^{-1}_{\lambda}$ exists for $p \in \mathcal{A}_{\alpha, \Theta}$, we have that 

\begin{lemma}[Uniform boundedness of Fourier coefficients from elements of $\mathcal{A}_{\alpha, \Theta}$ ]\label{lemma:boundeness_for_elements_of_dense_smooth_sub_algebra}
Let $s = (s_1, \cdots, s_d) \in \mathbb{Z}^d$ and  $x = (x_1, \cdots, x_d) \in \mathbb{N}^d$, denote
$$ s^x  := \prod_{j \leq d} s_j^{x_j},\; (is)^x  := \prod_{j \leq d} (is_j)^{x_j},\; |s|^x  := \prod_{j \leq d} |s_j|^{x_j},$$
if $p \in \mathcal{A}_{\alpha, \Theta}$ then for every $x \in \mathbb{N}^d$ there is $K_{x,p} < \infty$ such that
$$ \forall s \in \mathbb{Z}^s, \; |s|^x  \| \Phi_s (p) \| \leq K_{x,p}. $$
\end{lemma}
\begin{proof}
In \cref{lemma:bound_on_the_fourier_coefficients} we mentioned that $\| \Phi_s(p) \| \leq \|p \|$, so, by the previous discussion on the Fourier coefficients we have that
$$  \Psi_s \left(\frac{\partial\tau((e^{iy_1}, \cdots, e^{i y_d})) (p) }{ \partial^{x_1} y_1 \cdots \partial^{x_d} y_d } (\omega_1, \cdots, \omega_d) \right) =  (is)^x  \Psi_s\left(\tau((e^{i \omega_1}, \cdots, e^{i \omega_d})) (p) \right), $$
consequently,
\begin{align*}
    |s|^x \| \Phi_s (p) \| = \|  \Psi_s \left(\frac{\partial\tau((e^{iy_1}, \cdots, e^{i y_d})) (p) }{ \partial^{x_1} y_1 \cdots \partial^{x_d} y_d } (\omega_1, \cdots, \omega_d) \right) \| \\
    \leq  \| \frac{\partial\tau((e^{iy_1}, \cdots, e^{i y_d})) (p) }{ \partial^{x_1} y_1 \cdots \partial^{x_d} y_d } (\omega_1, \cdots, \omega_d) \|.
\end{align*}
Hence, 
$$K_{x,p} = \| \frac{\partial\tau((e^{iy_1}, \cdots, e^{i y_d})) (p) }{ \partial^{x_1} y_1 \cdots \partial^{x_d} y_d } (\omega_1, \cdots, \omega_d) \|.$$
\end{proof}

Due to the previous results, we evaluate the partial derivatives at the identity we get special transformations that map $\mathcal{A}_{\alpha, \Theta}$ into itself,
$$ p \mapsto  \frac{\partial\tau((e^{iy_1}, \cdots, e^{i y_d})) (p) }{ \partial^{x_1} y_1 \cdots \partial^{x_d} y_d } (1_{\mathbb{T}^d}).  $$ 
Any of those transformations can be obtained as the composition of $d$ of the partial derivatives with respect to each one of the $d$ canonical coordinates of $\mathbb{R}^d$, and those transformations are derivations over $\mathcal{A}_{\alpha, \Theta}$. Recall that a derivation\index{derivation} over $\mathcal{A}_{\alpha, \Theta}$ is a linear map from $\mathcal{A}_{\alpha, \Theta}$ into $\mathcal{A}_{\alpha, \Theta}$ that satisfies the Leibniz rule.

\begin{lemma}[Derivations over $\mathcal{A}_{\alpha, \Theta}$]
\label{lemma:derivations_over_smooth_sub_algebra_twisted_crossed_product}
Let 
$$ \partial_j : \mathcal{A}_{\alpha, \Theta} \to \mathcal{A}_{\alpha, \Theta}, \; \partial_j p :=  \frac{\partial\tau((e^{iy_1}, \cdots, e^{i y_d})) (p) }{ \partial y_j } (1_{\mathbb{T}^d}) ,$$
then $\partial_j$ is a derivation, moreover, $\partial_j \partial_l = \partial_l \partial_j$ for any $1 \leq j,l \leq d$. 
\end{lemma}
\begin{proof}
\begin{itemize}
    \item Given that $p \mapsto \tau(\lambda)(p)$ is a linear map and computing the derivative with respect to $y_j$ corresponds to computing a limit, we can use the fact that the limit of a sum if the sum of a limit to show that $\partial_j$ is a linear map.
    \item $\partial_j$ satisfies the Leibniz rule: take $p,q \in \mathcal{A}_{\alpha, \Theta}$, since $\tau(\lambda)$ is an algebra homomorphism we get that
    \begin{align*}
    \frac{ \partial \tau((e^{ix_1}, \cdots, e^{i x_d})) (pq)}{ \partial x_j } (y_1, \cdots, y_d) = \\
    \frac{ \partial \tau((e^{ix_1}, \cdots, e^{i x_d})) (p) \tau((e^{ix_1}, \cdots, e^{i x_d})) (q)}{ \partial x_j } (y_1, \cdots, y_d)    
    \end{align*}
    thus, we can use the proof for the Leibniz rule for differentiable maps into topological algebras ( \cref{remark:leibniz_rule_fn_values_in_topological_algebra}) to show that,
    \begin{align*}
    \frac{ \partial \tau((e^{ix_1}, \cdots, e^{i x_d})) (pq)}{\partial x_j} (y_1, \cdots, y_d) = \\
     \left(\frac{ \partial \tau((e^{ix_1}, \cdots, e^{i x_d})) (p) }{\partial x_j} (y_1, \cdots, y_d) \right) \tau((e^{iy_1}, \cdots, e^{i y_d})) (q) + \\
     \tau((e^{iy_1}, \cdots, e^{i y_d})) (p) \left(\frac{ \partial \tau((e^{ix_1}, \cdots, e^{i x_d})) (q) }{ \partial x_j} (y_1, \cdots, y_d) \right) \
    \end{align*}
    Which implies that $\partial_j (pq) = \left( \partial_j p \right) p + p \left( \partial_j q \right)$.    
    \item $\partial_j, \partial_l$ commute because every element is uniquely determined by its Fourier coefficients (\cref{definition:fourier_coefficients}) and we know that $\partial_j \partial_l$ produces the same action over the Fourier coefficients as $\partial_l \partial_j$ (\cref{remark:commutativity_of_partial_derivatives}).
\end{itemize}
\end{proof}

\begin{lemma}[$\mathcal{A}_{\alpha, \Theta}$ is a sub algebra of $A \rtimes_{\alpha,\Theta}\mathbb{Z}^d$]\label{lemma:smooth_elements_are_a_sub_algebra}
$\mathcal{A}_{\alpha, \Theta}$ is a sub algebra of $A \rtimes_{\alpha,\Theta}\mathbb{Z}^d$ and 
$$\mathcal{P}(\mathbb{Z}^d,A, \alpha,\Theta) \subset \mathcal{A}_{\alpha, \Theta},$$
with $\mathcal{P}(\mathbb{Z}^d,A, \alpha,\Theta)$ the *-algebra of generalized trigonometric polynomials (\cref{lemma:twisted_crossed_products_with_Z}).
\end{lemma}
\begin{proof}

The proof of the Leibniz rule in \cref{lemma:derivations_over_smooth_sub_algebra_twisted_crossed_product} also shows that if $p,q \in \mathcal{A}_{\alpha, \Theta}$ then 
$$ (x_1, \cdots, x_d) \mapsto \tau((e^{ix_1}, \cdots, e^{i x_d})) (pq), \text{ with } x_j \in (\theta_j - \pi, \theta_j + \pi)$$
is smooth, that is, all the partial derivatives 
$$ (y_1, \cdots, y_d) \mapsto  \frac{ \partial \tau((e^{ix_1}, \cdots, e^{i x_d})) (pq)}{ \partial^{l_1} x_1 \cdots \partial^{l_d}x_d} (y_1, \cdots, y_d) $$
exists and are continuous. Additionally, since the derivative is a linear map then $\alpha q + p \in \mathcal{A}_{\alpha, \Theta}$ for any $\alpha \in \mathbb{C}$. 

Also, any element of $A \rtimes_{\alpha,\Theta}\mathbb{Z}^d$ that takes the form $a u^s$ belongs to $\mathcal{A}_{\alpha, \Theta}$, because the map
$$ (x_1, \cdots, x_d) \mapsto \tau((e^{ix_1}, \cdots, e^{i x_d})) (a u^s) = \left( \prod_{j \leq d} e^{i s_j x_j} \right) a u^s $$
is smooth, thus, any finite polynomial belongs to $\mathcal{A}_{\alpha, \Theta}$, that is, 
$$\mathcal{P}(\mathbb{Z}^d,A, \alpha,\Theta) \subset \mathcal{A}_{\alpha, \Theta}.$$
\end{proof}

\begin{proposition}[Characterization of $\mathcal{A}_{\alpha, \Theta}$]\label{proposition:characterization_smooth_sub_algebra_twsited_crossed_product}
Let $p \in A \rtimes_{\alpha,\Theta}\mathbb{Z}^d$, then, the following statements are equivalent 
\begin{enumerate}
    \item  $p \in \mathcal{A}_{\alpha, \Theta}$.
    \item For every $x \in \mathbb{N}^d$ there is $K_{x,p} < \infty$ such that
$$ \forall s \in \mathbb{Z}^s, \; |s|^x  \| \Phi_s (p) \| \leq K_{x,p}. $$
    \item For every $x \in \mathbb{N}^d$ there is $\tilde{K}_{x,p} < \infty$ such that
$$ \sum_{s \in \mathbb{Z}^d} |s|^x  \| \Phi_s (p) \| \leq \tilde{K}_{x,p}. $$
\end{enumerate}
\end{proposition}
\begin{proof}
\begin{itemize}
    \item $1 \implies 2$: This was stablished in \cref{lemma:boundeness_for_elements_of_dense_smooth_sub_algebra}
    \item $3 \iff 2$: If 
    $$ \sum_{s \in \mathbb{Z}^d} |s|^x  \| \Phi_s (p) \| \leq \tilde{K}_{x,p}, $$
    then, all the elements of the series are bounded by $\tilde{K}_{x,p}$ since we are adding positive numbers. Let's prove the other implication, denote 
    $$|s|^{(x+n)} = \prod_{j \leq d} |s_j|^{x_j + n},$$
    then,
    $$ |s|^n |s|^{x} \| \Phi_s (p) \| =  |s|^{(x+n)} \| \Phi_s (p) \| \leq K_{(x+n),p}  $$
    therefore,
    $$ |s|^{x} \| \Phi_s (p) \| \leq \frac{ K_{(x+n),p}}{|s|^n}.$$
    Since $\sum_{l \in \mathbb{Z}} 1/l^2 \leq \frac{ \pi ^2}{ 3}$ (\citep{8353}), then,
    $$ \sum_{s \in \mathbb{Z}^d} |s|^{x} \| \Phi_s (p) \| \leq K_{(x+2),p} \sum_{s \in \mathbb{Z}^d} \frac{1}{|s|^2} = K_{(x+2),p} \prod_{j \leq d} \left( \sum_{l \in \mathbb{Z}} \frac{1}{l^2} \right) \leq K_{(x+2),p} \left( \frac{ \pi ^2}{ 3} \right)^d.$$
    Hence, for any $x \in \mathbb{N}^d$ the set of positive numbers $\{ |s|^x \| \Phi_s (p) \| \}_{s \in \mathbb{Z^d}}$ is summable, that is, $\sum_{s \in\mathbb{Z}^d} |s|^x \| \Phi_s (p) \| < \infty$ where the order of the sum does not matter.

    \item $3 \implies 1$: Denote $y_j = (0, \cdots, 1, \cdots, 0)$ with $1$ in the j$^{th}$ position, then 
    $$\sum_{s \in \mathbb{Z}^d} \phi_s (p) u^s = \lim_{n \to \infty} S^{(n)} (p) = p,$$
    also, $\sum_{s \in \mathbb{Z}^d} is^{y_j} \phi_s (p) u^s$ converges (absolutely), so, lets call it $q := \sum_{s \in \mathbb{Z}^d} is^{y_j} \phi_s (p) u^s $, then,
    $$\sum_{s \in \mathbb{Z}^d} is^{y_j} \phi_s (p) u^s = \lim_{n \to \infty} S^{(n)} (q) = q.$$
    
    Since $\tau(\lambda)$ is an automorphism we have that $\| a \| = \| \tau(\lambda)(a) \|$ for any $a \in A \rtimes_{\alpha,\Theta}\mathbb{Z}^d$ and $\lambda \in \mathbb{T}^d$, hence, if we define 
    \begin{align*}
    f_n : \mathbb{T}^d  \to A \rtimes_{\alpha,\Theta}\mathbb{Z}^d,& \; f_n(\lambda):= \tau(\lambda)(S^{n}(p)), \\
    g_n : \mathbb{T}^d  \to A \rtimes_{\alpha,\Theta}\mathbb{Z}^d,& \; f_n(\lambda):= \tau(\lambda)(S^{n}(q)),
    \end{align*}
    we end up with
    \begin{align*}
    \| f_n (\lambda) - \tau(\lambda)(p) \| = \| \sum_{s \in V_n} \gamma_s (\lambda) \phi_s (p) + \sum_{s \in \mathbb{Z}^d} \gamma_s (\lambda) \phi_s (p) \|  \leq \sum_{s \notin V_n} \| \phi_s (p) \|, \\
    \| g_n (\lambda) - \tau(\lambda)(q) \| = \| \sum_{s \in V_n} \gamma_s (\lambda) \phi_s (q) + \sum_{s \in \mathbb{Z}^d} \gamma_s (\lambda) \phi_s (q) \|  \leq \sum_{s \notin V_n} \| \phi_s (q) \|.
    \end{align*}
    Since both $ \sum_{s \in \mathbb{Z}^d} \phi_s (p), \sum_{s \in \mathbb{Z}^d} \phi_s (q)  $ converge absolutely, we have that $f_n \to \tau(\cdot)(p)$ uniformly over $\mathbb{T}^d$ and $g_n \to \tau(\cdot)(q)$ uniformly over $\mathbb{T}^d$.
    
    For the generalized trigonometric polynomials $ S^{(n)} (p), S^{(n)} (q) $ we have that
    $$ \Psi_s \left(\frac{\partial\tau((e^{iy_1}, \cdots, e^{i y_d})) (S^{(n)}(p)) }{ \partial y_j} (x_1, \cdots, x_d) \right) = i s^{y_j}  \Psi_s\left(\tau((e^{ix_1}, \cdots, e^{i x_d})) (S^{(n)}(p)) \right) $$
    and 
    $$ i s^{y_j}  \Psi_s\left(\tau((e^{ix_1}, \cdots, e^{i x_d})) (S^{(n)}(p)) \right) = \Psi_s\left(\tau((e^{ix_1}, \cdots, e^{i x_d})) (S^{(n)}(q)) \right),   $$
    thus, 
    $$ \frac{\partial f_n \circ \phi^{-1}_{\lambda} (y_1, \cdots, y_d)  }{ \partial y_j}  (x_1, \cdots, x_d) = g_n \circ \phi^{-1}_{\lambda} (x_1, \cdots, x_d) .$$
    
    Under this setting the third item of \cref{lemma:generalizations_of_some_calculus_results} implies that $ (x_1, \cdots, x_d) \mapsto f_n \circ \phi^{-1}_{\lambda} (x_1, \cdots, x_d)$ is differentiable with respect to the j$^{th}$ component, and
    $$ \frac{\partial f \circ \phi^{-1}_{\lambda} (y_1, \cdots, y_d)  }{ \partial y_j}  (x_1, \cdots, x_d) = g \circ \phi^{-1}_{\lambda} (x_1, \cdots, x_d),$$
    with
    $$ x_k \in  (\theta_k - \pi, \theta_k + \pi), \; \lambda_k = e^{i \theta_k}.$$
    
    Since $\sum_{s \in \mathbb{Z}^d} is^{y_j + y_l} \phi_s (p) u^s$ also converges (absolutely), the previous process can be iterated to show that
    $$ (x_1, \cdots, x_d) \mapsto  \frac{ \partial \tau((e^{i y_1}, \cdots, e^{i y_d})) (p)}{ \partial^{m_1} y_1 \cdots \partial^{m_d} y_d} (x_1, \cdots, x_d)   $$
    exists and is continuous for any $(m_1, \cdots, m_d) \in \mathbb{N}^d$. 
\end{itemize}
\end{proof}

\begin{corollary}\label{corollary:smooth_elements_are_a_sub_star_algebra}
$\mathcal{A}_{\alpha, \Theta}$ is a dense *-subalgebra of $A \rtimes_{\alpha,\Theta}\mathbb{Z}^d$, aditionally, for any $p \in \mathcal{A}_{\alpha, \Theta}$ and $1 \leq j \leq d$ we have that,
$$ (\partial_j p )^* = \partial_j (p^*), $$
\end{corollary}
\begin{proof}
Since $\mathcal{P}(\mathbb{Z}^d,A, \alpha,\Theta)$ is a subset of $\mathcal{A}_{\alpha, \Theta}$ (\cref{lemma:smooth_elements_are_a_sub_algebra}), and $\mathcal{P}(\mathbb{Z}^d,A, \alpha,\Theta)$ is dense in $A \rtimes_{\alpha,\Theta}\mathbb{Z}^d$ (\cref{lemma:twisted_crossed_products_with_Z}), we have that $\mathcal{A}_{\alpha, \Theta}$ is dense inside $A \rtimes_{\alpha,\Theta}\mathbb{Z}^d$.

From \cref{proposition:characterization_smooth_sub_algebra_twsited_crossed_product} we know that $p \in \mathcal{A}_{\alpha, \Theta}$ iff for every $x \in \mathbb{N}^d$ there is $K_{x,p} < \infty$ such that
$$ \forall s \in \mathbb{Z}^s, \; |s|^x  \| \Phi_s (p) \| \leq K_{x,p}. $$
Take $p \in \mathcal{A}_{\alpha, \Theta}$, from \cref{lemma:Fourier_coefficients_multiplication_and_involution} we know that
$$  \Phi_s(p^*) = e^{i s^t \Theta s } \alpha(s)(\Phi_{-s}(p)^*), $$
therefore, since $\alpha$ is norm decreasing by the virtue of being a C* homomorphism (\cref{proposition:automatic_continuity_C_star_algebras}), we get that 
$$ \forall s \in \mathbb{Z}^s, \; |s|^x  \| \Phi_s (p^*) \| \leq K_{x,p}, $$
consequently, if $p \in \mathcal{A}_{\alpha, \Theta}$ then $p^* \in \mathcal{A}_{\alpha, \Theta}$. In conclusion, $p \in \mathcal{A}_{\alpha, \Theta}$ is a sub *algebra of $A \rtimes_{\alpha,\Theta}\mathbb{Z}^d$.

Also,
\begin{align*}
    \Phi_s ( \partial_j (p^*) ) = i s_j e^{i s^t \Theta s } \alpha(s)(\Phi_{-s}(p)^*),\\
    \Phi_s ( (\partial_j p)^* ) = e^{i s^t \Theta s } \alpha(s)( (-i s_j \Phi_{-s}(p))^*),
\end{align*}
since $\alpha(s)$ is a linear map we get that
$$ \Phi_s ( \partial_j (p^*) ) = i s_j e^{i s^t \Theta s } \alpha(s)(\Phi_{-s}(p)^*) =  \Phi_s ( (\partial_j p)^* ) . $$
Given that an element of $A \rtimes_{\alpha,\Theta}\mathbb{Z}^d$ is uniquely determined by its Fourier coefficients (\cref{definition:fourier_coefficients}) we have that
$$ \partial_j (p^*) =  (\partial_j p)^*.  $$
\end{proof}

\begin{remark}[Alternative descriptions of $\mathcal{A}_{\alpha, \Theta}$]\label{remark:alternative_descriptions_of_smooth_sub_algebra}
We have characterized the sub *algebra $\mathcal{A}_{\alpha, \Theta}$ in terms of the decay of the Fourier coefficients of its elements, however, this characterization is not unique. For example, in \citep[Definition 12.7]{gracia-bondia_elements_2001} $\mathcal{A}_{\alpha, \Theta}$ is defined as the set of elements whose Fourier coefficients satisfy
$$ (1 + \sum_{j \leq d} s_j^2)^{k} \| \Phi_s (p) \|^2 \leq K_{k,p}, \; k \in \mathbb{N}, \; K_{k,p} < \infty, $$
which is also used in \citep[page 106]{carey_index_2014}. This definition is equivalent to the one we have given in \cref{proposition:characterization_smooth_sub_algebra_twsited_crossed_product}, and you can check it after doing some algebraic manipulations. 
\end{remark}

Recall that non-commutative C* algebras can be intuitively thought as algebras of continuous functions over "non-commutative topological spaces" (\cref{section:non_commutative_geometry_dictionary}), under this setting, $\mathcal{A}_{\alpha, \Theta}$ plays the role of algebra of smooth functions over the "non-commutative topological space" associated to $A \rtimes_{\alpha,\Theta}\mathbb{Z}^d$. Since $A \rtimes_{\alpha,\Theta}\mathbb{Z}^d$ arise as a generalization of the C* algebra $C(\mathbb{T})$, we have that $\mathcal{A}_{\alpha, \Theta}$ comes as generalization of the *algebra $\mathscr{S}(\mathbb{Z})$ , which is isomorphic to the *algebra of smooth functions over the torus through the Fourier transform (\cref{section:Fourier_transform_and_Frechet_algebras}). Our last step to complete this analogy is to provide $\mathcal{A}_{\alpha, \Theta}$ with the structure of a Fréchet algebra, which we proceed to do by copying the form of the seminorms over $C^{\infty}(\mathbb{T})$ (\cref{section:Fourier_transform_and_Frechet_algebras}).

\begin{lemma}[$\mathcal{A}_{\alpha, \Theta}$ is a Fréchet space\index{Fréchet space}]\label{lemma:algebra_of_smooth_elements_is_Frechet_space}
Let $x \in \mathbb{N}^d$ given by $x:= (x_1, \cdots, x_d)$, if $x_j \neq 0$ denote,
$$ \partial_j^{x_j}(p) := ( \underbrace{\partial_j \circ \cdots \circ \partial_j}_{x_j \text{ times }} )(p),$$
else, denote
$$ \partial_j^{x_j} ( p) := p,$$
so, for any $p \in  \mathcal{A}_{\alpha, \Theta}$ denote
$$  \partial^{x} p := (\partial_1^{x_1} \circ \cdots \circ \partial_d^{x_d}) p.$$
Under the set of seminorms
$$ \| p \|_{x} := \| \partial^{x} p \|, \; x \in \mathbb{N}^d  $$
the *-algebra $\mathcal{A}_{\alpha, \Theta}$ is complete. Therefore, $\mathcal{A}_{\alpha, \Theta}$ is a Fréchet space and for any $x\in \mathbb{Z}^s$ the map following map is continuous
$$ p \mapsto \partial^{x} p. $$
\end{lemma}
\begin{proof}
First we will prove that $\| p \|_{x} := \| \partial^{x} p \|, \; x \in \mathbb{N}^d$ is a seminorm, then, we will check that the the *-algebra $ \mathcal{A}_{\alpha, \Theta}$ provided with the topology generated by the seminorms $\| \cdot \|_{x}$ is a Fr\'echet space, finally, we will check that the maps $p \mapsto \partial^{x} p$ are continuous over the Fr\'echet space $ \mathcal{A}_{\alpha, \Theta}$.

\begin{itemize}
    \item Since each $\partial_j$ is linear (\cref{lemma:derivations_over_smooth_sub_algebra_twisted_crossed_product}) we have that,
    $$ \| \partial^{x} (p + q) \| = \| \partial^{x} p + \partial^{x} q \| \leq \| \partial^{x}p \| + \| \partial^{x} q \|   $$
    and 
    $$ \| \partial^{x} (\nu p) \| = | \nu | \| \partial^{x}p \|, \nu \in \mathbb{C},$$
    additionally, $\| 0 \|_{x} = 0$, therefore, $\| \cdot \|_{x}$ is a seminorm.
    
    \item Since the topology of $ \mathcal{A}_{\alpha, \Theta}$ is generated by a family of seminorms $\{ \| \cdot \|_{x}\| \}_{x \in \mathbb{N}^d}$ it becomes a locally convex space (\cref{theorem:characterization_of_lcs_using_seminorms}), additionally, since the family of seminorms is a countable set then $\mathcal{A}_{\alpha, \Theta}$ becomes a metrizable locally convex space (\cref{lemma:metrizable_lcs_and_family_of_seminorms}). Assume that $\{p_n\}_{n \in \mathbb{N}}$ is a Cauchy sequence under the topology induced by the family of seminorms $\{ \| \cdot \|_{x}\| \}_{x \in \mathbb{N}^d} $, then, from the definition of Cauchy sequence in a metrizable locally convex space (\cref{remark:equivalent_def_of_Cauchy_sequence_in_metrizable_LCS}) we have that, for any $P \subset \mathbb{N}^d$ with $|P|$ finite, and $\epsilon>0$, there is $N > 0$ such that if $n,m\geq N$, then, for all $x \in P$ 
    $$\| p_n - p_m \|_{x} \leq \epsilon.$$ 
    Given that,
    $$ \| |s|^{x} \Phi_s (p_n - p_m) \| \leq \| \partial^{x} ( p_n - p_m )\| = \| p_n - p_m \|_x, \; x \in \mathbb{N}^d, $$
    from the proof of \cref{proposition:characterization_smooth_sub_algebra_twsited_crossed_product}, we know that
    $$ \sum_{s \in \mathbb{Z}^d} |s|^{x} \| \Phi_s (p_n - p_m) \| \leq \| p_n - p_m \|_{x+2} \left( \frac{ \pi ^2}{ 3} \right)^d, . $$
    Since $\{p_n\}_{n \in \mathbb{N}}$ is a Cauchy sequence, the previous discussion on Cauchy sequences on $ \mathcal{A}_{\alpha, \Theta}$  assert us that, given $x \in \mathbb{N}^d$ and $\epsilon > 0$, there is $N$ such that $n,m > N$ then 
    $$ \| p_n - p_m \|_{x+2} < \epsilon \left( \frac{3}{ \pi ^2} \right)^d. $$
    Hence, knowing that 
    $$ \| \sum_{s \in \mathbb{Z}^d} (is)^{x}  \Phi_s(p_n) u^s - \sum_{s \in \mathbb{Z}^d} (is)^{x}  \Phi_s(p_m) u^s \| \leq   \sum_{s \in \mathbb{Z}^d} |s|^{x} \| \Phi_s (p_n - p_m) \|, $$
    we get that, if $n,m >N$ then
    $$ \| \sum_{s \in \mathbb{Z}^d} (is)^{x}  \Phi_s(p_n) u^s - \sum_{s \in \mathbb{Z}^d} (is)^{x}  \Phi_s(p_m) u^s \| \leq \epsilon,$$
    which implies that the following sequence is a Cauchy sequence of elements of $L^1(\mathbb{Z}^d,A)$,
    $$  \{ \sum_{s \in \mathbb{Z}^d} (is)^{x}  \Phi_s(p_n) u^s  \}_{n \in \mathbb{N}}.$$
    Since $L^1(\mathbb{Z}^d,A)$ is complete (\cref{example:L1_Bohner_spaces_and_convolution}) and the norm of $A \rtimes_{\alpha,\Theta}\mathbb{Z}^d$ is less than or equal to the norm on $L^1(\mathbb{Z}^d,A)$, then, for any $x \in \mathbb{N}^d$ the sequence $\{  \sum_{s \in \mathbb{Z}^d} (is)^x \Phi_s(p_n) u^s \}_{n \in \mathbb{N}}$ converges in $A \rtimes_{\alpha,\Theta}\mathbb{Z}^d$, moreover, the limit of $\{  \sum_{s \in \mathbb{Z}^d} (is)^x \Phi_s(p_n) u^s \}_{n \in \mathbb{N}}$ is also an element of $L^1(\mathbb{Z}^d,A)$.
    
    Denote by $p$ the limit of the sequence $\{  \sum_{s \in \mathbb{Z}^d} \Phi_s(p_n) u^s \}_{n \in \mathbb{N}}$, then, for any $x \in \mathbb{N}^d$ the continuity of the computation of the Fourier coefficients (\cref{lemma:bound_on_the_fourier_coefficients}) implies that 
    $$\lim_{n \to \infty} \Phi_s \left(  \sum_{s \in \mathbb{Z}^d} (is)^x \Phi_s(p_n) u^s \right) = \lim_{n \to \infty} (is)^x\Phi_s(p_n) = (is)^x\Phi_s(p),$$
    therefore, for any $x \in \mathbb{N}^d$, the series
    $$  \sum_{s \in \mathbb{Z}^d} (is)^x \Phi_s(p) u^s $$
    converges in $L^1(\mathbb{Z}^d,A)$. Using the characterization of the elements of $\mathcal{A}_{\alpha, \Theta}$ provided in \cref{proposition:characterization_smooth_sub_algebra_twsited_crossed_product} we get that $p \in \mathcal{A}_{\alpha, \Theta}$, as desired.
    \item Notice that $\| \partial^x p \|_y = \| \partial^{x+y} p \|$, so, take $p_n \to p$ in $\mathcal{A}_{\alpha,\Theta}$, then $\| p_n -p \|_s \to 0$ for all $s \in \mathbb{N}^d$ according to the definition of convergence in a metrizable locally convex space (\cref{proposition:convergence_metrizable_lcs}). Given that
    $$\| (\partial^x p_n) - (\partial^x p)\|_y = \| p_n - p \|_{x+y} $$
    and $\| p_n - p \|_{x+y} \to 0$ by assumption, the characterization of continuity of maps between metrizable locally convex spaces (\cref{proposition:continuity_metrizable_lcs}) implies that 
    $$\partial^x : \mathcal{A}_{\alpha,\Theta} \to \mathcal{A}_{\alpha,\Theta}, \; x \in \mathbb{N}^d$$ 
    are continuous maps.
\end{itemize}
\end{proof}

The results from \citep{bhatt_class_2013} imply that Fr\'echet $D^{*}_{\infty}$ subalgebras of $A \rtimes_{\alpha,\Theta}\mathbb{Z}^d$ are smooth sub algebras of $A \rtimes_{\alpha,\Theta}\mathbb{Z}^d$, hence, we will proof that $\mathcal{A}_{\alpha, \Theta}$ is a Fr\'echet $D^{*}_{\infty}$ subalgebra of $A \rtimes_{\alpha,\Theta}\mathbb{Z}^d$. A *-algebra $\mathcal{B}$ is called a Fr\'echet $D^{*}_{\infty}$ subalgebra of $A \rtimes_{\alpha,\Theta}\mathbb{Z}^d$ if (\cref{def:Frechet_d_infinity_subalgebras}) there is a sequence of seminorms $\{ \| \cdot \|_i: 0 \leq i < \infty \}$ such that $\| \cdot \|_0$ is the norm over $A \rtimes_{\alpha,\Theta}\mathbb{Z}^d$ and the following criteria are met:
\begin{itemize}
    \item $\mathcal{B}$ is a Fr\'echet algebra whose topology is defined by the sequence of seminorms $\{ \| \cdot \|_i: 0 \leq i < \infty \}$. 
    \item For all $i, 1 \leq i<\infty$, for all $x, y$ in $\mathcal{B}$, we have that $\|x y\|_i \leq\|x\|_i\|y\|_i,\left\|x^*\right\|_i=\|x\|_i$.
    \item For each $i, 1 \leq i<\infty$, there exists $D_i>0$ such that $\|x y\|_i \leq D_i\left(\|x\|_i\|y\|_{i-1}+\right.$ $\left.\|x\|_{i-1}\|y\|_i\right)$ holds for all $x, y$ in $\mathcal{B}$. $\left\{\|\cdot\|_i: 0 \leq i<\infty\right\}$. 
\end{itemize}
So, we will look for an additional set of seminorms that also generate the topology of $\mathcal{A}_{\alpha, \Theta}$, we take the explicit form of those seminorms from \citep[Examples 1.6]{bhatt_class_2013}, where they are exposed in the more general setting of an action of a Lie group $G$ on an unital C* algebra, which in our scenario corresponds to look at the action of $\mathbb{T}^d$ over the C* algebra $A \rtimes_{\alpha,\Theta}\mathbb{Z}^d$ (\cref{lemma:action_of_torus_over_twisted_crossed_product}). 


\begin{proposition}[$\mathcal{A}_{\alpha, \Theta}$ as a Fréchet $D^{*}_{\infty}$ subalgebra of $A \rtimes_{\alpha,\Theta}\mathbb{Z}^d$]\label{proposition:smooth_elements_as_D_infinity_Freche_subalgebra}
Define $\| p \|_0 := \| p \|$ for $p \in \mathcal{A}_{\alpha, \Theta}$. Let  $\left\{\partial_{i_1}, \partial_{i_2}, \ldots, \partial_{i_n}\right\}$ be an ordered $n$-tuple from $\left\{\partial_1, \partial_2, \ldots, \partial_d\right\}$ and define 
$$
\|p\|_n:=\|p\|_0+\sum_{k=1}^n \frac{1}{k!} \sum_{i_1, i_2, \ldots, i_k=1}^d \left\|\partial_{i_1} \partial_{i_2} \cdots \partial_{i_k} p\right\| , \; n \geq 1, 
$$
for $p \in \mathcal{A}_{\alpha, \Theta}$, where the sum 
$$ \sum_{i_1, i_2, \ldots, i_k=1}^d \left\|\partial_{i_1} \partial_{i_2} \cdots \partial_{i_k} p\right\| $$
runs over all the ordered $k$-tuples $\left\{\partial_{i_1}, \partial_{i_2}, \ldots, \partial_{i_k}\right\}$. Then, 
\begin{enumerate}
    \item The topology generated by $\{ \| \cdot \|_n \}_{n \in \mathbb{N}}$ is the Fréchet topology of $\mathcal{A}_{\alpha, \Theta}$, that is, the topology generated by $\{ \| \cdot \|_n \}_{n \in \mathbb{N}}$ is the same topology generated by the family of seminorms $\{ \| \cdot \|_{\alpha} \}_{\alpha \in \mathbb{N}^d}$ described in \cref{lemma:algebra_of_smooth_elements_is_Frechet_space}.
    \item Under the set of seminorms $\{ \| \cdot \|_n \}_{n \in \mathbb{N}}$, $\mathcal{A}_{\alpha, \Theta}$ becomes a Fréchet $D^{*}_{\infty}$ subalgebra of $A \rtimes_{\alpha,\Theta}\mathbb{Z}^d$. 
\end{enumerate}
\end{proposition}
\begin{proof}
\begin{enumerate}
    \item Let denote $T_1$ the topology generated by  $\{ \| \cdot \|_n \}_{n \in \mathbb{N}}$ and $T_2$ the topology generated by $\{ \| \cdot \|_{\alpha} \}_{\alpha \in \mathbb{N}^d}$, then, we need to show that given $p \in \mathcal{A}_{\alpha, \Theta}$
    \begin{itemize}
        \item If $O_1$ is an open neighborhood of $p$ in the topology $T_1$, then, there is an element of $T_2$, which we call $O_2$, such that $O_2 \subset O_1.$
        \item If $O_2$ is an open neighborhood of $p$ in the topology $T_2$, then, there is an element of $T_1$, which we call $O_1$, such that $O_1 \subset O_2.$
    \end{itemize}
    Both of these can be easily shown from the description of the seminorms, we are going to tackle one of them and the reader can perform the computations for the other. Assume that $p \in O_2$ and $O_2 \in T_2$, then, from the definition of the topology generated by a family of seminorms (\cref{def:topology_geenrated_by_seminorm}), we know that there is a $P \subset \mathbb{N}^d$ with $P$ finite and $\epsilon > 0$ such that if
    $$ \| p - q \|_{x} \leq \epsilon \; \forall x \in P, $$
    then, $q \in O_2$. Let $x:= (x_1, \cdots, x_d)$ and set
    $$ n = \max_{x \in P} \{ \sum_{j \leq d} x_j \}, $$
    then, if 
    $$ \| p - q \|_n \leq \frac{\epsilon}{n!}, $$
    we have that
    $$ \| p - q \|_{x} \leq \epsilon, $$
    thus, if 
    $$O_1 = \{ q \in \mathcal{A}_{\alpha, \Theta} : \| p -q \|_n \leq \frac{\epsilon}{n!} \} $$
    then $O_1 \subset O_2$.
    
    \item Following \cref{def:Frechet_d_infinity_subalgebras} we have to show that:
    \begin{itemize}
        \item For all $g$ with $1 \leq g<\infty$, if $p, q$ in $\mathcal{A}_{\alpha, \Theta}$ we have that $\|p q\|_g \leq\|p\|_g\|q\|_g$: let $p,q \in \mathcal{A}_{\alpha, \Theta}$, take the ordered $k$-tuple $\left\{\partial_{i_1}, \partial_{i_2}, \ldots, \partial_{i_k}\right\}$, then, the Leibniz rule for the derivations $\partial_1, ..., \partial_d$ can be used to express the element 
        $$\partial_{i_1} \partial_{i_2} \cdots \partial_{i_k} (pq)$$
        as a sum of $2^k$ elements each one taking the form 
        $$(\partial_{i_{l_1}} \cdots \partial_{i_{l_m}} p)(\partial_{i_{j_1}} \cdots \partial_{i_{j_n}} q), $$
        where $k = m+n$, $l_1 < l_2 < \cdots < l_m$, $j_1 < l_2 < \cdots < j_n$ and following the set equality
        $$ \{ l_1,  \cdots, l_m \} \cup \{ j_1, \cdots, j_n \} = \{ 1, \cdots, k \}.  $$
        Each one of the elements 
        $$(\partial_{i_{l_1}} \cdots \partial_{i_{l_m}} p)(\partial_{i_{j_1}} \cdots \partial_{i_{j_n}} q), $$
        is uniquely determined by the ordered tuples 
        $$\{ \partial_{i_{l_1}}, \cdots, \partial_{i_{l_m}}\} \text{ and } \{\partial_{i_{j_1}}, \cdots, \partial_{i_{j_n}}\},$$ 
        and those ordered tuples may be empty. Given the previous discussion, we use the notation
        $$ \partial_{i_1} \partial_{i_2} \cdots \partial_{i_k} (pq) = \sum_{m+n = k} (\partial_{i_{l_1}} \cdots \partial_{i_{l_m}} p)(\partial_{i_{j_1}} \cdots \partial_{i_{j_n}} q), $$
        where the last sum goes over all the elements that arise from applying the Leibniz rule iteratively, and as we mentioned, there are $2^k$ elements in that sum. Using the triangle inequality and the submultiplicativity of the C* norm of $A \rtimes_{\alpha,\Theta}\mathbb{Z}^d$ we get that
        $$ \frac{1}{k!} \|\partial_{i_1} \partial_{i_2} \cdots \partial_{i_k} (pq) \| \leq \frac{1}{k!} \sum_{m+n = k} \|(\partial_{i_{l_1}} \cdots \partial_{i_{l_m}} p)\| \| (\partial_{i_{j_1}} \cdots \partial_{i_{j_n}} q) \|. $$
        After applying the previous reasoning to each one of the $d^k$ elements of the sum 
        $$ \frac{1}{k!}\sum_{i_1, i_2, \ldots, i_k=1}^d \left\|\partial_{i_1} \partial_{i_2} \cdots \partial_{i_k} (pq)\right\|, $$
        we end up with an expression which we call $S_k$, and that expression is greater or equal to the sum on the previous statement. $S_k$ consists of the sum of $d^k 2^k$ elements of the form
        $$ \frac{1}{k!} \|(\partial_{i_{l_1}} \cdots \partial_{i_{l_m}} p)\| \| (\partial_{i_{j_1}} \cdots \partial_{i_{j_n}} q) \|, $$
        additionally, given the ordered tuples 
        $$\{ \partial_{i_{l_1}}, \cdots, \partial_{i_{l_m}}\} \text{ and } \{\partial_{i_{j_1}}, \cdots, \partial_{i_{j_n}}\},$$
        there are $\frac{k!}{n! m!}$ different ordered tuples $\left\{\partial_{i_1}, \partial_{i_2}, \ldots, \partial_{i_k}\right\}$ such that the element 
        $$\| \partial_{i_{l_1}} \cdots \partial_{i_{l_m}} p \| \| \partial_{i_{j_1}} \cdots \partial_{i_{j_n}} q \| $$
        appears in the sum 
        $$ \partial_{i_k} (pq) \| \leq \frac{1}{k!} \sum_{m+n = k} \|(\partial_{i_{l_1}} \cdots \partial_{i_{l_m}} p)\| \| (\partial_{i_{j_1}} \cdots \partial_{i_{j_n}} q) \| $$
        associated to 
        $$\frac{1}{k!} \|\partial_{i_1} \partial_{i_2} \cdots \partial_{i_k} (pq) \|,$$
        therefore, the term
        $$\frac{1}{n! m!} \|\partial_{i_{l_1}} \cdots \partial_{i_{l_m}} p \| \| \partial_{i_{j_1}} \cdots \partial_{i_{j_n}} q\|, $$
        appears in $S_k$ once it is simplified. Since $\| pq \|_g \leq \sum_{k \leq g} S_k$, we need to show that each element of $\sum_{k \leq g} S_k$ appears once in $\| p \|_g \| q \|_g$, this is straightforward from the definition of $\| \cdot \|_g$ as we proceed to check. The norm $\| p \|_g$ contains the term $\frac{1}{m!}\|\partial_{i_{l_1}} \cdots \partial_{i_{l_m}} p\|$ and the norm $\| q \|_g$ contains the term $\frac{1}{n!}\|\partial_{i_{j_1}} \cdots \partial_{i_{j_n}} q\|$, therefore, the multiplication $\| p \|_g \| q \|_g$ contains the term 
        $$ \frac{1}{m! n!}\|\partial_{i_{l_1}} \cdots \partial_{i_{l_m}} p\| \|\partial_{i_{j_1}} \cdots \partial_{i_{j_n}} q \|, $$
        which is the term that appears in $S_k$. The previous discussion assure us that $\| pq\|_g \leq \|p \|_g \|q\|_g$.
        \item For all $i$ with $1 \leq i<\infty$, we have that $\left\|p^*\right\|_i=\|p\|_i$. Given that,
            \begin{align*}
                \Phi_s (\partial_j (p ^*)) = i s_j e^{s^t \Theta s} \alpha(s) (\Phi_{-s} (p)^*), \\
                \Phi_s ((\partial_j p )^*) = e^{s^t \Theta s} \alpha(s) (i s_j \Phi_{-s} (p)^*),
            \end{align*}    
            and $\alpha$ is a linear transformation, we get that $\partial_j (p^*) = (\partial_j p)^*$, therefore, 
            $$ \| p \|_n = \| p^* \|_n, \; \forall n \in \mathbb{N}. $$
        \item For each $i$ with $ 1 \leq i<\infty$, there exists $D_i>0$ such that $\|p q\|_i \leq D_i\left(\|p\|_i\|q\|_{i-1}+\right.$ $\|p\|_{i-1}\|q\|_i$ ) holds for all $p, q$ in $\mathcal{A}_{\alpha, \Theta}$: In this case $D_i = 1$ and the inequality comes from the Leibniz rule for $\partial_j$.
        \item $\mathcal{A}_{\alpha, \Theta}$ is a Hausdorff Fréchet *-algebra with the topology defined by the seminorms $\left\{\|\cdot\|_i: 0 \leq i<\infty\right\}$: In \cref{lemma:algebra_of_smooth_elements_is_Frechet_space} we showed that $\mathcal{A}_{\alpha, \Theta}$ is a Fréchet space, and in the present proposition we showed that the seminorms $\{ \| \cdot \|_n \}_{n \in \mathbb{N}}$ are submultiplicative, therefore, \cref{lemma:joint_continuity_mutliplication_on_m_convex_frechet_sapce} implies that $\mathcal{A}_{\alpha, \Theta}$ is a Fr\'echet algebra, that is, the multiplication over $\mathcal{A}_{\alpha, \Theta}$ is jointly continuous (\cref{def:Frechet_algebra}). Additionally, we showed that $\| p^* \|_i = \| p \|_i$, therefore, it is a Fréchet *-algebra.
    \end{itemize}
    In conclussion, $\mathcal{A}_{\alpha, \Theta}$ is a Fréchet $D^{*}_{\infty}$ subalgebra of $A \rtimes_{\alpha,\Theta}\mathbb{Z}^d$. 
\end{enumerate}
\end{proof}

Let $A$ be a C* algebra and $\mathcal{A}$ a Fr\'echet $D^{*}_{\infty}$ subalgebra of $A$, according to \cref{proposition:invariance_under_holomorphic_func_calculus} we have that $\mathcal{A}$ is invariant with respect to the holomorphic functional calculus of $A$, under this setting, \cref{proposition:smooth_elements_as_D_infinity_Freche_subalgebra} implies that $\mathcal{A}_{\alpha, \Theta}$ is a invariant under the holomorphic functional calculus of $A \rtimes_{\alpha,\Theta}\mathbb{Z}^d$. Given that $\mathcal{A}_{\alpha, \Theta}$ is a dense sub *-algebra of $A \rtimes_{\alpha,\Theta}\mathbb{Z}^d$ (\cref{corollary:smooth_elements_are_a_sub_star_algebra}) and has a stronger topology than $A \rtimes_{\alpha,\Theta}\mathbb{Z}^d$, we have that $\mathcal{A}_{\alpha, \Theta}$ is a smooth sub algebra of the C* algebra $A \rtimes_{\alpha,\Theta}\mathbb{Z}^d$ (\cref{def:smooth_sub_algebra}).

In the following statement, we say that  $\mathcal{A}_{\alpha, \Theta}$ is a spectrally invariant with respect to $A \rtimes_{\alpha,\Theta}\mathbb{Z}^d$ if for any $a \in \mathcal{A}_{\alpha, \Theta}$, the spectrum of $a$ as an element of $\mathcal{A}_{\alpha, \Theta}$ is equal to the spectrum of $a$ as an element of $A \rtimes_{\alpha,\Theta}\mathbb{Z}^d$ (\cref{definition:spectrally_invariant}). Given the previous discussion, the following result is a consequence of \cref{lemma:automatic_continuity_spectrally_invariant_sub_alegbras}, \cref{proposition:invariance_under_holomorphic_func_calculus} and \cref{proposition:smooth_elements_as_D_infinity_Freche_subalgebra}.

\begin{corollary}[$\mathcal{A}_{\alpha, \Theta}$ is invariant under holomorphic calculus and $C^{\infty}$ calculus]\label{corollary:smooth_elements_are_invariant_under_holomoprhic_and_c_infty_calculus}
We have that:
\begin{enumerate}
    \item $\mathcal{A}_{\alpha, \Theta}$ is a smooth sub algebra (\cref{def:smooth_sub_algebra}) of $A \rtimes_{\alpha,\Theta}\mathbb{Z}^d$.
    \item $\mathcal{A}_{\alpha, \Theta}$ is a spectrally invariant with respect to $A \rtimes_{\alpha,\Theta}\mathbb{Z}^d$.
    \item $A \rtimes_{\alpha,\Theta}\mathbb{Z}^d$ is the enveloping C* algebra of $\mathcal{A}_{\alpha, \Theta}$, that is,
     $$ C^*(\mathcal{A}_{\alpha, \Theta}) \simeq A \rtimes_{\alpha,\Theta}\mathbb{Z}^d. $$
    \item $\mathcal{A}_{\alpha, \Theta}$ is invariant under the $C^{\infty}$ calculus on $A \rtimes_{\alpha,\Theta}\mathbb{Z}^d$ for self-adjoint elements, i.e. for any $a \in \mathcal{A}_{\alpha, \Theta}$ such that $a=a^*$ and $f \in C^{\infty}(\mathbb{R})$, we have that $f(a) \in \mathcal{A}_{\alpha, \Theta}$.
\end{enumerate}
\end{corollary}

\begin{remark}[Projective limits\index{projective limit} and $\mathcal{A}_{\alpha, \Theta}$]\label{remark:projective_limits_and_smooth_sub_algebras}
Let $A$ be a C* algebra and $\mathcal{A}$ be a Fréchet $D^{*}_{\infty}$ subalgebra of $A$ whose topology is defined by the family of seminorms $\{ \| \cdot \|_i \}_{i \in \mathbb{N}}$, in \citep[Theorem 3.1]{bhatt_class_2013} those are used to express $\mathcal{A}$ as a projective limit of Banach $\left(D_k^*\right)$-subalgebras of a $A$. Such projective limit is constructed as follows, take $n \in \mathbb{N}$  and denote by $\mathcal{A}_n$ the closure of $\mathcal{A}$ with respect to the norm $\| \cdot \|_n$, given that $\| a \|_n \leq \| a\|_{n+1}$ there are Banach alebra homomorphisms $\phi_{n}:\mathcal{A}_{n+1} \to \mathcal{A}_n: $ with dense range. The projective limit of $\{ \mathcal{A}_n \}_{n \in \mathbb{N}}$ consist of the elements $(a_n)_{n \in \mathbb{N}} $  of $\prod_{n \in \mathbb{N}} \mathcal{A}_n$ such that 
$$ \phi_n(a_{n+1}) = a_n, $$
this set is denoted by $\varprojlim \mathcal{A}_n$ and its topology is the subspace topology from $\prod_{n \in \mathbb{N}} \mathcal{A}_n $ (\citep[defintion in Section 3.3.1]{goldmann_uniform_1990}). In the setting of  $\mathcal{A}_{\alpha, \Theta}$, the Banach algebras $\mathcal{A}_n$ are a generalization of the Banach algebras of $k$-times continuously differentiable functions over $\mathbb{T}$ (\citep{4404313}). In \citep[Section 3.3.3]{prodan_bulk_2016} $\mathcal{A}_{\alpha, \Theta}$ is explicitly described as the intersection of those Banach $\left(D_k^*\right)$ algebras.
\end{remark}

\subsection{Equivalent descriptions of $A \rtimes_{\alpha,\Theta}\mathbb{Z}^d$}\label{section:equivalent_descriptions_of_twisted_crossed_product}

So far we have given four equivalent descriptions of $A \rtimes_{\alpha,\Theta}\mathbb{Z}^d$,
\begin{enumerate}
    \item $A \rtimes_{\alpha,\Theta}\mathbb{Z}^d \simeq C^*(L^1(\mathbb{Z}^d,A ; \Theta,\alpha))$, with these we mean that it is a twisted crossed product in the sense of \citep{packer_twisted_1989} (\cref{definition:twsited_crossed_product}).
    \item $C^*(\mathcal{P}(\mathbb{Z}^d,A, \alpha,\Theta)) \simeq A \rtimes_{\alpha,\Theta}\mathbb{Z}^d$ (\cref{lemma:twisted_crossed_products_with_Z})
    \item $A \rtimes_{\alpha,\Theta}\mathbb{Z}^d \simeq C^* (\mathcal{G}| \mathcal{R})$, where the generators $\mathcal{G}$ and relations $\mathcal{R}$ are described in (\cref{lemma:twisted_crossed_products_with_Z}).
    \item $C^*(\mathcal{A}_{\alpha, \Theta}) \simeq A \rtimes_{\alpha,\Theta}\mathbb{Z}^d$, that is, we can use the smooth sub algebra to define $A \rtimes_{\alpha,\Theta}\mathbb{Z}^d$ (\cref{corollary:smooth_elements_are_invariant_under_holomoprhic_and_c_infty_calculus}).
\end{enumerate}

Our starting point was the first definition, and we piggyback on various results on twisted crossed products to describe faithful representations of $A \rtimes_{\alpha,\Theta}\mathbb{Z}^d$, which led us to find a description of any element of $A \rtimes_{\alpha,\Theta}\mathbb{Z}^d$ as a countable set of elements of $A$, such that, the aforementioned description resembles the Fourier coefficients of continuous functions over $\mathbb{T}$. Once we had a set of generalized Fourier coefficients, we were able to use them to establish that last equivalence.

These results have been known by the community of non-commutative geometry for more than thirty years, however, it took us a while to understand them and write them in this thesis. In the literature, the C* algebra $A \rtimes_{\alpha,\Theta}\mathbb{Z}^d$ is presented in all of the previous ways, such that, each author chooses the description that better suits its purpose, and is very common to see authors jumping from one description to another. For example, \citep[Chapter 3]{prodan_bulk_2016} favors the description in terms of generators, while \citep[Definition 12.7]{gracia-bondia_elements_2001} favors the description in terms of smooth sub algebras.

Also, you may have noticed that once we had the results on twisted crossed products from \citep{packer_twisted_1989} and \citep{bedos_fourier_2015}, all we did was to take ideas from classical Fourier analysis over $\mathbb{T}$ and translate them into the context of $A \rtimes_{\alpha,\Theta}\mathbb{Z}^d$. Such translation is easily navigated by experts on the field but may be challenging for newcomers, and we believe that it may be useful to write these computations down to improve the reproducibility of the results presented in \citep{prodan_bulk_2016} and many other books that analyze twisted crossed products.

\section{Twisted transformation group C* algebras}
\label{sec:twsited_transformation_group_C_star_algebras}

In this section we will look into separable twisted dynamical systems $(A,G, \alpha, \zeta)$ where $A$ is a commutative C* algebra, by the Gelfand representation theorem (\cref{theorem:gelfand_representation_theorem}) this means that we will focus on the case $A = C_0(\Omega)$ with $\Omega$ a locally compact Hausdorff space, and the action\index{action of a group} of $G$ over $A$ comes from an action of $G$ over $\Omega$. Additionally, since we want $A$ to be separable we will assume that $\Omega$ is a second countable topological space, which implies that $C_0(\Omega)$ is a separable C* algebra (\cref{lemma:separability_functions_dewcaying_at_infinity}). These types of separable twisted dynamical systems are important for us because one of two components of the Non-Commutative Brillouin torus is a twisted crossed product associated to one of those separable twisted dynamical systems (\cref{sec:topological_algebras_for_disordered_crystals}).

\subsection{Introduction to Twisted transformation group C* algebras}
\label{sec:intro_twsited_transformation_group_C_star_algebras}

Let $\Omega$ be a locally compact Hausdorff space, then, we will denote $\text{Hom}(\Omega)$ the set of homeomorphisms of $\Omega$, recall that a function $f: \Omega \to \Omega$ is called a homeomorphism if it is a continuous bijection and has a continuous inverse.

\begin{definition}[Twisted transformation group C* algebras \index{Twisted transformation group C* algebra}, taken from Transformation group C* algebras: A selective survey \citep{doran_c-algebras_1994}]\label{definition:twisted_transformation_group_C_star_algebras}
Let $(A,G, \alpha, \zeta)$ be a separable twisted dynamical system (\cref{def:separable_twisted_dynamical_system}), assume that, $A=C_0(\Omega)$ with $\Omega$ a locally compact and second countable Hausdorff space, additionally, assume that,
\begin{itemize}
    \item $(\Omega, G, \varrho)$ is a topological dynamical system\index{topological dynamical system} (\citep[Section II.10]{blackadar_operator_2006}): there is a continuous action of $G$ over $\Omega$, which means that, there is a map $\varrho: G \to \text{Hom}(\Omega)$ such that for all $g, g' \in G$ we have that $\varrho(g) \cdot \varrho(g') = \varrho(g +  g')$, and the corresponding map from $\Omega \times G$ to $\Omega$ is continuous.
    \item The action of $G$ over $C_0(\Omega)$ takes the form,
    $$ \alpha : G \to \text{Aut}(C_0(\Omega)), \; \alpha(g)(f)(\omega):= (f \circ \varrho(-g) )(\omega), \; g \in G, \; \omega \in \Omega. $$
\end{itemize}
We will refer to the C* algebra 
$$ C_0(\Omega)\rtimes_{\alpha, \zeta} G $$
as a twisted transformation group C* algebra.
\end{definition}

\begin{remark}[Hilbert spaces where $C_0(\Omega)\rtimes_{\alpha, \zeta} G$ has a faithful representation]\label{remark:hilebrt_spaces_where_twsited_trans_group_C_star_alg_have_faithful_repre}
In \cref{definition:twisted_transformation_group_C_star_algebras} we have taken $\Omega$ second countable because that guaranties that $C_0(\Omega)$ is separable (\cref{lemma:separability_functions_dewcaying_at_infinity}). The C* algebra $C_0( \Omega)$ has a faithful representation on the Hilbert space $\sum_{\omega \in \Omega} \mathbb{C}$ (direct sum of Hilbert spaces) (\cref{sec:Continuous_functions_with_values_on_C_star_algebra}) given by multiplication operators i.e. if $h \in C_0(\Omega)$ and $h \in \sum_{\omega \in \Omega} \mathbb{C}$ then $f \cdot h (\omega) = f(\omega) h(\omega)$. Thus, from \cref{theorem:norm_of_twisted_crossed_products} we know that $C_0(\Omega) \rtimes_{\alpha, \Theta} G$ has a faithful representation on 
$$ L^2(G, \sum_{\omega \in \Omega} \mathbb{C}) \simeq L^2(G) \otimes ( \sum_{\omega \in \Omega} \mathbb{C}) \simeq  \sum_{\omega \in \Omega} L^2(G). $$
Moreover, from \cref{section:algebra_continuous_functions_locally_comp_space} if $\Omega$ has a Radon measure (\cref{definition:Radon_measure}) with full support (\cref{definition:measure_with_full_support}) we have that $C_0(\Omega)$ has a faithful representation on $L^2(\Omega)$ given by multiplication operators, hence, \cref{theorem:norm_of_twisted_crossed_products} implies that $C_0(\Omega) \rtimes_{\alpha, \Theta} G$ has a faithful representation over 
$$ L^2(G,L^2(\Omega)) \simeq L^2(G) \otimes L^2(\Omega) \simeq L^2(\Omega, L^2(G)).  $$
\end{remark}

We will focus on twisted transformation group C* algebras with $\mathbb{Z}^d$, in this case, the normalized 2 cocycle $\zeta$ takes the form (\cref{proposition:characterization_of_+cocycles_over_integers})
$$ \zeta(x,y) = \exp(i x^t \Theta y).$$
Recall that $C_0(\Omega) \rtimes_{\alpha, \Theta} \mathbb{Z}^d$ can be characterized as a C* algebra generated by elements of the form $f u_j$ with $f \in C_0(\Omega)$ which follow among others the commutation relations (\cref{lemma:twisted_crossed_products_with_Z}),
\begin{itemize}
    \item $ \alpha(e_j)(a) u_j = u_j a$.
    \item $u_j u_l = \exp \left( i\hat{\Theta}_{j,l} \right) u_l u_j$.
\end{itemize}
Hence, if we provide a set of bounded linear operators $\{ F_fU_j \}_{f \in C_0(\Omega), j \leq d}$ that satisfy these commutation relations, the universal property of enveloping C* algebras (\cref{proposition:factoring_representations_with_enveloping_C_star_algebras}) imply that there must be a C* homomorphism that maps $C_0(\Omega) \rtimes_{\alpha, \Theta} \mathbb{Z}^d$ into the C* algebra generated by the bounded linear operators 
$$\{ F_fU_j \}_{f \in C_0(\Omega), j \leq d}.$$ 
The following lemma is closely related to \citep[Proposition 3.1.2]{prodan_bulk_2016},

\begin{lemma}[Representation on $L^2(\mathbb{Z}^d)$]\label{lemma:representation_on_L2_Z}
Let $C_0(\Omega) \rtimes_{\alpha, \Theta} \mathbb{Z}^d$ be a twisted transformation group C* algebra, then, given $\omega \in \Omega$, there is a unique representation $\pi_{\omega}$ of $C_0(\Omega) \rtimes_{\alpha, \Theta} \mathbb{Z}^d$ over $L^2(\mathbb{Z}^d)$ that takes the following form on the generators of $C_0(\Omega) \rtimes_{\alpha, \Theta} \mathbb{Z}^d$:
$$ \pi_{\omega}(u_j) \ket{x} := \exp(i e_j^t \Theta x) \ket{x + e_j} \text{ with } x \in \mathbb{Z}^d, $$
$$ \pi_{\omega}(f) := \sum_{y \in \mathbb{Z}^d} \alpha(-y)(f)(\omega)  \ket{y}\bra{y} \text{ i.e. } \pi_{\omega}(f)\ket{x} = \alpha(-x)(f)(\omega)\ket{x}, $$
with  $x \in \mathbb{Z}^d$ and $f \in C_0(\Omega)$.
\end{lemma}
\begin{proof}
The representation will take as domain the C* algebra $C_0(\Omega) \rtimes_{\alpha, \Theta} \mathbb{Z}^d$ and as range the sub C* algebra of $B(L^2(\mathbb{Z}^d))$ generated by the bounded linear operators $\pi_{\omega}(f) \pi_{\omega}(u_j) $ with $f \in C_0(\Omega)$ and $1 \leq j \leq d$. For such C* homomorphism to exists we need to show that the operators $\pi_{\omega}(f) \pi_{\omega}(u_j) $ satisfy the commutation relations of the generators of $C_0(\Omega) \rtimes_{\alpha, \Theta} \mathbb{Z}^d$  (\cref{lemma:twisted_crossed_products_with_Z}). First of all, using the description of $\pi_{\omega}(u_j)$ and $\pi_{\omega}(f)$ it is easy to check that those are bounded linear operators over $L^2(\mathbb{Z}^d)$, therefore, we proceed to check that $\pi_{\omega}(f) \pi_{\omega}(u_j) $ satisfy the commutation relations of the generators of $\mathcal{P}( \mathbb{Z}^d, C_0(\Omega), \alpha, \zeta)$(\cref{def:algebra_generalized_trigonometric_polynomials}):
\begin{itemize}
    \item $\pi_{\omega}(u_k) \pi_{\omega}(u_j) = \exp(i e_j^t \hat{\Theta} e_k)  \pi_{\omega}(u_j) \pi_{\omega}(u_k)$: \\
    Take $x \in \mathbb{Z}^d$, then
    $$  \pi_{\omega}(u_j) \pi_{\omega}(u_k) \ket{x} =  \pi_{\omega}(u_j) \exp(i e_k^t \Theta x) \ket{x + e_k} $$
    $$= \exp(i (e_k^t \Theta x) + i(e_j^t \Theta (x+e_k)) ) \ket{x + e_k + e_j}$$
    and
    $$  \pi_{\omega}(u_k) \pi_{\omega}(u_j) \ket{x} =  \pi_{\omega}(u_k) \exp(i e_j^t \Theta x) \ket{x + e_j} $$
    $$= \exp(i (e_j^t \Theta x) + i(e_k^t \Theta (x+e_j)) ) \ket{x + e_k + e_j},$$
    therefore,
    $$  \pi_{\omega}(u_k) \pi_{\omega}(u_j) \ket{x} = \exp(i (e_k^t \Theta e_j) - i(e_j^t \Theta e_k))  \pi_{\omega}(u_j) \pi_{\omega}(u_k) \ket{x} .$$
    Since $\Theta$ is a lower triangular matrix, if $k \geq j$ then
    $$ \pi_{\omega}(u_k) \pi_{\omega}(u_j) \ket{x} = \exp(i e_k^t \Theta e_j)  \pi_{\omega}(u_j) \pi_{\omega}(u_k) \ket{x}, $$
    if $j \geq k$ then
    $$ \pi_{\omega}(u_k) \pi_{\omega}(u_j) \ket{x} = \exp(-i e_j^t \Theta e_k)  \pi_{\omega}(u_j) \pi_{\omega}(u_k) \ket{x}, $$
    so, following the notation for the commutation relations for the generators of  $C_0(\Omega) \rtimes_{\alpha, \Theta} \mathbb{Z}^d$ (\cref{lemma:twisted_crossed_products_with_Z}), we have that
    $$ \pi_{\omega}(u_k) \pi_{\omega}(u_j) \ket{x} = \exp(i e_j^t \hat{\Theta} e_k)  \pi_{\omega}(u_j) \pi_{\omega}(u_k) \ket{x}, $$
    with $\hat{\Theta} = \Theta - \Theta^t$. Given that $\{ \ket{x} \}_{x \in \mathbb{Z}^d}$ is an orthonormal basis for $L^2(\mathbb{Z}^d)$, we have that 
    $$ \pi_{\omega}(u_k) \pi_{\omega}(u_j) = \exp(i e_j^t \hat{\Theta} e_k)  \pi_{\omega}(u_j) \pi_{\omega}(u_k), $$
    additionally, $\pi_{\omega}(u_k)$ are bounded linear operators with 
    $$\pi_{\omega}(u_k) (\pi_{\omega}(u_k))^* = (\pi_{\omega}(u_k))^*\pi_{\omega}(u_k) = 1_{L^2(\mathbb{Z}^d)}.$$
    
    \item $ \pi_{\omega}(u_k) \pi_{\omega}(f) = \pi_{\omega}( \alpha(e_j) (f)) \pi_{\omega}(u_k) $:\\
    Let $x \in \mathbb{Z}^d$, then
    $$ \pi_{\omega}(f) \pi_{\omega}(u_k) \ket{x} = \pi_{\omega}(f) \exp(i e_k^t \Theta x) \ket{x + e_k} $$
    $$  \exp(i e_k^t \Theta x) \alpha(-x-e_k)(f)(\omega) \ket{x + e_k},  $$
    and 
    $$  \pi_{\omega}(u_k) \pi_{\omega}(f) \ket{x} = \pi_{\omega}(u_k) \alpha(-x)(f)(\omega) \ket{x} $$
    $$  \exp(i e_k^t \Theta x) \alpha(-x)(f)(\omega) \ket{x+e_k},  $$
    therefore,
    $$ \pi_{\omega}(u_k) \pi_{\omega}(f) \ket{x} = \exp(i e_k^t \Theta x) \alpha(-x-e_j)( \alpha(e_j)(f))(\omega) \ket{x+e_k} $$
    $$ \pi_{\omega}( \alpha(e_j) (f)) \pi_{\omega}(u_k) \ket{x}. $$
    Given that $\{ \ket{x} \}_{x \in \mathbb{Z}^d}$ is an orthonormal basis for $L^2(\mathbb{Z}^d)$, we have that 
    $$ \pi_{\omega}(u_k) \pi_{\omega}(f) = \pi_{\omega}( \alpha(e_j) (f)) \pi_{\omega}(u_k), $$
    additionally, $\pi_{\omega}(f)$ are bounded linear operators with
    $$ \pi_{\omega}(f + g) = \pi_{\omega}(f) + \pi_{\omega}(g), \; \pi_{\omega}(f^*) = (\pi_{\omega}(f))^*.  $$
    \item The relations $ \pi_{\omega}(u_k)^* \pi_{\omega}(f) = \pi_{\omega}( \alpha(-e_j) (f)) \pi_{\omega}(u_k)^* $ and $\pi_{\omega}(u_k)^* \pi_{\omega}(u_j) = \exp(-i e_j^t \hat{\Theta} e_k)  \pi_{\omega}(u_j) \pi_{\omega}(u_k)^*$ can be checked using computations similar to the ones in the previous items. 
\end{itemize}
The previous computations imply that the set 
$$\{ \pi_{\omega}(f) \pi_{\omega}(u_j) \}_{f \in C_0(\Omega), j \leq d}$$
satisfy the commutation relations of the generators of $C_0(\Omega) \rtimes_{\alpha, \Theta} \mathbb{Z}^d$, therefore, there is a C* homomorphism
$$ \pi_{\omega} : C_0(\Omega) \rtimes_{\alpha, \Theta} \mathbb{Z}^d \to A_{\omega}$$
where $A_{\omega}$ is the sub C* algebra of $B(L^2(\mathbb{Z}^d))$ generated by $\{ \pi_{\omega}(f) \pi_{\omega}(u_j) \}_{f \in C_0(\Omega), j \leq d}$. In this case, $\pi_{\omega}$ is a representation of $C_0(\Omega) \rtimes_{\alpha, \Theta} \mathbb{Z}^d$, and the automatic continuity of C* homomorphisms tell us that (\cref{proposition:automatic_continuity_C_star_algebras})
$$ \| \pi_{\omega}(p) \| \leq \|p\|, \; \forall p \in C_0(\Omega) \rtimes_{\alpha, \Theta} \mathbb{Z}^d. $$
\end{proof}


When $\Omega$ is compact, in the following lemma we will use the family of representations $\{ \pi_{\omega} \}_{\omega \in \Omega}$ to show that any $p \in C(\Omega) \rtimes_{\alpha, \Theta} \mathbb{Z}^d$ can be seen as a continuous function from $\Omega$ into $B(L^2(\mathbb{Z}^d))$. For the following lemma we need to take into account the following facts,
\begin{itemize}
    \item If $\Omega$ is compact, then $\Omega$ is a metric space (\cref{lemma:second_coutnalbe_locally_compact_hausdorff_space}), so, we denote by $d: \Omega \times \Omega \to \mathbb{R}^+$ a metric over $\Omega$ that generates the topology of $\Omega$.
    \item We denote by $L^2(\Omega, L^2(\mathbb{Z}^d))$ the set of equivalence classes of strongly $\mu$-measurable functions from $\Omega$ into $B(L^2(\mathbb{Z}^d))$ such that (\cref{def:Bochner_Lp_spaces})
    $$ \int_{\Omega} \| f(\omega)\|^2 d \mu(\omega) < \infty ,$$
    additionally, recall a strongly $\mu$-measurable function is a function from $\Omega$ into $B(L^2(\mathbb{Z}^d))$ such that there exists a sequence $\{ f_n\}_{n \in \mathbb{N}}$ of $\mu$-simple functions (\cref{def:mu_simple_func}) converging to $f$ almost everywhere (\cref{def:mu_strong_measu}).
\end{itemize}

\begin{definition}[Isometric action on a Twisted transformation group C* algebras]\label{definition:isometric_action_twisted_transf_group_C_star_algebras}
Let $C(\Omega) \rtimes_{\alpha, \Theta} \mathbb{Z}^d$ be a twisted transformation group C* algebra such that $\Omega$ is compact. If for any $x \in \mathbb{Z}^d$, $\omega_i \in \Omega, \; i = 1,2$ we have that the action $\varrho$ of $\mathbb{Z}^d$ over $\Omega$ that defines the transformation group C* algebra satisfies the relation 
$$ d(\omega_1, \omega_2) = d(  \varrho(x)(\omega_1), \varrho(x)(\omega_2) ),$$
then, we say that $\alpha$ is an isometric action\index{isometric action}.
\end{definition}


\begin{lemma}[Elements of $C(\Omega) \rtimes_{\alpha, \Theta} \mathbb{Z}^d$ as continuous functions of operators]\label{lemma:elements_of_twisted_transformation_groups_as_continuous_func}
Let $C(\Omega) \rtimes_{\alpha, \Theta} \mathbb{Z}^d$ be a twisted transformation group C* algebra such that $\Omega$ is compact, assume that there is a measure $\mu$ over $\Omega$ and $\Omega$ has finite measure i.e. $\mu(\Omega) < \infty$. If the $\alpha$ is an isometric action (\cref{definition:isometric_action_twisted_transf_group_C_star_algebras}), then,

\begin{enumerate}
    \item Given $f \in C(\Omega)$, then, the map $\omega \mapsto \pi_{\omega} (f)$ is continuous with respect to the norm topology on $B(L^2(\mathbb{Z}^d))$
    \item Let
    $$ \pi_1 : C(\Omega) \to B(L^2(\Omega, L^2(\mathbb{Z}^d))),$$
    by defined by 
    $$ \pi_1(g)(h)(\omega) := \pi_{\omega}(g)(h(\omega)),  $$
    where $h \in L^2(\Omega, L^2(\mathbb{Z}^d))$ and $g \in C(\Omega)$.
    Then, $\pi_1$ is a faithful representation of $C(\Omega)$, where $\pi_{\omega}$ is defined in \cref{lemma:representation_on_L2_Z}.
    \item Given $p \in C(\Omega) \rtimes_{\alpha, \Theta} \mathbb{Z}^d$, then, the map
    $$ \omega \mapsto \pi_{\omega} (p) $$
    is continuous.
    \item Given $p \in C(\Omega) \rtimes_{\alpha, \Theta} \mathbb{Z}^d$, then, the following defines a bounded linear operator over $L^2(\Omega, L^2(\mathbb{Z}^d))$,
    $$ \pi_1(p)(h)(\omega) := \pi_{\omega}(p)(h(\omega)), \; h \in L^2(\Omega, L^2(\mathbb{Z}^d)). $$
    Additionally, $\pi_1$ is a representation of $C(\Omega) \rtimes_{\alpha, \Theta} \mathbb{Z}^d$ over $L^2(\Omega, L^2(\mathbb{Z}^d))$. 
\end{enumerate}
\end{lemma}
\begin{proof}
\begin{enumerate}
    \item Take $f \in C(\Omega)$, since $f$ is a continuous function over a compact metric space, it is uniformly continuous \cite{nlab:continuous_metric_space_valued_function_on_compact_metric_space_is_uniformly_continuous} i.e. given $\epsilon > 0$ there is $\delta >0$ such that if $d(\omega_0, \omega_1) \leq \delta$ then
    $$ | f(\omega_0) - f(\omega_1) | \leq \epsilon, \text{ when } \omega_0, \omega_1 \in \Omega, $$
    additionally, by out assumption this implies/ that,
    $$ | f(\varrho(y)(\omega_0)) - f(\varrho(y)(\omega_1)) | \leq \epsilon  $$
    for all $y \in \mathbb{Z}^d$.
    By the definition of $\pi_{\omega}$ we have that
    $$\| \pi_{\omega}(f) \| \leq \sup_{s \in \mathbb{Z}^d} | f(\varrho(s)(\omega))|, $$
    therefore, 
    $$ \| \pi_{\omega}(f-g) \| \leq \sup_{s \in \mathbb{Z}^d} | f(\varrho(s)(\omega)) - g(\varrho(s)(\omega))|, $$
    so, if $d(\omega_0, \omega_1) \leq \delta$, then,
    $$\| \pi_{\omega}(f-g) \| \leq \epsilon,$$
    which in turns imply that the map
    $$ \omega \mapsto \pi_{\omega}(f)  $$
    is continuous.
    \item First we need to show that given $h \in L^2(\Omega, L^2(\mathbb{Z}^d))$ then the map $ \omega \mapsto \pi_{\omega}(g)(h(\omega))$ is an element of $L^2(\Omega, L^2(\mathbb{Z}^d))$, for this we need to show that the aforementioned map satisfies the following two criteria:
    \begin{itemize}
        \item It is $\mu$-strongly measurable.
        \item We have that $\int_{\Omega} \|\pi_{\omega}(g)(h(\omega))\|^2 d \mu \leq \infty$.
    \end{itemize}
    \cref{lemma:representation_on_L2_Z} establishes that $\pi_{\omega}$ is a representation of $C(\Omega)$, therefore,  $\pi_{\omega}(g)(h(\omega))$ is an element of $L^2(\mathbb{Z}^d)$, therefore, the map $ \omega \mapsto \pi_{\omega}(g)(h(\omega))$ goes from $\Omega$ into $L^2(\mathbb{Z}^d)$. By the previous item we know that the map $\omega \mapsto \pi_{\omega}(g)$ is continuous, so, it is also a $\mu$-strongly measurable map, that is, there is a sequence of functions $\{g_i\}_{i \in \mathbb{N}}$ converging almost everywhere to the map $\omega \mapsto \pi_{\omega}$ and each $g_i$ is a sequence of $\mu$-simple functions (\cref{example:Bochner_integral_continuos_funcion}). By assumption $h$ is an element of $L^2(\Omega, L^2(\mathbb{Z}^d))$, so, it can be approximated almost everywhere by a sequence of $\mu$-simple functions over $\Omega$ and taking values over $L^2(\mathbb{Z}^d)$, call that sequence $\{ f_i\}_{i \in \mathbb{N}}$. Given $f_i$ and $g_i$ denote by $g_i f_i$ the function that maps $\omega$ into $g_i(\omega)(f_i(\omega))$ for $\omega \in \Omega$, then, $g_i f_i$ is also a sum of $\mu$-simple functions by construction, so, since $g_i$ converges to $ \omega \mapsto \pi_{\omega}(g)$ almost everywhere and $f_i$ converges to $h$ almost everywhere we have that $g_if_i$ converges to $\pi_{\omega}(g)(h(\omega))$ almost everywhere, consequently, it is a $\mu$-strongly measurable map. 
    
    From the argument on the previous item we have that $\| \pi_{\omega}(g) \| \leq \sup_{\omega \in \Omega} |g(\omega)|$, hence, 
    $$\int_{\Omega} \|\pi_{\omega}(g)(h(\omega))\|^2 d \mu \leq (\sup_{\omega \in \Omega} |g(\omega)|)^2 \int_{\Omega} \|h(\omega)\|^2,$$
    so, given that we assumed that $\int_{\Omega} \|h(\omega)\|^2 < \infty$ we end up with
    $$\int_{\Omega} \|\pi_{\omega}(g)(h(\omega))\|^2 d \mu \leq \infty.$$
    
    It remains to prove that $\pi_1(g)$ is a bounded linear operator, because $\pi_1(g)$ is a linear operator by definition. Since $\Omega$ is compact, then, for $g \in C(\Omega)$, $|g|$ must attain its maximum valued at some $\omega \in \Omega$, lets us call it $\omega_0$ i.e.
    $$ \forall \omega \in \Omega, \; |g(\omega)| \leq |g(\omega_0)|.  $$
    By construction we have that
    $\| \pi_{\omega_0}(g) \| \geq |g(\omega_0)|,$
    also, since $\| \pi_{\omega_0}(g) \| \geq \sup_{\omega \in \Omega}|g(\omega)|$, we have that
    $$ \| \pi_{\omega_0}(g) \| = |g(\omega_0)|. $$
    Then, 
    $$\| f \| = \| \pi_{\omega_0}(f) \|,$$
    which implies that $\pi_1$ is faithful.
    \item Given $p \in C(\Omega) \rtimes_{\alpha, \Theta} \mathbb{Z}^d$ take $q \in \mathcal{P}(\mathbb{Z}^d,C(\Omega), \alpha,\Theta)$(\cref{def:algebra_generalized_trigonometric_polynomials}) such that 
    $$ \| q - p \| \leq \epsilon / 3, $$
    then, since C* homomorphisms are norm decreasing (\cref{proposition:automatic_continuity_C_star_algebras}), we have that
    $$ \forall \omega \in \Omega, \; \| \pi_{\omega}(q) - \pi_{\omega}(p) \| \leq \epsilon/3.  $$
    As an application of the triangle inequality we get
    $$ \| \pi_{\omega_0}(p) - \pi_{\omega_1}(p) \| \leq \| \pi_{\omega_0}(p) - \pi_{\omega_0}(q) \| + $$
    $$\| \pi_{\omega_0}(q) - \pi_{\omega_1}(q) \| + \| \pi_{\omega_1}(q) - \pi_{\omega_1}(p) \|, $$
    so, the first and third terms are bounded above by $\epsilon/3$, thus, to show that the map $ \omega \mapsto \pi_{\omega}(p) $ is continuous we need to show find $\delta$ such that, if $d(\omega_0, \omega_1) \leq \delta$ then 
    $$\| \pi_{\omega_0}(q) - \pi_{\omega_1}(q) \|.$$
    Since $\pi_{\omega}(u^s)$ does not depend on $\omega$ and $\pi_{\omega}(u^s)$ is an unitary operator we have that,
    $$ \| \pi_{\omega}(fu^s) - \pi_{\omega_0}(fu^s) \| = \| (\pi_{\omega}(f) - \pi_{\omega_0}(f)) \pi_{\omega}(u^s) \| \leq \| \pi_{\omega}(f) - \pi_{\omega_0}(f) \|,$$
    therefore, the map
    $$ \omega \mapsto \pi_{\omega}(f u^s) $$
    is continuous for any $s \in \mathbb{Z}^d$. Since $q$ is a finite sum of terms of the form $f u^s$ with $f \in C(\Omega)$ (\cref{def:algebra_generalized_trigonometric_polynomials}), we have the map
    $$ \omega \mapsto \pi_{\omega}(q) $$
    is also continuous, which tells us that the aforementioned $\delta$ exists.    
    \item Given that the map $\omega \mapsto \pi_{\omega}(p)$ is continuous, an argument similar to the one in the second item implies that, if $h \in L^2(\Omega, L^2(\mathbb{Z}^d))$ then the map $\omega \mapsto \pi_{\omega}(p)(h(\omega))$ is strongly $\mu$-measurable. Also, given that for every $\omega \in \Omega$ the map $\pi_{\omega}$ is a representation of $C(\Omega) \rtimes_{\alpha, \Theta} \mathbb{Z}^d$ (\cref{lemma:representation_on_L2_Z}), then, the automatic continuity of C* homomorphisms (\cref{proposition:automatic_continuity_C_star_algebras}) implies that $\| \pi_{\omega}(p) \| \leq \| p \|$, therefore, 
    $$ \int_{\Omega} \| \pi_{\omega}(p)(h (\omega)) d \mu \|^2 \leq \| p \|^2 \int_{\Omega} \| h (\omega) \|^2, $$
    hence, the map $\omega \mapsto \pi_{\omega}(p)(h(\omega))$ is an element of $L^2(\Omega, L^2(\mathbb{Z}^d))$. 
    
    Since $\pi_{\omega}$ is a representation of $C(\Omega) \rtimes_{\alpha, \Theta} \mathbb{Z}^d$ and $\pi_1(p)$ is an element of $B(L^2(\Omega, L^2(\mathbb{Z}^d)))$, then the map $\omega \mapsto \pi_1(p)$ is also a representation of $C(\Omega) \rtimes_{\alpha, \Theta} \mathbb{Z}^d$.
\end{enumerate}
\end{proof}

If $\Omega$ is compact, $\alpha$ is an isometric action (\cref{definition:isometric_action_twisted_transf_group_C_star_algebras}) and there is a measure $\mu$ over $\Omega$ such that $\mu(\Omega) < \infty$, \cref{lemma:elements_of_twisted_transformation_groups_as_continuous_func} provides a representation of $C(\Omega) \rtimes_{\alpha, \Theta} \mathbb{Z}^d$ over $L^2(\Omega, L^2(\mathbb{Z}^d))$ that is constructed by interpreting any $p \in C(\Omega) \rtimes_{\alpha, \Theta} \mathbb{Z}^d$ as a continuous function from $\Omega$ into $B(L^2(\mathbb{Z}^d))$. The representation of $C(\Omega) \rtimes_{\alpha, \Theta} \mathbb{Z}^d$ outlined in \cref{lemma:elements_of_twisted_transformation_groups_as_continuous_func} is inspired by the faithful representation of the elements of $C(\Omega, B(L^2(\mathbb{Z}^d)))$ over $L^2(\Omega, L^2(\mathbb{Z}^d))$  (\cref{lemma:faithful_representation_of_functions_decaying_with_values_on_algebra}), now, we proceed to check that if $\mu$ is a Radon measure (\cref{definition:Radon_measure}) with full support (\cref{definition:measure_with_full_support}), then, this representation is the same as left regular representation described in \cref{remark:reduced_crossed_products_and_left_regular_representation}

\begin{lemma}\label{lemma:left_regular_representation_of_C_0_Omega_rtimes_Z}
Let $C(\Omega) \rtimes_{\alpha, \Theta} \mathbb{Z}^d$ be a twisted transformation group C* algebra such that $\Omega$ is compact, assume that there is a Radon measure with full support $\mu$ over $\Omega$ and $\Omega$ has finite measure i.e. $\mu(\Omega) < \infty$, additionally, assume that $\alpha$ is an isometric action (\cref{definition:isometric_action_twisted_transf_group_C_star_algebras}).
Denote by 
$$\pi_l: C(\Omega) \rtimes_{\alpha, \Theta} \mathbb{Z}^d \to B(L^2(\mathbb{Z}^d, L^2(\Omega))) $$ 
the left regular representation of $C(\Omega) \rtimes_{\alpha, \Theta} \mathbb{Z}^d$ described in  (\cref{remark:reduced_crossed_products_and_left_regular_representation}), also, denote by
$$\pi_m : C(\Omega) \rtimes_{\alpha, \Theta} \mathbb{Z}^d \to B(L^2(\Omega, L^2(\mathbb{Z}^d)))$$
the representation of $C(\Omega) \rtimes_{\alpha, \Theta} \mathbb{Z}^d$ taking the following form,
$$ \pi_m(p)(h)(\omega) := \pi_{\omega}(p)(h(\omega)),$$
and described in the fourth item of \cref{lemma:elements_of_twisted_transformation_groups_as_continuous_func}, then, 
$$ \pi_l = \pi_m .$$
\end{lemma}
\begin{proof}
Taking into account the following,
\begin{itemize}
    \item From \cref{lemma:hilbert_space_and_L2_spaces_with_sigma_finite} we know that
    $$  L^2(\mathbb{Z}^d, L^2(\Omega)) \simeq L^2(\mathbb{Z}^d) \otimes L^2(\Omega) \simeq L^2(\Omega, L^2(\mathbb{Z}^d)),$$
    thus, using these canonical isomorphisms we can assume that both representations act on the same Hilbert space.
    \item From \cref{definition:tensor_product_hilbert_spaces} we know that $L^2(\mathbb{Z}^d) \odot L^2(\Omega)$ is dense in $L^2(\mathbb{Z}^d) \otimes L^2(\Omega)$, where $L^2(\mathbb{Z}^d) \odot L^2(\Omega)$ denotes the algebraic tensor product of $L^2(\mathbb{Z}^d)$ and $L^2(\Omega)$.
    \item In \cref{lemma:twisted_crossed_products_with_Z} we mentioned the the algebra of generalized trigonometric polynomials $ \mathcal{P}(C(\Omega),\mathbb{Z}^d; \alpha, \Theta )$ is dense in $C(\Omega) \rtimes_{\alpha,\Theta} \mathbb{Z}^d$.
    \item By \cref{lemma:faithful_representation_of_functions_decaying_at_infinity} $C(\Omega)$ has a faithful representation over $L^2(\Omega)$ given by the multiplication operators i.e. if $h \in C_0(\Omega)$ and $h \in \sum_{\omega \in \Omega} \mathbb{C}$ then $f \cdot h (\omega) = f(\omega) h(\omega)$. So, \cref{remark:reduced_crossed_products_and_left_regular_representation} tell us that $C(\Omega) \rtimes_{\alpha,\zeta} \mathbb{Z}^d$ has a faithful representation over $L^2(\mathbb{Z}^d, L^2(\Omega))$ given by the left regular representation.
\end{itemize}

If we show that 
$$\pi_l(f u^s)(h \otimes g) = \pi_m(f u^s)(h \otimes g)  $$
for any $h \in L^2(\mathbb{Z}^d), \; g \in L^2(\Omega)$, then, we would have that $\pi_l(f u^s) = \pi_m(f u^s)$ inside $L^2(\mathbb{Z}^d) \odot L^2(\Omega)$. Given that $L^2(\mathbb{Z}^d) \odot L^2(\Omega)$ is dense in $L^2(\mathbb{Z}^d) \otimes L^2(\Omega)$, the equality of $\pi_m(f u^s)$ and $\pi_l(f u^s)$ over $L^2(\mathbb{Z}^d) \odot L^2(\Omega)$ implies that $\pi_l(f u^s) = \pi_m(f u^s)$ in $L^2(\mathbb{Z}^d) \otimes L^2(\Omega)$.

Since a representation is a linear transformations, and $\mathcal{P}(\mathbb{Z}^d,C(\Omega), \alpha, \Theta )$ is linearly generated by elements of the form $f u^s$, if $\pi_l(f u^s)(h \otimes g) = \pi_m(f u^s)(h \otimes g)$, then, $\pi_l = \pi_m$ over $\mathcal{P}(\mathbb{Z}^d,C(\Omega), \alpha, \Theta )$. Given that $C(\Omega) \rtimes_{\alpha, \Theta} \mathbb{Z}^d$ is the enveloping C* algebra of $\mathcal{P}(\mathbb{Z}^d,C(\Omega), \alpha, \Theta )$, then, the universal property of C* algebras (\cref{proposition:factoring_representations_with_enveloping_C_star_algebras}) tells us that any representation of $C(\Omega) \rtimes_{\alpha, \Theta} \mathbb{Z}^d$ is uniquely determined by its value on $\mathcal{P}(\mathbb{Z}^d,C(\Omega), \alpha, \Theta )$, so, if $\pi_l = \pi_m$ over $\mathcal{P}(\mathbb{Z}^d,C(\Omega), \alpha, \Theta )$ then $\pi_l = \pi_m$ over $C(\Omega) \rtimes_{\alpha, \Theta} \mathbb{Z}^d$.

Since $\{ \ket{x} \}_{x \in \mathbb{Z}^d}$ is an orthonormal base for $L^2(\mathbb{Z}^d)$, in order to prove that $$\pi_l(f u^s)(h \otimes g) = \pi_m(f u^s)(h \otimes g)  $$
for any $h \in L^2(\mathbb{Z}^d), \; g \in L^2(\Omega)$, we only need to show that
$$\pi_l(f u^s)(h \otimes \ket{x}) = \pi_m(f u^s)(h \otimes \ket{x}), $$
where $h \in L^2(\Omega)$ and $h \otimes \ket{x}$ is the function in $L^2(\Omega, L^2(\mathbb{Z}^d))$ that maps $\omega$ into $h(\omega) \ket{x}$ (\cref{def:simple_func}). Under this setting, $\pi_m$ takes the form
$$ \pi_m(f u^s) (h \otimes \ket{x}) (\omega) = \pi_{\omega}(f u^s) (h(\omega) \ket{x}),  $$
which from \cref{lemma:representation_on_L2_Z} we get that results on
$$ \pi_m(f u^s) (h \otimes \ket{x}) (\omega) = \exp(i s^t \Theta x) \alpha(-x-s)(f)(\omega) h(\omega) \ket{x+s}. $$

Under the isomorphism $L^2(\mathbb{Z}^d) \otimes L^2(\Omega) \simeq L^2(\Omega, L^2(\mathbb{Z}^d))$, the map 
$$ \omega \mapsto  \exp(i s^t \Theta x) \alpha(-x-s)(f)(\omega) h(\omega) \ket{x+s} $$
corresponds to the tensor product
$$ \exp(i s^t \Theta x) \left( \pi(\alpha(-x-s)(f))(h)  \otimes \ket{x+s} \right),  $$
which is the result of applying $\pi_l(f u^s) $ to the vector $\ket{x} \otimes h$, thus, 
$$ \pi_l(f u^s)(\ket{x} \otimes h) = \pi_m(f u^s)(\ket{x} \otimes h). $$

Using the aforementioned equality along with the previous discussion we get that
$$  \pi_l = \pi_m.$$

\end{proof}

\begin{remark}[Faithful representation of $C(\Omega) \rtimes_{\alpha, \Theta} \mathbb{Z}^d$]\label{remark:faithful_representation_of_C_0_Omega_rtimes_Z}
Let $C(\Omega) \rtimes_{\alpha, \Theta} \mathbb{Z}^d$ be a twisted transformation group C* algebra such that $\Omega$ is compact, assume that there is a measure $\mu$ over $\Omega$ and $\Omega$ has finite measure i.e. $\mu(\Omega) < \infty$, additionally, assume that $\alpha$ is an isometric action (\cref{definition:isometric_action_twisted_transf_group_C_star_algebras}).In \cref{remark:reduced_crossed_products_and_left_regular_representation} it is mentioned that the left regular representation of $C(\Omega) \rtimes_{\alpha, \Theta} \mathbb{Z}^d$ is faithful, so, under the previous assumptions \cref{lemma:left_regular_representation_of_C_0_Omega_rtimes_Z} tells us that
$C(\Omega) \rtimes_{\alpha, \Theta} \mathbb{Z}^d$ is a sub C* algebra of $C(\Omega,B(L^2(\mathbb{Z}^d)))$.
\end{remark}

\subsection{Topological algebras for disordered crystals}
\label{sec:topological_algebras_for_disordered_crystals}

In the present section, we follow \citep[Section 2.4.1]{prodan_bulk_2016} for the construction of the C* algebra used for the tight binding models of homogeneous materials. The topological space $\Omega$ is termed the disorder space (\cref{sec:motivation_from_physics}), and is constructed as a countable product of compact and convex spaces, one for each position of the lattice $\mathbb{Z}^d$. These are the steps needed to construct $\Omega$:

\begin{enumerate}
    \item Let $\{ (E_s, \mu_s) \}_{s \in \mathbb{Z}^d}$ a set of tuples consisting of, $E_s$, a compact, convex (hence contractible) and second countable Hausdorff space, and $\mu_s$: a Radon probability measure (\cref{definition:Radon_measure}) over $E_s$ with full support (\cref{definition:measure_with_full_support}).
    \item Then, the Tychonoff's theorem\index{Tychonoff's theorem} (\citep[Theorem 1.29]{allan_introduction_2011}) tell us that $O = \prod_{s \in \mathbb{Z}^d} E_s $ is a compact space under the product topology, additionally, $O$ is a Hausdorff space (\citep[Proposition 1.28]{allan_introduction_2011}). Also, since each $E_s$ is second countable, and a base for the topology of $O$ is the set of all infinite cartesian products where only a finite amount of them is different from the whole space (\citep[Section 1.10]{allan_introduction_2011}), we have that $O$ is a second countable Hausdorff space. Additionally, $O$ is convex (hence contractible) because all of the $E_s$ are convex. 
    \item From \citep[Theorem 2.4.4]{tao_introduction_2011} we know that $d \mu := \prod_{s \in \mathbb{Z}} d \mu_s $ is a Radon measure defined in basic open sets by
    $$ \mu(\prod_{s \in \mathbb{Z}^d} U_s ) = \prod_{s | U_s \neq E_s} \mu_s (U_s), $$
    where $s \in \mathbb{Z}^d$ such that $ U_s \neq E_s$ for a finite amount of $s$. Additionally, given that a base for the topology of $\prod_{s \in \mathbb{Z}^d} E_s $ are infinite cartesian products with only a finite amount of them different from $E_s$, we can use the fact that each measure $\mu_s$ has full support to guarantee that $d \mu$ also has full support. 
\end{enumerate}

Since each $E_s$ is a compact and second countable space, each $E_s$ is a metric space (\cref{lemma:second_coutnalbe_locally_compact_hausdorff_space}), so, we denote by
$$ d_s : E_s \times E_s \to \mathbb{R}^+ $$
a metric that generates the topology of $E_s$.

\begin{lemma}[c.f. \citep{1326186} Metric over $\prod_{n \in \mathbb{N}} E_n$]\label{lemma:metric_over_countable_product_metric_spaces}
Let $\{ E_n \}_{n \in \mathbb{N}}$ be a sequence of Hausdorff compact spaces and denote by
$$ d_n : E_n \times E_n \to \mathbb{R}^+ $$
a metric that generates the topology of $E_n$, then, the topology over $O:=\prod_{n \in \mathbb{N}} E_n$ is the topology induced by the metric
$$ d_O : O \times O \to \mathbb{R}^+, \; d_O(x,y) = \sum_{n \in \mathbb{N}} 2^{-n} \frac{ d_{n}( x(n), y(n) ) }{ 1 + d_{n}( x(n), y(n) ) } $$
\end{lemma}
\begin{proof}
Given that $\sum_{n}2^{-n} < \infty$ and $\frac{ d_{n}( x(n), y(n) ) }{ 1 + d_{n}( x(n), y(n) ) } < 1$, then, $d_0$ is well defined, now, we check that it is a metric. Notice that the function $x \mapsto \frac{x}{1+x}$ is an increasing function for $x \neq -1$, so, since $d_{n}(x,z) \leq d(x,y) + d(y,z)$, if $x,y,z \in \prod_{n \in \mathbb{N}} E_n$ we  must have that
$$ \frac{ d_{n}( x(n), z(n) ) }{ 1 + d_{n}( x(n), z(n) ) } \leq \frac{ d_{n}( x(n), y(n) ) + d_{n}( y(n), z(n) ) }{ 1 + d_{n}( x(n), y(n) ) + d_{n}( y(n), z(n) ) } $$
$$\leq  \frac{ d_{n}( x(n), y(n) ) }{ 1 + d_{n}( x(n), y(n) ) } + \frac{ d_{n}( y(n), z(n) ) }{ 1 + d_{n}( y(n), z(n) ) }, $$
therefore, $d_O$ is a metric.
If $d(x_l , y) \to 0$ as $l \to \infty$, then, for any $n \in \mathbb{N}$ we must have that $d_n((x_l(n)), y(n)) \to \infty$, hence, if $\prod_{n \in \mathbb{N}} E_n$ is endowed with the topology generated by $d_O$, then, the projections $p_l : \prod_{n \in \mathbb{N}} E_n \to E_l, \; p(\{ x(n) \}_{n \in \mathbb{N}}) = x_l$ are continuous, meaning that the topology generated by $d_O$ is finer than the product topology over $\prod_{n \in \mathbb{N}} E_n$ (\citep[Product spaces, page 26]{allan_introduction_2011}) over $\prod_{n \in \mathbb{N}} E_n$ i.e. every open set in the product topology is an open set in the topology generate by $d_O$.

Now we check that every open set of the topology generated by $d_O$ is an open set of the product topology. Take $x \in \prod_{n \in \mathbb{N}} E_n$ and $\epsilon> 0$, let
$$ B_{\epsilon}(x) = \{ y \in \prod_{n \in \mathbb{N}} E_n | d_O(x,y) < \epsilon \}, $$
then, we need to show that there is a neighbourhood $U$ of $x$ under the product topology such that $U \subset B_{\epsilon}(x)$. Take $n_0 \in \mathbb{N}$ such that $\sum_{l \geq n_0} 2^{-l} < \epsilon/2$, let $\eta = \epsilon/4$ and set $B_{\eta}^{d_n}(x(n)) = \{ h \in E_n | d_n(h,x(n)) \leq \eta \}$, then, if we define 
$$ U:= \prod_{n = 0}^{n_0} B_{\eta}^{d_n}(x(n)) \times \prod_{n > n_0} E_n,$$
we have that $U$ is open neighborhood of $x$ in the product topology of $\prod_{n \in \mathbb{N}} E_n$ and $U \subset B_{\epsilon}(x)$.
\end{proof}

Given $\{ (E_s, \mu_s) \}_{s \in \mathbb{Z}^d}$ denote by $O$ the compact topological space $\prod_{s \in \mathbb{Z}^d} E_s$. Let $f: \mathbb{N} \to \mathbb{Z}^d$ be a bijective function, that is, $f$ is an enumeration of $\mathbb{Z}^d$, then, \cref{lemma:metric_over_countable_product_metric_spaces} tell us that the metric 
$$ d_O : O \times O \to \mathbb{R}^+, \; d_O(x,y) = \sum_{l \in \mathbb{N}} 2^{-l} \frac{ d_{f(l)}( x(f(l)), y(f(l)) ) }{ 1 + d_{f(l)}( x(f(l)), y(f(l)) ) } $$
generates the product topology over $O$, where, $x = \left( x_s \right)_{s \in \mathbb{Z}^d}, \; y = \left( y_s \right)_{s \in \mathbb{Z}^d} \in O$.

In the context of tight-binding models of homogeneous materials (\cref{sec:motivation_from_physics}), the space $E_l$ parametrizes the disorder assigned to applying the operator (\cref{lemma:representation_on_L2_Z})
$$U^l \in B(L^2(\mathbb{Z}^d)), \; U^l := S^l \exp(i l^t \Theta X), $$
with
$$ S^l \ket{x} = \ket{x + l} , \;  \exp(i l^t \Theta X) \ket{x} = \exp(i l^t \Theta x) \ket{x}.$$

In homogeneous materials (\citep[Chapter 2]{prodan_bulk_2016}), the disorder associated to applying the operator $U^l$ to the vectors $\ket{x}$ and $\ket{y}$ is the same for any $x,y \in \mathbb{Z}^d$, so, to take this into account we use a copy of $O$ for each one of the positions of the lattice $\mathbb{Z}^d$, which will lead us to use $\prod_{s \in \mathbb{Z}^d} O$ as our topological space of interest. Using similar arguments as the one exposed for the construction of $O$ is possible to check that   
\begin{itemize}
    \item $\Omega = \prod_{s \in \mathbb{Z}^d} O$ is a compact second countable and convex (hence contractible) Hausdorff space,
    \item $d \mathbb{P} = \prod_{s \in \mathbb{Z}^d} d \mu$ is a Radon measure over $\Omega$ with full support.
\end{itemize}

There is action of $\varrho$ of $\mathbb{Z}^d$ over $\Omega$ that takes the following form,
$$ \varrho(s)(\omega)(x) = \omega(x+s), \; \omega(x) \in O, $$
in particular, if $s := (s_1, \cdots, s_d)$ and $e_j := (0, \cdots, 1, \cdots, 0)$ with $1$ in the j$^{th}$ position, then
$$ \varrho(s) = \varrho(e_1)^{\circ s_1} \dots \varrho(e_d)^{\circ s_d}, $$
and we denote $\varrho_j := \varrho(e_j)$.

Under this setting, using the results from \cref{section:algebra_continuous_functions_locally_comp_space} we know that $C(\Omega)$ has a faithful representation over $L^2(\Omega, d\mathbb{P})$, also, from \cref{definition:twisted_transformation_group_C_star_algebras}, the following defines an action of $\mathbb{Z}^d$ over $C(\Omega)$
$$ \alpha(s)(f)(\omega) = f(\varrho(-s)(\omega)), \; \omega \in \Omega,\; s \in \mathbb{Z}^d, \; f \in C(\Omega). $$
Since $\Omega$ is a second countable compact Hausdorff space, \cref{lemma:separability_functions_dewcaying_at_infinity} tells us that $C(\Omega)$ is a separable C* algebra, also, we know that $\mathbb{Z}^d$ is a second countable locally compact abelian group, thus, $(C(\Omega), \mathbb{Z}^d, \alpha, \zeta)$ becomes a separable twisted dynamical system (\cref{def:separable_twisted_dynamical_system}), where $\zeta(x,y) = \exp(i x \Theta y^t)$ and $\Theta$ is a lower triangular matrix with zeros in the diagonal and entries on $[0,2 \pi)$.

Using the previously described setting we have a twisted crossed product $C(\Omega)\rtimes_{\alpha, \Theta} \mathbb{Z}^d$ along with a smooth sub algebra of $C(\Omega)\rtimes_{\alpha, \Theta} \mathbb{Z}^d$ which we call $\mathcal{C}(\Omega)_{\alpha, \Theta}$ (\cref{corollary:smooth_elements_are_invariant_under_holomoprhic_and_c_infty_calculus}). Additionally, the elements of $\mathcal{C}(\Omega)_{\alpha, \Theta}$ are have the following property, (\cref{lemma:algebra_of_smooth_elements_is_Frechet_space})
$$ p \in \mathcal{C}(\Omega)_{\alpha, \Theta}  \text{ iff } \forall n \in \mathbb{N} \exists K_n < \infty \text{ s.t. } |s|^n \| \phi_s (p) \|^n \leq K^n,  $$
and it has $d$ continuous maps $\{ \partial_j \}_{j \leq d}$ that commute and are derivations. By \cref{lemma:multiplicative_identity_of_twisted_crossed_product}, given that $C(\Omega)$ is a unital C* algebra, we have that $C(\Omega)\rtimes_{\alpha, \Theta} \mathbb{Z}^d$ is a unital C* algebra.

\begin{definition}[Non-commutative Brillouin Torus\index{Non-commutative Brillouin Torus}]\label{definition:non_commutative_brillouin_torus}
Given $\{ (E_s, \mu_s) \}_{s \in \mathbb{Z}^d}$ a set of tuples of the form, ( $E_s$: compact, convex (hence contractible) and second countable Hausdorff space, $\mu_s$: a Radon probability measure over $E_s$ with full support), denote by $O = \prod_{s \in \mathbb{Z}^d} E_s $ and set $\Omega = \prod_{s \in \mathbb{Z}^d} O$. Assume that $\Theta$ is a lower triangular matrix with entries in $[0, 2 \pi)$, and let $\alpha$ be the action of $\mathbb{Z}^d$ over $C(\Omega)$ described in the previous discussion, then, the pair of topological algebras
$$ (C(\Omega)\rtimes_{\alpha, \Theta} \mathbb{Z}^d, \mathcal{C}(\Omega)_{\alpha, \Theta}) $$\index{$\mathcal{C}(\Omega)_{\alpha, \Theta}$} \index{$C(\Omega)\rtimes_{\alpha, \Theta} \mathbb{Z}^d$}
will be called the \textbf{Non-commutative Brillouin Torus}.
\end{definition}

\begin{remark}\label{remark:non_commu_brilluouin_torus_nuclear_C_star_algebra}
Since the C* algebra $C(\Omega)$ is nuclear (\cref{example:nuclear_C_star_algebras}), by \cref{theorem:twsited_crossed_product_amenable_nuclear} we get that $C(\Omega)\rtimes_{\alpha, \Theta} \mathbb{Z}^d$ is a nuclear C* algebra.
\end{remark}

\begin{remark}\label{remark:non_commu_brilluouin_torus_separable_C_star_algebra}
Since $(C(\Omega), \mathbb{Z}^d, \alpha, \zeta)$ is a separable twisted dynamical system, \cref{lemma:twsited_crossed_product_is_separable} implies that $C(\Omega)\rtimes_{\alpha, \Theta} \mathbb{Z}^d$ is a separable C* algebra.
\end{remark}

 In \citep[Chapter 9]{prodan_computational_2017} and \citep[Chapter 1]{prodan_bulk_2016} you can find examples of more concrete \textbf{Non-Commutative Brillouin Torus}. If $E_s$ is a point for all $s \in \mathbb{Z}^d$ and $\Theta$ is the zero matrix, then the \textbf{Non-commutative Brillouin Torus} coincides with the \textbf{$\boldsymbol{d}$ non-commutative torus}\index{non-commutative torus} described in \citep[Section 3.1.1]{prodan_bulk_2016} and studied in \citep[Chapter 12]{gracia-bondia_elements_2001}.

\begin{remark}[Other structures over the Non-Commutative Brillouin torus]\label{remark:additional_structure_on_the_non_coommutative_brillouin_torus}
In \cref{definition:non_commutative_brillouin_torus} we have only taken into account the topological and smooth structure of the non-commutative smooth manifolds associated to the twisted crossed products $ C(\Omega)\rtimes_{\alpha, \Theta} \mathbb{Z}^d$, where, the topological information is stored in the C* algebra $ C(\Omega)\rtimes_{\alpha, \Theta} \mathbb{Z}^d $ and the smooth structures is stored in the smooth sub algebra  $ \mathcal{C}(\Omega)_{\alpha, \Theta})$. Is important to mention that those are not the only structures that can be studied over such non-commutative spaces, for example, is possible to explore their analytical structure by looking at Sobolev spaces (\citep[Section 3.6]{prodan_computational_2017}) and pseudo-differential operators (\citep{jimenez_pseudodifferential_2023}). The Sobolev spaces over the Non-Commutative Brillouin Torus are an important tool for studying topological invariants under strong disorder (\citep[Setion 6.5]{prodan_bulk_2016}).
\end{remark}

\chapter{K theory and the Non Commutative Brillouin Torus}
\label{chap:K_theory}

Let $(C(\Omega)\rtimes_{\alpha, \Theta} \mathbb{Z}^d, \mathcal{C}(\Omega)_{\alpha, \Theta})$ be The Non Commutative Brillouin Torus (\cref{definition:non_commutative_brillouin_torus}), in the present chapter we look into the steps necessary to prove the existence of the isomorphisms (\cref{remark:k_groups_and_the_Non_cOmmutative_brillouin_torus})
$$ K_0(C(\Omega)\rtimes_{\alpha, \Theta} \mathbb{Z}^d) \simeq K_0(\mathcal{C}(\Omega)_{\alpha, \Theta}), \; K_1(C(\Omega)\rtimes_{\alpha, \Theta} \mathbb{Z}^d) \simeq K_1(\mathcal{C}(\Omega)_{\alpha, \Theta}). $$
To prove the existence of the previous isomorphisms we will start from the following setup, assuming that $A$ is an unital C* algebra and $\mathcal{A}$ is a smooth sub algebra of $A$, then, we will prove the following statement,
$$ K_0(A) \simeq K_0(\mathcal{A}), \; K_1(A) \simeq K_1(\mathcal{A}), $$
which will be enough for our purposes because $\mathcal{C}(\Omega)_{\alpha, \Theta}$ is a smooth sub algebra of $C(\Omega)\rtimes_{\alpha, \Theta} \mathbb{Z}^d$ (\cref{corollary:smooth_elements_are_invariant_under_holomoprhic_and_c_infty_calculus}).

To devise a proof for the isomorphisms
$$ K_0(A) \simeq K_0(\mathcal{A}), \; K_1(A) \simeq K_1(\mathcal{A}), $$
we will take the following path,
\begin{enumerate}
    \item In \cref{sec:k_theory_for_C_star_algebras} we will define the groups $K_0(A), \; K_1(A)$ associated to a C* algebra $A$, these will be defined using equivalence relations on matrix algebras over $A$ and will relay on the topology of $A$. The maps $A \mapsto K_0(A)$ and $A \mapsto K_1(A)$ allow us to define a homology theory over the category of C* algebras (\citep[Chapter 11]{wegge-olsen_k-theory_1993}), we will not dive into these subjects but rather we will focus on the relation between the topological/analytical structure of $A$ and the groups $K_0(A), \; K_1(A)$.
    \item Let $A$ be a unital C* algebra and $\mathcal{A}$ be a smooth sub algebra of $A$, in \cref{sec:K_0_and_K_1_for_smooth_sub_algebras} we will define the pair of abelian groups $K_0(\mathcal{A}), \; K_1(\mathcal{A})$.
    \item In \cref{theorem:isomorphism_K_0_and_k_1_for_smooth_sub_algebras} we will proof the isomorphisms 
    $$ K_0(A) \simeq K_0(\mathcal{A}), \; K_1(A) \simeq K_1(\mathcal{A}). $$
\end{enumerate}

The group $K_0(A)$ is constructed taking into account homotopies of projections in matrix algebras of $A$ (\cref{sec:group_K_0}) while the group $K_1(A)$ is constructed taking into account homotopies of unitaries in matrix algebras of $A$ (\cref{sec:group_K_1}), and this construction can be extended into $\mathcal{A}$ because $\mathcal{A}$ is also a topological * algebra (\cref{sec:K_0_and_K_1_for_smooth_sub_algebras}). Under this context, the isomorphisms
$$ K_0(A) \simeq K_0(\mathcal{A}), \; K_1(A) \simeq K_1(\mathcal{A}), $$
are a direct consequence of two facts:
\begin{itemize}
    \item There is a one-to-one correspondence between the equivalence classes of homotopic unitaries over $A$ and the equivalence classes of homotopic unitaries over $\mathcal{A}$ (\cref{proposition:isomorphism_of_groups_of_unitaries_for_smooth_sub_algebras}).
    \item There is a one-to-one correspondence between the equivalence classes of homotopic projections over $A$ and the equivalence classes of homotopic projections over $\mathcal{A}$ (\cref{proposition:isomorphism_of_semigroups_of_projections_for_smooth_sub_algebras}).
\end{itemize}

 Our approach towards the isomorphisms of the K groups of $A$ and $\mathcal{A}$ takes an explicit look into the topology of $A$ and $\mathcal{A}$, this is relevant because the topology of $\mathcal{A}$ is stronger than the topology of $A$, which implies that we cannot use the topology of $A$ to guaranty that the a smooth path of projections over $A$ is also a smooth path of projections over $\mathcal{A}$ even if the path takes values inside $\mathcal{A}$ (\cref{remark:importance_funct_calculus_smooth_sub_algebras}). Under this context, to provide smooth paths of projections and unitaries with respect to the topology of $\mathcal{A}$ (\cref{proposition:cannonical_paths_idempotents_projections_smooth_sub_algebras}, \cref{proposition:cannonical_paths_invertibles_unitaries_smooth_sub_algebras}) we use the \textbf{holomorphic functional calculus} over $\mathcal{A}$ (\cref{theorem:holomorphic_funct_calculus_smooth_sub_algebras}), similarly, to provide smooth paths of projections and unitaries with respect to the topology of $A$ (\cref{proposition:canonical_paths_between_projections}, \cref{proposition:cannonical_smooth_paths_unitaries_and_invetibles_C_star_algebras}) we use the \textbf{holomorphic functional calculus} over $A$ (\cref{theorem:holomorphic_functional_calculus_banach_algebras}). For a brief discussion on alternative approaches towards the isomorphisms of the K groups of $A$ and $\mathcal{A}$, the reader can refer to \cref{sec:alternative_proofs_for_the_isomorphism_of_K_groups}.

The path taken in the present chapter can be intuitively thought of as a non-commutative version of the "smoothening" of vector bundles over compact smooth manifolds, that is, the equivalence classes of homotopic vector bundles over a compact smooth manifold are in one-to-one correspondence with the equivalence classes of homotopic smooth vector bundles over the same smooth compact manifold, this is mentioned in \citep[Introduction to Chapter 3]{connes_noncommutative_2014}. We do not delve into the previous statement, however, it is an interesting insight for our work, mostly because we are replacing Hausdorff compact spaces by C* algebras with unit and Hausdorff smooth compact manifolds by Fréchet algebras with unit.

\section{K theory for C* algebras}
\label{sec:k_theory_for_C_star_algebras}

\subsection{Equivalence relations}
\label{sec:equivalence_relations_C_star_algebras}

We will look into equivalence relations between various types of elements in a C* algebra $A$, so, lets us introduce some notation. Recall that $M_n(A)$ is a C* algebra (\cref{proposition:C_star_norm_on_matrix_algebras}), so, denote by $Q_n(A)$\index{$Q_n(A)$} the set of idempotents of $M_n(A)$ and by $P_n(A)$\index{$P_n(A)$} the set of projections of $M_n(A)$. If $A$ has a unit we denote by $G_n(A)$\index{$G_n(A)$} the group of invertible elements in $M_n(A)$ and $U_n(A)$\index{$U_n(A)$} for the group of unitary elements in $M_n(A)$. Following \cref{sec:unitization_of_C_star_algebras} we denote by $A^+$ the C* algebra whose underlying set is $\{ (a, \lambda): \; a \in A, \; \lambda \in \mathbb{C} \}$ provided with the following operations
\begin{itemize}
    \item $(a, \lambda) + (b , \gamma) := (a + b, \lambda + \gamma)$,
    \item $(a, \lambda)^* := (a^*, \lambda^*)$,
    \item $(a, \lambda) (b, \gamma) := (ab + \gamma a + \lambda b , \lambda \gamma)$.
\end{itemize}

\begin{definition}\label{definition:homotopy_equivalence_relations}
Let $A$ be a C* algebra, then, we introduce the following notation      
\begin{itemize}
    \item If $p,q \in P(A)$, we say that $p \sim_h q$ if there is a norm-continuous path of projections in $A$ from $p$ to $q$. (homotopy)\index{homotopy!of projections} \index{$\sim_h$ (homotopy)}
    \item If $p,q \in P(A)$, we say that $p \sim_u v$ if there is an unitary $u$ in $A^+$ with $u p u^{*}=q$. (unitary equivalence)\index{$\sim_u$ (unitary equivalence)}
    \item If $e,f \in Q(A)$, we say that $p \sim_h q$ if there is a norm-continuous path of idempotents in $A$ from $e$ to $f$. (homotopy)\index{homotopy!of idempotents}
    \item If $e,f \in Q(A)$, we say that $e \sim_s f$ if there is an invertible $z$ in $A^+$ with $z e z^{-1}=f$. (similarity)\index{$\sim_s$ (similarity)}
    \item If $x,y \in G(A)$, we say that $x \sim_h y$ if there is a norm-continuous path of invertible elements of $A$ from $x$ to $y$. (homotopy)\index{homotopy!of invertible elements}
    \item If $u,v \in U(A)$, we say that $u \sim_h v$ if there is a norm-continuous path of unitaries in $A$ from $u$ to $v$. (homotopy)\index{homotopy!of unitaries}
\end{itemize}
\end{definition}

It is easy to check that both $\sim_s$ and $\sim_h$ are equivalence relations between idempotents of $A$. It is easy to check that both $\sim_u$ and $\sim_h$ are equivalence relations between projection of $A$.

\begin{lemma}\label{lemma:equivalence_relations_invertibles_and_unitaries}
Let $A$ be a unital C* algebra, then, 
\begin{enumerate}
    \item If $e,f \in Q(A)$,then, $e \sim_s f$ iff there is $v \in G(A)$ such that $v e v^{-1}= f$.
    \item If $p,q \in P(A)$, then, $p \sim_u q$ iff there is $u \in G(A)$ such that $u e u^{*}= q$.
\end{enumerate}
\end{lemma}
\begin{proof}
In \cref{sec:unitization_of_C_star_algebras} is mentioned that if $A$ is unital, then, $A^+$ it is isomorphic to the C* algebra $A \oplus \mathbb{C}$ through the following mapping
$$ (a, \alpha) \mapsto a + \alpha f, \text{ with } (a, \alpha) \in A \oplus \mathbb{C} \text{ and } f = (-1_A, 1). $$
Given that an element $(a, \alpha)$ of $A \oplus \mathbb{C}$ is invertible iff both $a$ and $\alpha$ are invertibles, if $(v, \nu)$ is an invertible in $A^+$, then, $\nu$ is an invertible in $\mathbb{C}$ and $v + \nu 1_A$ is an invertible in $A$. Similarly, given that an element $(a, \alpha)$ of $A \oplus \mathbb{C}$ is unitary iff both $a$ and $\alpha$ are unitary, if $(u, \upsilon)$ is unitary in $A^+$, then, $\upsilon$ is unitary in $\mathbb{C}$ and $u + \upsilon 1_A$ is an unitary in $A$. 

\begin{enumerate}
    \item Since $A$ is a sub C* algebra of $A \oplus \mathbb{C}$ through the map $a \mapsto (a, 0)$, we have that, if $(v, \nu)$ is an invertible in $A^+$ and $(v, \nu) e (v \nu)^{-1} = f$, then, 
    $$(v + 1_A \nu ) e (v + 1_A \nu )^{-1} = f.$$ 
    Similarly, if $v \in G(A)$ and $v e v^{-1} = f$, then, $(v-1_A,1)$ is an invertible in $A^{+}$ and 
    $$(v - 1_A, 1 ) (e,0) (v - 1_A,1 )^{-1} = (f,0).$$ 
    \item Since $A$ is a sub C* algebra of $A \oplus \mathbb{C}$ through the map $a \mapsto (a, 0)$, we have that, if $(u, v)$ is an unitary in $A^+$ and $(u, v) p (u v)^{*} = p$, then, 
    $$(u + 1_A v ) p (u + 1_A v )^{*} = p.$$    
    Similarly, if $u \in U(A)$ and $u p u^{*} = q$, then, $(u-1_A,1)$ is an unitary in $A^{+}$ and 
    $$(u - 1_A, 1 ) (p,0) (u - 1_A,1 )^{*} = (p,0).$$ 
\end{enumerate}
\end{proof}

The main idea of the following sections is the fact that homotopies of invertibles, homotopies of unitaries, homotopies of idempotents and homotopies of projections have an appealing form based on exponentials, and this form comes as a consequence of the properties of the holomorphic functional calculus over C* algebras as explained in \cref{proposition:cannonical_smooth_paths_unitaries_and_invetibles_C_star_algebras}, \cref{proposition:canonical_paths_between_idempotents} and \cref{proposition:canonical_paths_between_projections}.

\subsection{Unitaries and invertibles}
\label{sec:unitaries_equivalence_relations}

Let $A$ be a unital C* algebra, from \cref{proposition:polar_decomposition_invertible_elements} we know that there is a map $\omega: G(A) \to U(A)$ that takes the following form $\omega(a) = |a|^{-1} a$ where $|a| = (a a^*)^{1/2}$, this implies that $a$ has a unique decomposition of the form $a = b u$ with $u$ and unitary in $A$ and $b$ a self adjoint element of $A$, where $b = |a|$  and $u=  \omega(a)$. The expresion $ a = |a| \omega(a)$ is called the polar decomposition of $a$ and plays an important role in the following proposition,

\begin{proposition}[$U(A)$ is a retraction of $G(A)$ (Proposition 2.1.8 \citep{rordam_introduction_2000})]\label{prop:U_A_retraction_of_G_A}

Let $A$ be a unital $C^*$-algebra.
\begin{enumerate}
    \item For every $a \in G(A)$ there is a unique decomposition $a = |a| \omega(a)$ with $|a| \in A_{\text{pos}}$, $|a| \in G(A)$ and $\omega(a) \in U(A)$
    \item The map $\omega$ : $G(A) \rightarrow \mathcal{U}(A)$ is continuous, $\omega(u)=u$ for every $u$ in $G(A)$, and $\omega(a) \sim_h a$ in $G(A)$ for every $a$ in $G(A)$.
    \item If $u, v$ are unitary elements in $U(A)$, and if $u \sim_h v$ in $G(A)$, then $u \sim_h v$ in $U(A)$. 
\end{enumerate}
\end{proposition}
\begin{proof}
\begin{enumerate}
    \item This is established by  \cref{proposition:polar_decomposition_invertible_elements}.
    \item From  \cref{proposition:polar_decomposition_invertible_elements} we know that $\omega(u) = u$ if $u \in U(A)$. Recall that $|a| = (a a^*)^{1/2}$ and $\omega(a) = |a|^{-1} a $, so
    \begin{itemize}
        \item the map $g : G(A) \to A_{\text{pos}} \cap G(A)$ given by $g(a) = a a^*$ is continuous because involution and multiplication are continuous, 
        \item the map $(\cdot)^{1/2}: A_{\text{pos}} \cap G(A) \to A_{\text{pos}}$ is continuous, because all the elements in $A_{\text{pos}} \cap G(A)$ have spectrum in $(0,\infty)$ and $f(z) = \sqrt{z}$ is an holomorphic function on $D$, thus \cref{proposition:properties_of_the_holomoprhic_functional_calculus} tell us that $a \to (a)^{1/2}$ is continuous.
        \item the map $(\cdot)^{-1} : G(A) \cap A_{\text{pos}} \to  G(A) \cap A_{\text{pos}} $ is continuous because the inversion on $G(A)$ is continuous,
    \end{itemize}
    therefore the maps $|\cdot|: G(A) \to G(A) \cap A_{\text{pos}}$ and $a \to |a|^{-1}$ are continuous, and we end up with the map $ \omega: G(A) \to U(A) $ given by $\omega(a) = |a|^{-1} a$ is also continuous.\\
    Take $a \in G(A)$ then set $\gamma(t) = \omega(a)(t|a| + (1-t)1_A)$ for $t \in [0,1]$, then $\gamma(0) = \omega(a)$ and $\gamma(1)=a$, also, there is $\lambda > 0$ such that $|a| \geq \lambda 1_A$. Notice that $t |a| + (1-t)1_A \geq 0$ because is the sum of two positive elements, moreover, 
    $$ t |a| + (1-t)1_A \geq t \lambda 1_A + (1-t)1_A = (1 - t(1 - \lambda))1_A \geq \lambda 1_A$$  
    and we end up with $t |a| + (1-t)1_A \in G(A)$, which implies that $\gamma(t) \in G(A)$ for $t \in [0,1]$. The path $\gamma$ gives us a homotopy of invertible elements connecting $a$ with $\omega(a)$, thus $a \sim_h \omega(a)$.
    
    \item Let $u,v \in U(A)$ and $v \sim_h u$ in $G(A)$ with $\gamma: [0,1] \to G(A)$, $\gamma(0) = u$ and $\gamma(1)=v$, then $\omega \circ \gamma : [0,1] \to U(A)$ is continuous map because if the composition of two continuous maps.
\end{enumerate}
\end{proof}

Given that $U(A)$ is a subset of $G(A)$, \cref{prop:U_A_retraction_of_G_A} tells us that the connected components of $G(A)$ are in one to one correspondence to the connected components of $U(A)$, hence, to analyze the connected components of $G(A)$ is equivalent to analyze the connected components of $U(A)$.

Let $\mathbb{R}^- = \{ x \in \mathbb{R} | x \leq 0 \}$ and $D = \mathbb{C} - \mathbb{R}^-$, for any $\theta \in [0, 2 \pi) $ denote by $D_{\theta}$ the set $\exp(\theta) D$. In \cref{sec:hol_cal_consecuences} it is shown how every invertible element with spectrum in $D_{\theta}, \; \theta \in \mathbb{C}$ is connected to the identity because it can be expressed as $\exp(a)$, in the next lemma we look into the consequences of this fact when we work with unitaries instead of invetibles,

\begin{lemma}[ c.f. Lemma 2.1.3 \citep{rordam_introduction_2000}]\label{lemma:connected_component_of_unitaries}

Let $A$ be a unital $C^*$-algebra, then
\begin{enumerate}
    \item For each self-adjoint element $h$ in $A, \exp (i h)$ belongs to $\mathcal{U}_0(A)$ with $\mathcal{U}_0(A)$ the connected component of $1_A$ inside $U(A)$.
    \item If $u$ is a unitary element in $A$ with $\operatorname{sp}(u) \neq \mathbb{T}$, then $u$ belongs to $\mathcal{U}_0(A)$ and $u = \exp(ih)$ with $h$ a self adjoint element of $A$.
    \item If $u, v$ are unitary elements in $A$ with $\|u-v\|<2$, then $u \sim_h v$ and there is a path connecting $u$ with $v$ taking the following form
    $$ \eta(t) = v \exp(t ih), $$
    where $h$ is a self adjoint element of $A$.
\end{enumerate}
\end{lemma}
\begin{proof}
\begin{enumerate}
    \item \cref{lemma:unitaries_from_exponential_of_self_adjoint} tells us that $\exp(ih)$ is an unitary element of $A$. Since the map $z \mapsto \exp(z)$ is an entire function, \cref{proposition:smoomth_paths_from_entire_functions} tells us that $t \to \exp(iht)$ is a continuous map, in particular, it is a homotopy of unitaries from $1_A$ to $\exp(ih)$.
    \item Recall that $\text{Sp}(u) \subseteq \mathbb{T}$ (\cref{lemma:espectrum_of_unitaries_c_star_algebra}). Since $\text{Sp}(u) \neq \mathbb{T}$, there is $\theta \in [0, 2 \pi)$ such that $\exp(i\theta) \notin \text{Sp}(u)$, which means that $\text{Sp}(u) \in D_{\theta}$ for some $\theta \in [0, 2 \pi)$ (\cref{remark:shifting_logarithms_and_n_roots}). Since $D_{\theta}$ is a simply connected domain of $\mathbb{C}$ there is a function called $\log$ on it (\cref{remark:shifting_logarithms_and_n_roots}) and we have that 
    $$ \exp(\log(u)) = u. $$
    The function $z \mapsto \log(z)$ is continuous over the spectrum of $u$, thus, given that continuous functional calculus and the holomorphic functional calculus coincide inside C* algebras (\cref{lemma:holomorphic_calculus_continuous_calculus_coincide}), the element $\log(u)$ can be computed with the continuous functional calculus of normal elements, and we get that $\log(u)$ commutes with $\log(u)^*$ because $\log(u)$ belongs to the sub C* algebra of $A$ generated by $1_A$ and $u$ (\cref{theorem:continuous_functional_calculus}). By the spectral mapping of the holomorphic functional calculus (\cref{theorem:holomorphic_functional_calculus_banach_algebras}) we have that $\text{Sp}(\log  (u)) \in i[ \theta, 2 \pi + \theta]$, thus $\log(u)/i$ has spectrum in $[ \theta, 2 \pi + \theta]$. Set $h = \log(u)/i$, then, $\exp(ih)=u$, additionally, $h$ is normal because $\log(u)$ is normal, and we get that $h$ is self adjoint because a normal element with spectrum inside $\mathbb{R}$ is a self adjoint element by \cref{lemma:characterization_normal_elements_with_their_spectrum}. The previous item implies that $u \in U_0(A)$, where $U_0(A)$ is the set of unitaries of $A$ that are homotopic to the identity.
    \item We have that,
    $$\| v^*u - 1_A \| = \|v^*(u-v)\| \leq \| v^* \| \|u -v \| \leq 2,$$
    thus, given that the spectral radius is bounded above by the norm (\cref{remark:upper_bound_on_spectral_radius}), we have that $\text{Sp}(v^* u - 1_A) \subset B(0,2)$. Since $v^*u = (v^*u - 1_A) + 1_A$, the spectral mapping of the holomorphic functional calculus (\cref{theorem:holomorphic_functional_calculus_banach_algebras}) implies that $\text{Sp}(v^* u) \subset B(1,2)$. The previous statement implies that $-1 \notin \text{Sp}(v^* u)$, thus $ Sp(v^* u) \subset D$, and we can compute the logarithm of $v^*u$ using the holomorphic functional calculus, so set
    $$ h = \log(v^* u)/i, $$
    as in the previous item $h$ is a self adjoint element by the continuous functional calculus, and $u = v \exp{ih}$. Since the map $z \mapsto \exp(z)$ is an entire function, \cref{proposition:smoomth_paths_from_entire_functions} tells us that $t \to \exp(iht)$ is a continuous map, thus, $\eta$ is a continuous path from $v$ to $u$, and we end up with $u \sim_h v$ as desire.
\end{enumerate}
\end{proof}

From \cref{theorem: description_of_GA} we know that any two homotopic invertibles $u,v$ of a unital C* algebra $A$ can be related through an expression like
$$u = v (\exp \left(a_1\right) \exp \left(a_2\right) \cdots \exp \left(a_k\right)),$$
in the next theorem we show how this translates into an continuous path of unitaries if both $u$ and $v$ are unitaries,

\begin{theorem}[Description of $U(A)$ (Proposition 2.1.6 \citep{rordam_introduction_2000})]\label{theorem:description_of_UA}
Let $A$ be a unital $C^*$-algebra, denote by $A_{\text{sa}}$\index{$A_{\text{sa}}$} the set of self-adjoint elements of $A$, denote by $U_0(A)$\index{$U_0(A)$} is the set of unitaries of $A$ that are homotopic to the identity, then,
\begin{enumerate}
    \item $U_0(A)$ is a normal subgroup of $U(A)$.
    \item $U_0(A)$ is open and closed relative to $U(A)$.
    \item An element $u$ in $A$ belongs to $U_0(A)$ if and only if
    $$
    u=\exp \left(i h_1\right) \cdot \exp \left(i h_2\right) \cdots \exp \left(i h_n\right)
    $$
    for some natural number $n$ and $h_1, h_2, \ldots, h_n \in A_{\text{sa}}$. 
    \item If $A$ is commutative then $U_0(A) = \exp{i A_{\text{sa}}}$
\end{enumerate}
\end{theorem}
\begin{proof}

\begin{enumerate}
    \item Given that the computing the inverse is a continuous from $G(A)$ into $G(A)$ and the multiplication is continuous in $G(A)$ (\cref{proposition:GA_topological_group_and_open_in_banach_algebra}), if we assume that $t \to u_t,\; t \to v_t$ are continuous paths untiaries from $1_A$ to $u$ and $v$ respectively, then, $t \to (u_t)^{-1}$ and $t \to v_t u_t (v_t)^{-1}$ are continuous paths of unitaries from $1_A$ to $u^*$ and $v u v^*$ respectively, which implies that $U_0 (A)$ is a normal sub group of $A$.
    \item Let $\beta$ be an accumulation point of $U_0 (A)$, then, there is an element $u \in U_0 (A)$ such that $\| a - \beta \| \leq 2$, which implies that $u \sim_h \beta$ by \cref{lemma:connected_component_of_unitaries}. If $t \to a_t$ is an homotopy connecting $1_A$ and $u$ then we can concatenate the homotopy $t \to u_t$ with the homotopy from $u$ to $\beta$ to get an homotopy of unitaries from $1_A$ to $\beta$, that is, $\beta \in U_0 (A)$. Therefore, $U_0 (A)$ is closed with respect to $U(A)$.
    Using \cref{lemma:connected_component_of_unitaries} you can show that if $u \in U_0 (A)$, then, any unitary close enought to $u$ are homotopic of $u$, thus $U_0 (A)$ is open with respect to $U(A)$.
    \item Let 
    $$E = \left\{\exp \left(i h_1\right) \exp \left(i h_2\right) \cdots \exp \left(i h_k\right): h_1, \ldots, h_k \in A_{\text{sa}}, k \in \mathbb{N}\right\},$$
    then $E$ is a subgroup of $U(A)$ and $E \in U_0 (A)$. Take 
    $$u = \exp \left(i h_1\right) \exp \left(i h_2\right) \cdots \exp \left(i h_k\right)$$
    then if $\| v - u \| \leq 2$ there is $x \in A_{\text{sa}}$ such that $v = u \exp{ih}$, thus $v \in E$ and $E$ is open. Also, every left coset of $E$ is open with respect to $U(A)$ because the mapping $u \to uv$ is a homeomorphism when $u \in E$, which implies that $G \setminus E$ is open with respect to $U(A)$, hence $E$ is closed with respect to $U(A)$. So, $E$ is a non-empty open and closed subset of $U_0 (A)$, which leaves no other alternative that $U_0(A) = E$.
    We end up having that every coset of $U_0 (A)$ in $U(A)$ is a connected open and closed subset of $U(A)$, and are the topological components of $U(A)$.
    \item If $A$ is commutative then 
    $$ \exp \left(i h_1\right) \exp \left(i h_2\right) \cdots \exp \left(i h_k\right) = \exp \left( i (h_1 + h_2 + \dots + h_k) \right) $$
    by \cref{proposition:logarithms_and_n_roots}, therefore $U_0 (A) = \exp{i A_{\text{sa}}}$.
\end{enumerate}
\end{proof}

In the following, if $A$ is a C* algebra and we have a function $\eta: [0,1] \to A$, we say that \textbf{$\eta$ is a smooth path}\index{smooth path} if $\eta$ is infinitely differentiable as a function from $[0,1]$ into $A$, where the derivative of such a function takes the usual definition as exposed in \cref{remark:leibniz_rule_fn_values_in_topological_algebra}.

\begin{lemma}\label{lemma:basci_smooth_paths_of_unitaries_and_invertibles}
Let $A$ be a unital C* algebra, then,
\begin{enumerate}
    \item If $a_1, \cdots, a_n$ are elements of $A$, then, the following is a smooth path of invertible elements of $A$
    $$ \eta(t) = \exp \left(t a_1\right) \exp \left(t a_2\right) \cdots \exp \left(t a_n\right).$$
    \item If $h_1, \cdots, h_n$ are self adjoint elements of $A$, then, the following is a smooth path of unitaries of $A$
    $$ \eta(t) = \exp \left(t i h_1\right) \exp \left(t i h_2\right) \cdots \exp \left(t ih_n\right).$$
\end{enumerate}
\end{lemma}
\begin{proof}
\begin{enumerate}
    \item From \cref{proposition:logarithms_and_n_roots} we know that $\exp \left(t a_j\right)$ is an invertible element for any $t \in \mathbb{C}$. Since the function $t \mapsto \exp(t)$ is an holomorphpic function, \cref{proposition:smoomth_paths_from_entire_functions} tells us that the map $t \mapsto \exp \left(t a_j\right)$ is a smooth path of elements of $A$, so, given that $\eta$ is a multiplication of such paths, the Leibniz rule (\cref{remark:leibniz_rule_fn_values_in_topological_algebra}) implies that $\eta$ is a smooth path of invertible elements of $A$.
    \item From \cref{theorem:description_of_UA} we know that $\exp \left(t i  h_j\right)$ is an unitary for any $t \in \mathbb{C}$. Since the function $t \mapsto \exp(t)$ is an holomorphpic function, \cref{proposition:smoomth_paths_from_entire_functions} tells us that the map $t \mapsto \exp \left(t i h_j\right)$ is a smooth path of elements of $A$, so, given that $\eta$ is a multiplication of such paths, the Leibniz rule (\cref{remark:leibniz_rule_fn_values_in_topological_algebra}) implies that $\eta$ is a smooth path of unitaries of $A$.
\end{enumerate}
\end{proof}

\begin{proposition}\label{proposition:cannonical_smooth_paths_unitaries_and_invetibles_C_star_algebras}
Let $A$ be a unital C* algebra, then,
\begin{enumerate}
    \item If $u,v \in G(A)$ and $u \sim_h v$, then, there is a smooth path of invertible elements $\eta$ from $u$ to $v$ taking the following form,
    $$ \eta(t) =  u \exp \left(ta_1\right) \exp \left(t a_2\right) \cdots \exp \left(ta_k\right), \; \eta(0)=u, \; \eta(1) = 1, $$
    where $a_1, \cdots, a_n$ are elements of $A$.
    \item If $u,v \in U(A)$ and $u \sim_h v$, then, there is a smooth path of unitaries  $\eta$ from $u$ to $v$ taking the following form,
    $$ \eta(t) =  v \exp \left(ti h_1\right) \exp \left(ti h_2\right) \cdots \exp \left(tih_k\right), \; \eta(0)=u, \; \eta(1) = 1, $$
    where $h_1, \cdots, h_n$ are self-adjoint elements of $A$.
\end{enumerate}
\end{proposition}
\begin{proof}
This proposition is a consequence of the properties of the holomorphic functional calculus over Banach algebras (\cref{sec:banach_alg_hol_func_cal}) as we proceed to check,
\begin{enumerate}
    \item From \cref{theorem: description_of_GA} we know that the connected components of the group $G(A)$ look like 
    $$u \exp \left(a_1\right) \exp \left(a_2\right) \cdots \exp \left(a_n\right)$$
    with $a_1, \ldots, a_n$ elements of $A$ and $u$ is a representative of a connected component of $G(A)$, therefore, if two invertible elements $u,v$ are homotopic we can construct a canonical path between them as 
    $$ \eta(t) = u \exp \left(t a_1\right) \exp \left(t a_2\right) \cdots \exp \left(t a_n\right)), \; \eta(0) = u , \; \gamma(1) = v.$$
    From \cref{lemma:basci_smooth_paths_of_unitaries_and_invertibles} we get that $\eta$ is a smooth path of invertible elements of $A$.
    \item From \cref{theorem:description_of_UA} we know that the connected components of the group $U(A)$ look like 
    $$u \exp \left(i h_1\right) \exp \left(i h_2\right) \cdots \exp \left(ih_n\right)$$
    with $h_1, \ldots, h_n$ self adjoint elements of $A$ and $u$ is a representative of a connected component of $U(A)$, therefore, if two unitaries $u,v$ are homotopic we can construct a canonical path between them as 
    $$ \eta(t) = u \exp \left(t i h_1\right) \exp \left(t i h_2\right) \cdots \exp \left(t i h_n\right)), \; \eta(0) = u , \; \gamma(1) = v.$$
    From \cref{lemma:basci_smooth_paths_of_unitaries_and_invertibles} we get that $\eta$ is a smooth path of unitaries inside $A$.
\end{enumerate}
\end{proof}

The specific form of the smooth paths in \cref{proposition:cannonical_smooth_paths_unitaries_and_invetibles_C_star_algebras} is be very convenient for the pairing between cyclic cohomology and K theory (\cref{sec:pairing_with_K1}), additionally, this type of smooth paths takes the same form in smooth sub algebras (\cref{proposition:cannonical_paths_invertibles_unitaries_smooth_sub_algebras}). 

\subsection{Projections and idempotents}
\label{sec:projections_equivalence_relations}

Idempotents and projections are the other ingredients of the K theory of C* algebras, so, in the present section we focus on some properties of those elements,

\begin{proposition}[$P(A)$ is a retraction of $Q(A)$ (Lemma 11.2.7. \citep{rordam_introduction_2000})]\label{proposition:P_A_retraction_of_Q_A}
Let $A$ be a $C^*$-algebra.
\begin{enumerate}
    \item For every idempotent element $e$ in $A, \eta(e)=e e^*\left(1_{A^+}+\left(e-e^*\right)\left(e^*-e\right)\right)^{-1}$ defines a projection in $A$. 
    \item The map $\eta: Q(A) \rightarrow P(A)$ is continuous, $\eta(p)=p$ for every projection $p$ in $A$, and $\eta(e) \sim_h e$ in $Q(A)$ for every idempotent $e$ in $A$.
    \item If $p$ and $q$ are projections in $A$ with $p \sim_h q$ in $Q(B)$, then $p \sim_h q$ in $P(A)$.
\end{enumerate}
\end{proposition}
\begin{proof}
We follow  \citep[Lemma 11.2.7.]{rordam_introduction_2000}. We reproduce the proof because we want to showcase how all the constructions needed in the proof only need of the holomorphic calculus, which makes them applicable to the context of smooth sub algebras.
\begin{enumerate}
    \item Let $e \in Q(A)$ and set $h = 1_{A^+} + (e - e^*)(e^* -e)$, then $h$ is positive because is the sum of the positive elements $1_{A^+}$ and $(e - e^*)(e^* -e)$ (\cref{proposition:cahracterization_of_positive_elements}). Since $h$ is the result of applying the function $x \mapsto 1 + x$, the spectral mapping of the continuous functional calculus  (\cref{theorem:continuous_functional_calculus}) tell us that $Sp(h) = 1 + Sp((e - e^*)(e^* -e))$, given that $(e - e^*)(e^* -e)$ is a positive element we know that $Sp((e - e^*)(e^* -e)) \subset \mathbb{R}^+$, therefore, 
    $$0 \notin \text{Sp}(1_{A^+} + (e - e^*)(e^* -e)),$$
    which tell us that $h$ is invertible. Let $\eta (e) = e e^* h^{-1}$, then $\eta(e) \in A$ because $A$ is an ideal of $A^+$. By definition $h$ is self adjoint, therefore $h^{-1}$ is also self adjoint, and doing algebraic manipulations (\citep[Lemma 11.2.7.]{rordam_introduction_2000}) is possible to check that $e e^* h = (e e^*)^2 = h e e^*$, $\eta(e) e = e$ and $e \eta(e) = \eta(e)$, which implies that $(e e^* h^{-1})^* = h^{-1} e e^* = e e^* h^{-1}$, thus $\eta(e)$ is self ajdoint. Also $\eta(e)^2 = \eta(e)$, thus $\eta(e) \in P(A)$.
    
    \item By definition, if $e \in P(A)$ then $\eta(e) = e e^* 1_B = e$, thus $P(A)$ is a fixed point of $\eta$. Recall that multiplication, inversion and addition are continuous on $A^+$, therefore $e \to \eta(e)$ is continuous. Set $\gamma(t) = 1_{A^+} - t(e - \eta(e))$ for $t \in [0,1]$, then $\gamma(t)$ is invertible with inverse $(\gamma(t))^{-1} = 1 + t(e - \eta(e))$. Therefore $(\gamma(t))^{-1} e \gamma(t) \in Q(A)$ and we have that 
    $$ e = (\gamma(0))^{-1} e \gamma(0) \sim_h (\gamma(1))^{-1} e \gamma(1) = \eta(e) \text{ in }  Q(A). $$
    
    \item Suppose that $t \to e_t$ is an homotopy of elements in $Q(A)$ from $e_0 = p$ to $e_1=q$ with $p,q \in P(A)$, then $t \to \eta(e_t)$ is a composition of two continuous maps, therefore is a continuous path in $P(A)$ from $\eta(e_0) = p$ to $\eta ( e_1 ) = q$.
\end{enumerate}
\end{proof}

Given that $P(A) \subset Q(A)$, the previous result tell us that the connected components of $Q(A)$ are in one to one correspondence to the connected components of $P(A)$, therefore, to analyze the connected components of $Q(A)$ is equivalent to study the connected components of $P(A)$.

In \cref{proposition:cannonical_smooth_paths_unitaries_and_invetibles_C_star_algebras} we showed how the holomorphic calculus provided us with canonical smooth paths between invertibles and unitaries, now we look in how this can be generalized into idempotents and projections.

\begin{lemma}[(Proposition 4.3.2 \citep{blackadar_k-theory_2012})]\label{lemma:close_idempotents_are_homotopic}
Let $A$ be a C* algebra and $e,f \in Q(A)$ with $\| e -f \| < 1/\|2e -1_{A^+}\|$, then $e \sim_s f$ i.e. there exists $x \in G(A)$ with $x^{-1} e x = f$, additionally, $x$ takes the following form $x=\exp(b) \in G_0(A)$. Additionally, we have that $e \sim_h f$ through the following smooth path of idempotents,
$$ \gamma(t) = \exp(-tb) e \exp(tb), \; \gamma(0) = e, \; \gamma(1) = f.  $$
\end{lemma}
\begin{proof}

Set $v = (2e - 1_{A^+})((2f - 1_{A^+})) + 1_{A^+}$, then 
$$ \| 1_{A^+} - v/2 \| \leq \|2e - 1_{A^+} \| \|e - f\| < 1 ,$$
therefore $v/2$ is invertible and can be expressed $\exp(b) = v/2$ with $b \in A^+$ (\cref{proposition:log_of_element_close_to_identity}). After algebraic manipulations you can check that $ ev = vf = 2ef $, therefore if $x = v/2$ we end up with 
$$ x^{-1} e x = f = \exp(-b) e \exp(b),$$
as desired, moreover, we can create a smooth path of idempotents as follows
$$ \gamma(t) = \exp(-tb) e \exp(tb), $$
with $\gamma(0)=e$ and $\gamma({1})=f$. \cref{lemma:basci_smooth_paths_of_unitaries_and_invertibles} tells us that $\gamma$ is a smooth path, additionally, lies inside $A$ because $e \in A$ and $A$ is an ideal of $A^+$ (\cref{sec:unitization_of_C_star_algebras}).
\end{proof}

\begin{proposition}[Canonical paths between idempotents (Proposition 4.3.2 \citep{blackadar_k-theory_2012})]\label{proposition:canonical_paths_between_idempotents}
Let $A$ be a C* algebra and $e,f \in Q(A)$ with $e \sim_h f$, then there is a smooth path of invertibles
$$ \eta: [0,1] \to G_0 (A^+) $$
such that $\eta(0) = 1_{A^+}$, and $(\eta(1))^{-1} e \eta(1) = f$. The path $\eta$ takes the following form,
$$ \eta(t) = \exp(tb_1) \dots \exp(tb_{n-1}) \exp(tb_n), \; b_1, \cdots, b_n \in A^+,$$
hence, we have that $e \sim_s f$. Additionally, $e \sim_h f$ through the smooth path of idempotents,
$$ \gamma(t) = \eta(t)^{-1} e \eta(t), \; \gamma(0) = e, \; \gamma(t) = f.$$
\end{proposition}
\begin{proof}

Let $\gamma: [0,1] \to Q(A)$ be a path of idempotents with $\gamma(0)=e$ and $\gamma(1)=f$, chose $k$ with $\| 2 \gamma(t) - 1_{A^+} \| < k$, then, the uniform continuity of the function 
$$ t \to \gamma(t) $$
assets us that we can find a partition $0 = t_0 \leq t_1 \leq \dots \leq t_n =1$ such that 
$$\| \gamma(t) - \gamma(s) \| < 1/k, \; t,s \in [t_i, t_{i+1}].$$
From \cref{lemma:close_idempotents_are_homotopic}, we know that there is $b \in A^+$ such that $s \to \exp(-sb_i) \gamma(t_{i-1}) \exp(sb)$ is a smooth path of idempotents from $\gamma(t_{i-1})$ to $\gamma(t_{i})$.

Since $\gamma(t_1) = \exp(-b_1) e \exp(b_1)$ and $\gamma(t_2) = \exp{-b_2} \gamma(t_1) \exp{b_2}$ then we have that 
$$ \gamma(t_2) = \exp(-b_2) \exp(-b_1) e \exp(b_1) \exp(b_2), $$
this process can be iterated to get
$$ f = \exp(-b_n) \dots \exp(-b_1) e \exp(b_1) \dots \exp(b_n). $$
Set $a = \exp{b_1} \dots \exp{b_n}$, then \cref{theorem: description_of_GA} tell us that $a \in G_0 (A^+)$, also, \cref{lemma:basci_smooth_paths_of_unitaries_and_invertibles} implies that
$$\eta: [0,1] \to Q(A), \; s \to \exp(-tb_n) \dots \exp(-tb_1) e \exp(tb_1) \dots \exp(tb_n)$$
is a smooth path of idempotents. $\eta(t)$ lies within $A$ because $e \in A$ and $A$ is an ideal of $A^+$ (\cref{sec:unitization_of_C_star_algebras}).
\end{proof}

We have constructed canonical paths between homotopic idempotents, these smooth paths are useful when in the realm of cyclic cohomology (\cref{sec:pairing with K theory}). Additionally, if we look into projections, we can construct these paths out of unitaries.

\begin{proposition}[Similar projections are unitary equivalent (Proposition 2.2.5 \citep{rordam_introduction_2000})]\label{lemma:similar_projetions_are_unitary_equivalent}
Let $a, b$ be self-adjoint elements in a unital $C^*$-algebra $A$, and suppose that $b=z^{-1} a z$ for some invertible element $z$ in $A$. Let $z=u|z|$ be the polar decomposition for $z$ with $u$ in $U(A)$ (\cref{proposition:polar_decomposition_invertible_elements}), then, $b=u^* a u$.
\end{proposition}
\begin{proof}
The equation $b=z^{-1} a z$ implies that $zb = az$, additionally, given that $a,b$ are self adjoint we get $b z^* = z^* a $. The previous claims imply that,
$$ |z|^2 a = (z z^*) a = z b z^* = a z z^* = a |z|^2,$$
thus, $a$ commutes with $z z^* = |z|^2$, and with any complex polynomial in the variable $z z^* = |z|^2$. Since $C^*(1_A, z z^*)$ is the sub C* algebra of $A$ where the polynoimals on $z z^*$ are dense, we have that $a$ commutes with any element of $C^*(1_A, z z^*)$. Since $z$ has an inverse, then $z^*$ is also invertible, thus, $z z^*$ is an invertible element of $A$, which implies that $0 \notin Sp(z z^*)$, additionally, $z z^*$ is a positive element of $A$ (\cref{proposition:cahracterization_of_positive_elements}), thus, $\text{Sp}(z z^*) \subset (0,\infty)$. Given that the function $\omega \to \omega^{-1/2}$ is holomorphic over the spectrum of $zz^*$, the holomorphic functional calculus tell us that $|z|^{-1} = (z z^*)^{-1/2}$, additionally, $|z|^{-1}$ can also be computed using the continuous functional calculus over $A$ (\cref{lemma:holomorphic_calculus_continuous_calculus_coincide}), thus, the argument layout at the beginning of the proof implies that $a$ commutes with $|z|^{-1}$. From the previous results it follows that
$$ u^* a u =  u^* a |z|^{-1} z  = u^* |z|^{-1} a z = u^* |z|^{-1} z b = u^* u b = b, $$
as desired.
\end{proof}

We can use \cref{lemma:similar_projetions_are_unitary_equivalent} to come up with canonical paths between projections using unitaries

\begin{proposition}[Close projections are unitary equivalent]\label{proposition:close_projetions_are_homotopic}
If $p,q \in P(A)$ and 
$$\| p-q\| \leq 1,$$
then $p \sim_u q$ with $u^{*} p u = q$ and $u \in U_0 (A^+)$, also $p \sim_h q$ through a smooth path of projections taking the following form, 
$$ \eta(t) = \exp \left(-i t h_n\right) \cdots \exp \left(-i t h_1\right) p \exp \left(i t h_1\right)\cdots \exp \left(i t h_n\right), \; \eta(0) = 0, \; \eta(1) = q ,$$
where $h_1, \cdots, h_n$ are self adjoint elements of $A^+$.
\end{proposition}
\begin{proof}
Notice that $2q -1_{A^+} \in U(A^+)$, thus $\| 2q -1_{A^+} \| =1$ and we can use \cref{lemma:close_idempotents_are_homotopic} to find $b \in G_0 (A^+)$ such that $b^{-1} p b =q$. Let $b = |b| u$ the polar decomposition of $b$, then \cref{lemma:similar_projetions_are_unitary_equivalent} tell us that $u^* p u = q$, moreover, since $U(A^+)$ is a retraction of $G(A^+)$ (\cref{prop:U_A_retraction_of_G_A}), we get that $u \sim_h 1_{A^+}$ in $U(A^+)$. 

By the characterization of $U(A^+)$ (\cref{theorem:description_of_UA}), we get that
$$ u = \exp \left(i h_1\right) \cdots \exp \left(i h_n\right), $$
where $\; h_1, \dots, h_n$ are self adjoint elements of $A^+$, also, \cref{lemma:basci_smooth_paths_of_unitaries_and_invertibles} tells us that the path
$$ \eta: [0,1] \to P(A), \; \eta(t) = \exp \left(-i t h_n\right) \cdots \exp \left(-i t h_1\right) p \exp \left(i t h_1\right)\cdots \exp \left(i t h_n\right) $$
is a smooth path of projections from $p = \eta(0)$ to $q = \eta(1)$. We have that $\eta$ lies within $A$ because $p \in A$ and $A$ is an ideal of $A^+$ (\cref{sec:unitization_of_C_star_algebras}).
\end{proof}

The previous proposition leads to,

\begin{proposition}[Canonical paths between projections]\label{proposition:canonical_paths_between_projections}
Let $A$ be a C* algebra and $p,q \in P(A)$ with $p \sim_h q$, then there is a smooth path of unitaries
$$ \eta: [0,1] \to U_0 (A^+) $$
such that $\eta(0) = 1_{A^+}$, and $(\eta(1))^{*} p \eta(1) = q$. The path $\eta$ takes the following form
$$ \eta(t) = \exp(tih_1) \dots \exp(tih_{n-1}) \exp(tih_n)$$
where $h_1, \cdots, h_n$ as self adjoint elements of $A^+$, hence, $p \sim_u q$. Additionally, $p \sim_h q$ through the smooth path 
$$ \gamma(t) = \eta(t)^* p \eta(t), \; \gamma(0)=p, \; \gamma(1)=q.$$
\end{proposition}
\begin{proof}
Let $\tilde{\gamma}: [0,1] \to P(A)$ be a path of projections with $\tilde{\gamma}(0)=p$ and $\tilde{\gamma}(1)=q$, then the uniform continuity of the function 
$$ t \to \tilde{\gamma}(t) $$
asset us that we can find a partition $0 = t_0 \leq t_1 \leq \dots \leq t_n =1$ such that 
$$\| \tilde{\gamma}(t) - \tilde{\gamma}(s) \| < 1, \; t,s \in [t_i, t_{i+1}].$$
From \cref{proposition:close_projetions_are_homotopic} we know that there are $u_i \in U_0(A^+)$ such that $\tilde{\gamma}(t_{i}) = u_i^* \tilde{\gamma}(t_{i-1}) u_i$, therefore, we have that
$$ q = u_n^* u_{n-1}^* \dots u_1^* p u_1 \dots u_{n-1} u_n .$$
Since $U_0 (A^+)$ is a subgroup of $U(A^+)$ we have that $v = u_1 \dots u_{n-1} u_n$ belongs to $U_0 (A^+)$, and by \cref{theorem:description_of_UA} we get that 
$$ v = \exp(i h_1) \cdots \exp(i h_n), $$
where $h_1, \dots, h_n$ are self adjoint elements of $A^+$. Consequently, we can define that path of projections
$$ \gamma : [0,1] \to P(A), \; \gamma(t) = \exp(- i t h_n) \cdots \exp(- i t h_1) p \exp(i t h_1) \cdots \exp(i t h_n), $$
and \cref{lemma:basci_smooth_paths_of_unitaries_and_invertibles} tells us that it is a smooth path. We have that $\gamma(t) \in A$ because $p \in A$ and $A$ is an ideal of $A^+$ (\cref{sec:unitization_of_C_star_algebras}).
\end{proof}

Given that matrix algebras with entries in C* algebras are C* algebras, the previous results also apply to matrix algebras.


\subsection{Groups $K_0$ and $K_1$}
\label{sec:groups_K_0_and_K_1}

For $M_n(A)$ there are canonical injective *-homomorphisms (\cref{sec:C_star_alg_matrix_alg})
$$
i_{n,m}: M_n(A) \rightarrow M_{m}(A): T \mapsto \left(\begin{array}{cc}
T & 0 \\
0 & 0_{m-n}
\end{array}\right),
$$
where we denote,
$$ T \oplus 0_{m-n} := \left(\begin{array}{cc}
T & 0 \\
0 & 0_{m-n}
\end{array}\right),  $$
therefore, we have the following maps
$$ i_{n,m}: Q_n(A) \rightarrow Q_{m}(A), \; i_{n,m}(e) = e \oplus 0_{m-n},$$ 
$$i_{n,m}: P_n(A) \rightarrow P_{m}(A), \; i_{n,m}(p) = p \oplus 0_{m-n},$$ 
because 
$$ (e \oplus 0_{m-n})(e \oplus 0_{m-n}) = e \oplus 0_{m-n}  \text{ iff } e^2 = e$$
and 
$$(p \oplus 0_{m-n})^* = p \oplus 0_{m-n} \text{ iff } p^* = p.$$ 

If $A$ has a unit, then for invertible and unitary matrices we have the following injective *-homomorphisms,
$$
j_{n,m}: G_n(A) \to G_{m}(A), \; j_{n,m}(v) = v \oplus 1_{M_{m-n}(A)} , \text{ with } v \oplus 1_{M_{m-n}(A)} := \left(\begin{array}{ll}
v & 0 \\
0 & 1_{M_{m-n}(A)}
\end{array}\right)
$$
and 
$$
j_{n,m}: U_n(A) \to U_{m}(A), \; j_{n,m}(u) = u \oplus 1_{M_{m-n}(A)}, \text{ with } u \oplus 1_{M_{m-n}(A)} :=  \left(\begin{array}{ll}
u & 0 \\
0 & 1_{M_{m-n}(A)}
\end{array}\right)
$$
because 
$$v \oplus 1_{M_m(A)} \in G_{m}(A) \text{ iff } v \in G_{n}(A)$$
and 
$$u \oplus 1_{M_m(A)} \in U_{m}(A) \text{ iff } u \in U_{n}(A).$$


Let $A$ be a C* algebra, then, the groups $K_0$ and $K_1$ are constructed using matrix algebras over $A$ with arbitrary dimension, therefore, we introduce the sets,
$$
M_{\infty}(A):=\bigcup_{n=1}^{\infty} M_n(A) ; \quad U_{\infty}(A):=\bigcup_{n=1}^{\infty} U_n(A) ; \quad G_{\infty}(A):=\bigcup_{n=1}^{\infty} G_n(A) ;$$
$$ \quad Q_{\infty}(A):=\bigcup_{n=1}^{\infty} Q_n(A) ;\quad P_{\infty}(A):=\bigcup_{n=1}^{\infty} P_n(A).
$$
On the aforementioned sets we will introduce the following equivalence relations,
\begin{itemize}
    \item If $e,f \in M_{\infty}(A)$ and $i_{n,m}(e) =f$, then $e$ and $f$ would represent the same element inside $M_{\infty}(A)$\index{$M_{\infty}(A)$}.
    \item If $e,f \in Q_{\infty}(A)$ and $i_{n,m}(e) =f$, then $e$ and $f$ would represent the same element inside $Q_{\infty}(A)$\index{$Q_{\infty}(A)$}.
    \item If $p,q \in P_{\infty}(A)$ and $i_{n,m}(p) =q$, then $p$ and $q$ would represent the same element inside $P_{\infty}(A)$\index{$P_{\infty}(A)$}.
    \item If $v,w \in G_{\infty}(A)$ and $j_{n,m}(v) =w$, then $v$ and $w$ would represent the same element inside $G_{\infty}(A)$\index{$G_{\infty}(A)$}.
    \item If $u,v \in U_{\infty}(A)$ and $j_{n,m}(u) =v$, then $u$ and $v$ would represent the same element inside $U_{\infty}(A)$\index{$U_{\infty}(A)$}.
\end{itemize}
From now on the notation $M_{\infty}(A)$, $G_{\infty}(A)$, $U_{\infty}(A)$, $Q_{\infty}(A)$ and $P_{\infty}(A)$ will refer to the sets obtained from identifying the elements using the equivalence relations coming from the injective *-homomorphisms $i_{n,m}$ and $j_{n,m}$.

\begin{remark}\label{remark:other_structure_over_M_infinity_A}
Some authors provide $M_{\infty}(A)$ with the structure of an inductive limit of C* algebras \citep[Section 3.3]{blackadar_k-theory_2012}, in which case is becomes a dense *-algebra of $\mathcal{K} \otimes A$ as is mentioned in \cref{section:stabilization_C_stal_algebra}. In this case the groups $K_0(A)$ and $K_1(A)$ can defined in terms of $Q(\mathcal{K} \otimes A)$ ($P(\mathcal{K} \otimes A)$) and $G(\mathcal{K} \otimes A)$ ($U(\mathcal{K} \otimes A)$), and their definition is going to be equivalent to the one we give in \cref{sec:group_K_0} and \cref{sec:group_K_1} (\citep[Section 8.1]{blackadar_k-theory_2012}, \citep[page 116]{wegge-olsen_k-theory_1993}). 
\end{remark}

\subsubsection{$K_0$}
\label{sec:group_K_0}

Let $A$ be a C* algebra, then, on the set $P_{\infty}(A)$, we can define the following equivalence relation, take $p \in P_{n}(A), \; q \in P_m (A)$, then,
$$ p \sim_h q \text{ if } \exists k > n,m \text{ s.t. } p \oplus 0_{k-n} \sim_h q \oplus 0_{k-m} \text{ over } P_{k}(A).$$
Denote 
$$V(A) := P_{\infty}(A) / \sim_h,$$
then, $V(A)$ becomes an abelian semigroup under the following addition (\citep[Proposition 6.1.3]{wegge-olsen_k-theory_1993}),
$$ [p] + [q] = [p \oplus q]  $$
and its neutral element is $0 = [0]$. Notice that \citep[Proposition 6.1.3]{wegge-olsen_k-theory_1993} does not use the equivalence relation $\sim_h$ to define $V(A)$, however, it is stated in \citep[Remark 6.1.2]{wegge-olsen_k-theory_1993} that $\sim_h$ define the same equivalence classes as the equivalence relation used in \citep[Proposition 6.1.3]{wegge-olsen_k-theory_1993}.

The mapping $A \mapsto V(A)$ is a covariant functor from the category of C* algebras into the category of abelian semigroups (\citep[Proposition 6.1.3]{wegge-olsen_k-theory_1993}), such that, if $\alpha: A \rightarrow B$ is a C* homomorphism, then the induced map $\alpha_*: V(A) \rightarrow V(B)$ given by
$$
\alpha_*\left(\left[\left(a_{i j}\right)\right]\right):=\left[\left(\alpha\left(a_{i j}\right)\right)\right]
$$
is a well-defined homomorphism of semigroups.

For the definition of $K_0(A)$ the following group is key, denote by $\mathcal{GT}(V(A))$ the Grothendieck group\index{Grothendieck group} of $V(A)$ (\citep[Appendix G]{wegge-olsen_k-theory_1993}), which can be seen as the set of formal differences of elements of $V(A)$, i.e.
$$ \mathcal{GT}(V(A))  = \{ [p] - [q] | [p], [q] \in V(A)   \}.$$  
We would like to emphasize that $[p] - [q]$ is a notation, but it does not refer to the application of substraction between elements of $V(A)$, in fact, $[p] - [q] = [p'] - [q]$ imply $[p] = [p']$ iff $V(A)$ has the cancelation property (\citep[Appendix G, proposition in page 296]{wegge-olsen_k-theory_1993}). We denote 
$$K_{00}(A) := \mathcal{GT}(V(A)),$$
and we have that $K_{00}(\mathbb{C}) = \mathbb{Z}$ \index{$K_{00}(\cdot)$} (\citep[Examples 6.1.4]{wegge-olsen_k-theory_1993}).

Let A be a C* algebra and let
$$
0 \longrightarrow A \stackrel{\iota}{\longrightarrow} A^{+} \underset{\lambda}{\stackrel{\pi}{\rightleftarrows}} \mathbb{C} \longrightarrow 0
$$
be the exact sequence of the unitization of $A$ (\cref{sec:unitization_of_C_star_algebras}), then, there is a group homomorphism (\citep[Appendix G, corollary in page 297]{wegge-olsen_k-theory_1993}) 
$$ \pi_* : K_{00}\left(A^{+}\right) \rightarrow K_{00}(\mathbb{C}), $$

\begin{definition}[group $K_0$]\label{definition:group_K_0}
Let $A$ be a C* algebra, then, set
$$
K_0(A):=\operatorname{Ker}\left(\pi_*: K_{00}\left(A^{+}\right) \rightarrow \mathbb{Z}\right) \subset K_{00}\left(A^{+}\right) .
$$\index{$K_0(\cdot)$}
\end{definition}

\begin{theorem}[Properties of $K_0$]\label{proposition:properties_of_K_0}
Let $A$ be a C* algebra, then:
\begin{itemize}
    \item \citep[Proposition 6.2.2]{wegge-olsen_k-theory_1993}: 
    \begin{itemize}
        \item if $A$ is unital then $K_0(A) = K_{00}(A)$
        \item $K_0(A^+) \simeq K_0(A) \oplus \mathbb{Z}$
    \end{itemize} 
    \item \citep[Proposition 6.2.7]{wegge-olsen_k-theory_1993}:
    \begin{itemize}
        \item For any $C^*$-algebra $A, K_0(A)$ is an abelian group.
        \item The elements of $K_0(A)$ can be visualized as formal differences
        $$
        [p]-[q]
        $$
        where $p$ and $q$ are projections in $M_k\left(A^{+}\right)$for some $k \in \mathbb{N}$ and $p-q \in$ $M_k(A)$. When $A$ is unital, $p$ and $q$ may be chosen in $M_k(A)$ rather than $M_k\left(A^{+}\right)$.
    \end{itemize}
    \item \citep[Proposition 6.2.4]{wegge-olsen_k-theory_1993}: $K_0$ is a covariant functor from the category of $C^*$ algebras to the category of abelian groups. Let $\alpha: A \rightarrow B$ be a homomorphism of C* algebras, then, there is an induced homomorphism of abelian groups given by,
    $$
    \alpha_*\left(\left[\left(x_{i j}\right)\right]-\left[\left(y_{i j}\right)\right]\right):=\left[\left(\alpha^{+} x_{i j}\right)\right]-\left[\left(\alpha^{+} y_{i j}\right)\right],
    $$
    where the matrices $\left(x_{i j}\right),\left(y_{i j}\right)$ are projections in $M_{\infty}\left(A^{+}\right)$, and $\alpha^{+}: A^{+} \rightarrow$ $B^{+}$is the unital morphism defined by $\alpha^{+}(a+\lambda):=\alpha(a)+\lambda$.

\end{itemize}
\end{theorem}

In \citep[Section 6.5]{wegge-olsen_k-theory_1993} and \citep[Table of K-groups]{rordam_introduction_2000} you can find a catalog of the commonly used $K_0$ groups.

\subsubsection{$K_1$}
\label{sec:group_K_1}

Let $A$ be a C* algebra, then, on the set $U_{\infty}(A^+)$ we can define the following equivalence relation, take $u \in U_{n}(A^+), \; v \in U_m (A^+)$, then,
$$ u \sim_h v \text{ if } \exists k > n,m \text{ s.t. } u \oplus 1_{k-n} \sim_h v \oplus 1_{k-m} \text{ over } U_{k}(A^+).$$
Under this setting, the Whitehead lemma (\cref{lemma:Whitehead_lemma_c_star_alg}) can be used to show that $U_{\infty}(A^+)/\sim_h$ becomes an abelian group (\citep[Proposition 7.1.2]{wegge-olsen_k-theory_1993}) under the following addition,
$$ [u] + [v] = [u \oplus v] .$$

\begin{definition}[group $K_1$]\label{definition:group_K_1}
Let A be a C* algebra, then, define
$$ K_1(A) = U_{\infty}(A^+) / \sim_h. $$\index{$K_1(\cdot)$}
When $u \in U_n(A^+)$, then, $[u] \in K_1(A)$ denotes the connected component containing $u \oplus 1_k$ for all $k \in \mathbb{N}$.
\end{definition}

\begin{theorem}[Properties of $K_1$]\label{proposition:properties_of_K1}
Let $A$ be a C* algebra, then
\begin{itemize}
    \item \citep[Proposition 7.1.2]{wegge-olsen_k-theory_1993}: $K_1(A)$ is a commutative group with addition given by $$ [u] + [v] := [u \oplus v] .$$
    \item \citep[Remarks 7.1.3]{wegge-olsen_k-theory_1993}: If $A$ is unital then
    $$ K_1(A) = U_{\infty}(A) / \sim_h. $$
    \item \citep[Proposition 7.1.6]{wegge-olsen_k-theory_1993}: $K_1$ is a covariant, homotopy invariant functor from the category of $C^*$ algebras to the category of abelian groups.
\end{itemize}

\end{theorem}

In \citep[Corollary 7.1.12]{wegge-olsen_k-theory_1993} and \citep[Table of K-groups]{rordam_introduction_2000} you can find various examples of $K_1$ groups.

\begin{remark}[$K_1(A)$ as equivalence of invertibles]\label{remark:K_1_equivalence_of_invertibles}
From propositon \cref{prop:U_A_retraction_of_G_A} we know that the connected components of $G(A)$ are in one to one correspondence to the connected components of $U(A)$, moreover, since $M_n(A)$ is also a unitary C* algebra we have that $U_n(A)$ is also a deformation retract of $G_n(A)$. Therefore, $u \sim_{h} v$ in $G_n(A)$ iff $u |u|^{-1} \sim_{h} v |v|^{-1}$ in $U_n(A)$, which in turn provides us with an isomorphism
$$ G_{\infty} (A) / \sim_{h} \;  \simeq \; U_{\infty} (A) / \sim_{h} \; = \; K_1(A). $$
\end{remark}

\begin{remark}[If $A$ is a separable C* algebras then $K_i(A)$ is at most countable]\label{remark:K_theory_of_separable_algebra_is_countble}
Let $A$ be a separable C* algebra, then
\begin{itemize}
    \item $K_0(A)$ is at most a countable group (\citep{2214789}),
    \item $K_1(A)$ is at most a countable group (\citep{3753144}).
\end{itemize}
\end{remark}

K groups of separable C* algebras are at most countable (\cref{remark:K_theory_of_separable_algebra_is_countble}), which makes them a great tool to assign indices to homotopy classes of unitaries and projections. Countable groups provide an intuitive framework to assign labels, and they have a natural topology that clearly differentiates points (discrete topology), whilst uncountable groups have no natural topology per se, and are way harder to deal with. For example, the C* algebras that contain the tight binding models for topological insulators are separable (\cref{remark:non_commu_brilluouin_torus_separable_C_star_algebra}), which implies that their K groups are countable (\cref{sec:top_inva_over_the_non_commutative_brillouim_torus}).

\section{$K_0$ and $K_1$ for smooth sub algebras}
\label{sec:K_0_and_K_1_for_smooth_sub_algebras}

Let $A$ be a unital C* algebra and $\mathcal{A}$ a smooth sub-algebra of $A$ (\cref{def:smooth_sub_algebra}), then we define $K_{0}(\mathcal{A})$ and $K_{1}(\mathcal{A})$ in the same way as for the C* algebra $A$, that is, using the homotopy equivalence relations over projections and unitaries on matrix algebras over $\mathcal{A}$.


\subsection{Equivalence relations on smooth sub algebras}
\label{sec:equivalence_relation_smooth_sub_algebras}

The characterization of unitaries and invertibles of C* algebras described in \cref{proposition:cannonical_smooth_paths_unitaries_and_invetibles_C_star_algebras} has a similar setup in unital smooth sub algebras because as is mentioned in \cref{sec:smooth_subalgebras}, many of the properties of C* algebras are present in smooth sub algebras, like the fact that the inversion is continuous (\cref{prop:pre_C_star_algebras_have_continuous_inversion}) and the set of invertible elements is open (\cref{lemma:some_properties_of_smooth_sub_algebras}). Following \cref{definition:unitization_smooth_sub_algebra} we denote by $\mathcal{A}^+$ the Fréchet algebra whose underlying set is $\{ (a, \lambda): \; a \in \mathcal{A}, \; \lambda \in \mathbb{C} \}$ provided with the following operations
\begin{itemize}
    \item $(a, \lambda) + (b , \gamma) := (a + b, \lambda + \gamma)$,
    \item $(a, \lambda)^* := (a^*, \lambda^*)$,
    \item $(a, \lambda) (b, \gamma) := (ab + \gamma a + \lambda b , \lambda \gamma)$.
\end{itemize}
Notice that if $A$ is a non unital C* algebra, then, $\mathcal{A}^+$ is a smooth sub algebra of $A^+$(\cref{corollary:unitization_preserves_smooth_sub_algebras}).

\begin{definition}\label{definition:homotopy_equivalence_relations_smooth_sub_algebra}
Let $A$ be a unital C* algebra and $\mathcal{A}$ a smooth sub algebra of $A$, then, we introduce the following notation      
\begin{itemize}
    \item If $p,q \in P(\mathcal{A})$, we say that $p \sim_h q$ inside $P(\mathcal{A})$ if there is a path of projections from $p$ to $q$ and that path is continuous with respect the topology of $\mathcal{A}$. (homotopy)\index{homotopy!of projections} \index{$\sim_h$ (homotopy)}
    \item If $e,f \in Q(\mathcal{A})$, we say that $p \sim_h q$ inside $Q(\mathcal{A})$ if there path of idempotents from $e$ to $f$ and that path is continuous with respect the topology of $\mathcal{A}$ . (homotopy)\index{homotopy!of idempotents}
    \item If $x,y \in G(\mathcal{A})$, we say that $x \sim_h y$ if there is a path of invertible elements of $\mathcal{A}$ from $x$ to $y$ and that path is continuous with respect the topology of $\mathcal{A}$. (homotopy)\index{homotopy!of invertible elements}
    \item If $u,v \in U(\mathcal{A})$, we say that $u \sim_h v$ if there is a path of unitaries in $\mathcal{A}$ from $u$ to $v$ and that path is continuos with respect the topology of $\mathcal{A}$. (homotopy)\index{homotopy!of unitaries}
\end{itemize}
\end{definition}

Smooth sub-algebras are well-behaved with respect to homotopic equivalence relations because we will be able to capture the structure of the path connected components of idempotents, projections, invertibles and unitaries of C* algebras through the connected components of the corresponding elements in smooth sub algebras.

\begin{proposition}[Invertibles and unitaries of smooth sub algebras are dense]\label{proposition:invertibles_and_untairies_of_smooth_algebras_are_dense}
Let $A$ be a unital C* algebra and $\mathcal{A}$ a smooth sub algebra of $A$ (\cref{sec:smooth_subalgebras}), then 
\begin{enumerate}
    \item $G(\mathcal{A})$ is dense in $G(A)$
    \item $U(\mathcal{A})$ is dense in $U(A)$
\end{enumerate}
\end{proposition}
\begin{proof}
\begin{enumerate}
    \item From \cref{proposition:GA_topological_group_and_open_in_banach_algebra} we know that $G(A)$ is open in $A$, thus, every element of $G(A)$ has a neighborhood $V_a$ of invertible elements. Since $\mathcal{A}$ is dense in $A$ (\cref{def:smooth_sub_algebra}) we know that there is an element $b$ of $\mathcal{A}$ that belongs to $V_a$, additionally, from \cref{lemma:some_properties_of_smooth_sub_algebras} we know that $G(\mathcal{A}) = \mathcal{A} \cap G(A)$, thus, $b \in G(\mathcal{A})$ and we get that $G(\mathcal{A})$ is dense in $G(A)$.
    \item From \cref{prop:U_A_retraction_of_G_A} we know that the map $\omega: G(A) \to U(A)$ given by the polar decomposition $z = |z| \omega(z)$ (\cref{proposition:polar_decomposition_invertible_elements}) is continuous and surjective. Since $G(\mathcal{A})$ is dense inside $G(A)$, for any $\epsilon > 0$ and any $u \in U(A) \subset G(A)$ we can find $x \in G(\mathcal{A})$ such that $\| \omega(x) - u \| \leq \epsilon$. From \cref{proposition:polar_decomposition_of_invertible_elements_smooth_sub_algebras} we know that $\omega(x) \in U(\mathcal{A})$, which implies that $U(\mathcal{A})$ is dense $U(A)$.
\end{enumerate}
\end{proof}

The following argument will be useful for the proof of \cref{proposition:idempotents_and_projections_of_smooth_sub_algebras_are_dense}. 

\begin{proposition}[Idempotents and projections of smooth sub algebras are dense]\label{proposition:idempotents_and_projections_of_smooth_sub_algebras_are_dense}
Let $A$ be a C* algebra and $\mathcal{A}$ a smooth sub algebra of $A$ (\cref{sec:smooth_subalgebras}), then 
\begin{enumerate}
    \item $Q(\mathcal{A})$ is dense in $Q(A)$
    \item $P(\mathcal{A})$ is dense in $P(A)$
\end{enumerate}
\end{proposition}
\begin{proof}
Before we outline the proofs for these statements we will look into a small result that will be helpful. Let $A$ be a Banach algebra without unit and $e \in Q(A^+)$ with $e = (a,\alpha)$, we have that
$$ e^2 = (a^2 + 2 \alpha a, \alpha^2) = (a, \alpha). $$
The only idempotents of $\mathbb{C}$ are $0,1$, thus we must have that $\alpha = 0$ or $\alpha =1$. If $\alpha = 1$ then we get that $a^2 = -a$, thus $ia \in Q(A)$, on the other side, if $\alpha = 0$ then $a^2 = a$, thus $a \in Q(A)$.

Now we proceed to proof for the results,
\begin{enumerate}
    \item Assume that $A$ has a unit and $e \in Q(A)$, in this case \cref{lemma:approximating_idempotents_with_idempotents} tell us that for every $\epsilon >0$ there is a $\delta >0$ such that, if $\| e - a \| \leq a$, then, there exists a function $g$ that is holomorphic on $\text{Sp}(a)$ such that $g(a) \in Q(A)$ and $\| g(a) - e \| \leq \epsilon$. Since $\mathcal{A}$ is dense in $A$ (\cref{def:smooth_sub_algebra}) we can chose $a \in \mathcal{A}$, and the invariance under holomorphic functional calculus of $\mathcal{A}$ tell us that $g(a) \in \mathcal{A}$, thus $Q(\mathcal{A})$ is dense in $Q(A)$. 
    
    If $A$ has no unit, then we use the previous argument to find an element $b \in Q(\mathcal{A}^+)$  such that $\| b - e \|_{A^+} \leq \epsilon$ when $e \in \mathcal{A}$, since $\mathcal{A}^+$ is invariant under the holomorphic calculus of $A^+$ by \cref{proposition:unitization_is_invariant_under_holomorphic_calculus}. Given that $b$ is an idempotent of $\mathcal{A}^+$, we could only have that $b = (b_0, \beta)$ with $\beta = 0$ of $\beta =1$, if $\beta = 0$ then $b \in \mathcal{A}$, also, if $\epsilon < 1$ then $\beta$ must be $0$ because $|\beta - 0 | \leq \| b - e \| \leq \epsilon $, so, taking $\epsilon <1$ guaranties that $b \in Q(\mathcal{A})$. 
    
     \item  From \cref{proposition:P_A_retraction_of_Q_A} we know that the map 
     $$e \to e e^*\left(1_{A^+}+\left(e-e^*\right)\left(e^*-e\right)\right)^{-1}$$
     takes an idempotent and returns a projection, such that $\eta(p) = p$ is $p$ is a projection. Also, the map $\eta : Q(A) \to P(A)$ is continuous, thus, if $p \in P(A)$ then there is $\delta > 0$ such that if $\| e - f \| \leq \delta$ then $\| \eta(e) - \eta(f) \| \leq \epsilon$. So, we just need to use the density of $Q(\mathcal{A})$ in $Q(A)$ to find a $e \in Q(\mathcal{A})$ such that $\| e - p \| \leq \delta$, which implies that $\| \eta(e) - p \| \leq \epsilon$. Since $\mathcal{A}$ is closed under the computation of involution (\cref{def:smooth_sub_algebra}) and the computation of inverse (\cref{lemma:some_properties_of_smooth_sub_algebras}), if we set $e \in Q(\mathcal{A})$ we get that $\eta(e) \in P(\mathcal{A})$, hence, $P(\mathcal{A})$ is dense in $P(A)$.
\end{enumerate}
\end{proof}

From \cref{proposition:idempotents_and_projections_of_smooth_sub_algebras_are_dense} and \cref{proposition:invertibles_and_untairies_of_smooth_algebras_are_dense} we get that smooth sub algebras allow us to arbitrarily approximate invertibles in their C* algebras with invertibles in the smooth sub algebra, and similar relations hold for unitaries, idempotents and projections. 

In the following, if $A$ is a C* algebra and $\mathcal{A}$ a smooth sub algebra of $A$, if we have a function $\eta: [0,1] \to \mathcal{A}$, we say that \textbf{$\eta$ is a smooth path}\index{smooth path} if $\eta$ is infinitely differentiable as a function from $[0,1]$ into $\mathcal{A}$, where $\mathcal{A}$ is understood as a Fréchet algebra (\cref{def:smooth_sub_algebra}) and the derivative of $\eta$ takes the usual definition as exposed in \cref{remark:leibniz_rule_fn_values_in_topological_algebra}.

\begin{lemma}\label{lemma:basci_smooth_paths_of_unitaries_and_invertibles_smooth_sub_algebra}
Let $A$ be a unital C* algebra and $\mathcal{A}$ a smooth sub algebra of $A$, then,
\begin{enumerate}
    \item If $a_1, \cdots, a_n$ are elements of $\mathcal{A}$, then, the following is a smooth path of invertible elements of $\mathcal{A}$
    $$ \eta(t) = \exp \left(t a_1\right) \exp \left(t a_2\right) \cdots \exp \left(t a_n\right).$$
    \item If $h_1, \cdots, h_n$ are self adjoint elements of $\mathcal{A}$, then, the following is a smooth path of unitaries of $\mathcal{A}$
    $$ \eta(t) = \exp \left(t i h_1\right) \exp \left(t i h_2\right) \cdots \exp \left(t ih_n\right).$$
\end{enumerate}
\end{lemma}
\begin{proof}
\begin{enumerate}
    \item Given that $a_j \in A$, from \cref{proposition:logarithms_and_n_roots} we know that $\exp \left(t a_j\right)$ is an invertible element for any $t \in \mathbb{C}$, additionally, since $\mathcal{A}$ is closed under the holomorphic functional calculus of $A$ (\cref{def:smooth_sub_algebra}), we have that $\exp \left(t a_j\right) \in \mathcal{A}$. Since the function $t \mapsto \exp(t)$ is an holomorphpic function, \cref{proposition:smoomth_paths_from_entire_functions_smooth_sub_algebras} tells us that the map $t \mapsto \exp \left(t a_j\right)$ is a smooth path of elements of $\mathcal{A}$, so, given that $\eta$ is a multiplication of such paths, the Leibniz rule (\cref{remark:leibniz_rule_fn_values_in_topological_algebra}) implies that $\eta$ is a smooth path of invertible elements of $\mathcal{A}$.
    \item  Given that $a_j \in A$, from \cref{theorem:description_of_UA} we know that $\exp \left(t i  h_j\right)$ is an unitary for any $t \in \mathbb{C}$, additionally, since $\mathcal{A}$ is closed under the holomorphic functional calculus of $A$ (\cref{def:smooth_sub_algebra}), we have that $\exp \left(t i  h_j\right) \in \mathcal{A}$. Since the function $t \mapsto \exp(t)$ is an holomorphpic function, \cref{proposition:smoomth_paths_from_entire_functions_smooth_sub_algebras} tells us that the map $t \mapsto \exp \left(t i h_j\right)$ is a smooth path of elements of $\mathcal{A}$, so, given that $\eta$ is a multiplication of such paths, the Leibniz rule (\cref{remark:leibniz_rule_fn_values_in_topological_algebra}) implies that $\eta$ is a smooth path of unitaries of $\mathcal{A}$.
\end{enumerate}
\end{proof}

\begin{remark}[The importance of the holomorphic functional calculus over smooth sub algebras]\label{remark:importance_funct_calculus_smooth_sub_algebras}
Let $A$ be a unital C* algebra and $\mathcal{A}$ a smooth sub algebra of $A$, take $a_1, \cdots, a_n$ as elements of $\mathcal{A}$. Since $\mathcal{A}$ is closed under the holomorphic functional calculus over $A$ (\cref{def:smooth_sub_algebra}), we can use the holomorphic functional calculus over $A$ (\cref{theorem:holomorphic_functional_calculus_banach_algebras}) to come up with a path, $\eta: [0,1] \to G(\mathcal{A})$ taking the following form 
$$ \eta(t) = \exp \left(t a_1\right) \exp \left(t a_2\right) \cdots \exp \left(t a_n\right),$$
and \cref{lemma:basci_smooth_paths_of_unitaries_and_invertibles} tell us that is is a smooth path from $[0,1]$ into $A$. Since the topology of $\mathcal{A}$ is stronger than the topology of $A$ (\cref{def:smooth_sub_algebra}), we cannot deduce that $\eta$ is a smooth path from $[0,1]$ into $\mathcal{A}$ using \cref{lemma:basci_smooth_paths_of_unitaries_and_invertibles}, thus, we need to use other tools. 

Under this setting, the holomorphic functional calculus over $\mathcal{A}$ (\cref{section:holomorphic_functional_calculus_for_smooth_sub_algebras}) provides us with the tools to prove that $\eta$ is a smooth path from $[0,1]$ into $\mathcal{A}$. Notice that the holomorphic functional calculus over $\mathcal{A}$ takes into account its topology as a Fréchet algebra, and this is the necessary setting to show that $\eta$ is a smooth function from $[0,1]$ into $\mathcal{A}$.
\end{remark}

\begin{proposition}(Canonical paths of idempotents and projections in smooth sub algebras)\label{proposition:cannonical_paths_idempotents_projections_smooth_sub_algebras}
Let $A$ be C* algebra and $\mathcal{A}$ a smooth sub algebra of $A$, then,
\begin{enumerate}
    \item Take $e,f \in Q(\mathcal{A})$, then $e \sim_h f$ in $Q(\mathcal{A})$ iff $e \sim_h f$ in $Q(A)$. Additionally, if $e,f \in Q(\mathcal{A})$ and $e \sim_h f$ inside $Q(\mathcal{A})$, there is a smooth path $\eta$ of idempotents inside $\mathcal{A}$ that connects $e$ with $f$ and takes the following form,
    $$ \eta(t) = \exp(-tb_n) \cdots \exp(-tb_1) e \exp(tb_1) \cdots \exp(tb_n), \; \eta(0) = e, \; \eta(1) = f,  $$
    where $b_1, \dots, b_n$ are elements of $\mathcal{A}^+$, and $b_1, \dots, b_n$ are elements of $\mathcal{A}$ is $A$ is a unital C* algebra.
    \item Take $p,q \in P(\mathcal{A})$, then $p \sim_h q$ in $P(\mathcal{A})$ iff $p \sim_h q$ in $P(A)$. Additionally, if $p,q \in P(\mathcal{A})$ and $p \sim_h q$ inside $P(\mathcal{A})$, there is a smooth path of projections inside $\mathcal{A}$ that connects $p$ with $q$ and takes the following form,
    $$ \eta(t) = \exp(-tih_n) \cdots \exp(-tih_1) p \exp(tih_1) \cdots \exp(tih_n), \; \eta(0) = p, \; \eta(1) = q, $$
    where $h_1, \dots, h_n$ are self adjoint elements of $\mathcal{A}^+$, and $h_1, \dots, h_n$ are elements of $\mathcal{A}$ is $A$ is a unital C* algebra.
\end{enumerate}
\end{proposition}
\begin{proof}
\begin{enumerate}
    \item If $e \sim_h f$ in $Q(\mathcal{A})$ then the definition of smooth sub algebra guaranties that $e \sim_h f$ in $Q(A)$, because any continuous path of idempotents in $\mathcal{A}$ is also a continuous path on idempotent in $A$.
    
    The converse implication is the important one, so, assume $e \sim_h f$ in $Q(A)$ with $\gamma: [0,1] \to Q(A)$ such that $\gamma(0)=e$ and $\gamma(1)=f$, then chose $k$ with $\| 2 \gamma(t) - 1_{A^+} \| < k$. The uniform continuity of the function 
    $$ t \to \gamma(t) $$
    asset us that we can find a partition $0 = t_0 \leq t_1 \leq \dots \leq t_n =1$ such that 
    $$\| \gamma(t) - \gamma(s) \| < 1/(2k), \; t,s \in [t_i, t_{i+1}].$$
    
    Use the density of $Q(\mathcal{A})$ on $Q(A)$ (\cref{proposition:idempotents_and_projections_of_smooth_sub_algebras_are_dense}) to find elements $\alpha_i \in Q(\mathcal{A})$ such that $\| \alpha_i - \gamma(t_i)\| \leq 1/(4k)$ and $\| 2 \alpha_i - 1_{A^+} \| < k$ for $1 \leq i \leq n-1$. Set $\alpha_0 = \gamma(0)$ and $\alpha_n = \gamma(1)$, then
    $$ \| \alpha_{i+1} - \alpha_i \| \leq \| \alpha_{i+1} - \gamma(t_{i+1})\| + \|\gamma(t_{i+1}) - \gamma(t_{i}) \| + \| \gamma(t_{i}) - \alpha_i \|$$
    $$ \leq 2/(4k) + 1/(2k) = 1/k ,$$
    and $\| 2 \alpha_i - 1_{A^+}\| \leq k$. Consequently
    $$ \| \alpha_{i+1} - \alpha_i \| \leq 1/k \leq 1/\|  2 \alpha_i - 1_{A^+} \| ,$$
    thus we can define $v_i = (2\alpha_i - 1_{A^+})((2\alpha_{i+1} - 1_{A^+})) + 1_{A^+}$ and we end up with
    $$ \| 1_{A^+} - v_i/2 \| \leq \|2\alpha_i - 1_{A^+} \| \|\alpha_i - \alpha_{i+1}\| \leq 1 ,$$
    therefore, $v_i /2$ is invertible and can be expressed $\exp(b_i) = v_i/2$ with $b_i \in A^+$ (\cref{proposition:log_of_element_close_to_identity}). Moreover, $b_i = \log(v_i/2)$ using functional calculus on $A^+$, thus, it belongs to $\mathcal{A}^+$ because $v_i/2 \in \mathcal{A}^+$, notice that also $\exp(b_i) \in \mathcal{A}^+$. Doing some algebraic manipulation we can check that $\alpha_i v_i/2 = (v_i/2) \alpha_{i+1} = \alpha_i \alpha_{i+1}$, hence 
    $$\exp(-b_i) \alpha_{i} \exp(b_i) = \alpha_{i+1}.$$
    
    We can compose these expressions for each $i$ to get
    $$ \alpha_{n} = \exp(-b_n) \exp(-b_{n-1}) \dots \exp(-b_1) \alpha_0 \exp(b_1) \dots \exp(b_{n-1}) \exp(b_n),$$
    which tell us that $e \sim_s f$ in $\mathcal{A}^+$. \cref{lemma:basci_smooth_paths_of_unitaries_and_invertibles_smooth_sub_algebra} assure us that the map $\eta: [0,1] \to Q(\mathcal{A}),$
    $$ \eta(t) = \left( \exp(-tb_n) \dots \exp(-tb_1) \right) e  \left( \exp(tb_1) \dots \exp(tb_n) \right) $$
    is a smooth homotopy from $e$ to $f$. As in C* algebras, $\eta$ takes values in $\mathcal{A}$ instead of $\mathcal{A}^+$ because $\mathcal{A}$ is an ideal of $\mathcal{A}^+$ (\cref{lemma:properties_of_unitization_of_smooth_sub_algebras}) and $e \in \mathcal{A}$. Thus, $e \sim_h f$ in $Q(\mathcal{A})$ as desire.
    If $A$ is a unital C* algebra then $\mathcal{A}$ has a unit (\cref{lemma:some_properties_of_smooth_sub_algebras}), thus, we can use the holomorphic functional calculus over $\mathcal{A}$ to find $b_1, \dots, b_n$ and we get that $b_1, \dots, b_n$ are elemenst of $\mathcal{A}$.
    
    \item You can follow a similar argument as the in the previous item, that is, use the constructions from \cref{proposition:close_projetions_are_homotopic} and \cref{proposition:canonical_paths_between_projections} to show that, if $p \sim_h q$ in $P(A)$, then $e \sim_u f$ in $\mathcal{A}^+$ and the there is a canonical path $\eta: [0,1] \to P(\mathcal{A}),$
    $$ \eta(t) = \left( \exp(-tih_n) \cdots \exp(-tih_1) \right) p \left( \exp(tih_1) \dots \exp(tih_n) \right) $$
    where $h_1, \dots, h_n $ self adjoint elements of $\mathcal{A}^+$, additionally $\eta$ is smooth.
    If $A$ is a unital C* algebra then $\mathcal{A}$ has a unit (\cref{lemma:some_properties_of_smooth_sub_algebras}), thus, we can use the holomorphic functional calculus over $\mathcal{A}$ to find $h_1, \dots, h_n $ and we get that $h_1, \dots, h_n $ are elemenst of $\mathcal{A}$.
\end{enumerate}
\end{proof}

\begin{proposition}(Canonical paths of invertibles and unitaries in smooth sub algebras)\label{proposition:cannonical_paths_invertibles_unitaries_smooth_sub_algebras}
Let $A$ be a unital C* algebra and $\mathcal{A}$ a smooth sub algebra of $A$, then,
\begin{enumerate}
    \item Take $x,y \in G(\mathcal{A})$, then $x \sim_h y$ in $G(\mathcal{A})$ iff $x \sim_h y$ in $G(A)$. Additionally, if $x,y \in G(\mathcal{A})$ and $x \sim_h y$ inside $G(\mathcal{A})$, there is a smooth path of invertible elements of $\mathcal{A}$ that connects $x$ to $y$ and takes the following form,
    $$ \eta(t) = x \exp(tb_1) \dots \exp(tb_{n-1}) \exp(tb_n), \; \eta(0) = x, \; \eta(1) = y,  $$
    where $b_1,\cdots, b_n $ are elements of $\mathcal{A}$.
    \item Take $u,v \in U(\mathcal{A})$, then $u \sim_h v$ in $U(\mathcal{A})$ iff $u \sim_h v$ in $U(A)$. Additionally, if $u,v \in U(\mathcal{A})$ and $u \sim_h v$ inside $U(\mathcal{A})$, there is a smooth path of unitaries inside $\mathcal{A}$ that connects $u$ and $v$ and takes the following form
    $$ \eta(t) = u \exp(tih_1) \dots \exp(tih_{n-1}) \exp(tih_n), \; \eta(0) = u, \; \eta(1) = v,  $$
    where $h_1, \cdots, h_n$ are self adjoint operators of $\mathcal{A}$.
\end{enumerate}
\end{proposition}
\begin{proof}
\begin{enumerate}
    \item If $x \sim_h y$ in $G(\mathcal{A})$ then $x \sim_h y$ in $G(A)$ because the inclusion $i: \mathcal{A} \to A$ is continuous.
    
    The converse implication is the important one, so, assume $x \sim_h y$ in $G(A)$ with $\gamma: [0,1] \to G(A)$ such that $\gamma(0)=x$ and $\gamma(1)=y$, then chose $k$ with $\| (\gamma(t))^{-1} \| < k$. The uniform continuity of the function 
    $$ t \to \gamma(t) $$
    asset us that we can find a partition $0 = t_0 \leq t_1 \leq \dots \leq t_n =1$ such that 
    $$\| \gamma(t) - \gamma(s) \| < 1/(2k), \; t,s \in [t_i, t_{i+1}].$$
    
    From the proof of \cref{theorem: description_of_GA} we have that if $\|b -a \| \leq \| a^{-1} \|^{-1} $ then $\log(z)$ is an holomorphic function on the spectrum of $a^{-1} b$, thus $b = a \exp(\log(a^{-1} b))$, so, if $a,b \in \mathcal{A}$ then $\log(a^{-1} b) \in \mathcal{A}$.
    
    Use the density of $G(\mathcal{A})$ on $G(A)$ (\cref{proposition:idempotents_and_projections_of_smooth_sub_algebras_are_dense}) to find elements $\alpha_i \in G(\mathcal{A})$ such that $\| \alpha_i - \gamma(t_i)\| \leq 1/(4k)$ and $\| (\alpha_i)^{-1} \| < k$ for $1 \leq i \leq n-1$. Set $\alpha_0 = \gamma(0)$ and $\alpha_n = \gamma(1)$, then
    $$ \| \alpha_{i+1} - \alpha_i \| \leq \| \alpha_{i+1} - \gamma(t_{i+1})\| + \|\gamma(t_{i+1}) - \gamma(t_{i}) \| + \| \gamma(t_{i}) - \alpha_i \|$$
    $$ \leq 2/(4k) + 1/(2k) = 1/k ,$$
    and $\|(\alpha_i)^{-1}\| \leq k$. Consequently
    $$ \| \alpha_{i+1} - \alpha_i \| \leq 1/k \leq 1/\| (\alpha_i)^{-1} \| ,$$
    thus we have that $\log(z)$ is holomorphic on $\alpha_i^{-1} \alpha_{i+1}$, so if $b_i = \log(\alpha_i^{-1} \alpha_{i+1})$ we get that $\alpha_{i+1} = \alpha_i \exp(b_i)$ with $b_i, \exp(b_i) \in \mathcal{A}$ because it is invariant under holomorphic calculus of $A$.
    
    We can compose these expressions for each $i$ to get
    $$\alpha_n = \alpha_0 \exp(b_1) \dots \exp(b_n), \; b_i \in \mathcal{A},$$
    and \cref{lemma:basci_smooth_paths_of_unitaries_and_invertibles_smooth_sub_algebra} assure us that the map $\eta: [0,1] \to G(\mathcal{A}),$
    $$ \eta(t) = x \exp(tb_1) \cdots \exp(tb_{n-1}) \exp(tb_n)  $$
    is a smooth homotopy from $x$ to $y$. Thus, $x \sim_h y$ in $G(\mathcal{A})$ as desire.
  
    \item You can follow a similar argument as the in the previous item, that is, use the constructions from \cref{lemma:connected_component_of_unitaries} and \cref{theorem:description_of_UA} to show that, if $u \sim_h v$ in $U(A)$, then $u \sim_h v$ in $U(\mathcal{A})$ through the canonical path $\eta: [0,1] \to U(\mathcal{A}),$
    $$ \eta(t) = u \exp(tih_1) \cdots \exp(tih_{n-1}) \exp(tih_n)  $$
    where $h_1, \dots, h_n$ self adjoint elements of $\mathcal{A}$, additionally $\eta$ is smooth.
\end{enumerate}
\end{proof}

\subsection{$K_0$ and $K_1$}
\label{sec:def_K_0_and_K_1_smooth_subalgebras}

Let $A$ be an unital C* algebra and $\mathcal{A}$ a smooth sub algebra of $A$, from \cref{prop:matrix_alegbras_of_pre_C_star_algebras} we know that for any $n \in \mathbb{N}$ the algebra $M_n(\mathcal{A})$ is a smooth sub algebra of $M_n(A)$. We introduce the followint notation,
\begin{enumerate}
    \item We denote by $Q_n(\mathcal{A})$ the set of idempotents of $M_n(\mathcal{A})$.
    \item We denote by $P_n(\mathcal{A})$ the set of projections of $M_n(\mathcal{A})$.
    \item We denote by $G_n(\mathcal{A})$ the set of invertible elements of of $M_n(\mathcal{A})$.
    \item We denote by $U_n(\mathcal{A})$ the set of unitaries of $M_n(\mathcal{A})$.
\end{enumerate}
The maps $i_{m,n}$ and $j_{m,n}$ will have the same form as in (\cref{sec:groups_K_0_and_K_1}), also, we introduce the sets,
$$
M_{\infty}(\mathcal{A}):=\bigcup_{n=1}^{\infty} M_n(\mathcal{A}) ; \quad U_{\infty}(\mathcal{A}):=\bigcup_{n=1}^{\infty} U_n(\mathcal{A}) ; \quad G_{\infty}(\mathcal{A}):=\bigcup_{n=1}^{\infty} G_n(\mathcal{A}) ;$$
$$ \quad Q_{\infty}(\mathcal{A}):=\bigcup_{n=1}^{\infty} Q_n(\mathcal{A}) ;\quad P_{\infty}(\mathcal{A}):=\bigcup_{n=1}^{\infty} P_n(\mathcal{A}).
$$
On the aforementioned sets we will introduce the following equivalence relations,
\begin{itemize}
    \item If $e,f \in M_{\infty}(\mathcal{A})$ and $i_{n,m}(e) =f$, then $e$ and $f$ would represent the same element inside $M_{\infty}(\mathcal{A})$.
    \item If $e,f \in Q_{\infty}(\mathcal{A})$ and $i_{n,m}(e) =f$, then $e$ and $f$ would represent the same element inside $Q_{\infty}(\mathcal{A})$.
    \item If $p,q \in P_{\infty}(\mathcal{A})$ and $i_{n,m}(p) =q$, then $p$ and $q$ would represent the same element inside $P_{\infty}(\mathcal{A})$.
    \item If $v,w \in G_{\infty}(\mathcal{A})$ and $j_{n,m}(v) =w$, then $v$ and $w$ would represent the same element inside $G_{\infty}(\mathcal{A})$.
    \item If $u,v \in U_{\infty}(A\mathcal{A}$ and $j_{n,m}(u) =v$, then $u$ and $v$ would represent the same element inside $U_{\infty}(\mathcal{A})$.
\end{itemize}
From now on the notation $M_{\infty}(\mathcal{A})$, $G_{\infty}(\mathcal{A})$, $U_{\infty}(\mathcal{A})$, $Q_{\infty}(\mathcal{A})$ and $P_{\infty}(\mathcal{A})$ will refer to the sets obtained from identifying the elements using the equivalence relations coming from the injective *-homomorphisms $i_{n,m}$ and $j_{n,m}$.

On the set $P_{\infty}(\mathcal{A})$, we can define the following equivalence relation, take $p \in P_{n}(\mathcal{A}), \; q \in P_m (\mathcal{A})$, then,
$$ p \sim_h q \text{ if } \exists k > n,m \text{ s.t. } p \oplus 0_{k-n} \sim_h q \oplus 0_{k-m} \text{ over } P_{k}(\mathcal{A}).$$
Denote 
$$V(\mathcal{A}) := P_{\infty}(\mathcal{A}) / \sim_h,$$
then, given that for any $n \in \mathbb{N}$ the algebra $M_n(\mathcal{A})$ is a smooth sub algebra of $M_n(A)$ (\cref{prop:matrix_alegbras_of_pre_C_star_algebras}), \cref{proposition:idempotents_and_projections_of_smooth_sub_algebras_are_dense} tells us that $P_n(\mathcal{A})$ is dense inside $P(A)$, therefore,\cref{proposition:cannonical_paths_idempotents_projections_smooth_sub_algebras} implies that the elements of $V(\mathcal{A})$ are in one to one correspondence with the elements of $V(A)$. Since the map
$$ (p,q) \mapsto [p \oplus q]  $$
maps elements of $P_{\infty}(\mathcal{A})$ into elements of $P_{\infty}(\mathcal{A})$, then, the fact that $V(A)$ is an abelian group (\cref{sec:group_K_0}) implies that $V(\mathcal{A})$ is also an abelian group under the addition 
$$ [p] + [q] = [p \oplus q] .$$

The previous discussion gives the following result, 

\begin{proposition}[Isomorphism of semigroups $P_{\infty}(A) / \sim_h$ and $P_{\infty}(\mathcal{A}) / \sim_h$]\label{proposition:isomorphism_of_semigroups_of_projections_for_smooth_sub_algebras}
Let $A$ be a unital C* algebra and $\mathcal{A}$ a smooth sub algebra of $A$ , then, we have the following isomorphism of abelian semi groups
$$P_{\infty}(A) / \sim_h \; \simeq \; P_{\infty}(\mathcal{A}) / \sim_h,$$
and the isomorphism of abelian groups
$$ V(\mathcal{A}) \simeq V(A).$$
\end{proposition}

On the set $U_{\infty}(\mathcal{A})$ we can define the following equivalence relation, take $u \in U_{n}(\mathcal{A}), \; v \in U_m (\mathcal{A})$, then,
$$ u \sim_h v \text{ if } \exists k > n,m \text{ s.t. } u \oplus 1_{k-n} \sim_h v \oplus 1_{k-m} \text{ over } U_{k}(\mathcal{A}).$$
Given that for any $n \in \mathbb{N}$ the algebra $M_n(\mathcal{A})$ is a smooth sub algebra of $M_n(A)$ (\cref{prop:matrix_alegbras_of_pre_C_star_algebras}), \cref{proposition:invertibles_and_untairies_of_smooth_algebras_are_dense} tells us that $U_n(\mathcal{A})$ is dense inside $U_n(A)$, therefore,  \cref{proposition:cannonical_paths_invertibles_unitaries_smooth_sub_algebras} implies that the elements of $U_{\infty}(\mathcal{A})/\sim_h$ are in one to one correspondence with the elements of $U_{\infty}(A)/\sim_h$. Since the map
$$ (u,v) \mapsto [u \oplus v]  $$
maps elements of $U_{\infty}(\mathcal{A})$ into elements of $U_{\infty}(\mathcal{A})$, then, the fact that $U_{\infty}(A)/\sim_h$ is an abelian group (\cref{sec:group_K_1}) implies that $U_{\infty}(\mathcal{A}) / \sim_h$ is also an abelian group under the addition 
$$ [u] + [v] = [u \oplus v] .$$

The previous discussion gives the following result, 

\begin{proposition}[Isomorphism of groups $U_{\infty}(A) / \sim_h$ and $U_{\infty}(\mathcal{A}) / \sim_h$]\label{proposition:isomorphism_of_groups_of_unitaries_for_smooth_sub_algebras}
Let $A$ be a unital C* algebra and $\mathcal{A}$ a smooth sub algebra of $A$ , then, we have the following isomorphism of abelian groups,
$$U_{\infty}(A) / \sim_h \; \simeq \; U_{\infty}(\mathcal{A}) / \sim_h.$$
\end{proposition}

\begin{definition}[K groups for smooth sub algebras]\label{definition:K_groups_smooth_sub_algebras}
Let $A$ be a unital C* algebra and $\mathcal{A}$ a smooth sub algebra of $A$, then we define,
$$ K_0(\mathcal{A}) = \mathcal{GT}(V(\mathcal{A})),$$\index{$K_0(\cdot)$}
and 
$$ K_1(\mathcal{A}) = U_{\infty}(\mathcal{A})/\sim_{h}.$$\index{$K_1(\cdot)$}
\end{definition}

The isomorphisms $K_0(\mathcal{A}) \simeq K_0(A)$  and $K_1(\mathcal{A}) \simeq K_1(A)$ are a direct consequence of the results on \cref{proposition:cannonical_paths_invertibles_unitaries_smooth_sub_algebras} and \cref{proposition:cannonical_paths_idempotents_projections_smooth_sub_algebras}, as we proceed to check

\begin{theorem}[$K_0(\mathcal{A}) \simeq K_0(A)$ ($K_1(\mathcal{A}) \simeq K_1(A)$)]\label{theorem:isomorphism_K_0_and_k_1_for_smooth_sub_algebras}
Let $A$ be a unital C* algebra and $\mathcal{A}$ a smooth sub algebra of $A$, then,
\begin{enumerate}
    \item $ K_0(\mathcal{A}) = K_0(A).$
    \item $ K_1(\mathcal{A}) = K_1(A).$    
\end{enumerate}
\end{theorem}
\begin{proof}
\begin{enumerate}
    \item Given that $K_0(A) = V(A)$ (\cref{proposition:properties_of_K_0}) and $K_0(\mathcal{A}) = V(\mathcal{A})$ \cref{definition:K_groups_smooth_sub_algebras}, this is a diret consequence of \cref{proposition:isomorphism_of_semigroups_of_projections_for_smooth_sub_algebras}.
    \item Given that $K_1(A) = U_{\infty}(A)/ \sim_h$ (\cref{proposition:properties_of_K1}) and $K_1(\mathcal{A}) = U_{\infty}(\mathcal{A}) / \sim_h$ \cref{definition:K_groups_smooth_sub_algebras}, this is a diret consequence of \cref{proposition:isomorphism_of_groups_of_unitaries_for_smooth_sub_algebras}.    
\end{enumerate}
\end{proof}

The isomorphisms $K_0(\mathcal{A}) \simeq K_0(A)$  and $K_1(\mathcal{A}) \simeq K_1(A)$ are well known in the literature, however, we were not able to find a proof that we could follow, and that is one of the main motivation behind the setup of this thesis. In \cref{sec:alternative_proofs_for_the_isomorphism_of_K_groups} you can find a brief discussion on the alternative proofs of \cref{theorem:isomorphism_K_0_and_k_1_for_smooth_sub_algebras} that we were able to find in the literature. Those isomorphisms come from the special relation established between the holomorphic functional calculus on both $\mathcal{A}$ and $A$, such relation is usually taken for granted when dealing with smooth sub algebras but rarely explicitly stated, which can make the literature hard to follow for newcomers. 

With \cref{theorem:isomorphism_K_0_and_k_1_for_smooth_sub_algebras} we have achieved the main purpose of the present chapter and we have looked into a key result of non-commutative differential topology whose applications have been exposed in \cref{section:mathematical_framework}.

\begin{remark}\label{remark:k_groups_and_the_Non_cOmmutative_brillouin_torus}
Let $ (C(\Omega)\rtimes_{\alpha, \Theta} \mathbb{Z}^d, \mathcal{C}(\Omega)_{\alpha, \Theta}) $ be The Non-commutative Brillouin Torus (\cref{definition:non_commutative_brillouin_torus}), since $C(\Omega)\rtimes_{\alpha, \Theta} \mathbb{Z}^d$ is an unital C* algebra and $\mathcal{C}(\Omega)_{\alpha, \Theta}$ is a smooth sub algebra of $C(\Omega)\rtimes_{\alpha, \Theta} \mathbb{Z}^d$ (\cref{corollary:smooth_elements_are_invariant_under_holomoprhic_and_c_infty_calculus}), from \cref{theorem:isomorphism_K_0_and_k_1_for_smooth_sub_algebras} we have the following isomorphisms,
$$ K_0(C(\Omega)\rtimes_{\alpha, \Theta} \mathbb{Z}^d) \simeq K_0 (\mathcal{C}(\Omega)_{\alpha, \Theta}), \;  K_1(C(\Omega)\rtimes_{\alpha, \Theta} \mathbb{Z}^d) \simeq K_1 (\mathcal{C}(\Omega)_{\alpha, \Theta}) .$$
\end{remark}

\chapter{Non Commutative Geometry and Topological invariants of Hamiltonians}
\label{chap:non_commutative_geometry_and_topological_invariants_for_hamiltonians}

In the present chapter we give a high-level description of the conceptual framework from \citep{prodan_bulk_2016} which is designed to study topological invariants of effective models for the dynamics of electrons on a crystal subject to weak disorder, we will focus on how the results of the present document fit within that framework, especially the results of \cref{chap:fourier_analysis} and \cref{chap:K_theory}.

Let $(C(\Omega)\rtimes_{\alpha, \Theta} \mathbb{Z}^d, \mathcal{C}(\Omega)_{\alpha, \Theta})$ be the Non Commutative Brillouin torus (\cref{definition:non_commutative_brillouin_torus}), the previously mentioned topological invariants come from the bilinear pairings (\cref{section:mathematical_framework}),
$$ \langle \cdot, \cdot \rangle : K_0(C(\Omega)\rtimes_{\alpha, \Theta} \mathbb{Z}^d) \times HC^{\text{ev}}_{\text{con}}( \mathcal{C}(\Omega)_{\alpha, \Theta}) \to \mathbb{C}, $$ 
$$ \langle \cdot ,\cdot \rangle : K_1(C(\Omega)\rtimes_{\alpha, \Theta} \mathbb{Z}^d) \times HC^{\text{odd}}_{\text{con}}( \mathcal{C}(\Omega)_{\alpha, \Theta}) \to \mathbb{C}, $$
where $HC^{\text{odd}}_{\text{con}}( \mathcal{C}(\Omega)_{\alpha, \Theta})$ and $HC^{\text{ev}}_{\text{con}}( \mathcal{C}(\Omega)_{\alpha, \Theta})$ are the components of the continuous cyclic cohomology of $ \mathcal{C}(\Omega)_{\alpha, \Theta}$ (\cref{sec:cyclic cohomology definition}). The aforementioned bilinear pairings take the following form (\cref{sec:even_cyclic_cocycles_non_commutative_brillouin_torus}, \cref{sec:odd_cyclic_cocycles_non_commutative_brillouin_torus}),
$$ \langle [P_{h, g}], [\xi_I] \rangle := \Lambda_n \sum_{\rho \in S_n} (-1)^{\rho} \int_{\Omega} \text{tr} \left( \Phi_0 \left( P_{h,g} (\partial_{\rho(1)} P_{h,g}) \cdots (\partial_{\rho(n)} P_{h,g}) \right)(\omega) \right) d \mu(\omega),   $$
and 
$$ \langle [ U_{h,0}], [\xi_I] \rangle := \Lambda_n \sum_{\rho \in S_n} (-1)^{\rho} \int_{\omega \in \Omega} \text{tr} \left( \Phi_0 \left(\prod_{l = 1}^{n} U_{h,0}^* (\partial_{\rho(l)} U_{h,0}) \right)(\omega) \right) d \mu(\omega),   $$
where $U_{h,0}$ is a unitary associated to a self-adjoint element of $M_n(\mathcal{C}(\Omega)_{\alpha, \Theta})$, $P_{h, g}$ is a projection associated to a self-adjoint element of $M_n(\mathcal{C}(\Omega)_{\alpha, \Theta})$, $\Phi_0(p)$ is the zeroth Fourier coefficient of $p$ (\cref{definition:fourier_coefficients}), $I$ is an ordered tuple of $n$ elements taken from $\{ 1, \cdots, d \}$, $\Omega$ is a compact topological space related the Non-Commutative Brillouin Torus and $\xi_I$ is an element of the continuous cyclic cohomology of $\mathcal{C}(\Omega)_{\alpha, \Theta}$. These pairings are a generalization of the even and odd Chern numbers.

In the limit of temperature near zero and infinite relaxation time, the pairings $\langle [P_{h, g}], [\xi_I] \rangle$ and $\langle [ U_{h,0}], [\xi_I] \rangle$ describe macroscopic properties of covariant families of Hamiltonians (\cref{sec:motivation_from_physics}, \citep[Section 7.2]{prodan_bulk_2016}). For example, if we work with $2$ dimensional systems and $I = \{ 1,2\}$, then, $\langle [P_{\text{Fermi}}], [\xi_I] \rangle$ describes the entry $(1,2)$ of the conductivity tensor, and we have that 
$$\langle [P_{\text{Fermi}}], [\xi_I] \rangle \subset \mathbb{Z},$$
where $P_{\text{Fermi}}$ is the Fermi projection of the physical system under study. The previously mentioned case corresponds to the Integer Quantum Hall Effect (\cref{sec:motivation_from_physics}). The topological invariants studied in \citep{prodan_bulk_2016} can be defined for physical systems with unitary symmetry or physical systems with chiral symmetry.

\section{Mathematics framework}
\label{section:mathematical_framework}

In this section, we will look at how the mathematical framework of \cref{chap:fourier_analysis} and \cref{chap:K_theory} fits within the analysis of bulk models of homogeneous materials (\cref{sec:motivation_from_physics}), which is a key part of the study of topological insulators and the bulk-boundary correspondence (\cite{prodan_bulk_2016}).

Let $A$ be a C* algebra, then, we use the term the K-theory of $A$ to refer to the pair of groups $K_0(A)$ (\cref{definition:group_K_0}) and $K_1(A)$ (\cref{definition:group_K_1}). Let $\mathcal{A}$ be a Frech\'et algebra, then, we use the term the continuous cyclic cohomology of $\mathcal{A}$ to refer to the modulus $HC^{*}_{\text{con}}(\mathcal{A})$ (\cref{definition:cyclic_cohomology}). Let $A$ be a C* algebra with an unit and $\mathcal{A}$ a smooth sub algebra of $A$, then, there is bilinear pairing between the K-theory of $A$ and the continuous Cyclic cohomology of $\mathcal{A}$ which is constructed as follows,
\begin{enumerate}
    \item From \cref{proposition:pairing_topological_k_0_even_ciclyc_cohomology} there is a bilinear pairing between $K_0 (\mathcal{A})$ and $HC^{\text{ev}}_{\text{con}}(\mathcal{A})$,
    $$ \langle \cdot, \cdot \rangle : K_0(\mathcal{A}) \times HC^{\text{ev}}_{\text{con}}(\mathcal{A}) \to \mathbb{C}$$
    and we have that
    $$ \langle [p], [\phi] \rangle \propto \phi(p, p, \cdots, p), \; \text{ for } \phi = HC^{2n}_{\text{con}}(\mathcal{A}), \; [p] \in K_0(\mathcal{A}) .$$
    From \cref{proposition:pairing_topological_k_1_odd_ciclyc_cohomology} there is a bilinear pairing between $K_1 (\mathcal{A})$ and $HC^{\text{odd}}_{\text{con}}(\mathcal{A})$, 
    $$ \langle \cdot, \cdot \rangle : K_1(\mathcal{A}) \times HC^{\text{odd}}_{\text{con}}(\mathcal{A}) \to \mathbb{C}, $$
    and we have that
    $$ \langle [u], [\phi] \rangle \propto \phi(u^*-1, u-1, u^*-1, \cdots, u^*-1, u-1),$$
     for  $\phi = HC^{2n+1}(\mathcal{A})_{\text{con}}, \; [u] \in K_1(\mathcal{A})$.
    \item \cref{theorem:isomorphism_K_0_and_k_1_for_smooth_sub_algebras} states that there are two isomorphisms $K_0 (A) \simeq K_0 (\mathcal{A})$ and $K_1 (A) \simeq K_1 (\mathcal{A})$,
    therefore, we can provide the bilinear pairings\index{bilinear pairing} (\cref{proposition:pairing_topological_k_0_even_ciclyc_cohomology_C_star_algebras}, \cref{proposition:pairing_topological_k_1_odd_ciclyc_cohomology_C_star_algebras}),
    $$ \langle \cdot, \cdot \rangle : K_0(A) \times HC^{\text{ev}}_{\text{con}}(\mathcal{A}) \to \mathbb{C}, \; \; \;  \langle \cdot, \cdot \rangle : K_1(A) \times HC^{\text{odd}}_{\text{con}}(\mathcal{A}) \to \mathbb{C}. $$
\end{enumerate}

Since the K-theory groups $\boldsymbol{K_i(A), \; i=0,1}$ are constructed using homotopy equivalent classes of projections/unitaries, this pairing becomes a \textbf{topological invariant}\index{topological invariant} and it is quite a powerful tool because it allows us to provide a test on whether two projections/unitaries are not homotopically equivalent. Under the Non-Commutative Geometry dictionary, the C* algebra $A$ represents the \textbf{topology of a non-commutative space}, whilst the Fréchet algebra $\mathcal{A}$ represents the \textbf{smooth structure of the same non-commutative space}, which in this context, makes it a \textbf{non-commutative smooth manifold}. Having said this, the pairing between $\boldsymbol{HC^*_{\text{con}}(\mathcal{A})}$ and $\boldsymbol{K_j(A), \; j=0,1}$ belongs to the field of \textbf{non-commutative differential topology} since we are using the smooth structure of a non-commutative smooth manifold to provide topological invariants to that non-commutative space.

The bilinear pairing 
$$ \langle \cdot, \cdot \rangle : K_0(A) \times HC^{\text{ev}}_{\text{con}}(\mathcal{A}) \to \mathbb{C}, \; \langle \cdot, \cdot \rangle : K_1(A) \times HC^{\text{odd}}_{\text{con}}(\mathcal{A}) \to \mathbb{C}, $$
behaves well with respect to the vector space structure of both $K_i(A)$ and $HC^m_{\text{con}}(\mathcal{A})$, with this we mean that:
\begin{itemize}
    \item Given that $\boldsymbol{K_i(A), \; i = 0,1}$ are abelian groups, we have that
    $$
    \begin{aligned}
        \langle [p] + [q] , \phi \rangle = \langle [p] , \phi \rangle + \langle [q] , \phi \rangle , \; [p], [q] \in K_0(A), \; \phi \in HC^{\text{ev}}_{\text{con}}(\mathcal{A}), \\
        \langle [v] + [u] , \phi \rangle = \langle [v] , \phi \rangle + \langle [u] , \phi \rangle , \; [v], [u] \in K_1(A), \; \phi \in HC^{\text{odd}}_{\text{con}}(\mathcal{A}).
    \end{aligned}
    $$ 
    \item Given that both $\boldsymbol{HC^{\text{ev}}_{\text{con}}(\mathcal{A}), \; HC^{\text{odd}}_{\text{con}}(\mathcal{A})}$ are modules, we have that
    $$
    \begin{aligned}
        \langle [p] , \phi + \psi \rangle = \langle [p] , \phi \rangle + \langle [p] , \psi \rangle , \; [p] \in K_0(A), \; \phi, \psi \in HC^{\text{ev}}_{\text{con}}(\mathcal{A}), \\
        \langle [v] , \phi \rangle = \langle [v] , \phi \rangle + \langle [v] , \psi \rangle , \; [v] \in K_1(A), \; \phi, \psi \in HC^{\text{odd}}_{\text{con}}(\mathcal{A}).
    \end{aligned}
    $$
\end{itemize}

\begin{remark}[Why to use smooth sub algebras]\label{remark:why_we_use_smooth_sub_algebras}
These are three reasons why smooth sub algebras are used,
\begin{itemize}
    \item \textbf{By necessity:} One of the components of the Non-commutative Brillouin Torus is the C* algebra $C(\Omega) \rtimes_{\alpha,\Theta} \mathbb{Z}^d$ (\cref{definition:non_commutative_brillouin_torus}), turns out that this C* algebra is nuclear (\cref{remark:non_commu_brilluouin_torus_nuclear_C_star_algebra}), and the continuous cyclic cohomology of nuclear C* algebras has trivial information (\cref{example:nuclear_C_star_algebras}). Since the continuous cyclic cohomology of $C(\Omega) \rtimes_{\alpha,\Theta} \mathbb{Z}^d$ has no relevant information, we need to resort to other objects, in this case we resort to the smooth sub algebra $\mathcal{C}(\Omega)_{\alpha, \Theta})$ motivated by the fact that $C^{\infty}(V)$ is a smooth sub algebra of $C(V)$ (\citep[Chapter 3, section 2]{connes_noncommutative_2014}).

    \item \textbf{By convenience:} Let $A$ be a C* algebra with unit and $\mathcal{A}$ be a smooth sub algebra of $A$, then, $\mathcal{A}$ has all the information needed to compute the K theory of $A$, that is, the homotopy classes of projections (unitaries) of $A$ are in one-to-one correspondence with the homotopy classes of projections (unitaries) of $\mathcal{A}$ (\cref{theorem:isomorphism_K_0_and_k_1_for_smooth_sub_algebras})
    
    \item \textbf{By insight:} As exposed in \cref{sec:smooth_algebras_non_commutative_smooth_manifolds}, it is sensible to look into non-commutative Fréchet sub algebras of a given C* algebra for a setup to generalize the algebras of smooth functions over smooth compact manifolds.
\end{itemize}

\end{remark}

\subsection{Topological invariants over the Non-commutative Brillouin Torus}
\label{sec:top_inva_over_the_non_commutative_brillouim_torus}

Recall that the Non-commutative Brillouin Torus (\cref{definition:non_commutative_brillouin_torus}) is the pair of topological algebras,
$$ (C(\Omega)\rtimes_{\alpha, \Theta} \mathbb{Z}^d, \mathcal{C}(\Omega)_{\alpha, \Theta}).$$
Using the description of $C(\Omega)\rtimes_{\alpha, \Theta} \mathbb{Z}^d$ as an iterated crossed product (\cref{lemma:iterated_crossed_product}), in \citep[Proposition 4.2.4]{prodan_bulk_2016} it is proven that the K theory of $C(\Omega)\rtimes_{\alpha, \Theta} \mathbb{Z}^d$ takes the form
$$ K_j(C(\Omega)\rtimes_{\alpha, \Theta} \mathbb{Z}^d) =  \mathbb{Z}^{2^{d-1}}, \; j = 0,1.$$

From the pairing between the K theory of $C(\Omega) \rtimes_{\alpha, \Theta} \mathbb{Z}^d$ and the cyclic cohomology of $\mathcal{C}(\Omega)_{\alpha, \Theta}$ (\cref{proposition:pairing_topological_k_0_even_ciclyc_cohomology_C_star_algebras}, \cref{proposition:pairing_topological_k_1_odd_ciclyc_cohomology_C_star_algebras}), we know that given two cyclic cocycle $\phi_0 \in HC^{2n}_{\text{con}}(\mathcal{C}(\Omega)_{\alpha, \Theta}), \; \phi_1 \in HC^{2n+1}_{\text{con}}(\mathcal{C}(\Omega)_{\alpha, \Theta})$, there are two group homomorphisms given by
$$ \eta_{\phi_0} : K_0(C(\Omega) \rtimes_{\alpha, \Theta} \mathbb{Z}^d) \to \mathbb{C}, \; \eta_{\phi_0}([p])= \langle [p],  [\phi_0] \rangle, \; [\phi_0] \in HC^{2n}_{\text{con}}(\mathcal{C}(\Omega)_{\alpha, \Theta})$$
$$ \eta_{\phi_1} : K_1(C(\Omega) \rtimes_{\alpha, \Theta} \mathbb{Z}^d) \to \mathbb{C}, \; \eta_{\phi_1}([u])= \langle [u],  [\phi_1] \rangle, \; [\phi_1] \in HC^{2n+1}_{\text{con}}(\mathcal{C}(\Omega)_{\alpha, \Theta}).$$

Some cyclic cocycles are of special importance because they come as a generalization of the cyclic cocycles over the $d$ non-commutative torus (\citep[equation 5.3]{nest_cyclic_1988}, \citep{nest_cyclic_1988_2}), these are defined using the derivations $\partial_j$ over $\mathcal{C}(\Omega)_{\alpha, \Theta}$, recall that the derivations $\partial_j$ are defined in \cref{lemma:derivations_over_smooth_sub_algebra_twisted_crossed_product}. To define such cyclic cocycles we need to have a continuous trace over $C(\Omega) \rtimes_{\alpha, \Theta} \mathbb{Z}^d$, in this case we can take $\mu$ to be the normalized Radon measure over $\Omega$ with full support which is described in \cref{sec:topological_algebras_for_disordered_crystals}, then, the map
$$ f \mapsto \int_{\Omega} f(\omega) d \mu(\omega)  $$
is a faithful continuous trace over $C(\Omega)$ (\cref{section:non_commutative_geometry_dictionary}).

Since $\mu$ is invariant under $\alpha(s)$ for any $s \in \mathbb{Z}$, we can define the following faithful continuous trace over $C(\Omega) \rtimes_{\alpha, \Theta} \mathbb{Z}^d$ (\cref{proposition:trace_over_non_commutative_brillouin_torus}),
$$ \mathscr{T}: C(\Omega) \rtimes_{\alpha, \Theta} \mathbb{Z}^d \to \mathbb{C}, \; \mathscr{T}(p) := \int_{\Omega} \Phi_{0}(p)(\omega) d \mu(\omega), $$
where $\Phi_0(p)$ is the zero Fourier coefficient of $p$, which is defined in \cref{definition:fourier_coefficients}. Since the inclusion $i: \mathcal{C}(\Omega)_{\alpha, \Theta} \to C(\Omega) \rtimes_{\alpha, \Theta} \mathbb{Z}^d$ is continuous, then, the map $\mathscr{T} \circ i$ is also continuous, thus, the following is also a faithful continuous trace over $\mathcal{C}(\Omega)_{\alpha, \Theta}$,
$$ \mathscr{T}: \mathcal{C}(\Omega)_{\alpha, \Theta} \to \mathbb{C}, \; \mathscr{T}(p) := \int_{\Omega} \Phi_{0}(p)(\omega) d \mu(\omega). $$
 
Under the setup of \cref{definition:non_commutative_brillouin_torus}, the continuous cyclic cocycles of interest for the study of topological invariants over $\mathcal{C}(\Omega)_{\alpha, \Theta}$ take the following form (\citep[Section 5.2]{prodan_bulk_2016}), $\xi_{I} : \mathcal{C}(\Omega)_{\alpha, \Theta} \times \cdots \times \mathcal{C}(\Omega)_{\alpha, \Theta} \to \mathbb{C}$,
$$ \xi_I (p_0, \cdots, p_n) := \Lambda_n \sum_{\rho \in S_n} (-1)^{\rho} \mathscr{T}(p_0 (\partial_{\rho(1)} p_1) \cdots (\partial_{\rho(n)} p_n) )  ,$$
and we use the following notation,
\begin{itemize}
    \item Denote by $I$ an ordered subset of $\{1, \cdots, d \}$ with $n$ elements and an order not necessarily induced by $\mathbb{Z}$, we refer to $I$ as a multi-index\index{multi-index}.
    \item Denote by $S_n$ the symmetric group of degree $n$, and $\rho \in S_n$ is understood as a bijection between $\{ 1, \cdots, n \}$ and $I$ with signature $(-1)^{\rho}$.
    \item Define the constants
    $$ \Lambda_n := \frac{(2 i \pi)^{\frac{n}{2}}}{\frac{n}{2}!} \text{ for n even, }  \Lambda_n := \frac{i(i \pi)^{\frac{n-1}{2}}}{n!!} \text{ for n odd, }  .$$
\end{itemize}
Notice that if $I = \emptyset$ we get $\xi_I = \mathscr{T}$.

Recall that every normalized 2-cocycle over $\mathbb{Z}^d$ is equivalent to a normalized 2-cocycle of the form (\cref{proposition:characterization_of_+cocycles_over_integers})
$$ \zeta(s,l) = \exp{i s^t \Theta l}, $$
with $\Theta$ a lower triangular matrix with zeros on the diagonal and entries in $[0, 2 \pi)$, so, denote $\tilde{\Theta} = \Theta - \Theta^t$, in which case we have that $\tilde{\Theta}$ is an antisymmetric real matrix.

\begin{theorem}[c.f. Theorem 5.7.1,Corollary 5.7.2 \citep{prodan_bulk_2016}]\label{theorem:range_of_the_non_comu_chern_numebrs}
Using the previously introduced notation, the following maps
$$ \eta_{\xi_I} : K_0(C(\Omega) \rtimes_{\alpha, \Theta} \mathbb{Z}^d) \to \mathbb{C}, \; \eta_{\xi_{I}}([p])= \langle [p],  [\xi_I] \rangle, \; [\xi_I] \in HC^{2n}_{\text{con}}(\mathcal{C}(\Omega)_{\alpha, \Theta}), \; |I| = 2m$$
$$ \eta_{\xi_I} : K_1(C(\Omega) \rtimes_{\alpha, \Theta} \mathbb{Z}^d) \to \mathbb{C}, \; \eta_{\xi_I}([u])= \langle [u],  [\xi_I] \rangle, \; [\xi_I] \in HC^{2n+1}_{\text{con}}(\mathcal{C}(\Omega)_{\alpha, \Theta}), \; |I|=2m+1$$
have the range
$$ \boldsymbol{\text{Ran}(\eta_{\xi_I}) = \mathbb{Z} + \sum_{I \subset J} (2 \pi)^{- \frac{1}{2} |J \setminus I |} \text{Pf}(\tilde{\Theta}_{J \setminus I}) \mathbb{Z} }, $$
where we use the following notation,
\begin{itemize}
    \item Given $J$ a multi-index, if all the elements of $I$ belong to $J$, denote by $J \setminus I$ the set of elements in $J$ that are not in $I$, notice that $J \setminus I$ is not an ordered set. 
    \item Denote by $|J \setminus I|$ the cardinality of the set $J \setminus I$.
    \item The sum $\sum_{I \subset J}$ goes over all multi indexes $J$ whose order coincides with the order induced by $\mathbb{Z}$ and satisfy that $|J \setminus I|$ is even.
    \item Denote by $\tilde{\Theta}_{J \setminus I}$ the square matrix obtained from $\tilde{\Theta}$ by restricting to only the rows and columns whose indices correspond to the elements of $J \setminus I$.
    \item Denote by $\text{Pf}(\tilde{\Theta}_{J \setminus I})$ the Pfaffian of the square matrix $\tilde{\Theta}_{J \setminus I}$.
\end{itemize}
\end{theorem}

Take $I$ such that $|I| = 2m$, if the range of $\eta_{\xi_I}$ is a discrete sub group of $\mathbb{C}$, then, for any $p \in P_{m}(C(\Omega) \rtimes_{\alpha, \Theta} \mathbb{Z}^d)$ there is an $\epsilon > 0$ such that there are no $q \in P_{m}(C(\Omega) \rtimes_{\alpha, \Theta} \mathbb{Z}^d) $ with 
$$| \langle [p],  [\xi_I] \rangle  - \langle [q],  [\xi_I] \rangle| < \epsilon.$$
In physical terms, the previous statement means that a precise enough measurement of the physical property defined by $\eta_{\xi_I}$ can be used to assess whether $p$ and $q$ are not homotopically equivalent projections. A similar argument can be layout for $u,v \in U_n(C(\Omega) \rtimes_{\alpha, \Theta} \mathbb{Z}^d)$ and $\eta_{\xi_I}$ with $|I|=2m+1$. In \cref{remark:list_of_interesting_cocycles_for_physics} we will list some cases that are relevant in the analysis of topological phases of Hamiltonians where the range of $\eta_{\xi_I}$ is discrete in $\mathbb{C}$ .


For applications in physics, our starting point will be a self-adjoint element of $M_n(C(\Omega)) \rtimes_{\alpha, \Theta} \mathbb{Z}^d$, then, we will use the continuous functional calculus over $M_n(C(\Omega)) \rtimes_{\alpha, \Theta} \mathbb{Z}^d$ to come up with projections or unitaries depending on whether we want to use even or odd cyclic cocycles. Since $M_n(C(\Omega)) \simeq C(\Omega) \otimes M_n(\mathbb{C})$ (\cref{proposition:C_star_norm_on_matrix_algebras}), according to \cref{proposition:isomorphism_tensor_product_twisted_crossed_product} we have that
$$ M_n(C(\Omega)) \rtimes_{\alpha, \Theta} \mathbb{Z}^d \simeq ( C(\Omega) \rtimes_{\alpha, \Theta} \mathbb{Z}^d ) \otimes M_n(\mathbb{C}) \simeq M_n(C(\Omega) \rtimes_{\alpha, \Theta} \mathbb{Z}^d). $$
From \cref{example:matrix_algebras_and_twisted_crossed_products} we know that $M_n(\mathcal{C}(\Omega)_{\alpha,\Theta})$ is a smooth subalgebra of 
$$M_n(C(\Omega)) \rtimes_{\alpha, \Theta} \mathbb{Z}^d,$$ 
and there is a canonical extension of the derivations $\partial_j$ into $M_n(\mathcal{C}(\Omega)_{\alpha,\Theta})$ which consists of applying the derivations to each one of the entries of the matrix. 

Given that $C(\Omega)) \rtimes_{\alpha, \Theta} \mathbb{Z}^d$ is a sub C* algebra of $C(\Omega, B(L^2(\mathbb{Z}^d)))$ (\cref{remark:faithful_representation_of_C_0_Omega_rtimes_Z}), we have that $M_n(C(\Omega)) \rtimes_{\alpha, \Theta} \mathbb{Z}^d$ is a sub C* algebra of $C(\Omega, M_n(B(L^2(\mathbb{Z}^d))))$, therefore, each element of $ M_n(C(\Omega)) \rtimes_{\alpha, \Theta} \mathbb{Z}^d$ can be expressed as $p = \{ p_{\omega} \}_{\omega \in \Omega}$, where the map $\omega \mapsto p_{\omega}$ is continuous and $p_{\omega} \in M_n(B(L^2(\mathbb{Z}^d)))$. Given that the map $f \mapsto \int_{\Omega} \text{tr}(f(\omega)) d \mu$ is a trace over $M_n(C(\Omega))$, we have that the following functions defines a trace over $M_n(C(\Omega)) \rtimes_{\alpha, \Theta} \mathbb{Z}^d$,
$$\mathscr{T}(p) = \int_{\Omega} \text{tr} \left( \Phi_0(p)(\omega) \right) d \mu(\omega). $$

Take $ h \in M_n(C(\Omega)) \rtimes_{\alpha, \Theta} \mathbb{Z}^d$  $p \in A \rtimes_{\alpha, \Theta} \mathbb{Z}^d$ a self adjoint elemet i.e. $h = h^*$, then, \cref{elmma:spectrum_of_elements_in_twisted_crossed_product} tells us that the spectrum of $h$ is the union of the spectrums of $h_{\omega}$,
$$ Sp(h) = \bigcup_{\omega \in \Omega} Sp(h_{\omega}),$$
hence, if $g \subset \mathbb{R}$ is a gap in the spectrum of $h_{\omega}$ for all $\omega \in \Omega$ (\cref{definition:spectral_gap}), we have that $g$ is also gap in the spectrum of $h$. We can use the continuous calculus over $M_n(C(\Omega)) \rtimes_{\alpha, \Theta} \mathbb{Z}^d$ to compute the projection
$$ \chi(h \leq \tilde{g}), \; \tilde{g} \in g, $$
where we set $\chi(x \leq \tilde{g}) = 1$ if $x \leq \tilde{g}$, else, $\chi(x \leq \tilde{g}) = 0$, and we get that (\cref{lemma:continuous_calculus_and_self_adjoint_elements})
$$ \chi(h \leq \tilde{g}) = \{ \chi(h_{\omega} \leq \tilde{g}) \}_{\omega \in \Omega}.$$
Since the element $\chi(h \leq \tilde{g})$ only depends on the value of the real valued function $x \mapsto \chi(x \leq \tilde{g})$ over the spectrum of $h$ (\cref{theorem:continuous_functional_calculus}), we have that $\chi(h \leq \tilde{g}) = \chi(h \leq \tilde{g}^{'})$ for any $\tilde{g}, \tilde{g}^{'} \in g$, therefore, we will use the notation $\chi(h \leq g)$ to refer to $\chi(h \leq \tilde{g})$ for any $\tilde{g} \in g$.

In \cref{lemma:decay_fourier_coefficients_projections_from_gapped_self_adjoint_elemetns} it is mentioned that the element $\chi(h \leq g)$ can be computed using a smooth function over $\mathbb{R}$, therefore, if $h \in M_n(\mathcal{C}(\Omega)_{\alpha, \Theta})$, then, 
$$ \chi(h \leq g) \in M_n(\mathcal{C}(\Omega)_{\alpha, \Theta}). $$
We will use the notation $ P_{h,g} : =\chi(h \leq g)$ and $P_{h,g,\omega} := \chi(h_{\omega} \leq g)$, thus, we get that, 
$$P_{h,g} = \{ P_{h,g, \omega} \}_{\omega \in \Omega}.$$

In the physics context, in the previous arguments, we asked for the self-adjoint element $h$ to satisfy the Bulk Gap Hypothesis (\cref{definition:bulk_gap_hypothesis}).

\subsubsection{Even cyclic cocycles}
\label{sec:even_cyclic_cocycles_non_commutative_brillouin_torus}

Recall that $I$ denotes an ordered subset of $\{ 1, \dots, d \}$ with $n$ elements and an order not necessarily induced by $\mathbb{Z}$, $S_n$ is the symmetric group of order $n$ and $\rho \in S_n$ is a bijection between $\{ 1, \dots, n \}$ and $I$. Take $|I| = 2l$, then, $\eta_{\xi_I}$ takes the form
$$ \eta_{\xi_I} ([P_{h,g}]) = \Lambda_n \sum_{\rho \in S_n} (-1)^{\rho} \mathscr{T}(P_{h,g} (\partial_{\rho(1)} P_{h,g}) \cdots (\partial_{\rho(n)} P_{h,g}) ).   $$
Taking into account that the trace $\mathscr{T}$ is constructed using a measure $\mu$ over $\Omega$, the pairing of $[P_{h, g}]$  with $[\xi_I]$ looks like
$$ \langle [P_{h, g}], [\xi_I] \rangle = \Lambda_n \sum_{\rho \in S_n} (-1)^{\rho} \int_{\Omega} \text{tr} \left( \Phi_0 ( P_{h,g} (\partial_{\rho(1)} P_{h,g}) \cdots (\partial_{\rho(n)} P_{h,g}) )(\omega) \right) d \mu(\omega),   $$
which comes as a generalization of the even Chern numbers\index{even Chern numbers} (\citep[Section 2.2.1]{prodan_bulk_2016}),
$$
\mathrm{Ch}_d\left(P_{\text{Fermi}}\right)=\frac{(2 \pi \mathrm{i})^{\frac{d}{2}}}{\left(\frac{d}{2}\right) !} \sum_{\rho \in \mathcal{S}_d}(-1)^\rho \int_{\mathbb{T}^d} \frac{d k}{(2 \pi)^d} \operatorname{tr}\left(P_{\text{Fermi}}(k) \prod_{j=1}^d \frac{\partial P_{\text{Fermi}}(k)}{\partial k_{\rho(j)}}\right),
$$\index{$\mathrm{Ch}_d$}
where
$$
P_{\text{Fermi}}(k):=\chi\left(H_k \leq g \right),
$$
$\{ H_k \}_{k \in \mathbb{T}^d}$ is the decomposable form of a self-adjoint operator $H$ over the Hilbert space
$$ L^2(\mathbb{T}^d, \mathbb{C}^N) \simeq \int_{\mathbb{T}^d}^{\oplus} \mathbb{C}^N. $$
Notice that $H$ can be understood as an operator over $L^2(\mathbb{Z}^d, \mathbb{C})$ if we identify $L^2(\mathbb{T}^d)$ with $L^2(\mathbb{\mathbb{Z}^d})$ through the Fourier transform (\cref{theorem:Plancherel_theorem})
$$ \mathcal{F}: L^2(\mathbb{T}^d) \to L^2(\mathbb{Z}^d).$$

When the ingredients which define the twisted crossed product $\alpha$, $\Theta$ (\cref{definition:twsited_crossed_product}) are trivial, and $\Omega$ is a point ($C(\Omega) \simeq \mathbb{C}$), we have that $(C(\Omega) \rtimes_{\alpha,\Theta} \mathbb{Z}^d ) \otimes M_n(\mathbb{C})$ falls back into $M_n(\mathbb{C})\otimes C(\mathbb{T}^d)$ (\cref{sec:trivial_twisting_actions}), in this case, the Fourier transform over $\mathbb{T}^d$ tell us that (\cref{example:Fourier_transform_and_dual_groups})
$$ \mathcal{F}(f)(0):= \Phi_0(f) = \int_{\mathbb{T}^d} f(k) \frac{d k}{(2 \pi)^k},$$
and we can define a faithful trace over $M_n(\mathbb{C})\otimes C(\mathbb{T}^d)$ as
$$\mathscr{T}(f) = \text{tr} \left( \int_{\mathbb{T}^d} f(k) \frac{d k}{(2 \pi)^k} \right).$$
Under this setting, the pairing $\langle [P_{h, g}], [\xi_I] \rangle$ falls back into the even Chern numbers $\mathrm{Ch}_d\left(P_{\text{Fermi}}\right)$.

\subsubsection{Odd cyclic cocycles}
\label{sec:odd_cyclic_cocycles_non_commutative_brillouin_torus}

Recall that $I$ denotes an ordered subset of $\{ 1, \dots, d \}$ with $n$ elements and an order not necessarily induced by $\mathbb{Z}$, $S_n$ is the symmetric group of order $n$ and $\rho \in S_n$ is a bijection between $\{ 1, \dots, n \}$ and $I$. As in the previous section, $h$ denotes a self adjoint element of $M_n(\mathcal{C}(\Omega)_{\alpha,\Theta})$, since $M_n(\mathcal{C}(\Omega)_{\alpha,\Theta})$ is a sub algebra of $M_n(C(\Omega)) \rtimes_{\alpha,\Theta} \mathbb{Z}^d$ we can view $h$ as an element of $C(\Omega, M_n(B(L^2(\mathbb{Z}^d))))$, i.e. $h := \{ h_{\omega} \}_{\omega \in \Omega}$ (\cref{remark:faithful_representation_of_C_0_Omega_rtimes_Z}), additionally, $P_{h,0}$ denotes the element 
$$ \chi(h \leq 0), $$
where we set $\chi(x \leq 0) = 1$ if $x \leq 0$, else, $\chi(x \leq 0) = 0$. Assume that every $h_{\omega}$ satisfies the chiral symmetry for $\omega \in \Omega$, that is, for every $\omega$
$$ (J \otimes \mathbb{1}) h_{\omega} (J \otimes \mathbb{1}) = -h_{\omega}, \text{ with } J = \begin{pmatrix}
 \mathbb{1}_n & 0 \\
 0 & - \mathbb{1}_n
\end{pmatrix},$$
additionally, assume that for every $\omega \in \Omega$ we have that $0 \notin Sp(h_{\omega})$, then, according to \citep[Section 2.3.1]{prodan_bulk_2016} $P_{h,0}$ can be computed using the smooth functional calculus of functions over $\mathbb{R}$, and takes the form
$$ P_{h,0} = \frac{1}{2} \begin{pmatrix}
 0 & -U_{h,0}^* \\
 -U_{h,0} & 0
\end{pmatrix}$$
with $U_h \in M_n( \mathcal{C}(\Omega)_{\alpha,\Theta} )$ an unitary. In this case, when $h \in M_{2n}(\mathcal{C}(\Omega)_{\alpha,\Theta})$ we can provide an homotopic invariant for $P_{h,0}$ using $U_{h,0}$ and the continuous cyclic cocycle $\xi_I$ with  $|I| = 2l+1$, resulting in
$$ \langle [U_{h,0}], [\xi_I] \rangle = \Lambda_n \sum_{\rho \in S_n} (-1)^{\rho} \mathscr{T}((U_{h,0}^* - \mathbb{1}) (\partial_{\rho(1)} U_{h,0}) \cdots (\partial_{\rho(n-1)} U_{h,0}^*) (\partial_{\rho(n)} U_{h,0}) ),$$
which, according to \citep[Example 3.45]{prodan_computational_2017} is equivalent to,
$$ \langle [U_{h,0}], [\xi_I] \rangle = \Lambda_n \sum_{\rho \in S_n} (-1)^{\rho} \mathscr{T}( \prod_{l = 1}^{n} U_{h,0}^* (\partial_{\rho(l)} U_{h,0})),$$
and takes the following form as an integral over $\Omega$
$$ \langle [ U_{h,0}], [\xi_I] \rangle = \Lambda_n \sum_{\rho \in S_n} (-1)^{\rho} \int_{\omega \in \Omega} \text{tr} \left( \Phi_0 \left(\prod_{l = 1}^{n} U_{h,0}^* (\partial_{\rho(l)} U_{h,0}) \right)(\omega) \right) d \mu(\omega).   $$
The aforementioned pairing comes as a generalization of the odd Chern numbers\index{odd Chern numbers} (\citep[Section 2.3.1]{prodan_bulk_2016}),
$$
\mathrm{Ch}_d\left(U_{\text{Fermi}}\right)=\frac{i (\pi \mathrm{i})^{\frac{d-1}{2}}}{d!!} \sum_{\rho \in \mathcal{S}_d}(-1)^\rho \int_{\mathbb{T}^d} \frac{d k}{(2 \pi)^d} \operatorname{tr}\left( \prod_{l = 1}^{d} U_{\text{Fermi}}^*(k) \frac{\partial U_{\text{Fermi}}}{\partial k_{\rho(l)}}(k) \right),
$$\index{$\mathrm{Ch}_d$}
where,
$$
P_{\text{Fermi}}(k)=\chi\left(H_k \leq g \right) = \frac{1}{2} \begin{pmatrix}
 0 & -U_{\text{Fermi}}(k)^* \\
 -U_{\text{Fermi}}(k) & 0
\end{pmatrix},
$$
$\{ H_k \}_{k \in \mathbb{T}^d}$ is the decomposable form of a self-adjoint operator $H$ over the Hilbert space
$$ L^2(\mathbb{T}^d, \mathbb{C}^N) \simeq \int_{\mathbb{T}^d}^{\oplus} \mathbb{C}^N. $$
Notice that $H$ can be understood as an operator over $L^2(\mathbb{Z}^d, \mathbb{C})$ if we identify $L^2(\mathbb{T}^d)$ with $L^2(\mathbb{\mathbb{Z}^d})$ through the Fourier transform (\cref{theorem:Plancherel_theorem})
$$ \mathcal{F}: L^2(\mathbb{T}^d) \to L^2(\mathbb{Z}^d).$$
As in the case of the even cyclic cocycles, $\langle [U_{h,0}], [\xi_I] \rangle$ falls back into $\mathrm{Ch}_d\left(U_{\text{Fermi}}\right)$ then $\alpha, \Theta$ are trivial and $\Omega$ is a point i.e. $C(\Omega) \simeq \mathbb{C}$.

\subsubsection{Generalized Chern numbers}\index{generalized Chern numbers}
\label{sec:generalized_chern_numbers}

Motivated by the fact that the pairing $\langle [P_{h,g}], [\xi_I] \rangle$ is a generalization of the even Chern numbers and the pairing $\langle [U_{h,0}], [\xi_I] \rangle$ is a generalization of the odd Chern numbers, the following notation is used in \citep[Sections 5.2 and 5.3]{prodan_bulk_2016},
\begin{itemize}
    \item Let $|I| = 2l$, assume $P_{h,g}$ belongs to $M_n(\mathcal{C}(\Omega)_{\alpha, \Theta})$, then define $\text{Ch}_I (P_{h,g}) := \langle [P_{h,g}], [\xi_I] \rangle$\index{$\text{Ch}_I$} and call $\text{Ch}_I$ the \textbf{even Non-Commutative Chern number associated to the multi index $I$}. Under this setting,
    $$ \text{Ch}_I (P_{h,g}) = \Lambda_{|I|} \sum_{\rho \in S_{|I|}} (-1)^{\rho} \int_{\Omega} \text{tr} \left( \Phi_0 \left( P_{h,g} (\partial_{\rho(1)} P_{h,g}) \cdots (\partial_{\rho(|I|)} P_{h,g}) \right)(\omega) \right) d \mu(\omega), $$
    \item Let $|I| = 2l+1$, assume $U_{h,0}$ belongs to $M_n(\mathcal{C}(\Omega)_{\alpha, \Theta})$, then, define $\text{Ch}_I (U_{h,0}) := \langle [U_{h,0}], [\xi_I] \rangle$ and call $\text{Ch}_I$ the \textbf{odd Non-commutative Chern number associated to $I$}. Under this setting
    $$ \text{Ch}_I (U_{h,0}) = \Lambda_{|I|} \sum_{\rho \in S_{|I|}} (-1)^{\rho} \int_{\Omega} \text{tr} \left( \Phi_0 \left(\prod_{l = 1}^{{|I|}} U_{h,0}^* (\partial_{\rho(l)} U_{h,0}) \right)(\omega) \right) d \mu(\omega), $$
    \item If $I=\{1, \cdots, d\}$, then, use the notation $\text{Ch}_d := \text{Ch}_I$.
\end{itemize}
Given the formula for the range of $\text{Ch}_I$ from \cref{theorem:range_of_the_non_comu_chern_numebrs}, in the following two cases the range of $\text{Ch}_I$ is discrete,
\begin{enumerate}
    \item If $I = \{ 1, \cdots, d \}$ with the order induced by $\mathbb{Z}$, then, $\text{Ran}(\text{Ch}_I) = \mathbb{Z}$.
    \item If $I = \{ 1, \cdots, d-1 \}$ with the order induced by $\mathbb{Z}$, then, $\text{Ran}(\text{Ch}_I) = \mathbb{Z}$.
\end{enumerate}
The case $I = \{ 1, \cdots, d \}$ is special because there is a way of expressing both $\text{Ch}_I(P_{h,g})$ and $\text{Ch}_I(U_{h,0})$ as the index of a Fredholm operator (\citep[Corollary 6.5.2]{prodan_bulk_2016}). When $I = \{ 1, \cdots, d \}$ the topological invariant $\text{Ch}_I$ is called a \textbf{strong Chern number}\index{strong Chern number}, for the other cases we refer to $\text{Ch}_I$ as a \textbf{weak Chern number}\index{weak Chern number} (\citep[Remark 5.3.5]{prodan_bulk_2016}).

\begin{remark}\label{remark:list_of_interesting_cocycles_for_physics}
From the formula for the range of $\eta_{\xi_I}$ a.k.a. $\text{Ch}_{I}$ (\cref{theorem:range_of_the_non_comu_chern_numebrs}), we get the following
\begin{itemize}
    \item \citep[Corollary 7.2.1]{prodan_bulk_2016}: If $d=2$, then, $\text{Ran}(\text{Ch}_2) = \mathbb{Z}$. This result is closely related to the Integer Quantum Hall effect in 2-dimensions.
    \item \citep[Corollary 7.2.2]{prodan_bulk_2016}: If $d=3$ and $|I|=2$, then, $\text{Ran}(\text{Ch}_I) = \mathbb{Z}$. This result is closely related to the Quantum Hall phases in 3-dimensions.
    \item \citep[Corollary 7.3.2]{prodan_bulk_2016}: If $d=1$ and $I = \{1\}$, then, $\text{Ran}(\text{Ch}_I) = \mathbb{Z}$. This result is closely related to the quantization of the chiral polarization of a 1-dimensional chiral symmetric solid-state system.
    \item \citep[Corollary 7.3.3]{prodan_bulk_2016}: If $d=2$ and $|I | = 1$, then, $\text{Ran}(\text{Ch}_I) = \mathbb{Z}$. This result is closely related to the quantization of the chiral polarization of a 2-dimensional chiral symmetric solid-state system.
\end{itemize}
For more context on the first two items refer to \cref{sec:motivation_from_physics}.
\end{remark}

\section{Physics framework}
\label{sec:motivation_from_physics}

In this section, from the point of view of physics we look at how the results exposed in \cref{sec:top_inva_over_the_non_commutative_brillouim_torus} can be used to study tight-binding models\index{tight-binding model} for homogeneous materials, in particular, we will look into effective models for infinite crystals with no boundary, also known as bulk models\index{bulk models}. We will describe the form of the effective models, starting from perfect crystals, then, adding the effect of a constant magnetic field over the crystals, and finally dealing with models that include disorder and describe physical observables invariant under disorder, also known as homogeneous materials.

Recall that a Hamiltonian\index{Hamiltonian} over $L^2(\mathbb{Z}^d)$ is a self-adjoint member of the C* algebra $B(L^2(\mathbb{Z}^d))$. As a first step, we are interested in effective models over lattices, that is, 
\begin{itemize}
    \item instead of working with Hamiltonians over $\boldsymbol{L^2(\mathbb{R}^d)}$ we will be working with Hamiltonians over $\boldsymbol{L^2(\mathbb{Z}^d)}$,
    \item instead of working with \textbf{many particle Hamiltonians} we will be working with \textbf{one particle Hamiltonians}.
\end{itemize}
Even though these may seem to be extreme assumptions that could hardly provide any useful physical information, under the appropriate settings, like temperatures close to zero or low electron density, the effective models provide accurate descriptions of the physical systems under study (\citep[Chapter 2]{prodan_bulk_2016}). So, what exactly are those effective models describing? 
\textbf{These effective models describe the dynamics of electrons on a crystal.}

\begin{figure}[h!]
\centering
\begin{tikzpicture}
    \node[anchor=south west,inner sep=0] (image1) at (0,0) {\includegraphics[width=0.9\textwidth]{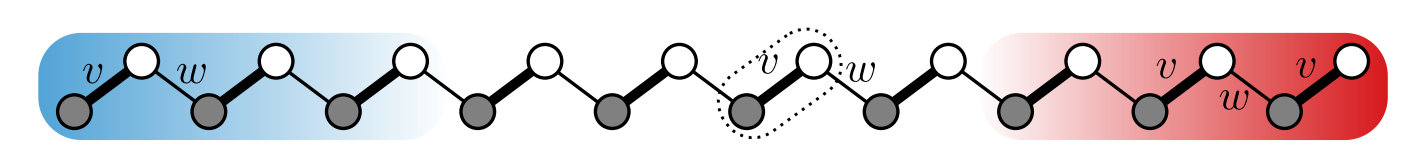}};
    \node[anchor=south west,inner sep=0] (image2) at (0,-5) {\includegraphics[width=0.9\textwidth]{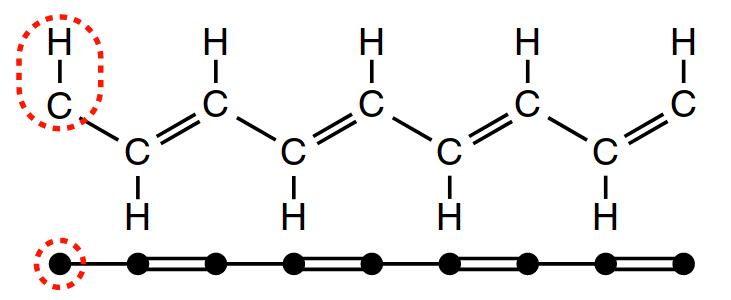}};
\end{tikzpicture}
\caption{Unit cells for the SSH model: Unit cell given by C-C bond  (up) (\citep[Figure 1.1]{asboth_short_2015}), Unit cell given by H-C bond (down) (\citep[Figure 1]{meier_observation_2016})}
\label{fig:unit_cells_in_ssh_model}
\end{figure}

Therefore, in a nutshell, after an elaborate process of approximation that involves using the maximally localized Wannier basis set  (\citep{marzari_maximally_2012}), physicists come up with a lattice model i.e. a Hamiltonian $H$ over $L^2(\mathbb{Z}^d) \otimes \mathbb{C}^N$ that describes the behavior of a possibly virtual particle, such that, $H$ contains the relevant information about the \textbf{dynamics of the particle on a crystal}. Each element of the lattice $\mathbb{Z}^d$ represents a unit cell of the crystal, that is, a set of atoms whose translation defines the complete crystal, for example, in \cref{fig:unit_cells_in_ssh_model} there are depicted two ways of defining unit cells in the SSH model. Given this setup, we focus on translation-invariant Hamiltonians, that is, models of the form (\citep[Section 2.2.1]{prodan_bulk_2016})
$$ H \in B(\mathbb{C}^N \otimes L^2(\mathbb{Z}^d) ), \; H = \sum_{y \in \mathbb{Z}^d} W_y \otimes S^{y}, $$
where $S^{y}$ is the shift operator by $y$ on $L^2(\mathbb{Z}^d)$ given by $S^y \ket{x} = \ket{ x + y}$, and the $ N \times N$ matrices $W_y$, called the hopping matrices, satisfy the constrain
$$ W_y^* = W_{-y}. $$

\begin{figure}[h!]
\centering
\includegraphics[width=0.9\textwidth]{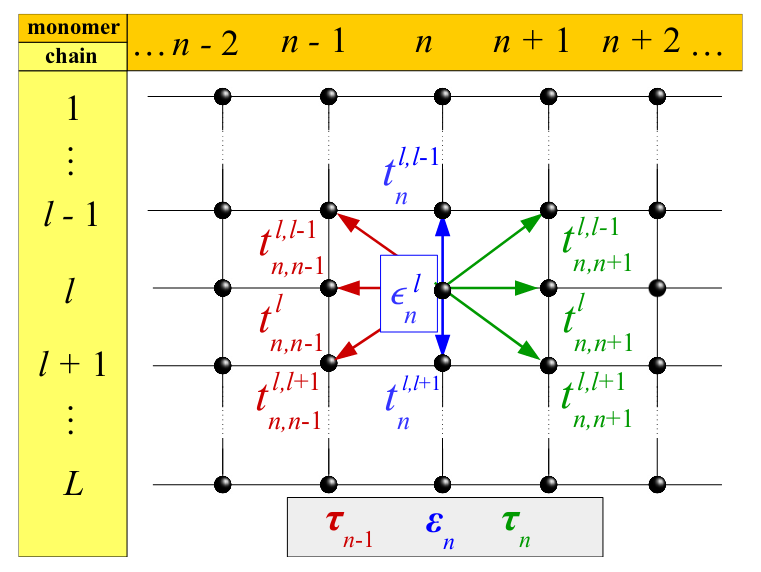}
\caption{Schematics of hopping matrices in a 1d model, on the horizontal axis are depicted the lattices sites and on the vertical axis are depicted the energy levels per site, notice that each site has the same energy levels because we are working with translation invariant systems. We have that $W_y \in M_L(\mathbb{C})$, such that, we are depicting the components $(l,l-1), (l, l), (l, l+1)$ (from top to bottom) of the following hopping matrices: in \textcolor{blue}{blue} we have depicted $W_0$ (inter-cell hopping terms), in \textcolor{red}{red} we have $W_{-1}$ (nearest neighbor on the left) and in \textcolor{green}{green} we have $W_1$ (nearest neighbor on the right) (\citep[Figure 1]{lambropoulos_tight-binding_2019}).}
\label{fig:energy_levels_and_hopping_matrices}
\end{figure}

In this setting, $\mathbb{C}^N$ represents the different energy levels of the unit cell that the electron can take, so, for instance, if we work in a one-dimensional system, like the SSH model, and $N = 2$, then, $W_2 \in M_2(\mathbb{C})$ and it describes the dynamics of a particle moving from the site $\ket{x}$ to the site $\ket{x+2}$, with $x \in \mathbb{Z}$. In \cref{fig:energy_levels_and_hopping_matrices} you can find a schematic of a 1d lattice with $L$ energy levels per site and a depiction of the hopping matrices.

For such Hamiltonians we have that $S^{y} H (S^y)^* = H $ for any $y \in \mathbb{Z}^d$, this is called \textbf{translation invariance}, and it allows to use the Fourier transform 
$$ \mathcal{F}: L^2(\mathbb{T}^d) \to L^2(\mathbb{Z}^d)$$
in order to express $H$ as a decomposable operator (\cref{def:decomposable_operator_direct_integral}) over the \textbf{direct integal} 
$$L^2(\mathbb{T}^d, \mathbb{C}^N) \simeq \int_{\mathbb{T}^d}^{\oplus} \mathbb{C}^N,$$
in which case
$$ \mathcal{F} H \mathcal{F}^* = \int_{\mathbb{T}^d}^{\oplus} dk H_k, \text{ with } H_k \in M_N(\mathbb{C}). $$
The space $\mathbb{T}^d$ is called the Brillouin Torus\index{Brillouin Torus}, and we have that for each $k \in \mathbb{T}^d$
$$ H_k : \mathbb{C}^N  \to \mathbb{C}^N, \; H_k = \sum_{y \in \mathcal{R}} e^{i\braket{y}{k}} W_y,$$ 
where, for simplicity, we take $\mathcal{R} \subset \mathbb{Z}^d$ finite, thus
$$ H = \sum_{y \in \mathcal{R}} W_y S^y. $$
Under this setting, the map $f : \mathbb{T}^d \to M_n(\mathbb{C}), \; f(k) := H_k$ is continuous and the set of operators $\{ H_y \}_{y \in \mathcal{R}}$ over $M_n(\mathbb{C})$ is called the Fourier coefficients of $f$, because, under the Fourier transform we get 
$$ W_y = \int_{\mathbb{T}^d} e^{i\braket{-y}{k}} H_k \frac{d \mu(k)}{(2 \pi)^d}. $$
In this setting $\mathbb{T}^d$ is called the \textbf{Brillouin Torus}.

The previously described systems have been widely studied, such that, there are even exact solutions to various of those models (\citep[Chapter 2]{prodan_bulk_2016}). Thus, the objective of Prodan et.al. in \citep{prodan_bulk_2016} is to present a generalization of the translation invariant models which can be analyzed using techniques from non-commutative geometry. The path towards those models consists of two steps (\citep[Chapter 1]{prodan_bulk_2016}),
\begin{enumerate}
    \item \textbf{Add the effect of an external magnetic constant field.} The presence of a uniform magnetic field can be modeled with the Peierls substitution\index{Peierls substitution} (\citep[Chapter 2]{prodan_bulk_2016}), which, takes into account the effect of the magnetic field through a real anti-symmetric matrix $B$ with entries in $[-\pi, \pi)$. Denote by $B_{+}$ the lower triangular matrix of $B$, also, set $B_{-} := B_{+}^T$, thus, $B = B_{+} - B_{-}$, then, in the Peierls substitution we need to replace the \textbf{shift operators} $S^y$ with the \textbf{dual magnetic translations} $U^y$, which, under the \textbf{Landau Gauge} looks like 
    $$  S^y \mapsto U^y, \; U^y := S^y e^{i \bra{y} B_{+} \ket{X}}, \text{ that is, } U^y \ket{x} = e^{i \bra{y} B_{+} \ket{x}} \ket{x + y}.  $$
    After replacing $S^y$ with $U^y$, the Hamiltonian becomes
    $$ \boldsymbol{H = \sum_{y \in \mathcal{R}} e^{\frac{i}{2} \bra{y} B_{+} \ket{y}} W_y \otimes U^y}. $$
    This Hamiltonian is no longer invariant under $S^y$, however, we can construct a new set of unitary transformations that generalize $S^y$ and preserve $H$, these are called the \textbf{direct magnetic translations},
    $$ V^y := e^{i \bra{X} B_{+} \ket{y}} S^y, \text{ that is, } \; V^y \ket{x} = e^{i \bra{x+y} B_{+} \ket{y}} \ket{x + y}. $$
    Given this setting, we have that
    $$ \boldsymbol{V^y H (V^y)^* = H}. $$
    The dual magnetic translations satisfy the commutation relation
    $$ \boldsymbol{U^x U^y = e^{i \bra{x} B_{+} \ket{y}} U^{x+y}}. $$
    Let $e_j = (0,\cdots, 1, \cdots, 0)$ where the position $j$ has a $1$ and it is zero elsewhere, then, the operators $\{ U_{e_j} \}_{1 \leq j \leq d}$ satisfy the commutation relations of the operators $u_j$ used to express twisted crossed products as universal c* algebras (\cref{lemma:twisted_crossed_products_with_Z}).
    
    \item \textbf{Work with random hopping matrices.} Let $\mathcal{H}$ be the Hilbert space where the Hamiltonians we are studying act on,
    $$\boldsymbol{\mathcal{H} = L^2(\mathbb{Z}^d) \otimes \mathbb{C}^N, \; N \in \mathbb{N}}.$$ When we work with random hopping elements the Hamiltonian is no longer translation invariant, nor invariant under the direct magnetic translations. In this scenario, we want to study physical properties that are invariant under the disorder of the physical system e.g. transport coefficients like the conductivity (\citep[Proposition 7.1.1]{prodan_bulk_2016}), the chiral polarization (\citep[Proposition 7.3.1]{prodan_bulk_2016}), so, we take the following steps,
    \begin{enumerate}
        \item Instead of working with a particular Hamiltonian $H$, we work with a family of Hamiltonians $\boldsymbol{\{H_{\omega}\}_{\omega \in \Omega} \subset B(\mathcal{H})}$.
        \item We ask for the family of Hamiltonians to have the \textbf{covariant property}, that is, we have a topological dynamical system (\citep[Section II.10]{blackadar_operator_2006}) and the following relation is satisfied,
        $$\boldsymbol{U^y H_{\omega} (U^y)^* = H_{\varrho(-y)(\omega)}, \; \forall y \in \mathbb{Z}^d}. $$
        Notice that the covariant property coincides with the fourth property of covariant representation of twisted dynamical systems (\cref{definition:representation_twsited_dynamical_system}), if we work with twisted dynamic systems where the C* algebra takes the form $C(\Omega)$ (\cref{definition:twisted_transformation_group_C_star_algebras}).
    \end{enumerate}
    The family of Hamiltonians $\boldsymbol{\{H_{\omega}\}_{\omega \in \Omega}}$ is called a \textbf{covariant family of Hamiltonians}\index{covariant family of Hamiltonians} and describes what is known as a \textbf{homogeneous material} (\citep[Section 1.1]{prodan_computational_2017}), also, the covariant families of Hamiltonians are well suited to study phenomena resilient to disorder, like the Integer Quantum Hall Effect (\citep{bellissard_non-commutative_1994}). In this setting, $\boldsymbol{\Omega}$ is a parametrization of the space of disordered configurations, it has a topology that makes the map 
    $$ \boldsymbol{\omega \mapsto H_{\omega}} $$
    continuous, and is a \textbf{compact space} (\citep[Definition 1.1]{prodan_computational_2017}).
    
    Denote $\boldsymbol{H = \{ H_{\omega} \}_{\omega \in \Omega} \subset B(\mathcal{H})}$ a covariant family of Hamiltonians, then, we are particularly interested in those Hamiltonians $H$ that come as a generalization of the Hamiltonian 
    $$\boldsymbol{\sum_{y \in \mathcal{R}} e^{\frac{i}{2} \bra{y} B_{+} \ket{y}} W_y \otimes U^y},$$ 
    one example of such a family is
    $$\boldsymbol{ \{ \sum_{y \in \mathcal{R}} \sum_{x \in \mathbb{Z}^d} e^{\frac{i}{2} \bra{y} B_{+} \ket{y}} W_y( \varrho(-x)(\omega)) \otimes \ket{x} \bra{x} U^y\}_{\omega \in \Omega}},$$ where $\boldsymbol{W_y : \Omega \to M_n(\mathbb{C})}$ are continuous functions, these are the covariant families of Hamiltonians that we care about. Notice that the operator
    $$\sum_{x \in \mathbb{Z}^d} e^{\frac{i}{2} \bra{y} B_{+} \ket{y}} W_y( \varrho(-x)(\omega)) \otimes \ket{x} \bra{x} U^y$$
    corresponds to the operator $\pi_{\omega}(W_y) \pi_{\omega}(u_y)$ comming from the map $\pi_{\omega}$ described in \cref{lemma:representation_on_L2_Z}.
        
    Let $\boldsymbol{H = \{ H_{\omega} \}_{\omega \in \Omega}} $ be a covariant family of Hamiltonians, then, the space $\boldsymbol{\Omega}$ is called the \textbf{Hull} of $\boldsymbol{H}$, is homeomorphic to the closure of 
    $$\boldsymbol{\{ U^y H (U^y)^* | y \in \mathbb{Z}^d \}}$$
    in the strong topology of $\boldsymbol{B(\mathcal{H})}$ (\citep[Definition 1.1]{prodan_computational_2017}).
\end{enumerate}

Let $\boldsymbol{A = \{ A_{\omega} \}_{\omega \in \Omega} \subset B(\mathcal{H})}$ be a covariant family of bounded linear operators, then, the set of all such $\boldsymbol{A}$ that take the form
$$\boldsymbol{ \{ \sum_{y \in \mathcal{R}} \sum_{x \in \mathbb{Z}^d} e^{\frac{i}{2} \bra{y} B_{+} \ket{y}} W_y( \varrho(-x)(\omega)) \otimes \ket{x} \bra{x} U^y\}_{\omega \in \Omega}},$$
with $\boldsymbol{W_y : \Omega \to M_n(\mathbb{C})}$ a continuous function, is a \textbf{*algebra} under \textbf{point wise multiplication, addition and involution}. We can provide a \textbf{C* norm} to this algebra as follows
$$ \boldsymbol{ \| A \| := \sup_{\omega \in \Omega} \| A_{\omega} \|_{B(\mathcal{H})}}, $$
and \textbf{its completion becomes C* algebra}. This C* algebra is isomorphic to the C* algebra
$$ \boldsymbol{M_n(C(\Omega) \rtimes_{\alpha, B_{+}} \mathbb{Z}^d))}, $$
where $\alpha$ is an action of $\mathbb{Z}^d$ over $C(\Omega)$ given by 
$$\alpha(y)(f)(\omega) = f(\varrho(-y)(\omega)),\; y \in \mathbb{Z}^d.$$
Notice that we have looked at some of the properties of the C* algebra $C(\Omega) \rtimes_{\alpha, B_{+}} \mathbb{Z}^d$ in \cref{chap:fourier_analysis}. We also have that
$$ \boldsymbol{(C(\Omega) \rtimes_{\alpha, B_{+}} \mathbb{Z}^d) \otimes M_n(\mathbb{C}) \simeq  M_n(C(\Omega)) \rtimes_{\alpha, B_{+}} \mathbb{Z}^d}.  $$

The C* algebra $\boldsymbol{M_n(C(\Omega)) \rtimes_{\alpha, B_{+}} \mathbb{Z}^d}$ contains the covariant families of Hamiltonians that describe \textbf{the dynamics of an electron in a homogeneous material with no boundaries} and $\boldsymbol{n}$ \textbf{energy levels per unit site in the lattice}. 

For practical applications we can provide $\boldsymbol{(\Omega, \varrho, \mathbb{Z}^d)}$ with more structure, which comes motivated from the disordered introduced in a pure crystal due to thermal fluctuations (\citep[Example 2.4]{prodan_computational_2017}), and takes the following form (\citep[Section 3.1]{prodan_computational_2017}),
$$\boldsymbol{ \Omega = \prod_{s \in \mathbb{Z}^d} \Omega_0, \; \varrho(y): \Omega \to \Omega, \; \varrho(y)\left((l \mapsto \omega_l)_{l \in \mathbb{Z}^d}\right) = (l \mapsto \omega_{l - y})_{l \in \mathbb{Z}^d}},$$
with $\Omega_0$ a compact, metrizable, and convex space. For example, if we want to work with Hamiltonians that include at most interactions with the closest neighbors inside the cube of side $2L$, we can take (\citep[Section 2.4.1]{prodan_bulk_2016})
$$\boldsymbol{\Omega_0 =  \prod_{y \in V_L} \Omega_0^y, \; V_L = \{ (y_1, \cdots, y_d) \in \mathbb{Z}^d | |y_i| \leq L \}},$$
with $\boldsymbol{\Omega_0^y}$ a compact, metrizable and convex space for all $\boldsymbol{y \in V_L}$. As explained in \cref{sec:topological_algebras_for_disordered_crystals}, with this particular choice for $\boldsymbol{\Omega}$, it becomes a \textbf{compact, metrizable and convex space} with a bounded Radon measure with full support, additionally, $\boldsymbol{\varrho(y)}$ becomes \textbf{homotopically equivalent to the identity map} on $\boldsymbol{\Omega}$ (\citep[Proposition 4.2.1]{prodan_bulk_2016}), and the space $\boldsymbol{\Omega_0^y}$ parametrizes the randomness associated to the shift $\boldsymbol{U^y}$. Notice that $M_n(C(\Omega) \rtimes_{\alpha, B_{+}} \mathbb{Z}^d)$ is the C* algebra of matrices with entries in the Non-Commutative Brillouin Torus (\cref{definition:non_commutative_brillouin_torus}).

Recall that the Fermi level\index{Fermi level} of $H$ is the highest energy level an electron can occupy at the absolute zero temperature in the crystal described by $H$. Each covariant family of Hamiltonians $\boldsymbol{H = \{ H_{\omega} \}_{\omega \in \Omega} \subset B(\mathcal{H})}$ describes the dynamics of electrons inside a homogeneous material, and, if we impose certain restrictions on the Hamiltonians $\boldsymbol{H}$ we can make relevant assessments about those materials.  In particular, we are going to focus on covariant families of Hamiltonians that satisfy the \textbf{Bulk Gap Hypothesis} (\citep[Section 2.4.2]{prodan_bulk_2016}), which states that 

\begin{definition}[Bulk Gap Hypothesis \index{Bulk Gap Hypothesis}]\label{definition:bulk_gap_hypothesis}
Let  $H = \{ H_{\omega} \}_{\omega \in \Omega} \subset B(\mathcal{H})$ be a covariant family of Hamiltonians, then, we say that $H$ satisfy the Bulk Gap Hypothesis if \textbf{the Fermi level $\boldsymbol{\mu} \in \mathbb{R}$ lies in a gap of the spectrum of $\boldsymbol{H_{\omega}}$ for all $\omega \in \Omega$.}   
\end{definition}

Recall that $M_n(\mathcal{C}(\Omega)_{\alpha, B_{+}})$ is a smooth sub algebra of $M_n(C(\Omega)) \rtimes_{\alpha, B_{+}} \mathbb{Z}^d$, thus, if $H \in M_n(\mathcal{C}(\Omega)_{\alpha, B_{+}})$, then, by \cref{lemma:decay_fourier_coefficients_projections_from_gapped_self_adjoint_elemetns} we have that
$$\chi(H \leq \mu) \in M_n(\mathcal{C}(\Omega)_{\alpha, B_{+}}).$$
For example, if $H$ is a finite polynomial
$$ \{ \sum_{y \in \mathcal{R}} \sum_{x \in \mathbb{Z}^d} e^{\frac{i}{2} \bra{y} B_{+} \ket{y}} W_y( \varrho(-x)(\omega)) \otimes \ket{x} \bra{x} U^y\}_{\omega \in \Omega}, $$
where, $\mathcal{R}\subset \mathbb{Z}^d$  and $|\mathcal{R}| \leq \infty$,
then, $\chi(H \leq \mu) \in M_n(\mathcal{C}(\Omega)_{\alpha, B_{+}})$.

In real-world applications the finite polynomials are used (\citep[Section 2.2.1]{prodan_bulk_2016}), so, let $P_{\text{Fermi}} = \chi(H \leq \mu)$\index{$P_{\text{Fermi}}$ (Fermi projection)} be the Fermi projection\index{Fermi projection}, then, in the limit of \textbf{temperature near zero} and \textbf{infinite relaxation time} the Fermi distribution becomes the Fermi projection (\citep[Section 4.2]{bellissard_non-commutative_1994}),
$$\lim _{\beta \uparrow \infty} f_{\beta, \mu}(H)=P_{\text{Fermi}},  \text{ with } f_{\beta, \mu}(H)=\frac{1}{1+\exp (\beta(H-\mu))}.$$
Also, the Kubo formula for the current density provides the following description of the non-diagonal elements of the conductivity tensor (\citep[Proposition 7.1.1]{prodan_bulk_2016})
$$ \sigma_{i,j} = \langle [P_{\text{Fermi}}], [\xi_{\{i,j\}}] \rangle = \text{Ch}_{\{i,j \}}(P_{\text{Fermi}}).  $$

From \cref{sec:top_inva_over_the_non_commutative_brillouim_torus} we know the specific form that takes the range of $\text{Ch}_{\{i,j \}}$ and how it depends on the dimension of the model under study, so, we can use \cref{remark:list_of_interesting_cocycles_for_physics} to make statements on the physical properties of homogeneous materials as described in \citep[Section 7.2]{prodan_bulk_2016}, 
\begin{itemize}
    \item \citep[Corollary 7.2.1]{prodan_bulk_2016}: If $d=2$, then, 
    $$\sigma_{1,2} = \text{Ch}_{2}(P_{\text{Fermi}}) \subset \mathbb{Z} \text{ and } \sigma_{2,1} = \text{Ch}_{\{2,1\}}(P_{\text{Fermi}}) \subset \mathbb{Z},$$
    such that, we get a proof for the \textbf{quantization of the transverse conductivity in the IQHE (Integer Quantum Hall Effect) in two dimensions}\index{Integer Quantum Hall Effect} and the \textbf{resilience of the quantization under disorder that does not close the spectral gap around the Fermi level}.
    \item \citep[Corollary 7.2.2]{prodan_bulk_2016}: If $d=3$, then, 
    $$\sigma_{i,j} = \text{Ch}_{\{ i,j\}}(P_{\text{Fermi}}) \subset \mathbb{Z} \text{ for any } 1 \leq i \neq j \leq 3,$$ 
    such that, we get a proof for the \textbf{quantization of the transverse conductivity in the IQHE (Integer Quantum Hall Effect) in three dimensions} and the \textbf{resilience of the quantization under disorder that does not close the spectral gap around the Fermi level}.
\end{itemize}
Both of the aforementioned results hold for any deformation of $\boldsymbol{H}$ that does not close the gap around the Fermi level $\boldsymbol{\mu}$. When a particular property holds as long as the spectral gap does not close around the Fermi level we say that it holds in the \textbf{regime of weak disorder}\index{weak disorder regime}, that is, when it holds under the Bulk Gap Hypothesis (\cref{definition:bulk_gap_hypothesis}).

\appendix
\newpage
\thispagestyle{empty}
\mbox{}
\newpage
\vspace*{\fill}
\begin{center}
\Huge \textbf{Appendices}
\end{center}
\vspace*{\fill}

\chapter{Comments on the appendices}
\label{chapter:ccomments_on_the_appendices}

There is a rich connection between the appendices and the main body of the document, the following is a list of the most important facts stated in the appendices that are useful for understanding the main body of the document:
\begin{itemize}
    \item The twisted crossed product $A \rtimes_{\alpha, \zeta} G$ is a C* algebra whose norm is closely related to the norm of representations of a $L^1$ Banach algebra, so, in \cref{sec:oper_valu_func} we provide a review on the representation of $L^1$ Banach algebras.
    \item The Fourier coefficients of an element of the twisted crossed product $A \rtimes_{\alpha, \zeta} \mathbb{Z}^d$ can be described as an integral of a C* algebra valued function, these type of integrals are called Bochner integrals and in \cref{sec:bochner_integral} we provide a review of such integrals.
    \item The twisted crossed product $A \rtimes_{\alpha, \zeta} \mathbb{Z}^d$ can be described as a universal C* algebra, so, in \cref{sec:Universal_C_star_algebras} we provide a review on universal C* algebras.
    \item Let $B$ be a C* algebra and $\mathcal{B}$ a smooth sub algebra of $B$, then, $\mathcal{B}$ is a Fréchet algebra, so, in \cref{chapter:frechet_algebras} we provide a review on Fréchet algebras and in \cref{sec:smooth_subalgebras} we provide a review on smooth sub algebras.
    \item The holomorphic functional calculus and the continuous functional calculus over a C* algebra $B$ are important for the definition of the groups $K_0(B)$ and $K_1(B)$, so, in \cref{sec:C_star_alg_cont_func_cal} we provide a review of the continuous functional calculus over C* algebras and in \cref{sec:banach_alg_hol_func_cal} we provide a review of the holomorphic functional calculus over Banach algebras.
    \item The holomorphic functional calculus over a smooth sub algebra $\mathcal{B}$ is important for the definition of the groups $K_0(\mathcal{B})$ and $K_1(\mathcal{B})$, so, in \cref{sec:smooth_sub_algebras_functional_calculus} we provide a review of the holomorphic functional calculus on smooth sub algebras.
    \item We will look at the twisted crossed product $A \rtimes_{\alpha, \zeta} \mathbb{Z}^d$ as an algebra of operators over a Hilbert space, so, in \cref{chapter:hilbert_spaces_section} we provide a review on the relevant results about Hilbert spaces used in this document.
    \item We will look at how some concepts from the Fourier analysis over $\mathbb{T}^d$ translate into the context of the twisted crossed product $A \rtimes_{\alpha, \zeta} \mathbb{Z}^d$, so, in \cref{chapter:review_of_Fourier_analysis} we provide a review on basic results of harmonic analysis over abelian locally compact Hausdorff groups and how these lead to relevant results on the Fourier analysis over $\mathbb{T}^d$.
    \item In \cref{chapter:cyclic_cohomology} we provide a review on the facts of cyclic cohomology that are relevant to the formulation of topological invariants for effective models that describe the dynamics of electrons on a crystal subject to weak disorder.
\end{itemize}

\chapter{Banach algebras}
\label{chapter:Banach_algebras}

A Banach algebra is a normed algebra that is complete in its norm, notice that a Banach algebra is by definition a Banach space. Let us dissect this definition. First, a normed algebra\index{normed algebra} is an associative algebra over a scalar field  $\mathbb{K}$ ($\mathbb{C}$ or $\mathbb{R}$) such that there is a mapping $a \mapsto\|a\|$ of $A$ into $\mathbb{R}^{+}$ with the following properties:

\begin{itemize}
    \item $(A ;\|\cdot\|)$ is a normed space over $\mathbb{K}$.
    \item $\|a b\| \leq\|a\|\|b\| \quad(a, b \in A)$ (submultiplicative).
    \item $\|a + b\| \leq \|a\| + \|b\| \quad(a, b \in A)$ (triangle inequality).
    \item $\|\lambda a\| = |\lambda| \|a\| \quad(a\in A, \: \lambda \in \mathbb{K}) $ (homogeneity).
\end{itemize}

\begin{remark}[Norm is continuous]\label{remark:norm_is_continuous}
In a normed space the norm is a continuous function, a fact that comes from the triangle inequality (\citep{265285}), so, we have that
$$ a \mapsto \|a\|  $$
is continuous.
\end{remark}

A Banach space\index{Banach space} is a normed vector space that is complete under its norm, that is, all the previously mentioned properties of the norm hold but not the sub-multiplicativity, since a Banach space does not require a multiplication operation to exist. Thus, a Banach algebra\index{Banach algebra} is a Banach space with a multiplication operation that is continuous with respect to the norm. Since the multiplication and the addition are continuous, we have that for any polynomial $p(x)$ the map $p: A \to A $ with $a \mapsto p(a)$ is continuous. 

Banach algebras are ubiquitous in functional analysis, just to mention some examples we have:

\begin{example}[Example of Banach algebras]\label{example:banach_algebras}

\begin{itemize}
    \item The field of the real numbers with the norm given by the absolute value is a Banach algebra i.e. $(\mathbb{R},*,+, \| \|)$
    
    \item The space $l^{1}$ of convergent series $a = \sum_{i \in \mathbb{N}} a_i$ such that $\sum_{i\in \mathbb{N}} \|a \| < \infty$ is Banach algebra, were the addition is given term by term and the multiplication is given by the convolution $(fg)(n) = \sum_{i \in \mathbb{N}}f(i-n)g(n)$. In $l^{1}$ the norm is $ \| a \| = (\sum_{i\in \mathbb{N}} \|a \|)$. This is a special case of the Banach spaces $l^{p}$ ($p \geq 1$) of convergent series $a = \sum_{i \in \mathbb{N}} a_i$ such that $\sum_{i\in \mathbb{N}} \|a \| ^{p} < \infty$, were the addition is given term by term, and the norm is $ \| a \|_p = (\sum_{i\in \mathbb{N}} \|a \| ^{p})^{1/p}$. Is important to mention that neither $l^p$ is a Banach algebra, for example, $l^2$ is not a Banach algebra since the convolution does not give an element in $l^2$ \citep{l2_not_banach_alg_stacx}.
    
    \item A generalization of the previous example: let $(X,\mathcal{A},\mu)$ be a measure space, then the space of all equivalence classes of measurable functions from $X$ to the complex numbers $\mathbb{C}$ such that $\int_{X} \|f(x)\| d\mu \leq \infty$  is a Banach space ($L^{1}(X,\mu)$). If $X$ is a locally compact group with $\mu$ the  left Haar meassure of $X$ (\citep[Theorem 2.1]{vinroot_haar_2008}), we can turn $L^{1}(X,\mu)$ into a Banach algebra by setting the following operations
    \begin{itemize}
        \item $(f+g)(x) = f(x) + g(x)$
        \item  $f^*(x)=\overline{f\left(x^{-1}\right)} \Delta(x)^{-1} \quad\left(f \in L^1(X), x \in X\right)$, with $ \Delta(x)$ the modular function of the Haar measure \citep[page 6]{vinroot_haar_2008}.
        \item $(f \star g)(x)=\int_G f(y) g\left(y^{-1} x\right) \mathrm{d} m(y) \; \left(f, g \in L^1(X), \; y, x \in X\right)$ such that $\| f \star g \| \leq \| f \| \| g \|$ (\citep[Theorem 1.6.2]{deitmar_principles_2009}), this operation is called the convolution\index{convolution}.
    \end{itemize}
    this is called the group algebra of $X$ \citep[Proposition 9.1.4 and Definition 9.1.5]{dales_introduction_2003}. $L^1(X, \mu)$ is very special, because in general we cannot make $L^2(X,\mu)$ into a Banach algebra for a general locally compact group, for example, there are functions in $L^{2}(\mathbb{R},\mu)$ such that their convolution is not in $L^2(\mathbb{R},\mu)$ \citep{L2_over_real_no_banach_alg_mathx} with $\mu$ the Lebesgue measure.
    
    \item Banach algebras usually arise as operator algebras, for example, let $A$ be a Banach space, the algebra of continuous linear operators from $A$ to $A$ ($\mathcal{L(A)}$) is also a Banach algebra with multiplication given by composition and norm given by the operator norm i.e. for $T: A \to B$ we have $\| T \| = \sup_{x \in A, \: \| x \|_{A} = 1 } \| T(x) \|_{A}$. Again, this is a special case of the Banach spaces $\mathcal{L}(A,B)$ of continuous linear maps from the Banach space $A$ to the Banach space $B$. 
\end{itemize}

For more examples and results on Banach algebras we highly recommend \citep{dales_introduction_2003} and \citep{allan_introduction_2011}. 

\end{example}

\section{Spectrum}
\label{ref:spec_banach_alg}

\textbf{From now on we will work with Banach algebras over the complex numbers.}\\

When we study operators an important property of them is their spectrum, which tell us when the operator $(T-\lambda I)$ has an inverse, where $I$ is the identity operator on the Banach space. Furthermore, for applications in quantum mechanics the spectrum of a self-adjoint operator has a great importance, the spectrum of a self-adjoint operator $T$ can be interpreted as the set of values that can be obtained from a physical measurement of its associated physical observable.

Let $A$ be a unital Banach algebra and $a \in A$, we will denote by $G(A)$\index{$G(A)$} the set of invertible elements of $A$, that is:

$$ G(A) = \{ a \in A | \exists b \in A \: \text{s.t.} \: ab = ba = 1  \}. $$

For $a \in A$, the resolvent set of $a$ is defined as,

$$  
R_A(a) = \{ \lambda \in \mathbb{C} | (\lambda 1 -a) \in G(A) \},
$$\index{$R_A(a)$ (resolvent set)}

and the spectrum of $a$ is defined as $Sp_A(a) = \mathbb{C} - R_A(a)$\index{$\text{Sp}_A(a)$}. When it is clear to which algebra we are referring we will drop the subscripts and just write $Sp(a)$ and $R(a)$\index{$R(a)$ (resolvent set)}. $Sp_A(a)$ is also referred to as the algebraic spectrum of $a$ with respect to $A$\index{algebraic spectrum}. 

The spectrum has many interesting properties both at the algebraic and topological level, for example, if $p$ is a complex polynomial then (\citep[Lemma 4.22]{allan_introduction_2011})
$$ Sp_A(p(a)) = \{ p(\lambda) | \; \lambda \in Sp_A (a) \}.  $$

\subsection{Unitization}
\label{sec:banach_alg_unitization}

If $A$ has no unit, we cannot define the spectrum of an element as we did in the previous paragraph, because the group $G(A)$ is not defined. However, we can easily construct a unital Banach algebra $A^{+}$ such that $A \subset A^{+}$ and $A$ is an ideal of $A^{+}$, to that algebra we will refer to as the unitization of $A$ \citep[4.3 page 161]{allan_introduction_2011}:

$$ A^{+} = A \oplus \mathbb{C} $$ \index{$A^+$}

 with:
 \begin{itemize}
     \item Addition: $(a,\lambda) + (b,\sigma) = (a+b, \lambda + \sigma)$ with addition identity $(0,0)$.
     
     \item Multiplication: $(a,\lambda) \cdot (b,\sigma) = (ab + \lambda b + \sigma a, \lambda \cdot \sigma)$, with multiplication identity $(0,1)$.
     
     \item Norm: $\| (a,\lambda) \| = \| a \| + | \lambda |$
 \end{itemize}

Using the submultiplicativity of the norm, it is possible to show that $A^{+}$ is a normed algebra, furthermore, is possible to show that $A^{+}$ is a complete iff $A$ is a complete. There are many ways to define a norm on $A^{+}$, the norm we have described is termed the $l_1$-norm and is the greatest regular norm that can be assigned to $A^{+}$ if the norm on $A$ is regular, that is, if $\|a \| = \sup \{ \| ax \|, \| xa \|: x \in A, \|x \| = 1 \}$ \citep{arhippainen_norms_2007}.

Now, for $A$ a non-unital Banach algebra and $a \in A$, define $Sp_A(a) = Sp_{A^{+}}(a)$ and $R_A(a) = R_{A^{+}}(a)$. Since $A$ is an ideal of $A^{+}$ and the unit of $A^{+}$ does not belong to $A$ we have that $0 \in Sp(a)$ i.e. $a$ is not invertible. Also, there is a canonical Banach algebra morphism that comes with the unitization,

$$ \pi: A^{+} \to \mathbb{C}, \; (a,\lambda) \to \lambda .$$

\subsection{Topology of the spectrum}
\label{sec:banach_alg_top_of_spec}

Take $A$ a Banach algebra and $a \in A$, for every $\lambda \in R(a)$ the resolvent mapping is defined as $r: R_a \to A$ such that $R_a(\lambda) = (\lambda 1 -a)^{-1}$. Note that the resolvent mapping is defined for all $\lambda > \| a \|$, since $(\lambda 1 -a)^{-1} = \lambda^{-1} + \sum_{k=1}^{\infty} \lambda^{-(k+1)}a^k$; this comes from the fact that for $b \in A$ with $\| b \| <1$ we have that $1 - b \in G(A)$ and $(1-b)^{-1} = 1 + \sum_{k=1}^{\infty}b^k$ \citep[Lemma 4.10]{allan_introduction_2011}.

\begin{proposition}\label{proposition:GA_topological_group_and_open_in_banach_algebra}
Let $A$ be a unital Banach algebra, then, 
\begin{enumerate}
    \item The set $G(A)$ is open in $A$ (\citep[Corollary 4.11]{allan_introduction_2011})
    \item $G(A)$ is a topological group, where its topology is inherited from the norm topology of $A$ , that is, 
    \begin{itemize}
        \item The multiplication $a \times b \mapsto ab$ is a continuous operation inside $G(A)$.
        \item The inversion $a \mapsto a^{-1}$ is a continuous operation inside $G(A)$ (\citep[Corollary 4.12]{allan_introduction_2011})
    \end{itemize}
\end{enumerate}
\end{proposition}

\begin{remark}\label{remark:resolvent_mapping_is_continuous_banach_algebra}
 Since the inversion operation is continuous, for any element $a \in A$ the resolvent mapping $R_a \to G(A)$ is a continuous.
\end{remark}

After extending the Cauchy integral formula to weakly holomorphic functions \citep[Theorem 3.11]{allan_introduction_2011} and extending the Liouville's theorem to vector valued functions \citep[Theorem 3.12]{allan_introduction_2011}, it is possible to give some nice properties of the topology of both the spectrum and the resolvent sets:

\begin{theorem}[\textbf{Theorem 4.17} \citep{allan_introduction_2011}]
\label{theo:spec_compact_non_empty}
Let $A$ be a Banach algebra, and let $a \in A$. Then $\mathrm{Sp}(a)$ is a nonempty, compact subset of the disc $\{\lambda \in \mathbb{C}:|\lambda| \leq\|a\|\}$.
\end{theorem}

The previous theorem also tell us that $R(a)$ is open. Now, since $Sp(a)$ is compact, we can ask what is the radius of the greatest disk that contains $Sp(a)$, this takes us to define the spectral radius of $a$,

$$ \rho(a) := \sup_{\lambda \in Sp(a)} \{ | \lambda | \} .$$\index{$\rho(a)$ (spectral radius)}

The spectral radius is related to the algebra norm through the Beurling-Gelfand spectral radius formula:

\begin{theorem}[\textbf{Theorem 4.23} \citep{allan_introduction_2011} Spectral radius formula]
\label{theo:spec_rad_form}
Let $A$ be a Banach algebra, and let $a \in A$. Then
$$
\rho_A(a)=\lim _{n \rightarrow \infty}\left\|a^n\right\|^{1 / n}=\inf _{n \in \mathbb{N}}\left\|a^n\right\|^{1 / n} .
$$
\end{theorem}

\begin{remark}\label{remark:upper_bound_on_spectral_radius}
Let $A$ be a Banach algebra and take $a \in A$, then, $\| a ^n \| \leq \| a \|^n$ due to the submultiplicativity property of Banach algebras, hence, \cref{theo:spec_rad_form} implies that $\rho_A(a) \leq \| a\|$. 
\end{remark}

As a final remark, for every unital Banach algebras $A$, we have that $G(A)$ is open \citep[Corollary 4.11]{allan_introduction_2011}; this fact is proven by showing that for every $a \in G(A)$ and $b$ such that $\| b -a \| < 1/ \| a ^{-1} \|$ then $b \in G(A)$. Denote by 
$$B(a, r) := \{ b \in A | \; \| b-a \| \leq r \},$$
then is possible to use the fact that $B(a,1/ \| a ^{-1} \|) \subset G(A)$ when $a \in G(A)$ to show the following useful results for $R_a(\lambda)$

\begin{proposition}[Lower bound for $R_a (\lambda)$(Corollary 6.9 \citep{abramovich_invitation_2002})]\label{proposition:bound_on_the_resolvent_norm}

Let $A$ be a Banach algebra and $a \in A$ then:
\begin{itemize}
    \item If $\lambda \in R_A (a)$, then $\|R_a(\lambda)\| \geq \frac{1}{d(\lambda, \text{Sp}_A (a) )}$, where $d(\lambda,  \text{Sp}_A (a))$ is the distance from $\lambda$ to the spectrum $\text{Sp}_A (a)$. 
    \item Assume that a sequence $\left\{\lambda_n\right\} \subseteq R_A (a)$ satisfies $\lambda_n \rightarrow \lambda_0 \in \mathbb{C}$. Then $\lambda_0 \in \text{Sp}_A (a)$ if and only if $\left\|R_a \left(\lambda_n\right)\right\| \rightarrow \infty$. 
\end{itemize}

\end{proposition}

There are some special elements of Banach algebras that will be useful in our study of K theory \cref{sec:equivalence_relations_C_star_algebras}, these are the idempotents, so, let $A$ be a Banach algebra and $e \in A$, $e$ is an idempotent\index{idempotent} if $e^{2} = e$. The sets of idempotents of the Banach algebra $A$ is denoted by $Q(A)$\index{$Q(A)$}. The spectrum of an idempotent is fairly simple: $Sp(e) \subset \{ 0,1 \}$ \citep[Proposition 4.16]{allan_introduction_2011}, also, the set of idempotents is close, this comes from the continuity of the multiplication on Banach algebras because if $a_n \to a$ and $a_n^2 = a_n$ then we must have that $a_n^2 \to a$, which together with $a_n^2 \to a^2$ gives us that $a^2 = a$. The set of idempotent elements of a Banach algebra $A$ is denoted by $Q(A)$. 

Another type of special elements that are usually studied in Banach algebras are the nilpotents\index{nilpotent} elements, these are $a \in A$ such that there is $n \in \mathbb{N}$ with $a^{n} = 0$, the spectrum of these elements is simpler that the spectrum of the idempotents since it consists of only the eigenvalue $0$  \citep[Page 166]{allan_introduction_2011}.

\subsection{Morphisms}
\label{sec:banach_alg_morph}

A morphism between two Banach algebras $A,B$ is a morphism of algebras $\phi: A \to B$ that is continuous with respect to the norm topologies of the algebras. For every algebraic morphism $\phi : A \to B$ between Banach algebras we have that $\phi(G(A)) \subseteq G(B)$ which implies that  $Sp_B(\phi(a)) \subseteq Sp_A(a)$.

A Banach algebra has two structures, the algebraic structure and the topological structure, which are bound by the continuity of the algebraic operations, nevertheless there is no a priori relation between the algebraic and topological structures of algebraic homorphisms between Banach algebras. Let 
$$\psi: A \to B$$
be an algebraic morphism between two algebras, were one or both of them are Banach algebras $A$ and $B$, we may ask if \textbf{$\psi$ is continuous?} This question is studied under the Banach algebra theory as \textit{automatic continuity}\index{automatic continuity} \citep[Part I Chapter 5]{dales_introduction_2003} and there are both positive and negative results for this question.

Automatic continuity is a rather complicated subject were many technicalities come into play, for example:

\begin{itemize}
    \item One of the first positive results on automatic continuity says that if $B$ is a semisimple Banach algebra then $\psi$ is automatically continuous \citep[Proposition 5.1.1]{dales_introduction_2003}.
    \item We also have negative results, like the theorem that tells us that for any infinite compact group $G$ there is a discontinuous homorphism from $L^2(G)$ into a Banach algebra \citep[Theorem 2]{runde_discontinuous_1996}.
    \item If we are concerned with $A$ and $B$ both being Banach algebras there are examples of discontinuous homorphisms were the domain is a Commutative Banach algebra ($C(K)$ with $K$ compact), and these constructions relay heavily on the Continuous Hypothesis \citep[Thereom 1.10]{dales_introduction_1987}, such that the existence of those homomorphisms cannot without the Continuous Hypothesis \citep[Automatic continuity of homomorphisms from C* -algebras, section 4]{bierstedt_functional_1984}, \citep[Part I Chapter 5 Additional note 4]{dales_introduction_2003}.
\end{itemize}  

Automatic continuity has a positive answer if $A$ and $B$ are both C* algebras ( \cref{proposition:automatic_continuity_C_star_algebras}), and this is a strong motivation for us to focus into C* algebras. For more information about automatic continuity of morphisms between Banach algebras we recommend \citep[Chapter 5.1]{dales_introduction_2003} and \citep[Chapter 5, section Automatic continuity]{allan_introduction_2011}.

\subsection{Spectrum continuity}
\label{sec:bana_alg_spec_conti}

In this document we work with functional calculus over our algebras, and the functional calculus depends on the spectrum of our elements, thus we would like to describe more carefully the spectrum.

Take $a \in A$ an element of a Banach algebra, as we have mentioned, its spectrum $Sp(a)$ is a compact subset of the complex numbers, thus, we can use the Hausdorff metric to compare the spectrum of two elements of the Banach algebra. Recall that for any complete metric space $(X,d)$ the set of closed and bounded subsets of $X$ ($\mathcal{K}(X)$) is a complete metric space under the Hausdorff metric \citep[Example 6.15]{barnsley_fractals_nodate}.

The Hausdorff metric has two equivalent definitions \citep[Proposition 2.1]{henrikson_completeness_1999}, on one side, take $A,B \in \mathcal{K}(X)$, then the Hausdorff metric takes the form,

$$
d(A, B)=\max \left\{\sup _{e \in A} d(e, B), \sup _{e \in B} d(e, A)\right\},
$$
where $d(e, B): X \times \mathcal{K}(X) \rightarrow \mathbf{R}$ is given by
$$
d(e, B)=\inf _{b \in B} d(e, b)
$$

On the other side, for $A$ in $\mathcal{K}(X)$\index{$\mathcal{K}(X)$} define its $\epsilon$-expansion as $ E_\epsilon(A)=\bigcup_{x \in A} B(x, \epsilon) $\index{$E_\epsilon(A)$}, which corresponds to the union of all $\epsilon$-open balls around points in $A$. Then $d(A, B)$ is defined as the "smallest" $\epsilon$ that allows the expansion of $A$ to cover $B$ and vice versa:
$$
d(A, B)=\inf \left\{\epsilon>0 \mid E_\epsilon(A) \supset B \text { and } E_\epsilon(B) \supset A\right\}
$$

\begin{figure}[H]
    \centering
    \includegraphics[width=0.7\textwidth]{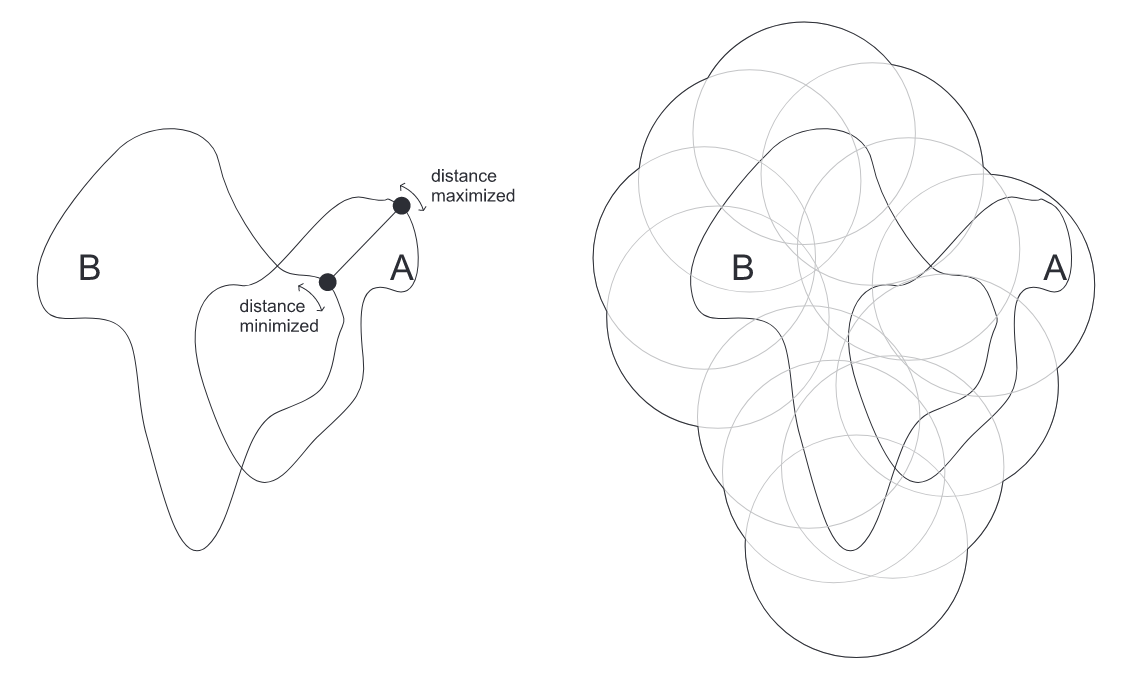}
    \caption{Visual representation of both definitions of the Hausdorff metric. Taken from \citep{henrikson_completeness_1999}.}
    \label{fig:hausd_metric}
\end{figure}

Many authors have studied the map $Sp: A \to \mathcal{K}(\mathbb{C}), \; a \mapsto 
Sp(a)$ when $A$ is a Banach algebra, if you wish to know more about that subject we recommend the article \citep{burlando_continuity_1994}. Continuity of the spectrum can be tackle for more general scenarios, for example the continuity of the spectrum of a field of self-adjoint operators $(H_{\alpha})_{\alpha \in \tau}$ \citep{beckus_continuity_2016}, which plays an important role in the study of aperiodic Schrodinger operators \citep{beckus_thesis_2016}. When fields of operators are considered is possible to use other topologies than the norm topology, and unexpected behaviours come into light, for example, there are examples of fields of self-adjoint operators $(H_{\alpha})_{\alpha \in \tau}$ that are strongly continuous in $\alpha$ and $ \alpha \mapsto Sp(H_{\alpha})$ is also continuous, but $\alpha \mapsto \| H_{\alpha} \|$ is not continuous \citep[Example 2.0.2]{beckus_thesis_2016}.

\begin{definition}\label{definition:upper_and_lower_semi_continuity_spectral_map}
Take $a_0 \in A$ with $A$ a Banach algebra, the map $Sp : A \mapsto \mathcal{K}(\mathbb{C})$ is called upper semi-continuous\index{upper semi-continuous map} at $a_0 \in A$ if for any $\epsilon>0$ there exists a neighbourhood $\Omega$ of $x_0$ in $A$ such that $Sp(a) \subset E_{\epsilon}(Sp(a_0))$. The map the map $Sp : A \mapsto \mathcal{K}(\mathbb{C})$ is called lower semi-continuous\index{lower semi-continuous map} at $a_0 \in A$ if for any $\epsilon>0$ there exists a neighbourhood $\Omega$ of $x_0$ in $A$ such that $Sp(a_0) \subset E_{\epsilon}(Sp(a))$ for any $a \in \Omega$.
\end{definition}

\begin{proposition}[c.f. Theorem 1.1 \citep{burlando_continuity_1994} and Proposition 4.21 \citep{allan_introduction_2011}]\label{proposition:upper_semi_continuity_spectrum_map}
Let $A$ be a Banach algebra and $a \in A$, then, the map $a \mapsto Sp(a)$ is upper semi-continuous.
\end{proposition}

For particular elements of a Banach algebra $A$ the map $Sp$ is continuous, for example, if $a \in A$ and $Sp(a)$ is a totally disconnected set, then the map $Sp$ is continuous at $a$ \citep[Theorem 1.6]{burlando_continuity_1994}. However, as we have mentioned, the map $a \mapsto Sp(a)$ is not in general continuous since there are Banach algebras were the map $Sp$ fails to be lower semi-continuous, for example, there is an example of nilpotent elements whose limit is an element with spectrum different from $\{0\}$ \citep[Page 282]{rickart_general_1974}.  

\section{Bochner integral}
\label{sec:bochner_integral}

In this section we will give a brief account of the results behind the integration of functions taking values in a Banach space, that is $f:S \to A$ with $A$ a Banach space and $(S,\mathcal{A})$ a measurable space. The integration of vector valued functions is designed to generalize the integration of complex (or real) valued functions to functions with values in Banach spaces, and as we would expect, the results exposed here fall back into the well-known results from complex valued functions. The integrals that we will define are widely used in Banach spaces theory and provide a powerful tool to perform computations on Banach spaces. Note that all the results exposed are valid for Banach algebras.

\subsection{Measurable functions}
\label{sec:measur_funct}

 Let $X$ be a Banach space, $(S, \mathscr{A})$ a measurable space, and $f : S \to X$ a function, then we want to study what it means for $f$ to be measurable. We want our results to fall back into the standard integration theory when $X = \mathbb{C}$. In $\mathbb{C}$ we have two widely used $\sigma$-algebras, the Lebesgue $\sigma$-algebra and the Borel $\sigma$-algebra, with the Lebesgue $\sigma$-algebra being complete. 
 
 We may want to generalize the Lebesgue $\sigma$-algebra\index{Lebesgue $\sigma$-algebra} of Euclidean spaces into the context of Banach spaces, however, there are some undesirable behaviors associated with Lebesgue measurable functions, for example the composition of Lebesgue measurable functions is not always Lebesgue measurable (\citep{283548}). Also, we will deal with the measurability of functions $x^{*} \circ f$ with $x^{*}$ a member of the continuous dual of $X$, thus we need to generalize a $\sigma$-algebra of $\mathbb{C}$ for our analysis.
 
 For simplicity we will use the Borel $\sigma$-algebra for $X$ and $\mathbb{C}$, that is, the Borel $\sigma$-algebra\index{Borel $\sigma$-algebra} $\mathscr{B}(X)$\index{$\mathscr{B}(X)$} is the $\sigma$-algebra generated by the open sets of $X$. More generally, let $(X,\tau)$ be a topological space, then $\mathscr{B}(X) = \sigma(\tau)$, where $\sigma(\tau)$ is the smallest sigma algebra containing the sets in $\tau$ i.e. the open sets of $Y$ (\citep[Theorem 3.4]{schilling_measures_2005}). The Borel $\sigma$-algebra has some nice properties with respect to measures, for example, if $X$ locally compact metric space then any locally finite measure over the Borel sets is automatically a regular measure \citep[Remark of Corollary 11.22]{knapp_basic_2005}. Recall that a topological space $X$ is said to be second countable\index{second countable topological space} is there exists a countable collection of open sets $U = \{ U_n \}_{n \in \mathbb{N}}$ such that any open set of $X$ is an union of elements of a sub familiy of $U$. Notice that if $(X,\tau)$ is second countable, then, there is a countable set of open sets that generate $\mathcal{B}(X)$, in this case we say that $\mathcal{B}(X)$ is a countably generated $\sigma$-algebra\index{countable generated $\sigma$-algebra}.

\begin{definition}[Measurable functions]\label{definition:measurable_functions}

For functions $f:S \to X$ we will concerned with two ways of defining measurability:

\begin{itemize}
    \item \textbf{Measurable function:}\index{measurable function} The function $f$ is called measurable if the pre-images
$$
f^{-1}(B):=\{f \in B\}:=\{s \in S: f(s) \in B \}
$$
are measurable for $B \in \mathcal{B}(X)$. Here we are concerned about the Borel  $\sigma$-algebra of the Banach algebra $X$.

\item \textbf{Weakly measurable function:}\index{weakly measurable function} The function $f$ is called measurable if the scalar-valued function $x^* \circ f$ is measurable for every functional $x^*$ in the dual space $X^*$. Here we use both the structure of the continuous dual of $X$ and the Borel $\sigma$-algebra of $\mathbb{C}$.
\end{itemize}

\end{definition}

In general, the set of weakly measurable functions is contained in the set of measurable functions, and there are Banach spaces were this inclusion is strict \citep[example 1.4.3]{hytonen_analysis_2016}, however, if $X$ is separable both sets coincide (\cref{theo:measu_vs_weak_measu}), recall that a topological space is said to be separable\index{separable topological space} if it has a dense countable set.

\begin{theorem}[\textbf{Corollary 1.1.2} \citep{hytonen_analysis_2016}]
\label{theo:measu_vs_weak_measu}
If $X$ is separable, then for a function $f: S \rightarrow X$ the following assertions are equivalent:
\begin{itemize}
    \item $f$ is measurable;
    \item $x^* \circ f$ is measurable for all $x^* \in X^*$ (f is weakly measurable). 
\end{itemize}
\end{theorem}

Thus, separable Banach spaces will play an important role in our constructions.

\textbf{What functions make sense to integrate?}

Indicator functions will play in important role in the integration theory of Banach space valued functions, as they do in the definition of Lebesgue integral. Therefore, we are motivated to define a new type of measurability related to finite sum of indicator functions. So, first, define an indicator function in this context:

\begin{definition}[Simple functions\index{simple function} \textbf{Definition 1.1.3} \citep{hytonen_analysis_2016}]
\label{def:simple_func}
A function $f: S \rightarrow X$ is called simple if it is of the form $f=\sum_{n=1}^N \mathbf{1}_{A_n} \otimes x_n$ with $A_n \in \mathscr{A}$ and $x_n \in X$ for all $1 \leqslant n \leqslant N$. Here $\mathbf{1}_A$ denotes the indicator function of the set $A$ and we use the notation
$$
(f \otimes x)(s):=f(s) x.
$$
\end{definition}

Thus, we are led to the definition of strongly measurable functions:

\begin{definition}[Strongly measurable function\index{strongly measurable function} \textbf{Definition 1.1.4} \citep{hytonen_analysis_2016}]
\label{def:stron_meas_func}
A function $f: S \rightarrow X$ is strongly measurable if there exists a sequence of simple functions $f_n: S \rightarrow X$ such that $\lim _{n \rightarrow \infty} f_n=f$ point wise on $S$.
\end{definition}

Since the point-wise limit of measurable functions is measurable \citep[Corollary 8.9]{schilling_measures_2005}, any strongly measurable function is measurable, so we may wonder if there are spaces where these definitions coincide, and as we would have expected, these coincide when $X$ is separable. Furthermore, if $X$ is not separable there are examples of measurable functions that are not strongly measurable \citep[Example 1.1.5]{hytonen_analysis_2016}. 

Since the separability of $X$ is such a key component in our analysis, let us generalize a bit and look into functions $f$ whose range is separable. 

\begin{definition}[Separably valued function\index{separably valued function} \textbf{Page 5} \citep{hytonen_analysis_2016}]
\label{def:separ_val_func}
A function $f: S \rightarrow X$ is called separably valued if there exists a separable closed subspace $X_0 \subseteq X$ such that $f(s) \in X_0$ for all $s \in S$.
\end{definition}

With this definition we can look at one of the cornerstone theorems of the integration theory of Banach space valued functions, the Pettis measurability theorem:

\begin{theorem}[\textbf{Theorem 1.1.6} \citep{hytonen_analysis_2016} Pettis measurability theorem\index{Pettis measurability}, first version]
\label{theo:Pettis_measu_theo_first_version}
Let $(S, \mathscr{A})$ be a measurable space and let $Y$ be a weak* dense subspace of $X^*$. For a function $f: S \rightarrow X$ the following assertions are equivalent:

\begin{itemize}
    \item $f$ is strongly measurable
    \item $f$ is separably valued and weakly measurable
    \item $f$ is separably valued and $\left\langle f, x^*\right\rangle$ is measurable for all $x^* \in Y$.
\end{itemize}
Moreover, if $f$ takes its values in a closed linear subspace $X_0$ of $X$, then $f$ is the pointwise limit of a sequence of $X_0$-valued simple functions.
\end{theorem}
\begin{remark}\label{remark:all_measurabilities_coincide_in_seperable_range}
If $X$ is separable all three definitions of measurability coincide, thus we can use them interchangeably.
\end{remark}

Additionally, we have a nice characterization of strongly measurable functions in terms of measurability and the separability of their range,

\begin{theorem}[Corollary 1.1.10. \citep{hytonen_analysis_2016}]
\label{coro:meas_vs_stron_measu_and_separ}
 For a function $f: S \rightarrow X$, the following assertions are equivalent:
 
 \begin{itemize}
     \item $f$ is strongly measurable
     \item $f$ is separably valued and measurable
 \end{itemize}
\end{theorem}

We can approximate strongly measurable functions as scalar valued measurable functions,

\begin{proposition}[Approximating strongly measurable functions]\label{proposition:approximating_strongly_measurable_functions}
For strongly measurable functions we have the following results:

\begin{itemize}
    \item \citep[Corollary 1.1.7]{hytonen_analysis_2016} If $f: S \rightarrow X$ is strongly measurable, then there exists a sequence of simple functions $\left(f_n\right)_{n \geqslant 1}$ such that
$\left\|f_n(x)\right\| \leqslant\|f(x)\|$ and $f_n(x) \rightarrow f(x)$ for all $x \in X$.
    \item \citep[Corollary 1.1.9]{hytonen_analysis_2016} The pointwise limit $f: S \rightarrow X$ of a sequence of strongly measurable functions $f_n: S \rightarrow X$ is strongly measurable.
    \item \citep[Corollary 1.1.11]{hytonen_analysis_2016} If $f: S \rightarrow X$ is strongly measurable and $\phi: X \rightarrow Y$ is measurable, where $Y$ is another Banach space, then $\phi \circ f$ is strongly measurable.
\end{itemize}
\end{proposition}

\subsection{Strong measurability with respect to a measure}
\label{sec:strong_measu}

Now we turn our attention to the measure space $(S,\mathcal{A},\mu)$ (Space, $\sigma$-algebra, measure). Since we are looking forward to generalize the $L^{p}(S,\mu)$ spaces, we want our integrals to have finite values, which motivates the definition of $\mu$-strongly measurable functions. First some definitions.

\begin{definition}[$\mu$-simple function\index{$\mu$-simple function} (\textbf{Definition 1.1.13} \citep{hytonen_analysis_2016})]
\label{def:mu_simple_func}
A $\mu$-simple function with values in $X$ is a function of the form $f=\sum_{n=1}^N \mathbf{1}_{A_n} \otimes x_n$, where $x_n \in X$ and the sets $A_n \in \mathscr{A}$ satisfy $\mu\left(A_n\right)<\infty$
\end{definition}

\begin{definition}[$\mu$-almost everywhere \textbf{Page 8} \citep{hytonen_analysis_2016}]
\label{def:mu_almost_everyw}

We say that a property holds $\mu$-almost everywhere\index{$\mu$-almost everywhere} if there exists a $\mu$-null set $N \in \mathscr{A}$ such that the property holds on the complement of $N$.

\end{definition}

This leads us to the definition of $\mu$-strongly measurable functions

\begin{definition}[Strongly $\mu$-measurable function\index{strongly $\mu$-measurable function} \textbf{Definition 1.1.14. } \citep{hytonen_analysis_2016}]
\label{def:mu_strong_measu}

A function $f: S \rightarrow X$ is strongly $\mu$-measurable if there exists a sequence $\left(f_n\right)_{n \geqslant 1}$ of $\mu$-simple functions converging to $f \mu$-almost everywhere.

\end{definition}

As you may expect, strongly measurable functions are not always $\mu$-strongly measurable, for example, the constant function $1$ is always strongly measurable but is only $\mu$-strongly measurable iff $\mu$ is $\sigma$-finite \citep[Example 1.1.17]{hytonen_analysis_2016}. So, let us look into what $\sigma$-finite means:

\begin{definition}[$\sigma$-finite measure\index{$\sigma$-finite measure} \textbf{Page 8} \citep{hytonen_analysis_2016}]
\label{def:sigma_finite_measu}

A measure $\mu$ is said to be $\sigma$-finite if it admits an exhausting sequence, i.e., an increasing sequence $S^{(1)} \subseteq S^{(2)} \subseteq \ldots$ of sets in $\mathscr{A}$ of finite $\mu$-measure such that $\bigcup_{n \geqslant 1} S^{(n)}=S$.

\end{definition}

Turns out that $\mu$-strongly measurable functions are $\mu$-essentially supported on $\sigma$-finite measure spaces \citep[Proposition 1.1.15]{hytonen_analysis_2016}, and $\mu$-strongly measurable functions are $\mu$-almost everywhere equal to strongly measurable functions if $\mu$ is $\sigma$-finite,

\begin{proposition}[Proposition 1.1.16. \citep{hytonen_analysis_2016}]
\label{prop:stron_meas_and_mu_stron_measu}
Consider a function $f: S \rightarrow X$.

\begin{itemize}
    \item If $f$ is strongly $\mu$-measurable, then $f$ is $\mu$-almost everywhere equal to a strongly measurable function.
    \item If $\mu$ is $\sigma$-finite and $f$ is $\mu$-almost everywhere equal to a strongly measurable function, then $f$ is strongly $\mu$-measurable. 
\end{itemize}

\end{proposition}

Also, there is a Pettis measurability theorem for $\mu$-strongly measurable functions which adds $\mu$ into the characterizatio given in \cref{theo:Pettis_measu_theo_first_version},

\begin{theorem}[Pettis measurability theorem, second version (Theorem 1.1.20 \citep{hytonen_analysis_2016})]\label{theorem:pettis_measurability_theorem_second_version}
An $X$-valued function $f$ is said to be $\mu$-essentially separably valued if there exists a closed separable subspace $X_0$ of $X$ such that $f(s) \in X_0$ for $\mu$-almost all $s \in S$, and weakly $\mu$-measurable if $\left\langle f, x^*\right\rangle$ is $\mu$-measurable for all $x^* \in X^*$.

For a function $f: S \rightarrow X$ the following assertions are equivalent:
\begin{itemize}
    \item $f$ is strongly $\mu$-measurable.
    \item $f$ is $\mu$-essentially separably valued and weakly $\mu$-measurable.
    \item  $f$ is $\mu$-essentially separably valued and there exists a weak* dense subspace $Y$ of $X^*$ such that $\left\langle f, x^*\right\rangle$ is $\mu$-measurable for all $x^* \in Y$.
\end{itemize}
Moreover, if $f$ takes its values $\mu$-almost everywhere in a closed linear subspace $X_0$ of $X$, then $f$ is the $\mu$-almost everywhere pointwise limit of a sequence of $X_0$-valued simple functions.
\end{theorem}

And is also possible to add the effect of $\mu$ into the results on \cref{proposition:approximating_strongly_measurable_functions}

\begin{proposition}[Approximating $\mu$-strongly measurable functions]\label{proposition:approximating_mu_strongly_measurable_functions}
For $\mu$-strongly measurable functions we have the following results:

\begin{itemize}
    \item \citep[Corollary 1.1.21]{hytonen_analysis_2016}  If $f: S \rightarrow X$ is strongly $\mu$-measurable, there exists a sequence of $\mu$-simple functions $\left(f_n\right)_{n \geqslant 1}$ such that
$\left\|f_n(x)\right\| \leqslant\|f(x)\|$ and $f_n(x) \rightarrow f(x)$ for $\mu$-almost all $x \in X$.
    \item \citep[Corollary 1.1.23]{hytonen_analysis_2016} The $\mu$-almost everywhere limit $f: S \rightarrow X$ of a sequence of strongly $\mu$-measurable functions $f_n: S \rightarrow X$ is strongly $\mu$-measurable.
    \item \citep[Corollary 1.1.24]{hytonen_analysis_2016} If $f: S \rightarrow X$ is strongly $\mu$-measurable and $\phi: X \rightarrow Y$ is measurable, where $Y$ is another Banach space, then $\phi \circ f$ is strongly $\mu$ measurable provided at least one of the following two conditions is satisfied:
    \begin{itemize}
        \item $\mu$ is $\sigma$-finite
        \item $\phi(0)=0$.
    \end{itemize}
\end{itemize}
\end{proposition}

\begin{remark}[Algebra of strongly ($\mu$-)measurable functions]\label{remark:algebra_of_strongly_mu_measurable_functions}

Let $X$ be a Banach algebra, and denote by $S_{X,\mu}$ ($S_X$) the set of $\mu$-strongly measurable (strongly measurable) functions, so if $f \in S_{X,\mu}$ ($f \in S_{X}$) there is a sequence of $\mu$-simple functions (simple functions) $\{ f_n \}_{n \in \mathbb{N}}$ such that $f_n \to f$ almost everywhere.

\begin{itemize}
    \item \textbf{Scalar multiplication:} Take $\lambda \in \mathbb{C}$ and $f \in S_{X,\mu} (S_X)$ with $f_n \to f$ a.e., set $(\lambda f)(x) = \lambda f(x)$, then $\lambda f_n$ are $\mu$-simple functions (simple functions) such that $\lambda f_n \to \lambda f_n$ a.e., thus $\lambda f \in S_{X,\mu} (S_X)$.
    \item \textbf{Addition:} Take $f,g \in  S_{X,\mu} (S_X)$ with $f_n \to f, \; g_n \to g$ a.e., set $(f+g)(x) = f(x) + g(x)$, then $f_n + g_n$ are $\mu$-simple functions (simple functions) such that $f_n + g_n \to f_g$ a.e., thus  $f+g \in  S_{X,\mu} (S_X)$.
    \item \textbf{Multiplication:} Take $f,g \in  S_{X,\mu} (S_X)$ with $f_n \to f, \; g_n \to g$ a.e., set $(fg)(x) = f(x) g(x)$, then $f_n g_n$ are $\mu$-simple functions (simple functions) such that $f_n g_n \to f_g$ a.e., thus  $fg \in  S_{X,\mu} (S_X)$.
\end{itemize}

The set of $\mu$-strongly measurable (strongly measurable) functions is an algebra under point-wise multiplication.

\end{remark}

\begin{remark}[$f: S \to X$ with $X$ separable]\label{remark:what_happend_if_X_is_separable}

Let $f: S \to X$ and $X$ separable, then our most of the definitions of measurability coincide:

\begin{itemize}
    \item $f$ is measurable iff $f$ is weakly measurable by \cref{theo:measu_vs_weak_measu}.
    \item $f$ is strongly measurable iff $f$ is measurable by \cref{coro:meas_vs_stron_measu_and_separ}.
\end{itemize}

Now, what can we say about strong $\mu$-measurability? Our best guess is to enforce $\sigma$-finitenes on $\mu$, 
\begin{itemize}
    \item if $\mu$ is $\sigma$-finite then $f$ is strongly measurable iff $f$ is $\mu$-strongly measurable by \cref{prop:stron_meas_and_mu_stron_measu},
\end{itemize}
otherwise a strongly measurable function may fail to be strongly $\mu$-measurable \citep[Example 1.1.17]{hytonen_analysis_2016}. 
\end{remark}

\begin{remark}[$\mu$-strongly measurable functions and composition with continuos functions]\label{remark:norm_of_mu_strongly_measurable_functions}
Also, if $f:S \to X$ is strongly $\mu$-measurable ($f_n \to f$ $\mu$-almost every where and $f_n$ are $\mu$-simple functions) and $g: X \to Y$ is continuous, then $g \circ f_n$ are $\mu$-simple functions converging to $g \circ f$ $\mu$-almost everywhere, therefore $g \circ f$ is strongly $\mu$-measurable. Consequently, if $f$ is strongly $\mu$-measurable then $s \mapsto \| f(s) \|$ is also strongly $\mu$-measurable, along with $s \mapsto \| f(s) \|^{p}$ for any $p > 0$. Notice that this fact also comes as a consequence of \cref{proposition:approximating_mu_strongly_measurable_functions} because $\|0\|^p = 0$.
\end{remark}

\subsection{Bochner integral}
\label{sec:Bochner_integral}

Now we turn our attention into defining the Bochner integral. Let $(S,\mathcal{A},\mu)$ be a measure space, then for a $\mu$-simple function $f=\sum_{n=1}^N \mathbf{1}_{A_n} \otimes x_n$ define
$$
\int_S f \mathrm{~d} \mu:=\sum_{n=1}^N \mu\left(A_n\right) x_n .
$$

For $\mu$-simple functions we have that:

\begin{itemize}
    \item $\left\|\int_S f \mathrm{~d} \mu\right\| \leqslant \int_S\|f\| \mathrm{d} \mu$
    \item $\int_S f \mathrm{~d} \mu+\int_S g \mathrm{~d} \mu=\int_S f+g \mathrm{~d} \mu$
\end{itemize}

Now, let us see what a Bochner integrable function is

\begin{definition}[Definition 1.2.1. \citep{hytonen_analysis_2016}]
\label{def:bochner_integ_func}
A strongly $\mu$-measurable function $f: S \rightarrow X$ is Bochner integrable\index{Bochner integrable} with respect to $\mu$ if there exists a sequence of $\mu$-simple functions $f_n: S \rightarrow X$ such that
$$
\lim _{n \rightarrow \infty} \int_S\left\|f-f_n\right\| \mathrm{d} \mu=0 .
$$

\end{definition}

For any $\mu$-strongly measurable function $f:S \to X$ the function $\|f \|: S \to \mathbb{R}$ is strongly $mu$-measurable by \cref{remark:norm_of_mu_strongly_measurable_functions}. Also, from
$$
\left\|\int_S f_n \mathrm{~d} \mu-\int_S f_m \mathrm{~d} \mu\right\| \leqslant \int_S\left\|f_n-f_m\right\| \mathrm{d} \mu \leqslant \int_S\left\|f_n-f\right\| \mathrm{d} \mu+\int_S\left\|f-f_m\right\| \mathrm{d} \mu
$$
we see that the integrals $\int_S f_n \mathrm{~d} \mu$ form a Cauchy sequence. By completeness, this sequence converges to an element of $X$. This limit is called the Bochner integral\index{Bochner integral} of $f$ with respect to $\mu$, notation

$$\int_S f \mathrm{~d} \mu:=\lim _{n \rightarrow \infty} \int_S f_n \mathrm{~d} \mu .$$

According to \citep{hytonen_analysis_2016} this definition does not depends on the choice of approximation sequence. Notice that for given a function $f: S \to X$ that is Bochner integrable, we have just assigned a new object inside $X$ to the function $f$, this will be a crucial tool in the development of the holomorphic functional calculus over Banach algebras (\cref{remark:holomorphic_calculus_is_a_Bochner_integral}). 

\begin{remark}[Bochner integral and multiplication]\label{remark:bochner_integral_and_multiplication}
Let $X$ be a Banach algebra and $f: S \to X$ a Bochner integral function, also, let $Q,P \in X$, since
$$ \| Pf(s)Q - P f_n(s)Q \| \leq \| P \| \| f(s) - f_n(s) \| \| Q \|,$$
we have that 
$\lim_{n \to \infty} \int_{S} \| P f(s) Q - Pf_n(s)Q \| d \mu =0.$
Since, each $P f_n Q$ is a $\mu$-simple function, and $P f_n Q \to P f Q$ almost everywhere, then, $PfQ$ is Bochner integrable. Additionally, since $\int_S f d\mu$ is corresponds to the limit finite sum, the continuity of multiplication on $X$ tells us that
$$ \int_{S} P f(s) Q d \mu  = P \left( \int_{S} f d \mu \right) Q. $$

A similar argument show us that if $f: S \to \mathbb{C}$ is a Bochner integrable function, then for any $P\in X$ we have that
$$ \int_{S} Pf  d \mu = \left( \int_{S} f d \mu \right) P.  $$
\end{remark}

As expected, a function $f$ is Bochner integrable iff its norm is integrable,

\begin{proposition}[Proposition 1.2.2. \citep{hytonen_analysis_2016}]
\label{prop:bochn_integra_condition}

$\mu$-measurable function $f: S \rightarrow X$ is Bochner integrable with respect to $\mu$ if and only if
$$
\int_S\|f\| \mathrm{d} \mu<\infty
$$
and in this case, we have
$$
\left\|\int_S f \mathrm{~d} \mu\right\| \leqslant \int_S\|f\| \mathrm{d} \mu
$$

\end{proposition}

\begin{example}[Continuous functions over a compact space]\label{example:Bochner_integral_continuos_funcion}
 If $(S,\mathcal{A},\mu)$ is a compact measure space with finite measure $\mu(S) < \infty$ such that $\mathcal{A} \subseteq \mathscr{B}(S)$, and $f: S \to A$ is a continuous function then $f$ is  Bochner integrable, let us check why this happens:
 
 \begin{itemize}
     \item f is strongly $\mu$-measurable: let us create an open cover of $S$ as follows, for $s\in S$ let $b_{x,\epsilon}$ be and open neighbourhood of $s$ such that if $y \in b_{x,\epsilon}$ then $\| f(y) - f(s) \| \leq \epsilon$, then $\cup_{s \in S}b_{s,\epsilon} = S$. Since $S$ is compact we can get a finite open cover $\cup_{i \leq n} b_{s_i, \epsilon} = X$, and based on that open cover we can create a simple function as follows:
     \begin{itemize}
         \item Set $a_1 = b_{s_1,\epsilon}$ and $a_j = b_{s_j,\epsilon} \cap (\cup_{i < j}a_i)^{c}$, then $a_j \in \mathcal{A}$ and $\cup a_i = S$.
         \item $f_{\epsilon}(s) = \sum_{i \leq n} 1_{a_i}(s)f(s_i)$ and $\mu(a_i)\leq \mu(S) < \infty$
        \item Then, $\| f(s) - f_{\epsilon}(s) \| \leq \epsilon$ for every $s \in S$
     \end{itemize}
     Hence, $\{ f_{1/n} \}_{n \in \mathbb{N}}$ is a sequence of $\mu$-simple functions that converge to $f$ poin-twise, which makes $f$ into strongly $\mu$-measurable, moreover, $f_n$ approximates $f$ uniformly 
     
    \item Let $\overline{a_i}$ the closure of $a_i$, then by the continuity of $f$ we know that $\|f(s) - f(s') \| \leq \epsilon$ for any $s \in \overline{a_i}$ and $s' \in a_i$. Since $a_i$ is closed then $\{\| f(s) \|\}_{s \in \overline{a_i}}$ has a minimum, denote it by $\hat{f_i}$ and $\hat{s_i} \in \overline{a_i}$ such that $\|f(s_i)\| = \hat{f_i}$, then set
     $$\overline{f}_{\epsilon}(s) = \sum_{i \leq n} 1_{a_i}(s) f(\hat{s_i}),$$
    then $\{ \overline{f}_{1/n} \}_{n \in \mathbb{N}}$ is a sequence of $\mu$-simple functions that approximate $f$ uniformly and $\|f_{1/n}(s) \| \leq \|f(s)\|$ for all $s$, thus, this is an example of the sequences that \cref{proposition:approximating_mu_strongly_measurable_functions} assure that exists, with the added valued that it approximates $f$ uniformly.  
     
     \item $f$ is Bochner integrable: We have that $s \mapsto \|f(s)\|$ is continuous, therefore is bounded because $S$ is compact, which in turns implies that
     $$ \int_S \|f(s)\| d \mu \leq  (\sup_{s \in S} \|f(s)\| ) \mu(S)< \infty,$$
     and by \cref{prop:bochn_integra_condition} we have that $f$ is Bochner mesurable.
 \end{itemize}
 
 This setup will come up in the formulation of the holomorphic functional calculus and in the definition of the Fourier Analysis over twisted crossed product C* algebras (\cref{sec:Fourier_analysis_twistted_crossed_product}).  
\end{example}

\begin{example}[$C_0(S;X)$ continuous functions decaying at infinity]\label{example:continuous_functions_decaying_at_infinty_are_strongly_measurable}
If $(S,\mathcal{A},\mu)$ is a locally compact Hausdorff space with a measure $\mu$ over a $\sigma$-algebra $\mathcal{A}$ such that $\mathcal{A} \subseteq \mathscr{B}(S)$, let $X$ be a Banach space then denote by $C_0(S;X)$ the vector space of continuous functions that decay at infinity, much like in \cref{sec:Continuous_functions_with_values_on_C_star_algebra}, that is, if $f \in C_0(S;X)$ then for any $\epsilon>0$ there is a $K \subset S$ compact such that $\|f_{K^c}(s)\| \leq \epsilon$.

Let $f\in C_0(S;X)$ and $\{K_n\}_{n \in \mathbb{N}}$ be a sequence of compact subsets of $S$ such that $\| f_{K_n^c} (s) \| \leq 1/n$, then from \citep{nlab:compact_subspaces_of_hausdorff_spaces_are_closed} we know that $K_n$ are closed, therefore $K_n \in \mathcal{A}$. Since the topology of $K_n$ is the inherited topology from $S$ as a subspace then from \cref{example:Bochner_integral_continuos_funcion} we know that there is $\mu$-simple function $f_n$ on $K_n$ such that $\|f_n(s) - f(s) \| \leq 1/n$ and $\| f_n(s)\| \leq \|f(s)\|$ for all $s \in K_n$, thus, extend it to $S$ by setting $f_n(s) =0$ if $s \notin K_n$, which again gives us a $\mu$-simple function. Then,
$$ \| f_n(s) - f(s) \| \leq 1/n, \; \| f_n(s)\| \leq \|f(s)\| \; s \in S,$$
thus, $f$ is $\mu$-strongly measurable, and can be approximated uniformly by $\mu$-simple functions.

Notice that $C_0(S) \odot X \subset C_0(S;X)$ because the finite sum of compact sets is a compact set.
\end{example}

\begin{example}[Continuous functions over a $\sigma$-finite space]\label{example:continuous_functions_on_sigmna_fintie_space}
Let $X$ be a $\sigma$-finite space, that is, $X = \cup_{n \in \mathbb{N}} K_n$ with $K_n \subseteq K_{n+1}$, $K_n$ compact and $\mu(K_n) < \infty$. Take $f \in C(X;A)$ with $A$ a Banach space, then 
$$ f = \lim_{n \to \infty} \mathbf{1}_{K_n}f $$ 
pointwise, and $\mathbf{1}_{K_n}f$ is a continuous function in a compact set. Under this setup we can follow a similar argument as in \cref{example:continuous_functions_decaying_at_infinty_are_strongly_measurable} to show that $f$ is the pointwise limit of $\mu$-strongly measruable functions, which in turn makes $f$ into a $\mu$-strongly measurable function.
\end{example}

The Bochner integral has the expected behavior with respect to linear continuous functionals and linear bounded transformations,

\begin{proposition}[Page 15 \citep{hytonen_analysis_2016}]
\label{prop:bochner_int_and_lin_bound_trans}

The Bochner integral has the following properties:

\begin{itemize}
    \item It $f:S \to X$ is Bochner integrable and $T \in \mathscr{L}(X, Y)$ then $Tf: S \to Y$ is Bochner integrable and 
    $$
    T \int_S f \mathrm{~d} \mu=\int_S T f \mathrm{~d} \mu .
    $$
    
    \item For all $x^* \in X^*$ we have that
    $$
    \left\langle\int_S f \mathrm{~d} \mu, x^*\right\rangle=\int_S\left\langle f, x^*\right\rangle \mathrm{d} \mu.
    $$
\end{itemize}

\end{proposition}

\begin{proposition}[Properties of the Bochner integral]\label{proposition:properties_of_Bochner_integral}
The results from Lebesgue integration carry over to the Bochner integral as long as the measures are not signed measures, to mention a few:
\begin{itemize}
    \item \textbf{Dominated convergence:}\index{Dominated convergence}  \citep[Proposition 1.2.5]{hytonen_analysis_2016} Let the functions $f_n: S \rightarrow X$ be Bochner integrable. If there exists a function $f: S \rightarrow X$ and a non-negative integrable function $g: S \rightarrow \mathbb{R}$ such that $\lim _{n \rightarrow \infty} f_n=f$ almost everywhere and $\left\|f_n\right\| \leqslant g$ almost everywhere, then $f$ is Bochner integrable and we have
$$
\lim _{n \rightarrow \infty} \int_S\left\|f_n-f\right\| \mathrm{d} \mu=0
$$
In particular,
$$
\lim _{n \rightarrow \infty} \int_S f_n \mathrm{~d} \mu=\int_S f \mathrm{~d} \mu .
$$

\item \textbf{Substitution theorem:} \citep[Proposition 1.2.6]{hytonen_analysis_2016} Let $(S, \mathscr{A}, \mu)$ be a measure space and let $(T, \mathscr{B})$ be a measurable space. Let $\phi: S \rightarrow T$ be measurable and let $f:$ $T \rightarrow X$ be strongly measurable. Let $\nu=\mu \circ \phi^{-1}$ be the image measure of $\mu$ under $\phi$. Then $f \circ \phi$ is Bochner integrable with respect to $\mu$ if and only if $f$ is Bochner integrable with respect to $\nu$, and in this situation
$$
\int_S f \circ \phi \mathrm{d} \mu=\int_T f \mathrm{~d} \nu
$$

    \item \textbf{Fubini theorem:}\index{Fubini theorem} \citep[Proposition 1.2.7]{hytonen_analysis_2016} Let $(S, \mathscr{A}, \mu)$ and $(T, \mathscr{B}, \nu)$ be $\sigma$-finite measure spaces and let $f: S \times T \rightarrow X$ be Bochner integrable, then
    
    \begin{itemize}
        \item For almost all $s \in S$ the function $t \mapsto f(s, t)$ is Bochner integrable.
        \item For almost all $t \in T$ the function $s \mapsto f(s, t)$ is Bochner integrable.
        \item The functions $s \mapsto \int_T f(s, t) \mathrm{d} \nu(t)$ and $t \mapsto \int_S f(s, t) \mathrm{d} \mu(s)$ are Bochner integrable and
$$
\int_{S \times T} f(s, t) \mathrm{d} \mu \times \nu(s, t)=\int_T \int_S f(s, t) \mathrm{d} \mu(s) \mathrm{d} \nu(t)=\int_S \int_T f(s, t) \mathrm{d} \nu(t) \mathrm{d} \mu(s) .
$$
    \end{itemize}

    
    \item \textbf{Fundamental theorem of calculus and more}\index{Fundamental theorem of calculus}:
    Various results from integro-differential calculus can be extended into the context of Bochner integrable functions if the domain of the function is a subset of $\mathbb{R}^d$, this happens in part because $\mathbb{R}^d$ is $\sigma$-finite and as we have seen, $\sigma$-finite measures have nice properties with respect to the Bochner integral e.g. \cref{proposition:approximating_mu_strongly_measurable_functions} and \cref{remark:what_happend_if_X_is_separable}.

    \begin{itemize}
        \item \textbf{Simple results:} The following are simple results that generalize the well-known results from basic integro-differential calculus. We say that a function $F$ is a \textit{primitive} of a function $f$, if $F'=f$ holds at each point, where we use the notation
        $$ F' (x) := \lim_{\delta x \to 0} \frac{ F(x + \delta x) - F(x)}{\delta x} .$$
        The proofs from the theorems applying to real-valued functions cannot be translated into the context of Banach space valued functions, because they relay on the mean value theorem, which ha no analog for the Banach valued functions, thus, other results are needed e.g. Heine-Borel covering theorem \citep[Chapter XIII, Section 2]{mikusinski_bochner_1978}.
        \begin{itemize}
            \item \citep[Chapter XIII, Corolary 1.1]{mikusinski_bochner_1978}: If $f$ is a Bochner integrable function on an interval $I \subset \mathbb{R}^1$ (closed or open) and $x_0 \in I$, then the indefinite integral $F(x)=\int_{x_0}^x f$ is continuous on $I$.  
            \item \citep[Chapter XIII, Theorem 2.1]{mikusinski_bochner_1978}: If the derivative $h'$ of a Banach value function $h$ vanishes everywhere, then $h$ is constant.  
            \item \citep[Chapter XIII, Theorem 2.2]{mikusinski_bochner_1978}: If $f$ is a continuous vector function in an interval $I \subset \mathbb{R}^1$, then the indefinite Bochner integral
                $$
                F(x)=\int_{x_0}^x f \quad\left(x_0 \in I\right)
                $$
            is, in $I$, a primitive function for $f$.
            \item \citep[Chapter XIII, Theorem 2.3]{mikusinski_bochner_1978}: If a function $f$ is continuous in an interval $(a, b)$ and $F$ is continuous on $[a, b]$ and has $f$ for its derivative in $(a, b)$, then the Bochner integral $\int_a^b f$ exists and we have
                $$
                \int_a^b f=F(b)-F(a)
                $$
        \end{itemize}
        \item \textbf{Generalization of absolute continuity:} The right setting for generalizing the notion of absolutely continuous functions is the setting of continuous local primitives of Banach valued functions. If $f : \mathbb{R} \to A$ with $A$ a Banach space, we say that $f$ is locally integrable if $f$ is Bochner integrable over any interval $[a,b]$ (\citep[Chapter XI section 1]{mikusinski_bochner_1978}). If $g$ is locally integrable then $f$ is called a locally derivative of $g$ if 
        $$
        \lim _{h \rightarrow 0} \int_a^b\left|\frac{1}{h}[f(x+h)-f(x)]-g(x)\right| d x=0
        $$
        holds for every bounded interval $(a, b)$ (\citep[Chapter XI section 3]{mikusinski_bochner_1978}).
        \begin{itemize}
            \item \citep[Chapter XIII theorem 3.3]{mikusinski_bochner_1978}: So, if $f$ is a locally integrable function, set 
                $$
                F(x):=\int_{x_0}^x f
                $$
                then $f$ is a local derivative of $F$, such that $F$ is continuous and
                $$
                \lim _{h \rightarrow 0} \frac{F(x+h)-F(x)}{h}=f(x)
                $$
                holds for almost every $x$, we denote $F' := f$. Similar results can be generalized into functions from $\mathbb{R}^d$ into a Banach algebra (\citep[Chapter XIII section 5]{mikusinski_bochner_1978}).
            \item If $f,g$ have local derivatives that are locally integrable e.g. if they are continuously derivable            functions, then we have the Leibniz rule (\citep[Chapter XIII theorem 6.1]{mikusinski_bochner_1978})
                $$ (fg)' = f' g +f g' ,$$
                 and we can perform integration by parts (\citep[Chapter XIII theorem 6.1]{mikusinski_bochner_1978})
                 $$ f(b)g(b) - f(a)g(a) = \int_{a}^b f' g + \int_{a}^bf g' .$$
            \item \citep[Chapter XIII theorem 3.4]{mikusinski_bochner_1978}: Continuous local primitives are derivable almost everywhere and their derivatives are equal to their local derivatives. More exactly, if $F$ is a continuous local primitive of $f$, then
            $$ \lim_{h \to 0} \frac{F(x + h) - F(x)}{h} = f(x)  $$
            holds for almost every $x$.
        \end{itemize}
    \end{itemize}
\end{itemize}
\end{proposition}

\begin{corollary}[Bochner integral and Riemann sums\index{Riemann sum}]\label{corollary:Bochner_integral_and_Riemann_sums}
Let $a,b \in \mathbb{R}$, let $f: [a,b] \to A$ be a continuous function into a Banach space $A$. Denote by $\text{part}$ the set of numbers $a = x_0 \leq x_1 \leq \dots \leq x_{\text{part}_{n}} = b$ and set $\delta(\text{part}) = \min_{1 \leq i \leq \text{part}_{n}} \{ |x_i - x_{i-1} |\}$, then,
$$ \int_{a}^{b} f(x) dx = \lim_{\delta(\text{part}) \to 0} \sum_{i = 1}^{\text{part}_{n}} f(\tilde{x}_i) (x_i - x_{i-1}),$$
where $\tilde{x}_i \in [x_{i-1}, x_{i}]$.
\end{corollary}
\begin{proof}
Since $f$ is continuous over a compact set it is uniformly continuous over $[a,b]$ (\citep{nlab:continuous_metric_space_valued_function_on_compact_metric_space_is_uniformly_continuous}), thus, the function 
$$f_{\text{part}_{n}}(x) =  \sum_{i = 1}^{\text{part}_{n}} f(\tilde{x}_i) 1_{[x_{i-1}, x_i)}(x)$$
converges uniformly to $f$ when $\delta(\text{part}) \to 0$, where $1_{[x_{i-1}, x_i)}(x)$ is the indicator function for the set $[x_{i-1}, x_i)$. Under this setting, the dominated convergence theorem (\cref{proposition:properties_of_Bochner_integral}) tells us that 
$$ \int_{a}^{b} f(x) d x = \lim_{\delta(\text{part}) \to 0} \int_{a}^{b} f_{\text{part}_{n}}(x) dx. $$
Since $f_{\text{part}_{n}}(x)$ is a $\mu$-simple function (\cref{def:mu_simple_func}), we have that
$$  \int_{a}^{b} f_{\text{part}_{n}}(x) dx = \sum_{i = 1}^{\text{part}_{n}} f(\tilde{x}_i) 1_{[x_{i-1}, x_i)}(x), $$
thus, we get that $\int_{a}^{b} f(x) dx$ is the limit of the Riemann sums $ \lim_{\delta(\text{part}) \to 0} \sum_{i = 1}^{\text{part}_{n}} f(\tilde{x}_i) (x_i - x_{i-1})$. 
\end{proof}

\begin{remark}[Approximating Bochner integrable functions by step functions]\label{remark:approximating_bochner_integrable_functions_with_step_functions}
Let $f$ be a Bochner integrable function, then, from \cref{proposition:approximating_mu_strongly_measurable_functions} we know that there is a sequence of $\mu$-simple functions $\{ f_n \}_{n \in \mathbb{N}}$ such that $f_n \to f$ and $\|f(s)\| \leq \|f(s)\|$ almost everywhere. So, from the dominated convergence theorem for Bochner integrable functions (\cref{proposition:properties_of_Bochner_integral}) we conclude that $\lim_{n \to \infty} \int_S f_n = \int_S f$ and $\lim_{n \to \infty} \int_{S} \| f_n - f \| = 0$.
\end{remark}

Since the Bochner integral behaves much like the Lebesgue integral when the measure space is $\sigma$-finite (\cref{proposition:properties_of_Bochner_integral}), we can translate various results from real (or complex) valued calculus into this setting. One example of this is the formulation of the holomorphic functional calculus on Banach algebras (\cref{theorem:holomorphic_functional_calculus_banach_algebras}), now we look into another set of examples:

\begin{lemma}[Generalizing some results from calculus]\label{lemma:generalizations_of_some_calculus_results}

These are some results that can be generalized into the context of Banach space valued functions, and by no means this is a comprehensive list of all the results that translate into this context:

\begin{enumerate}
    \item Suppose that $\left\{f_n\right\}$ is a sequence of functions which are continuous on $[a, b]$, and that $\left\{f_n\right\}$ converges uniformly on $[a, b]$ to $f$. Then $f$ is also continuous on $[a, b]$.

    \item Suppose that $\left\{f_n\right\}$ is a sequence of functions which are Bochner integrable on $[a, b]$, and that $\left\{f_n\right\}$ converges uniformly on $[a, b]$ to a function $f$ which is Bochner integrable on $[a, b]$. Then
$$
\int_a^b f=\lim _{n \rightarrow \infty} \int_a^b f_n
$$
    \item Suppose that $\left\{f_n\right\}$ is a sequence of functions which are differentiable on $[a, b]$, with Bochner integrable derivatives $f_n{ }^{\prime}$, and that $\left\{f_n\right\}$ converges (pointwise) to $f$. Suppose, moreover, that $\left\{f_n{ }^{\prime}\right\}$ converges uniformly on $[a, b]$ to some continuous function $g$. Then $f$ is differentiable and
$$
f^{\prime}(x)=\lim _{n \rightarrow \infty} f_n^{\prime}(x) .
$$

    \item Let $\sum_{n=1}^{\infty} f_n$ converge uniformly to $f$ on $[a, b]$.
    \begin{enumerate}
        \item If each $f_n$ is continuous on $[a, b]$, then $f$ is continuous on $[a, b]$.
        \item If $f$ and each $f_n$ is integrable on $[a, b]$, then
            $$
            \int_a^b f=\sum_{n=1}^{\infty} \int_a^b f_n .
            $$ 
        \item Moreover, if $\sum_{n=1}^{\infty} f_n$ converges (pointwise) to $f$ on $[a, b]$, each $f_n$ has an integrable      derivative $f_n{ }^{\prime}$ and $\sum_{n=1}^{\infty} f_n{ }^{\prime}$ converges uniformly on $[a, b]$ to some       continuous function, then
            $f^{\prime}(x)=\sum_{n=1}^{\infty} f_n^{\prime}(x) \quad$ for all $x$ in $[a, b]$.
    \end{enumerate}

\end{enumerate}
\end{lemma}
\begin{proof}
In this case, the definition of derivative is the usual one as exposed in \cref{remark:leibniz_rule_fn_values_in_topological_algebra}, 

\begin{enumerate}
    \item Follow the proof of \citep[Chapter 24, theorem 2]{spivak_calculus_1994} by replacing the absolute value by the norm of an element in a Banach space.

    \item Following the proof from \citep[Chapter 24, Theorem 1]{spivak_calculus_1994} : Let $\varepsilon>0$. There is some $N$ such that for all $n>N$ we have
    $$
    \left|f(x)-f_n(x)\right|<\varepsilon \quad \text { for all } x \text { in }[a, b] .
    $$
    Thus, if $n>N$ we can use \cref{prop:bochn_integra_condition} to get,
    $$
    \begin{aligned}
    \left|\int_a^b f(x) d x-\int_a^b f_n(x) d x\right| & =\left|\int_a^b\left[f(x)-f_n(x)\right] d x\right| \\
    & \leq \int_a^b\left|f(x)-f_n(x)\right| d x \\
    & \leq \int_a^b \varepsilon d x \\
    & =\varepsilon(b-a) .
    \end{aligned}
    $$
    Since this is true for any $\varepsilon>0$, it follows that
    $$
    \int_a^b f=\lim _{n \rightarrow \infty} \int_a^b f_n
    $$
    
    \item Following the proof from \citep[Chapter 24, Theorem 3]{spivak_calculus_1994}: Using the properties of the Bochner integral (\cref{proposition:properties_of_Bochner_integral}) and applying the previous result to the interval $[a, x]$, we see that for each $x$ we have
    $$
    \begin{aligned}
    \int_a^x g & =\lim _{n \rightarrow \infty} \int_a^x f_n{ }^{\prime} \\
    & =\lim _{n \rightarrow \infty}\left[f_n(x)-f_n(a)\right] \\
    & =f(x)-f(a) .
    \end{aligned}
    $$
    Since $g$ is continuous, it follows that $f^{\prime}(x)=g(x)=\lim _{n \rightarrow \infty} f_n^{\prime}(x)$ for all $x$ in the interval $[a, b]$.

    \item Following the proof from \citep[page 498, Corollay]{spivak_calculus_1994}: 
    \begin{enumerate}
        \item If each $f_n$ is continuous, then so is each $f_1+\cdots+f_n$, and $f$ is the uniform limit of the sequence $f_1, f_1+f_2, f_1+f_2+f_3, \ldots$, so $f$ is continuous by the first result on this lemma.
        \item Since $f_1, f_1+f_2, f_1+f_2+f_3, \ldots$ converges uniformly to $f$, it follows from the second result on this lemma that
        $$
        \begin{aligned}
        \int_a^b f & =\lim _{n \rightarrow \infty} \int_a^b\left(f_1+\cdots+f_n\right) \\
        & =\lim _{n \rightarrow \infty}\left(\int_a^b f_1+\cdots+\int_a^b f_n\right) \\
        & =\sum_{n=1}^{\infty} \int_a^b f_n .
        \end{aligned}
        $$ 
        \item Each function $f_1+\cdots+f_n$ is differentiable, with derivative $f_1{ }^{\prime}+\cdots+f_n{ }^{\prime}$, and     $f_1{ }^{\prime}, f_1{ }^{\prime}+f_2{ }^{\prime}, f_1{ }^{\prime}+f_2{ }^{\prime}+f_3{ }^{\prime}, \ldots$ converges uniformly to a continuous function, by hypothesis. It follows from the third results on this lemma that
            $$
            \begin{aligned}
            f^{\prime}(x) & =\lim _{n \rightarrow \infty}\left[f_1{ }^{\prime}(x)+\cdots+f_n{ }^{\prime}(x)\right] \\
            & =\sum_{n=1}^{\infty} f_n{ }^{\prime}(x) .
            \end{aligned}
            $$
    \end{enumerate}
\end{enumerate}
\end{proof}

\subsection{The Bochner spaces $L^p(S;X)$}
\label{sec:Bochner_L_p_spaces}

For $f,g : S \to X$ functions that are strongly $\mu$-measurable define the following relation, $f \sim g$ is $f(s)=g(s)$ $\mu$-almost everywhere, then $\sim$ is an equivalence relation. Take $f$ a strongly $\mu$ measurable function from $S$ into $X$, then, from \cref{remark:norm_of_mu_strongly_measurable_functions} we know that for any $1 \leq p < \infty$ the map $S \mapsto \| f(s) \|$ is also strongly $\mu$ measurable, therefore, is makes sense ask whether $\int_S\|f\|^p \mathrm{~d} \mu<\infty$ or not.

\begin{definition}[Definition 1.2.15. \citep{hytonen_analysis_2016}]
\label{def:Bochner_Lp_spaces}

 For $1 \leqslant p<\infty$ we define $L^p(S ; X)$ \index{$L^p(S ; X)$} as the linear space of all (equivalence classes of) strongly $\mu$-measurable functions $f: S \rightarrow X$ for which
$$
\int_S\|f\|^p \mathrm{~d} \mu<\infty,
$$

and endowed with the norms
$$
\|f\|_{L^p(S ; X)}:=\left(\int_S\|f\|^p \mathrm{~d} \mu\right)^{1 / p}.
$$        

If $p = \infty$ then we define $L^{\infty}(S;X)$ as the linear space of all (equivalent classes) of $\mu$-strongly measurable functions $f: S \to X$ for which there exists a real number $r \geq 0$ such that $\mu\{\| f\| > r \} = 0$. In this case we define
$$ \|f\|_{L^{\infty}(S ; X)}:= \inf \{  r \geq 0 : \mu\{\| f\| > r \} = 0 \}. $$
\end{definition}

\begin{remark}[Approximating $L^p$ functions by step functions]\label{remark:approximating_L_p_functions_by_step_functions}

Take $f \in L^p(S;X)$ and $p \in [0, \infty)$, thus $f$ and $s \mapsto \| f(s) \|^p$ are $\mu$-strongly measurable, so by \cref{proposition:approximating_mu_strongly_measurable_functions} we know that there is a sequence $\{ f_n \}_{n \in \mathbb{N}}$ of $\mu$-simple functions such that $\| f_n(s) \| \leq \|f(s) \|$ almost everywhere and $f_n \to f$ almost everywhere. From \cref{remark:norm_of_mu_strongly_measurable_functions} we know that $\| f_n \|^p \to \| f \|^p$ almost everywhere and is a sequence of $\mu$-strongly measurable functions, thus the dominated convergence theorem \cref{proposition:properties_of_Bochner_integral} tell us that 
$$\int_{S}\|f_n(s)\|^p d \mu(s) \to \int_S \|f(s)\|^p d \mu(s), \; \lim_{n \to \infty } \int_{S} | \|f_n(s)\|^p - \|f(s)\|^p | d \mu(s) = 0.$$

Since $| a -b |^p \leq | a^p - b^p|$ for $p \geq 1$ and $a,b \in \mathbb{R}^+$, then 
$$ \lim_{n \to \infty} \int_{S} \| f_n(s) - f(s) \|^p d \mu(s) \leq \lim_{n \to \infty} \int_{S} |\| f_n(s)\|^p - \|f(s) \|^p| d \mu(s) = 0, $$
thus $f_n \to f$ in $L^p(S;X)$.
\end{remark}

Various of the results of $L^p$ spaces translates into the Bochner $L^p$ spaces because their statements relay solely on the functions $s \to \| f(s)\|$, which are measurable by definition. The following are relevant to us

\begin{proposition}[Useful inequalities on Bochner $L^p$ spaces]\label{proposition:useful_inequalities_of_Bochner_L_p_spaces}

Let $(S, \mathscr{B}(S), \mu)$ be a measure space and $X$ a Banach space, if $r,p,q \in [1, \infty)$ then

\begin{itemize}
    \item \textbf{Minkowski inequality:}\index{Minkowski inequality} $\| f + g \|_p \leq \|f \|_p + \| g \|_p$ (\citep[Theorem 3.6]{Potter2014TheBI}).
    \item \textbf{Completeness:} The proof of completeness of $L^p(S,\mu)$ from complex valued functions \citep[Theorem 9.6]{knapp_basic_2005} (\citep[Chapter IV, section 3 theorem 7]{nagy_real_2019}) can be translated to show that $L^p(S;X)$ is a Banach space (\citep[Theorem 1.23]{driver_funcitonal_2020}). An important step in the proof of the completeness of $L^p(S;X)$ is the fact that given a a sequence $\{ h_i \}_{i \in \mathbb{N}}$ such that $h_i \to h$ in $L^p(S;X)$ then there is a sub-sequence $\{ h_{i(j)} \}_{j \in\mathbb{N}}$ such that $h_{i(j)} \to h$ almost everywhere (recall that convergence in $L^p$ does not imply convergence almost everywhere (\citep{cook_modes_2023}) ).
    \item \textbf{H\"older inequality:}\index{H\"older inequality} Assume $X$ is a Banach algebra and $f,g : S \to X$, if $f,g$ are $\mu$-strongly measurable then \cref{remark:algebra_of_strongly_mu_measurable_functions} tell us that $fg$ is $\mu$-strongly measurable, thus if we apply the H\"older inequality for $\mathbb{C}$ valued functions (\citep[Proposition 1.1]{bahouri_fourier_2011}) we get that
    $$\| fg \|_{r} \leq \| f \|_p \| g \|_q$$
    for $\frac{1}{r} = \frac{1}{p} + \frac{1}{q}$, $r,p,q \in [1, \infty]$ and $f \in L^p(S ; X), \; f \in L^q(S ; X)$ (\citep[Chapter 8, item 36]{pap_handbook_2002}).
    \item \textbf{Young inequality for inverse invariant Haar measures:}\index{Young inequality} Assume $X$ is a Banach algebra, and $S$ is a locally compact topological group with a Haar measure (\cref{definition:Haar_measure}) such that $\mu(A^{-1}) = \mu(A)$ for all $A \in \mathscr{S}$ e.g. unimodular groups (\cref{proposition:examples_unimodular_groups}), take $f \in L^p(S;X)$ and $g \in L^q(S;X)$, define 
    $$ (f \ast g)(x) = \int_G f(y) g(y^{-1} x) d \mu(y). $$
    Take $1 + \frac{1}{r} = \frac{1}{p} + \frac{1}{q}$ for $r,p,q \in [1, \infty]$, then $(f \ast g)$ converges almost everywhere, $f \ast g$ is $\mu$-strongly measurable and $\| f \| \in L^r(X;A)$, moreover  $\| f \ast g \|_r \leq \| f \|_p \| g \|_q$.
\end{itemize}
\end{proposition}
\begin{proof}
The only statement that requires a proof is the Young inequality. Notice that $y \to y^{-1}x$ is a continuous function, therefore is measurable, thus, by \cref{proposition:approximating_mu_strongly_measurable_functions} we know that $y \to g(y^{-1}x)$ is $\mu$-strongly measurable for any $x \in X$. If $r = \infty$ then the Young inequality comes from H\"older inequality since 
$$\| (f \ast g)(x)\|  \leq \int_G \| f(y)g(y^{-1} x)\| d \mu(x) \leq \| f\|_p \| g \|_q$$
almost everywhere, thus $(f \ast g) \in L^{\infty}(X;A)$. If $p = \infty$ then $r = \infty$ and $q = 1$ and we go back to H\"older inequality.

Assume $p,q,r \neq \infty$, we have that
$$ \|(f \star g)(x)\| \leq  \int_G \|f(y)\| \|g(y^{-1} x)\| d \mu(y),$$
so, from the Young inequality for scalar valued functions (\citep[Theorem 20.18]{hewitt_abstract_1963}) we get that $\int_G \|f(y)\| \|g(y^{-1} x)\| d \mu(y)$ converges almost everywhere, therefore, by \cref{prop:bochn_integra_condition} $y \to f(y^{-1} x) g(y)$ is Bochner integrable, that is, $(f \ast g)(x)$ exists almost everywhere. From \cref{lemma:L_p_bochner_spaces_for_locally_compact_hausdroff_spaces} we know that there are sequences $\{ f_n \}_{n \in \mathbb{N}}, \{ g_n \}_{n \in \mathbb{N}}$ of $\mu$-simple functions each one of those with support inside compact sets and $\| f_n(s) \| \leq \| f(s)\|, \; \| g_n(s) \| \leq \| g(s)\|$ almost everywhere, such that $f_n \to f, \; g_n \to g$ almost everywhere. Each $f_n, g_n$ are Bochner integrable, therefore $(f_n \ast g_n)(x)$ exists for all $x$, also, $y \mapsto f_n(y)g_n(y^{-1}x)$ converges to $y \mapsto f(y)g(y^{-1}x)$ almost everywhere, therefore, if $\int_G \|f(y)\| \|g(y^{-1} x)\| d \mu(y) < \infty$ then, by the dominated convergence theorem (\cref{proposition:properties_of_Bochner_integral}) we get that
$$ \lim_{n \to \infty} \int_G f_n(y) g_n(y^{-1} x) d \mu(y) \to \int_G f(y) g(y^{-1} x) d \mu(y). $$

We have that $f_n = \sum_{i \leq k}  \mathbf{1}_{P_i} a_i, \; g_n = \sum_{j \leq L}  \mathbf{1}_{Q_i} b_i$ and there is $K \subset G$ compact such that $P_i,Q_i \subseteq K$, so
$$ \int_G f_n(y) g_n(y^{-1} x) d \mu(y) = \sum_{i \leq k} \sum_{j \leq l} a_i b_j \int_G \mathbf{1}_{P_i}(y)\mathbf{1}_{Q_i}(y^{-1}x) d \mu(y). $$

Since indicator functions with finite measure belong to any $L^p$ then by \citep[Theorem 20.16]{hewitt_abstract_1963} we get that $x \mapsto \int_G \mathbf{1}_{P_i}(y)\mathbf{1}_{Q_i}(y^{-1}x) d \mu(y)$ belongs to $C_0(G)$, henceforth, $x \mapsto  \int_G f_n(y) g_n(y^{-1} x) d \mu(y)$ belongs to $C_0(G) \odot A \subset C_0(G;A)$, and from \cref{example:continuous_functions_decaying_at_infinty_are_strongly_measurable} we know that all the elements of $C_0(G;A)$ are $\mu$-strongly measurable. Also, we have shown that $x \mapsto \int_G f_n(y) g_n(y^{-1} x) d \mu(y)$ converges to $x \mapsto \int_G f(y) g(y^{-1} x) d \mu(y)$ for almost all $x \in G$, so, since $x \mapsto \int_G f_n(y) g_n(y^{-1} x) d \mu(y)$ is $\mu$-strongly measurable (\cref{lemma:L_p_bochner_spaces_for_locally_compact_hausdroff_spaces}) we get that $x \mapsto \int_G f(y) g(y^{-1} x) d \mu(y)$ is $\mu$-strongly measurable.

To check that $(f \ast g) \in L^r(G;X)$ you just need to follow the proof of the young inequality for scalar valued functions (\citep[Theorem 20.18]{hewitt_abstract_1963}) to check that 
$$ \| f \ast g \|_r \leq \| f\|_p \| g \|_q .$$
\end{proof}

\begin{lemma}[$L^p(S;A)$ for locally compact Hausdorff spaces]\label{lemma:L_p_bochner_spaces_for_locally_compact_hausdroff_spaces}
Let $(S, \mathscr{B}(S), \mu)$ with $S$ a locally compact Hausdorff space and $\mu$ a Radon measure (\cref{definition:Radon_measure}) i.e. it is finite on compact sets as in \cref{section:algebra_continuous_functions_locally_comp_space}, also, let $A$ be a Banach space and $ 1\leq p < \infty$. Denote by $C_c(S;A)$ the vector space of continuous functions with compact support, then
\begin{itemize}
    \item $C_c(S;A) \in L^P(S;A)$.
    \item $C_c(S;A)$ is dense in $L^p(S;A)$.
    \item If $f \in L^p(S;A)$ then $f$ can be approximated by $\mu$-simple functions $\{ f_n \}_{n \in \mathbb{N}}$ that are supported on compact sets i.e. for each $f_n$ there is $K_n \subset S$ compact such that $f_n(s) = 0$ if $s \notin K_n$. This sequence can be taken such that $\| f_n(s) \|_p \leq \| f(s) \|_p$ almost everywhere, and $\| f_n - f \|_p \leq 1/n$.
    \item If $f \in L^p(S;A)$ then for any $\epsilon > 0$ there exists $K \subset S$ compact such that $\|\mathbf{1}_{K^c} f \|_p \leq \epsilon $.
    \item If $A$ is a Banach algebra, if $f \in L^p(S;A)$ then for any $g \in C(S;A)$ $fg$ is $\mu$-strongly measurable. 
\end{itemize}
\end{lemma}
\begin{proof}
\begin{itemize}
    \item Take $f \in C_c(S;A)$ and $K \subset S$ a compact set such that $f(s) = 0$ if $s \notin K$, from \cref{section:algebra_continuous_functions_locally_comp_space} we know that any compact set of $S$ is closed, thus $K \in \mathscr{B}(S)$. Since the topology of $K$ is the inherited topology from $S$ we can use  \cref{example:Bochner_integral_continuos_funcion} to show that $f$ is $\mu$-strongly measurable, because $f|K$ is $\mu$-strongly measurable, moreover, $f \in L^1(S;A)$. Also, $f \in L^p(S;A)$ for any $p \in [1,\infty]$, since 
$$ \int_X \|f(x) \|^p \leq( \sup \{ \| f(x) \| \} )^p\mu(K) < \infty. $$
    \item From \cref{theprem:Lp_approximation_simple_functions} we know that $L^p(S) \odot A$ is dense in $L^p(S;A)$, also, given that $\mu$ is a regular Borel measure i.e. inner regular and outer regular (\cref{definition:Radon_measure}), from \citep[Proposition 11.21]{knapp_basic_2005} we know that $C_c(S)$ is dense in $L^p(S)$, therefore, $C_c(S) \odot A$ is dense in $L^p(S;A)$, since $C_c(S;A) \subset L^p(S;A)$ then $C_c(S;A)$ is dense in $L^p(S;A)$.
    \item Take $f \in L^p(S;A)$ and $g \in C_c(S;A)$ such that $\|f -g \|_p \leq  \epsilon/2$ and $g(s) = 0$ if $s \notin K$ with $K$ compact, also, take $h$ a $\mu$-simple function such that $\| g - h \|_p \leq \epsilon/2$, then $\hat{h} = \mathbf{1}_{K}h$ is a $\mu$-simple function and $\| \hat{h} - g \|_p \leq \epsilon/2$. Thus, $\| f - \hat{h} \|_p \leq \epsilon$, meaning that any function in $L^p(S;A)$ can be approximated by $\mu$-simple functions with support inside a compact set. Moreover, from \cref{remark:approximating_L_p_functions_by_step_functions} we know that $\{ f_n \}_{n \in \mathbb{N}}$ can be taken such that $\| f_n (s) \|_p \leq \|f(s) \|_p$ almost everywhere, also, we can chose $\epsilon - 1/n$ giving us a sequence of $\mu$-simple functions with $\|f_n - f\|_p \leq 1/n$.
    \item Take $f \in L^p(S;A)$ and $g$ a $\mu$-simple function with support inside a compact set $K \subset S$ such that $\|g - f \|_p \leq \epsilon$, then 
    $$\|\mathbf{1}_{K^c} f \|_p = \| \mathbf{1}_{K^c} (f-g) \|_p \leq \| f -g \|_p \leq \epsilon.$$
    \item Let $\{ f_n \}_{n \in \mathbb{N}}$ be a sequence of $\mu$-simple functions with $f_n$ supported inside a compact $K_n$, such that $\| f_n - f \|_p \leq 1/n$ and $\| f_n(s) \|_p \leq \| f(s) \|_p$ for all $s \in S$. We know that $\|\mathbf{1}_{K_n^c} f \|_p \leq 1/n$, also from \citep{nlab:compact_subspaces_of_hausdorff_spaces_are_closed} we know that $K_n$ are closed, thus $K_n \in \mathscr{B}(S)$. Since the topology on $K_n$ is the inherited topology from $S$ then we can use \cref{example:Bochner_integral_continuos_funcion} to find a $\mu$-simple function $g_n$ such that $\| g_n(s) \| \leq \| g(s) \|$ and 
    $$\sup \{ |g_n(s) - g(s)| : s \in K_n\} \leq 1/n.$$ 
    Extend $g_n$ to $S$ be setting it to zero outside $K_n$, thus it is still a $\mu$-simple function that approximates uniformly $g$, that is, $|g_n(s) - g(s)| \leq 1/n$ for all $n \in \mathbb{N}$. Then $f_n g_n \to fg$ for all $s \in \cup_{n \in \mathbb{N}} K_n$, so, by the definition of $K_n$ we must have $\int_S \mathbf{1}_{(\cup_{n \in \mathbb{N}} K_n)^c}\| f (s) \|^p d\mu(s) = 0$, meaning the either $(\cup_{n \in \mathbb{N}} K_n)^c$ has measure zero or $f(s) =0$ almost everywhere in $\cup_{n \in \mathbb{N}} K_n$, and in both cases we still have that $f_n g_n \to fg$ over $\cup_{n \in \mathbb{N}} K_n$, thus, $f_n g_n \to fg$ over $S$ and $fg$ is $\mu$-strongly measurable.
    \end{itemize}
\end{proof}

\begin{example}(Banach algebras $L^{1}(S;X)$) \index{$L^{1}(S;X)$}
\label{example:L1_Bohner_spaces_and_convolution}
 If $S$ is a locally compact unidomular group, that is, $\mu(A) = \mu(A^{-1})$ for $A \in \mathscr{B}(G)$, then $L^{1}(S,\mu)$ admits an convolution which can be translated into a multiplication on $L^{1}(S;X)$, and it makes $L^{1}(S;X)$ into a Banach algebra when $X$ is a Banach algebra. So, let $(G,\mathscr{B}(G),\mu)$ be a locally compact topological group with $\mu$ a left-invariant Haar measure such that $\forall A \in \mathscr{B}(G) \: \mu(A^{-1}) = \mu(A) $. From the Young inequality (\cref{proposition:useful_inequalities_of_Bochner_L_p_spaces}) we get that if $f,g \in L^1(G;X)$, then $\| g \ast f \| \leq \|g \| \|f \|$, therefore, $(g \ast f)(x) = \int_G g(y)f(y^{-1}x) d \mu (y)  \in L^{1}(G;x)$ and  $L^1(G;X)$ is a Banach algebra with multiplication given by convolution. An example of this setup is $S = \mathbb{R}^d$ with the Lebesgue measure, were if $f,g \in L^{1}(\mathbb{R}^d,X)$ then $(g \ast f )(x) = \int_{\mathbb{R}^d} g(y) f(x-y) dy \in L^{1}(\mathbb{R}^d,X)$ by \citep[Lemma 1.2.30]{hytonen_analysis_2016} and $\| f \ast g \| \leq \|f \| \|g \|$.
 
 If $X$ is a Banach *-algebra (\cref{def:Banach_star_algebra}), then we can define an involution on $L^1(S;X)$ as 
 $$*: L^1(S;X) \to L^1(S;X), \; (f^*)(s) = f(-s)^*,$$
 under this setup $L^1(S;X)$ becomes a Banach *-algebra, because $\| x^* \| = \|x\|$ for $x \in X$ and $\mu(A) = \mu(A^{-1})$.
\end{example}

\begin{example}[$L^p$ spaces for the counting measure\index{counting measure}]\label{example:L_p_sapces_for_counting_measure}

Let $\Lambda$ be a set and $\tau = \mathcal{P}(\Lambda)$ be the discrete topology on $\Lambda$, then $(\Lambda, \mathcal{P}(\Lambda), \mu)$ is a measure space with $\mu$ the counting measure (\citep{noauthor_counting_2023}), that is, $\mu(A) = |A|$ i.e. $|A| = \infty$ if $A$ is infinite else $|A| = n$ with n the amount of elements in $A$. If $X$ is a Banach space then $L^p(\Lambda;X)$ is a the Banach space of at most countable sums of elements that are indexed by $\Lambda$, or equivalently
$$ L^p(\Lambda;X)  = \left\{  \left(x_\lambda\right)_{\lambda \in \Lambda}, \; x_\lambda \in X \; : \sum_\lambda\left\|x_\lambda\right\|^p<\infty\right\}.$$
Notice that we the sums $\sum_\lambda\left\|x_\lambda\right\|^p$ cannot have a uncountable number of terms because that would imply that it diverges (\citep{rspuzio_uncountable_2013}), which is consistent with the properties of the counting measure over $\Lambda$.
\end{example}

Now we will look at some properties of the spaces $L^p(S;X)$.

\begin{definition}[Page 4 \citep{hytonen_analysis_2016}]\label{def:algebraic_tensor_product_functions_and_algebras}

If $F$ is a vector space of scalar-valued functions define
$$
F \odot X:=\left\{\sum_{n=1}^N f_n \otimes x_n: f_n \in F, x_n \in X, n=1, \ldots, N ; N=1,2, \ldots\right\},
$$

Notice that if $F$ is the vector space of ($\mu$-)simple function $f: S \to \mathbb{C}$ then $F \odot X$ is a vector space of ($\mu$-)simple functions $f: S \to X$.
\end{definition}

The $L^p(S;X)$ spaces have similar properties to the $L^p(S)$ spaces, for example, the $\mu$-simple functions are dense in $L^{p}(S;X)$

\begin{theorem}[Lemma 1.2.19 (Approximation by simple functions) \citep{hytonen_analysis_2016}]
\label{theprem:Lp_approximation_simple_functions}

 Let $1 \leqslant p < \infty$, then
 
 \begin{itemize}
     \item The $\mu$-simple functions are dense in $L^p(S ; X), 1 \leqslant p<\infty$. In particular, the algebraic tensor product $L^p(S) \odot X$ is dense in $L^p(S ; X), 1 \leqslant p<\infty$. 
 \end{itemize}

\end{theorem}

\begin{example}[$L^{p}(X;A)$ for second countable locally compact Hausdorff spaces]\label{example:Lp_second_countable_locally_compact_hausdorff_sapce}
Let $X$ be a second countable locally compact Hausdorff space, and $\mu$ a regular Borel measure in $X$, then from \cref{section:algebra_continuous_functions_locally_comp_space} we know that $C_{c}(X)$ is dense in $L^{p}(X,\mu)$. Also, if $A$ is a Banach algebra, then $L^{p}(X) \odot A$ is dense in $L^p(X ; A)$ and we can check that $\| f(x) \otimes a \| = \|a \| \|f\|_{p}$ for $a \in A$ and $f \in L^{p}(X)$, therefore, 
$$ C_{c}(X) \odot A $$
is dense in $L^p(X ; A)$. 
Additionally $C_{c}(X)$ is separable, so, if $A$ is separable then we have that $L^p(X ; A)$ is also separable. This can also be deduced using \cref{prop:separability_Lp_spaces} because $\mu$ is countably generated, which we mentioned in \cref{section:algebra_continuous_functions_locally_comp_space}.
\end{example}

In the $L^P(S;X)$ spaces there is even a continuous version of the Minkowski inequality \citep[Proposition 1.2.22]{hytonen_analysis_2016}, which reduces to the following isomorphism of Banach spaces:

\begin{proposition}[Corollary 1.2.23. \citep{hytonen_analysis_2016}]\label{prop:ismorphism_banch_spaces}

 Let $(S, \mathscr{A}, \mu)$ and $(T, \mathscr{B}, \nu)$ be measure spaces and let $X$ be a Banach space. For all $1 \leqslant p<\infty$ the mapping $\mathbf{1}_A \otimes\left(\mathbf{1}_B \otimes x\right) \mapsto \mathbf{1}_B \otimes\left(\mathbf{1}_A \otimes x\right)$ induces an isometric isomorphism
$$
L^p\left(S ; L^p(T ; X)\right) \approx L^p\left(T ; L^p(S ; X)\right) .
$$
From \cref{def:simple_func} recall that $\mathbf{1}_B \otimes x$ is a simple function that gives the value $x$ to any element $t \in B$ and $0$ elsewhere. 
\end{proposition}

Note that \cref{prop:ismorphism_banch_spaces} can be regarded as a Fubini theorem. Moreover, if $\mu$ and $\nu$ is are $\sigma$ finite then we have a more robust result that uses the product measure $\mu \times \nu$ that is also $\sigma$-finite,

\begin{proposition}[Proposition 1.2.24. \citep{hytonen_analysis_2016}]\label{prop:isomorphism_funcitons_over_sum_of_spaces}

 Let $(S, \mathscr{A}, \mu)$ and $(T, \mathscr{B}, \nu)$ be $\sigma$-finite measure spaces and let $1 \leqslant p<\infty$. The mapping $\mathbf{1}_A \otimes\left(\mathbf{1}_B \otimes x\right) \mapsto\left(\mathbf{1}_A \otimes \mathbf{1}_B\right) \otimes x$ extends uniquely to an isometric isomorphism
$$
\iota: L^p\left(S ; L^p(T ; X)\right) \simeq L^p(S \times T ; X) .
$$
If $\tilde{f}: S \times T \rightarrow X$ is a strongly $(\mu \times \nu)$-measurable function in the equivalence class of $f \in L^p(S \times T ; X)$, then for almost all $s \in S$ the function $\widetilde{f}(s, \cdot)$ belongs to $L^p(T ; X)$, the function $F: s \mapsto \widetilde{f}(s, \cdot)$ defines an element of $L^p\left(S ; L^p(T ; X)\right)$, and we have $\iota F=f$

\end{proposition}

Now we center in the separability of the $L^p(S;X)$ spaces. First let us establish some terminology,

\begin{definition}[Definition 1.2.27. \citep{hytonen_analysis_2016}]\label{def:mu_countably_geenrated}

A measure space $(S, \mathscr{A}, \mu)$ is called:

\begin{itemize}
    \item countably generated\index{countably generated}, if there exists a sequence $\left(S_n\right)_{n \geqslant 1}$ in $\mathscr{A}$ which generates $\mathscr{A}$, i.e. $\mathscr{A} = \sigma(\left(S_n\right)_{n \geqslant 1})$. 
    \item $\mu$-countably generated\index{$\mu$-countably generated}, if there exists a sequence $\left(S_n\right)_{n \geqslant 1}$ in $\mathscr{A}$, consisting of sets of finite $\mu$-measure, which $\mu$-essentially generates $\mathscr{A}$ in the sense that for all $A \in \mathscr{A}$ we can find a set $A^{\prime}$ in the $\sigma$-algebra generated by $\left(S_n\right)_{n \geqslant 1}$ such that $\mu\left(A \Delta A^{\prime}\right)=0$.
    \item purely infinite, if $\mu(A) \in\{0, \infty\}$ for all $A \in \mathscr{A}$.
\end{itemize}
Here, $A \Delta A^{\prime}=\left(A \backslash A^{\prime}\right) \cup\left(A^{\prime} \backslash A\right)=\left(A \cup A^{\prime}\right) \backslash\left(A \cap A^{\prime}\right)$ is the symmetric difference of $A$ and $A^{\prime}$.

\end{definition}

Given these definitions, we can give a characterization of the separability of $L^p(S ; X)$:

\begin{proposition}[(Separability of Bochner spaces) Proposition 1.2.29  \citep{hytonen_analysis_2016}]\label{prop:separability_Lp_spaces}

Let $(S, \mathscr{A}, \mu)$ be a measure space, let $1 \leqslant p<\infty$, and let $X$ be a Banach space. If $\operatorname{dim} L^p(S ; X) \geqslant$ 1 , the following assertions are equivalent:

\begin{itemize}
    \item $L^p(S ; X)$ is separable;
    \item $X$ is separable and we have a disjoint decomposition $S=S_0 \cup S_1$ in $\mathscr{A}$ such that $\left(S_0,\left.\mathscr{A}\right|_{S_0},\left.\mu\right|_{S_0}\right)$ is purely finite and $\left(S_1,\left.\mathscr{A}\right|_{S_1},\left.\mu\right|_{S_1}\right)$ is $\mu$-countably generated. 
\end{itemize}

If these equivalent conditions hold, then $\left(S_1,\left.\mathscr{A}\right|_{S_1},\left.\mu\right|_{S_1}\right)$ is $\sigma$-finite and we have $L^p(S ; X)=L^p\left(S_1 ; X\right)$ isometrically.
\end{proposition}

For further results on the Bochner integral and $L^p(S;X)$ spaces you can look into \citep{hytonen_analysis_2016}, \citep{driver_funcitonal_2020} \citep{mikusinski_bochner_1978}, 
\citep{pap_handbook_2002}, 
\citep{Potter2014TheBI}, \citep{mikusinski_integrals_2014}.

\subsection{Operator valued functions}
\label{sec:oper_valu_func}

From \cref{theorem:pettis_measurability_theorem_second_version} we know that an important feature of Bochner integrable functions is their separable range, this condition may not be satisfied if we work with operator valued functions. These types of functions will come up in the study of twisted crossed products (\cref{sec:twisted_crossed_products}), and are quite relevant to the study of the representations of locally compact Hausdorff topological groups (\citep[Chapter 8]{de_chiffre_haar_2011}, \citep[Chapter 6]{deitmar_principles_2009}).

Take $H$ a separable Hilbert space, then $B(H)$ is not separable under the norm topology (\citep{119273}), not even $U(H)$\index{$U(H)$} (the group of bounded unitary operators on $H$) is separable in the norm topology when $H$ is separable \citep[page 299]{neeb_theorem_1997}, recall that the norm topology on $B(H), U(H)$ is the topology of convergence in the operator norm i.e. $T_k \to T$ if $\| T_k - T\| \to 0$. As an example, take $H = L^2(\mathbb{R})$ and $T(s)\in B(H)$ for $s \in \mathbb{R}$ given by $$T(s)(f)(t) = f(t-s), \; f \in L^2(\mathbb{R}), \; t,s \in \mathbb{R},$$
each $T(s)$ is an unitary with $T^*(s) = T(-s)$, and, for any $s \leq s'$ we have that $\| T(s) - T(s') \|_{B(H)}=2$ (\citep[Section 1.1.c]{hytonen_analysis_2016}), consequently neither $B(H)$ nor $U(H)$ are separable in the operator norm topology. Thus, if $f \in C_c(\mathbb{R})$ with $f \neq 0$ we cannot use the Bochner integral to define an operator like $\int_{\mathbb{R}} f(x) T(x) d\mu(x)$.

If $H$ is a separable space, then $U(H)$ is a group under the multiplication of operators, and it becomes a topological group under the strong operator topology (\citep[Lemma 1.4]{espinoza_topological_2014}), that is, the topology of pointwise convergence, 
$$T_k \to T \text{ if for all } x \in H,\; T_k (x) \to T(x) \text{ in } H,$$
and we denote it by $U(H)_s$\index{$U(H)_s$}. $U(H)$ also becomes a topological group under the weak operator topology (\citep[Lemma 1.5]{espinoza_topological_2014}), that is, the topology of convergence in the continuous linear functionals, 
$$T_k \to T \text{ if for all } \hat{y} \in H^*, \; x \in H,\; \hat{y}(T_k(x)) \to \hat{y}(T(x)),$$
and we denote it by $U(H)_w$\index{$U(H)_w$}. From Riesz representation theorem\index{Riesz representation theorem} for continuous functionals on Hilbert spaces (\citep[Theorem 2.53]{allan_introduction_2011}) we have that if $\hat{y} \in H^*$ then there is a unique $y \in H$ such that $\langle x, y \rangle = \hat{y}(x)$ for all $x \in H$, thus, the weak convergence in $B(H)$ can also be formulated as, $T_k \to T$ if $\langle T_k (x) , y \rangle \to \langle T(x),y\rangle$ for $x,y \in H$. 

\begin{theorem}[Topology on $U(H)$ for $H$ separable (Theorem 1.2 \citep{espinoza_topological_2014})]\label{theorem:topology_if_unitary_group}
The topologies of $U(H)_s$, $U(H)_w$ agree, moreover, $U(H)$ endowed with any of those topologies becomes a Polish group i.e. a separable completely metrizable topological group. 
\end{theorem}

\begin{definition}[Strongly continuous function]\label{def:strongly_continuous_functions}
A function $f: G \to U(H)$ is called strongly continuous if $s \mapsto f(g)(y)$ is continuous for any $y \in H$
\end{definition}

\begin{lemma}\label{lemma:equivalence_of_strong_continuity_into_U(H)}
Let $f$ be a function $f: G \to U(H)$, then, $f$ is strongly continuous iff $g \mapsto \langle f(g)(x) , y \rangle,$ is continuous for any $x,y \in H$.
\end{lemma}
\begin{proof}
Since the topology of $U(H)_s$, $U(H)_w$ agree (\cref{theorem:topology_if_unitary_group}), then, strong continuity is equivalent to continuity of linear functionals.
\end{proof}

\begin{proposition}[Measurability and continuity on $U(H)$]\label{proposition:measurability_in_unitary_group}
Let $H$ be a separable Hilbert space and $(G, \mathcal{A})$ a mesurable space, then, a function $f: G \to U(H)_s$ is strongly measurable iff $g \mapsto f(g)(y),\; g \in G,$ is measurable for any $y \in H$, with $H$ under the Borel $\sigma$-algebra, equivalently, $f$ is strongly measurable iff $f: G \to U(H)_s$ is measurable iff $g \mapsto \langle f(g)(x), y \rangle, \; g \in G,$ is mesurable, with $\mathbb{C}$ endowed with the Borel $\sigma$-algebra.
\end{proposition}
\begin{proof}
This is a consequence of \cref{theorem:topology_if_unitary_group}.
\end{proof}

\begin{example}[Left regular representation\index{Left regular representation} of $G$]\label{example:left_regular_representation_of_G}
Let $G$ be a locally compact Hausdorff topological group and $\mu$ its Haar measure, then $L^2(G,\mu)$ is a Hilbert space under the inner product $\langle f,h \rangle = \int_G f(g) h(g)^* d \mu(g)$ for $f,h \in L^2(G,\mu)$ (\cref{proposition:inner_prodict_L_2_S_H}). The map 
$$\pi: G \to B(L^2(G,\mu)), \; \pi(g)(\xi)(x) = \xi(g^{-1} x), \; \xi \in L^2(G,\mu), \; x,g \in G, $$
is called the left regular representation of $G$ on $L^2(G,\mu)$.

The map $\pi$ has the following properties (\citep[Proposition 8.7]{de_chiffre_haar_2011}):
\begin{itemize}
    \item $\pi(g) \in U(L^2(G,\mu))$.
    \item $\pi$ is a group homomorphism, that is, $\pi(gj) = \pi(g)\pi(j)$ for $g,j \in G$.
    \item $\pi$ is continuous with respect to the strong operator topology, that is, if $g_k \to g$ then $\pi(g_k)(\xi) \to \pi(g)(\xi)$ in $L^2(G,\mu)$ for any $\xi \in L^2(G,\mu)$ and $g \in G$.
\end{itemize}

Let $H$ be a Hilbert space, then the left regular representation extends to $L^2(G;H)$, which we know that is isomorphic to the tensor product of Hilbert space $L^2(G)\otimes H$ (\cref{proposition:inner_prodict_L_2_S_H}), as follows
$$ (\pi \otimes Id_H)(g) (f \otimes h) = \pi(g)(f) \otimes h, \; f \in L^2(G), h \in H. $$
Since $L^2(G) \odot H$ is dense in $L^2(G;h)$ (\cref{theprem:Lp_approximation_simple_functions}), if $f \in L^2(G;H)$ then $(\pi \otimes Id_H)(g)(f)(j) = f(g^{-1}j)$ almost everywhere, thus, we can consider $(\pi \otimes Id_H)(g)(f)(j) := f(g^{-1}j)$ for all $j,g \in G$, moreover, $(\pi \otimes Id_H)(g) \in U(L^2(G;H))$ becomes it preserves the inner product on simple tensors, also $(\pi \otimes Id_H)(g) (\pi \otimes Id_H)(j) = (\pi \otimes Id_H)(gj)$.

Take $f \in L^2(G;H)$ and $f_n = \sum_{i \leq n} f_i \otimes h_i \in L^2(G) \odot H$ such that $\|f - f_n \|_2 \leq \epsilon/3$, then 
$$ \| (\pi \otimes Id_h)(g)(f) - (\pi \otimes Id_h)(j)(f) \|_2 $$
$$\leq  \| (\pi \otimes Id_h)(g)(f) - (\pi \otimes Id_h)(g)(f_n) \|_2 + \| (\pi \otimes Id_h)(g)(f_n) - (\pi \otimes Id_h)(j)(f_n) \|_2 + \dotsc$$
$$\leq \frac{2 \epsilon}{3} + \| (\pi \otimes Id_h)(g)(f_n) - (\pi \otimes Id_h)(j)(f_n) \|_2$$
$$ \leq  \frac{2 \epsilon}{3} +  \sum_{i \leq n} \|( \pi(g)(f_i) - \pi(j)(f_i) )\otimes h_i \|_2$$
$$ \leq \frac{2 \epsilon}{3} + \sum_{i \leq n} \|( \pi(g)(f_i) - \pi(j)(f_i) )\|_2 \| h_i \|.$$

So, take $U$ an open neighbourhood of $g$ such that if $j \in U$ then $\sum_{i \leq n} \|( \pi(g)(f_i) - \pi(j)(f_i) )\|_2 \| h_i \| \leq \epsilon/3$, which can be done because the left regular representation is strongly continuous on $L^2(G,\mu)$,  in this case  $\| (\pi \otimes Id_h)(g)(f) - (\pi \otimes Id_h)(j)(f) \|_2 \leq \epsilon$ and we have that $\pi \otimes Id_H$ is strongly continuous on $L^2(G;H)$.
\end{example}

Under the previous example, we have that $T(s) = \pi(s)$ for $G = \mathbb{R}$ and $s \in \mathbb{R}$. One of our objectives is to make sense of the operator $\int_{\mathbb{R}} f(x) T(x) d\mu(x)$, which requires to look into a weak version of the Bochner integral, and is defined using bounded sesquilinear forms.

\begin{definition}[Sesquilinear form \index{Sesquilinear form}(Section 2.14 \citep{allan_introduction_2011})]\label{definition:sesquilinear_bounded_functional}
A map $\phi: H \times H \to \mathbb{C}$ is called a sesquilinear form on $H \times H$ if
\begin{itemize}
    \item $\phi(a \xi, b \zeta) = a \phi(\xi,\zeta)b^*$ for $\xi, \zeta \in H$ and $a,b \in \mathbb{C}$.
    \item $\phi(\xi_1 + \xi_2, \zeta) = \phi(\xi_1, \zeta) +  \phi(\xi_2, \zeta)$ for $\xi_i, \zeta \in H$.
    \item $\phi(\xi_1, \zeta_1 + \zeta _2) = \phi(\xi, \zeta_1) +  \phi(\xi, \zeta_2)$ for $\xi, \zeta_i \in H$.
\end{itemize}
The form is called bounded if there is $M \geq 0$ such that $|\phi(\xi,\zeta)| \leq M \| \xi\| \|\zeta \| $ for $\xi, \zeta \in H$. If $\phi$ is bounded then we define
$$ \| \phi \| = \sup \{ |\phi(\xi,\zeta) | : \; \xi, \zeta \leq 1 \}. $$
\end{definition}

Let $T \in B(H)$, then $(x,y) \mapsto \langle T(x),y \rangle$ is a bounded sesquilinear form, which can be checked from the Cauchy-Schwartz inequality, $|\langle T(X) ,y \rangle | \leq \|T\| \| x\| \|y\|$. Set $\phi_T(x,y) =  \langle T(x),y \rangle$, then $\| \phi_T \| = \|T \|$ (\citep[page 76]{allan_introduction_2011}), moreover, every bounded sequilinear form arises this way,
 
\begin{proposition}[Correspondence between bounded linear operators and bounded sesquiliean forms (Theorem 2.55 \citep{allan_introduction_2011})]\label{proposition:correspondence_bounded_operators_and_sesquilinear_forms}
Let $\phi: H \times H \to \mathbb{C}$ be a bounded sesquilinear form, then there is a unique $T \in B(H)$ such that 
$$\phi(x,y) = \langle T(x), y \rangle, \; x,t \in H,$$
moreover, $\phi_T = \phi$ and $\|\phi\| = \|T\|$. 
\end{proposition}
 
Now, we look into an alternative definition of $L^1(X,B(H))$ that will be more general than the one given in \cref{sec:Bochner_L_p_spaces},

\begin{definition}[$L^1$ spaces of weakly integrable functions\index{weakly integrable functions} (Definition 8.10 \citep{de_chiffre_haar_2011})]\label{definition:L1_spaces_of_weakly_integrable_functions}
Let $(X, \mathcal{A}, \mu)$ be a measure space and let $H$ be a Hilbert space. By $\tilde{L}^1(X, \mathcal{B}(H))$ we denote the set of functions $f: X \rightarrow \mathcal{B}(H)$ satisfying the following two conditions:
\begin{itemize}
    \item For all $\xi, \eta \in H$, the function defined by
$$
x \mapsto\langle f(x) (\xi), \eta\rangle,
$$
for $x \in X$, is $\mu$-strongly measurable.
    \item The function defined by $x \mapsto\|f(x)\|$, for $x \in X$, belongs to $L^1(X, \mu)$.
\end{itemize}
\end{definition}

\begin{proposition}Strong measurability of $\|f\|$ on $\tilde{L}^1(X, \mathcal{B}(H))$ (Lemma 5.7 \citep{sundar_notes_202})]\label{proposition:strong_measurability_of_absolute_value_in_separable_hilbert_space}
Let $(X, \mathcal{A}, \mu)$ be a measure space and let $H$ be a separable Hilbert space, assume that $f \in \tilde{L}^1(X, \mathcal{B}(H))$, then,  $x \mapsto \| f(x) \|$ is $\mu$-strongly continuous.
\end{proposition}
\begin{proof}
Since $H$ is separable there is a dense countable set $E \subset B(0,1)_H$, thus from \cref{proposition:correspondence_bounded_operators_and_sesquilinear_forms}, 
$$\| f (x)\| = \sup_{\xi,\eta \in E} \{ | \langle f(x)(\xi), \eta \rangle \}.$$
Since $x \mapsto \langle f(x)(\xi), \eta \rangle$ is $\mu$-strongly measurable and $x \mapsto |x|$ is continuous, then \cref{remark:norm_of_mu_strongly_measurable_functions} tell us that $x \mapsto |\langle f(x)(\xi), \eta \rangle|$ is $\mu$-strongly measurable. Denote $E = \{ \xi_i \}_{i \in \mathbb{N}}$ and $E_n = \{ \xi_i \}_{i \leq n }$, then 
$$\| f (x)\| =  \lim_{n \to \infty} \sup_{\xi,\eta \in E_n} \{ | \langle f(x)(\xi), \eta \rangle \}.$$
Since the supremum of linear combinations of $\mu$-simple functions is still a linear combination of $\mu$-simple functions,  we have that $x \mapsto \sup_{\xi,\eta \in E_n} \{ | \langle f(x)(\xi), \eta \rangle \}$ can be approximated by $\mu$-simple functions almost everywhere, so, since the almost everywhere limit of $\mu$-strongly continuous functions is $\mu$-strongly continuous (\cref{proposition:approximating_mu_strongly_measurable_functions}), then $x \mapsto \| f(x) \|$ is $\mu$-strongly continuous.
\end{proof}

Given this setup up we can give meaning to the integral $\int_X f(x) d\mu(x)$ for $f \in \tilde{L}^1(X,B(H))$. Take $\xi, \eta \in H$, then define
$$ \phi(\xi,\eta) = \int_X \langle f(x)(\xi), \eta \rangle d \mu(x), $$
then by the Cauchy-Schwartz inequality on $H$, $\phi(\xi,\eta) \leq \| f\|_1 \| \xi \| \|\eta \|$. Moreover, the properties of the inner product on $H$ imply that $\phi$ is a bounded sesquilinear form, therefore, \cref{proposition:correspondence_bounded_operators_and_sesquilinear_forms} guaranties that there is a unique $T \in B(H)$ such that $\langle T(\xi), \eta \rangle = \phi(\xi,\eta)$, this operator is denotes by $\int_X f(x) d \mu(x)$. Also, if $f \in \tilde{L}^1(X,B(H))$ then $(f^*)(x):=f(x)^*$ also belongs to $\tilde{L}^1(X;B(H))$ because 
$$x \mapsto \langle f^*(x)(\xi), \eta \rangle = (\langle f(x)(\eta), \xi)^*,$$
thus, since the involution on $\mathbb{C}$ is continuous and thus measurable, then $x \mapsto \langle f^*(x)(\xi), \eta \rangle$ is $\mu$-strongly measurable (\cref{proposition:approximating_mu_strongly_measurable_functions}).

\begin{proposition}[Integration of elements in $\tilde{L}^1(X,B(H))$ \index{$\tilde{L}^1(X,B(H))$}]\label{proposition:integration_of_elements_in_weak_L1}

Let $(X, \mathcal{A}, \mu)$ be a measure space and let $H$ be a Hilbert space, if $f \in \tilde{L}^1(X,B(H))$ then
\begin{itemize}
    \item \citep[Proposition 8.11]{de_chiffre_haar_2011} There is a unique element $\int_X f(x) d \mu(x) \in B(H)$ such that 
    $$ \langle (\int_X f(x) d \mu(x))(\xi), \eta\rangle = \int_X \langle f(x)(\xi), \eta \rangle d \mu(x), \; \xi, \eta \in H$$
    and
    $$ \|\int_X f(x) d \mu(x))(\xi) \| \leq \|f\|_1.$$
    \item \citep[Proposition 8.12]{de_chiffre_haar_2011} The vector-valued integral is linear in the sense that if $f, g \in L^1(X, \mathcal{B}(H))$ and $A, B \in \mathcal{B}(H)$, then $(A f B+g) \in L^1(X, \mathcal{B}(H))$ and
$$
\int A f(x) B+g(x) \mathrm{d} \mu(x)=A\left(\int f(x) \mathrm{d} \mu(x)\right) B+\int g(x) \mu(x) .
$$
    \item \citep[Proposition 5.8 item 3]{sundar_notes_202} $\int_X f^*(x) d \mu(x) = \left( \int_X f(x) d \mu(x) \right)^*$.
    \item \citep[Proposition 8.12]{de_chiffre_haar_2011} If $E \subseteq X$ is measurable of finite measure, then $1_E \cdot Id_H \in L^1(X, \mathcal{B}(H))$ and
$$
\int 1_E(x) \mathrm{d} \mu(x)=\mu(E) Id_H .
$$
    \item \citep[Proposition 8.12]{de_chiffre_haar_2011} If $f \in L^1(X, \mathcal{B}(H))$ and $f(x) \geq 0$ for all $x \in X$, then $\int f(x) \mathrm{d} \mu(x) \geq 0$.
\end{itemize}
\end{proposition}

\begin{proposition}[$\tilde{L}^1(X,B(H))$ and Bochner integrable functions]\label{proposition:weak_L1_and_Bochner_integrable_functions}
Let $(X, \mathcal{A}, \mu)$ be a measure space and let $H$ be a Hilbert space, take $f \in L^1(X, B(H))$, then, $f \in \tilde{L}^1(X,B(H))$ and the Bochner integral $\int_X f$ is the same operator as the one constructed in \cref{proposition:integration_of_elements_in_weak_L1} called $\int_X f(s) d \mu(x)$. 
\end{proposition}
\begin{proof}
If $f \in L^1(X, B(H))$, then $f$ is $\mu$-strongly measurable and $\int_X \|f\| d \mu(x) \leq \infty$, thus by \cref{prop:bochn_integra_condition} we have that $\int_X f(x) d \mu(x)$ is the Bochner integral of $f$. From \cref{prop:bochner_int_and_lin_bound_trans} we have that if $g^* \in B(H)^*$ then $g^*(\int_X f(x) d \mu(x)) = \int_X g^*(f(x)) d \mu(x)$, since, $\phi_{\xi,\eta}(T) = \langle T(\xi), \eta \rangle$ is a member of the continuous dual of $B(H)$, which we denoted by $B(H)^*$ we have that 
$$  \langle (\int_X f(x) d \mu(x))(\xi), \eta\rangle = \int_X \langle f(x)(\xi), \eta \rangle d \mu(X), \; \xi, \eta \in H.$$

By \cref{theorem:pettis_measurability_theorem_second_version} we know that $f$ is weak $\mu$-measurable, which means that $x \mapsto \langle f(x)(\xi), \eta \rangle$ is $\mu$-strongly measurable for $\xi, \eta \in H$, also $f \in L^1(X)$ by definition, thus \cref{proposition:integration_of_elements_in_weak_L1} gives us an alternative definition of the operator $\int_X f(X) d \mu(x)$, which coincides with the Bochner integral of $f$ because they generate the same sesquilinear bounded form
$$ \phi(x,y) =  \int_X \langle f(x)(\xi), \eta \rangle d \mu(X), \; \xi, \eta \in H.$$
\end{proof}

An example of operator-valued functions that are integrable are the continuous functions over a compact topological Hausdorff space (\cref{example:Bochner_integral_continuos_funcion}).

\begin{remark}[Weak convergence of partial sums vs norm convergence convergence of partial sums]
If $f \in \tilde{L}^1(G;B(H))$, the operator $\int_G  f(g) d\mu(g)$ can be understood as the limit of sums of the form $\sum_{i \leq n} f(g_i) \mu(S_i)$ under the weak operator topology \citep[page 515]{busby_representations_1970}, which is a parallel setting to the one given by the Bochner integral, that was understood as the limit of the sums $\int_G e(g) f(g) d\mu(g)$ under the norm operator topology \cref{sec:Bochner_integral}. Since the norm operator topology is stronger than the weak operator topology we found that any Bochner integral function can be weakly integrated \cref{proposition:weak_L1_and_Bochner_integrable_functions}, while the converse those not holds, as we saw with the left regular representation of the group $\mathbb{R}$ at the beginning of this section.
\end{remark}

\begin{remark}[Unitary representations and *-representations]\label{remark:untary_representations_and_star_representations}
Take $H = L^2(G)$ and $f \in L^1(G)$ with $G$ a second countable locally compact group, then $\mu$ is the Haar measure which is a Radon measure (\cref{definition:Radon_measure}), then $\mu$ is $\sigma$-finite (\cref{section:algebra_continuous_functions_locally_comp_space}). Denote by $\pi: G \to B(H)$ the left regular representation of $G$ (\cref{example:left_regular_representation_of_G}), then $g \mapsto \langle \pi(g)(\xi), \eta \rangle$ is measurable because it is continuous, moreover, it is $\mu$-strongly measurable due to \cref{remark:what_happend_if_X_is_separable}. The function $g \mapsto f(g) \langle \pi(g)(\xi), \eta \rangle$ is also $\mu$-strongly measurable (\cref{remark:algebra_of_strongly_mu_measurable_functions}), therefore, $g \mapsto f(g)\pi(g)$ is an element of $\tilde{L}^1(G;H)$, which implies that there is an element $\int_G f(g) \pi(g) d \mu(g) \in B(H)$ that satisfies the properties from \cref{proposition:integration_of_elements_in_weak_L1}. Let $\hat{H}$ be a Hilbert space, then a similar argument can be used to construct $\int_G f(g) (\pi \otimes Id_{\hat{H}}) d \mu(g) \in B(L^2(G;\hat{H}))$.

Turns out that the mapping $f \mapsto \int_G f(g) \pi(g) d \mu(g)$ is a representation of $L^1(G)$ on $L^2(G)$ (\citep[Proposition 6.2.1]{deitmar_principles_2009}), where $L^1(G)$ is provided with the structure of an algebra through the convolution $(f \ast e)(j) := \int_G f(g)e(g^{-1}j ) d \mu(g) $ (\citep[Theorem 1.6.2]{deitmar_principles_2009}). If $\mu$ is unimodular then we know that $L^1(G)$ a Banach *-algebra (\cref{example:L1_Bohner_spaces_and_convolution}), if $\mu$ is not unimodular then $L^1(G)$ can also become a Banach *-algebra by modifying the involution (\citep[Lemma 2.6.2]{deitmar_principles_2009}), and in both cases $\hat{\pi(f)} := f \mapsto \int_G f(g) \pi(g) d \mu(g)$ becomes a *-representation, that is, $\hat{\pi}(f \ast e) = \hat{\pi}(f) \hat{\pi}(e)$ and $\hat{\pi}(f^*) = \hat{\pi}(f)^*$ for $f,e \in L^1(G)$. As a matter of fact, there is a one-to-one between unitary representations of $G$ and *-representations of $L^1(G)$ (\citep[Proposition 6.2.1]{deitmar_principles_2009}, \citep[Proposition 6.2.3]{deitmar_principles_2009}).
\end{remark}

Now that we have been able to define operators $\int_G f(g) (\pi \otimes Id_H)(g) d \mu(x) \in B(L^2(G;H))$ for $G$ a second countable locally compact group and $f \in L^1(g)$ (\cref{remark:untary_representations_and_star_representations}), we now look into making $f(g)$ an operator valued function instead of a scalar-valued function. When we make $f$ into an operator valued function we get the appropriate setting for the representation of twisted crossed products (\cref{sec:twisted_crossed_products}).

Take $f: S \to X$ strongly measurable and $g: S \to \mathscr{L}(X,Y)$ a function from $S$ to the Banach space of continuous linear maps from $X$ to $Y$, this set up allow us to create a new function $gf$ as follows,

$$ gf(s) := g(s)(f(s)),$$

so let us look at some sufficient conditions for $gf$ to be strongly measurable.

$\mathscr{L}(X,Y)$ is not necessarily separable even if $X$ and $Y$ are separable, for example, if $X$ is a separable Hilbert space then $\mathscr{L}(X,X)$ is not separable \citep{119273}, thus we modify the definition of strongly measurable for functions of the type $g: S \to \mathscr{L}(X,Y)$,

\begin{definition}[Definition 1.1.27. \citep{hytonen_analysis_2016}]
\label{def:oper_valu_strong_measu}
A function $f: S \rightarrow \mathscr{L}(X, Y)$ is called strongly measurable (respectively, strongly $\mu$-measurable) if for all $x \in X$ the $Y$-valued function $f x: s \mapsto f(s) x$ is strongly measurable (respectively, strongly $\mu$ measurable).

\end{definition}

Notice that in the previous definition, we are taking into account the strong topology of $\mathscr{L}(X, Y)$ instead of the norm topology of $\mathscr{L}(X, Y)$. As we would expect, the strongly ($\mu$-) measurable functions $g: S \to \mathscr{L}(X,Y)$ allow us to create strongly ($\mu$-) measurable functions $gf : S \to Y$,

\begin{proposition}[Proposition 1.1.28. \citep{hytonen_analysis_2016}]
\label{prop:oper_valu_new_strong_measu_funcs}
Let $(S, \mathscr{A})$ be a measurable space (respectively, $(S, \mathscr{A}, \mu)$ a measure space) and let $X$ and $Y$ be Banach spaces.
\begin{itemize}
    \item \citep[Proposition 1.1.28.]{hytonen_analysis_2016} If $f: S \rightarrow X$ and $g: S \rightarrow \mathscr{L}(X, Y)$ are strongly ( $\mu$-)measurable, then $g f: S \rightarrow Y$ is strongly $(\mu-)$ measurable.
    \item \citep[Corollary 1.1.29.]{hytonen_analysis_2016} Let $X, Y, Z$ be Banach spaces. If $f: S \rightarrow \mathscr{L}(X, Y)$ and $g: S \rightarrow \mathscr{L}(Y, Z)$ are strongly $(\mu-)$ measurable, then $g \circ f: S \rightarrow \mathscr{L}(X, Z)$ is strongly $(\mu$-)measurable.
\end{itemize}
\end{proposition}

If $Y$ is separable the setup is simpler and we have the following result which will come in handy later:

\begin{proposition}
\label{prop:oper_valu_new_strong_measu_funcs_in_separable_case}
Let $(S, \mathscr{A})$ be a measurable space (respectively, $(S, \mathscr{A}, \mu)$ a measure space), let $X$ and $Y$ be Banach spaces with $Y$ separable. If $f: S \rightarrow X$ is strongly measurable and  $g: S \rightarrow \mathscr{L}(X, Y)$ is measurable with respect to $\mathcal{B}(\mathscr{L}(X, Y))$ (Borel $\sigma$ algebra of the norm topology), then $gf: S \to Y$ is strongly measurable. Furthermore, if $\mu$ is $\sigma$-finite then $gf: S \to Y$ is strongly $\mu$-measurable.
\end{proposition}

\begin{proof}
For every $x \in X$ the evaluation map

$$ ev_x : \mathscr{L}(X, Y) \to Y, \; ev_x (L) = L(x) $$

is continuous because $\| ev_x (L) - ev_x(T) \| = \| T(x) - L(x) \| \geq \| x \| \| T-L \|$, which is equivalent to saying that the inverse image of any open set in $Y$ is an open set in $\mathscr{L}(X, Y)$. Since we are using the $\sigma$-algebras $\mathcal{B}(\mathscr{L}(X, Y))$ and $\mathcal{B}(Y)$, the previous statement implies that the evaluation map is measurable. 

The map $gx : s \mapsto g(s)(x)$ is the composition of the evaluation map and the map $g$, thus it is measurable because the composition of measurable functions is measurable \citep[Lemma 3.31]{heil_introduction_2011}. Since $Y$ is separable we have that $gx$ is separably valued, and \cref{coro:meas_vs_stron_measu_and_separ} tells us that $gx$ is strongly measurable. Now, we can use \cref{prop:oper_valu_new_strong_measu_funcs} to show that $gf$ is strongly measurable.

Furthermore, if $\mu$ is $\sigma$-finite we can use \cref{prop:stron_meas_and_mu_stron_measu} to show that $gf$ is strongly $\mu$-measurable. If $g$ is $\mu$-almost everywhere equal to a measurable function, then $gf$ is $\mu$-almost everywhere to a strongly measurable function, which in turn implies that is $\mu$-strongly measurable because $\mu$ is $\sigma$-finite and $Y$ is separable (\cref{remark:what_happend_if_X_is_separable}).
\end{proof}

\begin{proposition}[Functions of $\tilde{L}^1(G;H)$ from composition]\label{proposition:creating_operators_on_L_2_from_composition_of_measurealbe_functions}
Let $G$ be second countable locally compact group with $\mu$ its Haar measure and $H$ a separable Hilbert space, assume that $f: G \to U(H)_s$ measurable and $e \in L^1(G;B(H))$, then, the map $s \mapsto e(s)(f(s))$ is an element of $\tilde{L}^1(G;H)$ and has an associated bounded operator $ \int_G e(g)f(g) d \mu(g) \in B(H)$ over $L^2(G,H)$ satisfying
$$ \langle (\int_G e(g) f(g) d\mu(g) )(\xi), \eta \rangle = \int_G \langle (e(g)f(g))(\xi), \eta \rangle d \mu(g), \; \xi, \eta \in H .$$
\end{proposition}
\begin{proof}
From \cref{proposition:measurability_in_unitary_group} we know that $g \mapsto U(g)(\xi)$ and $g \mapsto \langle U(g)(\xi), \eta \rangle$ are both measurable, moreover, since $\mu$ is $\sigma$-finite (\cref{section:algebra_continuous_functions_locally_comp_space}) both maps are $\mu$-strongly measurable (\cref{remark:what_happend_if_X_is_separable}).
 
Take $e \in L^1(G;B(H))$ i.e. a Bochner integrable function, then $e$ is equal to a measurable function almost everywhere, under this setup, \cref{prop:oper_valu_new_strong_measu_funcs_in_separable_case} tell us that $g \mapsto e(g)(f(g)(\xi))$ is $\mu$-strongly measurable for any $\xi \in H$. Since the inner product is a continues map, then, \cref{remark:norm_of_mu_strongly_measurable_functions} implies that the map $g \mapsto \langle e(g)(f(g)(\xi)), \eta \rangle$ is $\mu$-strongly measurable for any $\eta \in H$. Also, from \cref{proposition:strong_measurability_of_absolute_value_in_separable_hilbert_space} we know that $g \mapsto \|e(g)f(g) \|$ is $\mu$-strongly measurable, additionally, since $f(g)$ is unitary we have that $\|e(g)f(g) \| \leq \| e(g) \|$ which implies that $g \mapsto \|e(g)f(g) \| $ belongs to $L^1(G,\mu)$. The previous statements imply that the map $g \mapsto e(g)f(g)$ is an element of $\tilde{L}^1(G;H)$ (\cref{definition:L1_spaces_of_weakly_integrable_functions}), therefore, by \cref{proposition:integration_of_elements_in_weak_L1} we know that there is an element
$$ \int_G e(g)f(g) d \mu(g) \in B(H), $$
such that
$$ \langle (\int_G e(g) f(g) d\mu(g) )(\xi), \eta \rangle = \int_G \langle (e(g)f(g))(\xi), \eta \rangle d \mu(g), \; \xi, \eta \in H .$$
\end{proof}

\begin{example}[Functions of $\tilde{L}^1(G;H)$ from composition]\label{example:weak_L1_functions_from_composition}
Let $G$ be second countable locally compact group with $\mu$ its Haar measure and $H$ a separable Hilbert space, take $f: G \to U(H)_s$ measurable, then from \cref{proposition:measurability_in_unitary_group} we know that $g \mapsto U(g)(\xi)$ and $g \mapsto \langle U(g)(\xi), \eta \rangle$ are both measurable, moreover, since $\mu$ is $\sigma$-finite (\cref{section:algebra_continuous_functions_locally_comp_space}) both maps are $\mu$-strongly measurable (\cref{remark:what_happend_if_X_is_separable}). 

Take $e \in L^1(G;B(H))$ i.e. a Bochner integrable function, then $e$ is equal to a measurable function almost everywhere, under this setup, \cref{prop:oper_valu_new_strong_measu_funcs_in_separable_case} tell us that $g \mapsto e(g)(f(g)(\xi))$ is $\mu$-strongly measurable for any $\xi \in H$. Also, from \cref{proposition:strong_measurability_of_absolute_value_in_separable_hilbert_space} we know that $g \mapsto \|e(g)f(g) \|$ is $\mu$-strongly measurable, since, $f(g)$ is unitary we have that $\|e(g)f(g) \| \leq \| e(g) \|$, therefore, by \cref{proposition:integration_of_elements_in_weak_L1} we know that there is an element
$$ \int_G e(g)f(g) d \mu(g) \in B(H), $$
such that
$$ \langle (\int_G e(g) f(g) d\mu(g) )(\xi), \eta \rangle = \int_G \langle (e(g)f(g))(\xi), \eta \rangle d \mu(g), \; \xi, \eta \in H .$$
\end{example}

\section{Holomorphic functional calculus}
\label{sec:banach_alg_hol_func_cal}

The holomorphic functional calculus provides a powerful computational tool for unital Banach algebras and will be useful on our characterization of unitaries and projections of C* algebras. In principle, it is defined as a topological algebra homorphism, that goes from the algebra of holomorphic functions over an open set and ends up in an unital Banach algebra, that is, let $A$ be a unital Banach algebra and $a\in A$, as we have seen $sp(a)$ is compact, thus let $\Omega$ be an open neighbourhood of $Sp(a)$ that is a Cauchy domain \citep[Defitinion 5.2]{allan_spectral_1965} with smooth boundary, we will have the following homomorphism,

$$ \gamma : \text{Hol}(\Omega) \to A , \; f \mapsto \int_{\eta} f(z) (z1 -a)^{-1} dz, $$

were $\eta: [0,1] \oplus ... \oplus [0,1] \to \mathbb{C}$ is a piece-wise $C^{1}$ curve surrounding $sp(a)$, which corresponds to the boundary of $\Omega$. The integral is defined as a Bochner integral and we know that it exists due to \cref{example:Bochner_integral_continuos_funcion}. This homorphism is also a continuous maps of topological spaces, therefore we will take a brief detour into the topology of $\mathscr{Hol}(\Omega)$ which in turns will help us understand better the homomorphism $\gamma$.

\subsection{Algebra of holomorphic functions}
\label{section:algebra_of_holomorphic_functions}
Let $\Omega$ be an open set of the complex numbers, let $\text{Hol}(\Omega)$\index{$\text{Hol}(\Omega)$} by the algebra of holomorphic functions on $\Omega$, this algebra becomes a Fréchet m-convex algebra (\cref{example:holomorphic_functions_over_open_set}) under the compact-open topology \citep[Example 4.7]{allan_introduction_2011}. The compact-open topology is the topology of uniform convergence on compacts subsets of $\Omega$, and is the topology given to $\text{Hol}(\Omega)$ because for every sequence $\{ f_n \}$ that converges uniformly in every compact subset of $\Omega$ we have that $f = \lim_{n \to \infty} f_n$ is holomorphic in $\Omega$ \citep[Theorem 7.2]{gilman_complex_2007}.

The compact-open topology is the topology generated by the sub-multiplicative semi norms
$$ p_{K} (f)= \sup_{k \in K} \{| f(k) | \text{ were } K \subset \Omega \text{ and } K \text{ is compact,} \}. \; p_{K} (fg) \leq p_{K} (f)p_{K} (g),$$

thus with this topology $\text{Hol}(\Omega)$ is a complete locally m-convex algebra, that is, a locally convex vector space with continuous multiplication. To check that $Hol(\Omega)$ is a Fréchet algebra, we need to provide it with an upper directed countable family of semi norms that provide an equivalent topology (\cref{theorem:metrizable_lcs_and_seminorms}), fortunately we can do this because $\mathbb{C}$ is second countable locally compact topological spaces. We follow \citep[Section 7.2]{gilman_complex_2007}, so, for each rational $z \in \Omega$ i.e. $z=x+\imath y$ with both $x$ and $y \in \mathbb{Q}$, take all the closed balls with rational radii that are contained in $\Omega$, that is, $\{ y | |y-z| \leq r\} \subset \Omega$. There are a countable number of such closed balls, and they cover $\Omega$. Call that collection of closed balls $\{ D_i \}_{i \in \mathbb{N}}$, and set 
$$
K_n=\bigcup_{i \leq n} D_i ,
$$
then, each $K_n$ is compact and 
$$\bigcup_{i \in \mathbb{N}} K_i = \Omega.$$ 
So, if $P \subset \Omega$ is compact then $P \subset K_l$ for some $l \in \mathbb{N}$, so, if $\{ f_n \}_{n \in \mathbb{N}}$ is a sequence of holomorphic functions, then $f_n \to f$ uniformly in every compact subset of $\Omega$ iff $f_n \to f$ uniformly in each $K_i$. 

\begin{definition}[Compact exhaustion\index{compact exhaustion}]\label{def:compact_exhaustion}
Let $X$ be a Hausdorff topological space. Assume that there is a sequence $\{ K_n \}_{n \in \mathbb{N}}$ of compact sets of $X$ such that, $K_n \subseteq K_{n+1}$, $K_n \subset K_{n+1}^{o}$ and $\cup_{n \in \mathbb{N}} K_n = X$, where $ K_{n+1}^{o}$ denotes th interior of $ K_{n+1}$. We call the sequence $\{ K_n \}_{n \in \mathbb{N}}$ a compact exhaustion of $X$.
\end{definition}
 
$\text{Hol}(\Omega)$ becomes a Fréchet m-convex algebra under the semi norms
$$ p_{i}(f) = \sup_{\omega \in K_i} | f(\omega) |, \; i \in \mathbb{N},$$
and this topology is equivalent to the topology generated by the translation invariant metric
$$ d(f,g) = \sum_{i \in \mathbb{N}} 2^{-i} \frac{p_i (f-g)}{1 + p_i (f-g)},$$
by \cref{theorem:metrizable_lcs_and_seminorms}. The previous construction can be performed with any compact exhaustion $\{ K_n \}_{n \in \mathbb{N}}$ of $\mathbb{C}$.

\subsection{Definition and properties}
\label{sec:holom_calc_def_and_prop}

The holomorphic functional calculus comes as generalization of the polynomial calculus \citep[Lemma 4.22]{allan_introduction_2011}, which for a given element $a$ of a Banach algebra $A$ and for every polynomial $p$ in one variable with complex coefficients assigns an element $p(a)\in A$ we have that 
$$ \text{Sp}(p(a)) = p(\text{Sp}(a)). $$

The holomorphic functional calculus is constructed using line integrals over $A$, so, let us define over what type of contours we will perform an integration on. 

\begin{definition}[Cauchy domain\index{Cauchy domain} (definition 5.2 \citep{allan_spectral_1965})]\label{definition:Cauchy_domain}
A subset $D$ of $\mathbb{C}$ is called a Cauchy domain if 
\begin{itemize}
    \item $D$ is open
    \item $D$ has a finite number of components, the closures of which are pairwise disjoint
    \item The boundary $\partial D$ of $D$ is a subset of $\mathbb{C}$, consisting of a finite positive number of closed rectifiable Jordan curves, no two of which intersect.
\end{itemize}
\end{definition}

For simplicity, we will work with Cauchy Domains with $C^1$ boundary, that is, 
$$\partial D = \bigcup_{i \leq l} \gamma_i, \; \gamma_i : [a,b] \to \mathbb{C}, \; \frac{ d \gamma_i}{dt} \text{ is continuous},$$
as exposed in \citep[section 1.11]{allan_introduction_2011}. For a continuous function $F: \partial D \to A$ with $A$ a Banach algebra, define 
$$ \int_{\partial D} F := \int_{\partial D} F(z) dz := \sum_{i \leq l} \int_{a_i}^{b_i} F( \gamma_i (t')) \frac{d \gamma_i}{dt}(t') dt'.  $$

Let $K \subset \mathbb{C}$ be compact, and $D$ an open neighbourhood of $K$ that is Cauchy domain with $C^1$ boundary such that $\partial D = \bigcup_{i \leq l} \gamma_i$, then for every $z \in K$ define 
$$ g_i(t):= \frac{1}{2 \pi} \text{arg}(\gamma_i (t) - z) $$
taking $g_i(t) \in [0, 2 \pi)$ and varying continuous with respect to $t$, then we denote
$n(\gamma_i ; z) := \int_{\gamma_i} g_i \in \mathbb{N}$ 
and it is called the winding number of the contour $\bigcup_{i \leq l} \gamma_i$ (\citep[section 1.11]{allan_introduction_2011}). 

We focus on special types of open neighbourhoods,

\begin{definition}[Contour\index{contour}]\label{definition:contoru_compact_set}

Let $K \in \mathbb{C}$ be a compact set and denote by $\partial D$ the boundary of $A$, then $\partial D$ with $\partial D = \bigcup_{i \leq l} \gamma_i$ is called a contour of $K$ if
\begin{itemize}
    \item $D$ is Cauchy domain with $C^1$ boundary and is an open neighbourhood of $K$. 
    \item For every $k \in K$ we have that 
    $$n(\partial D ; k) = \sum_{i \leq l} n(\gamma_i;k) = 1.$$ 
\end{itemize}
\end{definition}

For every $K \subset \mathbb{C}$ compact there is a contour \citep[Proposition 1.31]{allan_introduction_2011}, and contours are special because they \textit{surround} $K$. 

Given an $U \subset \mathbb{C}$, denote by $R(U)$ the set of rational functions that are holomorphic on $U$, that is, 
$$ R(u) = \{ p(x) / q(x) , \; \text{if }q(z) = 0 \text{ then } z \notin U \}.$$
Take $a \in A$ and $\text{Sp}(a) \subset U$ open, then you can check that
\begin{itemize}
    \item $R_{\lambda_0}(a) R_{\lambda_1}(a) = R_{\lambda_1}(a) R_{\lambda_0}(a)$ for $\lambda_0, \lambda_1 \in U$.
    \item $p(a) R_{\lambda}(a) = R_{\lambda}(a) p(a)$ for $\lambda \in U$ and $p(x)$ any polynomial in one variable with complex coefficients.
\end{itemize}
So, it can be shown that for $a \in A$ and $\text{Sp}(a) \subset U$ open with $\partial U$ a contour of $\text{Sp}(a)$ there is a continuous homorphism $\eta : R(U) \to A$ such that $\eta(r) = r(a)$ \citep[Lemma 4.88]{allan_introduction_2011}. This homorphism is extended into $\text{Hol}(U)$ using the fact that $R(U)$ is dense in $\text{Hol}(U)$ \citep[Corollary 4.86]{allan_introduction_2011}, which altogether gives the holomorphic functional calculus,

\begin{theorem}[Holomorphic functional calculus\index{Banach algebra!holomorphic functional calculus} (Theorem 4.89 \citep{allan_introduction_2011})]\label{theorem:holomorphic_functional_calculus_banach_algebras}
Let $A$ be an unital Banach algebra, let $a \in A$, and let $U$ be an open neighborhood of $\text{Sp} (a)$ in $\mathbb{C}$ that is a Cauchy domain. Then there is a unique continuous, unital homomorphism $\Theta_a: \text{Hol}(U) \rightarrow A$ such that $\Theta_a(Z)=a$, and has the following properties:
\begin{itemize}
    \item For every $\partial D \subset U$ a contour of $\text{Sp}(a)$ (\cref{definition:contoru_compact_set}) such that $D \subset U$, and every $f \in \text{Hol}(U)$, we have
    $$
    \Theta_a(f)=\frac{1}{2 \pi \mathrm{i}} \int_\gamma f(\lambda)(\lambda 1_A-a)^{-1} \mathrm{~d} \lambda
    $$
    \item $\Theta_a(r)=r(a)(r \in R(U))$
    \item $\Theta_a (\text{Hol}(D))$ belongs to the smallest commutative algebra containing $a$ ($\Theta_a (\text{Hol}(D)) \subseteq\{a\}^c \cap\{a\}^{c c}$)
    \item $\varphi\left(\Theta_a(f)\right)=f(\varphi(a)) \quad\left(f \in \text{Hol}(U), \varphi : \{a\}^{c c} \to \mathbb{C} \right)$, where $\{a\}^{c c}$ is the bi-commutant of $a$ \citep[page 180]{allan_introduction_2011}.
    \item \textbf{Spectral mapping:}\index{spectral mapping} $\text{Sp}\left(\Theta_a(f)\right)=f(\text{Sp} (a))(f \in \text{Hol}(U))$
\end{itemize}

To easy the notation we denote $\Theta_a(f)$ by $f(a)$. 
\end{theorem}

\begin{remark}[Holomorphic calculus is a Bochner integral]\label{remark:holomorphic_calculus_is_a_Bochner_integral}

We have defined the holomorphic calculus as a map that comes from an integral, and as you may have guessed, this is a Bochner integral, so, how do we know that $f(\lambda)(\lambda 1-a)^{-1}$ is Bochner integrable? This is a fact easy to check, so, let $\gamma = \bigcup_{i \leq n} \gamma_i$ be a contour of $\text{Sp}(a)$, then $\gamma_i : [0,1] \to \mathbb{C}$ is $C^1$ and the function $\hat{f} \circ \gamma_i : [0,1] \to A $ given by
$$  (\hat{f} \circ \gamma_i)(t) = f(\gamma_i (t))(\gamma_i (t) 1_A-a)^{-1} , \; \hat{f}(\lambda) = f(\lambda)(\lambda 1-a)^{-1}$$
is continuous because $R_a (\lambda) = (\lambda 1-a)^{-1}$ is a continuous when $\lambda \in \text{Sp}(a)$. We have assumed that the function 
$$t' \to \frac{d \gamma_i}{dt} (t'),$$
is continuous, thus the mapping
$$ \tilde{f}: [0,1] \to A, \tilde{f}(t') =  f(\gamma_i (t'))(\gamma_i (t') 1_A-a)^{-1} \frac{d \gamma_i}{dt} (t')$$
is continuous on $[0,1]$ and \cref{example:Bochner_integral_continuos_funcion} tell us that by being $[0,1]$ a compact measure space $\tilde{f}$ is Bochner integrable. Therefore, $f(a)$ exists because it is the sum of Bochner integrable functions. Additionally, since $\tilde{f}$ is a continuous function \cref{corollary:Bochner_integral_and_Riemann_sums} tell us that the element 
$$\frac{1}{2 \pi \mathrm{i}} \int_\gamma f(\lambda)(\lambda 1_A-a)^{-1} \mathrm{~d} \lambda$$
is the limit of Riemann sums.
\end{remark}

Let $\{ K_i \}_{i \in \mathbb{N}}$ be a compact exhaustion of $U$ (\cref{def:compact_exhaustion}), then, the continuity of $\Theta_a$ tell us that if  $\{ f_n \}_{n \in \mathbb{N}}$ is a sequence of holomorphic functions in $U$ such that $f_n \to f$ uniformly over all $K_i$ then $f_n(a) \to f(a)$ in the norm topology of $A$. 

\begin{itemize}
    \item The topology of $\text{Hol}(U)$ is important to understand the holomorphic calculus as a topological algebras homomorphism, for example, if we take $a \in A$, every contour $\gamma \subset U$ of $\text{Sp}(a)$ is a compact subset of $U$, and the convergence of $f_n \to f$ in the topology of $\text{Hol}(U)$ is the key to guaranty that the Banach valued integrals
    $$ \int_{\gamma}f_n(\lambda)(\lambda 1_A - a)^{-1} d \lambda. $$
    By \cref{prop:bochn_integra_condition} we have that
    $$ \| \int_{\gamma}f_n(\lambda)(\lambda 1_A - a)^{-1} d \lambda - \int_{\gamma}f(\lambda)(\lambda 1_A - a)^{-1} d \lambda \| $$ 
    $$ \leq \int_{\gamma} |f_n(\lambda)- f(\lambda)| \| (\lambda 1_A - a)^{-1} \| d \lambda,  $$
    thus, if we can make $|f_n(\lambda) -f(\lambda)|$ arbitrarily small in any compact subset of $U$, we can make it arbitrary small in any contour, and we get the convergence of $f_n(a) \to f(a)$ in the topology of the Banach algebra.
    \item Hence, if $f(z) = \sum_{k \in \mathbb{N}} \alpha_k (z - \omega)^k$ inside $\text{Sp}(a)$, then, $f_n = \sum_{i \leq n} \alpha_k (z - \omega)^k$ is a sequence of functions that converge to $f$ in the compact-open topology, and therefore $f(a) = \lim_{n \to \infty} f_n (a) = \sum_{k \in \mathbb{N}} \alpha_k (a - \omega 1_{A})^k$. When $U = \mathbb{C}$ we get a functional calculus for the entire functions, as expected.
\end{itemize}

The previous discussion provides a proof for the following fact,

\begin{lemma}\label{lemma:banach_algebra_holomorphic_func_calculus_entire_functions}
Let $A$ be a unital Banach algebra and $a \in A$, let $f$ be a an entire function with a series expansion given by $f(z) = \sum_{i \in \mathbb{N}} \alpha_i (z - \omega)^i$, then, $\Theta_a(f) = \sum_{k \in \mathbb{N}} \alpha_k (a - \omega 1_{A})^k$, where the right side is convergent series inside $A$.
\end{lemma}

Now, we describe a few properties of the holomorphic functional calculus that will be useful later:

\begin{proposition}[Properties of the holomorphic functional calculus]\label{proposition:properties_of_the_holomoprhic_functional_calculus}
Denote by $\mathbb{O}_{a}$ the algebra of functions that are holomorphic on an open neighbourhood $U$ of $\text{Sp}(a)$ such that $U$ is a Cauchy domain, notice that $\mathbb{O}_{a}$ depends on both $a$ and $U$, we are not taking into account $U$ in the notation because it will not be explicitly used in the following statements. The homomorphism $\Theta_a$ defined in \cref{theorem:holomorphic_functional_calculus_banach_algebras} has the following properties:

\begin{itemize}
    \item It is continuous as an $A$ valued function \citep[Proposition 4.93]{allan_introduction_2011}, that is, if $a_n \to a$ in $A$ and $f \in \mathbb{O}_{a}$ then $f (a_n) \to f(a)$ in $A$. Since we are dealing with a normed space then convergence of every sequence is equivalent to convergence on the normed space i.e. given $\epsilon > 0$ then there is  $\delta >0$ such that if $\| b - a\| \leq \delta$ then $\| f(a) - f(b) \| \leq \epsilon$. 
    \item It behaves well with respect to unital Banach homorphisms \citep[Proposition 4.94]{allan_introduction_2011}, that is, if $A, B$ are Banach algebras and $T_A \to B$ is a continuous unital homomorphism, then for $f \in \mathbb{O}_{a}$, we have that 
    $$T(f(a)) = f(T(a)).$$
    Notice that $f(T(a))$ is well defined because $\text{Sp}(T(a)) \subset Sp(a)$.
    \item It behaves well with respect to composition of holomorphic functions \citep[Proposition 4.95]{allan_introduction_2011}, that is, if $f \in \mathbb{O}_{a}$ and $g \in \mathbb{O}_{g(a)}$ then
    $$ (g \circ f)(a) = g(f(a)). $$
\end{itemize}

\end{proposition}

\begin{remark}[Spectrum upper semi-continuity and function evaluation]\label{remark:spectrum_upper_semi_continuity_and_function_evaluation}

Let $U$ be an open neighbourhood of $\text{Sp}(a)$ and $f$ holmorphic on $U$, notice $f(b)$ may not be defined for all elements of $A$, thus, what do we mean by, if $b \to a$ then $f(b) \to f(a)$?
In this claim we are using the upper semi-continuity of the spectrum on Banach algebras (\cref{proposition:upper_semi_continuity_spectrum_map}) as follows, let $\gamma$ be contour of $\text{Sp}(a)$ then $d(\gamma, \text{Sp}(a)) > 0$, take $0 <\epsilon < d(\gamma, \text{Sp}(a))$ then 
$$ E_{\epsilon}(\text{Sp}(a)) \subset U, $$
and $\gamma$ is also a contour of the clousure of $E_{\epsilon}(\text{Sp}(a))$. Then, there is a $\delta > 0$ such that $\| a -b \| \leq \delta$ implies that $\text{Sp}(b) \subset E_{\epsilon}(\text{Sp}(b))$, and this in turn implies that $f$ is also holomorphic on $\text{Sp}(a)$, or put in another terms, $f(b)$ exists by the holomorphic functional calculus. So, for elements close enough to $a$ we can apply $f$ and to those elements we refer in the claim $f(b) \to f(a)$. 
\end{remark}

So, we will look into some results that are a consequence of the holomorphic functional calculus, and that are helpful in our subsequent analysis of K theory.

\subsection{Consecuences}
\label{sec:hol_cal_consecuences}

First, is possible to define a holomorphic functional calculus if $A$ has no unit? It is possible, by taking advantage of the fact that there is a canonical homorphism from a Banach into its unitization
$$ i: A \to A^{+}, $$
so, for $a$ in $A$ and $f \in \mathbb{O}_{a}$, we define $f(a) = f(i(a)) = f(a,0)$. Also, notice that $a \in A$ iff $i(a) = 0$. On other side, we can use the other homorphism of Banach algebras $\pi: A^{+} \to \mathbb{C}$ to check that $\pi(f(\hat{a})) = f(\pi(\hat{a}))$for $\hat{a} \in A$ by \cref{proposition:properties_of_the_holomoprhic_functional_calculus}, therefore, if $a \in A$ then $f(a) = f(i(a)) \in A$ iff $\pi(f(i(a))) = f(\pi(a)) = f(0) = 0$.

Now, we turn into a couple of functions that are special for us, the logarithm and the n-rooths. Recall that neither the logarithm nor the n-rooths are entire functions, so, we can only define them when the spectrum of an element lays in the domain of holomorphicity of those functions. There are many conditions on $\text{Sp}(a)$ to be in the domain of holomorphicity of a branch of the complex logarithm and complex n roots, the most commonly used is for $\text{Sp}(a) \subset D$ with $D$ and open simply connected domain of $\mathbb{C}$ \citep[Remark 2.13]{timoney_complex_2004}, which fits good our purposes. So, set $\mathbb{R}^- = \{ x \in \mathbb{R} | x \leq 0 \}$ , $D = \mathbb{C} - \mathbb{R}^-$ and $\Pi=\{z=x+\mathrm{i} y \in \mathbb{C}: x>0\}$, then $D$ is a simply connected domain of $\mathbb{C}$. 

For any element $a \in A$ such that $\text{Sp}(a) \subset D$ we can use the holomorphic functional calculus to create elements $\log(a)$ and $a^{1/n}$, that satisfy $\exp(\log(a)) = a$ and $(a^{1/n})^n =a$. Also, notice that $f(z) = \exp(z)$ and $g(z) = \exp(-z)$ are inverse in $\mathbb{O}_a$ for every $a$, meaning that $\exp(a)^{-1} = \exp(-a)$ for every $a \in A$. We list these facts and a couple more for future reference,

\begin{lemma}\label{lemma:compute_inverse_with_holomorphic_functional_calculus}
Let $A$ be a unital Banach algebra and $a \in G(A)$, then, $a^{-1}$ is the result of applying the function $f(z) = z^{-1}$ with the holomorphic functional calculus.
\end{lemma}
\begin{proof}
Since $a \in G(A)$ we have that $0 \notin Sp(a)$, thus, the function $f(z) = 1/z$ is holomorphic over the spectrum of $a$, then, the holomorphic functional calculus (\cref{theorem:holomorphic_functional_calculus_banach_algebras}) tells us that $f(a)a = a f(a) = 1_A$ i.e. $f(a) =  a^{-1}$.  
\end{proof}

\begin{proposition}[Logarithms and n-roots]\label{proposition:logarithms_and_n_roots}
Let $A$ be a unital Banach algebra and $a \in A$, if $\text{Sp}(a) \subset D$, then
\begin{itemize}
    \item There is $b \in A$ such that $\exp(b)=a$, $b$ is the result of applying the holomorphic functional calculus over $a$ with the function $\log$. \citep[Proposition 4.100]{allan_introduction_2011}.
    \item There is a unique element $b \in A$ such that $b^2 =a$ and $\text{Sp}(b) \in \Pi$ \citep[Proposition 4.99]{allan_introduction_2011}. Moreover, for any $n \in \mathbb{N}$ there is an element $c$ such that $a = c^n$, to construct this element take the principal branch of the holomorphic function $f(z) = z^{1/n}$ over $D$ and apply it to $a$ using the holomorphic calculus, and $\text{Sp}(a^{1/n}) \subset \Pi$ due to the spectral mapping of the holomorphic calculus.
    \item If $ab = ba$, then $\exp(a + b) = \exp(a) \exp(b)$ \citep[Proposition 4.98]{allan_introduction_2011}.
    \item We have that $\exp(a) \in G(A)$ with $\exp(a)^{-1} = \exp(-a)$ \citep[Proposition 4.98]{allan_introduction_2011}.
\end{itemize}
\end{proposition}

\begin{corollary}\label{corolllary:computing_minus_n_root_holomorphic_funcitonal_calculus}
Let $A$ be a unital Banach algebra and $a \in A$, if $\text{Sp}(a) \subset D$, then, for any $n \in \mathbb{N}$ there is an element $b$ such that $b^{-1} = a^{1/n}$, and $b$ is computed using the function $f(z) = n^{-1/n}$.
\end{corollary}
\begin{proof}
From \cref{proposition:logarithms_and_n_roots} we know that $a^{1/n}$ can be computed using the function $g(z) = z^{1/n}$, additionally, $0 \notin in Sp(a^{1/n})$, thus, $a^{1/n}$ is an invertible element of $A$. Under this setting, \cref{lemma:compute_inverse_with_holomorphic_functional_calculus} tells us that $(a^{1/n})^{-1}$ is the result of applying the function $h(z) = 1/z$ to the element $a^{1/n}$, so, \cref{proposition:properties_of_the_holomoprhic_functional_calculus} tell us that the holomorphic functional calculus respects the composition of holomorphic functions i.e. $h \circ g (a) = f(a)$ with $f(z) = n^{-1/n}$.
\end{proof}

\begin{proposition}\label{proposition:log_of_element_close_to_identity}
Let $A$ be a unital Banach algebra and $a \in A$,
\begin{enumerate}
    \item  If $\rho(1_A-a) < 1$, then,
        $$ \log (a) = - \sum_{n = 1}^{\infty} \frac{(1-a)^n}{n}, $$
        and $a = \exp (\log(a)).$
    \item If if $\| 1_A -a \| \leq 1$, then, 
        $$ \log (a) = - \sum_{n = 1}^{\infty} \frac{(1-a)^n}{n}, $$
        and $a = \exp (\log(a)).$
    \item The open ball $\{a \in A:\|1-a\|<1\}$ belongs to $\exp(A)$, where, $\exp(A) = \{ \exp(a) | a \in A \}$ (\citep[Proposition 4.101]{allan_introduction_2011}).
\end{enumerate}
\end{proposition}
\begin{proof}
\begin{enumerate}
    \item Recall that $\rho(1_A-a) < 1$ is the spectral radius of $1_A-a$, so, if $\rho(1_A-a) < 1$, then, $\text{Sp}(1_A -a) \subset B(0,1)$. Also, we have that $a = -(1_A - a) + 1_A$, therefore, the spectral mapping of the holomorphic functional calculus (\cref{theorem:holomorphic_functional_calculus_banach_algebras}) implies that $\text{Sp}(a) \subset - B(0,1) + 1 = B(1,1)$, which means tha t$\text{Sp}(a) \subset D$. \cref{proposition:logarithms_and_n_roots} tell us that $\log(a)$ exists, additionally, given that the natural logarithm has the series expansion $log(z) = - \sum_{n = 1}^{\infty} \frac{(1-a)^n}{n}$ for all $z$ such that $|1-z| < 1$, we have that
    $$ \log (a) = - \sum_{n = 1}^{\infty} \frac{(1-a)^n}{n}. $$
    The fact that $a = \exp (\log(a))$ is a consequence of the composition of holomorphic functions (\cref{proposition:properties_of_the_holomoprhic_functional_calculus}).
    \item If $\| 1_A -a \| \leq 1$, we have that $\rho(1_A-a) < 1$ (\cref{remark:upper_bound_on_spectral_radius}), therefore, the previous item implies that
    $$ \log (a) = - \sum_{n = 1}^{\infty} \frac{(1-a)^n}{n}. $$
    The fact that $a = \exp (\log(a))$ is a consequence of the composition of holomorphic functions (\cref{proposition:properties_of_the_holomoprhic_functional_calculus}).
    \item This is a consequence of the previous item.    
\end{enumerate}
\end{proof}

\begin{remark}[Shifting the logarithm and the n-roots]\label{remark:shifting_logarithms_and_n_roots}
Denote $D_{\theta} = D \exp(i\theta)$ for $\theta \in [0, 2 \pi)$, then $D_{\theta}$ is also a simply connected domain of $\mathbb{C}$, thus for any $a\in A$ with $\text{Sp}(a)$ there are elements $\log(a)$ and $a^{1/n}$ which are computed with the principal branches of $f(z) = \log(z)$ and $g(z) = z^{1/n}$ on $D_{\theta}$ and the holomorphic functional calculus. 
\end{remark}

We can use the holomorphic calculus to create smooth paths on $A$ as follows

\begin{proposition}[Smooth paths from entire functions]\label{proposition:smoomth_paths_from_entire_functions}
For any $a \in A$ and $f \in \text{Hol}(\mathbb{C})$\index{$\text{Hol}(\mathbb{C})$} the function
$$ \gamma_{f,a} : \mathbb{C} \to A, \; z \mapsto f(za) $$
is infinitely continuously differentiable i.e $\gamma_{f,a} \in C^{\infty}(\mathbb{C},A)$. 
\end{proposition}
\begin{proof}
The proof of these facts relays on the continuity of the map $a \to \|(\lambda 1_A -a)^{-1}\|, \; \lambda \notin Sp(a) $ on Banach algebras (\citep[Proposition 4.93]{allan_introduction_2011}).

\begin{itemize}
    \item \textbf{Continuity:} 
    First, recall $f \in \mathbb{O}_a$ because is holomorphic over $\mathbb{C}$, also, note that $z \to za$ is a continuous functions for every $z \in \mathbb{C}$, thus by \cref{proposition:properties_of_the_holomoprhic_functional_calculus} the function $z \to f(za)$ is continuous for every $z$, because is the composition of two continuous functions
    $$ f(za) = (f \circ h_a) (a), \; f: A \to A, \; h_a : \mathbb{C} \to A \text{ s.t. } h_a(z) = za.  $$

    \item \textbf{Differentiable paths:} Given that the map $a \mapsto Sp(a)$ is upper semi-continuous (\cref{proposition:upper_semi_continuity_spectrum_map}), according to \cref{remark:spectrum_upper_semi_continuity_and_function_evaluation} if $D$ is a Cauchy domain of $Sp(za)$ and $\gamma$ is the boundary of $D$, there must be $\eta >0$ such that, if $|z- z'| \leq \eta$ then $D$ is a Cauchy domain of $Sp(z'a)$. 
    
    Set $g_z(a) = za$, since $f(za) = (f \circ g_z)(a)$, we have that
    $$ \frac{\left( f(za) - f(z'a) \right)}{z - z'} =  \frac{\left( (f \circ g_z)(a) - (f \circ g_{z'})(a) \right)}{z - z'}  $$
    $$ = \frac{1}{2 \pi \mathrm{i}} \left(  \int_\gamma (f \circ g_z)(\lambda)(\lambda 1-a)^{-1} \mathrm{~d} \lambda - \int_\gamma (f \circ g_{z'})(\lambda)(\lambda 1-a)^{-1} \mathrm{~d} \lambda  \right) \frac{1}{z - z'}$$
    $$ = \frac{1}{2 \pi \mathrm{i}} \int_\gamma \left( \frac{(f \circ g_z)(\lambda) - (f \circ g_{z'})(\lambda)}{z - z'}  \right) (\lambda 1-a)^{-1} \mathrm{~d} \lambda ,$$
    moreover, if we use the fundamental theorem of calculus for complex integrals we get,
    $$ \frac{\left( f(za) - f(z'a) \right)}{z - z'}  =  \frac{1}{2 \pi \mathrm{i}} \int_\gamma \lambda \left( \frac{\int_{\lambda z'}^{\lambda z} f^{(1)}(\omega) d \omega }{\lambda z - \lambda z'}  \right) (\lambda 1-a)^{-1} \mathrm{~d} \lambda ,$$
    with $f^{(1)}(\omega) = \frac{d f}{dz} (\omega)$.
    Since $f^{(1)}$ is a continuous function over $\mathbb{C}$, the fundamental theorem of calculus for complex valued integrals tells us that for any $\lambda \in \mathbb{C}$, if $z' \to z$ then
    $$ \frac{\int_{\lambda z'}^{\lambda z} f^{(1)}(\omega) d \omega }{\lambda z - \lambda z'} \to  f^{(1)}(\lambda z). $$
    Since $f$ is an entire function, we know that $f^{(1)}$ is an holomorphic function over $\mathbb{C}$, therefore, it is uniformly continuous over compact sets. 
    
    Here comes the key part of the demonstration, notice that  the function $\lambda \mapsto f_{z,z'}(\lambda)$ given by
    $$ f_{z,z'}(\lambda) = \left( \frac{(f \circ g_z)(\lambda) - (f \circ g_{z'})(\lambda)}{z - z'}  \right) =  \lambda  \frac{\int_{\lambda z'}^{\lambda z} f^{(1)}(\omega) d \omega }{\lambda z - \lambda z'}$$
    is an holomorphic function over $\mathbb{C}$ when $z \neq z'$, so, we will ask for the limit of $f_{z,z'}(\cdot)$ as elements of the topological algebra $\text{Hol}(\mathbb{C})$ when $z' \to z$, which will turn out to be the function $\lambda \mapsto \lambda f^{(1)}(\lambda z)$, in which case the continuity of the holomorphic functional calculus over $A$ (\cref{theorem:holomorphic_functional_calculus_banach_algebras}) would imply that 
    $$ \frac{d f(z'a)}{ dz'} |_{z} = \lim_{z' \to z} \frac{1}{2 \pi \mathrm{i}} \int_\gamma  f_{z,z'}(\lambda) (\lambda 1-a)^{-1} \mathrm{~d} \lambda $$
    $$= \frac{1}{2 \pi \mathrm{i}} \int_\gamma \lambda f^{(1)}(\lambda z) (\lambda 1-a)^{-1} \mathrm{~d} \lambda.$$
    
    Now we will look into how to show that 
    $$ (\lambda \mapsto f_{z,z'}(\lambda)) \to  (\lambda \mapsto \lambda f^{(1)}(\lambda z))$$
    inside $\text{Hol}(\mathbb{C})$. For $i \in \mathbb{N}$ set $K_i = \overline{B(0, i)}$, then, $\{ K_i \}_{i \in \mathbb{N}}$ is a compact exhaustion of $\mathbb{C}$ (\cref{def:compact_exhaustion}), then, the topology of $\text{Hol}(\mathbb{C})$ is given by the set of seminorms (\cref{example:holomorphic_functions_over_open_set})
    $$ p_{i}(f) = \sup_{\omega \in K_i} | f(\omega) |, \; i \in \mathbb{N},$$
    thus, by the content of \cref{proposition:convergence_metrizable_lcs} we need to check that, given $n < \infty$ and $ \epsilon > 0$, we can find $\delta > 0$ such that if $| z' - z | \leq \delta$, then 
    $$p_{n}\left( (\lambda \mapsto f_{z,z'}(\lambda)) -  (\lambda \mapsto \lambda f^{(1)}(\lambda z)) \right) \leq \epsilon.$$ 
    
    For convenience set $|z - z'| \leq 1$, under this setting, we have that, if $\lambda \in K_n$, then $z' \lambda \in B(0,n(|z| + 1))$. Chose $\delta' > 0$ such that, if $ \omega, \omega' \in \overline{B(0,n(|z| + 1))}$ and $|\omega - \omega'| \leq \delta '$,  then $|f^{(1)}(\omega) - f^{(1)}(\omega ') | \leq \epsilon/n$, under this setting, if we set
    $$ |z - z' | \leq \frac{\delta'}{n}, $$
    we have that
    $$ | \lambda z - \lambda z' | \leq |\lambda| |z - z'| \leq \delta ',  $$
    which guaranties that, for all $\lambda \in K_n$ the following holds 
    $$|f^{(1)}(z \lambda) - f^{(1)}(z' \lambda)| \leq \epsilon.$$ 
    Given that $\overline{B(0,n(|z| + 1))}$ is a convex set of $\mathbb{C}$, the line connecting $\lambda z$ with $\lambda z'$ lies inside $\overline{B(0,n(|z| + 1))}$, therefore, if $|z - z'| \leq \delta / n '$ we have that for any $0 \leq c \leq 1$ the following holds,
    $$|f^{(1)}(c z \lambda + (1-c) z' \lambda) - f^{(1)}(z \lambda)| \leq \epsilon/n.$$
    The fundamental theorem of calculus for complex valued functions tells us that,
    $$  \lambda f^{(1)}(\lambda z) = \lambda  \int_{\lambda z'}^{\lambda z} \frac{f^{(1)}(\lambda z)}{\lambda z - \lambda z'} d \omega, $$
    therefore,
    $$ \left| \lambda \frac{\int_{\lambda z'}^{\lambda z} f^{(1)}(\omega) d \omega }{\lambda z - \lambda z'} - \lambda f^{(1)}(\lambda z) \right| \leq \left( \sup_{\lambda \in K_n} |\lambda| \right)  \frac{\int_{\lambda z'}^{\lambda z} | f^{(1)}(\omega) - f^{(1)}(\lambda z) |d \omega }{ |\lambda z - \lambda z'|} .$$
    Since,
    $$ \int_{\lambda z'}^{\lambda z} | f^{(1)}(\omega) - f^{(1)}(\lambda z) |d \omega \leq \left( \sup_{0 \leq c \leq 1} |f^{(1)}(c z \lambda + (1-c) z' \lambda) - f^{(1)}(z \lambda)| \right) |\lambda z - \lambda z'| .$$
    if we use the bound on the values of $| f^{(1)}(\omega) - f^{(1)}(\lambda z) |$ when $\omega$ lies between $\lambda z'$ and $\lambda z$ we get that, if $\lambda \in K_n$, then
    $$ \left| \lambda \frac{\int_{\lambda z'}^{\lambda z} f^{(1)}(\omega) d \omega }{\lambda z - \lambda z'} - \lambda f^{(1)}(\lambda z) \right| \leq \left( \sup_{\lambda \in K_n} |\lambda| \right) \frac{\epsilon |\lambda z - \lambda z'|}{n} \frac{1}{|\lambda z - \lambda z'|} \leq \epsilon .$$
    The previous statement implies that if $z' \to z$, then $f_{z,z'}(\lambda) \to \lambda f^{(1)}(\lambda z)$ uniformly on $K_n$, under this setting, \cref{proposition:convergence_metrizable_lcs} tells us that $f_{z,z'}(\lambda) \to \lambda f^{(1)}(\lambda z)$ as elements of  $\text{Hol}(\mathbb{C})$, hence,
    $$ \frac{d f(z'a)}{ dz'} |_{z} = \frac{1}{2 \pi \mathrm{i}} \int_\gamma \lambda f^{(1)}(\lambda z) (\lambda 1-a)^{-1} \mathrm{~d} \lambda.$$
    We know that $f^{(1)} \circ g_z$ is holomorphic over $\mathbb{C}$, also, the function $\text{id}(\lambda) = \lambda$ is holomorphic over $\mathbb{C}$, from \cref{theorem:holomorphic_functional_calculus_banach_algebras} we know that $\text{id}(a) = a$, so, given that the holomorphic functional calculus over $A$ is an homomorphism of algebras (\cref{theorem:holomorphic_functional_calculus_banach_algebras}), we get that $ (\text{id} f^{(1)} \circ g_z)(a) = a f^{(1)}(a), $ which implies that 
    $$ \frac{d f(z'a)}{ dz'} |_{z} = a f^{(1)}(za). $$
    Notice that the map $z \mapsto a f^{(1)}(za)$ is continuous because $f^{(1)}$ is a continuous function over $\mathbb{C}$. Since the function $\lambda \mapsto \lambda f^{(1)}(\lambda z)$ is holomorphic on $\mathbb{C}$, we can iterate the previous argument to show that
    $$ \frac{d f(z'a)}{ dz'} |_{z} = \frac{1}{2 \pi \mathrm{i}} \int_\gamma \lambda^n f^{(n)}(\lambda z) (\lambda 1-a)^{-1} \mathrm{~d} \lambda = a^{n} f^{(n)}(za),$$
    which implies that the function $z \mapsto f(za)$ is smooth.
\end{itemize}
\end{proof}

\begin{remark}[Homotopy from exponential function]\label{remark:homotopy_from_exponential_functions}
Using \cref{proposition:smoomth_paths_from_entire_functions} if we take $f(x) = \exp(x)$ we have smooth homotopies of invertible elements of the form
$$ t \to \exp(ta), \; t \in [0,1], \; a \in A ,$$
henceforth, $\exp(A) \subset G_0(A)$, where $G_0(A)$\index{$G_0(A)$} is the connected component of the identity of the topological group $G(A)$.
If we were to be more careful with the conditions on \cref{proposition:smoomth_paths_from_entire_functions} we could deal with homotopies that arise from holomorhic functions in subsets of $\mathbb{C}$, for example, if we restrict to $z \in \Pi$ we could have come up with a smooth maps
$$ \gamma_{f,a} : (0,\infty) \to A, \; z \mapsto f(za), \; \text{ when } \text{Sp}(a) \subset \Pi. $$
This map is not relevant for our analysis, however, you may encounter it in the literature, for example, it is used as a key step in the study of the semi-group of homotopic projections of a C* algebra in \citep[Lemma 3.43]{gracia-bondia_elements_2001} with $f(z) = z^{1/2}$ (the principal branch of the complex square root).
\end{remark}

All of this study of the analytical structure of a Banach algebra will provide a nice description of the topological group $G(A)$ when $A$ has a unit, a description that will be fundamental for the study of K Theory of C* algebras from a topological point of view.

\begin{theorem}[Description of $G(A)$ (Theorem 4.105 \citep{allan_introduction_2011})]\label{theorem: description_of_GA}
Let $A$ be an unital Banach algebra, then:

\begin{itemize}
    \item $G_0(A)$ is an open-and-closed, normal subgroup of $G(A)$, and the connected components of $G(A)$ are precisely the cosets of $G_0(A)$ in $G(A)$
    \item $G_0(A)$ consists of all finite products of exponentials, so that
$$
G_0(A)=\left\{\exp \left(a_1\right) \exp \left(a_2\right) \cdots \exp \left(a_k\right): a_1, \ldots, a_k \in A, k \in \mathbb{N}\right\} ;
$$
\item in the case where $A$ is commutative, $G_0(A)=\exp(A)$
\end{itemize}
\end{theorem}

\begin{proof}
We denote by $G_0(A)$ the set of all invertibles of $A$ that are homotopic to $1_A$ through a path on invertibles, and we set $G_0 = G_0(A)$, also $G = G(A)$.
We are going to walk you through the proof given in \citep[Theorem 4.105]{allan_introduction_2011}, so before we start, we must state an important fact about $\exp(A)$, for $a \in G(A)$ and $b\in A$ such that $\|b -a \| \leq \| a^{-1} \|^{-1} $ we have that $a^{-1}b \in \exp(A)$, because
$$\| a^{-1} b - 1_A \| = \| a^{-1}(b -a) \| \leq \|a^{-1}\| \| b -a \| \leq 1  $$
and \cref{proposition:log_of_element_close_to_identity} tell us that $a^{-1}b \in \exp(A)$. Notice that the previous claim is a good way of characterizing close enough elements of $G(A)$ as $b = a \exp(\alpha)$ for $\alpha \in A.$ Denote $G_0 = G_0(A)$ and $G = G(A)$, then:
\begin{itemize}
    \item Take $a,b \in G_0$, then $b^{-1} \in G_0$ and $ab \in G_0$, thus $G_0$ is a subgruoup of $G$. Take $c \in G$, then $\alpha_c :  a \to c^{-1} a c$ is a continuous map and $\alpha_c (1_A) = 1_A$, which implies that $c^{-1} a c$ is homotopic to $1_A$ in $G$, thus, $G_0$ is a normal subgroup of $G$. \\
    
    Now, let $\beta$ be an accumulation point of $G_0$ with respect to $G$, then for any $a \in G_0$ such that $\| a - \beta \| \leq \| \beta^{-1} \|^{-1}$ we have that 
    $$\| 1_A - a \beta^{-1} \| \leq \| \beta^{-1}\| \| a - \beta \| \leq 1,$$
    therefore $a \in D$ and \cref{proposition:log_of_element_close_to_identity} tell us that there is $b \in A$ such that $a \beta^{-1} = \exp(b)$, thus $\beta = a \exp(-b)$. We now from \cref{remark:homotopy_from_exponential_functions} that $\exp(-b) \in G_0$, then we have that $\beta \in G_0$, that is, $G_0$ is closed with respect to $G$. Using a similar argument we can show that for any $a \in G_0$ and $b \in G_0$ with $\|b -a \| \leq \| a^{-1} \|^{-1} $ we have that $b \in G_0$, meaning that $G_0$ is open with respect to G.
    
    \item Let 
    $$E = \left\{\exp \left(a_1\right) \exp \left(a_2\right) \cdots \exp \left(a_k\right): a_1, \ldots, a_k \in A, k \in \mathbb{N}\right\},$$
    then $E$ is a subgroup of $G$ and $E \in G_0$. Take $a = \exp \left(a_1\right) \exp \left(a_2\right) \cdots \exp \left(a_k\right)$ then if $\| b - a \| \leq \| a ^{-1} \|^{-1}$ there is $c \in A$ such that $b = a \exp(c)$, thus $b \in E$ and $E$ is open. Also, every left coset of $E$ is open because the mapping $a \to ab$ is a homeomorphism when $b \in E$, which implies that $G \setminus E$ is open, hence $E$ is closed in $G$. So, $E$ is a non-empty open and closed subset of $G_0$, which leaves no other alternative that $G_0 = E$.\\
    
    We end up having that every coset of $G_0$ in $G$ is a connected open and closed subset of $G$, and are the topological components of $G$.
    \item If $A$ is commutative then 
    $$ \exp \left(a_1\right) \exp \left(a_2\right) \cdots \exp \left(a_k\right) = \exp \left( a_1 + a_2 + \dots + a_k \right) $$
    by \cref{proposition:logarithms_and_n_roots}, therefore $G_0 = \exp(A)$.
\end{itemize}

\end{proof}

\begin{remark}[Leibniz rule for functions taking values in topological algebras\index{Leibniz rule}]\label{remark:leibniz_rule_fn_values_in_topological_algebra}
The term topological algebra refers to an algebra that is a topological space where the multiplication and addition are continuous. Assume that $f,g : \mathbb{R} \to A$ are differentiable functions, then, the arguments of calculus for $\mathbb{R}$ valued functions can be translated into this context to show that both $g,f$ are continuous functions (\citep[Chapter 9, Theorem 1]{spivak_calculus_1994}). Under this setting we have that,
$$
\begin{aligned}
\frac{d (fg) }{dt}(x) & =\lim _{\Delta x \rightarrow 0} \frac{(fg)(x+\Delta x)-(fg)(x)}{\Delta x} \\
& =\lim _{\Delta x \rightarrow 0} \frac{f(x+\Delta x) g(x+\Delta x)-f(x) g(x)}{\Delta x} \\
& =\lim _{\Delta x \rightarrow 0} \frac{f(x+\Delta x) g(x+\Delta x)-f(x) g(x+\Delta x)+f(x) g(x+\Delta x)-f(x) g(x)}{\Delta x} \\
& =\lim _{\Delta x \rightarrow 0} \frac{[f(x+\Delta x)-f(x)] \cdot g(x+\Delta x)+f(x) \cdot[g(x+\Delta x)-g(x)]}{\Delta x} \\
& =\lim _{\Delta x \rightarrow 0} \frac{f(x+\Delta x)-f(x)}{\Delta x} \cdot \lim _{\Delta x \rightarrow 0} g(x+\Delta x) + \\
& \lim _{\Delta x \rightarrow 0} f(x) \cdot \lim _{\Delta x \rightarrow 0} \frac{g(x+\Delta x)-g(x)}{\Delta x} \\
& = \frac{df}{dt}(x) g(x)+f(x) \frac{dg}{dt}(x) ,
\end{aligned}
$$
The previous calculation implies that $fg$ is a differentiable function and the Leibniz rule is satisfied.
\end{remark}

\begin{remark}[Canonical homotopies between invertibles in Banach algebras]\label{remark:canonical_homotopies_between_invertibles_Banach_algebras}
From \cref{theorem: description_of_GA} we know that the connected components of the group $G(A)$ look like 
$$b (\exp \left(a_1\right) \exp \left(a_2\right) \cdots \exp \left(a_k\right))$$ 
with $a_1, \ldots, a_k \in A, k \in \mathbb{N}$ where $b$ is a representative of a connected component of $G(A)$, therefore, if two invertibles $x,y \in G(A)$ are homotopic we can construct a canonical path between them as 
$$ \gamma(t) = x (\exp \left(t a_1\right) \exp \left(t a_2\right) \cdots \exp \left(t a_k\right)), \; \gamma(0) = x , \; \gamma(1) = y.$$
The path $\gamma$ is a multiplication of continuous paths by \cref{remark:homotopy_from_exponential_functions}, therefore is continuous i.e. a homotopy. Additionally, the Lebniz rule for functions taking values in topological algebras (\cref{remark:leibniz_rule_fn_values_in_topological_algebra}) implies that $\gamma$ is a differentiable path since it is the multiplication of differentiable paths.

The specific form of this smooth paths will be very convenient for the pairing between cyclic cohomology and K theory, because it will also appear in smooth sub algebras.
\end{remark}

We end this section with a clever usage of the holomorphic calculus to come up with idempotents, and create convergent sequences of idempotents, which will be important when we want to relate idempotents of smooth sub algebras with the idempotent of C* algebras in \cref{proposition:idempotents_and_projections_of_smooth_sub_algebras_are_dense}.

\begin{lemma}[Approximating idempotents with idempotents]\label{lemma:approximating_idempotents_with_idempotents}

Let $A$ be a unital Banach algebra and $e$ an idempotent of $A$ i.e. $e \in Q(A)$, then $B = B(0,1/4) \cup B(1,1/4)$ is an open neighbourhood of $\text{Sp}(e)$ (recall $\text{Sp}(e) \subset \{ 0,1 \}$ ), so, let $g(z) = 0$ if $z \in B(0,1/4)$ and $g(z)=1$ if $z \in B(1,1/4)$, then $g$ is holomorphic on $B$ and
$$ e = \frac{1}{ 2 \pi i} \int_{\gamma} g(\lambda)(\lambda 1_A - e)^{-1} d \lambda $$
with $\gamma \subset B$ a contour of $\text{Sp}(a)$.

Also, for every $\epsilon >0$ there is a $\delta > 0$ such that if $\| e - a \| \leq \delta$ then 
$$\| g(a) - g(e) \| \leq \epsilon $$
and $g(a)$ is an idempotent.
\end{lemma}
\begin{proof}

Let $A$ be a unital Banach algebra and $a \in A$ with $a^2 =a$, then $\text{Sp}(a) \subset \{ 0,1\}$ and following the calculations on \citep{705540} we know that if $\lambda \notin \text{Sp}(a)$ then
$$ (\lambda 1_A - a)^{-1} = \frac{1}{1-\lambda} a-\frac{1}{\lambda}(1-a). $$
Let $B = B(0,1/4) \cup B(1,1/4)$, then $B$ is an open neighbourhood of $\text{Sp}(a)$ and $\gamma = \gamma_{\alpha,0} \cup \gamma_{1,\alpha}$ is a contour of $\text{Sp}(a)$ with 
$$ \gamma_{\alpha,0}[t] = \alpha \exp(it),  \;  \gamma_{\alpha,1}[t] = 1 + \alpha \exp(it), \; t \in [0, 2 \pi], \;0<\alpha<1/4. $$
Therefore, the function $g(z) = 0$ if $z \in B(0,1/4)$ and $g(z) = 1$ if $z \in B(1,1/4)$ is holomorphic on $B$ and the residue theorem together with Cauchy integral theorem for complex integrals tell us that
$$ a = a \left( \frac{1}{2 \pi i} \int_{\gamma_{\alpha,1}}(1 - \lambda)^{-1} d\lambda \right) =  \frac{1}{2 \pi i} \int_{\gamma_{\alpha,1}} g(\lambda) (\lambda 1_A - a)^{-1} =  \frac{1}{2 \pi i} \int_{\gamma} g(\lambda)(\lambda1_A -a)^{-1} d \lambda$$
which implies that $a = g(a)$.

Following \cref{remark:spectrum_upper_semi_continuity_and_function_evaluation} the upper semi-continuity of the map $a \mapsto Sp(a)$ tells us that there is a $\delta>0$ such that, if $\| b-a\| \leq \delta$, then $\text{Sp}(b) \subset B$, and finally \cref{proposition:properties_of_the_holomoprhic_functional_calculus} tell us that given $\epsilon >0$ there is $0<\delta_1 < \delta$ such that if $\| b -a \| \leq \delta_1$ then $\|g(b) - g(a) \| \leq \epsilon$. The holomorphic calculus also tells us that
$$ g(b)g(b) = (g^2)(b) = g(b)$$
therefore $g(b)$ is an idempotent.
\end{proof}

\chapter{Hilbert spaces}
\label{chapter:hilbert_spaces_section}

Hilbert spaces are a key tool of non-commutative geometry, in particular of operator algebras, because more often than none those algebras come as a sub algebra of $B(H)$, for example, Von Neumman algebras are sub algebras of $B(H)$ that are close under the weak operator topology, and C* algebras come as sub algebras of $B(H)$ that are close under the operator norm topology (\cref{theorem:faithfull_universal_representation_c_star_algebras}). The study of Hilbert spaces is by no means a small task, for example, the study of representation of Von Neumann algebras on Direct integrals of Hilbert spaces was an essential part of the \href{https://www.mathunion.org/imu-awards/fields-medal/fields-medals-1982}{Fields medal given to Connes in 1982}.

\section{Direct sums  of Hibert spaces }
\label{sec:direct_sums_of_Hilbert_spaces}
Now, we will introduce you to arguably the most useful Hilbert space to analyze C* algebras, it is the direct sum of Hilbert spaces, and is a simpler relative of the direct integral of Hilbert spaces (\cref{sec:direct_integral}).

\begin{definition}[Direct sum of Hilbert spaces \index{Hilbert space!direct sum} (Section 2.6 page 123 \citep{kadison_fundamentals_1983_V1})]\label{defintion:direct_sum_hilbert_spaces}

Let $H_{\lambda} \; (\lambda \in \Lambda)$ be a family of Hilbert spaces, finite or infinite and assume that $\mu$ is the counting  measure over $\Lambda$, then, define $\sum_{\lambda \in \Lambda} H_{\lambda} := \int_{\Lambda} H_{\lambda} d\mu(\lambda)$.  $\sum_{\lambda \in \Lambda} H_{\lambda}$, or equivalently 
$$ \sum_{\lambda \in \Lambda} H_{\lambda}  = \left\{\left(x_\lambda\right) \in \prod_{\lambda \in \Lambda} H_\lambda: \sum_\lambda\left\|x_\lambda\right\|_\lambda^2<\infty\right\}$$
is a Hilbert space, where the addition and scalar multiplication given coordinate-wise. The inner product on $\sum_{\lambda \in \Lambda} H_{\lambda}$ is given by
$$ \langle x, y\rangle=\sum_{\lambda \in \Lambda}\left\langle x_\lambda, y_\lambda\right\rangle_\lambda \quad\left(x=\sum_{\lambda \in\Lambda} x_{\lambda}, y=\sum_{\lambda \in\Lambda} y_{\lambda}\right). $$

Notice that even though $\Lambda$ may have an arbitrary cardinality if $(x_{\lambda}) \in \sum_{\lambda \in \Lambda} H_{\lambda}$, then, the sum $\sum_\lambda\left\|x_\lambda\right\|_\lambda^2$ needs to have at most a countable amount of elements, because if it were uncountable then the norm of the element would be infinite (\citep[Proposition 4.5]{timoney_functional_nodate}).
\end{definition}

Using the Zorn's lemma is possible to proof that any Hilbert space $H$ has an orthonormal basis $B = \{e_i \}_{i \in I}$, also called complete orthonormal systems, whether it is separable or not \citep[Corollary 4.6]{timoney_functional_nodate}, and that basis has the following properties

\begin{lemma}[Properties of orthonormal basis of Hilbert spaces\index{Hilbert space!orthonomal basis} ]\label{lemma:properties_orthonormal_basis_hilbert_spaces}
Let $H$ be a Hilbert space and $B = \{e_i \}_{i \in I}$ an orthonormal basis of $H$, then
\begin{itemize}
    \item $\langle x, e_i \rangle$ is nonzero for at most a countable number of $e_i$ \citep[Proposition 4.5]{timoney_functional_nodate}, as it happened in \cref{example:L_p_sapces_for_counting_measure}.
    \item \textbf{Bessel inequality:}\index{Bessel inequality} If $S \subseteq I$ then $\sum_{\phi \in S} | \langle x, \phi \rangle |^2 \leq \| x \|^2$ \citep[Proposition 4.5]{timoney_functional_nodate}.
    \item \textbf{Generalized Fourier expansion:}\index{Generalized Fourier expansion} $x = \sum_{i \in I} \langle x, e_i \rangle e_i$, \citep[Corollary 4.6]{timoney_functional_nodate}.
    \item \textbf{Bessel identity:}\index{Bessel identity} $\sum_{i \in I} | \langle x, e_i \rangle |^2 = \|x\|^2$ \citep[Corollary 4.8]{timoney_functional_nodate}.
    \item \textbf{Parseval identity:}\index{Parseval identity} $\langle x,y \rangle = \sum_{i \in I} \langle x, e_i \rangle \overline{ \langle y, e_i \rangle}$ \citep[Corollary 4.6]{timoney_functional_nodate}.
\end{itemize}
In the previous statements when infinite sums are presented these are to be understood as the limit taken over the set of countable elements such that $\langle x, e_i \rangle$ is nonzero, for example,
$$x = \lim_{n \to \infty} \sum_{j = 1}^{n} \langle x, e_{i_j} \rangle e_{i_j},$$
and the sum can be provided any ordering \citep[Corollary 4.6]{timoney_functional_nodate}.
\end{lemma}

Using the definition of direct sums of Hilbert spaces and the properties of orthonormal basis of Hilbert spaces we get the following result,

\begin{lemma}[Hilbert spaces as direct sums (cf. Theorem 4.10  \citep{timoney_functional_nodate})]\label{lemma:hilbert_space_as_direct_sums}
Let $H$ be a Hilbert space and $B = \{e_i \}_{i \in I}$ an orthonormal basis of $H$, then the Hilbert space $H$ is isomorphic to $\sum_{i \in I} H_i, \; H_i = \mathbb{C}$, which is isomorphic to $l^2(I)$, and the isomorphism preservers the inner product.
\end{lemma}

Consequently, all Hilbert spaces can be studied as direct sums of Hilbert spaces. You can consult \citep[Section 8.3]{kantorovitz_introduction_2003} for an exposition of the aforementioned results. Given that a Hilbert space is a topological space, we can translate concepts from topological spaces into the realm of Hilbert spaces

\begin{definition}[Separable Hilbert space\index{Hilbert space!separable}]\label{def:separable_hilbert_space}
A Hilbert space is separable if it is separable as a topological space.
\end{definition}

\begin{lemma}[Separability of Hilbert spaces and orthonormal basis (Theorem 4.9 \citep{timoney_functional_nodate}, \citep{1939088})]\label{lemma:separability_Hilbert_spaces_and_orthonormla_basis}
Hilbert space is separable iff it has a countable orthonormal basis.
\end{lemma}

The orthonormal basis $B = \{e_i \}_{i \in I}$ is not unique, actually, any linearly independent set of $H$ whose finite linear combinations are dense in $H$ generates a complete orthonormal basis of $H$ through the Gram-Schmidt construction (\citep[Theorem 4.12]{timoney_functional_nodate}).

\section{Infinite matrices and bounded operators}
\label{sec:infinite_mattrices_and_bounded_operators}

Let $\xi = \left\{\xi_i\right\}_{i \in I}$ be a complete orthonormal basis of $H$, then for $S \in B(H)$ each vector $S(\xi_j)$ has a generalized Fourier expansion as
$$ S(\xi_j) = \sum_{i \in I} s_{i,j} \xi_i $$
with the generalized Fourier coefficients given by
$$ s_{i,j} = \langle S(\xi_j), \xi_i \rangle. $$ 
Therefore, to $S$ we associate a complex matrix $[ s_{i,j}]_{i,j \in I}$ relative to the complete orthonormal basis $\xi$.

Take $x = \sum_{i \in I} \langle x, \xi_i \rangle \xi_i$, then 
$$S(x) = \sum_{i \in I} \langle x, \xi_i \rangle S(\xi_i) = \sum_{i \in I} \langle x, \xi_i \rangle (\sum_{k \in I} s_{k,i} \xi_k) =  \sum_{k \in I} (\sum_{i \in I}  \langle x, \xi_i \rangle s_{k,i} ) \xi_k,$$
which is the generalized Fourier expansion of $S(x)$, thus we only need to know how $S$ acts on $\xi$ to compute how it acts in any element of $H$.

\begin{definition}[Infinite matrix form of a Bounded operator\index{infinite matrix}]\label{def:infinite_matrix_form_of_boundede_operator}
Let $H$ be a Hilbert space and $\xi = \left\{\xi_i\right\}_{i \in I}$ be a complete orthonormal basis of $H$, take $S$ a bounded linear operator over $H$, we call the ordered set of values $[ \langle S(\xi_j), \xi_i \rangle]_{i,j \in I}$ the matrix representation of $S$.
\end{definition}

\begin{proposition}[Properties of bounded operators as infinite matrices]\label{proposition:properties_bouiinded_operators_as_infinite_matrices}
Let $H$ be a Hilbert space and $\xi = \left\{\xi_i\right\}_{i \in I}$ be a complete orthonormal basis of $H$, let $S, T \in B(H)$ with $[ s_{i,j}]_{i,j \in I}$ the matrix representation of $S$ and  $[ t_{i,j}]_{i,j \in I}$ the matrix representation of $T$, then:
\begin{enumerate}
    \item $S(x) = \sum_{i \in I} \langle x, \xi_i \rangle S(\xi_i) = \sum_{i \in I} \langle x, \xi_i \rangle (\sum_{k \in I} s_{k,i} \xi_k) =  \sum_{k \in I} (\sum_{i \in I}  \langle x, \xi_i \rangle s_{k,i} ) \xi_k$.
    \item $ \| S(\xi_j) \|^2 = \sum_{i \in I} | s_{i,j} |^2 \leq \| S \|, \; \|S^*(\xi_i) \|^2 = \sum_{j \in I} | s_{i,j} |^2 \leq \| S \|$.
    \item $ S + T = [s_{i,j} +  t_{i,j}]_{i,j \in I}$.
    \item If $S^* = [a_{i,j}]_{i,j \in I}$ then $a_{i,j} = s_{j,i}^*$. 
    \item $ \sum_{c \in I} |s_{i,c} t_{c,j}| \leq \left( \sum_{c \in I} |s_{i,c}|^2 \right)^{1/2} \left( \sum_{c \in I} |t_{c,j}|^2 \right)^{1/2} = \| (s_{i,c})_{i \in I} \|_{l^2(I)} \| (t_{c,j})_{j \in I} \|_{l^2(I)}$.
    \item $ST = [ \sum_{c \in I} s_{i,c} t_{c,j} ]_{i,j \in I} $.
\end{enumerate}
\end{proposition}
\begin{proof}
\begin{enumerate}
    \item This is a consequence of the properties of complete orthonormal basis over Hilbert spaces mentioned in \cref{lemma:properties_orthonormal_basis_hilbert_spaces}. 
    \item This is a consequence of the Bessel identity mentioned in \cref{lemma:properties_orthonormal_basis_hilbert_spaces}.
    \item This is a consequence of the linearity of the inner product.
    \item This is a consequence of the linearity of involution.
    \item This is a consequence of the Holder inequality mentioned in \cref{proposition:useful_inequalities_of_Bochner_L_p_spaces}
    \item This is a consequence of the Parseval identity mentioned in \cref{lemma:properties_orthonormal_basis_hilbert_spaces}.
\end{enumerate}
\end{proof}

Notice that if $U \in B(H)$ is a unitary operator it preserves the inner product, then $U(\xi) = \{ U(\xi) \}_{i \in I}$ is a complete orthonormal basis of $H$. Let $T \in B(H)$, then the matrix representation of $T$ under the orthogonal basis $U(\xi)$ is the same as the matrix representation of the operator $U^* T U$ under the basis $\xi$, so, it may be a different matrix but it still represents the same linear transformation over $H$.

\begin{proposition}[cf. Proposition 2.6.13 \citep{kadison_fundamentals_1983_V1}]\label{proposition:characterization_of_bounded_operators_throught_infinite_matrices}
Let $H$ be a Hilbert space and $\xi = \left\{\xi_i\right\}_{i \in I}$ be a complete orthonormal basis of $H$, then,
\begin{itemize}
    \item Let $[a_{i,j}]_{i,j \in I}$ be a mapping from $I \times I \to \mathbb{C}$, then, $[a_{i,j}]_{i,j \in I}$ is the infinite matrix of a bounded operator $A$ over $H$ iff 
    $$ \sup \{ \| A(J) \| : J \subseteq I, \; J \text{ finite} \} < \infty $$
    with $A(J)$ the operator given by $[a_{i,k}]_{i,k \in J}$ acting over $l^2(J) \simeq \mathbb{C}^{|J|}$
    \item If $A$ is a bounded operator over $H$ with an infinite matrix representation given by $[a_{i,j}]_{i,j \in I}$, then,  $\| A(J) \| \leq \| T(L)\|$ for $J \subseteq L$ with both $J,L$ finite, where $A(J)$ is the operator given by $[a_{i,k}]_{i,k \in J}$ acting over $l^2(J) \simeq \mathbb{C}^{|J|}$. Also, $|a_{i,j} | \leq  \| A \| \leq \sum_{i,j \in I} |a_{i,j}|$.
\end{itemize}
\end{proposition}

\section{Tensor product of Hilbert spaces}
\label{sec:tensor_product_of_hilbert_spaces}

\begin{definition}[Tensor product Hilbert spaces\index{Hilbert space!tensor product} ( Remark 2.6.7 \citep{kadison_fundamentals_1983_V1})]\label{definition:tensor_product_hilbert_spaces}

Let $H_1$, $H_2$ be two Hilbert spaces, then $H_1 \otimes H_2$ is the unique Hilbert space obtained as the completion of $H_1 \odot H_2$ with respect to the unique inner product on $H_1 \odot H_2$ that satisfies 
$$ \langle x_1 \otimes y_1 , x_2 \otimes y_2 \rangle_{H_1 \otimes H_2} = \langle x_1, x_2 \rangle_{H_1} \langle y_1, y_2 \rangle_{H_2}, \; x_i \in H_1, \; y_i \in H_2. $$
\end{definition}

The tensor product of Hilbert spaces has many interesting properties, to mention a few

\begin{proposition}[Properties of tensor product of Hilbert space]\label{proposition:properties_tensor_product_Hilbert_spaces}

Let $H_1, H_2, H_3$ be two Hilbert spaces, then

\begin{itemize}
    \item \textbf{Commutativity:} $H_1 \otimes H_2 \simeq H_2 \otimes H_1$.
    \item \textbf{Associativity:} $(H_1 \otimes H_2) \otimes H_3 \simeq H_1 \otimes (H_2 \otimes H_3)$ (\citep[Proposition 2.6.5]{kadison_fundamentals_1983_V1}).
    \item \textbf{Basis:} IF $\{ \xi_i \}_{i\in I}$ is an ortonormal basis of $H_1$ and $\{ \eta_j \}_{j \in J}$ is an ortonormal basis of $H_2$ then 
    $$ \{ \xi_i \otimes \eta_j | \; i \in I , \; j \in J \} $$
    is an ortonormal basis of $H_1 \otimes H_2$ (\citep[Theorem 2.6.4]{kadison_fundamentals_1983_V1}).
    \item \textbf{Dimension:} $\dim(H_1 \otimes H_2) = \dim (H_1) \dim (H_2)$ (\citep[Theorem 2.6.4]{kadison_fundamentals_1983_V1}).
    \item \textbf{Direct sum:} Let $H_1, H_2$ be Hilbert spaces and $\{ \xi_i \}_{i \in I}$ be a complete orthonormal base for $H_2$, then $H_2 \simeq l^2(I)$ (\cref{lemma:hilbert_space_as_direct_sums}), set $H_{1,i} = H_1$, then from \citep[Remark 2.6.8]{kadison_fundamentals_1983_V1} we get that $H_1 \otimes H_2 \simeq \sum_{i \in I} H_{i}$ with the isomorphism given by
    $$ \phi: \sum_{i \in I}H_{1,i} \to H_1 \otimes l^2(I), \; \phi(\sum_{i \in I}h_i ) = \sum_{i \in I} \xi_i \otimes h_i, \; h_i \in H_{1,i}. $$
    \item \textbf{Hilbert-Schmidt functionals:} $H_1 \otimes H_2$ is isomorphic to sub space of the space of all Hilbert-Schmidt functionals on $\overline{H_1} \times \overline{H_2}$, where $ \overline{H_i}$ is the conjugate Hilbert space of $H_i$ (\citep[Theorem 2.6.4]{kadison_fundamentals_1983_V1}).
\end{itemize}

\end{proposition}

From the previous proposition we get that the tensor product of Hilbert spaces exists, because it is a sub space of the space of functionals over a Hilbert space. 

Since the elements of a C* algebra can be seen as bounded operators over a Hilbert space, we now look on some properties of operators over tensor products of Hilbert spaces

\begin{proposition}[Operators on tensor product of Hilbert spaces (Proposition 2.6.12 \citep{kadison_fundamentals_1983_V1})]\label{proposition:opeators_tensor_product_hilbert_spaces}

Let $H_1,\;H_2,\;H_3$ be Hilbert spaces and $A_i, B_i \in B(H_i)$, then
\begin{itemize}
    \item \textbf{Existence:} There is a unique operator $A \in B(H_1 \otimes H_2)$ such that
    $$ A(x_1 \otimes x_2) = (A_1 (x_1) ) \otimes (A_2(x_2))$$
    for all $x_i \in H_i$, this operator is denoted by $A_1 \otimes A_2$.
    \item \textbf{Linearity:} 
    $ (A_1 \otimes A_2)(B_1 \otimes B_2) = (A_1 B_1 ) \otimes (A_2 B_2),$
    and $$(A_1 + A_2) \otimes B_1 = (A_1 \otimes B_1) + (A_2 \otimes B_1).$$
    \item \textbf{Involution:} $(A_1 \otimes A_2)^* = (A_1^*) \otimes (A_2)^*$
    \item \textbf{Norm:} $\| A_1 \otimes A_2 \| = \| A_1 \| \| A_2 \|$
    \item \textbf{Associativity:} $(A_1 \otimes A_2) \otimes A_3 = A_1 \otimes (A_2 \otimes A_3)$ as operators over the isomorphic Hilbert spaces $(H_1 \otimes H_2) \otimes H_3 \simeq H_1 \otimes (H_2 \otimes H_3)$.
    \item \textbf{Inclusion:} The mapping 
    $$i: B(H_1) \to B(H_1 \otimes H_2), \; i(A) = A \otimes Id$$
    with $Id$ the identity operator on $H_2$, preserves products, adjoints and norms, thus it is an injective *-homomorphism of C* algebras. In \cref{sec:tensor_product_of_bounded_opeartors_os_infinite_matrices} we will mention how does $A \otimes I$ looks like over a direct sum of Hilbert spaces.
    \item $B(H_1) \odot B(H_2)$ is not dense in $B(H_1 \otimes H_2)$  (\citep[Remark T4.4]{wegge-olsen_k-theory_1993})
\end{itemize}
\end{proposition}

\subsubsection{Tensor product of bounded operators as infinite matrices}
\label{sec:tensor_product_of_bounded_opeartors_os_infinite_matrices}

The results of this section are taken from \citep[Section 2.6, Matrix representations]{kadison_fundamentals_1983_V1}. Following a similar exposition as in \cref{sec:infinite_mattrices_and_bounded_operators}, if $H_1, H_2$ are Hilbert spaces, if $\{ \xi_i \}_{i \in I}$ is a complete orthonormal basis of $H_2$ then from \cref{proposition:properties_tensor_product_Hilbert_spaces} we know that $H_1 \otimes H_2 \simeq \sum_{i \in I} H_{i}$ with the isomorphism given by
    $$ \phi: \sum_{i \in I}H_{1,i} \to H_1 \otimes H_2, \; \phi(\sum_{i \in I}h_i ) = \sum_{i \in I} h_i \otimes \xi_i, \; H_{1,i} = H_1, \; h_i \in H_{1,i} .$$

For $a \in I$ we introduce the bounded linear operators $U_a,\; V_a$,
$$ U_a : H_1 \to \sum_{i \in I}H_{1,i}, \; U_a(x)=  \sum_{i \in I} x_i, \; x_i = x \text{ if } i = a \text{ else } x_i = 0,$$
and 
$$ V_a : \sum_{i \in I} H_{1,i} \to H_1, \;  V_a (\sum_{i \in I} x_i) = x_a.$$
Using $U_a, V_a$, given $T \in B(\sum_{i \in I} H_{1,i})$ we create a matrix $[T_{i,j}]_{i,j \in I}$ with entries $T_{i,j} \in B(H_{1,i})$ as follows
$$ T_{i,j} = V_{i}TU_{j}. $$
Let $x  =  \sum_{i \in I} x_i \in \sum_{i \in I} H_{1,i}$, then $T(x) \in \sum_{i \in I} H_{1,i}$ and 
$$ T(x) = \sum_{i \in I} y_i, \; \text{with } y_i = \sum_{j \in I} T_{i,j}(x_{j}) $$
where the sum giving $y_i$ can take any ordering and have at most a countable amount of terms because $\sum_{i \in I} x_i$ has at most a countable amount of terms (\cref{lemma:properties_orthonormal_basis_hilbert_spaces}). Notice that $T U_j$ is the restriction of $T$ to the sub space of vectors taking the form $\xi_i \otimes h$ with $h \in H_1$, and $V_{i}TU_{j}$ fro $i \in I$ gives us the description of $T(\xi_i \otimes h)$ as an element of $\sum_{i \in I} H_{1,i}$.

\begin{definition}[Infinite matrix form of bounded operators over tensor products]\label{def:infinite_matrix_bounde_operator_tensor_produc}
Let $H_2$ be a Hilbert space with a complete orthonormal basis $\{ \zeta_i \}_{i \in I}$ and $H$ a Hilbert space, take $T \in B(H_2 \otimes H)$, we call the ordered set $[V_i T U_j]_{i,j \in I}$ the matrix representation of $T$.
\end{definition}

\begin{proposition}[cf. Section 2.6, Matrix representations \citep{kadison_fundamentals_1983_V1}]\label{proposition:properties_infinite_matrices_bounded_operator_tensor_product}
Let $H_2$ be a Hilbert space with a complete orthonormal basis $\{ \zeta_i \}_{i \in I}$ and $H$ a Hilbert space, take $T, S \in B(H_2 \otimes H)$ with $[T_{i,j}]_{i,j \in I}$ the matrix representation of $T$ and $[S_{i,j}]_{i,j \in I}$ the matrix representation of $S$, then
\begin{enumerate}
    \item $ T(x) = \sum_{i \in I} y_i, \; \text{with } y_i = \sum_{j \in I} T_{i,j}(x_{j}) $
    \item $(T+S)_{i,j} = T_{i,j} + S_{i,j}$
    \item $T^* = [A_{i,j}]_{i,j \in I}$ then $A_{i,j} = T_{j,i}^*$
    \item $ R_{i,j} = \sum_{c \in I} S_{i,c} T_{c,j} $ with the sum converging in the strong operator topology in $I$ is infinite
\end{enumerate}
\end{proposition}

In \cref{section:convergence_of_the_Fourier_Series_twisted_crossed_products} we look at an example where $R_{i,j} = \sum_{c \in I} S_{i,c} T_{c,j}$ does not holds in the operator norm topology but it holds in the strong operator topology i.e. the sum converges when evaluated in an element of $\sum_{i \in I} H_{1,i}$.

If $A \in B(H)$ then $A \otimes Id_{H_2} \in H_1 \otimes H_2$ is the direct sum of operators $\sum_{i \in I} A$ and has a "diagonal matrix" given by $[\delta_{i,j}A]_{i,j \in I}$. Also, if $S \in B(H_2)$ with matrix representation $[s_{i,j}]_{i,j \in I}$ then $ Id_{H_1} \otimes S$ has a matrix representation given by $[s_{i,j}I_{H_1}]_{i,j \in I}$. Similarly, $A \otimes S$ has the matrix representation $[s_{i,j} ((1 - \delta_{i,j})Id_{H_1} + \delta_{i,j}A )]_{i,j \in I}$.

\begin{proposition}[cf. Proposition 2.6.13 \citep{kadison_fundamentals_1983_V1}]\label{proposition:infinite_matrices_over_tensor_products}
Let $H_2$ be a Hilbert space with an orthonormal basis given by $\{\zeta_i\}_{i \in I}$, then, 
\begin{enumerate}
    \item Let $H$ be a Hilbert space and $[T_{i,j}]_{i,j \in I}$ be a mapping from $I \times I \to B(H)$, then, $[T_{i,j}]_{i,j \in I}$ is the infinite matrix of a bounded operator over $H_2 \times H$ iff
    $$ \sup \{ \| T(J) \| : J \subseteq I, \; J \text{ finite} \} < \infty $$
    with $T(J)$ the operator given by $[T_{i,k}]_{i,k \in J}$ acting over $\sum_{i \in J} H_{1}$.
    \item If $T$ is a bounded operator over $H_2 \times H$ and has an infinite matrix representation given by $[T_{i,k}]_{i,k \in I}$, then, $\| T(J) \| \leq \| T(L)\|$ if $J \subseteq L$ with $J,L$ finite, where $T(J)$ is the operator given by $[T_{i,k}]_{i,k \in J}$ acting over $\sum_{i \in J} H_{1}$. Also, $ \| T_{i,j} \| \leq \| T \| \leq \sum_{i,j} \| T_{i,j} \|$.
\end{enumerate}
\end{proposition}

\subsubsection{$L^2$ Hilbert spaces}
\label{sec:L_2_Hilbert_spaces}

Let $\mathcal{H},\hat{H}$ be a Hilbert spaces and $(S,\mathcal{A},\mu)$ and $(T,\mathcal{B},\nu)$ measure spaces, then from \cref{prop:ismorphism_banch_spaces} and \cref{theprem:Lp_approximation_simple_functions} we have that
\begin{itemize}
    \item $L^{2}(S;L^2(T,\mathcal{H})) \simeq  L^{2}(T;L^2(S,\mathcal{H}))$
    \item $L^{2}(S)\odot L^2(T,\mathcal{H})$ is dense in $L^{2}(S;L^2(T,\mathcal{H}))$ and $L^{2}(T)\odot L^2(S,\mathcal{H})$ is dense in $L^{2}(T;L^2(S,\mathcal{H}))$
\end{itemize}

Now let us dive into the Hilbert space structure of $L^{2}(S;L^2(T,\mathcal{H}))$

\begin{proposition}[Inner product on $L^2(S,\mathcal{H})$]\label{proposition:inner_prodict_L_2_S_H}
Let $H$ be a Hilbert space and $(S,\mathcal{A},\mu)$ a measure space, then, $L^2(S,\mathcal{H})$ is a Hilbert space under the inner product $ \langle f, g\rangle = \int_S \langle f(s), g(s) \rangle_{\mathcal{H}} d \mu (s)$
\end{proposition}
\begin{proof}
From the polarization identity we have that 
$$\langle f(s), g(s) \rangle_{\mathcal{H}} =\frac{1}{4} (\| f(s) + g(s) \|^{2} - \|f(s) - g(s) \|^{2}$$
$$ + i( \| f(s) + ig(s) \|^{2} - \|f(s) - ig(s) \|^{2})),$$
since $s \to \| f(s) \pm (i) g(s) \|$ is strongly $\mu$-measurable by \cref{remark:norm_of_mu_strongly_measurable_functions} then $s \to \langle f(s), g(s) \rangle_{\mathcal{H}}$ is also strongly $\mu$-measurable. Also 
$$| \langle f(s), g(s) \rangle_{\mathcal{H}} | \leq \| f(s) + g(s) \|^{2} + \|f(s) - g(s) \|^{2} + $$
$$ \| f(s) + ig(s) \|^{2} + \|f(s) - ig(s) \|^{2},$$ 
therefore, $\int_S  |\langle f(s), g(s) \rangle_{\mathcal{H}}| d \mu (s) \leq \infty$ , which implies that 
$$\int_S \langle f(s), g(s) \rangle_{\mathcal{H}} d \mu (s) \in\mathbb{C},$$ i.e. the inner product is well defined. Furthermore, $\|f\|^2 = \langle f,f\rangle \geq 0$ thus $\langle f,f \rangle =0$ iff $f=0$ $\mu$-almost everywhere.
\end{proof}

\begin{proposition}\label{proposition:tensor_product_and_L_2_functs}
Let $H$ be a Hilbert space and $(S,\mathcal{A},\mu)$ a measure space, then, 
$$L^{2}(S)\otimes \mathcal{H} \simeq L^{2}(S;\mathcal{H}).$$
\end{proposition}
\begin{proof}
Take $\sum_{i \leq n} f_i (s) \otimes X_i \in L^{2}(S;\mathcal{H})$  and $\sum_{i \leq m} g_i (s) \otimes Y_i \in L^{2}(S;\mathcal{H})$, then 
$$
\begin{aligned}
\langle \sum_{i \leq n} f_i (s) \otimes X_i, \sum_{i \leq m} g_i (s) \otimes Y_i \rangle_{L^{2}(S;\mathcal{H})} & = & \int_S \langle \sum_{i \leq n} f_i (s) X_i , \sum_{i \leq m} g_i (s) Y_i \rangle_{\mathcal{H}} d \mu (s) \\
& = & \sum_{i\leq n, j \leq m} (\int_S f_i(s) \overline{g}_j(s) d \mu (s)) \langle X_i, Y_j \rangle_{\mathcal{H}} \\
& = & \sum_{i\leq n, j \leq m} \langle f_i,g_j \rangle_{L^2(S)} \langle X_i, Y_j\rangle_{\mathcal{H}}
\end{aligned}
$$
Denote by $L^{2}(S)\otimes \mathcal{H}$ the tensor product of Hilbert spaces following \cref{proposition:properties_tensor_product_Hilbert_spaces}, then $L^{2}(S)\otimes \mathcal{H}$ is the completition of $L^{2}(S)\odot \mathcal{H}$ with respect to the unique inner product satisfying 
$$\langle f_1 \otimes X_1 , f_2 \otimes X_2 \rangle_{L^{2}(S)\otimes \mathcal{H}} = \langle f_1,f_2\rangle_{L^{2}(S)} \langle X_1,X_2\rangle_{\mathcal{H}}.$$  
Since $L^{2}(S)\odot \mathcal{H}$ is dense in both $L^{2}(S)\otimes \mathcal{H}$ and $L^{2}(S;\mathcal{H})$ and has the same norm on both Hilbert spaces, we can establish the isomorphism of Hilbert spaces 
$$L^{2}(S)\otimes \mathcal{H} \simeq L^{2}(S;\mathcal{H}).$$ 
\end{proof}

\begin{lemma}\label{lemma:hilbert_space_and_L2_spaces_with_sigma_finite}
Let $H, \hat{H}$ be Hilbert spaces, let $(S,\mathcal{A},\mu)$ and $(T,\mathcal{B},\nu)$ $\sigma$-fintie measure spaces, then, $$L^{2}(S)\otimes L^2(T,\hat{H}) \simeq L^{2}(S;L^2(T,\hat{H})) \simeq L^2(T \times S; \hat{H}) \simeq L^{2}(T;L^2(S,\hat{H})),$$
and
$$L^{2}(S)\otimes L^2(T) \otimes \hat{H} \simeq L^{2}(S)\otimes L^2(T,\hat{H}).$$
\end{lemma}
\begin{proof}
Take if $\mathcal{H} = L^2(T;\hat{H})$, then, the result is a consequence of \cref{proposition:tensor_product_and_L_2_functs} and  \cref{prop:isomorphism_funcitons_over_sum_of_spaces}.
\end{proof}

\begin{proposition}\label{proposition:cauchy_schwarts_inequality}
Let $H$ be a Hilbert space and $(S,\mathcal{A},\mu)$ a measure space, then, $\int_S \langle f(s), g(s) \rangle d \mu(s)$ is absolutely integrable and
$$ | \int_{S} \langle f(s), g(s) \rangle  d\mu(s) | \leq \left( \int_{S} \|f(s) \|^2 d\mu(s)\right)^{1/2} \left( \int_{S} \|g(s) \|^2 d\mu(s)\right)^{1/2}$$
\end{proposition}
\begin{proof}
The Cauchy-Schwartz inequality on $H$ i.e $ |\langle x,y \rangle_{\mathcal{H}} | \leq \| x \|_{\mathcal{H}} \| y \|_{\mathcal{H}}$ and the Holder inequality on $L^2(S)$ (\cref{proposition:useful_inequalities_of_Bochner_L_p_spaces}) implies that
$$ | \int_{S} \langle f(s), g(s) \rangle  d\mu(s) | \leq \int_{S} | \langle f(s), g(s) \rangle | d\mu(s)$$
$$\leq \int_S \| f(s) \| \| g(s) \| d\mu(s) \leq \left( \int_{S} \|f(s) \|^2 d\mu(s)\right)^{1/2} \left( \int_{S} \|g(s) \|^2 d\mu(s)\right)^{1/2},$$
so, not only we have the Cauchy Schwartz inequality on the Hilbert space $L^2(S;H)$ but also $\int_S \langle f(s), g(s) \rangle d \mu(s)$ is absolutely integrable.
\end{proof}

\begin{remark}[Square integrable functions and direct sum]\label{remark:square_integrable_funct_and_direct_sum}
 If $\mu$ is the counting measure over $S$ (\cref{example:L_p_sapces_for_counting_measure}) then $L^2(S;H)$ is the Hilbert space of at most countable sequences of elements taking values in $H$ and indexed by $S$, where the addition and scalar multiplication given coordinatewise, and the inner is given by
$$ \langle x, y\rangle=\sum_{\lambda \in \Lambda}\left\langle x_\lambda, y_\lambda\right\rangle_\lambda \quad\left(x=\sum_{\lambda \in\Lambda} x_{\lambda}, y=\sum_{\lambda \in\Lambda} y_{\lambda}\right), \; x_\lambda, y_\lambda \in H. $$
If $H_{\lambda} = \mathbb{C}$ for all $\lambda \in \Lambda$, then we denote $l^2(\Lambda) :=\sum_{\lambda \in \Lambda} H_{\lambda}$, that is, the Hilbert space of all square summable sequences over $\Lambda$. In this setup, $L^2(S;H)$ corresponds to the direct sum of Hilbert spaces $H$ indexed by $S$ explained in \cref{defintion:direct_sum_hilbert_spaces}.
\end{remark}

\section{Direct integral}       
\label{sec:direct_integral}

Most of the content of this section comes from \citep[Section 14.1]{kadison_fundamentals_1983_V2}.
       
\begin{definition}[Direct integral\index{direct integral} ]\label{def:direct_integral}
Let $X$ be a $\sigma$-compact locally compact space and $\mu$ the completion of a measure over $\mathcal{B}(X)$ such that $\mu(K) < \infty$ if $K$ is compact, let $\{ H_x \}_{x \in X}$ a family of separable Hilbert spaces then a Hilbert space $H$ is the direct integral of $\{ H_x \}_{x \in X}$ over $(X,\mu)$ when for each $u \in H$ there is a function $f: X \to \cup_{x \in X}H_x$ such that $f(x) \in H_x$ and
\begin{itemize}
    \item If $f,g \in H$ then  $x \mapsto \langle f(x), g(x) \rangle_{H_x} $ is Bochner inegrable and
    $$ \langle f,g \rangle_{H} = \int_{X}\langle f(x), g(x) \rangle_{H_x} d\mu(x). $$
    \item If $f: X \to \cup_{x \in X}H_x$ such that $f(x) \in H_x$ and $x \mapsto \langle f(x) , g(x) \rangle_{H_x}$ is Bochner integrable for each $g \in H$ then there is $\hat{f} \in H$ such that $f = \hat{f}$ almost everywhere.
\end{itemize}
We say that  $ H = \int_{X}^{\oplus} H_x d\mu(x)$ is the direct integral decomposition of $H$.
\end{definition}
       
There is an alternative description of a direct integral in terms of coherences of functions taking values in a Hilbert space (\citep[Chapter 2 Section 6]{nielsen_direct_2017}), which is equivalent to the description mentioned (\citep[Lemma 8.2]{nielsen_direct_2017}).
Notice that if $X$ is $\sigma$-compact then $X = \cup_{i \in \mathbb{N}} Y_i$ with $X_n$ compact, so, since the finite union of compact sets is compact we have that $X_n := \cup_{i \leq n} Y_i$ is compact for every $n$. Under this setup $X = \cup_{n \in \mathbb{N}} X_n$ and $X_n \subseteq X_{n+1}$, thus $\mu$ is $\sigma$-finite.

Let $H$ be a separable Hilbert space, $X$ a $\sigma$-compact space and $\mu$ such that $\mu(K) < \infty$ for $K$ compact, then in \cref{sec:L_2_Hilbert_spaces} we have shown that $s \to \langle f(s), g(s) \rangle_{\mathcal{H}}$ is Bochner integrable for $f,g \in L^2(S,\mathcal{H})$, which is the first condition for $L^2(S,\mathcal{H})$ to be a direct integral. 

Since $\mu$ is $\sigma$-finite then we can write $X = \cup_{i \in \mathbb{N}}X_i$ with $X_i$ compact with $\mu(X_i) < \infty$ and $X_i \subseteq X_{i+1}$. Denote, $\hat{X}_n = \cup_{i \leq n} X_i$, then $\hat{X}_n$ is compact because a finite union of compact sets is compact, moreover, take $u: S \to \mathcal{H}$, then $u$ is the pointwise limit of the functions $\mathbf{1}_{\hat{X}_n} u$. Assume that $s \mapsto \langle u(s), g(s) \rangle_{\mathcal{H}}$ is Bochner integrable for every $g \in L^2(S,\mathcal{H})$, for any $y \in \mathcal{H}$ the function $\mathbf{1}_{\hat{X}_n}y \in L^2(S,\mathcal{H})$, thus we have that
$$ \phi_{y,n}(s) = \langle u(s), \mathbf{1}_{\hat{X}_n}(s)y \rangle = \langle \mathbf{1}_{\hat{X}_n}(s) u(s), y \rangle $$
is Bochner integrable. By the Riesz representation theorem (\citep[Theorem 2.53]{allan_introduction_2011}) all the continuous functionals on $\mathcal{H}$ come in the form $\phi(x) = \langle x,y \rangle$ with $y \in \mathcal{H}$, consequently, $\mathbf{1}_{\hat{X}_n} u$ is a $\mu$-weakly measurable function, so, since we are dealing with functions in a separable space and a $\sigma$-finite measure we have that $\mathbf{1}_{\hat{X}_n} u$ is $\mu$-strongly measurable (\cref{remark:what_happend_if_X_is_separable}). Since the $u = \lim_{n \to \infty} \mathbf{1}_{\hat{X}_n} u$, that is, $u$ is a $\mu$-almost everywhere limit of $\mu$-strongly functions, then $u$ is $\mu$-strongly continuous (\cref{proposition:approximating_mu_strongly_measurable_functions}).

Take $f \in L^2(S)$ and define 
$$\hat{f}(s) = \frac{f^*(s) u(s)}{ \| u(s) \|} \text{ if }  u \neq 0 \text{ else } \hat{f}(s) =0,$$
$\hat{f}(s)$ is $\mu$-strongly continuous because is the multiplication of $\mu$-strongly continuous functions, also $\int_{S} \| \hat{f}(s) \|^2 d \mu(s) = \int_{S} \| f(s) \|^2 d \mu(s) < \infty$, thus $\hat{f} \in L^2(S;\mathcal{H})$. So, by hypothesis we must have that 
$$ \psi(s) = \langle u(s), \hat{f}(s) \rangle = \| u(s)\|f(s), $$
is Bochner integrable, or equivalently, $\psi \in L^1(S)$. Since $f$ was arbitrary, then $s \mapsto \|u(s) \|$ can be shown to generate a continuous linear functional on $L^2(S)$, which in turn implies that it belongs to $L^2(S)$ (\citep{61549}), consequently $u \in L^2(G;H)$, that is, we have fulfilled the second condition of the direct integral.

As in \cref{sec:L_2_Hilbert_spaces}, the Cauchy-Shwartz inquality on $H_x$ plus the Holder inequality on $L^2(S)$ implies that 
$$ |\langle f,g \rangle_{H}| \leq \int_{X}|\langle f(x), g(x) \rangle_{H_x}| d\mu(x) \leq \int_{X} \| f(x) \|_{H_x} \|g(x) \|_{H_x} d\mu(x) \leq \| f \|_H^2 \| f \|_H^2, $$
therefore, $\int_{X}\langle f(x), g(x) \rangle_{H_x} d\mu(x)$ is absolutely integrable.

\begin{definition}[Decomposable operator over a direct integral\index{direct integral!decomposable operator} (Definition 14.1.6 \citep{kadison_fundamentals_1983_V2})]\label{def:decomposable_operator_direct_integral}
Let $H$ be the direct integral of $\{ H_x \}_{x \in X}$ over $(X,\mu)$, then, $T \in B(H)$ is said to be decomposable if there is a function $x \mapsto T(x)$ on $X$ such that $T(x) \in B(H_x)$ and, for each $h \in H$, $T(x)(h(x)) = T(h)(x)$ for almost every $x \in X$.
\end{definition}

\chapter{C* algebras}
\label{chapter:Banach_star_algebras_and_C_star_algebras}

Now we will add a new operation to our Banach algebras, this operation is the involution. The involution\index{involution} is an idempotent operation and has the following properties,

\begin{definition}[Involution Section 6.1\citep{allan_introduction_2011}]\label{def:involution}
Let $A$ be an algebra over $\mathbb{C}$. Then an involution on $A$ is a mapping
$$
*: a \mapsto a^*, \quad A \rightarrow A,
$$
such that:

\begin{itemize}
    \item $a^{* *}=a$ for all $a \in A$.
    \item $(\lambda a+\mu b)^*=\bar{\lambda} a^*+\bar{\mu} b^*$ for all $a, b \in A$ and $\lambda, \mu \in \mathbb{C}$
    \item $(a b)^*=b^* a^*$ for all $a, b \in A$ 
\end{itemize}
An algebra with an involution is a *-algebra\index{* algebra}. 
\end{definition}

From the properties of the involution, we get that, $0^*=0$ and, if $A$ has an identity $1$, then $1^*=1$. If $A$ has no identity then the involution on $A^{+}$ is defined as $(a,\lambda)^* = (a^*,\overline{\lambda})$. Also, we will say that a homomorphism $\theta: A \rightarrow B$ between $*$-algebras is a *-homomorphism\index{*-homomorphism} if $(\theta a)^*=\theta\left(a^*\right)(a \in A)$, that is, the homorphism preserves the involution. If we have a *-homomorphism $\theta: A \rightarrow B(H)$ between a Banach algebra $A$ and the C* algebra of bounded operators over a Hilbert space $H$ ($B(H)$\index{$B(H)$}), we use the term *-representation\index{*-representation}. 

The involution is not a priori linked to the topology of the Banach algebra, for example, there are Banach algebras with discontinuous involutions for which there is an element such that $exp(a^*) \neq (exp(a))^*$ \citep[Notes of section 6.1]{allan_introduction_2011} \citep[Thereom 5.6.83]{dales_banach_2000}. This is an undesired behavior, therefore we will restrict to those Banach algebras where the involution is an isometry,

\begin{definition}[Banach $*$-algebra Section 6.1 \citep{allan_introduction_2011}]\label{def:Banach_star_algebra}
The algebra $(A ; *)$ is a Banach $*$-algebra\index{Banach * algebra} if $*$ is an isometric involution on $A$, that is, 
$$\| a^* \| = \|a \|, \; \forall a \in A.$$
\end{definition}

In Banach *-algebras we get that $\| a a^* \| \leq \|a\|^2 $. Also, we have that $\text{Sp}(a^*) = (\text{Sp}(a))^*$ where
$$ (\text{Sp}(a))^* = \{ \lambda^* | \lambda \in \text{Sp}(a) \}, $$
and $a$ is invertible iff $a^*$ is invertible, in which case $(a^*)^{-1} = (a^{-1})^*$. Notice that $\rho(a) = \rho(a^*)$.

Holomorphic functional on Banach *-algebras couples nicely with entire functions, 

$$ (f(a))^*  = ( \lim_{n \to \infty} \sum_{i \leq n}f_i(a - \beta 1_A)^i )^* =  \lim_{n \to \infty} \sum_{i \leq n} (f_i(a - \beta 1_A)^i )^* = \lim_{n \to \infty} \sum_{i \leq n} f_i^*(a^* - \beta^* 1_A)^i ,$$

since $g(z) = \lim_{n \to \infty} \sum_{i \leq n} f_i^*(z - \beta^* 1_A)^i$ is an entire function because if has in infinite radius of converge then $(f(a))^* = g(a^*)$, in particular we have that $(\exp{a})^* = \exp{a^*}$.

To mention a few examples of Banach *-algebras, 

\begin{example}[Example Banach *-algebras]\label{example:banach_star_algebras}
\begin{itemize}
    \item $\mathbb{C}$ with the involution given by the adjoint
    \item Let $\Gamma$ a commutative locally compact group with a Haar measure $\mu$ (\citep[Theorem 2.1]{vinroot_haar_2008}) such that $\mu(A) = \mu(A^{-1})$ for $A \subset \Gamma$ measurable, the $L^1(\Gamma)$ is a Banach *-algebra with the involution given by $a: \Gamma \to \mathbb{C}, \; a^*(g) = \overline{a(g^{-1})}$. In \cref{example:banach_algebras} we explain how to make $L^1(\Gamma)$ into a Banach algebra, thus we only need to show that $\| a \| = \| a^* \|$, which can be seen from 
    $$\| a^* \| = \int_{\Gamma} |\overline{a(g^{-1})}| dg = \int_{\Gamma}  | \overline{a(g)}| dg,$$  
    that is a consequence of asking that $\mu(A) = \mu(A^{-1})$ for $A$ measurable. If $G$ is an abelian group then $\mu(A) = \mu(A^{-1})$ (\cref{remark:convolution_abelian_groups}). 
    We do not necessarily have that $\| a a^{*} \| = \|a \|^{2}$, for example, in $l^1(\mathbb{Z})$ that equality does not hold, moreover, there is no complete norm on $l^1(\mathbb{Z})$ such that $\| a a^{*} \| = \|a \|^{2}$ holds and the topology induced coincides with its Banach algebra topology \citep{95130}.
\end{itemize}

\end{example}

Now let's look at a special type of Banach *-algebras, those were the equality $\| aa^*\| = \| a \|^{2}$ holds, those are termed as C* algebras, 

\begin{definition}[C* algebra (Section 6.4 \citep{allan_introduction_2011})]\label{def:C_star_algebra}
A C* algebra\index{C* algebra} is a Banach *-algebra whose norm satisfies the C* property\index{C* property}:
$$ \| a a^* \| = \|a\|^2 .$$
\end{definition}

\begin{remark}\label{remark:isometry_involution_and_an_interesting_example}
The C* property is stronger than the involution being an isometry, because $\| a a^* \| = \|a\|^2$ implies $\|a \| =\|a^*\|$ \citep[section 6.4]{allan_introduction_2011}, while the converse statement does not holds, for example, $l^1 (\mathbb{Z})$ is a commutative Banach *-algebra that fails to be a C* algebra \citep{95130}.
\end{remark}

Given that a C* algebra is a topological space, is possible to translate concepts from the real of topological spaces into the realm of C* algebras,

\begin{definition}[Separable C* algebra (page 1 of \citep{rordam_introduction_2000})]\label{def:separable_c_star_algebra}
Let A be a C* algebra, then, A is called to be separable\index{C* algebra!separable} if it is separable as a topological space.
\end{definition}

Now we will look at some examples of C* algebras,

\begin{example}[Examples of C* algebra]\label{example:C_star_algebras}

\begin{itemize}
    \item An elementary example of a C* algebra is the algebra of bounded operators over a Hilbert space $H$, that is, $B(H)$, were the involution is given by the adjoint operator \citep{hkbst_6197_bounded_2013}. Moreover, any closed *-algebra of $B(H)$ is a C* algebra.
    \item Another example of C* algebra if $C_0(X)$ were $X$ is a locally compact Hausdorff space, notice that this is a commutative C* algebra and has unit when $X$ is compact. Recall that a locally compact Hausdorff space\index{locally compact Hausdorff space} is a Hausdorff topological space where every point has a neighborhood with compact closure.
\end{itemize}
We will see in \cref{sec:C_star_alg_repres} that every C* algebra is isomorphic to a closed *-algebra of $B(H)$, thus, we will have a nice characterization of all possible C* algebras.
\end{example}

The involution brings new types of elements,

\begin{definition}[Types of elements defined through the involution]\label{definition:elements_defined_through_the_involution}
\begin{itemize}
    \item if $u \in A$ and $u^* u = u u^* =1_A$ then $u$ is called an \textbf{unitary}\index{unitary}, the set of unitary elementes of a Banach *-algebra $A$ is denoted by $U(A)$\index{$U(A)$};
    \item if $x x^* = x^* x$ then $x$ is called \textbf{normal}\index{normal}
    \item if $x = x^*$ then $x$ is called \textbf{self-adjoint}\index{self-adjoint}, the set of self-adjoint elementes of a Banach *-algebra $A$ is denoted by $A_{\text{sa}}$;\index{$A_{\text{sa}}$}
    \item if $x^2 =x$ and $x^* = x$ then $x$ is called a \textbf{projection}\index{projection} i.e. a self-adjoint idempotent, the set of projections of a Banach *-algebra $A$ is denoted by $P(A)$\index{$P(A)$};
\end{itemize}
\end{definition}

We say that a C* algebra is an unital C* algebra\index{C* algebra!unital} if it has a unit, otherwise, we say that it is a non-unital C* algebra\index{C* algebra!non-unital}.

\begin{lemma}\label{lemma:espectrum_of_unitaries_c_star_algebra}
Let $A$ be a unital C* algebra and $u$ an unitary element of $A$, then,
\begin{enumerate}
    \item $\| u\| = 1$
    \item If $|\lambda| \neq 1$, then $(u - \lambda 1_A) \in G(A)$
    \item $\text{Sp}(u) \subseteq \mathbb{T}$
\end{enumerate}
\end{lemma}
\begin{proof}
\begin{enumerate}
    \item This is a consequence of the C* property, $\| u^2 \| = \| u^* u \| = \|1\| =1$.
    \item The proof is in \citep{2648203}.
    \item The proof is in \citep{2648203}.
\end{enumerate}
\end{proof}

\begin{lemma}\label{lemma:unitaries_from_exponential_of_self_adjoint}
Let $A$ be a unital C* algebra and let $h$ be a self adjoint element of $A$, then, 
\begin{enumerate}
    \item $(\exp{ih})^*=\exp{-ih}=(\exp{ih})^{-1}$, thus, $\exp(ih)$ is a uniraty of $A$.
    \item The spectrum of $\exp{ih}$ is contained in $\mathbb{T}$
\end{enumerate}
\end{lemma}
\begin{proof}
\begin{enumerate}
    \item Let $h$ be a self adjoint element of $A$, then, we have that 
    $$\exp(ih) = \sum_{k = 0} \frac{(ih)^k}{k!},$$ 
    where the right side is a convergent series of elements of $A$ (\cref{lemma:banach_algebra_holomorphic_func_calculus_entire_functions}). Since the involution is a continuous map in C* algebras (\cref{remark:isometry_involution_and_an_interesting_example}), we have that $(\exp(ih))^* = \exp((ih)^*)$, also, we know that $(ih)^* = -i h$, therefore, $(\exp(ih))^* = \exp(-i h)$, additionally, we know that $\exp(a)^{-1} = \exp(-a)$ (\cref{proposition:logarithms_and_n_roots}), thus, we get that 
    $$ (\exp(ih))^* = \exp(-i h) ,$$
    which means that $\exp(i h)$ is an unitary
    \item This is a consequence of the spectral mapping of the holomorphic functional calculus (\cref{theorem:holomorphic_functional_calculus_banach_algebras}).
\end{enumerate}
\end{proof}

\begin{remark}[Normal, self-adjoint and unitary operators are closed sets]\label{remark:normal_self_adjoint_unitary_are_closed}
In a Banach *-algebra the sets of normal operators, self adjoint and unitary operators operators are all close, because if $a_n \to a$ and $a_n^* a_n = a_n a_n^*$ ($a_n=a_n^*$, $a_n a_n^* = a_n^* a_n =1_A$) then one must have that $a a^* = a^* a$ ($a=a^*$, $a a^* = a^* a =1_A$) because both the multiplication and the involution are continuous operations.
\end{remark}

\begin{definition}[C* homomorphism\index{C* homomorphism}]\label{definition:C_star_homomorphism}

Let $A,B$ be two C* algebras, then a map $\theta : A \to B$ is called a C* homomorphism if:

\begin{itemize}
    \item $\theta$ is a *-algebra homomorphism
    \item $\theta$ is a continuous map between Banach algebras
\end{itemize}

$\theta$ is called a C* isomorphism, or isomorphism\index{C* isomorphism} when there is no ambiguity, if it is a one-to-one C* homomorphism. If $\theta$ is a map from $A$ to $A$ and is an isomorphisms it is called a C* automorphism\index{C* automorphism}.
\end{definition}

The C* homomorphisms are the natural maps between C* algebras. Let $A$ be a C* algebra, then, a representation\index{C* algebra!representation} $T$ of $A$ on a Hilbert space $H$ is a *-algebra homomorphism from $A$ into the C* algebra of bounded operators over a Hilbert space $H$ ($B(H)$), $T$ is said to be faithful\index{C* algebra!faithful representation} if it is injective. Notice that the concept of representation of a C* algebra coincides with the concept of *-representation of a Banach *-algebra when we use C* algebras.

If $A$ is a unital C* algebra, the definitions of resolvent, spectrum and spectral radius are the same as for Banach algebras, and all the results from unital Banach algebras still holds for $A$. If $A$ is non-unital then we face with a problem, because $\|a\| + |\lambda|$ is not a C* norm on $A^{+}$, thus, we will need to look for a C* norm on $A^{+}$, a task that is non trivial. Once we have provided a C* norm on $A^{+}$ such that $A^{+}$ is complete, we proceed to define the spectrum of $a \in A$ as in the case of Banach algebras, that is $\text{Sp}(a)_A = \text{Sp}(a,0)_{A^{+}}$. For the next result will assume that $A^{+}$ is a C* algebra and $\text{Sp}(a)_A = \text{Sp}(a,0)_{A^{+}}$, you can get more information about the C* structure of $A^{+}$ on \cref{sec:unitization_of_C_star_algebras}.

\begin{proposition}[Automatic continuity\index{automatic continuity} and a couple more of results]\label{proposition:automatic_continuity_C_star_algebras}

Let $A,B$ be a C* algebras, then

\begin{itemize}
    \item Let $a \in A$ be normal, then $\|a \| = \rho(a)$ \citep[Lemma 6.17]{allan_introduction_2011}
    \item \textbf{Automatic continuity} Let $\theta : A \to B$ be a *-homomorphism of *-algebras, then $\theta$ is continuous and $\| \theta(a) \| \leq \|a \|, \; a \in A$ \citep[Corollary 6.19]{allan_introduction_2011}
    \item Let $B \subset A$ with $A$ a unital C* algebra with $1_A \in B$ then $\text{Sp}_B (b ) = \text{Sp}_A (b)$ for $b \in B$ \citep[Proposition 6.23]{allan_introduction_2011}
\end{itemize}

\end{proposition}

\begin{remark}\label{remark:representation_C_star_alg_is_continuous}
Representations of C* algebras\index{C* algebra!representation} are continuous due to \cref{proposition:automatic_continuity_C_star_algebras}.    
\end{remark}

This proposition is by no means a small result, it is telling us that the analytical structure of the C* algebras is completely stored in its algebraic structure, but why is that you may ask, well, notice that $(a^* a)^* = a^* a$ thus $a^* a$ is self adjoint for every $a \in A$, then, the C* property tell us that $\| a^* a \| = \| a \|^2$, which ultimately lead us to
$$ \| a \| = \sqrt{\| a^* a \|} = \sqrt{ \rho(a^* a) } = \sqrt{ \sup \{ |\lambda| \text{ s.t. } (\lambda1 - a^* a ) \notin G(A) \}}, $$
and we can recover the norm of any element from a purely algebraic construction. Since the norm of a C* algebra end up being uniquely determined by its algebraic structure, therefore, is unique.

The automatic continuity of *-homomorphisms can be a misleading for newcomers, because you may incorrectly assume that any *-homomorphism densely defined between C* algebras can be extended to a *-homomorphisms on the whole C* algebras, which is not true,

\begin{lemma}[Extending *-homomorphisms into C* algebra homomorphisms]\label{lemma:extending_star_homomorphisms_into_C_star_homomorphisms}

Let $A,B$ be C* algebra and $C \subset A$ a dense *-algebra, suppose that there is a *-algebra homomorphism
$$ \phi: C \to B, $$
then, there is a C*-homomorphism $\psi:A \to B$ such that $\psi|_{C} = \phi$ iff 
$$ \| \phi(c) \| \leq \| c\|, \; \forall c \in C .$$   

\end{lemma}

\begin{proof}

We just need to ask one question, if $c_n \to c$ and $c \notin C$ ($c \in A$) then to which element of $B$ we map $c$? the condition on the norm allows us to compute this element easily. So, assume $\| \phi(c) \| \leq \| c\|, \; \forall c \in C$ then if $c_n \to c$ we have that $\{ c_n \}_{n \in \mathbb{N}}$ is Cauchy sequence, and since $\phi$ is a contraction then $\{ \phi(c_n) \}_{n \in \mathbb{N}}$ must be also a Cauchy sequence, which implies that it converges in $B$, thus set $\psi(c) = \lim_{n \to \infty} \phi(c_n)$.
Notice that $\phi(C)$ is a sub *-algebra of $B$, thus $c_n^* \in C$ and the isometric property of the involution tell us that $c_n \to c$ iff $c_n^* \to c^*$ in $A$, and we have that $\psi(c^*) = \lim_{n \to \infty} \phi(c_n^*) = \lim_{n \to \infty} \phi(c_n)^* = \psi(c)^* $. Using the continuity of the product and addition on both $A$ and $B$ is possible to show that if $a_n \to a$, $c_n \to c$ with $a_n, c_n \in C$ then
\begin{itemize}
    \item $\lim_{n \to \infty} \phi(a_n c_n) = (\lim_{n \to \infty} \phi(a_n))(\lim_{n \to \infty} \phi(c_n))$
    \item $\lim_{n \to \infty} \phi(a_n + c_n) = (\lim_{n \to \infty} \phi(a_n)) + (\lim_{n \to \infty} \phi(c_n))$
    \item $\lim_{n \to \infty} \phi(\lambda a_n) = \lambda \lim_{n \to \infty} \phi(a_n)$
\end{itemize}
therefore, $\phi$ is a *-homomorphism from $A$ to $B$ and as such is automatically continuous.

The converse implication is easier to check, because if there is $c \in C$ such that $\| \phi(c) \| > \| c\|$ and there were a *-homomorphism from $A$ to $B$ that extends $\phi$ that would contradict \cref{proposition:automatic_continuity_C_star_algebras}, therefore such extension must not exists.
\end{proof}

As an interesting side note, there is way of characterizing C* algebras among *-algebras using solely algebraic constructions and one completeness condition,

\begin{remark}[Characterizing C* algebras among *-algebras]\label{remark:characterize_C_star_algebras_among_star_algebras}

We can characterize C* algebras among all *-algebras using algebraic constructions and one analytical condition. We follow \citep{414904_mathoverflow} and \citep{bader_group_2020}. Suppose that $A$ is a unital *-algebra, if $A$ were non unital you can take its algebraic unitiation $A^{+}$ since $A^{+}$ is a C* algebra iff $A$ is a C* algebra (\cref{sec:unitization_of_C_star_algebras}). So, define
$$
A_{+}=\left\{\sum_{i=1}^n x_i^* x_i \mid n \in \mathbb{N}, x_1, \ldots, x_n \in A\right\}
$$\index{$A_{+}$}
which is a convex cone in $A$ \citep[page 4]{bader_group_2020}, and we can define a partial order on $A_+$ by
$$
x \leq y \quad \Longleftrightarrow \quad y-x \in A_{+}.
$$
Now we define a pseudo norm for $x \in A$ as,
$$
\|x\|_{0}=\sqrt{\inf \left\{\alpha \in \mathbb{R}_{+} \mid x^* x \leq \alpha\right\}} \in[0, \infty]
$$
(using the conventions $\inf \emptyset=\infty$ and $\sqrt{\infty}=\infty$ ).

The norm $\| \cdot \|_{0}$ is a C* norm by \citep[Theorem 2]{bader_group_2020}, moreover, as we will see in \cref{sec:cont_func_calc_consec} $a^*a$ is positive for every C* algebra and $\|a\|_{0}^{2}$ is the spectral radius of $a^* a$, so $\| a \| = \| a \|_{0}$ if $A$ is a C* algebra. Now, we characterize a C* algebra in terms of $\|\cdot \|_{0}$, as follows, a *-algebra $A$ is a C* algebra iff these three conditions are satisfied

\begin{itemize}
    \item \textbf{Archimedian property} \citep[Definition 4]{bader_group_2020}:Every element of $A$ is bounded, that is, $\| a\|_{0} < \infty$ for $a \in A$.
    \item \textbf{No non zero infinitesimals} \citep[page 4]{bader_group_2020}: and infinitesimal is an element such that $\| a \|_{0} =0$, so we are asking that $\|a \|_{0} = 0$ iff $a=0$.
    \item \textbf{Completeness}: $A$ is a complete algebra under the norm $\| \cdot \|_{0}$.
\end{itemize}

\end{remark}

Turns out that \cref{proposition:automatic_continuity_C_star_algebras} gives us two more pieces of information, 
\begin{itemize}
    \item A C* algebra has a unique norm because, to chekc this take $\| \cdot \|_1, \; \| \cdot \|_2$ two norms over $A$, then, given that both $id: A \to A$ and $id^{-1} = id : A \to A$ must be norm decreasing we have that $\| a \|_1 \geq \| id(a) \|_2$ and $\| a \|_2 \geq \| id^{-1}(a)\|_1$, which altoguehter implies that $\| a \|_1 = \| a \|_2$.
    \item Let $\phi: A \to B$ a *-homomorphism between C* algebras, then $\text{Ker}(\phi)$ is a sub *-algebra of $A$ and is closed because $\phi$ is continuous, moreover, $\text{Ker}(\phi)$ is a two sided ideal of $A$.
\end{itemize}

The study of ideals in C* algebras brings many useful results,

\begin{proposition}[Ideals on C* algebras]\label{proposition:ideal_on_C_star_algebras}
Let $A$ be a C* algebra and $I \subset A$ a two sided closed ideal, then
\begin{itemize}
    \item $I$ is a sub C* algebra of $A$  \citep[Theorem 3.1.3]{murphy_c-algebras_1990}
    \item $A/I$ becomes a C* algebra under the quotient norm, that is, the following defines a C* norm on $A/I$
$$ \|a + I \| = \inf \{ \| a+x \| : x \in I \},  $$
and $A/I$ is complete under this norm \citep[Theorem 3.1.4]{murphy_c-algebras_1990}
    \item The canonical mapping 
$$ \pi : A \to A/I, \; \pi(a) = a+I $$
becomes a C* homomorphism.
    \item If $\phi: A \to B$ is a C* momorphism then $\psi : A/\text{Ker}(\phi) \to B$ given by $\psi(a + \text{Ker}(\phi) ) = \phi(a)$ is an injective C* homomorphism, and $\phi(A) = \phi(A/\text{Ker}(\phi))$, thus as a consequence of the continuous funcitonal calculus on C* algebras (\cref{proposition:properties_continuous_functional_calculus}) you can show that $\phi(A)$ is a sub C* algebra of $B$ and the mapping $\psi$ is an isometry.
    \item Every C* algebra admits an approximate identity\index{approximate identity} \citep[Theorem 3.1.1]{murphy_c-algebras_1990}, and if $A$ is a separable algebra then it has an approximate identity that is a sequence, that is, there is $\{ a_n \}_{n \in \mathbb{N}} \subset A$ such that $\lim_{n \to \infty} \| b - ba_n \| =0$ for all $b \in A$ \citep[Remark 3.1.1]{murphy_c-algebras_1990}.
    \item Let $B$ and $I$ be respectively a $C^*$-subalgebra and a closed ideal in a $C^*$-algebra $A$. Then $B+I$ is a $C^*$-subalgebra of $A$ \citep[Theorem 3.1.7]{murphy_c-algebras_1990}.
\end{itemize}

\end{proposition}

C* algebras are a special among Banach algebras because their analytical and algebraic structures are tightly bound together, and these will make them the home of one of the most important results on non commutative geometry and arguably the one result that brought it to life, the categorical duality between of locally compact Hausdorff spaces and commutative C* algebras (\cref{section:non_commutative_geometry_dictionary}).

\section{Exact sequences of C* algebras}
\label{sec:exact_sequences_of_C_star_algebras}
A finite (or infinite) sequence of $C^*$-algebras and *-homomorphisms
$$
\cdots \longrightarrow A_n \stackrel{\varphi_n}{\longrightarrow} A_{n+1} \stackrel{\varphi_{n+1}}{\longrightarrow} A_{n+2} \longrightarrow \cdots
$$
is said to be \textbf{exact} if $\operatorname{Im}\left(\varphi_n\right)=\operatorname{Ker}\left(\varphi_{n+1}\right)$ for all $n$. An exact sequence of the form
$$
0 \longrightarrow I \stackrel{\phi}{\longrightarrow} A \stackrel{\psi}{\longrightarrow} B \longrightarrow 0
$$
is called \textbf{short exact}. In a short exact sequence of C* algebras\index{C* algebra!exact sequence} $\phi$ is injective because $\text{Ker}(\phi) = 0$, this implies that $\phi$ is isometric by \cref{proposition:properties_continuous_functional_calculus}, thus $\phi(I) \simeq I$. Also, $\psi$ is surjective, and by \cref{proposition:ideal_on_C_star_algebras} $\hat{\psi}: A/\text{Ker}(\psi) \to B$ is an injective *-homomorphism, therefore $B \simeq A/\text{Ker}(\psi)$. So, we get that any short exact sequence of C* algebras is equivalent to a short exact sequence of the form
$$
0 \longrightarrow I \stackrel{i}{\longrightarrow} A \stackrel{\pi}{\longrightarrow} A/I \longrightarrow 0,
$$
with $I$ a two sided closed ideal of $A$. Consequently, we can understand $A$ as an extension of $B$ by $I$.

\section{Unitization of C* algebras}
\label{sec:unitization_of_C_star_algebras}

We can construct new C* algebras from existing ones, for example, one of the simplest way of creating a new C* algebra is through their sum as vector spaces:

\begin{definition}[Sum of C* algebras\index{C* algebra!sum}]\label{definition:sum_of_C_star_algebras}
Let $A$, $B$ be two C* algebras, then the set $A \times B$ can be provided with a C* norm as follows
$$ \| (a,b)\| = \max \{ \|a\|_{A}, \|b \|_{B} \}, $$
and becomes a C* algebra under the poin-twise multiplication, addition, and involution. This C* algebra is called $A \oplus B$ and is unital iff both $A$ and $B$ are unital \citep[Proposition 2.1.5]{wegge-olsen_k-theory_1993}.
\end{definition}

In the context of C* algebras, the pursuit of unitizations\index{C* algebra!unitization} gets harder than in the case of Banach algebras, because the $l_1$-norm that we have previously defined for a Banach algebra does not define a C* norm in general. When dealing with C* algebras the proper way of assigning a C* norm to $A^+$\index{$A^+$} is though the operator norm, which is motivated by the fact that $A$ acts over itself as multiplication operators i.e. $b \mapsto ab$ with $a,b \in A$. There are many equivalent definitions of the operator norm, for example, is possible to provide an implicit definition of the operator norm using a representation of $A^{+}$ as operators on the Banach algebra $B(A)$ (bounded operators on $A$) \citep[Proposition 2.1.3]{wegge-olsen_k-theory_1993}, on the other side, is possible to explicitly provide a formula for the operator norm using the following formula \citep[page 4]{courtney_notes_nodate}:

$$ \| (a, \omega) \|_{A^{+}}= \sup_{b \in A, \|b\| \leq 1} \|ab + \omega b\|_{A}.$$

The norm $\| \cdot \|_{A^{+}}$ turns $A^{+}$ into a unital C* algebra \citep[Proposition 1.15]{courtney_notes_nodate}, and is unique, because the norm on a C* algebra is unique (\cref{proposition:automatic_continuity_C_star_algebras}), moreover, we have that $\|(a,\omega)\|_{A^{+}} \geq \sup \{ \|a \|_A, |\omega| \}$. Also, $A^{+}$ and $A$ are part of a short exact sequence of C* algebras,

$$
0 \longrightarrow A \stackrel{\iota}{\longrightarrow} A^{+} \underset{\lambda}{\stackrel{\pi}{\rightleftarrows}} \mathbb{C} \longrightarrow 0
$$
 were $\pi: A^{+} \rightarrow \mathbb{C}$ is the quotient mapping ($\pi (a,\omega) = \omega$), and let $\lambda: \mathbb{C} \rightarrow \widetilde{A}$ is defined by $\lambda(\omega)=\omega 1_{A^{+}}$. Notice that the form of the operator norm tell us that $\| (a,0) \|_{A^{+}} = \| a \|_{A}$ since $b = a^* / \|a\|_{A}$ has norm $1$, to check this notice that $\| (a,0) \|_{A^{+}} \leq \|a\|_{A}$ and the supremum is achieved in $b$ due to the C* property of $\| \cdot \|_{A}$. Is also possible to check that $\| (0,\omega) \|_{A^{+}} = | \omega|$. Additionally, $i(A) \simeq A$ because $i$ is an injective *-homomorphism (by \cref{proposition:properties_continuous_functional_calculus}), and $i(A) = \text{Ker}(\pi)$ is a closed two sided ideal of $A^{+}$.
 
 The algebra $A^{+}$ can also be defined for a unital C* algebra, and it has the property that $A$ is unital iff $A^{+} \simeq A \oplus \mathbb{C}$. Suppose $A$ is unital, then $f = 1_{A^{+}} - 1_A = (-1_A,1)$ is a projection in $A^{+}$ and 
 $$ A^{+} = \{ a + \alpha f : a \in A, \alpha \in \mathbb{C}\}, $$
 thus the map $A \oplus \mathbb{C} \to A^{+}$ given by $(a,\alpha) \mapsto a + \alpha f$ is a *-isomorphism of algebras, and becomes a C* isomorphism because there is only one norm in a C* algebra (\cref{proposition:automatic_continuity_C_star_algebras}). In case $A$ is not unital then $f$ cannot be defined, and $A \oplus \mathbb{C}$ has no unit while $A^{+}$ has a unit, therefore those algebras are not isomorphic \citep[Section 1.1.6]{rordam_introduction_2000}. Notice that if $A$ is unital and $a + \alpha f$ is invertible with inverse $b + \beta f$ then $a b = b a = 1_A$ and $\alpha \beta = 1$, because $A^{+} \simeq A \oplus \mathbb{C}$.
 
$A^{+}$ is a functorial construction in the sense that every C* algebra morphism $\alpha: A \to B$ induces a C* algebra morphism $\alpha^{+} : A^{+} \to B^+$ sending $(a,\omega)$ to $(\alpha^{+} (a), \omega)$, and is surjective [injective] iff $\alpha$ is surjective [injective] \citep[Proposition 2.1.7]{wegge-olsen_k-theory_1993}.

\begin{remark}[Multiplier algebras\index{multiplier C* algebras}]\label{remark:multiplier_algebras}
Wegge Olsen defines a unitization of a C* algebra $A$ as finding a unital C* algebras $B$ were $A$ can be embedded as an essential ideal \citep[Definition 2.1.1]{wegge-olsen_k-theory_1993}, thus, there could be more than one unitization of a C* algebra. Among those unitizations two stand out, the one point unitization $A^{+}$ and the multiplier algebra $\mathcal{M}(A)$, where $A^{+}$ is the smallest one and $\mathcal{M}(A)$\index{$\mathcal{M}(A)$} is the biggest one (\citep[Proposition 2.2.14]{wegge-olsen_k-theory_1993}), that is, $A \subset A^+ \subset \mathcal{M}(A)$ (\citep[Proposition 2.2.5]{wegge-olsen_k-theory_1993}). 

In the case of a commutative C* algebra $A$, $A \simeq C_0(\Omega)$ for a locally compact Hausdorff space (\cref{sec:C_star_alg_repres}), and we have that,
\begin{itemize}
    \item $A^{+} \simeq C(\Omega^{+})$, where $\Omega^{+}$ is the one point compatification of $\Omega$ (\citep[Remarks 2.1.8]{wegge-olsen_k-theory_1993}),
    \item $M(A) \simeq C(\beta X)$ where $\beta X$ is the Stone-Čeck compactificaion of $X$ (\citep[Chapter 12]{blackadar_k-theory_2012}), moreover, $\mathcal{M}(A) \simeq C_b(\Omega)$,
\end{itemize}
where $C_b(\Omega)$ is the C* algebra of bounded continuous functions over $\Omega$ (\citep[Examples 2.2.4]{wegge-olsen_k-theory_1993}). This is part of the dictionary between commutative C* algebras and locally compact Hausdorff spaces which we mentioned in the introduction and \cref{section:non_commutative_geometry_dictionary}. Given a faithful representation $\pi : A \to B(H)$, then, $\mathcal{M}(A)$ takes the following form (\citep[Definition 2.2.2]{wegge-olsen_k-theory_1993})
$$ \mathcal{M}(A) := \{ x \in B(H) | x \pi(A) \subseteq \pi(A) \text{ and } \pi(A) x \subseteq A \}, $$
and $\mathcal{M}(A)$ is independent of the particular faithful representation chosen (\citep[Definition 2.2.11]{wegge-olsen_k-theory_1993}).
\end{remark}

We have mentioned that $A^+$ has C* norm under which it is complete, also, we know that the norm on a C* algebra is unique (\cref{proposition:automatic_continuity_C_star_algebras}), therefore, we get that $A^+$ is a sub C* algebra of any unital C* algebra $B$ such that $A \subset B$. 

\begin{lemma}\label{lemma:uniqueness_of_unitization}
We have that,
\begin{itemize}
    \item Let $A$ be a non unital C* algebra and $B$ be a unital C* algebra such that $A \subset B$, then,
    $$ A^+ \simeq \{ (a, \lambda 1_B) | a \in A, \lambda \in \mathbb{C} \}. $$
    \item Let $A$ be a unital C* algebra and $B$ be a unital C* algebra such that $A \subset B$ and $1_B \neq 1_A$, then,
    $$ A \oplus \mathbb{C} \simeq \{ (a, \lambda (1_B - 1_A)) | a \in A, \lambda \in \mathbb{C} \}. $$
\end{itemize}
\end{lemma}
\begin{proof}
\begin{itemize}
    \item $A^+$ and $\{(a, \lambda 1_B) | a \in A, \lambda \in \mathbb{C} \}$ are canonically isomorphic as *algebras, so, since the C* norm is unique on $A^+$, they must be isomorphic as C* algebras.
    \item $A \oplus \mathbb{C}$ and $\{(a, \lambda (1_B - 1_A)) | a \in A, \lambda \in \mathbb{C} \}$ are isomorphic as *algebras, so, since the C* norm is unique on $A \oplus \mathbb{C}$, they must be isomorphic as C* algebras.
\end{itemize}
\end{proof}

 On \cref{section:unitization_universal_C_star_algebra} we will elaborate more on how this result is related to universal C* algebras.

\section{Representations and duality}
\label{sec:C_star_alg_repres}

We have talked about the nice properties about the C* algebras in terms of the relations between their analytical and algebraic structures, but how much these properties are intertwined? Actually pretty much, lo let's see how this develop

\subsection{Gelfand transform}
\label{section:Gelfand_transform}

One of the first results on automatic continuity of the theory of Banach algebras over $\mathbb{C}$ comes in the form characters. A character on a Banach algebra $A$ is a non-zero algebra homomorphism $\phi: A \to \mathbb{C}$, and the set of all characters of $A$ is denoted by $\Phi_A$,

\begin{theorem}[Automatic continuity\index{automatic continuity} of characters (Theorem 4.43 \citep{allan_introduction_2011})]\label{theorem:automatic_continuity_of_characters}
Let $A$ be a Banach algebra, and let $\varphi \in \Phi_A$. Then $\varphi$ is continuous and $\|\varphi\| \leq 1$. Suppose that $A$ is unital. Then $\|\varphi\|=1$.
\end{theorem}

In commutative unital Banach algebras $\Phi_A$ is non-empty and it is in one-to-one correspondence to the set of maximal ideals \citep[Theorem 4.46]{allan_introduction_2011}, also there are nice relations between the spectrum of an element of $A$, $G(A)$ and $\Phi_A$ \citep[Corollary 4.47]{allan_introduction_2011}. When $A=C(K)$, with $K$ a non-empty compact Hausdorff space, we end up with a commutative unital Banach algebra $A$, such that for every $x \in K$ there is the character $\varepsilon_x$, defined by
$$
\varepsilon_x(f)=f(x) \quad(f \in C(K))
$$
i.e. $\varepsilon_x$ is 'evaluation at $x$ ', with corresponding maximal ideal
$$
M_x=\{f \in C(K): f(x)=0\} .
$$
Interestingly, $\varepsilon_x$ are the only characters of $A$ and all the maximal ideals of $A$ take the form $M_x$ \citep[Example 4.48]{allan_introduction_2011}. So, we can see that characters are quite special.

We can provide $\Phi_A$ with a topology that comes directly from the algebraic structure of $A$, that is the final topology given by the homomorphisms $\hat{a} : \Phi_A \to \mathbb{C}, \; \hat{a}(f) = f(a)$ for $a \in A$, which is also known as the weak-* topology on $\Phi_A$, and is denoted as $\sigma(\Phi_A , A)$. The topological space $(\Phi_A; \sigma(\Phi_A , A))$ is called the character space of $A$, and also referred to as spectrum of $A$, for reasons we will see in upcoming paragraphs.

If $A$ is a commutative unital Banach algebra, then $(\Phi_A; \sigma(\Phi_A , A))$ is a compact Hausdorff space \citep[Theorem 4.54]{allan_introduction_2011}, and if $A$ is non unital then $\Phi_{A^{+}}$ can be identified with the one point compactification of $\Phi_A$ which implies that $(\Phi_A; \sigma(\Phi_A , A))$ is a locally compact space, possibly with $\Phi_A = \emptyset$. If we assume $\Phi_A \neq \emptyset$ then for $a \in A$ is possible to define the function $\widehat{a}: \Phi_A \rightarrow \mathbb{C}$ by
$$
\widehat{a}(\varphi)=\varphi(a) \quad\left(\varphi \in \Phi_A\right) ,
$$ 
and the definition of $\sigma(\Phi_A , A)$ tell us that $\widehat{a} \in C_0(\Phi_A)$. This mapping is called the Gelfand tansform\index{Gelfand transform}, and may look innocent but it is a powerful tool,

\begin{theorem}[Gelfand representation theorem\index{Gelfand representation theorem} (Theorem 4.54 \citep{allan_introduction_2011})]\label{theorem:gelfand_representation_theorem}

Let $A$ be a unital Banach algebra. Then the Gelfand transform
$$
\mathcal{G}: a \mapsto \widehat{a}, \quad A \rightarrow C\left(\Phi_A\right)
$$
is a continuous, unital homomorphism. Suppose that $A$ is commutative, and take $a \in A$. Then $a \in G(A)$ if and only if $\widehat{a}(\varphi) \neq 0$ for all $\varphi \in \Phi_A$, and
$$
\operatorname{Sp}_A (a)=\operatorname{Sp}_{C\left(\Phi_A\right)} (\widehat{a})=\left\{\widehat{a}(\varphi): \varphi \in \Phi_A\right\} .
$$
Let $A$ be a Banach algebra without an identity, so that $\Phi_A$ is a locally compact space. Suppose that $\Phi_A \neq \varnothing$. Then $\widehat{a} \in C_0\left(\Phi_A\right)$, and that the map
$$
\mathcal{G}: a \mapsto \widehat{a},(A ;\|\cdot\|) \rightarrow\left(C_0\left(\Phi_A\right) ;|\cdot|_{\Phi_A}\right)
$$
is a continuous homomorphism with $\|\mathcal{G}\| \leq 1$.
\end{theorem}

So, the Gelfand transform allows us to recover the spectrum of elements and it is a contractive (continuous) homomorphism, but is not always injective nor surjective, so we may wonder if it plays any role beyond another tool on the Banach algebras arsenal, well, the answer is positive and it may surprise you: when $A = L^1(G,\mu)$ with $G$ a locally compact abelian group and $\mu$ its Haar measure, amusingly the Gelfand transform on $L^1(G,\mu)$ is the Fourier transform on $L^1(G)$ and that the topology of $\hat{G}$ is the Gelfand topology on $\Phi_{L^1(G)}$ (take a look at \cref{remark:Gelfand_transform_and_Fourier_transform}).

We have gone really deep into the Gelfand transform and we have not talked about C* algebras, well, this is the time, turns out that if $A$ is a commutative C* algebra then $\Phi_A \neq \emptyset$, and

\begin{theorem}[Commutative Gelfand-Naimark theorem (Theorem 6.24 \citep{allan_introduction_2011})]\label{theorem:commutative_gelfand_naimark_theorem}
Let A be a commutative, unital $C^*$-algebra. Then the Gelfand representation of $A$ is an isometric *-isomorphism of $A$ onto $C\left(\Phi_A\right)$.\\
In the case where $A$ is a commutative, non-unital $C^*$-algebra then $\mathcal{G}: A \rightarrow C_0\left(\Phi_A\right)$ is an isometric $*$-isomorphism, i.e. $\mathcal{G}$ is a C* isomorphism.
\end{theorem}

The last piece of the duality between the category of locally compact Hausdorff spaces and the category of C* algebras, which has as functors $C(\cdot)$ and $\Phi_{\cdot}$, comes from the fact that if $A = C_0(\Omega)$ with $\Omega$ a locally compact Hausdorff space then $ \Psi : \Omega \to \Phi_A, \; \Psi(\omega) = \varepsilon_{\omega}$ is a bijective homemorphism of locally compact Hausdorff space that preserves the points at infinity ( \citep[Lemma 4.55]{allan_introduction_2011}, \citep[Theorem II.2.2.6]{blackadar_operator_2006}). In conclusion, we have look at results that establish:
\begin{itemize}
    \item We can capture the whole structure of a commutative C* algebra in a locally compact Hausdorff space space using the Gelfand transform.
    \item We can capture the whole structures locally compact Hausdorff space spaces in C* algebras using the space of characters.
\end{itemize}
These are deep dualities and we will talk more about them in \cref{section:non_commutative_geometry_dictionary}.

\subsection{Gelfand-Naimark theorem}
\label{section:Gelfand_Naimakr_theorem}

Given a C* algebra $A$ the direct sum of Hilbert spaces (\cref{defintion:direct_sum_hilbert_spaces}) is the key in the construction of $H$ such that $A$ is isomorphic to a close sub algebra of $B(H)$, and the missing piece in this construction is what family of Hilbert spaces is used. Here is where another result on automatic continuity comes into play, $H$ is constructed from positive states on $A$. We will work only with unital C* algebras, because if $A$ is not unital then it is a sub C* algebra of $A^{+}$, and if we have an isomorphic of $A^{+}$ into $B(H)$ then we can restrict it to get an isomorphism of $A$ into $B(H)$.

Let $A$ be a $*$-algebra, define the set
$$
A_{+}=\left\{\sum_{\jmath=1}^n a_{\jmath}^* a_{\jmath}: a_1, \ldots, a_n \in A, n \in \mathbb{N}\right\}.
$$
A linear functional $f$ on a $*$-algebra $A$ is said to be \textbf{positive} if
$$
f\left(a^* a\right) \geq 0 \quad(a \in A),
$$
therefore $f(A_{+}) \subset \mathbb{R}^+$. If $A$ has a unit, then $f$ is a \textbf{state}\index{C* algebra!state} if $f(1)=1$, and the \textbf{set of states} is denoted by $S(A)$. The \textbf{norm of a linear functional} $f$ is defined as
$$\| f \| = \sup \{ |f(a)| : a \in A, \| a\| \leq 1 \}$$
which is its operator norm, notice that $f$ is continuous iff $\| f \| \leq \infty$. Having stated this definitions, we can look at an interesting result on the automatic continuity\index{automatic continuity} of states,

\begin{proposition}[Automatic continuity of states]\label{propostion:automatic_continuity_of_states}
Let $A$ be a unital Banach *-algebra, and let $f$ be a positive linear functional on $A$. Then:
\begin{itemize}
    \item $f$ is continuous, with $\|f\|=f(1)$, and $\overline{ f\left(A^{+}\right)} \subset \mathbb{R}^{+}$ \citep[Proposition 6.8]{allan_introduction_2011}.
\end{itemize}

Also, by \citep[Corollary 6.9]{allan_introduction_2011}, we can characterize the space of states $S(A)$ in the following way,  
$$S(A)=\left\{f \in A^*: f\left(A_{+}\right) \subseteq \mathbb{R}^{+},\|f\|=f(1)=1\right\},$$
with $A^*$ the Banach dual of $A$.
\end{proposition}

We have come to the pinnacle and last step of this section, the GNS construction \citep[Section 6.3]{allan_introduction_2011}. If $f$ is a state, then $L_f = \{ a \in A : f(a^* a) = 0 \}$ is a closed left ideal of A (\citep[page 264]{allan_introduction_2011}), so, denote by $\pi_f$ the canonical projection map $\pi_f: A \to A / L_f$, then the following formula
$$ \langle \pi_f (a), \pi_f (b) \rangle_f = f(b^* a) \; a,b \in A $$
defines a inner product on $A / L_f$, and the norm is given by 
$$\| \pi_f (a) \|_f^2 = f(a^* a) \subset \mathbb{R}^+.$$
Denote by $H_f$ the completion of $A / L_f$ with respect to $\| \cdot \|_{f}$, then $H_f$ is a Hilbert space. For each $a \in A$ there is a mapping $T_{f,a} : A / L_f \to A / L_f$ given by $$ T_{f,a} (\pi_f (b)) = \pi_f (ab) , \; b \in A$$
such that the mapping $T_f : A \to B(H_f), \; a \to T_{f,a}$ is a unital *-homomorphism with $\| T_{f,a} \| \leq \| a\|$ and $\text{Ker}(T_f) = L_f$ (\citep[Lemma 6.14]{allan_introduction_2011}).

Let $A$ be a unital $*$-algebra, let $T$ be a representation of $A$ on a Hilbert space $H$, $T$ is said to universal if $T$ is unital and each state on $A$ has the form $a \mapsto\langle T a(x), x\rangle$ for some $x \in H$, with $\|x\|=1$.

\begin{theorem}[Gelfand-Naimark theorem\index{Gelfand Naimark theorem}]\label{theorem:faithfull_universal_representation_c_star_algebras}
Let $A$ be a C* algebra, and set 
$$H = \sum_{f \in S(A)} H_f,$$
then the *-homomorphism
$$ \pi: A \to  B(\sum_{f \in S(A)} H_f)$$
given by $\pi(a)(\sum_{f \in S(A)}x_f) = \sum_{f \in S(A)} T_{f,a}(h_f)$ is a faithfull universal *-representation $T$ of $A$ on $H$ \citep[Theorem 6.47]{allan_introduction_2011} , and 
$$S(A) = \{ f \in A^* : \| f \| = f(1) = 1 \}$$
by \citep[Corollary 6.44]{allan_introduction_2011}. 

If $A$ is separable then there is a separable Hilbert space $H$ where $A$ has a faithful representation \citep[Corollary II.6.4.10]{blackadar_operator_2006}.
\end{theorem} 

The previous result has a generalization into the realm of Banach *-algebras that are *-semisimple \citep[Theorem 6.15]{allan_introduction_2011}. Also, if $A$ is separable then $A$ has a faithful representation in a Hilbert space with a countable ortonormal basis, because for Hilbert spaces having a countable ortonormal basis and being separable are equivalent conditions (\cref{lemma:separability_Hilbert_spaces_and_orthonormla_basis}).


In \cref{sec:infinite_mattrices_and_bounded_operators} we discussed on how elements of $B(H)$ can be understood as "infinite matrices" once a complete orthonormal basis for $H$ is chosen, therefore, C* algebras can be studied as algebras of infinite matrices if we chose a particular orthonormal base for the Hilbert space where they have a faithful representation. If $A$ is a C* algebra that has a faithful representation over $H$, the elements of $a$ may have more than one way of seeing them as infinite matrices since there could be more than one orthonormal basis of $H$, moreover, if we look into the case of $A$ not finitely generated there are cases where $A$ can have faithful representations over Hilbert spaces with different cardinality e.g. countable vs not countable, for an example of this you can refer to \cref{section:algebra_continuous_functions_locally_comp_space} and \cref{section:faithful_representation_of_torus}. This representation of elements of C* algebras as infinite matrices will be a recurrent theme on this document, because most of the C* algebras will be studied through faithful representations and the action of their elements over a fixed orthonormal basis.

\subsection{Non-commutative geometry dictionary}
\label{section:non_commutative_geometry_dictionary}

The Gelfand transform gives the backbone duality of non-commutative geometry, the correspondence of two categories:

\begin{theorem}[Duality between locally compact Hausdorff space spaces and commutative C* algebras (Theorem II.2.2.8 \citep{blackadar_operator_2006})]\label{theorem:duality_LCH_spaces_commutative_S_star_algebras}
The correspondence $(X, *) \leftrightarrow C_{\mathrm{o}}(X \backslash\{*\})$ is a contravariant category equivalence between the category of pointed compact Hausdorff spaces and basepoint-preserving continuous maps and the category of commutative $\mathrm{C}^*$-algebras and ${ }^*$-homomorphisms.
\end{theorem}

Under this duality of categories non-commutative C* algebras can be intuitively though as algebras of functions over \textit{non-commutative locally compact Hausdorff space spaces}, and we would like that if $A$ is a non commutative C* algebra then $\Phi_A$ would be the non-commutative space, however, the range of the characters is a commutative C* algebra ($\mathbb{C}$) which implies that we are potentially loosing information of the non-commutative structure of $A$ if we restrict to the space of characters, for example, if $A = M_n(\mathbb{C})$ then $\Phi_A = \emptyset$ \citep{3930961}. Therefore, we embrace the duality and proceed to study C* algebras instead of locally compact Hausdorff space spaces, such that we provide a dictionary that gives algebraic terms that capture the topological ones as follow

\begin{figure}[H]
\centering
\includegraphics[width=0.7\textwidth]{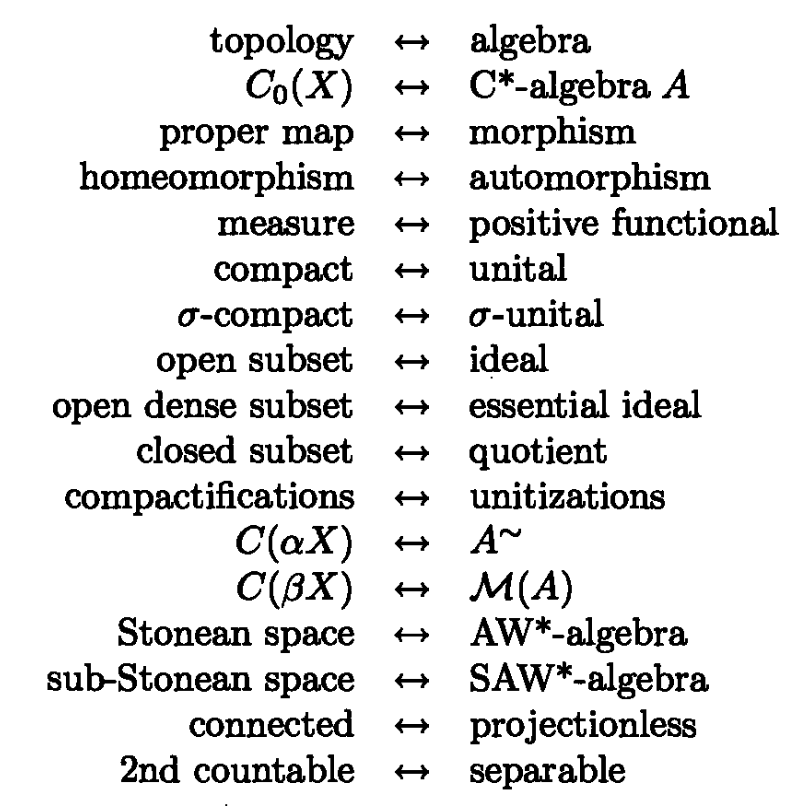}
\caption{List of dualities between topology and algebra for locally compact Hausdorff spaces and commutative C* algebras \citep[page 24]{wegge-olsen_k-theory_1993}.}
\label{fig:topology-algebra_dictionary}
\end{figure}

These are the most common dualities from non-commutative geometry, and by no means the previous list is exhaustive. Behind each duality there are highly non trivial results on analysis and algebra, and those are scattered on the vast bibliography of non-commutative geometry and operator algebras theory, in this document we will encounter some of those dualities and we will expand a little on them when those appear, for example, in \cref{remark:multiplier_algebras} we mention the relation between topological compactifications and unitization of C* algebras.

\begin{example}[Open/close sets vs ideals (Theorem 1.2.3 \citep{rordam_introduction_2000})]\label{example:open_close_sets_vs_ideals}

Let $X$ be a locally compact Hausdorff space space, then if $U \subseteq X$ is open then $U, U^c$ are locally compact Hausdorff space spaces \citep{243170_mathstack}, also, we have that $C_0(U)$ is the algebra of continuous functions over $X$ that decay at infinity and vanish at $U^c$, therefore $C_0(U)$ is an ideal of $C_0(X)$ (by setting $f(x)=0$ for $x \in U^c$ and $f \in C_0(U)$), moreover, any ideal of $C_0(X)$ comes in that form (\citep[Exercise 1.9.1]{putnam_lecture_nodate}). There is *-homomorphism $\pi: C_0(X) \rightarrow C_0\left(U^{\mathrm{c}}\right)$ given by restriction $\left.f \mapsto f\right|_{U^c}$ which is surjective, such that if $f = g+h$ with $h \in C_0(U)$ then $\phi(f) = \phi(g)$ and we end up with a short exact sequence of C* algebras:
$$
0 \longrightarrow C_0(U) \longrightarrow C_0(X) \longrightarrow C_0\left(U^{\mathrm{c}}\right) \longrightarrow 0 .
$$
The map $C_0(U) \rightarrow C_0(X)$ is given by extending a function $f$ in $C_0(U)$ to $X$ by giving it the value $0$ on $U^{\mathrm{c}}$.

\end{example}

The analysis realm also has its space in the non-commutative geometry, for example, integration gets its generalization in the form of linear functionals, which is motivated by the following results:
\begin{itemize}
    \item The Riesz representation theorem\index{Riesz representation theorem}  \citep[Theorem 11.1]{knapp_basic_2005} gives a one to one correspondence between positive functionals on $C_0(\Omega)$ and bounded Radon measures i.e. bounded regular Borel measures (\cref{definition:Radon_measure}) on $\Omega$. To avoid confusions, the term Borel measure will mean a measure defined on the Borel sigma algebra that is locally finite, that is, $\mu(K)\leq \infty$ for all $K$ compact.
    \item If $K$ is a compact Hausdorff space there is a one to correspondence between linear functionals on $C(K)$ and the space of regular Borel complex (locally finite) measures on $K$ \citep[Theorem 11.28]{knapp_basic_2005}
\end{itemize}
Therefore, we consider \textbf{linear functionals on C* algebras as generalizations of integration of functions against complex measures}, and in particular \textbf{positive linear functionals are generalization of integration against bounded Radon measures i.e. bounded regular Borel measures}. 

Let $A = C_0(\Omega)$ with $\Omega$ a locally compact Hausdorff space, then
$$ \text{Sp}(a) = \{ a(\omega) \in \mathbb{C} : \omega \in \Omega \} $$
and we can check that it fits all the results of C* algebras we have given, for example, if $\Omega$ is compact and $0 \notin \text{Sp}(a)$ then the fact that $a(\Omega) \subset \mathbb{C}$ is compact guaranties that $a^{-1}$ given by $a^{-1}(\omega = (a(\omega))^{-1}$ is a bounded continuous function over $\Omega$, thus $a$ is invertible. We can check that also if $\Omega$ is not compact then $\hat{a} \in A^{+} \simeq C_0 (\Omega^{+})$ is such that $\hat{a}(\infty) = 0$, meaning that $\hat{a}$ is not invertible in $A^{+}$. Therefore, in a non-commutative C* algebra \textbf{the spectrum of an element plays the role of its range its range if it is considered as a function over a non-commutative space}.\\

\textbf{C* algebras and the ring of complex numbers}\\

Depending on the values that takes a function we can perform certain operations with it, for example, if $A = C(\Omega)$ with $\Omega$ compact, $a \in A$ and $\text{Sp}(a) \subset D$ ($D = \mathbb{C} - \mathbb{R}^-$), then the holomorphic functional calculus (\cref{proposition:logarithms_and_n_roots}) tell us that there are elements $\log(a)$ and $a^{1/2}$ in $A$, and both belong to the smallest commutative Banach algebra that contains $a$. We can take this one step further and bring the involution into play, such that, we look for elements of C* algebras that behave almost like complex valued functions, since $\mathbb{C}$ is commutative we will ask for $x x^* = x^* x$, these are the normal elements as we already mentioned. For normal elements we will be able to define a functional calculus that comes from continuous functions on $\text{Sp} (a)$ \cref{sec:C_star_alg_cont_func_cal}.

Normal elements are quite special, and among them \textbf{self adjoint elements stand out because they are would be our generalization of functions that take real values}. We wonder if we can take this one step further and look for positive valued functions, and the answer is positive, these elements are termed \textbf{positive elements}, and have two properties, so $a \in A$ is a positive element if
\begin{itemize}
    \item $a^* = a$,
    \item $\text{Sp}(a) \subset \mathbb{R}^+$ (with the convention that $0 \in \mathbb{R}^+$).
\end{itemize}
The set of positive elements of $A$ is denoted by $A_{\text{pos}}$\index{$A_{\text{pos}}$}. Surprisingly $a$ is positive iff $a = b^* b$ with $b \in A$ , and if $g,f$ are positive then $g+f$ is positive with $\rho(g+f) \geq \max\{ \rho(f) + \rho(g)\}$ \cref{proposition:cahracterization_of_positive_elements} so, positive elements behave much like positive valued functions. Interestingly enough, the set of positive elements $A_{\text{pos}}$ coincides with the set $A_{+}$ defined for the GNS construction (\cref{proposition:cahracterization_of_positive_elements}), and an element is positive iff it can be written as $a a^*$ with $a \in A$. Moreover, in unital C* algebras the set of positive elements together with with the set of unitaries to provide a unique polar decomposition of invertible elements (\cref{proposition:polar_decomposition_invertible_elements}), that resembles the expression of complex numbers as $z = |z| \exp(i \arg (z))$ with $|z| > 0$ and $\exp(i \arg (z)) \in \mathbb{T}$. Thus, the set of unitaries can be intuitively though as a generalization of the set of unit length complex numbers.

In sections \cref{sec:cont_func_calc_consec} and \cref{sec:unitaries_equivalence_relations} we look on properties of C* algebras that makes them similar to the ring $\mathbb{C}$, where the set $A_{\text{pos}}$ plays the role of real positive numbers, the set $A_{\text{sa}}$ plays the role of the real numbers and $U(A)$ plays the role of the unit lenght numbers, and $G(A)$ plays the role of complex numbers that are non zero, one of these results is the polar decomposition of invertible elements. These nice properties of C* algebras was one of the motivations for studying Hilbert C* modules, which are generalization of the concept of Hilbert space where $\mathbb{C}$ is changed by a C* algebra.

A Hilbert C* module is a module over a C* algebra $A$ where there is a inner product taking values on $A$ such that it is complete as a complex vector space with respect to the norm induced by the inner product \citep{nlab:hilbert_module}. Hilbert C* modules are a great tool to study more sophisticated structures that involve C* algebras, like families of C* algebras, or vector bundles of Hilbert spaces, we won't deal with Hilbert C* modules in this document but we believe is good you now of their existence because sooner than latter you will come across them when studying non-commutative geometry, for example \citep[Chapter 2 (Non commutative topology: vector bundles)]{gracia-bondia_elements_2001}.\\

\textbf{Measures}\\

Take $\mu$ as a bounded Radon measure (\cref{definition:Radon_measure}) with full support (\cref{definition:measure_with_full_support}) on a locally compact Hausdorff space space $\Omega$, then,  
$$\int_{\Omega} f^* f d \mu = \int_{\Omega} |f|^2 d \mu  > 0 \text{ iff } f \neq 0.$$
The concept of full support measure is generalized in to the concept of faithful positive linear functional, so, we say that a positive linear functional $\phi: A \to \mathbb{C}$ is \textbf{faithful} if $f(a^* a) = 0$ iff $a = 0$.  The "faithfull" property is defined equally for any functional whose range contains $A_{\text{pos}}$, for example, a \textbf{faithful state} is a state $f$ such that $f(a^* a) = 0$ iff $a=0$. Notice that if $A$ is a unital C* algebra, then every positive linear functional can be scaled into a state as follows, if $f$ is a positive linear functional then $f/f(1)$ is a state, which corresponds to normalizing measures on $K$ a compact Hausdorff space.

When $A = C_0(\Omega)$ we can measure the \textit{size} of elements of $A$ with regular Borel measures, and we only need to resort to the elements of the form $a^* a$ to give a sense of the size of the elements of $A$, so, we can relax the linear functional over $A$ requirement and focus only on linear functionals that are defined on $A_{\text{pos}}$, these are referred to as weights and traces. A \textbf{weight}\index{C* algebra!weight} \citep[Section II.6.7]{blackadar_operator_2006} on a $\mathrm{C}^*$-algebra $A$ is a function $\phi: A_{\text{pos}} \rightarrow$ $[0, \infty]$ such that 
\begin{itemize}
    \item $\phi(0)=0, \phi(\lambda a)=\lambda \phi(a)$ for $a \in A_{\text{pos}}$ and $\lambda>0$,
    \item $\phi(a+b)= \phi(a)+\phi(b)$ for all $a, b \in A_{\text{pos}}$.
\end{itemize}
The weight $\phi$ is densely defined if $\left\{x \in A_{\text{pos}} \right.$: $\phi(x)<\infty\}$ is dense in $A_{\text{pos}}$, and $\phi$ is \textbf{faithful} if $\phi(x)=0$ implies $x=0$. Any finite valued weight comes from a positive linear functional, thus, for $\phi$ to not be continuous there must be an element $a$ such that $\phi(a) = \infty$, which will happen in the commutative case if we consider unbounded measures (\citep[Examples II.6.7.2, i]{blackadar_operator_2006}), also, in the commutative case every lower semicontinuous weight comes from Borel measures (\citep[Examples II.6.7.2, v]{blackadar_operator_2006}). 

In our examples (\cref{section:algebra_of_compact_operators}) a special type of weight will come up, it is called a \textbf{trace}\index{C* algebra!trace} and is characterized by $\tau\left(x^* x\right)=\tau\left(x x^*\right)$ for all $x \in$ A, that is, $\tau(a)=\tau(a^*)$ for any $a \in A_{\text{pos}}$ (\citep[Definition II.6.8.1]{blackadar_operator_2006}). Also, traces are unitary equivalent, that is, $\tau(u^* x u) = \tau(x)$ for any $x \in A_{\text{pos}}$ and $u \in A_{\text{pos}}$. In the commutative case, all weights are traces, therefore, all traces com from Borel measures. A tracial state\index{C* algebra!tracial state} (or normalized trace) is a state which is a trace.

\begin{remark}[Weights and continuity]\label{remark:weights_and_continuity}
We have mentioned that a weight that is finite everywhere comes from a positive linear functional, so, for weights to be discontinuous there must be an element $a \in A_{\text{pos}}$ such that $\phi(a) = \infty$. In this context, there is a special kind of weights, called lower semicontinuous weights\index{C* algebra!lower semicontinuous weight}, and has the property that if $a_n \to a$ with $\{ a_n \}_{n \in \mathbb{N}} \subset A_{\text{pos}}$ and $a \in A_{\text{pos}}$, then
$$ \phi(a) \leq \lim_{n \to \infty} \left( \inf_{m \geq n} \phi(a_m) \right) .$$
In \cref{section:algebra_of_compact_operators} there is an example of a lower semicontinuous weight, which is actually a trace.

Take $a,b \in A_{\text{pos}}$ such that $a \geq b$, then $b-a \in A_{\text{pos}}$ (\cref{remark:partial_ordering_in_self_adjoint_elements}), hence 
$$ \phi(b-a) + \phi(a) = \phi(b) \Rightarrow  \phi(a) \leq \phi(b),$$
that is, the weight $\phi$ is \textbf{monotonous} with respect to the order of $A_{\text{pos}}$. Also, if $a,b$ are arbitrary positive elements of $A$ we have that $a \leq a+b, b \leq a+b$ (\cref{proposition:cahracterization_of_positive_elements}), therefore, 
$$ \phi(a) \leq \phi(a +b), \; \phi(b) \leq \phi(a+b). $$
So, take $\{a_n \}_{n \in \mathbb{N}} \subset A_{\text{pos}}$ with $a_n \to a$  and $a_n \leq a_{n+1}$, from \cref{remark:partial_ordering_in_self_adjoint_elements} we know that $a_n \leq a$ for all $n \in \mathbb{N}$ and $a \in A_{\text{pos}}$, therefore,
$$ \phi(a_n) \leq \phi(a_l), \; n \leq l, \text{ and } \phi(a_n) \leq \phi(a).$$

Since $\{ \phi(a_n)\}_{n \in \mathbb{N}}$ is a set of increasing positive numbers it can do one of two things:
\begin{itemize}
    \item $\{ \phi(a_n)\}_{n \in \mathbb{N}}$ converges to a positive number, which we will call $c$,
    \item $\{ \phi(a_n)\}_{n \in \mathbb{N}}$ diverges, that is, any $K >0$ there is an $N \in \mathbb{N}$ such that, if $n \geq N$ then $\phi(a_n) > K$.
\end{itemize}

Assume that $\phi$ is a lower semicontinuous weight, then,
\begin{itemize}
    \item if $\phi(a) < \infty$ then $\phi(a) \leq \lim_{n \to \infty} \left( \inf_{m \geq n} \phi(a_m) \right)$, so, since $\phi(a) \geq c$ we must have that $\{ \phi(a_n)\}_{n \in \mathbb{N}}$ converges and $\phi(a) = c$,
    \item if $\phi(a) = \infty$ we get that $\phi(a) \leq \lim_{n \to \infty} \left( \inf_{m \geq n} \phi(a_m) \right)$, thus, we must have that $\phi(a) \leq \lim_{n \to \infty} \left( \inf_{m \geq n} \phi(a_m) \right) = \infty$, thus, $\{ \phi(a_n)\}_{n \in \mathbb{N}}$ diverges.
\end{itemize}

We have shown how a lower semicontinuous weight generalizes the concept of lower semicontinuous function, and in this context, taking an increasing sequence of positive elements amounts to computing the limit of the function from the left.
\end{remark}

\section{Continuous functional calculus}
\label{sec:C_star_alg_cont_func_cal}

Following the exposition on \cref{section:non_commutative_geometry_dictionary}, if $A$ is a C* algebra then the normal elements of $A$ are a good generalization of complex valued functions, and as such, we will be able to evaluate continuous functions on them. Notice that if $a a^* = a^* a$ then
\begin{itemize}
    \item if $p(z), q(z)$ are complex polynomials then $p(a)q(a^*) = q(a^*) p(a)$,
\end{itemize}
thus it makes sense to define mappings $a \to p(a,a^*)$ for $p(z,z^*)$ a polynomial on $z,z^*$ when $a$ is normal. Let $A$ be unital, then $C = C^*(1_A,a)$ (the sub C* algebra of $A$ generated by $1_A$ and $a$) is a commutative C* algebra, where $C^*(1_A,a)$ is the closure in $A$ of the algebra of complex polynomials on $z,z^*$ evaluated on $a$. It turns out that the Gelfand transform (\cref{theorem:commutative_gelfand_naimark_theorem}) establishes an isomorphism between $C$ and $C(\Phi_C) \simeq C(\text{Sp}(a))$, which gives us the continuous functional calculus

\begin{theorem}[Continuous functional calculus\index{C* algebra!continuous functional calculus} (Theorem 6.26 \citep{allan_introduction_2011})]\label{theorem:continuous_functional_calculus}

Let $A$ be a unital $C^*$-algebra, and let a be a normal element of $A$. Then there is a unique unital *-homomorphism $\Theta_a: C(\operatorname{Sp} a) \rightarrow A$ such that $\Theta_a((f(z) = z, \; z \in \text{Sp}(a)))=a$. Moreover:
\begin{itemize}
    \item $\Theta_a$ is isometric
    \item $\operatorname{im} \Theta_a=C^*(1_A,a)$
    \item $a \Theta_a(g)=\Theta_a(g) a$ and $\operatorname{Sp}\left(\Theta_a(g)\right)=g(\operatorname{Sp} a)$ for every $g \in C(\operatorname{Sp} a)$
\end{itemize}
We denote $\Theta_a(f) = f(a)$.
\end{theorem}

Notice that if $b \in C^*(a,1_A)$, then $bb^* = b^*b$ because it is the limit of normal elements, and the set of normal elements is closed (\cref{remark:normal_self_adjoint_unitary_are_closed}), moreover, if $a=a^*$ then $b=b^*$ because the set of self adjoint elements is closed (\cref{remark:normal_self_adjoint_unitary_are_closed}).

\begin{lemma}\label{lemma:holomorphic_calculus_continuous_calculus_coincide}
Let $A$ be a unital C* algebra and let $a$ be a normal element of $A$. Let $f$ be an holomorphic function on a neighborhood of the spectrum of $a$, denote by $\widehat{f}(a)$ the element of $A$ obtained by applying the continuous functional calculus over $A$ (\cref{theorem:continuous_functional_calculus}) and denote by $\tilde{f}(a)$ the element of $A$ obtained by applying the holomorphic functional calculus over $A$ (\cref{theorem:holomorphic_functional_calculus_banach_algebras}), then
$$ \tilde{f}(a) = \widehat{f}(a). $$
\end{lemma}
\begin{proof}
Let $A$ be a unital C* algebra and $a\in A$ normal, if $p(z) = \sum_{i=0}^{n}p_i z^i$ then in the functional calculus we have that
$$p(a) = \sum_{i=0}^{n}p_i a^i $$
with $a^0 = 1_A$. If $p(z)\neq 0$ when $z \in \text{Sp}(a)$ then $p^{-1}(z) = \left( \sum_{i=0}^{n}p_i z^i \right)^{-1}$ is an holomorphic function on $\text{Sp}(a)$, in particular it is continuous, and we get in the continuous calculus
$$ (p p^{-1})(a) = id(a) = 1_A = p(a)p^{-1}(a), $$
which implies that $p^{-1}(a) = (p(a))^{-1}$ in $A$. Therefore, if $f(z) = p(z)/q(z)$ is holomorphic on $\text{Sp}(a)$ then the continuous functional calculus gives
$$ f(a) = p(a)(q(a))^{-1}, $$
and we get that the holomorphic functional calculus and the continuous functional calculus coincide on the rational functions.

Let $U$ be a open neighbourhood of $\text{Sp}(a)$, we mentioned in \cref{sec:holom_calc_def_and_prop} that $R(U)$ is dense in $\text{Hol}(U)$ in the compact-open topology i.e. in the topology of uniform convergence in compact sets. Since $\text{Sp}(a)\subset U$ is compact then if $r_n \to f$ in the compact-open topology then $r_n \to f$ uniformly in $\text{Sp}(a)$, which is equivalent to saying that $r_n \to f$ in $C(\text{Sp}(a))$. Denote by $\tilde{f}(a) \in A$ the object obtained from the holomorphic calculus, and $\widehat{f}(a) \in A$ the object obtained from the continuous calculus, then the continuity of both functional calculus assert that
$$\tilde{f}(a) =   \lim_{n \to \infty} \tilde{r_n}(a) =  \lim_{n \to \infty} \widehat{r_n}(a) = \widehat{f}(a),$$
and we get that the continuous and holomorphic functional calculus coincide.
\end{proof}

If $B,A$ are unital C* algebras and $B \subseteq A$ with $1_A \subset B$ then $\text{Sp}_B (b) = \text{Sp}_A (b)$ for very $b \in B$ \citep[Proposition 6.23]{allan_introduction_2011}, this is an important property for the formulation of the functional calculus. 

\begin{proposition}[Properties of the continuous functional calculus]\label{proposition:properties_continuous_functional_calculus}

Let $A,B$ be unital C* algebras, a be normal element of $A$ and $\phi : A \to B$ a unital *-homomorphism, then

\begin{itemize}
    \item \citep[Corollary 6.27]{allan_introduction_2011} If $f \in C(\operatorname{Sp} (a))$ and $g \in C(\text{Sp}(f(a)))$, then $f(a)$ is normal and $(g \circ f)(a)=g(f(a))$.
    \item \citep[remarks in page 273]{allan_introduction_2011} If $f \in C(\operatorname{Sp} a)$ then $\| f(a) \| = \sup\{ |g(\alpha)| : \alpha \in \text{Sp}(a) \}$, which we denote by $|f|_{\text{Sp}(a)}$ 
    \item \citep[Proposition 6.29]{allan_introduction_2011}Recall that $\text{Sp}(\phi(a)) \subseteq \text{Sp}(a)$, so we have that $f(\phi (a))=\phi(f(a))\left(f \in C\left(\text{Sp}_A (a)\right)\right)$
    \item \citep[Proposition 6.29]{allan_introduction_2011} If $\phi$ is injective, $\mathrm{Sp}_B (\phi (a))=\mathrm{Sp}_A(a)$, the map $\phi$ is isometric, and $\phi(A)$ is a $C^*$-subalgebra of $B$.
\end{itemize}

\end{proposition}

From \cref{sec:unitization_of_C_star_algebras} we know that if $A,B$ are C* algebras (unital or not) and there is a C* algebra homomorphism $\phi: A \to B$, then there is an associated C* algebra homomorphism $\phi^+ : A^{+} \to B^+$ that is injective iff $\phi$ is injective, moreover, let $i_A: A \to A^{+}$ ($i_B: B \to B^+$) then $\| i_A (a) \| = \| a \|$ ($\| i_B (b) \| = \| b \|$). So, if $\phi$ is injective then $\phi^+$ is an isometry, which in turn implies that $\phi$ is an isometry.

If $A$ is a non-unital C* algebra and $a\in A$,  for $f \in C(\text{Sp}(a))$ we define $f(a) = f(i(a)) = f(a,0)$ as in the holomorphic calculus, then the commutativity of the continuous calculus with unital algebra *-homomorphism implies that $f(a) \in A$ iff $f(0) = 0$ and the argument is a copy of the one exposed in \cref{sec:hol_cal_consecuences}.

\begin{proposition}[Continuity of function evaluation]\label{proposition:continuity_of_function_evaluation}

Let $K$ be a non-empty compact subset of $\mathbb{C}$ and let $f : K \to \mathbb{C}$ be a continuous function. Let $A$ be a unital C* algebra and 
$$\Omega_K = \{ a \in A :\; a^* a = a a^*, \; \text{Sp}(a) \subseteq K\}$$
then the mapping 
$$\hat{f}: \Omega_K \to A, \; \hat{f}(a) = f(a) $$
is continuous.
\end{proposition}
\begin{proof}
This is small generalization of  \citep[Lemma 1.2.5]{rordam_introduction_2000}, so we follow mostly the same steps.

From the Stone-Weierstrass theorem for complex valued functions on an arbitrary compact set \citep[Theorem 11.7.16]{lebl_basic_2022} (or \citep[Corollary 2.35 (ii)]{allan_introduction_2011}) we know that the complex polynomials in the variables $z, z^*$ are dense in $C(K,\mathbb{C})$ when $K \subset \mathbb{C}$ compact, this means that for each $\epsilon > 0$ and each $f \in C(K,\mathbb{C})$ there is a polynomial $p_{\epsilon}$ in the variable $z, z^*$ such that $\sup_{z \in K}(\| p_{\epsilon}(z,z^*) - f(z) \|) < \epsilon$. Also, given a polynomial $p$ in the variables $z,z^*$ then the mapping $a \mapsto p(a,a^*)$ is well defined for a normal element, and is continuous, because is obtained by composing continuous maps i.e. involution, multiplication, addition, scalar multiplication.

Let $f: K \rightarrow \mathbb{C}$ be any continuous function, $a$ be an element in $\Omega_K$, and $\epsilon>0$. Use the Stone-Weierstrass theorem to find a complex polynomial $g$ such that $|f(z)-g(z)| \leqslant \varepsilon / 3$ for every $z$ in $K$. Since $g$ is a polynomial, we now that the mapping $a \mapsto p(a,a^*)$ is continuous, thus there is a $\delta>0$ such that $\|g(a)-g(b)\| \leqslant \epsilon / 3$ whenever $b$ is an element in $A$ with $\|a-b\| \leqslant \delta$. By \cref{proposition:properties_continuous_functional_calculus} we know that
$$
\|f(c)-g(c)\|=\|(f-g)(c)\|=\sup \{|(f-g)(z)|: z \in \text{Sp}(c)\} \leqslant \varepsilon / 3, 
$$
for all $c$ in $\Omega_K$, it follows that 
$$\|f(a)-f(b)\| \leq \| f(a) - g(a)\| + \| g(a) - g(b) \| + \| f(b) - g(b) \| \leqslant \varepsilon$$
for all $b$ in $\Omega_K$ with $\|a-b\| \leqslant \delta$.

\end{proof}

Notice that, if we chose $K$ with a non-empty interior and $\text{Sp}(a) \subset K^o$, then the upper semi-continuity of the map $a \mapsto \text{Sp}(a)$ (\cref{proposition:upper_semi_continuity_spectrum_map}) guaranties that there is $\delta > 0$ such that if $\|b-a\| \leq \delta$ then $\text{Sp}(b) \subset K^o$, so $f(b)$ exists when $b$ is close enough to $a$, and the mapping $b \mapsto f(b)$ is continuous (this requires a similar argument as \cref{remark:spectrum_upper_semi_continuity_and_function_evaluation}).

\subsection{Consequences}
\label{sec:cont_func_calc_consec}

First we can use the continuous calculus to provide a characterization of normal elements according to its spectrum

\begin{lemma}[Characterization of normal elements with their spectrum]\label{lemma:characterization_normal_elements_with_their_spectrum}
Let $A$ be a unital C* algebra and $a \in A$ with $a^* a = a a^*$, then
    \begin{itemize}
        \item $a \in A_{\text{sa}}$ iff its spectrum is real.
        \item $a \in U(A)$ iff its spectrum is contained in the unit circle
        \item $a \in P(A)$ iff its spectrum is contained in $\{0,1\}$
        \item $a \in A_{\text{pos}}$ iff its spectrum is contained in $\mathbb{R}^+$ 
    \end{itemize}
\end{lemma}
\begin{proof}
The continuous calculus establishes an isomorphism $\phi: C(\text{Sp}(a)) \to C^*(1_A,a)$ such that $\phi(id) = a$ where $id(z) = z$, so, 
\begin{itemize}
    \item $id$ is self adjoint iff $\text{Sp}(a) \subset \mathbb{R}$
    \item $id$ is unitary iff $\text{Sp}(a) \subseteq \mathbb{S}$
    \item $id$ is self adjoint and idempotent iff $\text{Sp}(a) \subseteq \{ 0,1\}$
    \item By the first item $a$ is self adjoint, also $\text{Sp}(a) \subset \mathbb{R}^+$ thus $a$ is positive.
\end{itemize}
\end{proof}

Also, we can compute the norm of the resolvent of a normal element using the continuous calculus, providing a stronger result than \cref{proposition:bound_on_the_resolvent_norm} when we deal with normal elements

\begin{lemma}[Norm of resolvent of normal element]\label{lemma:norm_of_resolvent_of_normal_element}

Let $A$ be a unital C* algebra and $a \in A$ a normal element, then for $\lambda \in R_A (a)$ we have that
$$ \| R_a (\lambda) \| = \| (\lambda 1 - a)^{-1} \| = \sup_{x \in sp(a)} \{ | (\lambda - x)^{-1} | \} = \frac{1}{d(\lambda, \text{Sp}(a))},$$
where $d(\lambda,\text{Sp}(a))$ is the Hausdorff distance between $\lambda$ and $\text{Sp}(a)$.

\end{lemma}
\begin{proof}
The function $f_{\lambda}(z) = (\lambda - z)^{-1}$ is continuous on $\text{Sp}(a)$, then \cref{proposition:properties_continuous_functional_calculus} tell us that
$$ \| (\lambda 1_A - a)^{-1} \| = \sup \{ | (\lambda - z)^{-1} | : z \in \text{Sp}(a) \}, $$
also we have 
$$ \sup \{ | (\lambda - z)^{-1} | : z \in \text{Sp}(a) \} = \sup \{ | \lambda - z|^{-1}  : z \in \text{Sp}(a) \}  = 1/ \inf \{ | \lambda - z| : z \in \text{Sp}(a) \}.$$
Since $\inf \{ | \lambda - z| : z \in \text{Sp}(a) \} = d(\lambda, \text{Sp}(a))$ we get the desire result.
\end{proof}

The continuous calculus also gives us a nice characterization of the positive elements\index{C* algebra!positive element}, so, if $a \in A_{\text{pos}}$ then $f(x) = \sqrt{x}$ is continuous on $\text{Sp}(a)$ and if $b = \sqrt{a}$ then $b^* = b$, $b^2 = b^* b = a$. We have that any positive element comes in the form $b^* b$, now, if we use the continuous function $g(z) = z^* z$ we can use the continuous functional calculus to show that $a^* a= g(a)$ for every $a \in A$ normal and more importantly 
$$ \text{Sp}(a^* a) = g(\text{Sp}(a)) \subset g(\mathbb{C}) = \mathbb{R}^+, $$
thus $a^* a$ is a positive element. We look into a generalization of this result and a couple more of interesting properties of positive elements in C* algebras

\begin{proposition}[Characterization of positive elements]\label{proposition:cahracterization_of_positive_elements}

Let $A$ be a C* algebra then

\begin{itemize}
    \item $a \in A_{\text{pos}}$ iff $a = b^2 = b^* b$ for $b \in A_{\text{sa}}$ \citep[Theorem 6.30]{allan_introduction_2011}
    \item An operator $T \in B(H)$ is a positive operator iff $T \in B(H)_{\text{pos}}$ \citep[Proposition 6.32]{allan_introduction_2011}
    \item If $a,b \in A_{\text{pos}}$ then $a+b \in A_{\text{pos}}$ \citep[Lemma 6.36]{allan_introduction_2011}. In the reference this fact is proven for unital C* algebras, however, is easy to generalize to any C* algebra is one see $A$ as a sub algebra of $A^{+}$.
    \item Let $A$ be a C* algebra, then $a^* a \in A_{\text{pos}}$ for every $a \in A$ \citep[Theorem 6.38]{allan_introduction_2011}
    \item $A_{\text{pos}}$ is closed in $A$ and $A_{+} = A_{\text{pos}}$ \citep[Corollary 6.39]{allan_introduction_2011}
    \item If $a,b \in A_{\text{pos}}$ then $\rho(a+b) \geq \max \{ \rho(a), \rho(b) \}$, or equivalently, $\|a+b\| \geq \max \{ \|a\|, \|b\| \}$. This implies that if $a \geq b$ then $\| a \| \geq \|b\|$.
\end{itemize}
\end{proposition}
\begin{proof}
All but the last item are provided a reference that contains their demonstration, thus we proceed to proof the last item.

From \cref{theorem:faithfull_universal_representation_c_star_algebras} we know that there is a Hilbert space $H$ and an injective *-homomorphism $\phi: A \to B(H)$, also, from \cref{proposition:properties_continuous_functional_calculus} we know that $\text{ Sp}(\phi(a)) = \text{Sp}(a)$. Take $a \in A_{\text{pos}}$, then $\phi(a)$ is a positive operator on $H$, so in particular $\phi(a)$ is self adjoint and its norm is given by \citep[Lemma 8.26]{hunter_aplied_2005} 
$$ \| \phi(a) \| = \sup \{ | \langle \phi(a)(x), x \rangle | : \; x \in H, \; \| x \| = 1  \} . $$
Take $a,b \in A_{\text{pos}}$ then
$$ \rho(a+b) = \| a+b \| = \| \phi(a+b) \| = \sup \{ | \langle \phi(a)(x), x \rangle + \langle \phi(b)(x), x \rangle| : \; x \in H, \; \| x \| = 1  \} ,$$
since both $\phi(a), \phi(b)$ are positive then $\langle \phi(a)(x) , x \rangle \geq 0$ and $\langle \phi(b)(x) , x \rangle \geq 0$, therefore
$$ \| \phi(a+b) \| = \geq  \sup \{ \langle \phi(a)(x), x \rangle : \; x \in H, \; \| x \| = 1  \} = \|\phi(a) \|$$
and similarly $\| \phi(a+b) \| \geq \| \phi(b) \|$. And we conclude that $\| a + b \| \geq \max \{ \|a\|, \|b\| \}$.
\end{proof}

\begin{remark}[Partial ordering in $A_{\text{sa}}$]\label{remark:partial_ordering_in_self_adjoint_elements}
The set $A_{\text{pos}}$ allows to set a partial ordering on the set $A_{\text{sa}}$, so, we denote $a \geq b$ if $a -b \in A_{\text{pos}}$, and $a > b$ if $a -b \in A_{\text{pos}}$ and $a \neq b$. Notice that if $a \leq b$ and $b \leq c$ we have that $a \leq c$ because $a-c = (a-b) + (b -c)$ and the set of positive elements is close under addition, also, $a \geq 0$ iff $a \in A_{\text{pos}}$ thus we use $a \geq 0$ and $a \in A_{\text{pos}}$ interchangeably. When $a\geq 0$ and $a \in G(A)$ then there is $\lambda > 0$ such that $\text{Sp}(a) > \lambda$, thus, the spectral mapping of the continuous functional calculus tell us that $\text{Sp}(a-\lambda1_A) \subset (0,\infty)$ ($(a-\lambda1_A) \in G(A)$) and $(a-\lambda1_A) \in A_{sa}$, consequently $a \geq \lambda1_A$. The previous argument tell us that for $a \in A_{\text{pos}}$ we have, $a \in G(A)$ iff there is $\lambda > 0$ such that $a \geq \lambda 1_A$. Notice that if $a \geq b$ then for every $c \in A_{\text{sa}}$ we have $a + c \geq b+c$.

Let $\{ a_n \}_{n \in \mathbb{N}} \subset A_{\text{sa}}$, such that $a_n \leq a_{n+1}$ and $a_n \to a$ in $A$. From \cref{remark:normal_self_adjoint_unitary_are_closed} we know that $a \in A_{\text{sa}}$, moreover, from \cref{proposition:cahracterization_of_positive_elements} we get that $a \in A_{\text{pos}}$. From \cref{sec:c_star_alg_spec_cont} we know that the map $a \mapsto sp(a)$ is continuous in the set of normal elements, so, if $b \in A_{\text{sa}}$ we have that
$$ sp(a_n -b) \to sp(a - b). $$
In particular, if we take $b = a_j$ for $j \in \mathbb{N}$ we get that
$$ sp(a_n - a_j) \xrightarrow[n \to \infty]{} sp(b - a_j) ,$$
therefore, $sp(b - a_j) \subset \mathbb{R}^+$ since $sp(a_l - a_j) \subset \mathbb{R}^+$ when $l \geq j$. This is one of many properties that translate from positive elements in $\mathbb{C}$ into the set of positive elements in a C* algebra. 

Additionally, from \citep[Corollary 6.31]{allan_introduction_2011} we have that, if $\phi: A \to B$ is an isomorphism of unital C* algebras, then, 
\begin{itemize}
    \item $\phi(A_{\text{sa}}) = B_{\text{sa}}$,
    \item $\phi|_{A_{\text{sa}}} : A_{\text{sa}} \to B_{\text{sa}}$ is an order preserving, isometric, real-linear isomorphism,
    \item if $A$ is commutative, $\phi|_{A_{\text{sa}}} : A_{\text{sa}} \to B_{\text{sa}}$ is a real algebra isomorphisms.
\end{itemize}
\end{remark}

As $A_{\text{pos}}$ plays the role of positive valued functions we have that $U(A)$ will plays the role of functions taking values on $\mathbb{S}$ and we will have a polar decomposition of invertible elements much like every complex number can be expressed as 
$$z = |z| \exp{i \text{arg}(z)}$$
with $|z| \in \mathbb{R}^+$ and $ \exp{i \text{arg}(z)} \in \mathbb{S}$.

\begin{proposition}[Polar decomposition\index{Polar decomposition} of invertible elements]\label{proposition:polar_decomposition_invertible_elements}

Let $A$ be a unital C* algebra, then
\begin{itemize}
    \item If $a \geq 0$ with $a \in G(A)$, then there exists $b \in A_{\text{sa}}$ such that $\exp{b}=a$ \citep[Proposition 6.33]{allan_introduction_2011}
    \item If $a \in G(A)$, then there is a unique decomposition $a=b u$ with $b \in A_{\text{pos}}$, and $u \in U(A)$. Moreover, $b=\left(a a^*\right)^{1 / 2}$. \citep[Corollary 6.40]{allan_introduction_2011}
    \item If $a \in U(A)$ then $b = 1_A$ and $u=a$, that is, the unitaries are a fixed point of the map $a \to u$.
\end{itemize}
We call $b$ the absolute value of $a$ and we denote by $|a|$, also, since $u$ is unique we denote it by $\omega(a)$.
\end{proposition}
\begin{proof}
We will use the functional calculus we have looked into so far,

\begin{itemize}
    \item Since $a \in G(A)$ then $0 \notin \text{Sp}(a)$ thus $\text{Sp}(a) \subset (0,\infty)$ and we have that $\text{Sp}(a) \subset D$ with $D$ defined as in \cref{sec:hol_cal_consecuences}. Therefore, $\log(z)$ is holomorphic on $\text{Sp}(a)$ and there is an element $b = \log(a)$ such that $\exp{b} = a$ by the holomorphic calculus, moreover, $\text{Sp}(\log (a)) \subset \mathbb{R}$ and \cref{lemma:characterization_normal_elements_with_their_spectrum} tell us that $\log(a)$ is a self-adjoint element as desired.
    \item Notice that $a^* \in G(A)$ which implies that $aa^{*} \in G(A)$ and $a a^* \in A_{\text{pos}}$ by \cref{proposition:cahracterization_of_positive_elements}. Let $b = (a a^*)^{1/2}$ which from \cref{proposition:cahracterization_of_positive_elements} we know that is self adjoint (and can be computed using the holomorphic calculus as in \cref{proposition:logarithms_and_n_roots}), and the spectral mapping tell us that $0 \notin \text{Sp}(b)$ (otherwise $0 \in \text{Sp}(a a^*)$ because for $t \geq 0$ $\sqrt{t} = t$ iff $t =0$), that is, $b \in G(A)$. 
    Let $u = b^{-1} a$ ($b^{-1}$ can also be computed with the holomophic calculus beacuse $z \to 1/z$ is holomorphic on $\text{Sp}(b)$), then $b \in G(A)$ and 
    $$u u^* = b^{-1} a a^* (b^{-1})^* = b^{-1} a a^* b^{-1} = b^{-1} b^2 b^{-1} = 1_A, $$
    thus $u^{-1} = u^*$ and $u \in U(A)$. 
    Let $a = c v$ with $c \in A_{\text{pos}}$ and $v \in U(A)$ then $aa^* = c v v^* c = c^2$, thus the uniquenes statement on the square root given in \cref{proposition:logarithms_and_n_roots} tell us that $c=b$ and $v=b^{-1}a=u$.
    \item If $a \in U(A)$ then $a a^* = 1_A$ and we get $b=1_A$, thus $u = a$.
\end{itemize}

Notice that the calculations are computed using the holomorphic calculus, and the properties of those elements are deduced from the continuous functional calculus. This procedure will be replicated in smooth sub algebras.
\end{proof}

The polar decomposition of invertible elements is the first step to show that $U(A)$ is a deformation retraction of $G(A)$ (\cref{prop:U_A_retraction_of_G_A}), which will play an important role in the various equivalent descriptions of the K theory of a C* algebra (\cref{remark:K_1_equivalence_of_invertibles}). 

Given a C* algebra $A$, if $A$ had no unit then $Sp(a)$ with $a \in A$ is computed over $A^+$, however, there is not a unique unital C* algebra where $A$ is a sub C* algebra, thus, we may wonder whether the spectrum of an element depends on the ambient algebra. This is actually the case, nonetheless, the variation of the spectrum is rather simple, because the continuous functional calculus gives us a nice result on which C* algebra contains the inverse of an element.

\begin{lemma}[Inverse and $C^*(a)$]\label{lemma:inverse_and_C_a}
Let $A$ be a unital C* algebra and $a \in G(A)$, then, 
$$a^{-1}, 1_A \in C^*(a).$$
\end{lemma}
\begin{proof}
We follow \citep[Exercise 1.6]{rordam_introduction_2000}. We do this in two steps, first we prove it for $a$ a normal element, then, we prove it for $a$ arbitrary.
\begin{itemize}
    \item \textbf{Assume $a$ is normal.} We follow \citep{2405957}. Let $B = C^*(1_A, a)$, then, the continuous functional calculus tell us that there is a unital C* isomorphism
    $$ \phi : C(Sp(a)) \to C^*(1_A, a), $$
    such that $\phi(\text{id}) = a$ with $\text{id}(z) = z$ over $Sp(a)$. Since $a \in G(A)$ then $0 \notin Sp(a)$, therefore, $\text{id}: C(Sp(a)) \to \mathbb{C}$ is a function that separates points and vanishes nowhere, so, the Stone-Weierstrass theorem \citep[Theorem 11.7.16]{lebl_basic_2022} (or \citep[Corollary 2.35 (ii)]{allan_introduction_2011}) tell us that 
    $$ C(Sp(a)) = C^*(\text{id}), $$
    with $C^*(\text{id})$ the closure of the *algebra generated by $\text{id}$. Since $\phi$ is a C* isomorphism we also have that 
    $$  C^*(1_A,a) \simeq C(Sp(a)) \simeq C^*(\text{id}) \simeq C^*(\phi(\text{id})) \simeq C^*(a), $$
    therefore, $a^{-1}, 1_A \in C^*(a)$.
    
    Another way of looking at this from the stand point of view of the Stone-Weierstrass theorem is to realize that the functions
    $$ f(z)= 1/z, \; g(z) = 1,  $$
    are continuous on $Sp(a)$, so, since the *algebra generated by $\text{id}$ separates points and vanishes no where on $Sp(a)$, we have that both $f,g$ can be uniformly approximated by polynomials on the variables $z, z^*$. This tell us that $\phi(f) = a^{-1}$ and $\phi(g) = 1_A$ belong to $C^*(\phi(\text{id})) = C^*(a)$.
    \item \textbf{Take $a$ an arbitrary element.} From \citep[Proposition 6.2]{allan_introduction_2011} we have that both $a a^*$ and $a^* a$ are invertible in $A$, additionally, $a^*$ is also invertible with $(a^*)^{-1} = (a^{-1})^*$. We have that 
    $$a^{-1} = (a^* a)^{-1} a, \; a^{-1} = a^* (a a^*)^{-1},$$
    so, since $a a^* \in A_{\text{sa}}$, by the previous item we have that $(a a^*)^{-1} \in C^*(a a^*)$, additionally, since $C^*(a a^*) \subseteq C^*(a)$ by definition of those C* algebras, we end up with
    $$ (a a^*)^{-1} a \in C^*(a), $$
    thus, $a^{-1}, 1_A \in C^*(a)$.
\end{itemize}

\end{proof}

\cref{lemma:inverse_and_C_a} gives us another way of proving \citep[Proposition 6.23]{allan_introduction_2011}, which will be useful for our analysis

\begin{corollary}[Spectrum and ambient C* algebra]\label{corollary:spectrum_and_ambient_C_star_algebra}
Let $A$ be a C* algebra and $B$ a sub C* algebra of $A$, then
\begin{enumerate}
    \item If $A$ has a unit $1_A$ and $1_A \in B$, then, 
    $$Sp_{A}(b) = Sp_{B}(b)$$
    for all $b \in B$.
    \item If $A$ has a unit and $B$ has no unit, then,
    $$ Sp_A(b) = Sp_{B}(b) $$
    for all $b \in B$.
    \item If $A$ has a unit $1_A$ and $B$ has a unit $1_B$, and $1_A \neq 1_B$, then
    $$ Sp_{A} (b) = Sp_{B}(b) \cup \{ 0 \} $$
    for all $b \in B$.
\end{enumerate}
\end{corollary}
\begin{proof}
\begin{enumerate}
    \item From \cref{lemma:inverse_and_C_a} we know that for $b \in G(A)$ and $b \in B$, 
    $$ b^{-1} \in C^*(b) \subset B, $$
    therefore, 
    $$ (\lambda 1_A - b) \in G(A) \text{ iff } (\lambda 1_A - b) \in G(B).  $$
    From the definition of spectrum (\cref{ref:spec_banach_alg}) we get that
    $$Sp_{A}(b) = Sp_{B}(b)$$
    for all $b \in B$.
    \item From \cref{lemma:uniqueness_of_unitization} we know that $B^+ \simeq C^*(\{1_A\} \cup B)$, and from \cref{lemma:inverse_and_C_a} we know that for $b \in G(A)$ and $b \in B^+$, 
    $$ b^{-1} \in C^*(b) \subset B^+, $$
    therefore, 
    $$ (\lambda 1_A - b) \in G(A) \text{ iff } (\lambda 1_A - b) \in G(B^+).  $$
    From the definition of spectrum (\cref{ref:spec_banach_alg}) we get that
    $$Sp_{A}(b) = Sp_{B^+}(b)$$
    for all $b \in B$. Since $Sp_B (b) := Sp_{B^+}(b,0)$ then $B$ is not unital (\cref{chapter:Banach_star_algebras_and_C_star_algebras}), we get that 
    $$ Sp_{A}(b) = Sp_{B}(b). $$
    \item We have that $B \oplus \mathbb{C}$ is unital. From \cref{lemma:uniqueness_of_unitization} we know that $B \oplus \mathbb{C} \simeq C^*(\{1_A\} \cup B)$, and from \cref{lemma:inverse_and_C_a} we know that for $b \in G(A)$ and $b \in B \oplus \mathbb{C}$,
    $$ b^{-1} \in C^*(b) \subset B \oplus \mathbb{C}, $$
    therefore,
    $$ (\lambda 1_A - b) \in G(A) \text{ iff } (\lambda 1_A - b) \in G(B \oplus \mathbb{C}). $$
    Any element of $B \oplus \mathbb{C}$ can be written as
    $$ (b, z(1_A - 1_B)), \; z \in \mathbb{C}, $$
    therefore, for $\lambda \in \mathbb{C}$ then
    $$ (b, z(1_A, - 1_B)) - \lambda 1_A = (b - \lambda 1_b, (z - \lambda)(1_A - 1_B)), $$
    so, 
    $$ (b, z(1_A, - 1_B)) - \lambda 1_A) \in G(A) \text{ iff } b - \lambda 1_B \in G(B), \; (z - \lambda) \neq 0. $$
    The previous equality tell along with the definition of spectrum (\cref{ref:spec_banach_alg}) tell us that
    $$ Sp_A(b) = Sp(b) \cup \{0\}. $$
\end{enumerate}

\end{proof}

The continuous calculus allow us to set many more properties of C* algebras and C* homomorphisms, to mention a few,
\begin{itemize}
    \item We used the continuous functional calculus to show that the image of C* homomorphisms are C* algebras (\cref{proposition:ideal_on_C_star_algebras}), whose proof involved also using properties of closed two sided ideals in C* algebras.
    \item The continuous calculus allow us to find suitable liftings of elements under C* homomorphism, that is, given a surjetive C* homomorphism $\phi: A \to B$ then given $b\in B$ with certain property, can we find an element $a$ such that $\phi(a) =b$ and $a$ has the same property? The answer is positive, literally, because if $b \geq 0$ then you can find $a \geq 0$ and $\|b\|_B = \|a\|_A$, this can also be made for self adjoint elements \citep[Section 2.2.10]{rordam_introduction_2000}, but not for normal nor unitary elements. These is a nice property is tighly related to the fact that $\text{ker}(\phi)$ is a two sided close ideal of $A$ and $B \simeq A / \text{ker}(\phi)$.
\end{itemize}
 That fact can be proven using solely the continuous functional calculus, for an explanation on this you can look at \citep{435105}. Also,

\section{Spectrum continuity}
\label{sec:c_star_alg_spec_cont}

\begin{definition}[Spectral gap\index{spectral gap}]\label{definition:spectral_gap}
Let $A$ be a C* algebra and $a$ a self adjoint element of $A$, then, $g$ is called a spectral gap of $a$ if the following conditions are met,
\begin{itemize}
    \item $g$ is a bounded and convex subset of $\mathbb{R}$ i.e. $g = (l,u)$ with $l,u \in \mathbb{R}$
    \item $g \cap Sp(a) = \emptyset$
    \item there is an element $p \in Sp(a)$ such that, for all $s \in g, \; s < p$
    \item there is an element $p \in Sp(a)$ such that, for all $s \in g, \; s > p$
\end{itemize}
\end{definition}

When we work with C* algebras the map 
$$Sp: A \to \mathcal{K}(\mathbb{C}), \; a \mapsto Sp(a) $$ 
described in \cref{sec:bana_alg_spec_conti} is continuous on normal elements \citep{Newburgh1951TheVO}. Notice that in the special case of self-adjoint operators we will be dealing with subsets of $\mathbb{R}$, thus, the edges of the spectral gaps will also vary continuously with respect to norm of the self-adjoint operator. In particular, we will have that if $a_n \to a$ and $a_n$ are normal elements then we know that $a$ is normal, and we will have that $$ \rho(a_n) = \| a_n \| \to \rho(a) = \| a \|, \text{ and } \text{Sp}(a_n) \to \text{Sp}(a)$$ 
in the topology $\mathcal{K}(\mathbb{C})$, which is described in \cref{sec:bana_alg_spec_conti}. 

In our constructions we will deal with a continuous field of self-adjoint operators over the same Hilbert space (\cref{sec:twsited_transformation_group_C_star_algebras}), meaning that the spectrum of operators varies continuously with respect to the parameter of the field of self-adjoint operators. 

\begin{remark}[Spectral gap and continuous field of self-adjoint operators]\label{remark:spectral_gap_continuous_field_self_adjoint_operators}
Let $\Omega$ be a locally compact Hausdorff space and $f: \Omega \to A_{\text{sa}}$ a continuous function, where $S_{\text{sa}}$ is the set of self-adjoint operators of the C* algebra $A$, let $ g \subset \mathbb{R}$ with $g$ open such that
$$ g \cap Sp(f(\omega)) = \emptyset, \; \forall \omega \in \Omega. $$
Then, the continuity of the spectrum for normal elements tell us that 
$$ g \cap \overline{ \left( \cup_{\omega \in \Omega} Sp(f(\omega)) \right)} = \emptyset. $$
\end{remark}

\section{Universal C* algebras}
\label{sec:Universal_C_star_algebras}

We have seen that *-homomorphisms of C* algebras are automatically continuous, that is, the algebraic component of *-homomorphisms are tighly linked to their analytical structure. Now, we dive into the question of whether we can define a C* algebra giving only algebraic relations, or in more precise terms, do universal C* algebras exists? The general answer to this question is negative, however, there are special cases where those universal C* algebras\index{universal C* algebra} exist, like the torus (\cref{section:C_star_generated_by_a_unitary}) and the compact operators on separable Hilbert spaces (\cref{section:algebra_of_compact_operators}).  

\begin{example}\label{example:bad_universal_C_star_algebras}
The following are examples of ill defined universal C* algebras

\begin{itemize}
    \item \textbf{Heisenberg relation:}\index{Heisenberg relation} there is no unital C* algebra $A$ with elements $a,b$ that satisfy $ab - ba = 1_A$ (\citep{151951}, \citep{sundar_notes_202}). The relation $ab -ba =1$ is a well known relation in physics, because it is the relation between the momentum and position operators, however, the momentum operator is unbounded, thus it cannot belong to a C* algebra.
    
    \item \textbf{Algebra of polynomials:}\index{algebra of polynomials} Let $a$ be variable and set 
    $$B = \{ \sum_{i =0}^{n} \alpha_i a^i : \; \alpha \in \mathbb{C} \}$$
    with $a^0 = 1$, then $B$ is a *-algebra with the involution given by $ (\sum_{i =0}^{i \leq n} \alpha_i a^i )^* = \sum_{i =0}^{i \leq n} \alpha_i^* a^i$. The Weierstrass theorem (\citep[Corollary 2.35 (ii)]{allan_introduction_2011}) tell us that $B$ is a dense *-algebra of $C(K)$ with $K$ a compatc subset of $\mathbb{R}$, therefore, the continuous calculus (\cref{theorem:continuous_functional_calculus}) tell us that $B$ is a dense *-algebra of $C^*(1_A,a)$, with $a$ being a self adjoint operator of a unital C* algebra $A$. 
    
    Since the norm of $\sum_{i =0}^{i \leq n} \alpha_i a^i$ on $C(K)$ depends on $K$, given that $K$ can be any compact subset of $\mathbb{R}$ we have that the norm of $\sum_{i =0}^{i \leq n} \alpha_i a^i$ can take any value in the range of $|\sum_{i =0}^{i \leq n} \alpha_i a^i|$ over $\mathbb{R}$, where $\sum_{i =0}^{i \leq n} \alpha_i a^i$ is considered as a polynomial in one variable over $\mathbb{R}$, therefore, we can assign many nonequivalent norms to $B$ that make them into a dense *-algebra of a C* algebra. Consequently, $B$ does not define uniquely a C* algebra.
    
    Also, this setup gives an example where \cref{lemma:extending_star_homomorphisms_into_C_star_homomorphisms} cannot be applied, because the norm of $\sum_{i =0}^{i \leq n} \alpha_i a^i$ depends on the C* algebra it lies in. So, if $K_0 \subset K_1$ with both compact sets then 
    $$\| \sum_{i =0}^{i \leq n} \alpha_i a^{i} \|_{C(K_0)} \leq \| \sum_{i =0}^{i \leq n} \alpha_i a^{i} \|_{C(K_1)},$$
    and there are elements where the equality does not holds, thus, we can only set a C* algebra homomorphism $\phi: C(K_0) \to C(K_1)$ which corresponds to the identity on the polynomials if $K_0 \subset K_1$.
\end{itemize}
\end{example}

As we have seen from \cref{example:bad_universal_C_star_algebras}, there are many nonequivalent C* algebras with isomorphic dense *-algebras, therefore, we have no trivial way of assigning a C* norm to a *-algebra given by a set of generators and relations between those generators. At first sight this appears to go against what is mentioned in some literature, specially physics like literature, were authors refer to C* algebras generated by a given set of generators with some set of relations, so, how de we make sense to those claims? There are many approaches to defining C* algebras by relations, one could go full category theory mode and follow the approach of Loring \citep{loring_c-algebra_2010}, or we could take a simpler route and follow the approach of Blackadar \citep[Section II.8.3]{blackadar_operator_2006} and \citep{blackadar_shape_1985}. I this text, we will focus on the former approach.

The core idea behind Blackadar approach is to explicitly give bounds on the norm of the elements of the *-algebra, so, we need two ingredients
\begin{itemize}
    \item $\mathcal{G} = \{ x_i : i \in \Omega \} $ a set of generators
    \item $\mathcal{R}$ a set of relations that implicitly or explicitly set a bound on the norm of the generators, the most common types of relations used are polynomial relations,
    $$\left\|p\left(x_{i_1}, \cdots, x_{i_n}, x_{i_1}^*, \cdots, x_{i_n}^*\right)\right\| \leq \eta$$
    where $p$ is a polynomial in $2n$ noncommuting variables with complex coefficients and $\eta \geq 0$. Also, the relation must be realizable among bounded operators on a Hilbert space and place a bound on the norm of each generators as an operator.
\end{itemize}

Then, a representation of $(\mathcal{G} | \mathcal{R})$ is a set of bounded operators $\{ T_i : i \in \Omega \}$ on a Hilbert space satisfying the relations $\mathcal{R}$. Each representation of $(\mathcal{G} | \mathcal{R})$ defines a *-representation of the free algebra $\mathcal{A}$ on the set $\mathcal{G}$, thus, we can use them to define a C* semi norm on $\mathcal{A}$ as follows
$$ \| x \| = \sup\{ \| \pi(x) \| : \pi \text{ is a representation of } (\mathcal{G} | \mathcal{R})\}, \text{ for } x \in \mathcal{A}. $$

Since the relations $\mathcal{G}$ set a bound on $\pi(x)$ then $\| x\|$ is finite, set 
$$ I = \{ x \in \mathcal{A} : \|x \| = 0 \},$$
then $I$ is an ideal on $\mathcal{A}$. The semi-norm descends to a C*-norm on $\mathcal{A} / I$, and the completion of $\mathcal{A} / I$ with respect to this norm is called the \textbf{universal C* algebra on }  $(\mathcal{G} | \mathcal{R})$ (\citep[Section II.8.3]{blackadar_operator_2006}), and is denoted by
$$ C^* (\mathcal{G} |  \mathcal{R}). $$
Notice that if you add a generator called $1$ with relations $1 = 1^* = 1^2$ and $1x=x1$ for each element of $\mathcal{G}$ you get a unital C* algebra. Also, any C* algebra can be described as a universal C* algebra, with the set of generators given by all its elements, and the relation given by all the algebraic relations between its elements.

If all the relations in $\mathcal{R}$ take the form 
$$\left\|p\left(x_{i_1}, \cdots, x_{i_n}, x_{i_1}^*, \cdots, x_{i_n}^*\right)\right\| = 0,$$
denote by $\mathcal{I} = \{ x \in A : x = p\left(x_{i_1}, \cdots, x_{i_n}, x_{i_1}^*, \cdots, x_{i_n}^*\right) ,\; p \in \mathcal{R} \}$, then we can use $I = \mathcal{A}\mathcal{I} + \mathcal{I} \mathcal{A} + \mathcal{A} \mathcal{I} \mathcal{A}$ to define a new *-algebra 
$$\hat{\mathcal{A}} = \mathcal{A} / I. $$
Then, if $C^*(\mathcal{G} | \mathcal{R})$ exists we call it the \textbf{enveloping C* algebra\index{enveloping C* algebra} of} $\hat{A}$ and we denote it by $C^*(\hat{\mathcal{A}})$ \citep[chapter 2]{sundar_notes_202}. Notice that the concept of enveloping C* algebra is a special case of a universal C* algebra, and the automatic continuity of C* homomorphism assure us that any C* algebra is isomorphic to its enveloping C* algebra. A more interesting example of enveloping C* algebras comes in the form of Banach *-algebras, where we encounter an example of automatic continuity that ensures the existence of the enveloping C* algebra of a Banach *-algebra,

\begin{proposition}[Automatic continuity\index{automatic continuity} between Banach *-algebras and C* algebras (Proposition 5.2 \citep{takesaki_theory_2003})]\label{proposition:automatic_continuity_banach_star_algebras}

If $\pi$ is a *-homomorphism of a Banach *-algebra $A$ into a $C^*$-algebra $B$, then
$$
\|\pi(x)\| \leq\|x\|, \quad x \in A
$$

\end{proposition}

This results is fundamental for the representation theory of Banach *-algebras and the representation theory of locally compact groups.

Let $A$ be a *-algebra and assume that $C^*(A)$ exists, then any representation of $A$ on a Hilbert space factors through $C^*(A)$,

\begin{proposition}[Universal property of universal of universal C* algebras\index{universal C* algebra!universal property} (Proposition 2.2 \citep{sundar_notes_202})]\label{proposition:factoring_representations_with_enveloping_C_star_algebras}
Let $\mathcal{A}$ be a *-algebra, assume that $C^*(\mathcal{A})$ exists, let be $B$ a $C^*$-algebra and $\pi: \mathcal{A} \rightarrow B$ be a *-homomorphism. Then, there exists a unique *-homomorphism $\widetilde{\pi}: C^*(\mathcal{A}) \rightarrow B$ such that $\widetilde{\pi}(x)=\pi(x)$ for every $x \in \mathcal{A}$.
\end{proposition}

On \cref{chap:fourier_analysis} we will look into the enveloping C* algebra of a Banach *-algebra, it is called the twisted crossed product and plays an important role in our analysis. Enveloping C* algebras is one case where *-homomorphisms defined over a dense *-algebra can be extended to C* homomorphisms on the whole C* algebra because by definition we know that all the representations of the dense *-algebra have a bounded norm. The sub-algebras of C* algebras whose representations are norm decreasing are called \textbf{core subalgebras}\index{core subalgebras} (\citep[section 4]{exel_envelope_2008}), and are pretty interesting because we still have automatic continuity over those algebras. In \cref{sec:smooth_subalgebras} we will encounter another example of core subalgebras, and they will be core sub algebra because they are spectrally invariant with respect to their C* algebras, which implies automatic continuity of *-homomorphisms (\cref{lemma:automatic_continuity_spectrally_invariant_sub_alegbras}).

\begin{remark}\label{remark:bound_in_the_generators_guaranties_existence_of_enveloping_c_star_algebra}
Let $(\mathcal{G},\mathcal{R})$ be a set of generators an relations among the generators, then if $\mathcal{R}$ set bounds on the norm of each generator then $C^*(\mathcal{G} | \mathcal{R})$ exists.
\end{remark}

\subsection{Examples}
\label{section:universal_C_star_algebras_examples}

Universal C* algebras and enveloping C* algebras are commonly used in applications of non-commutative geometry, in part because their description in terms of generators is very similar to other algebraic descriptions where there is no topology. We need to be cautious, because even if in some cases the algebraic description of the algebras may suggest that any element of the algebra can be described as an infinite series that is often not the case, as is exemplified by the case of continuous functions over the torus (\cref{section:convergence_of_the_Fourier_series}).

\textit{From now on every-time we include the identity ($1$) among the generators of a *-algebra we assume that commutation relations with all the elements of the *-algebra are taken for granted unless otherwise stated.}

In \citep[Examples II.8.3.3]{blackadar_operator_2006} you can look at examples of universal C* algebras, we will study one of those examples, the Non-Commutative Brilluoin Torus torus (\cref{definition:non_commutative_brillouin_torus}), which we will analyze as the enveloping C* algebra of a Banach *-algebra in \cref{sec:twisted_crossed_products}.

\subsubsection{Untization}
\label{section:unitization_universal_C_star_algebra}

Let $A$ be a C* algebra without a unit, then, from \cref{lemma:uniqueness_of_unitization} we know that $A^+$ is isomorphic to $C^*(A, 1_B)$ for any unital C* algebra $B$ such that $A \subset B$, thus, we must that $A^+$ is a universal C* algebra
$$ A^+ \simeq C^*(\mathcal{G} | \mathcal{R}), \; \mathcal{G} = A \cup \{1 \}, \; \mathcal{R} = \{ 1 a = a 1 = a, 1 = 1^* \}. $$
In a similar way, if $A$ is unital, we have that $A \oplus \mathbb{C}$ has the following description as a universal C* algebra
$$ A \oplus \mathbb{C} \simeq C^*(\mathcal{G} | \mathcal{R}), \; \mathcal{G} = A \cup \{\hat{1}\}, \; \mathcal{R} = \{ \hat{1} a = a \hat{1} = 0, \hat{1} = \hat{1}^* \}. $$

There is no universal description for $\mathcal{M}(A)$ that we are aware of, nonetheless, it has a nice description in terms of double centralizers for $A$. A double centralizer for $A$, is a pair $(L,R)$ of functions $L,R: A \to A$, satisfying
$$R(x) y = x L(y),  $$
for all $x,y \in A$. Turns out that $\mathcal{M}(A)$ is isomorphic to the *algebra of double centralizers of $A$ (\citep[Definition 2.2.11]{wegge-olsen_k-theory_1993}).

\subsubsection{Algebra generated by a self-adjoint element}
\label{section:algebra_generated_by_a_self_adjoint}
The following arguments are in \citep[Examples II.8.3.2]{blackadar_operator_2006}. In \cref{example:bad_universal_C_star_algebras} we saw that there is no enveloping C* algebra of the algebra of complex polynomials in one self-adjoint variable, however, if we set a restriction on the norm of the self adjoint element we can define a universal C* algebra. Set $\mathcal{G} = \{ x, 1\}$ and $\mathcal{R} = \{x =x^*, 1=1^*=1^2, 1 x = x1 = x, \| x \| \leq 1 \}$, then the continuous functional calculus (\cref{theorem:continuous_functional_calculus}) tell us that
$$C^*(1,x) = C(\text{Sp}(x)),$$
also, from \cref{proposition:automatic_continuity_C_star_algebras} we now that $\|x\| = \rho(x)$ and since $x$ is self adjoint we get that $\text{Sp}(x) \subset \mathbb{R}$. Therefore, $\text{Sp}(x) \subseteq [-1,1]$, thus
$$ C^*(\mathcal{G} | \mathcal{R}) \simeq C([-1,1]), $$
notice that the continuous functional calculus tell us that $C^*(\mathcal{G} | \mathcal{R}) \subseteq C([-1,1])$, and the equality is obtained with the self-adjoint element $f(x) = x$ for $f \in C([-1,1])$, thus we get the isomorphism.

If we want a C* algebra without a unit, then we need to take the unit out of $C([-1,1])$ and having it contained any polynomial in the variable $x$. Since the algebra of polynomials in one variable over $[-1,1]$ and no constant term is a self-adjoint sub-algebra of $C([-1,1])$ that separates points and vanishes nowhere, by the Stone-Weierstrass theorem in the locally compact version (\citep[Stone-Weierstrass notes, Corollary 6]{quigg_real_2005}) we know that is is dense in $\{ f \in C([-1,1]) \; | \; f(0)=0 \} = C_0([-1,0) \cup (0,1])$, therefore,
$$C^*(x| x= x^*, \|x\|\leq 1) \simeq C_0([-1,0) \cup (0,1]) .$$

If instead we consider a positive element i.e. $\mathcal{G} = \{ x, 1\}$ and $\mathcal{R} = \{x =x^*, 1=1^*=1^2, 1 x = x1 = x, \| x \| \leq 1 , \text{Sp}(x) \subset \mathbb{R}^+ \}$ we get that 
$$ C^*(\mathcal{G} | \mathcal{R}) \simeq C([0,1]), $$
and the C* algebra generated by a positive element with norm less then one and without a unit is isomorhic to
$$ C_0((0,1]). $$
If we deal with a normal element with norm less than one we get a unital C* algebra
$$ C_0(\mathbb{D}) ,$$
 with $\mathbb{D} = \{ c \in \mathbb{C} : \; |c|\leq 1 \}$ and a non unital C* algebra
$$ C_0(\mathbb{D}-\{ 0 \}). $$

\subsubsection{Universal C* algebra generated by a unitary}
\label{section:C_star_generated_by_a_unitary}

The following argument is in \citep[Proposition 2.4]{sundar_notes_202}. This is one of the most important C* algebras of this document, because we will study some of its non-commutative analogous. If $\mathbb{G} = \{ u, 1 \}$ and $\mathcal{R} = \{ u u^* = u^* u =1 \}$ then
$$ C^*(\mathcal{G} | \mathcal{R}) = C(\mathbb{T}).$$\index{$C(\mathbb{T})$}

First, recall that $\| u \| = 1$, by \cref{remark:bound_in_the_generators_guaranties_existence_of_enveloping_c_star_algebra} $C^*(\mathcal{G} | \mathcal{R})$ exists. Recall that the continuous functional calculus (\cref{theorem:continuous_functional_calculus}) tells us that there is a *-homomorphism 
$\phi : \text{Sp}(u) \to C^*(1,u) = C^*(\mathcal{G} | \mathcal{R})$,
that sends $f(z) =z$ to $u$. Since $\text{Sp}(u) \subseteq \mathbb{T}$, we can compose the previous C* homomorphism with the restriction $i^*: C(\mathbb{T}) \to C(\text{Sp}(u))$ obtained from the inclusion of compact Hausdorff spaces $i: \text{Sp} \to \mathbb{T}$, which gives us a *-homomorphism
$$ \phi \circ i^* : C(\mathbb{T}) \to C^*(\mathcal{G} | \mathcal{R}),$$
that maps $f(z)=z$ into $u$.
On the other side, from the Weierstrass theorem (\citep[Corollary 2.35 (ii)]{allan_introduction_2011}), we know that $C(\mathbb{T}) \simeq \overline{\mathcal{A}_u }$
with $\mathcal{A}_u$ the *-algebra of polynomials in one unitary variable given by $u:= f(z) =z$ for $z \in \mathbb{T}$, because $\mathcal{A}_u$ separates points on $\mathbb{T}$ and is close under involution (\citep[Exercise 11.7.8]{lebl_basic_2022}). Therefore, maximality of the norm of enveloping C* algebras (\cref{proposition:factoring_representations_with_enveloping_C_star_algebras}) tell us that there is a *-homomorphism 
$$\pi: C^*(\mathcal{G} | \mathcal{R}) \to C(\mathbb{T})$$ 
that sends $u$ to $f(z) = z$. Set $\eta =  \phi \circ i^*$, then $(\eta \circ \pi) (u) = u$ and $( \pi \circ \eta) (f(z) = z) = (f(z) = z)$, so, $\| \eta(p(u,u^*)) \| = \|p(u,u^*)\|$ for any complex polynomial in the variables $u, u^*$ because C* homomorphisms are norm decreasing. Since the elements of the form $p(u,u^*)$ and $\eta(p(u,u^*))$ are dense in $C^*(\mathcal{G} | \mathcal{R})$ and $C(\mathbb{T})$ respectively we get that $$ C^*(\mathcal{G} | \mathcal{R}) \simeq C(\mathbb{T}). $$      

In \cref{section:algebra_continuous_functions_locally_comp_space} we look into an important representation of $C(\mathbb{T})$.

\subsubsection{Matrices}
\label{example:matrices_gen_C_star_algebras}

The following example shows us how a commonly used C* algebra can be described using generators.
Set $\mathcal{G} = \{ e_{ij} : 1 \leq i,j \leq j \}$ with 
$$
\mathcal{R}=\left\{e_{i j}^*=e_{j i}, e_{i j} e_{k l}=\delta_{j k} e_{i l}: 1 \leq i, j, k, l \leq n\right\},
$$
where $\delta_{j k}$ is the Kronecker symbol. The relations tell us that $e_{jj}$ are projections, thus have norm equal to one, and also $\| e_{i,j} \|^2 = \| e_{ji}e_{i,j}\| = \| e_{jj}\| =1$, thus all the generators have bounded norm, consequently $C(\mathcal{G} | \mathcal{R})$ exists. Turns our that $C(\mathcal{G} | \mathcal{R}) \simeq M_n (\mathbb{C})$ (\citep[Examples II.8.3.2 item iv]{blackadar_operator_2006}), with $M_n (\mathbb{C})$ the algebra of matrices of size $n \times n$ with complex entries, moreover $M_n (\mathbb{C})$ is simple \citep[Lemma 1.2]{sundar_notes_202}. Recall that a C* algebra is simple if it has no nontrivial to sided ideals, or equivalently, for every non-zero representation $\pi$ we have that $\| \pi(a) \| = \|a\|$, for any element of the C* algebra (\citep[Lemma 1.3]{sundar_notes_202}).

If $a \in M_n(\mathbb{C})$\index{$M_n(\mathbb{C})$}, then $a = (a_{ij})_{1 \leq i,j \leq n}$ and $a_{ij} \in \mathbb{C}$, these are the coefficients of the elements of $\mathcal{G}$, thus $a$ can be seen as the polynomial
$$ a = \sum_{1 \leq i,j \leq n} a_{ij} e_{ij}.$$
Also, the following formula
$$ \text{tr}(a) \sum_{1 \leq i \leq n} a_{ii} $$
defines a tracial continuous functional on $M_n(\mathbb{C})$ i.e. $\text{tr}(ab) = \text{tr}(ba)$.

The algebra of matrices is special because any finite-dimensional C* algebra is isomorphic to a direct sum $\oplus_{r=1}^{m} M_{n_r} (\mathbb{C})$ (\citep[Examples II.8.3.2 item iv]{blackadar_operator_2006}). 

\subsubsection{Algebra of compact operators}
\label{section:algebra_of_compact_operators}

Most of the following arguments are in \citep[Chapter 1]{sundar_notes_202}. Now we look into a C* algebra that has a deep connection with the matrix algebras, this is the C* algebra of compact operators on a separable Hilbert space. Let $H$ be a separable Hilbert space, then, $H$ has a countable basis $\xi = \left\{\xi_i\right\}_{i \in \mathbb{N}}$ (\cref{lemma:separability_Hilbert_spaces_and_orthonormla_basis}), and $S \in B(H)$ can be seen as an infinite matrix $S = [s_{i,j}]_{i,j \mathbb{N}}$ (\cref{sec:infinite_mattrices_and_bounded_operators}).

The infinite matrix representation of a bounded operator is intuitive, however, checking that a given infinite matrix satisfies certain properties is non-trivial, for instance, there are examples of criteria that generate compact operators e.g. infinte matrices whose entries come from an absolutely summable sequence (\citep{1243780}) or matrices whose entries are square summable  (\citep{553928}), on the otherside, there are counter examples of matrices that one could naively consider as compact operators but actually are not compact operators e.g. matrices whose associated operator satisfy the relation  (\citep{4470683}) 
$$ \lim_{i \to \infty} \| T(\xi_i) \| = 0.$$ 

Let $H$ be a separable Hilbert space, then an operator on $H$ is compact iff it is the limit of a norm-convergent sequence of finite rank operators \citep[Chapter 3 Proposition 29]{melrose_introduction_2014}. denote by $\mathcal{K}(H)$\index{$\mathcal{K}(H)$} the *-algebra of compact operators on $H$ and $F(H)$\index{$F(H)$} the *-algebra of finite rank operators\index{*-algebra of finite rank operators} on $H$, we have that $\mathcal{K}(H)$ is a normed closed two sided ideal in $B(H)$ that is closed under taking adjoints, thus, $\mathcal{K}(H)$ is a C* algebra and $F(H)$ is a dense *-algebra inside $\mathcal{K}(H)$ (\citep[Chapter 1 introduction]{sundar_notes_202}).

Given the ortonormal basis $\xi = \left\{\xi_i\right\}_{i \in \mathbb{N}}$ of the Hilbert space $H$, let $E_{i j} \in B(H)$ be the operators defined on the basis by by $E_{ij}(\gamma) = \xi_i \langle \gamma, \xi_j \rangle$. Observe that
$$
\begin{aligned}
E_{i j} E_{k l} & =\delta_{j k} E_{i l} \\
E_{i j}^* & =E_{j i}
\end{aligned}
$$
for $i,j,k,l \in \mathbb{N}$, this a generalization of the set of generators of the algebra of matrices with entries in $\mathbb{C}$ in \cref{example:matrices_gen_C_star_algebras}, and by the universality of $M_n(\mathbb{C})$ we have a *-homomorphism $\pi_n : M_n(\mathbb{C}) \to A_n, \; \pi_n(e_{ij}) = E_{ij}$ with $A_n$ the linear span of $\{ E_{ij} : 1 \leq i,j \leq n  \} $, moreover, $A_n \subset \mathcal{K}(H)$ because each $E_{ij}$ has finite rank. 

If $L \in F(H)$ there is an ortonormal set $\{ e_k \}_{1 \leq k \leq n} \subset H$ such that (\citep[Chapter 3 Lemma 26]{melrose_introduction_2014}) 
$$ L(h) = \sum_{1 \leq i,j \leq n} c_{i,j}\langle h, e_{j} \rangle e_i. $$
For any $\epsilon > 0$ there is $m \in \mathbb{N}$ such that each of the elements of $\{ e_k \}_{1 \leq k \leq n}$ can be approximated by linear combinations of $\{ \xi_0, \cdots, \xi_n \}$ with a precision of $\epsilon$, therefore, we can construct a sequence of operators on $\cup_{n\in \mathbb{N}} A_n$ that are converge in norm to $L$. The previous statement tell us that $\cup_{n\in \mathbb{N}} A_n$ is dense in $F(H)$, consequently it is dense in $\mathcal{K}(H)$.

The density of $\cup_{n \geq 1} A_n$ in $\mathcal{K}(H)$ along with the isomorphisms $\pi_n : M_n(\mathbb{C}) \to A_n, \; \pi_n(e_{ij}) = E_{ij}$ can be used to show that $\mathcal{K}(H)$ is a simple C* algebra (\citep[Theorem 1.1]{sundar_notes_202}), meaning that any non-zero *-homomorphism starting from $\mathcal{K}(A)$ is an isometry (injective).

Hence, there is a description of $\mathcal{K}(H)$ as a universal C* algebra that resembles the description of $M_n(\mathbb{C})$:

\begin{proposition}[Proposition 1.4 \citep{sundar_notes_202}]\label{proposition_universal_description_C_star_algebra_compact_operators}
Let $A$ be a $C^*$-algebra. Suppose there exists a system of matrix units $\left\{e_{i j}: i, j \in \mathbb{N}\right\}$ in $A$, i.e. the set $\left\{e_{i j}: i, j \in \mathbb{N}\right\}$ satisfies the following relations.
$$
\begin{gathered}
e_{i j} e_{k l}=\delta_{j k} e_{i l} \\
e_{i j}^*=e_{j i}
\end{gathered}
$$
for $i, j, k, l \in \mathbb{N}$. Then, there exists a unique *-homomorphim $\pi: \mathcal{K}(H) \rightarrow A$ such that for $i, j \in \mathbb{N}, \pi\left(E_{i j}\right)=e_{i j}$. 
\end{proposition}

For any two separable Hilbert spaces $H_1, H_2$ we have $\mathcal{K}(H_1) \simeq \mathcal{K}(H_2)$, so, we denote with $\mathcal{K}$\index{$\mathcal{K}$} the C* algebra of compact operators\index{C* algebra!of compact operators} on a separable Hilbert space. From \cref{proposition_universal_description_C_star_algebra_compact_operators} we can characterize $\mathcal{K}$ as the C* algebra generated by the set $\mathcal{G} = \{ e_{ij} : i,j \in \mathbb{N} \}$ with 
$$
\mathcal{R}=\left\{e_{i j}^*=e_{j i}, e_{i j} e_{k l}=\delta_{j k} e_{i l}: 1 \leq i, j, k, l \in \mathbb{N} \right\}.
$$

Also, any non-degenerate representation of $\mathcal{K}(H)$ is unitarily equivalent to the identity representation on $H$ (\citep[Theorem 1.6]{sundar_notes_202}).

$\mathcal{K}(H)$ can be intuitively understood as the C* algebra that lies between the finite matrices and the rest of bounded linear operators, because we are able to approximate the operator with finite matrices, instead of just approximating its norm. So, let $a = (a_{i,j})_{i,j \in \mathbb{N}}$ a compact operator, we know that it is the norm limit of finite matrices but we should not interpret the notation $a = (a_{i,j})_{i,j \in \mathbb{N}}$ as meaning that in the operator norm we have $a = \lim_{n \to \infty} a_n$ with $a_n = (a_{i,j})_{i,j \leq n}$. Instead we should interpret the notation $a = (a_{i,j})_{i,j \in \mathbb{N}}$ as giving us the operator over $\sum_{i \in\mathbb{N}} H_i, \; H_i = \mathbb{C}$ that acts as
$$  a(h) = (\eta_i)_{i \in \mathbb{N}}, \; \eta_i = \sum_{j \in \mathbb{N}} a_{i,j}h_j.   $$

Let $H$ be a separable Hilbert space and $T \geq 0$ and $T \in B(H)$, if $\{ \xi_i \}_{i \in \mathbb{N}}$ an orthonormal basis for $H$, then the following defines a trace on $B(H)_{\text{pos}}$ (\citep[Section I.8.5]{blackadar_operator_2006}),
$$ \text{Tr}(T) = \sum_{i \in \mathbb{N}} \langle T \xi_i, \xi_i \rangle \in [0,\infty].$$\index{$\text{Tr}(\cdot)$}
This trace has the following properties (\citep[Item I.8.5.1]{blackadar_operator_2006}):
\begin{itemize}
    \item $\text{Tr}(T)$ is independent of the orthonormal basis
    \item $\text{Tr}(T) = \text{Tr}(U^* T U)$ for any unitary $U$
    \item $\text{Tr}(T) \geq \| T \|$
    \item If $S \leq S \leq T$ then $\text{Tr}(S) \leq \text{Tr}(T)$
    \item $\text{Tr}(T) < \infty$ iff $T$ is compact and $\sum_{n \in \mathbb{N}}\mu_n (T) < \infty$, in which case $\text{Tr}(T) = \sum_{n \in \mathbb{N}}\mu_n (T)$. The values $\{\mu_n (T) \}_{n \in \mathbb{N}}$ are the eigenvalues of $|T|$.
\end{itemize}
 
Since we have described $\mathcal{K}(H)$ using the projections $E_{ij}$, we can check that 
$$ \text{Tr}(a) = \sum_{i \in \mathbb{N}} \langle a \xi_i, \xi_i \rangle = \sum_{n \in \mathbb{N}} a_{nn}, $$
 for $a \in \mathcal{K}(H)$ written as $a = (a_{ij})_{i,j \in \mathbb{N}}$. $\text{Tr}$ is not a continuous trace over $\mathcal{K}_{\text{pos}}$ because it is not a bounded trace, that is, the set of compact operators is bigger than the set of trace class operators \citep{1822286}. However, $\text{Tr}$ is lower semi-continuous (\citep[Item I.8.5.2]{blackadar_operator_2006}). In \cref{example:infinte_matrices_with_rapid_decay}, we will look at a *-algebra that is dense in $\mathcal{K}(H)$ and is spectrally invariant and where $\text{Tr}$ is continuous with respect to a stronger topology than the norm topology.

\subsection{Toeplitz algebra}\index{Toeplitz algebra}
\label{section:Toeplitz_algebra}

Most of the following arguments are in \citep[Chapter 3]{sundar_notes_202}. Let $\mathcal{G} = \{ v,1\}$ and $\mathcal{R} = \{ v^* v = 1 \}$, then $\| \pi(v) \| = \sqrt{\|\pi(v)^* \pi(v)\|} = 1$, therefore $\|v\| =1$ and $C(\mathcal{G}| \mathcal{H})$ exists, it is denoted by $\mathcal{T}$. We have assumed that $\pi(v)$ exists, so, is this the case? to answer this question positively we need to provide a Hilbert space $H$ and an element $b \in B(H)$ such that $b^* b = 1_H$, in which case we would have a *-homomorphism
$$\phi: \mathcal{T} \to C^*(b,1_H)$$
where $C^*(b,1_H)$ denotes the sub C* algebra of $H$ that is generated by $b,1_H$, i.e. the closure of the *-algebra of complex polynomials on variables $b,b^*,1_H$ under the norm of $B(H)$.

Let $H = l^2(\mathbb{N})$, $\left\{\delta_n\right\}_{n \geq 1}$ be the standard orthonormal basis for $l^2(\mathbb{N})$, and $S: l^2(\mathbb{N}) \rightarrow \ell^2(\mathbb{N})$ be the unique operator such that $S\left(\delta_n\right)=\delta_{n-1}$, that is, the shift operator. Then $S^*$ is the left shift in $l^2(\mathbb{N})$ (\citep[Section 9a example 7.1]{garret_real_2020}) i.e. $S\left(\delta_{n+1}\right)=\delta_{n}$, and $S^* S = 1_H$, thus the universal property of $\mathcal{T}$ (\cref{proposition:factoring_representations_with_enveloping_C_star_algebras}) guaranties that there is a unique *-homomorphism
$$ \pi: \mathcal{T} \to C^*(S,1), \; \pi(v) =  S. $$

Turns out that $\pi$ is an isomorphism, and that result is by no means trivial, it is called Coburn's theorem (\citep[Theorem 3.3]{sundar_notes_202}) and its proof goes through establishing the short exact sequence of C* algebras
$$0 \longrightarrow \mathcal{K} \longrightarrow C^*(S,1) \longrightarrow C(\mathbb{T}) \longrightarrow 0.$$
Let us take a close look at this sequence of C* algebras. Let $P = 1 - s s^*$, then $P$ is a projection, using the notation $E_{0,0} = P$ and $E_{m,n} = S^{m} P (S^*)^n$ you can check that 
$$ (E_{m,n})^* = E_{n,m}, \;  S^* E_{0,0} = E_{0,0} S = 0, $$
which implies that $E_{j,i} E_{k,l} = \delta_{j k} E_{i, l} 1 \leq i, j, k, l \in \mathbb{N}$, therefore there is *-homomorphism $\phi: \mathcal{K} \to C^*(S,1)$ which is injective because $\mathcal{K}$ is simple. Moreover, $\phi(\mathcal{K})$ is a two sided closed ideal of $C^*(S,1)$, it is closed because the image of a C* homomorphism is closed (\cref{proposition:ideal_on_C_star_algebras}), and the following algebraic relations assure us that it is a two-sided ideal
$$ S^k E_{m,n}(S^*)^l = E_{m+k,n+l}, \; (S^*)^k E_{m,n} S^l = \delta(m\geq k) \delta(n \geq l) E_{m-k,n-l},$$
with $\delta(m \geq k) = 1$ if $m \geq k$ else $\delta(m \geq k) = 0$. You may wonder how the aforementioned relations gives us that information, in such argument we used the universality of both $\mathcal{K}$ and $C^*(S,1)$ as follows, to show that $ab \in \phi(\mathcal{K}), \; ba \in \phi(\mathcal{K})$ when $a \in \phi(\mathcal{K}), \; b \in C^*(S,1)$ we only need to check it in finite polynomials on $E_{i,j}$ (those are dense in $\phi(\mathcal{K})$), and finite polynomials on $S, S^*$ (those are dense in $C^*(S,1)$), because $\phi(\mathcal{K})$ is closed in $C^*(S,1)$.

The second part of the exact sequence comes in the form of quotients of two-sided closed ideals of C* algebras, so, we have shown that $\phi(\mathcal{K})$ is a two-sided close ideal of $C^*(S,1)$, therefore the map $\psi : C^*(S,1) \to C^*(S,1)/ \phi(\mathcal{K})$ is a surjective C*-homomorphism by \cref{proposition:ideal_on_C_star_algebras}, whose kernel is $\phi(\mathcal{K}) \simeq \mathcal{K}$ by definition, thus we have the following short exact sequence of C* algebras
$$0 \longrightarrow \mathcal{K} \longrightarrow C^*(S,1) \longrightarrow C^*(S,1) / \phi(\mathcal{K})  \longrightarrow 0.$$
The final step is to show that $C^*(S,1) / \phi(\mathcal{K}) \simeq C(\mathbb{T})$, which is equivalent to $$C^*(S,1) / \phi(\mathcal{K}) \simeq C(u, u^*u = u u^* = 1)$$
by \cref{section:C_star_generated_by_a_unitary}. Notice that $\psi(S^* S)= \psi(S)^* \psi(S) = \psi(1) =1$, also, $\psi(S S^*) = \psi(S) \psi(S)^* = \psi(1-P)$, so, since $P \in \phi(\mathcal{K})$ we have that $\psi(S S^*) = 1$, meaning that $\phi(S)$ is a unitary in $C^*(S,1) / \phi(\mathcal{K})$. Also, the *-algebra of polynomials in $\psi(S), \psi(S)^*$ is dense in $C^*(S,1) / \phi(\mathcal{K})$ because the polynomials in $S, S^*$ are dense in $C^*(S,1)$ and the C* homomorphisms are continuous (\cref{proposition:automatic_continuity_C_star_algebras}). The desired isomorphism comes from \citep[Lemma 3.2]{sundar_notes_202} that establishes $\text{Sp}(\psi(S)) = \mathbb{T}$, thus the continuous functional calculus (\cref{theorem:continuous_functional_calculus}) says that 
$$C(\mathbb{T}) = C(\text{Sp}(\psi(S))) \simeq C^*(\psi(S)).$$

The Toepliltz algebra and its associated short exact sequence has been thoroughly studied, it is the starting point of the short exact sequence of C* algebras that we will use to describe topological insulators (\cref{sec:motivation_from_physics}), also, the Toeplitz extension and has a description in terms of the Hardy space (\citep[Section 2.1]{arici_toeplitz_2020}) that is well fitted to generalize it into higher dimensions (\citep[Section 2.2]{arici_toeplitz_2020}). It has also been studied from a more general point of view by Cuntz \citep[Theorem 4.2]{cuntz_topological_2007}.

\section{Tensor products}\index{C* algebra!tensor product}
\label{sec:C_star_alg_tensor_prod}

In \cref{sec:Universal_C_star_algebras} we have looked at examples where a *-algebra did and did not uniquely define a C* algebra as its completion, a similar discussion arises when working with tensor product of C* algebras, because the algebraic tensor product of two C* algebras may have many equivalent C* norms. However, there are a special kind of C* algebras where the algebraic tensor product has a unique C* norm, and as expected it coincides with the norm as operators on the tensor product of very specific Hilbert spaces. The study of tensor products of C* algebras is a technical subject and now we give a brief review of the main results that are relevant for us, we follow mainly the exposition on \citep[appendix T]{wegge-olsen_k-theory_1993}.

\begin{definition}[Algebraic tensor product\index{algebraic tensor product}]\label{definition:algebraic_tensor_product}

Let $A,B$ be two *-algebras over $\mathbb{C}$, then we denote by $A \odot B$\index{$A \odot B$} the set of elements
$$ \sum_{k = 1}^n a_k \otimes b_k, \; a_k \in A , b_k \in B$$
with the operations
\begin{itemize}
    \item \textbf{Addition:} $(a_0 \otimes b_0) + (a_1 \otimes b_0) = (a_0 + a_1) \otimes (b_0)$
    \item \textbf{Multiplication:} $(a_0 \otimes b_0) (a_1 \otimes b_1) = (a_0 a_1) \otimes (b_0 b_1)$
    \item \textbf{Involution:} $(a_0 \otimes b_0)^* = (a_1^* \otimes b_1^*)$
    \item \textbf{Scalar multiplication:} $\lambda (a_0 \otimes b_0) = (\lambda a_0( \otimes b_0 = a_0 \otimes (\lambda b_0)$ for $\lambda \in \mathbb{C}$
\end{itemize}
and satisfying the relations
$$ (a_0 + a_1)\otimes b = (a_0 \otimes b) + (a_1 \otimes b), \; a \otimes (b_0 + b_1) = (a \otimes b)_0 + (a \otimes b_1), \; a,a_0,a_1 \in A, \; b,b_0,b_1 \in B.$$
The addition on $A \odot B$ is distributive with respect to the multiplication. Under this conditions $A \odot B$ becomes a *-algebra over $\mathbb{C}$.
\end{definition}

Now we mention some of the properties of $A \odot B$,

\begin{proposition}[Properties of algebraic tensor product]\label{proposition:properties_algebraic_tensor_product}

\begin{itemize}
    \item If we consider $A,B$ as vector spaces, that is, there is no multiplication nor involution, then the vector space defined in \cref{definition:algebraic_tensor_product} is the unique vector space where bilinear maps on $A \times B$ becomes a linear map on $A \odot B$ (\citep[Proposition T.2.4]{wegge-olsen_k-theory_1993}).
    \item The description $t= \sum_{i=1}^n a_i \otimes b_i, \; t \in A \odot B$ is not unique, however, if $\{ \beta_i \}_{i \in I}$ is a basis for $B$ then every element of $A \odot B$ has a unique representation as a finite sum of the form $\sum_{i \in I} a_i \otimes \beta_i$ (\citep[Proposition T.2.8]{wegge-olsen_k-theory_1993}).
    \item If $\psi: A \times B \to C$ is a bilinear, multiplicative and *-preserving map then there is unique map $\phi: A \odot B \to C$ that extends $\psi$.
    \item If $\phi_A: A \to C$ and $\phi_B : B \to C$ are linear maps and $C$ is an algebra then 
    $$ (\phi_A \odot \phi_B)(a \otimes b) = \phi_A(a) \phi_B(b) $$
    is a linear map that is multiplicative if $\phi_A, \phi_B$ commute (\citep[Section T.2.12]{wegge-olsen_k-theory_1993}).
    \item The algebraic tensor product is canonically commutative i.e. $A \odot B \simeq B \odot A$.
\end{itemize}
\end{proposition}

The algebraic tensor product of *-algebras can be iterated to get three fold, four fold, etc, algebraic tensor products.

Now we look into the topological side. There are various ways to assign a C* norm to the tensor product $A \odot B$, with this we mean that there are potentially many norms $\alpha : A \odot B \to \mathbb{R}^+$ that satisfy the following properties
\begin{itemize}
    \item \textbf{Cross property:} $ \alpha( a \otimes b ) = \| a \|_A \| b \|_B $ for all $a \in A, \; b \in B$.
    \item \textbf{C* property:} $\alpha(c )^2 = \alpha( c c^* )$ for $c \in A \odot B$.  
\end{itemize}
Notice that the completion of $A \odot B$ with respect to $\alpha$ satisfying the above properties will gives us a C* algebra. In particular, if $A,B$ are C* algebras then, the set of all C* norms on $A \odot B$ has a minimum element and a maximum, those are the spatial norm denoted by $\sigma$ and the maximal C* norm denoted by $\mu$. Is important to mention that $A \odot B$ in general is not complete with respect to a C* norm, for an example look at \cref{sec:Continuous_functions_with_values_on_C_star_algebra}.

\begin{theorem}[(Theorem T.6.21 \citep{wegge-olsen_k-theory_1993})]\label{theoren:maximum_and_minimum_norms_C_star_algebras}
Every C* norm on $A \odot B$ has the cross property, additionally
$$ \sigma \leq \alpha \leq \mu $$
for every C* norm $\alpha$ on $A \odot B$.
\end{theorem}

The C* norm $\mu$ is defined to be the maximum norm over all possible C* semi-norms over $A \odot B$ (\citep[Definition T.6.6]{wegge-olsen_k-theory_1993}), more precisely, 
$$ \mu(t) = \sup \{ \beta(t) | \beta \text{ is a C* seminorm on } A \odot B \},$$
and it has an intuitive description as the supremum of $\|\pi(t) \|$ over all representations $\pi$ of $A \odot B$ that come from commuting representations of $A,B$ over the same Hilbert space $H$ i.e. $\pi_A(a) \pi_B (b) = \pi_B (b) \pi_A(a)$ as operators on $H$ (\citep[Definition T.6.8]{wegge-olsen_k-theory_1993}).

The spatial tensor product of two C* algebras $A,B$, which will be a C* algebra were $A \odot B$ is dense, has a norm that arises as the norm of operator over a Hilbert space obtained from the tensor product of two Hilbert spaces (\cref{definition:tensor_product_hilbert_spaces}), more precisely,

\begin{theorem}[Spatial tensor product\index{spatial tensor product} (Definition T.5.16 \citep{wegge-olsen_k-theory_1993})]\label{theorem:spatial_tensor_product_C_star_algebra}

Let $A_1, A_2$ be two C* algebras, and $\pi_i : A_i \to B(H_i)$ a faithful representation of the C* algebras (we know that it exists by \cref{theorem:faithfull_universal_representation_c_star_algebras}), define
$$ \sigma(t) = \| (\pi_1 \odot \pi_2)(t) \|, \; t \in A_1 \odot A_2,$$
with $\pi_1 \odot \pi_2$ the unique injective algebraic representation of $A_1 \odot A_2$ on $B(H_1 \otimes H_2)$ (\citep[Proposition T.5.1]{wegge-olsen_k-theory_1993}) satisfying
$$ (\pi_1 \odot \pi_2 )(a_1 \otimes a_2) = \pi_1(a_1) \otimes \pi_2(a_2) \in B(H_1 \otimes H_2). $$

The definition of $\sigma$ is a good definition because \cref{theorem:spatial_tensor_product_C_star_algebra} it is independent of the faithful representation of $A_1, A_2$ (\citep[Proposition T.5.15]{wegge-olsen_k-theory_1993}), moreover, the spatial norm on $A_1 \odot A_2$ can be expressed using only the set of states of $A_1$ and $A_2$ (\citep[Proposition T.5.14]{wegge-olsen_k-theory_1993}).
\end{theorem}

Let $\alpha$ be a C* norm on $A \odot B$, then we denote by $A \otimes_{\alpha} B$ the completion of $A \odot B$ with respect to $\alpha$, so, $A \otimes_{\alpha} B$ is a C* algebra. We have many possible C* norms on $A \odot B$, so, which one to chose? Fortunately there are a type of C* algebras where this question takes a simple form, and we will use them in our analysis

\begin{definition}[c.f. Definition T.6.18 \citep{wegge-olsen_k-theory_1993} (Nuclear C* algebra\index{C* algebra!nuclear})]\label{definition:Nuclear_C_star_algebra}
A C* algebra $A$ is nuclear when, for every C* algebra $B$, there is only one C* norm on $A \odot B$. In this case $\sigma = \mu$ and we denote by $A \otimes B$\index{$A \otimes B$} the unique C* algebra obtained as the completion of $A \odot B$ with respect to $\sigma$.
\end{definition}

For examples of non nuclear C* algebras you can look into \citep[Example 11.3.14]{kadison_fundamentals_1983_V2}, also in \citep[Chapter 11]{dales_introduction_2003} there are interesting results on non nuclear C* algebras.

\begin{remark}[Nuclear C* algebras and enveloping C* algebras]\label{remark:nuclear_C_star_algebra_enveloping_C_star_algebra}
Notice that if $A$ is a nuclear C* algebra then we have that
$$ A \otimes B \simeq C^*(A \odot B) $$
for any C* algebra $B$, because there is only one C* norm on $A \odot B$. 
\end{remark}

\begin{remark}[Nuclearity of the tensor product and associativity]\label{label:nuclearity_of_tensor_product_and_associativity}
Let $A,B$ be nuclear C* algebras, then 
\begin{itemize}
    \item \textbf{Tensor product:} $A \otimes B$ is also a nuclear C* algebra (\citep[Remark 4.9]{pisier_tensor_2021}), thus, for any C* algebra $C$ we have that $(A \times B) \otimes C \simeq A \otimes (B \otimes C)$.
    \item \textbf{Sum of C* algebras:} $A \oplus B$ is also nuclear (\citep[Remark 4.9]{pisier_tensor_2021}) and for any C* algebra we have that $(A \otimes B) \oplus C \simeq (A \otimes C) \oplus (B \otimes C)$.
    \item \textbf{Commutativity:} Let $A,B$ be two nuclear C* algebras, then $A \otimes B \simeq B \otimes A$ (\citep[Equation 4.8]{pisier_tensor_2021}).
\end{itemize}

\end{remark}

\begin{lemma}[Tensor product and generated C* algebras]\label{lemma:tensor_products_and_generated_C_star_algebras}
Let $A$ be a nuclear C* algebra and $B$ a C* algebra such that $A = C^*(\mathcal{A})$ and $B = C^*(\mathcal{B})$, then $C^*(\mathcal{A} \odot \mathcal{B})$ exits and
$$ A \otimes B \simeq  C^*(\mathcal{A} \odot \mathcal{B}).$$
\end{lemma}
\begin{proof}
If $A = C^*(\mathcal{A})$, $B = C^*(\mathcal{B})$ then $\mathcal{A} \odot \mathcal{B}$ is a dense *-algebra of $A \otimes B$. If we denote
$$ \eta(t) = \sup \{  \| \pi(t) \| : \; \pi: \mathcal{A}\odot \mathcal{B} \to B(H) \text{ a representation}\} $$
then $\eta(t) \geq \| t\|_{A \otimes B}$ for $t \in \mathcal{A} \odot \mathcal{B}$, so, if $\eta(t) < \infty$ for all $t \in \mathcal{A} \odot \mathcal{B}$ then $\eta$ is the C* norm of $C^*(\mathcal{A} \odot \mathcal{B})$ (\cref{sec:Universal_C_star_algebras}). Assume that $C^*(\mathcal{A} \odot \mathcal{B})$ exists, then by \cref{lemma:extending_star_homomorphisms_into_C_star_homomorphisms} there must be a C* algebra homomorphism
$$ \phi : C^*(\mathcal{A} \odot \mathcal{B}) \to A \otimes B. $$
Assume that $\mathcal{A}, \mathcal{B}$ have units, otherwise add a unit to them as in \cref{sec:unitization_of_C_star_algebras} and call them $\mathcal{A}^+, \mathcal{B}^+$, so, in this case $C^*(\mathcal{A})^+ = C^*(\mathcal{A}^+)$ and $C^*(\mathcal{B})^+ = C^*(\mathcal{B}^+)$ which implies that $C^*(\mathcal{A})$ is a sub C* algebra of $C^*(\mathcal{A}^+)$.

If $\alpha$ is a C* semi-norm over $\mathcal{A} \odot \mathcal{B}$ then 
$$\alpha(a \odot b) = \alpha((a \odot 1_\mathcal{B})\odot (1_\mathcal{A} \odot a)) \leq \alpha(a \odot 1_\mathcal{B}) \alpha(1_\mathcal{A} \odot b),$$
also, $\alpha_{\mathcal{A}}(a) = \alpha(a \odot 1_\mathcal{B})$ is a C* semi-norm over $\mathcal{A}$ and $\alpha_{\mathcal{B}}(b) = \alpha(1_\mathcal{A} \odot b)$ is a C* semi-norm over $\mathcal{B}$. Since both $C^*(\mathcal{A}), C^*(\mathcal{B})$ exists then $\alpha_{\mathcal{B}}(b) \leq \| b \|_{C^*(\mathcal{B})}$ and $\alpha_{\mathcal{A}}(a) \leq \| a \|_{C^*(\mathcal{A})}$, therefore, $\eta(a \otimes b) \leq \| a\|_A \|b \|_b = \| a \odot b \|_{A \otimes B}$ for $a \in \mathcal{A}, \; b \in \mathcal{B}$, which in turn implies that  $\eta(t) < \infty$ for all $t \in \mathcal{A} \odot \mathcal{B}$ and by \cref{remark:bound_in_the_generators_guaranties_existence_of_enveloping_c_star_algebra} we get that $C^*(\mathcal{A} \odot \mathcal{B})$ exists. Also, $\| a \odot b \|_{A \otimes B} \leq \eta(a \odot b)$ by definition of $\eta$, consequently, $\| a\|_{A } \| b \|_{B} = \eta(a \odot b) $ for $a \in \mathcal{A}, \; b \in \mathcal{B}$, in particular 
$$\|a \odot 1_{\mathcal{B}}\|_{C^*(\mathcal{A} \odot \mathcal{B})} = \eta(a\odot 1_{\mathcal{B}}) = \| a \|_A, \; a \in \mathcal{A}$$
along with 
$$\| 1_{\mathcal{A}} \odot b\|_{C^*(\mathcal{A} \odot \mathcal{B})} = \eta( 1_{\mathcal{A}} \odot b) = \| b \|_B, \; b \in \mathcal{B}.$$

By the universality of $C^*(\mathcal{A})$ (\cref{proposition:factoring_representations_with_enveloping_C_star_algebras}) we get that there is a C* homomorphism
$$ \phi_A: C^*(\mathcal{A}) \to C^*(\mathcal{A} \odot \mathcal{B}), \;  \phi_B: C^*(\mathcal{B}) \to C^*(\mathcal{A} \odot \mathcal{B})$$
which is injective (isometric) due to the previous computation, and takes the following form 
$$ \phi_A(a) = a \odot 1_{\mathcal{B}}, a \in C^*(\mathcal{A}), \; \phi_B(b) = 1_{\mathcal{A}} \odot b, b \in C^*(\mathcal{B}).$$
If $\lim_{n \to \infty}a_n =a$ with $a \in C^*(\mathcal{A})$ and $a_n \in \mathcal{A}$ and $\lim_{n \to \infty}b_n =b$ with $b \in C^*(\mathcal{B})$ and $b_n \in \mathcal{B}$ then the equality $\eta( a_n \odot b_n) = \| a_n\| \|b_n\|$ implies that $\lim_{n \to \infty}(a_n \otimes 1_{\mathcal{B}})(1_{\mathcal{A}} \otimes b_n)$ has a one-to-one correspondence with $a \otimes b$, thus, 
$$A \odot B \subset C^*(\mathcal{A} \odot \mathcal{B}).$$  
Since there is only one C* norm on $A \odot B$ (\cref{remark:nuclear_C_star_algebra_enveloping_C_star_algebra}) and $A \odot B$ is dense in $C^*(\mathcal{A} \odot \mathcal{B})$ then $C^*(\mathcal{A} \odot \mathcal{B}) \simeq A \otimes B$ as desired.
\end{proof}

Nuclear C* algebras will be a key tool of our analysis because they not only have a good behaviour with respect to C* tensor products, they also have a simple behaviour with respect to bounded derivations as we mention on \cref{sec:when_is_cyclic_cohomology_interesting}.

\begin{example}[Examples of nuclear C* algebras]\label{example:nuclear_C_star_algebras}

The following C* algebras are nuclear

\begin{itemize}
    \item Commutative C* algebras (\citep[Theorem T.6.20]{wegge-olsen_k-theory_1993}), we will discuss a little more of these C* algebra on \cref{sec:Continuous_functions_with_values_on_C_star_algebra}.
    \item Matrix algebras $M_n(\mathbb{C})$ (\citep[Theorem T.5.20]{wegge-olsen_k-theory_1993}), more generally, all finite dimensional C* algebras (\cref{example:matrices_gen_C_star_algebras}). We will discuss a little more of these C* algebra on \cref{sec:C_star_alg_matrix_alg}.
    \item C* algebra of compact operators on a separable Hilbert space, also known as $\mathcal{K}$ (\citep[Exercise T.M]{wegge-olsen_k-theory_1993}), we will discuss a little more of these C* algebra on \cref{section:stabilization_C_stal_algebra}.
\end{itemize}
\end{example}

C* homomorphisms can be extended into tensor product of C* algebras

\begin{proposition}[Extending C* homomorphisms into tensor products]\label{proposition:extending_C_star_homomorphisms_into_tensor_products}
Let $A_k, B_k$ be $C^*$-algebras with *-homomorphisms $\psi_k: A_k \rightarrow B_k, k=1,2,$ with one of $A_k$ nuclear and one of $B_k$ nuclear, then

\begin{itemize}
    \item \citep[Corollary T.5.19]{wegge-olsen_k-theory_1993}: $\psi_1 \odot \psi_2$ extends by continuity to a *-homomorphism $\psi_1 \otimes \psi_2: A_1 \otimes A_2 \rightarrow B_1 \otimes B_2$. If $\psi_1$ and $\psi_2$ are both injective, so is $\psi_1 \otimes \psi_2$.
    \item \citep[Corollary T.5.15]{wegge-olsen_k-theory_1993}: If $\psi_1, \psi_2$ are isomorphisms then $\psi_1 \odot \psi_2$ is an isomorphism at the level of $A_1 \odot A_2$, thus, it extends to an isomorphism 
    $$\psi_1 \otimes \psi_2: A_1 \otimes A_2 \to B_1 \otimes B_2.$$
    \item \citep[Proposition T.6.23]{wegge-olsen_k-theory_1993}: If $\phi_k: A_k \rightarrow C$ are commuting *-homomorphisms then $\phi_1 \odot \phi_2: A_1 \odot A_2 \rightarrow C$ is continuous with respect to $\sigma$ and there is a unique *-homomorphism 
    $$\phi_1 \otimes \phi_2 : A_1 \otimes A_2 \to C$$
    that extends $\phi_1 \odot \phi_2$. If $f_k \in A_k^*$, then
    $$
\left|\left(f_1 \odot f_2\right)(t)\right| \leq \left\|f_1\right\|\left\|f_2\right\| \sigma(t).
$$
In particular the tensor product of states always extends to a state on $A_1 \otimes A_2$.
\end{itemize}
\end{proposition}

Also, we can extend short exact sequence using tensor products

\begin{theorem}[Extending short exact sequences with tensor products (Theorem T.6.26 \citep{wegge-olsen_k-theory_1993})]\label{theorem:extending_short_exact_sequences_with_tensor_product}

Let $0 \rightarrow A \stackrel{\phi_A}{\longrightarrow} B \stackrel{\phi_B}{\longrightarrow} C \rightarrow 0$ be an exact sequence of $C^*$-algebras and $D$ another $C^*$-algebra.
When $D$ is nuclear, there is an exact sequence
$$
0 \rightarrow A \otimes D \rightarrow B \otimes D \rightarrow C \otimes D \rightarrow 0.
$$
If $C$ is nuclear, there is an exact sequence
$$
0 \rightarrow A \otimes_\sigma D \rightarrow B \otimes_\sigma D \rightarrow C \otimes D \rightarrow 0.
$$
\end{theorem}

And the extension of a nuclear C* algebra by a nuclear C* algebra is a nuclear C* algebra

\begin{theorem}[Extension nuclear C* algebra (Theorem T.6.27 \citep{wegge-olsen_k-theory_1993})]\label{theorem:extension_nuclear_C_star_algebra_with_nuclear_C_star_algebra}
The extension of a nuclear $C^*$-algebra by a nuclear $C^*$ algebra is nuclear. To be precise: if $0 \rightarrow A \rightarrow B \rightarrow C \rightarrow 0$ is exact and $A$ and $C$ are nuclear, then so is $B$.
\end{theorem}

\begin{remark}[The Toeplitz algebra is nuclear]\label{remark:Toeplitz_algebra_is_nuclear}

From \cref{section:Toeplitz_algebra} we have the following short exact sequence
$$0 \longrightarrow \mathcal{K} \longrightarrow C^*(S,1) \longrightarrow C(\mathbb{T}) \longrightarrow 0,$$
since both $C(\mathbb{T})$ and $\mathcal{K}$ are nuclear by \cref{example:nuclear_C_star_algebras}, then \cref{theorem:extension_nuclear_C_star_algebra_with_nuclear_C_star_algebra} implies that $C^*(S,1 | S^*S = 1)$ is also a nuclear C* algebra. 
\end{remark}

The term "nuclear" has a similar meaning as the notion of nuclear space in Grothendieck's theory of tensor product of topological vector spaces, but they do not coincide in general for C* algebras \citep[Section II.9.4]{blackadar_operator_2006}. 

\begin{remark}[Tensor product of separable C* algebras]\label{remark:tensor_product_separable_C_star_algebras}
Let $A,B$ be separable nuclear C* algebras, then $A \otimes B$ is separable. This is an easy fact to check, let $\mathcal{A}, \mathcal{B}$ be dense countable subsets of $A, B$ respectively, then
$$ v = \{ \sum_{i \leq n} \alpha_i \otimes \beta_i , \; i \in \mathbb{N}, \; \alpha \in \mathcal{A}, \; \beta \in \mathcal{B} \}$$
is a dense subset of $A \odot B$ under the norm of $A \otimes B$. Also, $\mathcal{A} \odot \mathcal{B}$ is a countable set, thus $A \otimes B$ is separable.
\end{remark}

\subsection{Matrix algebras ($M_n(A)$)}\index{$M_n(A)$}
\label{sec:C_star_alg_matrix_alg}
In \cref{example:nuclear_C_star_algebras} we mentioned that $M_n(\mathbb{C})$ is a nuclear C* algebra. Let $\{ e_{ij} \}_{1 \leq i,j \leq n}$ the set of generators of $M_n(\mathbb{C})$, since the generators are linearly independent then by \cref{proposition:properties_algebraic_tensor_product} we have any element of $A \odot M_n (\mathbb{C})$ can be written uniquely as
$$ t = \sum_{1 \leq i,j \leq n} a_{i,j} \otimes e_{i,j}.$$
The map 
$$ \phi: A \odot M_n (\mathbb{C}) \to M_n(A), \; \phi(t) = (a_{ij})_{1 \leq i,j \leq 1} $$
is a bijective *-algebra homomorphism. The multiplication, addition and involution on $M_n(A)$ take the usual entry-wise form from $M_n(\mathbb{C})$, in particular
$$
\left(\begin{array}{cccc}
a_{11} & a_{12} & \cdots & a_{1 n} \\
a_{21} & a_{22} & \cdots & a_{2 n} \\
\vdots & \vdots & \ddots & \vdots \\
a_{n 1} & a_{n 2} & \cdots & a_{n n}
\end{array}\right)^*=\left(\begin{array}{cccc}
a_{11}^* & a_{21}^* & \cdots & a_{n 1}^* \\
a_{12}^* & a_{22}^* & \cdots & a_{n 2}^* \\
\vdots & \vdots & \ddots & \vdots \\
a_{1 n}^* & a_{2 n}^* & \cdots & a_{n n}^*
\end{array}\right)
$$

Additionally, $M_n(A)$ is complete under any C* norm, thus the spatial norm is the only C* norm on $M_n(A)$,

\begin{proposition}[C* norm on matrix algebras (Proposition T.5.20 \citep{wegge-olsen_k-theory_1993})]\label{proposition:C_star_norm_on_matrix_algebras}
The spatial norm is the only $C^*$-norm on $A \odot \mathbb{M}_n(\mathbb{C})$, and
$$
\begin{aligned}
A \otimes \mathbb{M}_n(\mathbb{C}) & \simeq \mathbb{M}_n(A) \\
\mathbb{M}_n(\mathbb{C}) \otimes \mathbb{M}_m(\mathbb{C}) & \simeq \mathbb{M}_{n m}(\mathbb{C})
\end{aligned}
$$
\end{proposition}

From the description of an element of $M_n(A)$ as a sum of elementary tensor products we have that
$$ \| t \| = \| \sum_{1 \leq i,j \leq n} a_{i,j} \otimes e_{i,j} \| \leq \sum_{1 \leq i,j \leq n} \|a_{i,j}\|\| e_{i,j}\| =  \sum_{1 \leq i,j \leq n} \|a_{i,j}\|. $$
From the description of the spatial norm be know that $M_n(A)$ has a faithful representation on $H_A \otimes \mathbb{C}^n$, with $\pi_A : A \to H_A$ a faithful representation of $A$. From the description of a tensor product of Hilbert spaces as a direct sum (\cref{proposition:properties_tensor_product_Hilbert_spaces}) we know that $H_A \otimes \mathbb{C}^n \simeq \sum_{1 \leq i \leq n} H_A$, therefore, any element of $H_A \otimes \mathbb{C}^n$ can be understood as an n tuple of elements of $H_A$ and its norm is given by $\sqrt{\sum_{1 \leq i \leq n} \| h_i \|^2}$ with $h_i \in H_A$. So, using Pythegoras theorem on $H_A \otimes \mathbb{C}^n$ (\citep[Proposition T.5.20]{wegge-olsen_k-theory_1993}) is possible to check that
$$ \| a_{i,j} \| = \|(\pi_A \odot \pi_0)(a_{i,j} \otimes e_{i,j}) \| \leq \| \sum_{1 \leq i,j \leq n} a_{i,j} \otimes e_{i,j} \|. $$

If $\eta \in M_n(\mathbb{C})$ and $\phi: A \to B$ a *-homomorphism between C* algebras, then by \cref{theorem:extending_short_exact_sequences_with_tensor_product} there is a map
$\phi \otimes \eta : M_n(A) \to M_n(B)$
that is injetive if both $\phi$ and $\eta$ are injective . If $\eta = id_{M_n(\mathbb{C})}$ then $\phi_n = \phi \otimes id_{M_n(\mathbb{C})}$ looks like
$$\phi_n\left(\begin{array}{ccc}
a_{11} & \cdots & a_{1 n} \\
\vdots & \ddots & \vdots \\
a_{n 1} & \cdots & a_{n n}
\end{array}\right)=\left(\begin{array}{ccc}
\phi\left(a_{11}\right) & \cdots & \phi\left(a_{1 n}\right) \\
\vdots & \ddots & \vdots \\
\phi\left(a_{n 1}\right) & \cdots & \phi\left(a_{n n}\right)
\end{array}\right).
$$

If $\alpha$ is a continuous linear functional on $A$ then by \cref{theorem:extending_short_exact_sequences_with_tensor_product} $\alpha \otimes \text{tr}$ is a continuous linear functional on $M_n(A)$ and takes the following form
$$\alpha \otimes \text{tr} : A \otimes M_n(\mathbb{C}) \to \mathbb{C}, \; (\alpha \otimes \text{tr}) (\sum_{1 \leq i,j \leq n} a_{i,j} \otimes e_{i,j}) = \sum_{1 \leq i \leq n} \text{tr}(a_{i,i}) .$$

Let $\sum_{1 \leq i,j \leq n} a_{i,j} \otimes e_{i,j} \in M_n(A)_{\text{pos}}$, then we can show that $a_{i,i}$ are positive elements of $A$, first notice that $a_{i,i}^* = a_{i,i}$ ($a_{i,i} \in A_{\text{sa}}$), also, if $\hat{h_i} = (0, \dots, h_i , \dots, 0)$ then
$$\langle \hat{h_i}  , (\pi_A \otimes \pi_0)(a_{i,i} \otimes e_{i,i}) \hat{h_i} \rangle = \langle h_i, \pi_A(a_{i,i}) h_i \rangle \geq 0 , \; \forall h_i \in H_A.$$
Thus, $\pi_A(a_{i,i})$ is a positive operator, which is equivalent to $\pi_A(a_{i,i})$ being a positive element of the C* algebra $B(H_A)$ (\cref{proposition:cahracterization_of_positive_elements}), and, since $\pi_A$ is an isomorphism then $a_{i,i}$ is a positive operator.

So, if $\alpha$ is a weight on $A$ then $\alpha \otimes \text{tr}$ defined by 
$$ \alpha \otimes \text{tr} : M_n(A)_{\text{pos}} \to [0, \infty], \;  (\alpha \otimes \text{tr})(\sum_{1 \leq i,j \leq n} a_{i,j} \otimes e_{i,j}) = \sum_{1 \leq i \leq n} \alpha(a_{i,i})$$
is also a weight on $M_n(A)$.

\subsection{Stabilization}\index{C* algebra!stabilization}
\label{section:stabilization_C_stal_algebra}

In \cref{example:nuclear_C_star_algebras} we mentioned that $\mathcal{K}$ is a nuclear C* algebra. Recall that $\mathcal{K} \simeq \mathcal{K}(l^2(\mathbb{N}))$, let $A$ be an arbitrary C* algebra (nuclear or not) with a faithful representation $\pi_A : A \to H_A$, then $\mathcal{K} \otimes A$ has a faithful representation on $H_A \otimes l^2(\mathbb{N})$. 

Since $l^2(\mathbb{N})$ has an ortonormal basis $\{ \xi_i \}_{i \in \mathbb{N}}$, then $l^2(\mathbb{N}) \simeq \sum_{i \in \mathbb{N}} H_i$ with $H_i = \langle \xi_i \rangle$ i.e  the closed sub space generated by $\xi_i$, which is isomorphic to $\mathbb{C}$. So, $H_A \otimes l^2(\mathbb{N}) \simeq \sum_{i \in \mathbb{N}}H_{A,i}$ with $H_{A,i} = H_A$ (\cref{proposition:properties_tensor_product_Hilbert_spaces}), with the isomorphism given by
$$ \phi: \sum_{i \in \mathbb{N}}H_{A,i} \to H_A \otimes l^2(\mathbb{N}), \; \phi(\sum_{i \in \mathbb{N}}h_i ) = \sum_{i \in \mathbb{N}} \xi_i \otimes h_i $$
(\citep[Remark 2.6.8]{kadison_fundamentals_1983_V1}). Thus any $T \in B(\sum_{i \in \mathbb{N}}H_{A,i})$ can be seen as an infinite matrix with entries in $B(H_A)$, i.e $T = [t_{i,j}]_{i,j \in \mathbb{N}}$ with $t_{i,j} \in B(H_A)$ (\citep[Page 149]{kadison_fundamentals_1983_V1}), such that if $h = \sum_{i \in \mathbb{N}} h_i, \; h_i \in H_{A,i}$ then $T(h) = \sum_{i \in \mathbb{N}} (\sum_{j \in \mathbb{N}} t_{i,j}(h_j))$. The representation of $T$ as an infinite matrix with entries in $B(H_A)$ comes with involution and addition as in the case of infinite matrices over $\mathbb{C}$ from \cref{section:algebra_of_compact_operators}, and the multiplication given by an infinite sum that converges in the strong operator topology, not necessarily in the norm topology (in \cref{section:faithful_representation_of_torus} we would like an example where ths sum does not converge in the operator norm). Also, the norm of the operator $T$ can be computed as the supremum of all the norms of finite sub matrices of $T$ (\citep[Proposition 2.6.13]{kadison_fundamentals_1983_V1})

Since $A \otimes \mathcal{K}$ has a faithful representation on $B(\sum_{i \in \mathbb{N}}H_{A,i})$ then any $a \in A \otimes \mathcal{K}$ can be seen as an infinite matrix $a = [a_{i,j}]_{i,j \in \mathbb{N}}$ with $a_{i,j} \in A$. Moreover, we can use the generators of $\mathcal{K}$ to write that matrix as follows
$$ a = \sum_{i,j \in \mathbb{N}} a_{i,j} \otimes E_{i,j}, $$
since $a_{i,j} \otimes E_{i,j}$ is the operator that contributes with $\pi_A(a_{i,j})(h_{j})$ to the $i^{th}$ entry of $\pi(a)(h)$ with $\pi: A \otimes \mathcal{K} \to B(\sum_{i \in \mathbb{N}} H_{A,i})$ and $h = \sum_{i \in \mathbb{N}}h_i \in \sum_{i \in \mathbb{N}} H_{A,i}$. Moreover, by \cref{proposition:extending_C_star_homomorphisms_into_tensor_products} there is an injective *-homomorphism
$$ id \otimes i : A \otimes M_n(\mathbb{C}) \to A \otimes \mathcal{K}, \; (id \otimes i)(\sum_{1 \leq i,j \leq n} a_{ij} \otimes e_{ij}) = \sum_{1 \leq i,j \leq n} a_{ij} \otimes E_{ij},$$
so, $A \otimes \mathcal{K}$ is the clousure of the *-algebra $\pi(\cup_{i=1}^{\infty} M_i(A))$ in $B(\sum_{i \in \mathbb{N}}H_{A,i})$.

As in the case of infinite matrix algebras over $\mathbb{C}$ (\cref{section:algebra_of_compact_operators}), we can impose restrictions on the coefficients $a = [a_{i,j}]_{i,j \in \mathbb{N}}$ with $a_{i,j} \in A$ to ensure that $a \in A \otimes \mathcal{K}$, for example, if $\sum_{i,j \in \mathbb{N}} \| a_{i,j} \|^2 < \infty$. This is an easy fat to check, take $h = \sum_{i \in \mathbb{N}} h_i, \; h_i \in H_A$ with norm one, then 
$$ | \pi(a)(h) - \pi(\sum_{i,j \leq n} a_{i,j} \otimes E_{i,j})(h) |^2 = | \pi(\sum_{i,j > n} a_{i,j} \otimes E_{i,j})(h) |^2 = \sum_{i,j > n} \| a_{i,j}(h_j) \|^{2} \leq  \sum_{i,j > n} \| a_{i,j} \|^{2}. $$
Since $\| a -  \sum_{i,j \leq n} a_{i,j} \otimes E_{i,j} \|$ is the supremum of the previous computation over all unit vectors, if we chose $n$ large enough such that $ \sum_{i,j > n} \| a_{i,j} \|^{2} \leq \epsilon$ then $\| a -  \sum_{i,j \leq n} a_{i,j} \otimes E_{i,j} \| \leq \epsilon$, meaning that $a \in A \otimes \mathcal{K}$.

As in the case of matrix algebras, if $a \in (A \otimes \mathcal{K})_{\text{pos}}$ then we must have that $a_{i,i} = A_{\text{pos}}$ for $i \in \mathbb{N}$, thus, if $\alpha$ is a weight (trace) on $A$ we can define a weight (trace)  
$$ \alpha \otimes \text{Tr} : (A \otimes \mathcal{K})_{\text{pos}} \to [0, \infty], \;  (\alpha \otimes \text{Tr})(\sum_{i,j \in \mathbb{N}} a_{i,j} \otimes E_{i,j}) = \sum_{i \in \mathbb{N}} \alpha(a_{i,i}).$$

Moreover, take $\{ a_n \}_{n \in \mathbb{N}} \in (A \otimes \mathcal{K})_{\text{pos}}$ such that $a_n \to a$ over $(A \otimes \mathcal{K})_{\text{pos}}$, then, $(a_n)_{i,j} \to (a_n)_{i,j}$ for all $i,j \in \mathbb{N}$. From \cref{remark:weights_and_continuity} we know that $a \in (A \otimes \mathcal{K})_{\text{pos}}$, therefore, is $\alpha$ is a lower semicontinuous weight (trace), we have that
$$ \alpha ( a_{i,i} ) \leq \lim_{k \to \infty} \left( \inf_{l \geq k} \alpha( (a_l)_{i,i} ) \right).$$
Since $\alpha$ takes only positive values, then $\sum_{j \in \mathbb{N}} \alpha(a_{j,j})$ either converges absolutely or diverges, and in both cases we have that
$$ \sum_{j \in \mathbb{N}} \alpha(a_{j,j}) \leq \sum_{j \in \mathbb{N}} \left( \lim_{k \to \infty} \left( \inf_{l \geq k} \alpha( (a_l)_{j,j} ) \right) \right) \leq \lim_{k \to \infty} \left( \inf_{l \geq k} \left(  \sum_{j \in \mathbb{N}}  \alpha( (a_l)_{j,j} ) \right) \right) , $$
therefore, $\alpha \otimes \text{Tr}$ is also a lower semicontinuous weight (trace).

\subsection{Continuous functions with values on C* algebras $C_0(X,A)$}\index{$C_0(X,A)$}
\label{sec:Continuous_functions_with_values_on_C_star_algebra}

In \cref{example:nuclear_C_star_algebras} we mentioned that any commutative C* algebra is nuclear, or equivalently, $C_0(X)$ is nuclear for $X$ a locally compact Hausdorff space.

Let X be a locally compact Hausdorff space and $A$ a Banach space, denote by $C_b(X;A)$ the vector space of bounded continuous functions from $X$ to $A$, then $C_b(X;A)$ is a Banach space under the uniform norm (\citep{1026265}) 
$$\| f \| = \sup_{x\in X} \| f(x) \|,$$
thus, it becomes a C* algebra under point wise multiplication, addition and involution. Denote by $C_0(X;A)$ the space of functions that decay at infinity, then $C_0(X;A)$ is also closed under the supremum norm (\citep{1315426}). Moreover, if $A$ is C* algebra, then $C_0(X,A)$ is a C* algebra with the following operations
\begin{itemize}
    \item $(f+g)(x) = f(x) + g(x)$
    \item $(f \cdot g)(x) = f(x) g(x)$
\end{itemize}
 and is a generalization of $C_0(X)$. Moreover, it has a simple description in terms of crossed products of C* algebras, because $C_0(X)$ is nuclear. So, taking $C_0(X) \odot A$ to be a *-subalgbera of $C_0(X,A)$ following \cref{def:algebraic_tensor_product_functions_and_algebras}, we can check that $C_0(X) \odot A$
 is dense in $C_0(X,A)$.
 
\begin{lemma}\label{lemma:C_0_x_times_A_dense_in_C_0(X,A)}
 Let $X$ be a locally compact Hausdorff space and $A$ a Banach space, then $C_0(X) \odot A$ is dense in $C_0(X;A)$
\end{lemma}
\begin{proof}
\begin{itemize}
    \item Take $f \in C_0(X;A)$ and $K$ compact with $\| f(x) \| \leq \epsilon$ for $x \in K$.
    \item Let $\{ b_i \}_{i \leq n}$ be finite open cover of $K$ such that if $x,y \in b_i$ then $\| f(x) - f(y) \| \leq \epsilon$. This cover exists because $K$ is compact.
    \item Let $\{ g_i \}_{i \leq n}$ be a partition of unity for $K$ and subordinate to the open cover $\{ b_i \}_{i \leq n}$, that is, $g_i : X \to [0,1]$, $supp(g_i)\subset b_i$, $\sum_{i \leq n} g_i(x) \in [0,1]$ and $\sum_{i \leq n} g_i(x) = 1$ if $x \in K$ by \citep[Propostion 4.41]{folland_real_1999}.
    \item Then, let $x_i \in b_i$, and set $f_{\epsilon}(x) = \sum_{i \leq n} g_i(x)f(x_i)$. The function $f_{\epsilon}$ is an approximation of $f$, to check this note that $f(x) = \sum_{i \leq n} (g_i(x)f(x))$ if $x \in K$, therefore,
    $$\| f(x) - f_{\epsilon}(x) \| \leq \sum_{i \leq n } \| f(x) - f(x_i) \| g_i(x) , \; x \in K .$$
    The sum on the right has non zero terms when $g_i(x) \geq 0$, which happens inside $b_i$, but we have taken $b_i$ such that $\| f(x) - f(x_i) \| \leq \epsilon$ if $g_i (x) \geq 0$, hence we end up having
     $$\| f(x) - f_{\epsilon}(x) \|\leq \epsilon , x \in K .$$
    Also, if $x \notin K$, $f_{\epsilon}(x)$ then
    $$\| f(x) - f_{\epsilon}(x) \| \leq \| f(x) (1 - \sum_{i \leq n} g_i(x)) \| +  \sum_{i \leq n } \| f(x) - f(x_i) \| g_i(x),$$
    which in turn implies that if $x \notin supp(\sum_{i \leq n} g_i$ then $\| f(x) - f_{\epsilon}(x) \| = \| f(x) \| \leq \epsilon$ and if $x \in supp(\sum_{i \leq n} g_i)$ then $\| f(x) - f_{\epsilon}(x) \| \leq 2 \epsilon$. Thus, $C_0(X) \odot A$ is dense in $C_0(X,A)$.
\end{itemize}  
\end{proof}

If $X$ is a locally compact Hausdorff space then there is a faithful representation of both $C_0(X), C_b(X)$ on the Hilbert space $l^2(X)$, which, as mentioned in \cref{defintion:direct_sum_hilbert_spaces}, is the direct sum of the family of Hilbert spaces $\{ H_x \}_{x \in X}$ with $H_x = \mathbb{C}$ (\cref{defintion:direct_sum_hilbert_spaces}). The representation is given as multiplication operators
$$ (\pi(f)(g))(x) = f(x)g(x), \; f \in C_0(X), \; g \in l^2 (X), $$
and is easy to check that $\| \pi(f) \| = \sup \{ |f(x) | : x \in X \}$, therefore, it is faithful. The representation $\pi$ along with the dense inclusion of $C_0(X) \odot A$ in $C_0(X,A)$ are key to prof the following result,

\begin{proposition}[Isomorphism of tensor products of commutative C* algebras (Proposition T.5.21 \citep{wegge-olsen_k-theory_1993})]\label{proposition:isomorphism_tensor_product_of_commutative_C_star_algebras}
The completion of $A \odot C_0(X)$ with respect to the spatial norm is isomorphic to the $C^*$-algebra $C_0(X \rightarrow A)$, moreover :
$$
\begin{aligned}
A \otimes_\sigma C_0(X) & \simeq C_0(X ; A), \\
C_0(X) \otimes_\sigma C_0(Y) & \simeq C_0(X \times Y) .
\end{aligned}
$$
\end{proposition}

Since $C_0(X)$ is nuclear, then we can drop the index $\sigma$ of the tensor product.

\begin{example}[Suspension and cone exact sequence]\label{example:Suspension_and_cone_exact_sequence}

 There is a short exact sequence
 $$ 0 \longrightarrow C_0((0,1)) \stackrel{\iota}{\longrightarrow} C_0((0,1]) \stackrel{\pi}{\longrightarrow} \mathbb{C} \longrightarrow 0, $$
where $\iota$ is the inclusion mapping and $\pi(f)=f(1)$. Since all the terms of the short exact sequence are nuclear we can use \cref{theorem:extending_short_exact_sequences_with_tensor_product} to get the exact sequence
$$ 0 \longrightarrow C_0((0,1] ; A) \stackrel{\iota}{\longrightarrow} C_0 ((0,1]; A) \stackrel{\pi}{\longrightarrow} A \longrightarrow 0, $$
for any C* algebra $A$ from \citep[Example 4.1.5]{rordam_introduction_2000}. The C* algebra $C_0((0,1] ; A)$ is named the cone of $A$ and denoted by $CA$, while the C* algebra $C_0 ((0,1); A)$ is named the suspension of $A$ ans denoted by $SA$. This short exact sequence is a key step in the description of the K theory of a C* algebras in terms of the suspension (\citep[Section 5.4]{prodan_bulk_2016}).

If there is an exact sequence of C* algebras 
$$
0 \rightarrow A \otimes D \rightarrow B \otimes D \rightarrow C \otimes D \rightarrow 0,
$$
then \cref{theorem:extending_short_exact_sequences_with_tensor_product} gives us a short exact sequence
$$
0 \rightarrow SA \rightarrow SB \rightarrow SC \rightarrow 0,
$$
thus, the functor $S: A \mapsto SA$ from the category of C* algebras into the category of C* algebras is exact.
\end{example}

\section{Examples}
\label{sec:C_star_algebras_examples}

\subsection{Continuous functions over a locally compact Hausdorff space}
\label{section:algebra_continuous_functions_locally_comp_space}

The algebras of continuous functions are ubiquitous in the analysis of C* algebras and give us various results that will help us to guide our study of non-commutative C* algebras, so, let us study some examples that are commonly used.

\begin{lemma}[Second countable Locally compact Hausdorff spaces\index{locally compact Hausdorff space!second countable}]\label{lemma:second_coutnalbe_locally_compact_hausdorff_space}
If $X$ is a second countable locally compact Hausdorff space then:
    \begin{enumerate}
        \item Our space $X$ is $\sigma$-compact\index{$\sigma$-compact space}, that means that there is a sequence of compact set $\{ a_n \}_{n \in \mathbb{N}}$ such that $\cup a_n = X$ .
        \item  $X$ has compact exhaustion (\cref{def:compact_exhaustion}).
        \item $X$ is a metrizable space, i.e., there is a metric $d : X \times X \to \mathbb{R}^+$ that induces the same topology of $X$.
    \end{enumerate}
\end{lemma}
\begin{proof}
    \begin{enumerate}
    \item This is proven in \citep{nlab:locally_compact_and_second-countable_spaces_are_sigma-compact}.
    \item Since $X$ is $\sigma$-compact, \citep[Proposition 10.25]{knapp_basic_2005} implies us that $X$ has compact exhaustion (\cref{def:compact_exhaustion}).
    \item Given that X is a locally compact Hausdorff space, \citep[Corollary 10.21]{knapp_basic_2005} tells us that $X$ is a regular topological space, additionally, $X$ is separable because every second countable spaces is separable; recall that a regular topological space\index{regular topological space} is one where for every closed set $C$ and a point $p$ not in $C$, there are two disjoint open set $U_0, \; U_1$ with $U_0$ a neighborhood of $C$ and $U_1$ a neighborhood of $p$. Since $X$ is a separable regular space, \citep[Theorem 10.45 (Urysohn Metrization Theorem)]{knapp_basic_2005}\index{Urysohn Metrization Theorem} tell us that $X$ is a metrizable space\index{metrizable space}, i.e., there is a metric $d : X \times X \to \mathbb{R}^+$ that induces the same topology of $X$.
    \end{enumerate}
\end{proof}

\begin{definition}[c.f. \citep{nlab:radon_measure} (Radon measure)\index{Radon measure}]\label{definition:Radon_measure}
Let $X$ be a locally compact, then, $\mu$ is called a Radon measure over $X$ if it satisfies the following conditions:
\begin{itemize}
    \item $\mu$ is a Borel measure\index{Borel measure} i.e. $\mu$ is a measure over the Borel $\sigma$-algebra of $X$.
    \item $\mu(K)$ is finite for any $K$ a compact subset of $X$.
    \item $\mu$ is inner regular\index{inner regular measure}: for any element $E$ of the Borel $\sigma$-algebra of $X$ we have that
    $$\mu(E)=\sup _{\substack{K \subseteq E, K \text { compact }}} \{\mu(K)\}.$$
    \item $\mu$ is outer regular\index{outer regular measure}: for any element $E$ of the Borel $\sigma$-algebra of $X$ we have that
    $$\mu(E)=\inf _{\substack{U \supseteq E, U \text { open } }} \{\mu(U)\}.$$
\end{itemize}
We say that $\mu$ is a bounded Radon measure if $\mu(X) < \infty$, also, we say that $\mu$ is a Radon probability measure if $\mu(X)=1$.
\end{definition}

\begin{definition}[Measure with full support\index{measure with full support}]\label{definition:measure_with_full_support}
Let $X$ be a locally compact Hausdorff space and $\mu$ a measure over $X$, then, we say that $\mu$ has full support if $\mu(G) > 0$ for every open set of $X$.
\end{definition}

For the following discussions, recall that $C_{c}(X)$\index{$C_{c}(X)$} denotes the algebra of functions with compact support over $X$.. 

\begin{proposition}[c.f. Theorem 3.18 \citep{kantorovitz_introduction_2003} and Theorem 11.1 \citep{knapp_basic_2005}]\label{proposition:riesz-markov-representation-theorem}
Let $X$ is a second countable locally compact Hausdorff space, then, there is a one to one correspondence between positive linear continuous functionals over $C_c(X)$ and Radon measures (\cref{definition:Radon_measure}) over $X$, the correspondence is given as follows
    $$ \phi(f) = \int_X f d \mu ,\; f \in C_c(X), $$
    $\mu$ a Radon measure and $\mathcal{A}$ is associated $\sigma$-algebra. The measure has the following properties
    \begin{itemize}
        \item $\mathscr{B}(X) \subset \mathcal{A}$
        \item $\mu(K) < \infty$ for $K\subset X$ compact
        \item $\mu$ is a regular measure:
        \begin{itemize}
        \item $\mu$ is inner regular: for any $E \in \mathcal{A}$ we have that $\mu(E)=\sup _{\substack{K \subseteq E, K \text { compact }}} \{\mu(K)\}$
        
        \item $\mu$ is outer regular: for any $E \in \mathcal{A}$ we have that $\mu(E)=\inf _{\substack{U \supseteq E, U \text { open } }} \{\mu(U)\}$
        \item Since $\phi$ is a positive functional then $\mu$ is an unsigned measure i.e. $\mu(A) \geq 0$ for all $A \in \mathcal{B}(X)$, where $\mathcal{B}(X)$ is the Borel $\sigma$-algebra over $X$.
        \end{itemize}
    \end{itemize}
\end{proposition}

\begin{lemma}\label{lemma:LCS_with_mu_radon_measure_mu_signa_finite_and_coutnablye_generated}
Let $X$ is a second countable locally compact Hausdorff space, if $\mu$ is a Radon measure (\cref{definition:Radon_measure}) over $X$ then,
\begin{itemize}
    \item $\mu$ is $\sigma$-finite (\cref{def:sigma_finite_measu}). 
    \item $(X, \mathcal{B}(X),\mu)$ is $\mu$-countably generated (\cref{def:mu_countably_geenrated}), where $\mathcal{B}(X)$ is the Borel $\sigma$-algebra over $X$. 
    \item If $X$ is second countable then $L^P(X,\mu)$ for $1 \leq p < \infty$ is separable (\citep[Corollary 11.22]{knapp_basic_2005}).
    \item  $L^2(X,\mu)$ has a countable orthonormal basis.
\end{itemize}
\end{lemma}
\begin{proof}
\begin{itemize}
    \item Given that $X$ has a compact exhaustion (\cref{lemma:second_coutnalbe_locally_compact_hausdorff_space}), then, $X = \cup_{n \in \mathbb{N}} K_n$ with $K_n$ compact and $\mu(K_n) < \infty$ with $K_i \subseteq K_{i+1}$ as desired.
    \item We need to check that there is a sequence $\{ s_n\}_{n \in \mathbb{N}}$ with $\mu(s_n) < \infty$ and $\mathcal{B}(X) = \sigma(\{ s_n\}_{n \in \mathbb{N}})$. To check this let $N = \{ N_x \}_{x \in X}$ a set of compact neighbourhoods for all the points in $X$, and let $ A = \{a_n \}_{n \in \mathbb{N}}$ a countable base for the topology of $X$, then set $S = \{ a \in A | \exists n \in N \text{ s.t. } a \subset n\}$. Since $S$ is a subset of $A$ we have that $S$ is countable,it is also a base for the topology of $X$ and $\mu(s) < \infty$ for all $s \in S$, so $S$ is our set of sets with finite measure that generate $\mathcal{B}(X)$.
    \item From \cref{lemma:second_coutnalbe_locally_compact_hausdorff_space} we have that $X$ is a locally compact separable metric space, additionally, by assumption, $\mu$ is a Borel measure (\cref{definition:Radon_measure}), therefore, this is guaranteed by \citep[Corollary 11.22]{knapp_basic_2005}. Notice that this fact also comes from \cref{prop:separability_Lp_spaces}, because $\mu$ is $\sigma$-finite and $\mathbb{C}$ is a separable Banach space.
    \item Given that $L^2(X,\mu)$ is separable and is a Hilbert space (\cref{proposition:inner_prodict_L_2_S_H}), then, this statement is a consequence of \cref{lemma:separability_Hilbert_spaces_and_orthonormla_basis}.
\end{itemize}
\end{proof}

\begin{remark}\label{remark:LCS_second_countable_and_regular_borel_measure}
Let $X$ be a second countable Locally compact Hausdorff space, then, as we have previously mentioned, $X$ is a locally compact Hausdorff space metric space (\cref{lemma:second_coutnalbe_locally_compact_hausdorff_space}), thus, from \citep[Remark of Corollary 11.22]{knapp_basic_2005} we know that we only need $\mu$ to be a Borel measure (measure over the Borel sets that is finite on compact sets) in order for $\mu$ to be a regular Borel measure. 
\end{remark}

One of the algebras we care about is the algebra of continuous functions that decay to infinity $C_0(X)$, this algebra has a nice set or properties:

\begin{lemma}[Separability of $C_0(X)$]\label{lemma:separability_functions_dewcaying_at_infinity}
If $X$ is a second countable locally compact Hausdorff space then $C_0(X)$ is separable.
\end{lemma}
\begin{proof}
\begin{itemize}
    \item Denote by $C_c(X)$ the algebra of continuous functions on $X$ with compact support, then $C_{c}(X)$ is dense in $C_0(X)$ \citep[Proposition 4.35]{folland_real_1999}.
    \item In a separable locally compact Hausdorff space metric $Y$ space, the algebra $C_{c}(Y)$ is separable as a normed space under the supremum norm \citep[Corollary 11.22]{knapp_basic_2005}. Recall that under our assumptions $X$ is metrizable and separable, therefore $C_{c}(X)$ is separable under the supremum norm.
    \item Since $C_{c}(X)$ is separable and $C_{c}(X)$ is dense in $C_0(X)$ by \citep[Proposition 4.35]{folland_real_1999} we have that $C_0(X)$ is also separable.
\end{itemize}
\end{proof}

\begin{lemma}[Approximate identity of $C_0(X)$]\label{lemma:approximate_identity_of_functions_decayinh_at_infinity}
If $X$ is a second countable locally compact Hausdorff space then $C_0(X)$ has an approximate identity
\end{lemma}
\begin{proof}
\begin{itemize}
    \item If $X$ is compact then $C_0(X) = C(X)$ and ti has an identity, thus now we look into the case of $X$ not compact.
    \item $X$ is second countable, thus there is a countable set of open sets $a =\{ a_n\}_{n \in \mathbb{N}}$ that generate the topology of $X$. Also, X is locally compact, thus for each $x \in X$ there is an compact neighbourhood $b_x$ of x. Set $ \hat{a} = \{ \alpha \in a | \exists b_x \text{ s.t. } \alpha \in b_x \}$, then $\hat{a}$ is still a countable base for the topology of $X$.
    \item Denote by $\overline{a}$ the closure of an open set, then $\overline{\alpha}$ is compact for every $\alpha\in \hat{a}$. Now, define inductively $\beta_n$ as follows:
    \begin{itemize}
        \item $\beta_1 = \overline{\alpha}$
        \item $b_n = \cup_{i \leq n}\beta_i$.
        \item Since $b_n$ is compact, by \citep[Corollary 10.23]{knapp_basic_2005} there is a compact $K_n$ such that $b_n \subset K_n^{o}$
        \item Set $\beta_n = b_n \cup K_n$
        \item Then $\beta_n \subset \beta_{n+1}^{o}$, $\beta_i$ is compact, $\sup \beta_i = X$ and for every $P$ compact in $X$ there $\beta_j$ such that $P \subset \beta_j$.
    \end{itemize}

    \item By the Urysohn's lemma (\citep[Proposition 4.32]{folland_real_1999}) there is a continuous function $f_n: X \to [0,1]$ such that $f_n|_{\beta_n} = 1$ and $\text{support}(f_n) = \beta_{n+1}$. Let's see why $\{ f_n \}$ is an approximate identity of $C_0(X)$:
    
    \begin{itemize}
        \item Take $g \in C_0(X)$, then there is $K$ compact such that $g|_{K} < 1/n$. Take $\beta_i$ such that $K \subset \beta_i$, then $(gf_i) (x) = (f_i g) (x) = g(x)$ for $x \in K$ and $\| (gf_i)(x) \| \leq \| g(x) \| \|f_i(x)\| \leq 1/n$ for $x \notin K$, which implies that
        $$\| (gf_i) - g \|_{K} = \| (f_i g) - g \|_{K} = 0 $$ 
        and 
        $$\| (gf_i) - g \|_{K^{c}} = \| (f_i g) - g \|_{K^{c}} \leq 1/n. $$
        Also, for each $j \geq i$ the same inequalities still hold, which implies that 
        $$\lim_{n \to \infty} f_i g = \lim_{n \to \infty} f_i g = g. $$
    \end{itemize}
\end{itemize}
\end{proof}

\begin{lemma}[Faithful representation of $C_0(X)$]\label{lemma:faithful_representation_of_functions_decaying_at_infinity}
If $X$ is a locally compact Hausdorff space and $\mu$ is Radon measure (\cref{definition:Radon_measure}) over $X$ with full support (\cref{definition:measure_with_full_support}), then, $C_0(X)$ has a faithful representation on $L^{2}(X,\mu)$ given by the multiplication operator 
    $$ \pi : C_0(X) \to B(L^{2}(X,\mu)), \: \pi(f)(h)(x) \mapsto f(x)h(x),$$
where $f \in C_0(X)$, $h \in L^{2}(X,\mu)$.
\end{lemma}
\begin{proof}
\begin{itemize}
    \item Every $f \in C_0(X)$ is measurable because  as we mentioned previously $\mathscr{B}(X) \subset \mathcal{A}$ and we are working with the Borel $\sigma$-algebra on $\mathbb{C}$, moreover, by \cref{example:continuous_functions_decaying_at_infinty_are_strongly_measurable} we know that $f$ is $\mu$-strongly measurable. Since the $\mu$-strongly measurable functions are an algebra (\cref{remark:algebra_of_strongly_mu_measurable_functions}) then for $h \in L^{2}(X,\mu)$ the function $fh$ is $\mu$-strongly measurable. The only requirement left to check is that $fh$ has a finite norm on $L^{2}(X,\mu)$ which is done in the next item.
     
    \item $\pi(f) \in B(L^{2}(X,\mu)):$ We have that $$ \| \pi(f)(h) \|^{2} = \int_{X} | f(x)h(x)|^{2} d \mu(x) \leq \| f \|^2 \| h \|^{2}, $$
    $\pi(f)(\lambda h_1 + h_2) = \lambda \pi(f)( h_1) +  \pi(f) (h_2)$, and $\pi(fg) = \pi(f)\pi(g)$.
    \item $\pi$ is faithful: for $f \in C_0(X)$ ($f \in C_b(X)$) we have that there is $x \in X$ such that $| f(x) | = \| f \|$, then, take $\epsilon > 0$ and $N_{x,\epsilon}$ an open neighborhood of $x$ with finite measure, which we know that exists because for every $x\in X$ there is an open neighbourhood with compact closure. Construct the indicator function $1_{N_{x,\epsilon}}$, then we have that $1_{N_{x,\epsilon}} \in L^{2}(X,\mu)$ and,

    $$ \| \pi(f)(h) \|^{2} = \int_{X} | f(y)1_{N_{x,\epsilon}}(y)|^{2} d \mu(y) \geq (\| f \|-\epsilon)^2 \| h \|^{2}. $$
    
    Since $\epsilon$ is arbitrary we end up having that $\| \pi(f) \| = \| f \|$. Using a similar argument is easy to check that if $f \neq g$ then there is $x \in L^{2}(X,\mu)$ such that $ \| \pi(f-g)(x) \| > 0$, which implies that $\pi(f) \neq \pi(g)$.

\end{itemize}
\end{proof}

\begin{lemma}[Faithful representation of $C_b(X)$]\label{lemma:faithful_representation_of_functions_bounded}
If we take $X$ to be a second countable then \cref{lemma:LCS_with_mu_radon_measure_mu_signa_finite_and_coutnablye_generated} tell us that $\mu$ is $\sigma$-finite, therefore, we can use \cref{example:continuous_functions_on_sigmna_fintie_space} to show that any $f \in C_b(X)$ is $\mu$-strongly measurable. Under this circumstances we can replicate the proof on \cref{lemma:faithful_representation_of_functions_decaying_at_infinity} to show that $C_b(X)$ has a faithful representation on $L^{2}(X,\mu)$ given by the multiplication operator 
    $$ \pi : C_b(X) \to B(L^{2}(X,\mu)), \: \pi(f)(h)(x) \mapsto f(x)h(x),$$ 
where $f \in C_b(X)$, $h \in L^{2}(X,\mu)$.
\end{lemma}

If we take $X$ to be second countable, then, from our previous discussion both $C_0(X), C_b(X)$ have a representation on a separable Hilbert space $H = L^2(X,\mu)$, also, since $H$ is separable it has a countable orthonormal basis  $\xi = \{ \xi_i \}_{ i \in \mathbb{N}} $ (\cref{lemma:separability_Hilbert_spaces_and_orthonormla_basis}) and under this basis if $f \in C_0(X)$ ($f \in C_b(X)$) we can express $\pi(f)$ as an infinite matrix with a countable set of entries (\cref{sec:infinite_mattrices_and_bounded_operators}). Also, in \cref{sec:Continuous_functions_with_values_on_C_star_algebra} we mentioned that both $C_0(X), C_b(X)$ have a faithful representation on the Hilbert space $l^2(X)$, which is a Hilbert space with a dimension equal to the cardinality of $X$, thus, if $X$ is not countable then we can give two representation of $f$ as infinite matrices each one with different cardinality but equivalent. In our context we prefer the representation over the separable Hilbert space, because the countable orthonormal basis will be intrinsically related to the Fourier analysis when $X$ is a second countable locally compact abelian group, more precisely, $X = \mathbb{T}$ (\cref{section:faithful_representation_of_torus}) or $\mathbb{X} = \mathbb{Z}^d$ when we look into the twisted crossed products with $\mathbb{Z}^d$ (\cref{sec:twisted_crossed_products}).

Let $A$ be a C* algebra with a faithful representation $\pi_A : A \to H_A$, then from \cref{sec:Continuous_functions_with_values_on_C_star_algebra} we get the isomorphism
$$ C_0(X) \otimes A \simeq C_0(X,A), $$
and from \cref{theorem:spatial_tensor_product_C_star_algebra} we know that $C_0(X,A)$ has a faithful representation on $L^2(X,\mu) \otimes H_A$ given by 
$$ \pi \otimes \pi_A: C_0(X,A) \to B(L^2(X,\mu) \otimes H_A), $$
that takes the following form on simple tensor
$$ (\pi \otimes \pi_A)(f\otimes a)(h \otimes h_a) = \pi(f)(h) \otimes \pi_A(a)(h_a), \; h \otimes h_a \in L^2(X,\mu) \otimes H_A .  $$
From \cref{proposition:tensor_product_and_L_2_functs} we know that $L^2(X,\mu) \otimes H_A \simeq L^2(X,H_A; \mu)$, with the isomorphism taking the following form on simple tensors
$$ h(x) \otimes h_a \mapsto (x \to h_a h(x)), $$
also, from \cref{theprem:Lp_approximation_simple_functions} we know that linear combinations of simple tensors are dense in $L^2(X,H_A)$, thus for $h \in L^2(X,H_A)$ there is a sequence 
$$ \{ h_i \}_{i \in \mathbb{N}} = \{ \sum_{1 \leq l \leq n_i} h_{l_i}(x) \otimes h_{a_{l_i}} \}_{i \in \mathbb{N}} $$
that converges to $h$ on $L^2(X,H_A)$. From \cref{sec:Continuous_functions_with_values_on_C_star_algebra} we know that finite sums of simple tensors $f(x) \otimes a$ are dense in $C_0(X,A)$, so, take $f \in C_0(X,A)$ and 
$$\{ f_i \}_{i \in \mathbb{N}} = \{ \sum_{1 \leq k \leq m_i} f_{k_i}(x) \otimes a_{k_i} \}_{i \in \mathbb{N}} $$
a sequence of functions converging to $f$, thus $(\pi \otimes \pi_A)(f_i) \to (\pi \otimes \pi_A)(f)$ in $B(L^2(X;H_A))$, then
$$ \lim_{i \to \infty}  (\pi \otimes \pi_A)(f_i)(h_i) = (\pi \otimes \pi_A)(f)(h),$$
in $L^2(X;H_A)$. 

From \cref{proposition:useful_inequalities_of_Bochner_L_p_spaces} we know that given $h_i \to h$ in $L^2(X,H_A)$ then there is a sub-sequence $\{ h_{i(j)} \}_{j \in \mathbb{N}}$ such that $ h_{i(j)} \to h$ almost everywhere, and for this sequence we still have
$$  \lim_{j \to \infty}  (\pi \otimes \pi_A)(f_{i(j)})(h_{i(j)}) = (\pi \otimes \pi_A)(f)(h),$$
in $L^2(X;H_A)$. Denote $\zeta =  (\pi \otimes \pi_A)(f)(h)$ and $ \zeta_j = (\pi \otimes \pi_A)(f_{i(j)})(h_{i(j)}) $, then $\zeta_j \to \zeta$, thus, let $\{ \zeta_{j(k)} \}_{k \in \mathbb{N}}$ a sub sequence such that $\zeta_{j(k)} \to \zeta$ almost everywhere, then
$$ \lim_{k \to \infty}  (\pi \otimes \pi_A)(f_{i(j(k))})(h_{i(j(k))})(x) = (\pi \otimes \pi_A)(f)(h)(x) \; \text{a.e.},$$
also,
$$ (\pi \otimes \pi_A)(f_{i(j(k))})(h_{i(j(k))})(x) = $$
$$(\pi \otimes \pi_A)\left(\sum_{1 \leq p \leq m_{i(j(k))}} f_{p_{i(j(k))}}(x) \otimes a_{p_{i(j(k))}}\right)\left(\sum_{1 \leq q \leq n_{i(j(k))}} h_{q_{i(j(k))}}(x) \otimes h_{a_{q_{i(j(k))}}}\right)$$
$$ =\sum_{1 \leq p \leq m_{i(j(k))}} \sum_{1 \leq q \leq n_{i(j(k))}} (\pi \otimes \pi_A)\left(f_{p_{i(j(k))}}(x) \otimes a_{p_{i(j(k))}}\right)\left( h_{q_{i(j(k))}}(x) \otimes h_{a_{q_{i(j(k))}}}\right)$$
$$ =\sum_{1 \leq p \leq m_{i(j(k))}} \sum_{1 \leq q \leq n_{i(j(k))}} f_{p_{i(j(k))}}(x)  h_{q_{i(j(k))}}(x) \pi_A(a_{p_{i(j(k))}})\left( h_{a_{q_{i(j(k))}}}\right)$$
$$ =   \pi_A\left( \sum_{1 \leq p \leq m_{i(j(k))}} f_{p_{i(j(k))}}(x) a_{p_{i(j(k))}}\right)\left( \sum_{1 \leq q \leq n_{i(j(k))}}   h_{q_{i(j(k))}}(x) h_{a_{q_{i(j(k))}}}\right).$$

Take $x \in X$ such that $ h_{i(j(k))}(x) \to h(x)$ and $\zeta_{j(k)}(x) \to \zeta(x)$, then from the definition of the sub sequences we know that the set where one or both of the previous convergences does not happen has zero measure. So, take $x \in X$ such that $ h_{i(j(k))}(x) \to h(x)$ and $\zeta_{j(k)}(x) \to \zeta(x)$, then 
$$(\pi \otimes \pi_A)(f_{i(j(k))})(h_{i(j(k))})(x) \to \pi_A(f(x))(h(x))\; a.e.,  $$
and
$$(\pi \otimes \pi_A)(f_{i(j(k))})(h_{i(j(k))})(x) \to (\pi \otimes \pi_A)(f)(h)(x) \; a.e., $$ 
therefore
$$\pi_A(f(x))(h(x)) =(\pi \otimes \pi_A)(f)(h) \; a.e..$$
So, $(\pi \otimes \pi_A)(f)$ is a generalization of the multiplication as follows
$$(\pi \otimes \pi_A)(f)(h)(x) = \pi_A(f(x))(h(x)), \; x \in X, h \in L^2(X;H_A), \; f \in C_0(X,A).$$
Since $\pi \otimes \pi_A \in B(L^2(X;H_A))$, the previous statement is telling us that $x \mapsto \pi_A(f(x))(h(x))$ is a $\mu$-strongly measurable function which lies in $L^2(X;H_A)$. Additionally, if $X$ is second countable then from \cref{remark:tensor_product_separable_C_star_algebras} if $A$ is separable, then $C_0(X;A)$ is separable, and as we have seen, it has a faithful representation on the separable Hilbert space $L^2(X,H_A)$.

\begin{lemma}[Faithful representation of $C_0(X;A)$]\label{lemma:faithful_representation_of_functions_decaying_with_values_on_algebra}
Let $A$ be a C* algebra with a faithful representation $\pi_A : A \to H_A$, let $X$ be a locally compact Hausdorff space and $\mu$ a Radon measure (\cref{definition:Radon_measure}) over $X$ with full support (\cref{definition:measure_with_full_support}), then, $C_0(X,A)$ has a faithful representation over $L^2(X,H_A)$ given by
$$(\pi \otimes \pi_A)(f)(h)(x) = \pi_A(f(x))(h(x)), \; x \in X, h \in L^2(X;H_A), \; f \in C_0(X,A).$$
\end{lemma}
\begin{proof}
\cref{sec:Continuous_functions_with_values_on_C_star_algebra} tell us that we have the following isomorphism of C* algebras,
$$ C_0(X) \otimes A \simeq C_0(X,A).$$
From \cref{lemma:faithful_representation_of_functions_decaying_at_infinity} we know that $C_0(X)$ has a faithful representation over $L^2(X;\mu)$, so, using \cref{theorem:spatial_tensor_product_C_star_algebra} we know that $C_0(X,A)$ has a faithful representation on $L^2(X,\mu) \otimes H_A$ given by 
$$ \pi \otimes \pi_A: C_0(X,A) \to B(L^2(X,\mu) \otimes H_A), $$
that takes the following form on simple tensor
$$ (\pi \otimes \pi_A)(f\otimes a)(h \otimes h_a) = \pi(f)(h) \otimes \pi_A(a)(h_a), \; h \otimes h_a \in L^2(X,\mu) \otimes H_A .  $$

Moreover, since $L^2(X,\mu) \otimes H_A \simeq L^2(X,H_A; \mu)$ (\cref{proposition:tensor_product_and_L_2_functs}), with the isomorphism taking the following form on simple tensors
$$ h(x) \otimes h_a \mapsto (x \to h_a h(x)), $$
we have that, given $h \in L^2(X,H_A; \mu)$ then
$$\pi_A(f(x))(h(x)) =(\pi \otimes \pi_A)(f)(h) \; a.e..$$
So, $(\pi \otimes \pi_A)(f)$ is a generalization of the multiplication as follows
$$(\pi \otimes \pi_A)(f)(h)(x) = \pi_A(f(x))(h(x)), \; x \in X, h \in L^2(X;H_A), \; f \in C_0(X,A).$$
\end{proof}

We could argue that 
$$ x \mapsto \pi_A(f(x))(h(x)) $$
is a sensible generalization for the representation of functions in $C_0(X)$ as multiplication operators. So, instead of starting with the representation of the tensor product of C* algebras $C_0(X) \otimes A$, and then proving that it gives us the expected result when we go from $L^2(X) \otimes H_A$ to $L^2(X;H_A)$, we could have just prove directly that 
$$ f \mapsto (x \mapsto \pi_A(f(x))(h(x))) $$
is a faithful representation of $C_0(X,A)$. This second approach is more natural from the point of view of generalizing the multiplication operator representation, however, we would have need to assume that $A$ is separable and $\mu$ is $\sigma$-compact in order to show that $x \to \pi_A(f(x))(h(x))$ is $\mu$-strongly measurable, and would need to use the results from \cref{sec:oper_valu_func}. Let's take this approach for educational purposes.

\begin{lemma}[Faithful representation of $C_0(X,A)$ with $A$ separable and $\mu$ $\sigma$-finite]\label{lemma:faithful_representation_C_0_X_A_separable_mu_sigma_finite}
Let $A$ be a separable C* algebra, let $X$ be a second countable locally compact Hausdorff space, and $\mu$ a $\sigma$-compact measure over $X$. For $f \in C_0(X,A)$, then, for all $h \in L^2(X,H_A)$ we have that
$$ \widehat{\pi_A} = \pi \otimes \pi_A ,$$
where we set $\widehat{\pi_A}(f)(h)(x) := \pi_A(f(x))(h(x)).$
\end{lemma}
\begin{proof}
Given that $A$ is separable, according to \cref{theorem:faithfull_universal_representation_c_star_algebras} we can choose $H_A$ to be separable. Notice that $x \mapsto \pi_A(f(x))$ is a continuous function, which implies that it is a measurable function with respect to the Borel $\sigma$ algebra of the norm topology on $B(H_A)$, additionally, if $\mu$ is $\sigma$-finite, we can use \cref{prop:oper_valu_new_strong_measu_funcs_in_separable_case} to show that:
\begin{itemize}
    \item if $x \to h(x)$ is $\mu$-strongly measurable then $x \to \pi_A(f(x))(h(x))$ is $\mu$-strongly measurable.
\end{itemize}
This argument assure us that $x \to \pi_A(f(x))(h(x))$ belongs to $L^2(X,H_A)$ when $h \in L^2(X,H_A)$.

Since $f \in C_0(X,A)$ then it is bounded, thus,
$$ \int_{X} \| \pi_A(f(x))(h(x)) \|^2 d \mu(x) \leq \| f\|^2 \int_X \| h(x) \|^2 d \mu(x), $$
and we have that $(x \mapsto \pi_A(f(x))(h(x))) \in B(L^2(X, H_A))$. Lets denote
$$ \widehat{\pi_A}(f)(h)(x) := \pi_A(f(x))(h(x)), $$
then, we need to check that $\widehat{\pi_A}$ is a faithful representation of $C_0(X,A)$. We have already shown that $\|\widehat{\pi_A}(f) \| \leq \| f\|$, so, we need to show that for any $\epsilon > 0$ $\|\widehat{\pi_A}(f) \| \geq \| f \| + \epsilon$, and we are done. Let's copy the argument from the scalar-valued case, take $x \in X$ such that $\| f(x) \| = \|f\|$ and let $U_x \subset X$ and compact neighbourhood of $x$ such that
$$ \text{ if } y \in U_x \text{ then } \| f(x) - f(y)\| \leq \epsilon/2. $$
Let $h \in H_A$ be a normalized vector such that
$$ | \| f(x) \| - \| \pi_A(f(x))(h) \| | \leq \epsilon/2,$$
then, by the triangle inequality, we have that
$$ | \| \pi_A (f(y))(h) \| - \| \pi_A (f(x))(h) \| | \leq \epsilon/2,  $$
consequently, for all $y \in U_x,$
$$ \| f \| - \epsilon \leq \| \pi_A (f(y))(h) \|. $$
So, if we define $g := 1_{U_x}h/\mu(U_x)$ we have that
$$ (\|f\| - \epsilon )^2 \leq \int_X \| 1_{U_x} \pi_A (f(y))(h) /\mu(U_x) \|^2 d \mu(y), $$
which implies that $\| \widehat{\pi_A}(f)\| =  \|f\|$, that is, $\widehat{\pi_A}$ is a faithful representation of $C_0(X,A)$. Additionally, is easy to check that 
$$
\begin{aligned}
\widehat{\pi_A}(f \otimes a)(x \mapsto h_a h(x)) = \pi \otimes \pi_A(f \otimes a) ( (x \mapsto h(x)) \otimes h_a), \\
\forall (x \mapsto h(x)) \in L^2(X), \; h_a \in H_A, \; f \in C_0(X), \; a \in A,
\end{aligned}
$$
therefore, since $L^2(X) \odot H_A$ is dense in $L^2(X) \otimes H_A$ (\cref{definition:tensor_product_hilbert_spaces}) we have that $\widehat{\pi_A}$ and $\pi \otimes \pi_A$ coincide over $C_0(X) \odot A$. Since $C_0(X) \odot A$ is dense in $C_0(X,A)$ (\cref{sec:Continuous_functions_with_values_on_C_star_algebra}) we end up with the desired equality
$$ \widehat{\pi_A} = \pi \otimes \pi_A .$$
\end{proof}

\begin{lemma}[Faithful representation of $C_b(X,A)$]\label{lemma:faithful_representation_C_b_X_A}
The previous argument can be generalised to the C* algebra $C_b(X,A)$, given that, when $\mu$ is $\sigma$-finite the elements of $C_b(X,A)$ are $\mu$-strongly measurable (\cref{example:continuous_functions_on_sigmna_fintie_space}), so, given $\pi_A : A \to B(H_A)$ a faithful representation of $A$, with $H_A$ a separable Hilbert space, then,
$$ \widehat{\pi_A} : C_b(X,A) \to L^2(X,H_A), \; \widehat{\pi_A}(f)(h)(x) := \pi_A(f(x))(h(x), $$
$$ f \in C_b(X,A), \; h \in L^2(X,H_A)  $$
is a faithful representation. Notice that this is giving us a generalization of the representation of $C_b (X)$ on $L^2(X)$.
\end{lemma}

Let $A$ be a unital C* algebra, notice that $C_0(X,A)$ is a sub C* algebra of $C_b(X,A)$, and $C_b(X,A)$ comes as a generalization of $C_b(X)$. Take $f \in C_0(X,A), \; g \in C_b(X,A)$, then $fg \in C_0(X,A)$, thus, from the definition of multiplier C* algebra (\citep[Definition 2.2.2]{wegge-olsen_k-theory_1993}) we have that
$$ C_b(X,A) \subseteq \mathcal{M}(C_0(X,A)),$$
we do not know if the reverse inclusion holds, but it seems reasonable for that to happen, since $C_b(X) \simeq \mathcal{M}(C_0(X))$ (\cref{remark:multiplier_algebras}). 

The C* algebra $C_0(X,A)$ is a continuous field of C* algebras (\citep[Example IV.1.6.5]{blackadar_operator_2006}), it is called the constant field. The C* algebra $C_b(X,A)$ behaves very much like a continuous filed of C* algebras, with the caveat that $(x \mapsto \| f(a) \|) $ does not necessarily belongs to $C_0(X)$ (\citep[Definition IV.1.6.1]{blackadar_operator_2006}). So, let $\pi_A : A \mapsto B(H_A)$ be a faithful representation, then, the following defines a family of representations of both $C_b(X,A), C_0(X,A)$
$$ \pi_x : C_b(X,A) \to B(H_A), \; \pi_x (f) := \pi_A(f(x)), $$
$$ \pi_x : C_b(X,A) \to B(H_A), \; \pi_x (f) := \pi_A(f(x)). $$
Since each one of the $\pi_x$ is an algebra homomorphis, from \cref{sec:banach_alg_morph} we know that
$$ Sp(\pi_x(f)) \subseteq Sp(f), $$
therefore,
$$  \bigcup_{x \in X} Sp(\pi_x (f)) \subseteq Sp(f), $$
in fact, we have the following

\begin{lemma}[Spectrum of elements in $C_b(X,A), C_0(X,A)$]\label{lemma:spectrum_elements_C_b_X_A}
Let $f \in C_0(X,A)$ ($C_b(X,A)$), then
\begin{itemize}
    \item if $X$ is compact then
    $$ Sp(f) =  \bigcup_{x \in X} Sp(\pi_x (f)),$$
    \item if $X$ is not compact then
    $$\overline{ \left( \bigcup_{x \in X} Sp(\pi_x (f))  \right)} \subseteq Sp(f),$$
    and if $f$ is normal then
    $$ \overline{ \left( \bigcup_{x \in X} Sp(\pi_x (f))  \right)} = Sp(f). $$
\end{itemize}
\end{lemma}
\begin{proof}
Assume that $A$ is a unital C* algebra, in case it is not unital replace $A$ by $A^+$ and the following arguments would still hold, since the spectrum of an element in a non unital C* algebra is defined as its spectrum in the unitization of the algebra (\cref{sec:unitization_of_C_star_algebras}). 

$C_0(X,A)$ is a sub C* algebra of $C_b(X,A)$ and has no unit, hence, from \cref{corollary:spectrum_and_ambient_C_star_algebra} we know that for $f \in C_0(X,A)$ 
$$Sp_{C_0(X,A)}(f) = Sp_{C_b(X,A)}(a).$$

Since 
$$\bigcup_{x \in X} Sp(\pi_x (f)) \subseteq Sp(f),$$
then, if $a - \lambda 1 \in G(C_b(X,A))$ we must have that $\pi_x(a) - \lambda 1 \in G(B(H_A))$, thus, assume that $\pi_x(a) - \lambda 1 \in G(B(H_A))$ for all $x \in X$. Since the inversion is a continuous map (\cref{sec:banach_alg_top_of_spec}), we have that the map
$$ x \mapsto (\pi_x(a) - \lambda 1)^{-1} $$
is also continuous, moreover, since the norm is a continuous map (\cref{remark:norm_is_continuous}), then, the map
$$ x \mapsto \| (\pi_x(a) - \lambda 1)^{-1} \|  $$
is also continuous. So, in order for $\lambda \in Sp(f)$ we must have that
$$ x \mapsto \| (\pi_x(a) - \lambda 1)^{-1} \|$$
is an unbounded map.

Since any continuous function on a compact space has a maximum, if $X$ is compact then 
$$ x \mapsto \| (\pi_x(a) - \lambda 1)^{-1} \| $$
is bounded, thus
$$ Sp(f) = \bigcup_{x \in X} Sp(\pi_x (f)).$$

Let's look into the case where $X$ is non compact. 
\begin{itemize}
    \item Assume that $\lambda \in \overline{ \left( \bigcup_{x \in X} Sp(\pi_x (f))  \right)}$ and $\lambda \notin \bigcup_{x \in X} Sp(\pi_x (f))$, then, for any $\epsilon >0$ there is $x \in X$ such that $d(\lambda, Sp(\pi_x(f))) \leq \epsilon$, so, from \cref{proposition:bound_on_the_resolvent_norm} we have that 
    $$ \| (\pi_x(a) - \lambda 1)^{-1} \| \geq \frac{1}{\epsilon}, $$
    therefore, $\lambda \in Sp(f)$.
    \item If $f$ is normal then we have a tider bound on the norm of th resolvent at $\lambda$ given by \cref{lemma:norm_of_resolvent_of_normal_element}, which tell us that if $d(\lambda, Sp(\pi_x (a))) = l$ then
    $$ \| (\pi_x(a) - \lambda 1)^{-1} \| = 1/l, $$
    therefore, $x \mapsto (\pi_x(a) - \lambda 1)^{-1} \|$ is unbounded iff 
    $$ \lambda \in \overline{ \left( \bigcup_{x \in X} Sp(\pi_x (f))  \right)}. $$
\end{itemize}
\end{proof}

\subsection{Algebra of continuous functions over the torus}
\label{sec:algerba_continuous_functions_over_torus}

\subsubsection{Faithful representation of the Torus C* algebra}
\label{section:faithful_representation_of_torus}

The torus is a separable compact Hausdorff space, and is homeomorphic to $[0, 2\pi], \; 0 \sim 2 \pi$, thus, we can use \cref{section:algebra_continuous_functions_locally_comp_space} to get useful information:

\begin{itemize}
    \item $\mathbb{T}$ is locally compact abelian group, thus we will take $\mu$ to be the Haar measure over $\mathbb{T}$. Since the Haar measure is a Radon measure (\cref{definition:Haar_measure}), $L^2(\mathbb{T})$ with the Haar measure is a separable Hilbert space, and has a countable complete orthonormal basis (\cref{lemma:separability_Hilbert_spaces_and_orthonormla_basis}).
    \item $C(\mathbb{T})$ is separable and is dense in $L^2(\mathbb{T})$
    \item $C(\mathbb{T})$\index{$C(\mathbb{T})$} has a faithful representation on $L^2(\mathbb{T})$ as a multiplication operator.
\end{itemize}

As we mentioned in \cref{chapter:hilbert_spaces_section} the ortonormal basis for $H=L^2[0, 2 \pi]$ is not unique, thus, we will focus on a particular orthonormal basis that comes from the Fourier analysis (\cref{section:Fourier_analysis_and_ortonormal_basis}). First, recall that Fourier analysis tell us that there is an isomorphism of Hilbert spaces
$$ \mathcal{F} :  l^2(\mathbb{Z}) \to L^2(\mathbb{T})  , \; \mathcal{F}((a_n)_{n \in \mathbb{Z}})(\exp(i \alpha)) = \sum_{n \in \mathbb{Z}} a_{i} \exp(- i \alpha n), \; \alpha \in [0,2 \pi)$$
and gives us the complete orthonormal basis $\{ \exp(i \alpha) \mapsto \exp(i n \alpha) \}_{n \in \mathbb{Z}}$ over $L^2(\mathbb{T})$, we are taking the Haar measure as normalized over $\mathbb{T}$ such that the Haar measure over its dual group $\mathbb{Z}$ is the counting measure (\cref{definition:dual_group}). Under this basis any function of $L^2(\mathbb{T})$ can be expressed as $\sum_{n \in \mathbb{Z}} a_{i} \exp(i \alpha n)$ that converges on $L^2(\mathbb{T})$, or in other terms, the Fourier series of $f$ converges to $f$ in the $L^2$ norm, moreover, if $f \in C(\mathbb{T})$ then the Cesaro sum of the Fourier series of $f$ converges to $f$ in the uniform topology i.e. in the topology of $C(\mathbb{T})$  (\cref{section:convergence_of_the_Fourier_series}).

Let $f$ in $L^2(\mathbb{T})$, then 
$$\mathcal{F}^{-1}(f)(n) = \frac{1}{2 \pi} \int_{0}^{2 \pi} f(\exp(i\alpha)) \exp(-i n \alpha) d \alpha ,$$
and from Plancherel theorem (\cref{theorem:Plancherel_theorem}) we know that
$$ \mathcal{F}^{-1} (fg) = \mathcal{F}^{-1}(f) \ast \mathcal{F}^{-1}(g).$$
Let $f \in C(\mathbb{T})$, $\pi: C(\mathbb{T}) \to B(L^2(\mathbb{T}))$ the faithful representation given by multiplication operators and $g \in L^2(\mathbb{T})$. For $h \in L^2(\mathbb{T})$ denote $\hat{h} := \mathcal{F}^{-1}(f)$ then    $$\pi(f)(g)(\exp(i \alpha)) = \sum_{n \in \mathbb{Z}} \widehat{fg}(n) \exp(in \alpha),$$
with 
$$ \widehat{fg}(n) = (\hat{f} \ast \hat{g})(n) = \sum_{k \in\mathbb{Z}} \hat{f}(n - k)\hat{g}(k).$$

Comparing the previous formula with the description of operators on $L^2(\mathbb{T})$ as infinite matrices (\cref{sec:infinite_mattrices_and_bounded_operators}) we conclude that the representation of $\pi(f)$ as an infinite matrix is 
$$ \pi(f) = [\hat{f}(l-m)]_{l,m \in \mathbb{Z}}, $$
which is a matrix with entries constant in the diagonals,
$$\pi(f) = \begin{bmatrix}
 & \vdots & \vdots & \vdots &   \\ 
\cdots & \hat{f}(0) & \hat{f}(1) & \hat{f}(2) \cdots & \\ 
\cdots & \hat{f}(-1) & \hat{f}(0) & \hat{f}(1)  &  \cdots\\ 
\cdots & \hat{f}(-2) & \hat{f}(-1) & \hat{f}(0) & \cdots\\ 
 & \vdots & \vdots & \vdots & 
\end{bmatrix} .$$

This matrix representation can also be computed with as indicated in \cref{sec:infinite_mattrices_and_bounded_operators}, this is, if $\pi(f) = [f_{i,j}]_{i,j \in \mathbb{Z}}$ then
$$ f_{l,m} = \mathcal{F}^{-1}(\pi(f)(\exp(i m \alpha)))(l) = \frac{1}{2 \pi} \int_{0}^{2 \pi} ( \exp(i m \alpha) f(\exp(i \alpha)) )\exp(-i l \alpha)$$
$$ = \frac{1}{2 \pi} \int_{0}^{2 \pi} f(\exp(i \alpha)) \exp(-i (l -m) \alpha)  = \hat{f}(l-m). $$

Denote $\{S_n(f)\}_{n \in \mathbb{N}}$ the Fourier series of $f$ (\cref{section:Fourier_analysis_and_ortonormal_basis}), then
$$\pi(S_N(f)) = [\hat{f}(l-m) \text{ if } |l-m| \leq N \text{ else } 0]_{l,m \in \mathbb{Z}},$$
and $\pi(S_N(f))$ converges to $\pi(f)$ in the strong topology, however, it does not necessarily converges in the supremum topology because there is a continuous function whose Fourier series diverges at a point (\cref{section:convergence_of_the_Fourier_series}). However, the Cesaro sum of the Fourier series, which we denote by $\{K_n \ast f\}_{N \in \mathbb{N}}$ with $\{K_N\}_{N \in \mathbb{N}}$ the Fejér kernel, converge in the supremum norm to $f$, and has a matrix representation given by 
$$ \pi(K_n \ast f) = [(1 - \frac{|l-m|}{N+1})\hat{f}(l-m) \text{ if } |l-m| \leq N \text{ else } 0]_{l,m \in \mathbb{Z}}.$$

Let $f \in C(\mathbb{T})$ with $f(z) = z$ i.e. $f(\exp(i \alpha)) = \exp(i \alpha)$ for $\alpha \in [0, 2 \pi]$, thus, if $S \in B(l^2(\mathbb{Z}))$ given by $S(e_j) = e_{j-1}$ (the right shift), then $\mathcal{F}(S(h)) = \pi(f)(\mathcal{F}(h))$ for any $h \in l^2(\mathbb{Z})$, and this extends to any polynomial on $S, S^*$ (notice that $S^*$ is the left shift over $l^2(\
\mathbb{Z})$ and $S S^* = S^* S = Id_{l^2(\mathbb{Z})}$). Since the Fourier transform is an isomorphism of Hilbert spaces, the C* algebra generated by $S$ on $B(l^2(\mathbb{Z}))$ is isomorphic to the C* algebra generated by $f$ on $B(L^2(\mathbb{T}))$, which is isomorphic to $C(\mathbb{T})$ by the Weierstrass theorem (\cref{section:C_star_generated_by_a_unitary}), consequently, $Sp(S) = Sp(f) = \mathbb{T}$, which is a fact that can be obtained from a pure analysis of the shift operator (\citep[Section 9a example 6.3]{garret_real_2020}).

Under the previous discussion the Toeplitz short exact sequence (\cref{section:Toeplitz_algebra}) becomes,
$$0 \longrightarrow \mathcal{K}(l^2(\mathbb{N})) \longrightarrow C^*(S_{l^2(\mathbb{N})}) \longrightarrow C^*(S_{l^2(\mathbb{Z})}) \longrightarrow 0.$$

Notice that $\pi(f)$ is not a compact operator for any $f \in C(\mathbb{T})$, because its matrix representation consists of infinite diagonal blocks, therefore, there is no finite range approximation of $\pi(f)$ under the operator norm. The aforementioned exact sequence of C* algebras is telling us how the Toeplitz algebra is the extension of an algebra of non-compact operators by an algebra of compact operators.

Using the matrix representation of $\pi(f)$ and the inequalities on the norm of $\pi(f)$ from \cref{sec:infinite_mattrices_and_bounded_operators} we also get that 
$$\sup_{z \in \mathbb{T}}\{ |f(z)| \} = \| \pi(f) \|_{C(\mathbb{T})} \geq |\hat{f}(n) |$$
for any $n \in \mathbb{Z}$, so, if $f_l \to f$ in $C(\mathbb{T})$ then $\hat{f_l}(n) \to \hat{f}(n)$ and 
$$| \hat{f_l}(n) - \hat{f}(n)| \leq \| f_n -f \|_{C(\mathbb{T})}.$$

\subsubsection{N Torus}
\label{section:n_torus}

Denote $A = C^*(\mathcal{G} | \mathcal{R})$ with $\mathcal{G} = \{ u_1, u_2\}$ and $\mathcal{R} = \{u_i u_i^* = u_i^* u_i =1, \text{ for } i =1,2, \; u_1 u_2 = u_2 u_1 \}$, then $\| u_i \|=1$ because those are unitaries, so by \cref{remark:bound_in_the_generators_guaranties_existence_of_enveloping_c_star_algebra} $A$ exists. If we denote $u_i^* = u_i^{-1}$ then the *-algebra
$$ \mathcal{A} = \{ \sum_{l,m \in P} \alpha_{l,m} u_1^l u_2^m, \; \alpha_{l,m} \in \mathbb{C}, \; P \subset \mathbb{Z}^2 \text{ and } |P| \text{ finite}\}, $$
then $\mathcal{A}$ is dense inside $A$ by definition, and we can check that $\mathcal{A}$ is isomorphic to 
$$ \mathcal{A}_1 \odot \mathcal{A}_2 $$
with 
$$\mathcal{A}_i = \{ \sum_{l\in P} \alpha_{l} u_i^l, \; \alpha_{l} \in \mathbb{C}, \; P \subset \mathbb{Z} \text{ and } |P| \text{ finite} \}.$$

From \cref{lemma:tensor_products_and_generated_C_star_algebras} we get that 
$$ C^*(\mathcal{A}_1) \otimes C^*(\mathcal{A}_2) \simeq C^*(\mathcal{A}_1 \odot \mathcal{A}_2),$$
so, since \cref{section:C_star_generated_by_a_unitary} tell us that $\mathcal{A}_i = C(\mathbb{T}), \; i = 1,2,$ then 
$$ A \simeq C(\mathbb{T})\otimes C(\mathbb{T}).  $$
Additionally, from \cref{sec:Continuous_functions_with_values_on_C_star_algebra} we know that 
$$  A \simeq C(\mathbb{T} \times \mathbb{T}).$$

$A$ has a faithful representation over $L^2(\mathbb{T}^2)$, which in turn is isomorphic to $l^2(\mathbb{Z}^2)$ by the Fourier analysis (\cref{section:Fourier_analysis_and_ortonormal_basis}), and $L^2(\mathbb{T}^2)$ has a complete orthonomal basis given by 
$$  \{(\alpha_1, \alpha_2) \mapsto \exp(i \alpha_1 n) \exp(i \alpha_2m): \; \alpha_i \in [0, 2 \pi] \}_{(n,m) \in \mathbb{Z}^2}.$$

As in \cref{section:faithful_representation_of_torus} every $f \in C(\mathbb{T}^2)$ has a representation as an infinite matrix under the complete orthonormal basis given by the Fourier analysis
$$ \pi(f) = [\hat{f}( \overrightarrow{n} - \overrightarrow{m})]_{\overrightarrow{n}, \overrightarrow{m} \in \mathbb{Z}^2}, $$
which can be deduced from the formula of convolution and multiplication of functions for the Fourier transform. Also, $\| f \| \geq |\hat{f}(\overrightarrow{n})|$ for any $\overrightarrow{n} \in \mathbb{Z}^2$. 

This procedure can be iterated to get\index{$C(\mathbb{T}^d)$} 
$$ C^*(u_1, \dotsc, u_d | u_i u_i^* = u_i^*u_i = 1, u_i u_j = u_j u_i) \simeq C(\mathbb{T}^d),$$
along with the faithful representation
$$ \pi(f) = [\hat{f}( \overrightarrow{n} - \overrightarrow{m})]_{\overrightarrow{n}, \overrightarrow{m} \in \mathbb{Z}^d}, $$
with
$$ \hat{f}(\overrightarrow{s}) = \int_{\mathbb{T}^d} \exp(-i \overrightarrow{s} \cdot \overrightarrow{\lambda}) d \mu(\lambda). $$

Since the Fejér kernel can be used to get a sequence of trigonometric polynomials over $\mathbb{T}^d$ converging uniformly to $f$ for every $f \in C(\mathbb{T}^d)$ (\cref{section:convergence_of_the_Fourier_series}), we get that $f$ is uniquely determined by $\hat{f}$. Also, from \cref{proposition:isomorphism_tensor_product_of_commutative_C_star_algebras} we get that
$$ C(\mathbb{T}^d) \simeq C(\mathbb{T}) \otimes \cdots \otimes C(\mathbb{T}^d) ,$$
such that, $C(\mathbb{T}^l)$ is a sub C* algebra of $C(\mathbb{T}^d)$ for every $1 \leq l \leq d$.

If $C$ is a C* algebra then \cref{remark:nuclear_C_star_algebra_enveloping_C_star_algebra} tell us that 
$$ C \otimes C(\mathbb{T}^d) \simeq C^*(C \odot \underbrace{\mathcal{A}_1 \odot \cdots \odot \mathcal{A}_d}_{d \text{ times }}), $$
therefore, $C \otimes C(\mathbb{T}^d)$ is the C* algebra generated by the set of elements of the form $c u_j$ with $j=1, \cdots, d$ and
$$ cu_j = u_j c, \;  u_i u_i^* = u_i^*u_i = 1, \; u_i u_j = u_j u_i.$$
Also, if $C^*(\mathcal{C})$ with $\mathcal{C}$ a *-algebra then \cref{remark:nuclear_C_star_algebra_enveloping_C_star_algebra} tell us that 
$$ C \otimes C(\mathbb{T}^d) \simeq C^*(\mathcal{C} \odot \underbrace{\mathcal{A}_1 \odot \cdots \odot \mathcal{A}_d}_{d \text{ times }}), $$
so, we only need to take into account the elements of $\mathcal{C}$ to define $C \otimes C(\mathbb{T}^d)$ as an enveloping C* algebra.

In \cref{definition:Haar_measure} is mentioned that the Haar measure $\mu$ over the torus is a Radon measure with full support, thus from \cref{section:non_commutative_geometry_dictionary} we get that the map
$$ f \mapsto \int_{\mathbb{T}^d} f(\lambda) d \mu(\lambda) $$
is a continuous trace over $C(\mathbb{T}^d)$, which happens to be faithful because $\mu$ has full support (\cref{definition:measure_with_full_support}).

In \cref{chap:fourier_analysis} and \cref{chap:K_theory} you can look for other examples of C* algebras.

\chapter{Fréchet algebras}
\label{chapter:frechet_algebras}

Fréchet algebras are a common tool for Non-Commutative Geometry and physics, these usually arise are dense sub-algebra of C* algebras (  \citep[Introduction]{gramsch_relative_1984}, \citep[Introduction]{fragoulopoulou_topological_2005}) and provide a well behaved setting for dealing with an algebraic approach to differential geometry over smooth manifolds (\citep{kainz_c-algebras_1987}, \citep{michor_characterizing_1994}). Fréchet algebras (or locally convex algebras in general) often arise as objects that capture unboundedness, for example, dense Fréchet sub *-algebras of C* algebras often arise as algebras were unbounded close derivations are well defined and continuous (\cref{sec:Frechet_d_infinity_subalgebras}). Additionally, locally convex algebras can arise as algebras of unbounded operators over a Hilbert space, which provided useful tools in the analysis of systems on quantum mechanics \citep[Applications of Generalized B*-Algebras to Quantum Mechanics]{kikianty_positivity_2021}, among those algebras the GB* algebras stand out as a generalization of Banach *-algebras \citep{fragoulopoulou_generalized_2022}.

In this section, we will focus on Fréchet algebras, since those are the most common setting for dealing with smooth sub algebras of C* algebras (\cref{sec:smooth_subalgebras}). Additionally, Fréchet algebras provide a useful setting for working with cyclic cohomology when dealing with nuclear C* algebras (\cref{section:mathematical_framework}). When dealing with C* algebras for homogeneous materials a.k.a. the Non Commutative Brilluoin torus (\cref{definition:non_commutative_brillouin_torus}), Fréchet algebras are the right setting for providing physical interpretations for the topological invariants that come up in the K theory of the Non-Commutative Brillouin Torus (\cref{sec:motivation_from_physics})

\begin{definition}[Topological vector space]\label{definition:topological_vector_space}
$A$ is a Topological Vector Space, so, it has the following properties (\citep[Definition in page 67]{narici_topological_2011}):
\begin{itemize}
    \item $A$ is a vector space over $\mathbb{C}$ (you can use other fields)
    \item $A$ has a topology where the addition and inverse additive are continuous functions, that is, the maps $+ : A \times A \to A,\; a + b \mapsto a + b$ and $^{-1} : A \to A, \; a \mapsto -a $ are continuous
    \item The scalar multiplication is continuous, that is, the map $\text{sm} : \mathbb{C} \times A \to A, \; sm(\lambda,a) = \lambda a$ is continuous.
\end{itemize}
\end{definition}

\begin{remark}\label{remark:translation_invariance_topology_in_tvs}
Let $A$ be a topological vector space, then, it has a topology that is translation invariant, which means that if $S$ is a subset of $A$ that contains $a$, then, $S$ is a neighbourhood of $a$ iff $x + S$ is a neighbourhood of $a + x$ (\citep[Page 14]{schaefer_topological_1999}).
\end{remark}

\begin{definition}[Locally convex space\index{locally convex space} (Page 122 \citep{narici_topological_2011}) ]\label{def:locally_convex_space}
$A$ is considered a locally convex space if it is a Topological Vector Space with a locally convex topology. In the context of topological vector spaces being locally convex is equivalent to the existence of a neighbourhood base at $0$ of convex sets (\citep[Definition 4.4.5]{narici_topological_2011}). A set $S$ is called convex\index{convex set} if $ta + (1-t)b$ belongs to $S$ whenever $a,b \in S$ and $t \in [0,1]$. If $A$ is a locally convex space then we will refer to $A$ as a LCS\index{LCS}.
\end{definition}

\begin{definition}[Absolutely convex set\index{locally convex space!absolutely convex set} (Theorem 4.2.8 \citep{narici_topological_2011})]\label{def:absolutely_convex_set}
Let $A$ be a LCS, then, a subset $D \subset A$ is called absolutely convex or disked if for all scalars $a,b$ such that $|a| + |b| \leq 1$ then $aD + bD \subset D$.
\end{definition}

The description given in \cref{def:locally_convex_space} gives us an overall structure of the topology of Locally Convex Space, however, is quite hard to perform computations with such a definition, fortunately, there is an analytical characterization of locally convex spaces using seminorms. Before we dive into this characterization we introduce some notation and definitions.

\begin{definition}[Seminorm \index{seminorm}]\label{def:seminorm}
Let $A$ be a vector space over $\mathbb{C}$, then a real valued function $p : A \to \mathbb{R}^{+}$ is called a seminorm if
\begin{itemize}
    \item Triangle inequality: $p(a + b) \leq p(a) + p(b)$
    \item Absolute homogeneity: $p(\lambda a ) = | \lambda | p (a)$ for $\lambda \in \mathbb{C}$ and $a \in A$ 
\end{itemize}
\end{definition}

One key aspect of LCS are topologies generated by seminorms, there are defined as follow,

\begin{definition}[Topology generated by seminorms \index{topology generated by seminorms} ( Example 4.5.4 \citep{narici_topological_2011})]\label{def:topology_geenrated_by_seminorm}
Let $A$ be a vector space and $P$ a family of seminorms over $A$, then, the topology generated by $P$ on $A$ is the weakest topology with respect to which each $p \in P$ is continuous, equivalently, this topology has as a subbasis the following sets
$$ \{ x,y \in A | p(x-y) \leq \epsilon \}, \; \epsilon > 0, p \in P .$$
\end{definition}

Now we are in position to give the characterization of locally convex spaces in terms of seminorms,

\begin{theorem}[Characterization of locally convex spaces using seminorms (cf. Theorem 5.5.2 \citep{narici_topological_2011})]\label{theorem:characterization_of_lcs_using_seminorms}
$A$ is a LCS iff it is a topological vector space whose topology is generated by a family of seminorms.
\end{theorem}

\begin{definition}[Equivalent families of seminorms \index{equivalent families of seminorms} (page 96 \citep{horvath_topological_1966})]\label{def:equivalent_families_of_seminorms}
Let $A$ be a LCS whose topology is generated by the family of seminorms $P$, let $F$ by another family of seminorms over $A$. We say that $P$ is equivalent to $F$ if $P$ and $F$ generate the same locally convex topology over $A$. 
\end{definition}

\begin{definition}[Saturated family of seminorms\index{saturated family of seminorms} (c.f. Definition 5.5.5 \citep{narici_topological_2011}, page 96 \citep{horvath_topological_1966})]\label{def:saturated_family_of_seminorms}
Let $A$ be a LCS whose topology is generated by the family of seminorms $P$, we say that $P$ is saturated family of seminorms if for any $I$ finite subset of $P$ we have that $r(x) = \max_{p \in I} p(x)$ belongs to $P$. If $P$ is countable then $P$ is called upper directed if $p_n \leq p_m$ when $m > n$.\index{upper directed family of seminorms} 
\end{definition}

It turns out that any family of seminorms is equivalent to a saturated family of seminorms, moreover, if the family of seminorms is countable then it is equivalent to an upper-directed family of seminorms,

\begin{lemma}[Equivalent saturated family of seminorms (page 96 \citep{horvath_topological_1966})]\label{lemma:equivalent_satureted_family_of_seminorms}
Let $A$ be an LCS whose topology is generated by the family of seminorms $P$, then
\begin{itemize}
    \item $P$ is equivalent to the family of seminorms $\tilde{P}$ whose elements are the seminorms
    $$ r_I(x) = \max_{p \in I} p(x), $$
    where $I$ runs over all finite subsets of $P$. The family of seminorms $\tilde{P}$ is saturated.
    \item Let $P$ be a countable family of seminorms, then, $P$ is equivalent to the family of seminorms $\tilde{P}$ whose elements are
    $$ r_m(x) = \max_{n \leq m} p_n (x), m \in \mathbb{N}. $$
    The family of seminorms $\tilde{P}$ is saturated.
\end{itemize}
\end{lemma}
\begin{proof}
Let $A$ be an LCS with a topology generated by the family of seminorms $P$ and $\tilde{P}$ be its saturated family of seminorms described in the statement. Take $I$ a finite subset of $P$, then, the definition of the topology generated by a family of seminorms (\cref{def:topology_geenrated_by_seminorm}) tells us that the set
$$ \{ x,y \in A | r_I(x-y) \leq \epsilon \}$$
is open in the topology generated by $\tilde{P}$, and corresponds to the intersection of the sets 
$$ \{ x,y \in A | p(x-y) \leq \epsilon \}, \; p \in I $$
which in turn are open sets in the topology generated by $P$. Given that the sets of the form 
$$ \{ x,y \in A | r_I(x-y) \leq \epsilon \}$$
are a subbasis of the topology generated by $\tilde{P}$, we have that the topology generated by $P$ is stronger than the topology generated by $\tilde{P}$. Also, any set of the form $\{ x,y \in A | p(x-y) \leq \epsilon \}$ belongs to the topology generated by $\tilde{P}$ because $I = \{ p \}$ is a finite subset of $P$, so, given that the set of the form
$$\{ x,y \in A | p(x-y) \leq \epsilon \}$$
are a subbasis of the topology generated by $P$, we have that the topology generated by $\tilde{P}$ is stronger than the topology generated by $P$. The previous two statements imply that the topology generated by $P$ is equal to the topology generated by $\tilde{P}$.     
\end{proof}

For the purposes of the current document we are interested in LCS that are metrizable, this restriction comes to be useful because it narrows down our scope of work into the realm of countable families of seminorms (\cref{theorem:metrizable_lcs_and_seminorms}).

\begin{definition}[Metrizable topological space (page 110 \citep{horvath_topological_1966})]\label{def:metrizable_topological_space}
If $A$ is a topological space then $A$ is said to be metrizable if there exists a metric $d$ over $A$ such that the topology defined by $d$ coincides with the topology of $A$.
\end{definition}

\begin{theorem}[Metrizable LCS\index{metrizable LCS} and upper directed family of seminorms (cf. Theorem 5.6.1 of \citep{horvath_topological_1966} and Chapter 2, Section 6 Proposition 2 of \citep{horvath_topological_1966} )]\label{theorem:metrizable_lcs_and_seminorms}
A LCS $A$ is metrizable iff its topology is generated by an upper directed sequence $\{ p_n \}_{n \in \mathbb{N}}$ of seminorms. In this case the topology of $A$ is the topology defined by the metric $\delta(x,y) = |x-y|$, where the map $x \mapsto |x|$ from $A$ into $\mathbb{R}^{+}$ is defined by:
$$ |x| = \sum_{n = 0}^{\infty} \frac{1}{2^n} \frac{p_n (x)}{ 1 + p_n (x)}. $$
The map $x \mapsto |x|$ has the following properties:
\begin{itemize}
    \item $|x|=0$ iff $x = 0$
    \item $|x| = |-x|$
    \item $|x + y| \leq |x| + |y|$
    \item $|\lambda| \leq 1$ implies $|\lambda x | \leq |x|$
    \item $\lambda \to 0$ implies $|\lambda x | \to 0$ for every $x \in A$
\end{itemize}
\end{theorem}

\begin{lemma}[Metrizable LCS and family of seminorms]\label{lemma:metrizable_lcs_and_family_of_seminorms}
A LCS $A$ is metrizable iff its topology is generated by a countable family of seminorms $\{ p_n \}_{n \in \mathbb{N}}$.
\end{lemma}
\begin{proof}
If $A$ is a metrizable LCS, then, its topology is generated by a countable family of seminorms $\{ p_n \}_{n \in \mathbb{N}}$ according to \cref{theorem:metrizable_lcs_and_seminorms}. On the other side, if the topology of $A$ is generate by a countable family of seminorms $\{ p_n \}_{n \in \mathbb{N}}$, then, by \cref{lemma:equivalent_satureted_family_of_seminorms} the family of seminorms $\{ r_m \}_{m \in \mathbb{N}}$ with
$$ r_m(x) = \max_{n \leq m} p_n (x), m \in \mathbb{N} $$
defines the same topology over $A$ as the family of seminorms  $\{ p_n \}_{n \in \mathbb{N}}$, since the fmaily of seminorms $\{ r_m \}_{m \in \mathbb{N}}$ is upper directed, by \cref{theorem:metrizable_lcs_and_seminorms} we get that $A$ is a metrizable LCS.
\end{proof}

Given that a metrizable LCS is a metric space then the continuity of a map becomes equivalent to the sequential continuity of a map,

\begin{definition}[Sequential continuity of a map]\label{def:sequential_continuity}
Let $A, B$ be topological spaces, then, $f: A \to B$ is called sequentiallly continuous at $a \in A$ is for any sequence $\{a_n \}_{n \in \mathbb{N}}$ converging to $a$ we have that $f(a_n)$ converges to $f(a)$ in $B$.
\end{definition}

\begin{proposition}[Continuity equivalent to sequential continuity]\label{proposition:continuity_equivalent_continuity}
Let $A,B$ be metrizable LCS, let $f: A \to B$, then $f$ is continuous at $x \in A$ iff $f$ is sequentially continuous at $a.$
\end{proposition}
\begin{proof}
Since $A$ and $B$ are metric spaces, this proposition is a consequence of \citep[Theorem 3.1]{analysis_course:intro_metric_spaces}.
\end{proof}

\begin{lemma}\label{lemma:characterization_of_convergence_in_metrizable_LCS}
Let $A$ be a metric space and $B$ be a metrizable LCS whose topology is generated by the sequence $\{ p_n \}_{n \in \mathbb{N}}$ of seminorms, let $f: A \to B$, then, $f$ is continuous at $a \in A$ iff for any sequence $\{ a_n \}_{n \in \mathbb{N}}$ converging to $a$ and any $m \in \mathbb{N}$ we have that $p_m(f(a_n) -f(a) )$ converges to $0$ as $n$ goes to infinity.
\end{lemma}
\begin{proof}
Continuity of the map $f$ at $a$ is equivalent to sequentially continuity at $a$ by \cref{proposition:continuity_equivalent_continuity}, then, the desired result comes from \citep[item b of Theorem 5.14]{folland_real_1999} (or \citep[item a of Theorem 5.7.2]{narici_topological_2011}) which asserts that $f(a_n) \to f(a)$ iff for any $m \in \mathbb{N}$ we have that $p_m(f(a_n) - f(a)) \to 0$ as $n$ goes to infinity.
\end{proof}

Assessing the completeness of an arbitrary LCS requires the usage of Cauchy nets, however, by \cref{theorem:metrizable_lcs_and_seminorms} if $A$ is a Metrizable LCS then each point $x$ has a countable neighborhoob basis given by $\{ \{ y \in A | |y-x| \leq 1/n \} \}_{n \in \mathbb{N}}$, so, under this setting \citep[page 167]{folland_real_1999} tells us that $A$ is said to be complete if every Cauchy sequence converges, so, let us look into what is the meaning of a Cauchy sequence in a metrizable LCS.

\begin{definition}[Cauchy sequence in a metrizable LCS]\label{def:cauchy_sequence_for_metrizable_LCS}
Let $A$ be a metrizable LCS whose topology is given by an upper directed family of seminorms $\{ p_m \}_{m \in \mathbb{N}}$, let $\{ x_n\}_{n \in \mathbb{N}}$ be a sequence of elements in $A$, then, $\{ x_n\}_{n \in \mathbb{N}}$ is said to be Cauchy if for any $\epsilon > 0$ there is $N \in \mathbb{N}$ such that if $j,k > N$ then $| x_j - x_k | \leq \epsilon$ with 
$$ |x| = \sum_{n = 0}^{\infty} \frac{1}{2^n} \frac{p_n (x)}{ 1 + p_n (x)}. $$
\end{definition}

\begin{remark}[Equivalent definition of Cauchy sequence in metrizable LCS]\label{remark:equivalent_def_of_Cauchy_sequence_in_metrizable_LCS}
Let $A$ be a metrizable LCS whose topology is given by the family of seminorms $\{ p_m \}_{m \in \mathbb{N}}$, then, using \citep[Theorem 5.14]{folland_real_1999} the definition of Cauchy sequence in $A$ is equivalent to the following statement:\\
$\{ x_n\}_{n \in \mathbb{N}}$ is a Cauchy sequence of $A$ if for any $p_m$ the sequence $\{ x_n\}_{n \in \mathbb{N}}$ is Cauchy with respect to $p_m$, or more precisely, given $p_m$ then for any $\epsilon >0$ there is a $N \in \mathbb{N}$ such that if $j,k > N$ we have that $p_m(x_j - x_k) \leq \epsilon$.
\end{remark}

\begin{definition}[Complete metrizable LCS]\label{def:complete_metrizable_lcs}
A metrizable LCS $A$ is complete is every Cauchy sequence in $A$ converges.
\end{definition}

Now we have the necessary ingredients to define a Fréchet space,

\begin{definition}[Definition of Fréchet space\index{Fréchet space} (Chapter 2, section 9, Definition 4 \citep{horvath_topological_1966})]\label{def:Frechet_space}
A Fréchet space is locally convex space that is metrizable and complete.
\end{definition}

We will proceed to give some definitions and facts over metrizable LCS. Let us look into the definition of convergence in a metrizable LCS, this definition is closely related to the definiton of Cauchy sequence (\cref{def:cauchy_sequence_for_metrizable_LCS}) and comes from \citep[Theorem 5.14]{folland_real_1999}

\begin{proposition}[Convergence in metrizable LCS]\label{proposition:convergence_metrizable_lcs}
Let $A$ be a metrizable LCS whose topology is given by the family of seminorms $\{ p_m \}_{m \in \mathbb{N}}$, then, $x_n \to x$ if $p_m (x_n - x) \to 0$ for all $m \in \mathbb{N}$. Equivalently, $x_n \to x$ if 
$|x_n - x | \to 0$, where
$$ |x| = \sum_{n = 0}^{\infty} \frac{1}{2^n} \frac{\tilde{p}_n (x)}{ 1 + \tilde{p}_n (x)}, $$
and $\tilde{p}_n (x) = \max_{j \leq n} p_j (x)$.
\end{proposition}
\begin{proof}
The first part of the proposition is a consequence of \cref{lemma:characterization_of_convergence_in_metrizable_LCS}. Additionally, since the family of seminorms $\{ \tilde{p}_n \}_{n \in \mathbb{N}}$ is equivalent to the family of seminorms $\{ p_n \}_{n \in \mathbb{N}}$ (\cref{lemma:equivalent_satureted_family_of_seminorms}), the second part of the proposition is a consequence of (\cref{def:cauchy_sequence_for_metrizable_LCS}). 
\end{proof}

We can also provide a characterization of the continuity of maps between two metrizable LCS,

\begin{proposition}[Continuity of maps between metrizable LCS (Proposition 5.15 \citep{folland_real_1999})]\label{proposition:continuity_metrizable_lcs}
Assume that $X$ and $Y$ are vector spaces with topologies defined, respectively, by the families $\left\{p_\alpha\right\}_{\alpha \in \mathbb{N}}$ and $\left\{q_\beta\right\}_{\beta \in \mathbb{N}}$ of seminorms (not necessarily upper directed). Let $T: X \rightarrow Y$ is a linear map, then, $T$ is continuous iff for each $\beta \in \mathbb{N}$ there exist $\alpha_1, \ldots, \alpha_k \in A$ and $C>0$ such that $q_\beta(T x) \leq C \sum_1^k p_{\alpha_j}(x)$ . If the family of seminorms is upper directed then this simplifies into, for each $\beta \in \mathbb{N}$ there exist $ \alpha \geq \beta$ and $C>0$ such that $q_\beta(T x) \leq C p_{\alpha}(x)$.
\end{proposition}

In the present document, we would like to work with Fréchet algebras. Before we dive into Fréchet algebras there are some facts that need to be stated.

\begin{definition}[Separately continuous multiplication\index{Separately continuous multiplication}]\label{def:separately_continuous_multiplication_Frechet_space}
Let $A$ be a Fréchet space with a multiplication defined over it, the multiplication is said to be separately continuous if for any convergent sequence $a_k \to a$ and any $b \in A$ we have that $\; a_k \to a$ then $b_k a_k \to ba$.
\end{definition}

\begin{definition}[Jointly continuous multiplication\index{Jointly continuous multiplication}]\label{def:jointly_continuous_multiplication_Frechet_space}
Let $A$ be a Fréchet space with a multiplication defined over it, the multiplication is said to be jointly continuous if for any convergent sequences $a_k \to a$ and $b_k \to b$ we have that $a_k b_k \to ab$.
\end{definition}

\begin{lemma}[Joint continuity of multiplication in Fréchet spaces (Chapter VII, Propostion 1 \citep{waelbroeck_topological_1971})]\label{lemma:joint_continuity_multiplication_in_Frechet_spaces}
Let $A$ be a Fréchet space with a separately continuous multiplication, then, the multiplication is jointly continuous.
\end{lemma}

\begin{definition}[Locally convex algebra\index{locally convex algebra}]\label{def:locally_convex_algebra}
Let $A$ be an algebra, then, $A$ is a locally convex algebra if it is a locally convex space where the multiplication is jointly continuous.
\end{definition}

\begin{definition}[Fréchet algebra\index{Fréchet algebra} (Definition 1.6 \citep{fragoulopoulou_topological_2005})]\label{def:Frechet_algebra}
Let $A$ be an algebra, then, $A$ is a Fréchet algebra if it is a Fréchet space where the multiplication is jointly continuous.
\end{definition}

\begin{lemma}[Characterization of continuity of multiplication]\label{lemma:joint_continuity_and_seminorms}
Assume that $A$ be a Fréchet space whose topology is given by the upper directed family of seminorms $\{ p_m \}_{m \in \mathbb{N}}$, assume that $A$ has a multiplication, then, the multiplication over $A$ is jointly continuous if for any $n \in \mathbb{N}$ there is a $m \geq n$ and a $C_n > 0$ such that $p_n(ab) \leq C_n p_m(a) p_m(b)$ for all $a,b \in A$ 
\end{lemma}
\begin{proof}
For any $m \in \mathbb{N}$ we have that,
$$ \| a_k b_k - ab \|n   $$
$$ = \| a_k b_k - a b_k + a b_k - ab \|n   $$
$$ \leq \| a_k b_k - a b_k \|_n + \| a b_k - a b \|_n   $$
$$ \leq C_n \left( \| a_k - a \|_m \| b_k \|_m +  \| a \|_m\| b_k - b \|_m  \right)  $$
$$ \leq C_n \left( \| a_k - a \|_m \| b \|_m +  \| a_k - a \|_m \| b_k - b \|_m + \| a \|_m\| b_k - b \|_m  \right).  $$
Since $\| a_k - a \|_m \to 0$ and $\| b_k - b \|_m \to 0$ as $k \to \infty$, using \cref{lemma:characterization_of_convergence_in_metrizable_LCS} we get that the previous inequality implies that $a_k b_k \to ab$ as $k \to \infty$, that is, the multiplication is jointly continuous.
\end{proof}

According to \citep{wiki:frechet_algebra} the previous statement is a double implication, however, we only proceed to prove the implication used in the current document.

As a direct application of \cref{lemma:joint_continuity_and_seminorms} we get the following lemma

\begin{lemma}\label{lemma:joint_continuity_mutliplication_on_m_convex_frechet_sapce}
Let $A$ Fréchet space whose topology is given by the upper directed family of seminorms $\{ p_m \}_{m \in \mathbb{N}}$, if $p_n(ab) \leq p_n(a) p_n(b)$ for any $a,b \in A$ and $n \in \mathbb{N}$, then, $A$ is a Fréchet algebra.
\end{lemma}

\begin{definition}[m-convex Fréchet algebra\index{Fréchet algebra!m-convex} (Definition 2.1 \citep{fragoulopoulou_topological_2005})]\label{def:m_convex_Frechet_algebra}
Let $A$ be a Fréchet algebra whose topology is given by the upper directed family of seminorms $\{ p_m \}_{m \in \mathbb{N}}$, then, $A$ is said to be m-convex if If $p_n(xy) \leq p_{n}(x)p_{n}(y)$ for every $n \in \mathbb{N}$. 
\end{definition}

There is one additional definition that will be needed in the present document,

\begin{definition}[Bounded sequence in a Fréchet space\index{Fréchet space!bounded subset} (Theorem 6.1.5 \citep{narici_topological_2011}) ]\label{def:bounded_sequence_Frechet_space}
Let $A$ be a Fréchet algebra whose topology is given by the family of seminorms $\{ p_m \}_{m \in \mathbb{N}}$. Let $\{ x_n \}_{n \in \mathbb{N}}$ a sequence of elements of $A$, then, $\{ x_n \}_{n \in \mathbb{N}}$ is said to be bounded if for every $m \in \mathbb{N}$ there is a $K_m$ such that 
$$ \forall n \in \mathbb{N}, \;  p_m (x_n) \leq k_m .$$
\end{definition}

\begin{remark}[Bounded set]\label{remark:bounded_set_in_frechet_algebras}
Notice that for Fréchet spaces, the notion of a bounded set on a metric space is not equivalent to the notion of a bounded set of a locally convex spaces, actually, it is a weaker notion \citep[Example 6.1.3]{narici_topological_2011}. In non normable Fréchet spaces any ball around the origin is not normable, and many of the examples of Fréchet algebras we have given are non normable Fréchet spaces (\citep{446294}).
\end{remark}

It is possible to define various types of tensor products on Fréchet algebras (\citep[Part III]{treves_topological_2006}), we will focus on the projective tensor product,

\begin{definition}[Projective tensor product of Fréchet algebras \index{Fréchet algebra!Projective tensor product} (Definition 43.2 \citep{treves_topological_2006})]\label{definition:projective_tensor_product_of_frechet_algebras}
Let $A,B$ be to Fréchet algebra whose topology is given by the families of seminorms $\{ \alpha_n \}_{n \in \mathbb{N}}$, $\{ \beta_n \}_{n \in \mathbb{N}}$ respectively, then, the projective topology on $A \odot B$ is the strongest locally convex topology on this vector space for which the canonical bilinear mapping $(x, y) \sim x \otimes y$ of $A \times B$ into $A \otimes B$ is continuous. Provided with it, the space $A \otimes B$ will be denoted by $A \otimes_\pi B$\index{$A \otimes_\pi B$}.
\end{definition}

Let's mention some of the properties of the projective tensor product, which will be helpful later

\begin{proposition}[Properties of the projective tensor product]\label{proposition:properties_projective_tensor_product}
Let $A,B$ be to Fréchet algebra whose topology is given by the families of seminorms $\{ \alpha_n \}_{n \in \mathbb{N}}$, $\{ \beta_n \}_{n \in \mathbb{N}}$ respectively, then,
\begin{itemize}
    \item The projective topology over $A \odot B$ is given by the family of seminorms (\citep[Equation 43.1]{treves_topological_2006})
    $$ \alpha_n \otimes \beta_m (x) = \inf \{ \sum \alpha_n (x_{j,A}) \beta(x_{j,B}) | x = \sum x_{j,A}\otimes x_{j,B}, \text{ a finite sum }\},   $$
    thus, the seminorms $\alpha_n \otimes \beta_m$ are called the product seminorms. $A \otimes_{\pi}B$ is a locally convex space whose topology is generated by the familiy of seminorsm $\{ \alpha_n \otimes \beta_m \}{n,m \in \mathbb{N}^2}$, additionally, the completition of $A \otimes_{\pi}B$ is a Fréchet space, and we denote it by $A \widehat{\otimes}_{\pi} B$.
    \item $A \otimes_{\pi}B$ is Hausdorff iff both $A$ and $B$ are Hausdorff (\citep[Proposition 43.3]{treves_topological_2006}).
    \item The dual of $A \otimes_{\pi} B$ is canonically isomorphic to $B(A,B)$, the space of continuous bilinear forms on $A \times B$ (\citep[corollary of proposition 43.4]{treves_topological_2006}).
    \item The projective tensor product of two Fréchet algebras is a Fréchet algebra, that is, the multiplication is continuous over $A \widehat{\otimes}_{\pi} B$. 
    \item If $\eta_1: A_1 \mapsto B_1$ and $\eta_2: A_2 \mapsto B_2$ are continuous Fréchet algebras homorphisms, then 
    $$\eta_1 \otimes_{\pi} \eta_2: A_1 \otimes_{\pi} A_2 \to B_1 \otimes B_2, \; \eta_1 \otimes_{\pi} \eta_2 (a_1 \otimes a_2) = \eta_1(a_1) \otimes \eta_2(a_2),$$
    is a continuous algebra homomorphis. Since $\eta_1 \otimes_{\pi} \eta_2$ is continuous, it can be extended into $A_1 \hat{\otimes}_{\pi} A_2$, we call its extension $\eta_1 \widehat{\otimes}_{\pi} \eta_2$. If $\eta_1(A_1), \; \eta_2(A_2)$ are dense in $B_1$ and $B_2$ respectively, then $\eta_1 \widehat{\otimes}_{\pi} \eta_2$ is a surjective algebra homomorphism.
    \item If $\eta_1: A_1 \mapsto \mathbb{C}$ and $\eta_2: A_2 \mapsto \mathbb{C}$ are continuous linear maps, then, $\eta_1 \widehat{\otimes}_{\pi} \eta_2$ is a continuous linear map, where $\eta_1 \widehat{\otimes}_{\pi} \eta_2 (x_1 \otimes x_2) = \eta_1(x_1) \eta_2(x_2)$ (\citep[corollary of proposition 43.6]{treves_topological_2006}).
    \item If $x \in A \widehat{\otimes}_{\pi} B$ then $x$ has a representation as an absolutelly convergent series (\citep[Theorem 6.4]{schaefer_topological_1999})
    $$ x = \sum_{i =1}^{\infty} \lambda_i x_1 \otimes y_i,  $$
    with $\sum_{i \in \mathbb{N}} | \lambda_i | \leq \infty$ and $\{ x_i \}_{i \in \mathbb{N}}, \; \{ y_i \}_{i \in \mathbb{N}}$ are sequence converging to $0$.
\end{itemize}
\end{proposition}
\begin{proof}
We provide references for most of these facts, so, let's check on those that do not have references
\begin{itemize}
    \item We provide references for this fact.
    \item We provide references for this fact.
    \item We provide references for this fact.
    \item Take $\{ \alpha_n \}_{n \in \mathbb{N}}$, $\{ \beta_n \}_{n \in \mathbb{N}}$ sets of seminorms that generate the topologies of $A,B$ respectively, moreover, following \cref{lemma:equivalent_satureted_family_of_seminorms} taken them upper directed i.e  $ \alpha_n \leq \alpha_m $ and $\beta_n \leq \beta_m $ when $n \leq m$, given this setup, if $n_1 \leq n_2$ and $m_1 \leq m_2$ then $\alpha_{n_1} \widehat{\otimes}_{\pi} \beta_{m_1} \leq \alpha_{n_2} \widehat{\otimes}_{\pi} \beta_{m_2} $. Provide the set $\{ \alpha_n \widehat{\otimes}_{\pi} \beta_m \}_{n,m \in \mathbb{N}^2}$ with an ordering, so, lets call if $\{ p_i \}_{i \in \mathbb{N}}$, then, according to \citep[page 96]{horvath_topological_1966} the set of seminorms $\{ \sup_{j \leq i} p_j \}_{i \in \mathbb{N}}$ generate the topology of $A \widehat{\otimes}_{\pi} B$ and are a upper directed set of seminorms. 
    
    Take $i \in \mathbb{N}$ and denote by $N,M$ the set of indices such that $p_j = \alpha_n \widehat{\otimes}_{\pi} \beta_m, \; n \in N, \; m \in M$ for $j \leq i$, then, for each $n \in N$ and $m \in M$ denote by $t(n), t(m)$ the index such that $\alpha_n(xy) \leq C_n \alpha_{t(n)}(x) \alpha_{t(n)}(y) $ , $\beta_m(xy) \leq C_m \alpha_{t(m)}(x) \alpha_{t(m)}(y) $. Since both $A$ and $B$ are Fréchet algebras, there must be $C_i$ such that $\alpha_n \widehat{\otimes}_{\pi} \beta_m \leq C_i \alpha_{t(n)} \widehat{\otimes}_{\pi} \beta_{t(m)} $, thus, take $l \in \mathbb{N}$ such that 
    $$\{  \alpha_{t(n)} \widehat{\otimes}_{\pi} \beta_{t(m)}, \alpha_{n} \widehat{\otimes}_{\pi} \beta_{m} \}_{n \in N, m \in M}$$
    is a subset of $\{ p_k \}_{k \leq l}$, then, we have that $p_i(xy) \leq C_i p_l(x) p_l(y)$.  
    \item From \citep[corollary of proposition 43.6]{treves_topological_2006} we know that $\eta_1 \otimes_{\pi} \eta_2$ is a continuous linear map, moreover, since both $\eta_1$ and $\eta_2$ are algebra homomorphisms, then, $\eta_1 \otimes_{\pi} \eta_2$ is also an algebra homomorphism, thus, it is a Fréchet algebra homomorphism. The extension of $\eta_1 \otimes_{\pi} \eta_2$ into $A_1 \widehat{\otimes}_{\pi} A_2$ is called $\eta_1 \widehat{\otimes}_{\pi} \eta_2$ (\citep[Definition 42.6]{treves_topological_2006}) and, if $\eta_1(A_1), \; \eta_2(A_2)$ are dense in $B_1$ and $B_2$ respectively, then $\eta_1 \widehat{\otimes}_{\pi} \eta_2$ is a surjective algebra homomorphism (\citep[Definition 43.9]{treves_topological_2006}).
    \item We provide references for this fact.
    \item We provide references for this fact.
\end{itemize}

\end{proof}

\subsubsection{Examples}
\label{sec:Frechet_algebras_examples}

\begin{example}[Holomorphic functions over an open set]\label{example:holomorphic_functions_over_open_set}
Let $\Omega$ be an open set of $\mathbb{C}$ and $\{ K_n \}_{n \in \mathbb{N}} $ a compact exhaustion of $\Omega$ (\cref{def:compact_exhaustion}), we have that the set
$$  \text{Hol}(\Omega) = \{ f \text{ is holomorphic over } \Omega \} $$\index{$\text{Hol}(\Omega)$}
becomes a Fréchet algebra under the set of seminorms
$$ p_n(f) = \sup_{x \in K_n} |f(x)|. $$
This is called the compact-open topology, and we have already encounter with this algebra in \cref{section:algebra_of_holomorphic_functions}
\end{example}

\begin{example}[Schwartz space\index{Schwartz space}]\label{example:schwartz_space}
The Schwartz space ($ \mathcal{S}$) is the space of smooth functions over $\mathbb{R}^d$ which, together with all their derivatives, vanish at infinity faster than any power of $|x|$. Thus, let
$$
\|f\|_{(N, \alpha)}=\sup _{x \in \mathbb{R}^n}(1+|x|)^N\left|\partial^\alpha f(x)\right|
$$
then
$$
\mathcal{S}(\mathbb{R}^n)=\left\{f \in C^{\infty}:\|f\|_{(N, \alpha)}<\infty \text { for all } N, \alpha\right\} .
$$
The Schwartz space is a Freceht space \citep[Proposition 8.2]{folland_real_1999}, and becomes a Fréchet algebra with the convolution as multiplication \citep[Proposition 8.11]{folland_real_1999}.
\end{example}

The projective tensor product is a useful tool for studying Fréchet algebras, the following is a simple example of the application of the projective tensor product

\begin{example}[Matrix algebras\index{Matrix algebra}]\label{example:Frechet_algebras_matrix_algebras}
Let $A$ be a Fréchet algebra whose topology is given by the saturated family of seminorms $\{ \alpha_n \}_{n \in \mathbb{N}}$, then, $$M_n(A) := M_n(\mathbb{C}) \otimes_{\pi} A = M_n(\mathbb{C}) \widehat{\otimes}_{\pi} A .$$\index{$M_n(A)$}

From \cref{example:matrices_gen_C_star_algebras} we know that $M_n(\mathbb{C})$ is a Fréchet algebra whose topology is given by one norm $\| \cdot \|$, because it is a Banach algebra, therefore, the topology on $M_n(\mathbb{C}) \otimes_{\pi} A$ is given by the family of seminorms $\{ \| \cdot \| \otimes \alpha_n \}_{n \in \mathbb{N}}$. Take $(b_{i,j})_{i,j \leq n} \in M_n(\mathbb{C})$, then, from \cref{sec:C_star_alg_matrix_alg} we know that
\begin{itemize}
    \item $\forall i,j \leq n, \; \| b_{i,j} \| \leq \|  (b_{i,j})_{i,j \leq n} \|,$
    therefore,
    $$\forall i,j \leq n, \; \|\cdot \| \otimes \alpha_n ((b_{i,j})_{i,j \leq n} \otimes a) \geq \| b_{i,j} \alpha_n (a), $$
    \item any $x \in M_n(A)$ can be written as $x = \sum_{i,j \leq n} \delta_{i,j} \otimes a$, thus, 
    $$ \| \cdot \| \otimes \alpha_n (x) \leq \sum_{i,j \leq n} \| \delta_{i,j} \| \alpha_n (a) = \sum_{i,j \leq n} \alpha_n (a). $$
\end{itemize}
The previous two statements imply that if $\{x_n \}_{n \in \mathbb{N}}$ with $x_n = (x_n^{i,j})_{i,j \leq n}$ is a Cauchy sequence on $M_n(A)$ we must have that $\{ x_n^{i,j}\}$ is a Cauchy sequence, so, since $A$ is a Fréchet algebra there must be $x \in M_n(A)$ with $x_n \to x$ and $x_n^{i,j} \to x^{i,j}$ for all $i,j \leq n$.
The aforementioned argument similar to the one shown to prove that $M_n(B)$ is complete when $B$ is a C* algebra (\citep[Proposition T.5.20]{wegge-olsen_k-theory_1993}). The previous argument implies that for any $m \in \mathbb{N}$ the map $p_{i,j}: M_m(A) \to A, \; p_{i,j}(\sum_{k,l \leq m} a_{k,l} \otimes \delta_{k,l}) = a_{i,j}$ is continuous.
\end{example}

In Fréchet algebras is possible to define various topological tensor products (\citep[Theorem 44.1]{treves_topological_2006}), which are in general not equivalent, interestingly, the concept of nuclear C* algebras translates into Fréchet algebras, such that, the nuclear Fréchet algebras are the Fréchet algebras where all the admisible topological tensor products coincide (\citep[Theorem 50.1]{treves_topological_2006}). The following Fréchet algebras are nuclear,
\begin{itemize}
    \item \citep[Theorem 51.1]{treves_topological_2006}: $\mathbb{C}^s$ with $s$ and arbitrary set.
    \item \citep[Theorem 51.5]{treves_topological_2006}: the Fréchet algebra of rapidly decreasing sequences $\mathscr{S}(\mathbb{Z}^d)$.
    \item \citep[Corollary of Theorem 51.5]{treves_topological_2006}: the Schwartz algebra, the algebra of holomoprhic function on an open set, and some more.
\end{itemize}

There are canonical isomorphism of Fréchet algebras $ \mathscr{S}(\mathbb{R}^n) \widehat{\otimes}_{\pi} \mathscr{S}(\mathbb{R}^m) \simeq \mathscr{S}(\mathbb{R}^{n+m}) $ and $\text{Hol}(\Omega_1) \widehat{\otimes}_{\pi} \text{Hol}(\Omega_2) \simeq \text{Hol}(\Omega_1 \times \Omega_2)$ (\citep[Corollary of Theorem 51.6]{treves_topological_2006}). Also, if $A$ is a complete locally convex topological vector space and $X$ is an open set of $\mathbb{R}^d$, then, the algebra $C^{m}(X,E)$ is isomorphic to the projective tensor product $C^{m}(X) \widehat{\otimes}_{\pi} A$ (\citep[Theorem 44.1]{treves_topological_2006}), also, $C^{\infty}(X) \odot A$ is dense in $C^{m}(X,A)$ (\citep[Proposition 44.2]{treves_topological_2006}).

\section{Smooth sub algebras}
\label{sec:smooth_subalgebras}

Smooth subalgebras come into play to generalize the notion of a smooth manifold, and we define them as sub *-algebras of C* algebras much like in the spirit of $C^{\infty}([a,b]) \subset C([a,b])$.

\begin{definition}[Smooth sub algebra \index{smooth sub algebra} (Definition 4.3.3 \citep{schulz-baldes_harmonic_2022})]\label{def:smooth_sub_algebra}
Let $A$ be a C* algebra, we say that $\mathcal{A} \subset A$ is a smooth sub algebra of $A$ if:
\begin{itemize}
    \item \textbf{Dense sub algebra:} $\mathcal{A}$ is an *-algebra dense in $A$.
    \item \textbf{Fréchet algebra:} $\mathcal{A}$ is a Fréchet algebra with a topology that is stronger than the norm topology inherited from $\| \cdot \|_{A}$, that is, the topology of $\mathcal{A}$ as a Fréchet algebra is stronger than the topology of $\mathcal{A}$ as a subspace of $A$.
    \item \textbf{Holomophic calculus invariance\index{Holomophic calculus invariance}:} $\mathcal{A}$ is invariant under the holomorphic calculus of $A$. This means that for $a \in \mathcal{A}$ and $f$ holomorphic on a neighborhood of $sp(a)$ then $f(a) \in \mathcal{A}$. If $A$ has no unit then $f(a) \in \mathcal{A}$ when $f(0) = 0$. The element $f(a)$ is computed using the holomorphic functional calculus over $A$ (\cref{theorem:holomorphic_functional_calculus_banach_algebras}).
\end{itemize}
\end{definition}

\begin{remark}[Other terms for smooth sub algebras]\label{remark:other_terms_for_smooth_sub_algebras}
Many terms have been used to refer to smooth sub algebras, the most common term is Pre C* algebras \citep[Chapter 3 section 8]{gracia-bondia_elements_2001}, while some authors prefer to use the term local C* algebras \citep[Chapter 3 section 1]{blackadar_k-theory_2012}. However, there is no common consensus on what those terms mean, for example, the term local algebra has also been used to refer to smooth sub algebras with local units \citep[Defintion 3]{rennie_smoothness_2003}, and plays an important role in the study non unital spectral triplets. Since our objective is to use results of cyclic cohomology for unital algebras (\cref{sec:pairing with K theory}) we can avoid such technicalities.
\end{remark}

\begin{lemma}[Some properties of smooth sub algebras]\label{lemma:some_properties_of_smooth_sub_algebras}
\begin{enumerate}
    \item We can assume that the norm of $A$ ($\| \cdot \|_{A}$) is one of the seminorms that define the topology of $\mathcal{A}$
    \item If $A$ is unital then $\mathcal{A}$ is also unital.
    \item \textbf{Q property:}\index{Q property} If $A$ is unital $i^{-1}(G(A)) \subset G(\mathcal{A})$ is open in $\mathcal{A}$, also, 
     $$i^{-1}(G(A)) = G(\mathcal{A}) = \mathcal{A} \cap G(A).$$
    \item \textbf{Spectrally invariant:} Take $A$ a unital C* algebra and $\mathcal{A}$ a smooth sub algebra of $A$, let 
    $$ sp(a)_{\mathcal{A}} = \{ \lambda \in \mathbb{C} | \nexists b \in \mathcal{A} \text{ s.t. } b (\lambda 1_{\mathcal{A}} - a) = (\lambda 1_{\mathcal{A}} - a) b = 1_{\mathcal{A}} \},$$
    then, for $a \in \mathcal{A}$ we have that $sp(a)_{\mathcal{A}} = sp(a)_{A}$. 
\end{enumerate}
\end{lemma}
\begin{proof}
\begin{enumerate}
    \item Since the inclusion $i : \mathcal{A} \to A$ is continuous, if $\{ p_n \}_{n \in \mathbb{N}}$ is an upper directed family of seminorms defining the topology of $\mathcal{A}$, then, according to \cref{proposition:continuity_metrizable_lcs} there are $i_1, ..., i_k$ and $C>0$ such that, $\| a\| \leq C \sum_{j \leq k} p_{i_j}(a)$. Take $\epsilon > 0$, then, the open set of $\mathcal{A}$ given by 
    $$ \cap_{j \leq k} \{ p_{i_j}(a) \leq \frac{\epsilon}{kC} \}$$
    is a subset of the open ball of $A$ given by  $\{ a| \|a \| \leq \epsilon \}$, therefore, if we add the norm $\| \cdot \|$ as one of the seminorms of $\mathcal{A}$ we will end up with the same topology over $\mathcal{A}$, that is, the family of seminorms $\{ \| \cdot \| \} \cup \{ p_n \}_{n \in \mathbb{N}}$ is equivalent to the family of seminorms $\{ p_n \}_{n \in \mathbb{N}}$.
    \item This happens because the constant function $f(x)=1$ is holomorphic on $sp(a)$ and $f(a)=1_A$ for every $a \in \mathcal{A}$. 
    \item Take $a \in \mathcal{A}$ such that $a \in G(A)$, since $a^{-1}$ can be computed using the holomorphic functional calculeus over $A$ (\cref{lemma:compute_inverse_with_holomorphic_functional_calculus}), we have that $G(\mathcal{A}) = i^{-1}(G(A))$. Given that $G(A)$ is open in $A$ (\cref{proposition:GA_topological_group_and_open_in_banach_algebra}), the fact that the inclusion is continuous implies that $G(\mathcal{A})$ is open in $\mathcal{A}$.
    \item Since $1_A \in \mathcal{A}$, we get that any element of the form $(\lambda 1_A - a)$ with $a \in \mathcal{A}$ is invertible in $A$ iff is invertible in $\mathcal{A}$, therefore, for $a \in \mathcal{A}$ we have that $sp(a)_{\mathcal{A}} = sp(a)_{A}$ .
\end{enumerate}
\end{proof}

\begin{remark}[Q property]\label{remark:Q_property}
In \cref{lemma:some_properties_of_smooth_sub_algebras} we refer to the Q property as $G(\mathcal{A})$ to be open, however, some texts refer to the Q property as related to the group of quasi-invertibles e.g. \citep[Definition 6.1]{fragoulopoulou_topological_2005}.  
\end{remark}

We saw that when $A$ is unital many good properties of the spectrum and invertibles is recovered in $\mathcal{A}$, so what about the non unital case? in contrast to C* algebras, $\mathcal{A}^+$ can be assigned many topologies that turn it into a Fréchet algebra, and those may or may not make it into a smooth sub algebra of $A^{+}$, fortunately, there is a standard way of generating a Fréchet algebra that is a dense sub *-algebra of $A^{+}$,

\begin{definition}[Unitization of a smooth sub algebra\index{smooth sub algebra!unitization}]\label{definition:unitization_smooth_sub_algebra}

Let $\mathcal{A}$ be a smooth sub algebra of a non unital C* algebra $A$, and $\{ p_n \}_{n \in\mathbb{N}}$ the seminorms giving the topology of $\mathcal{A}$. By the discussion on \cref{lemma:some_properties_of_smooth_sub_algebras} we can assume that $p_0 = \| \cdot \|_A $, thus we define $\mathcal{A}^{+}$\index{$A^+$} as the set $\mathcal{A} \times \mathbb{C}$ with the following operations,

\begin{itemize}
     \item Addition: $(a,\lambda) + (b,\sigma) = (a+b, \lambda + \sigma)$ with addition identity $(0,0)$.
     
     \item Multiplication: $(a,\lambda) \cdot (b,\sigma) = (ab + \lambda b + \sigma a, \lambda \cdot \sigma)$, with multiplication identity $(0,1)$.
     
     \item involution: $(a,\lambda)^* = (a^*, \overline{\lambda})$
 \end{itemize}
 
 and topology given by the family of seminorms
 
 \begin{itemize}
     \item $\hat{p_0}(a,\lambda) = \| (a,\lambda) \|_{A^{+}}$ (operator norm i.e. norm of $(a,\lambda)$ on the C* algebra $A^{+}$)
     \item $\hat{p_n}(a,\lambda) = p_n(a) + |\lambda|$ for $n \geq 1$.
 \end{itemize}

\end{definition}

\begin{lemma}[Properties of the unitization of a smooth sub algebra]\label{lemma:properties_of_unitization_of_smooth_sub_algebras}
Let $A$ be a non unital C* algebra and $\mathcal{A}$ a smooth sub algebra of $A$, then,
\begin{enumerate}
    \item $\mathcal{A}^+$ is a Fréchet algebra, such that, the involution is continuous on $\mathcal{A}^+$ iff is continuous on $\mathcal{A}$.
    \item $i: \mathcal{A} \to \mathcal{A}^+, \; i(a) = (a,0)$ is an injective homomorphism of Fréchet algebras (continuous) and $i(\mathcal{A})$ is a closed two sided ideal of $\mathcal{A}^+$.
\end{enumerate}
\end{lemma}
\begin{proof}
\begin{enumerate}
    \item From \cref{definition:unitization_smooth_sub_algebra} we know that $\mathcal{A}^+$ is a dense *-algebra of $A^{+}$ and has a topology stronger than the topology inhereted from $\| \cdot \|_{A^{+}}$. Moreover, the form of $\{\hat{p_n} \}_{n \in \mathbb{N}}$ guaranties that $\{ (a_n, \lambda_n)\}_{n \in \mathbb{N}}$ is a Cauchy sequence iff both $\{ a_n\}_{n \in \mathbb{N}}$ and $\{ \lambda_n\}_{n \in \mathbb{N}}$ are Cauchy sequences, thus $\mathcal{A}^+$ is complete because both $\mathcal{A}$ and $\mathbb{C}$ are complete. Using similar arguments you can check that the multiplication is continuous on $\mathcal{A}^+$, which altogether implies that $\mathcal{A}^+$ is a Fréchet algebra, such that, the involution is continuous on $\mathcal{A}^+$ iff is continuous on $\mathcal{A}$.
    \item From \cref{definition:unitization_smooth_sub_algebra} we know that $\mathcal{A}$ is a closed two sided ideal of $\mathcal{A}^+$, additionally, the form of $\{\hat{p_n} \}_{n \in \mathbb{N}}$ guaranties that $i: \mathcal{A} \to \mathcal{A}^+, \; i(a) = (a,0)$ is an injective homomorphism of Fréchet algebras i.e. it is a continuous homorphisms.
\end{enumerate}
\end{proof}

\begin{remark}[seminorms of unitization]\label{remark:seminorms_of_unitization}
The objective of $\mathcal{A}^+$ is to provide an unital algebra whose inclusion on $A^{+}$ is continuous. We chose to set $\hat{p_0} (a,\lambda) = \| (a ,\lambda) \|_{A^{+}}$ to ensure this, however, we may have set another norm that is equivalent to  $\| \cdot \|_{A^{+}}$. For example, we could have chosen $\hat{p_0}(a,\lambda) = \| a \|_{A} + |\lambda|$ i.e. the $l_1$ norm, because it generates the same topology as the operator norm, which can be deduced from the inequalities $\| a \|_{A} + |\lambda| \leq 6 e \| (a ,\lambda) \|_{A^{+}}$ \citep[Lemma 2.4]{bhatt_class_2013} and $\| (a ,\lambda) \|_{A^{+}} \leq \| a \|_{A} + |\lambda|$. Using a unitization with only $l_1$ norms is standard in the literature (\citep[Section 3.3]{fragoulopoulou_topological_2005}), however, we choose to use the operator norm for the zero$^{th}$ norm in order to emphasize that we are looking for a smooth sub algebra of $A^+$ and not an arbitrary Fréchet algebra inside $A^+$.
\end{remark}

In \cref{lemma:some_properties_of_smooth_sub_algebras} we saw how, for pair ($A$ unital C* algebra, $\mathcal{A}$: smooth sub algebra of $A$), invariance under holomorphic functional calculus implies invariance of the spectrum. The invariance of the spectrum of dense *-algebras is a nice property that will allow us to establish automatic continuity results (\cref{lemma:automatic_continuity_spectrally_invariant_sub_alegbras}), and is tightly connected to invariance under holomorphic functional calculus if the dense algebra is a Fréchet algebra (\cref{proposition:unitization_is_invariant_under_holomorphic_calculus}). We present a definition of spectral invariance that also takes into account non unital smooth sub algebras,

\begin{definition}[Spectrally invariant dense sub *-algebra\index{spectrally invariant dense sub *-algebra}]\label{definition:spectrally_invariant}
Let $\mathcal{A}$ a dense sub *-algebra of a unital C* algebra $A$, and denote by 
$$ Sp_{\mathcal{A}}(a) = \{ \lambda \in \mathbb{C} | \nexists b \in \mathcal{A} \text{ s.t. } b (\lambda 1_{\mathcal{A}} - a) = (\lambda 1_{\mathcal{A}} - a) b = 1_{\mathcal{A}} \},$$ 
then we say that $\mathcal{A}$ is spectrally invariant if $sp(a)_{\mathcal{A}} = sp_{A}(a)$ for all $a \in \mathcal{A}$, where $sp_{A}(a)$ is the spectrum of $a$ on the C* algebra $A$ (\cref{ref:spec_banach_alg}).

If $\mathcal{A}$ is non unital then we denote by $\mathcal{A}^+$ the set $\mathcal{A} \times \mathbb{C}$ with the same algebraic operations as in \cref{definition:unitization_smooth_sub_algebra}, then we say that $\mathcal{A}$ is spectrally invariant if $sp(a,0)_{\mathcal{A}^+} = sp_{A^{+}}(a,0)$ for all $a \in \mathcal{A}$, where $sp_{A^{+}}(a,0)$ is the spectrum of $(a,0)$ on the C* algebra $A^{+}$.
\end{definition}

Now, we look into some interesting results about spectrally invariance and invariance under holomorphic functional calculus. We have mentioned that unital smooth sub algebras are spectrally invariant, so what about non unital ones? That will actually be the case, furthermore, a dense sub *-algebra of a C* algebra that is a Fréchet algebra and has a topology stronger than the topology of the C* algebra is invariant under holomorphic functional calculus iff is spectrally invariant. To prove this we need to use results that are a consequence of the holomorphic functional calculus on Fréchet algebras e.g. \cref{prop:invariace_holom_calculus_equal_spectral_invariance}, which we cover in \cref{sec:smooth_sub_algebras_functional_calculus}. 

\begin{proposition}\label{proposition:unitization_is_invariant_under_holomorphic_calculus}

Let $\mathcal{A}$ be a dense *-algebra of a C* algebra $A$, then 
\begin{enumerate}
    \item If $\mathcal{A}$ is non unital and invariant under the holomorphic calculus of $A$, then, $\mathcal{A}^+$ is invariant under the holomorphic calculus on $A^{+}$.
    \item If $\mathcal{A}$ is invariant under holomorphic calculus of $A$, then $\mathcal{A}$ is spectrally invariant.
    \item If $\mathcal{A} \subset A$ is a dense *-algebra of $A$ with a topology stronger than $A$ and is a Fréchet algebra, then, $\mathcal{A}$ is spectrally invariant iff is invariant under holmorphic functional calculus.
\end{enumerate}
\end{proposition}
\begin{proof}
\begin{enumerate}
    \item We have that if $a \in \mathcal{A}^+$ then $a = (\alpha, \lambda)$ with $\alpha \in \mathcal{A}$ and $\lambda \in \mathbb{C}$, from \cref{ref:spec_banach_alg} this algebraic relation we get that
    $$ \text{Sp}_{\mathcal{A}^+}(\alpha, \lambda) =  \text{Sp}_{\mathcal{A}^+}(\alpha, 0) + \lambda.$$
    Hence, a function $z \to f(z)$ is holomorphic on a neighbourhood of $\text{Sp}_{\mathcal{A}^+}(\alpha, \lambda)$ iff the function $z \to f(z + \lambda)$ is holomorphic on a neighbourhood of $\text{Sp}_{\mathcal{A}^+}(\alpha, 0)$. Set $g(z) = f(z + \lambda)$, we know that $0 \in \text{Sp}_{\mathcal{A}^+}(\alpha, 0)$, thus $h(z) = g(z) - f(\lambda)$ is an holomorphic function on a neighbourhood of $\text{Sp}_{\mathcal{A}^+}(\alpha, 0)$ with $h(0) =0$, and the definition of invariance under holomorphic functional calculus (\cref{def:smooth_sub_algebra}) tell us that $h(\alpha,0) \in \mathcal{A}$. Then, the properties of the holomorphic calculus on $A^{+}$ (\cref{proposition:properties_of_the_holomoprhic_functional_calculus}) tell us that 
    $$ f(\alpha,\lambda) = (f\circ l_{\lambda} )(\alpha,0) = g(\alpha,0)$$
    with $l_{\lambda}(z) = z + \lambda$, so, given that $g(z) = h(z) - f(\lambda)$, we have that
    $$f(\alpha, \lambda) = h(\alpha,0) + 1_{A^{+}}f(\lambda).$$
    Since $1_{A^+} \in \mathcal{A}^+$ by definition (\cref{definition:unitization_smooth_sub_algebra}), we have that $f(\alpha,\lambda) \in \mathcal{A}^+$, as desire.
    
    \item If $\mathcal{A}$ is unital then this comes from the last item on \cref{lemma:some_properties_of_smooth_sub_algebras}. If $\mathcal{A}$ is non unital then we can use the previous item to show that $\mathcal{A}^+$ is invariant under holomorphic functional calculus on $A^{+}$, which in turn implies that $\mathcal{A}^+$ is spectrally invariant. Since the spectrum of the elements on $\mathcal{A}$ is defined as their spectrum as elements on $\mathcal{A}^+$ (\cref{definition:spectrally_invariant}) we get the desire result.
    
    \item We have seen that invariance under holomorphic calculus implies spectrally invariant, now we look in the converse. If $\mathcal{A}$ is unital then proposition  \cref{prop:invariace_holom_calculus_equal_spectral_invariance} tell us spectrally invariance is equivalent to invariance under holomorphic functional calculus on $A$. Now, if $\mathcal{A}$ is non unital then the behaviour of the spectrum with respect to polynomials (\cref{ref:spec_banach_alg}) tells us that
    $$ Sp_{\mathcal{A}^+}(a, \lambda) = Sp_{\mathcal{A}^+}(a, 0) + \lambda = Sp_{A^+}(a,0) + \lambda =  Sp_{A^+}(a,\lambda),  $$
    which in turn implies that $\mathcal{A}^+$ is invariant under holomorphic calculus on $A^{+}$. So, the holomorphic calculus on $A^{+}$ (\cref{sec:hol_cal_consecuences}) tell us that $f(a) \in A$ if $f(0) = 0$, therefore, for $a \in \mathcal{A}$ we have that $f(a) \in \mathcal{A} \subset A$ if $f(0)=0$.
    
\end{enumerate}
\end{proof} 

\begin{corollary}\label{corollary:unitization_preserves_smooth_sub_algebras}
Let $A$ be a non-unital C* algebra and $\mathcal{A}$ a smooth sub algebra of $A$, then, $\mathcal{A}^+$ is a smooth sub algebra of $A^+$.
\end{corollary}
\begin{proof}
From \cref{lemma:properties_of_unitization_of_smooth_sub_algebras} we know that $\mathcal{A}^+$ is a sub *-algebra if $A^+$ and $\mathcal{A}^+$ is a Fréchet algebra with a topology stronger than the topology of $A^+$. Since $\mathcal{A}$ is invariant under the holomorphic funtional calculus of $A$, from \cref{proposition:unitization_is_invariant_under_holomorphic_calculus} we know that $\mathcal{A}^+$ is invariant under the holomorphic functional calculus of $A^+$. The previous two statements imply that $\mathcal{A}^+$ is a smooth sub algebra of $A^+$ (\cref{def:smooth_sub_algebra}).
\end{proof}

From the discussion after \cref{proposition:automatic_continuity_C_star_algebras} we now that all the analytical information of a C* algebra is stored in the spectrum of their elements, because we have that
$$ \| a\| = \sqrt{ \rho(a a^*) } = \sqrt{ \sup \{ |\lambda| : \lambda \in \text{Sp}(a) \}}$$
for any element of a C* algebra $A$. Since smooth sub algebras are spectrally invariant, we can recover the norm of an element of a smooth sub algebra solely from the structure of the smooth sub algebra, which implies that a smooth sub algebra has a unique C* algebra where it is dense and invariant under the holomorphic functional calculus. This discussion lets us to 

\begin{lemma}\label{lemma:smooth_sub_algebra_has_unique_C_star_algebra}
Let $\mathcal{A}$ be a smooth sub algebra of a C* algebra $A$, then $A$ is unique and
$$ C^*(\mathcal{A}) \simeq A.  $$
\end{lemma}

If $\phi: A \to B$ is a *-homormophism of algebras then we now that $\text{Sp}(\phi(a)) \subseteq \text{Sp}(a)$, thus if $A$ and $B$ are C* algebras and $C \subset A$ is a sub *-algebra that is spectrally invariant ($\text{Sp}_C(c) = \text{Sp}_A(c)$) then for every *-homomorphism $\psi: C \to B$ we have that
$$ \| \phi(c) \|_B = \sqrt{ \rho( \phi(c) \phi(c^*) )} = \sqrt{ \rho_B(\phi(c c^*))} \leq \sqrt{ \rho_C(c c^*)}  = \sqrt{ \rho_A (c c^*)} = \| c \|_A .$$
Therefore, any we have automatic continuity of *-homomorphisms that go from spectrally invariant sub algebras in C* algebras. If $C$ is dense in $A$ then by \cref{lemma:extending_star_homomorphisms_into_C_star_homomorphisms} we can extend $\psi$ into a C* algebra homomorphism 
$$ \hat{\psi}: A \to B \; \hat{\psi}|_{C} = \psi.$$

We condense this discussion on the following lemma

\begin{lemma}\label{lemma:automatic_continuity_spectrally_invariant_sub_alegbras}
Let $A, B$ be C* algebras and $C \subset A$ a *-algebra that is spectrally invariant i.e.
$$ \text{Sp}_A(c) = \text{Sp}_C (c), \; \forall c \in C. $$
Then, any *-homomorphism $\phi: C \to B$ is automatically continuous with respect to the norms of $A$ and $B$ i.e.
$$ \| \phi(c) \|_B \leq \| c \|_A. $$
If $C$ is dense in $A$ then there is a unique C* homomorphism $\hat{\phi} : A \to B$ extending $\phi$ i.e.
$$\hat{\phi}|_C = \phi,$$
and $C$ becomes a core subalgebra of $A$ (\citep[section 4]{exel_envelope_2008}).
\end{lemma}

We have seen how dense spectrally invariant *-algebras allow us to recover the norm of their C* algebras and recover C* homomorphisms, these is just the tip of the iceberg regarding what can be achieved with those algebras, you can actually recover the results of the K theory of C* algebras using dense spectrally invariant *-algebras that behaved well with respect to matrix algebras, which referred to as local C* algebras on \citep{blackadar_k-theory_2012}. Notice that \cref{lemma:automatic_continuity_spectrally_invariant_sub_alegbras} tell us that dense spectrally invariant *-algebras are a class of core subalgebras (\citep[section 4]{exel_envelope_2008}), moreover, \cref{lemma:smooth_sub_algebra_has_unique_C_star_algebra} tell us that they are *-algebras that admit an enveloping C* algebra, because, if $A$ is a C* algebra and $\mathcal{A} \subset A$ a dense *-algebra spectrally invariant then for every representation of $\pi: \mathcal{A} \to B(H)$ we have that (\cref{sec:banach_alg_morph})
$$ Sp_{B(H)}(\pi(A)) \subseteq Sp_{\mathcal{A}}(a),  $$
which implies that
$$\| a \|_A = \sup \{ \| \pi(a) \| \}.  $$
These and many more properties of smooth sub algebras will make them a great tool in non-commutative geometry.

\begin{remark}[Enveloping C* algebra of a smooth sub algebra]\label{remark:enveloping_C_star_algebra_of_smooth_sub_algebra}
In \cref{proposition:automatic_continuity_banach_star_algebras} is mentioned that the enveloping C* algebra of a Banach *-algebra capture the representation theory of the Banach *-algebra, and it was maximal in the sense of having the greatest norm over all the images of representations of the Banach *-algebra. This concept can be generalized into the context of certain types of Fréchet algebras, these are called m *-convex Fréchet algebras and have the following properties,
\begin{itemize}
    \item $A$ is a Fréchet algebra with the topology given by a family of seminorms $\{ p_i \}_{i \in I}$
    \item The seminorms are sub multiplicative $p_i(yx)\leq p_i(x) p_i(y)$
    \item The ivolution is an isometry in the seminorms i.e. $p_i(x) = p_i(x^{*})$
\end{itemize}
To every Fréchet m*-convex algebra you can assign a special type of Fréchet algebra that plays the role of the enveloping C* algebra \citep[Defintion 18.14]{fragoulopoulou_topological_2005}, and captures the representationm theory of the Fréchet m*-convex algebra \citep[Chapter IV]{fragoulopoulou_topological_2005}, these are called locally C* algebras and have the special property of having seminorms with the C* property $p_i(x^{*} x) = p_i(x)^{2}$ \citep[Defintion 7.5]{fragoulopoulou_topological_2005}. 

It turns out that if a Fréchet m*-convex algebra $F$ has an open set of invertibles (Q property) then its enveloping locally C* algebra is isomorphic to a C* algebra $\hat{F}$ \citep[Corollary 18.16]{fragoulopoulou_topological_2005}, that is, there is only one C* algebra were $F$ is a dense *-algebra, which fits well the concept of smooth sub algebra, because the set of invertibles of a unital smooth sub algebra is open. 

Fréchet m*-convex Q-algebras appear as a very important type of smooth subalgebras, the Fréchet $D^{*}_{\infty}$-subalgebras (\cref{sec:Frechet_d_infinity_subalgebras}).

\end{remark}

The following is a key property of smooth sub algebras that will come into play when working with the holomorphic functional calculus on smooth sub algebras

\begin{proposition}[Smooth sub algebras have continuous inversion]\label{prop:pre_C_star_algebras_have_continuous_inversion}
Let $A$ be a unital C* algebra and $\mathcal{A}$ a smooth sub algebra of $A$, then, the map 
$$ (\cdot )^{-1} : G(\mathcal{A}) \to G(\mathcal{A}), \; x (x)^{-1} = (x)^{-1} x = 1_\mathcal{A}, $$
is continuous.
\end{proposition}
\begin{proof}
Given that the inclusion is continuous $i: \mathcal{A} \to A$, \cref{lemma:some_properties_of_smooth_sub_algebras} tells us that $G(\mathcal{A}) = i^{-1}(G(A))$ is open. \citep[Corollary page 115]{waelbroeck_topological_1971} tell us that if the set of invertibles in a Fréchet algebra is open, then the inversion is continuous, so, given that $\mathcal{A}$ is a Fréchet algebra (\cref{def:smooth_sub_algebra}), we have that the inversion in $\mathcal{A}$ is continuous.
\end{proof}

\subsection{Functional calculus}
\label{sec:smooth_sub_algebras_functional_calculus}

The functional calculus plays an important role in the K theory of both C* algebras and smooth sub algebras (\cref{chap:K_theory}). The functional calculus is useful because it provides a nice characterization of paths of unitaries and projections on unital C* algebras and unital smooth sub algebras (\cref{sec:equivalence_relations_C_star_algebras}), that is why we shift our attention towards a formulation of a holomorphic functional calculus on Fréchet algebras. 

The holomorphic functional calculus depends on the spectrum of an element, so. we look at the spectrum of an element in a Fréchet algebra, which is a generalization of the algebraic spectrum we have been using for Banach algebras and C* algebras (\cref{ref:spec_banach_alg}). We refer to the new spectrum as the topological spectrum, since it will take into account the topology of the algebra (\cref{definition:topological_spectrum_frechet_algebras}), so, turns out that the algebraic and topological spectrum coincides for smooth sub algebras (\cref{corollary:topological_and_algebraic_spectrum_of_smooth_sub_algebras}).

\begin{definition}[Bounded element\index{Fréchet algebra!bounded element} (Definition 2.2.1 \citep{fragoulopoulou_generalized_2022})]\label{definition:bounded_element}
Let $\mathcal{A}$ be a Fréchet algebra. An element $x$ of $\mathcal{A}$ is bounded if and only if, for some non-zero complex number $\lambda$, the set $\left\{(\lambda x)^n: n=1,2, \ldots\right\}$ is a bounded subset of $\mathcal{A}$ (\cref{def:bounded_sequence_Frechet_space}). The set of all bounded elements of $\mathcal{A}$ will be denoted by $\mathcal{A}_0$\index{$\mathcal{A}_0$ (set of bounded elements of a Fréchet algebra)}.
\end{definition}

The bounded property can be translated into the concept of operators over locally convex spaces, where it is referred to as tameness, and is important to provide a holomorphic functional calculus on those operators (\citep{arikan_holomorphic_2003}). The notion of a bounded element seems to be a little unintuitive, however, it has a simpler interpretation in terms of limits i.e. an element $x$ is bounded iff there is $\lambda \in \mathbb{C}$ such that $(\lambda x)^n \to 0$ as $n \to \infty$ (\citep[Proposition 2.14]{allan_spectral_1965}). 

\begin{definition}[Convex and bounded sets (Definition 2.2.2 \citep{fragoulopoulou_generalized_2022})]\label{definition:absolutely_convex_and_bounded_sets}
For a Fréchet algebra $\mathcal{A}$, let $\mathfrak{B}_0$\index{$\mathfrak{B}_0$ (convex and bounded sets of a Fréchet algebra)} denote the collection of all subsets $B$ of $\mathcal{A}$, which fulfill the following properties:
\begin{itemize}
    \item $B$ is absolutely convex (\cref{def:absolutely_convex_set}) and $B^2 \subset B$;
    \item $B$ is bounded and closed.
\end{itemize}
\end{definition}

 Denote by $\mathcal{A}[B]$\index{$\mathcal{A}[B]$ (algebra generated by a convex bounded set)} the subalgebra of $\mathcal{A}$ generated by $B \in \mathfrak{B}_0$, which, based on the properties of $B$, takes the following form
 $$ \mathcal{A}[B] = \{ \lambda x : \; \lambda \in \mathbb{C}, \; x \in B \},$$
then, the Minkowski functional
$$  \| x \|_B = \inf \{ t > 0 : \; x \in tB \}, \; x \in \mathcal{A}[B] $$
is a norm in $\mathcal{A}[B]$ (\citep[Equation 2.2.2]{fragoulopoulou_generalized_2022}). The topology generated on $\mathcal{A}[B]$ by the norm $\| \cdot \|_B$ is stronger that the inherited topology from $\mathcal{A}$ (\citep[page 12]{fragoulopoulou_generalized_2022}). According to \citep[Section 2.2]{fragoulopoulou_generalized_2022}, the previous statements hold in the general setting of locally convex algebras, so, among all the locally convex algebras we are interested in the following,

\begin{definition}[Pseudo-complete locally convex algebra\index{locally convex algebra!pseudo-complete }]\label{definition:pseudo_complete_locally_convex_algebra}
Let $\mathcal{A}$ be a locally convex algebra, then, $\mathcal{A}$ is called pseudo-complete if $\mathcal{A}[B]$ is a Banach algebra for any $B \in \mathfrak{B}_0.$
\end{definition}

\begin{lemma}[Fréchet algebras are pseudo-complete]\label{lemma:frechet_algebras_are_pseudo_complete}
Let $\mathcal{A}$ be a Fréchet algebra, then $\mathcal{A}$ is pseudo-complete.
\end{lemma}
\begin{proof}
By definition $\mathcal{A}$ is a complete metrizalbe LCS (\cref{def:Frechet_space}), so, from the definition of completeness of a metrizable LCS we get that $\mathcal{A}$ is sequentially complete (\cref{def:complete_metrizable_lcs}). Under this setting, the content of \citep[Proposition 2.2.5]{fragoulopoulou_generalized_2022} tells us that $\mathcal{A}$ is pseudo-complete.
\end{proof}

Pseudo-complete algebras behave well with respect to unitization, because, $\mathcal{A}$ is pseudo-complete iff $\mathcal{A}^+$ is pseudo-complete (\citep[Proposition 2.2.8]{fragoulopoulou_generalized_2022}), therefore, if $\mathcal{A}$ is non-unital the holomorphic functional calculus will be defined through its unitization, as in done for Banach algebras (\cref{sec:hol_cal_consecuences}).

\begin{definition}[Radius of boundedness]\label{definition:radius_of_boundedness}
Let $\mathcal{A}$ be a Fréchet algebra, then, for $x \in \mathcal{A}$ define
$$ \beta(x) :=  \inf  \{ \lambda > 0 : \{ \left( \frac{1}{\lambda} x \right)^n  \} \text{ is bounded} \}, \; \inf \emptyset = + \infty $$
so, $\beta(x)$ is called the radius of boundedness of $x$.
\end{definition}

By definition of $\mathcal{A}_0$ we have that $\beta(x)$ iff $x \in \mathcal{A}_0$ (\citep[Proposition 2.2.14]{fragoulopoulou_generalized_2022}), additionally, if $\{ p_n \}_{n \in \mathbb{N}}$ is a upper directed family of seminorms generating the topology of the Fréchet algebra $\mathcal{A}$, we have that (\citep[Proposition 2.2.18]{fragoulopoulou_generalized_2022}),
$$ \beta(x) =  \sup \{ \underset{m \to \infty}{\lim \sup} | p_n(x^m) |^{\frac{1}{m}} : \; n \in \mathbb{N} \}, \; x \in \mathcal{A}. $$

We can use the set $\mathcal{A}_0$ to define a topological spectrum for the elements of $\mathcal{A}$, which we consider to be a subset of the one-point compactification of $\mathbb{C}$. Denote by $\mathbb{C}^*:= \mathbb{C} \cup \{ \infty \}$ the one-point compactification of $\mathbb{C}$, then 

\begin{definition}[Topological spectrum\index{topological spectrum} of Fréchet algebras (Definition 2.3.1 \citep{fragoulopoulou_generalized_2022})]\label{definition:topological_spectrum_frechet_algebras}
Let $\mathcal{A}$ be a Fréchet algebra, then, denote
$$  \sigma_{\mathcal{A}} (x) := \{ \lambda \in \mathbb{C} : \; \lambda 1_{\mathcal{A}} - x \text{ has no inverse in } \mathcal{A}_0 \} \cup \{ \infty \text{ iff } x \notin \mathcal{A}_0 \}, \; \sigma_{\mathcal{A}}(x) \subset \mathbb{C}^*. $$\index{$ \sigma_{\mathcal{A} } (\cdot)$ (topological spectrum)}
In case $\mathcal{A}$ has no identity then $\sigma_{\mathcal{A}}(x) := \sigma_{\mathcal{A}^+}(x,0)$.
\end{definition}

The topological spectrum may not be bounded, so, if it is not bounded we say that $\infty \in \sigma_{\mathcal{A}}(x)$. If we chose $\mathcal{A}$ to be pseudo-complete, the following two statements are valid
\begin{enumerate}
    \item $\sigma_{\mathcal{A}}(x)$ is a close subset of $\mathbb{C}^*$ (\citep[Corollary 2.3.8]{fragoulopoulou_generalized_2022}).
    \item $\sigma_{\mathcal{A}}(x)$ is bounded iff $x \in \mathcal{A}_0$ (\citep[Corollary 2.3.14]{fragoulopoulou_generalized_2022}).
\end{enumerate}

\begin{definition}[c.f. Definition 2.4.2 \citep{fragoulopoulou_generalized_2022} (Cauchy domain Fréchet algebra)]\label{definition:cauchy_domain_frechet_algebra}
A subset $D$ of $\mathbb{C}^*$ is called a Cauchy domain\index{Cauchy domain} if it fulfills the following conditions, 
\begin{enumerate}
    \item $D$ is open
    \item $D$ has a finite number of components, whose closures are pairwise disjoint
    \item the boundary $\partial D$ of $A$ is a subset of $\mathbb{C}$, which consists of a finite number of rectifiable Jordan curves no two of which intersects
\end{enumerate}
\end{definition}

\begin{remark}\label{remark:cauchy_domain_and_bounded_elements_psedo_complete_alg}
    Notice that if $\mathcal{A}$ is pseudo-complete and $a \in \mathcal{A}_0$, then $D$ can be taken to be a compact subset of $\mathbb{C}$ because $\sigma_{\mathcal{A}}(a)$ is a bounded subset of $\mathbb{C}$ (\citep[Corollary 2.3.14]{fragoulopoulou_generalized_2022}) and it is also a close subset of $\mathbb{C}^*$ (\citep[Corollary 2.3.8]{fragoulopoulou_generalized_2022}).
\end{remark}

 If $ \infty \in \sigma_{\mathcal{A}}(x)$ we say that a function is holomorphic over $\sigma_{\mathcal{A}}(x)$ if $f$ is holomorphic at any $x \neq \infty$, $f(1/ \lambda)$ has a limit at $0$ and is holomorphic at $\lambda = 0$ (\citep[page 21]{fragoulopoulou_generalized_2022}). According to \citep[Lemma 2.4.3]{fragoulopoulou_generalized_2022} for every $x \in \mathcal{A}$ and $f$ holomorphic on $\sigma_{\mathcal{A}}(x)$, there is a Cauchy Domain $D$ satisfying the following two conditions,
\begin{enumerate}
    \item $\sigma_{\mathcal{A}}(x) \subset D$
    \item $f$ is holomorphic on the closure of $D$ and the integral 
    $$ \int_{\partial D}  f(\lambda) (\lambda 1_{\mathcal{A}} - x)^{-1} d \lambda $$
    defines an element of $\mathcal{A}_0$ independent of the choice of the Cauchy domain satisfying the previously mentioned conditions.
\end{enumerate}

The element $ \int_{\partial D}  f(\lambda) (\lambda 1_{\mathcal{A}} - x)^{-1} d \lambda $ is constructed as follows. Since $\partial D$ is a compact subset of $D$, according to \citep[Lemma 2.3.10]{fragoulopoulou_generalized_2022} there is a $B \in \mathfrak{B}_0 $ such that $(\lambda 1_{\mathcal{A}} - x)^{-1} \in \mathcal{A}[B]$ for all $\lambda \in \partial D$, thus, as an application of the holomorphic functional calculus for Banach algebras (\cref{theorem:holomorphic_functional_calculus_banach_algebras}) we have that the Bochner integral
$$ \int_{\partial D} f(\lambda) (\lambda 1_{\mathcal{A}} - x)^{-1} d \lambda  $$
exists and belongs to $\mathcal{A}[B] \subset \mathcal{A}_0$ (\citep[Proposition 2.4.1]{fragoulopoulou_generalized_2022}). To prove that $ \int_{\partial D} f(\lambda) (\lambda 1_{\mathcal{A}} - x)^{-1} d \lambda  $ is independent of the choice of $D$ assume that you have another Cauchy domain $D'$, then, given that $\partial D \cup \partial D'$ is a compact subset of $\mathbb{C}$, according to \citep[Lemma 2.3.10]{fragoulopoulou_generalized_2022} there must be a $\tilde{B} \in  \mathfrak{B}_0$ such that $(\lambda 1_{\mathcal{A}} - x)^{-1} \in \mathcal{A}[\tilde{B}]$ for all $\lambda \in \partial D \cup \partial D'$, under this setting the holomorphic functional calculus over the Banach algebra $\mathcal{A}[\tilde{B}]$ tells us that $ \int_{\gamma} f(\lambda) (\lambda 1_{\mathcal{A}} - x)^{-1} d \lambda  $ does not depend on whether we used $\gamma = \partial D$ or $\gamma = \partial D'$.

Since Fréchet algebras are pseudo-complete (\cref{lemma:frechet_algebras_are_pseudo_complete}), \citep[Theorem 2.4.4]{fragoulopoulou_generalized_2022} gives us the following homomorphism of algebras based on the results explained in the previous paragraphs,

\begin{theorem}[c.f. Theorem 2.4.4 \citep{fragoulopoulou_generalized_2022}]\label{theorme:holomorphic_funcitonal_calculus_Frechet_algebras}
Let $\mathcal{A}$ be a Fréchet algebra\index{Fréchet algebra!holomorphic functional calculus} and $x \in \mathcal{A}$, denote by $F_x$ the set of holomorphic functions over $\sigma_{\mathcal{A}}(x)$. If $\mathcal{A}$ has a unit set $e = 1_{\mathcal{A}}$, else, set $e = 1_{\mathcal{A}^+}$. Then, there is an algebra homomorphism $\widehat{\Theta}_x: F_x \to \mathcal{A}$, which is given by the following formulae,
\begin{itemize}
    \item if $x \in \mathcal{A}_0$, then 
    $$ \widehat{\Theta}_x(f) = \frac{1}{2 \pi i} \int_{\partial D'} f(\lambda) (\lambda 1_{\mathcal{A}} - x)^{-1} d \lambda,  $$
    \item if $x \notin \mathcal{A}_0$ and $\sigma_{\mathcal{A}}(x) \neq \mathbb{C}$ then
    $$  \widehat{\Theta}_x(f) = f(\infty) e + \frac{1}{2 \pi i} \int_{\partial D'} f(\lambda) (\lambda 1_{\mathcal{A}} - x)^{-1} d \lambda, $$
    \item if $\sigma_{\mathcal{A}}(x) = \mathbb{C}$ then $F_x$ contains only constant functions, so, if $f(\lambda) = c$ then $\widehat{\Theta}_x(f) = c e$.
\end{itemize}

Furthermore, for all cases, if $\mathcal{C}$ is a maximal commutative subalgebra of $\mathcal{A}$, which contains $x$, then $f(x) \in \mathcal{A}_0 \cap \mathcal{C}$. Additionally, denote by $u_0, \; u_1$ the complex functions $u_0(\lambda)  =1, \; u_1(\lambda) = \lambda$, then,
\begin{itemize}
    \item $u_0 \in F_x$ and $\widehat{\Theta}_x(u_0) = e$
    \item If $x \in \mathcal{A}_0$ we have that $u_1 \in F_x$ and $ \widehat{\Theta}_x(u_1) = x$
\end{itemize}
\end{theorem}

\begin{corollary}[Continuous holomorphic functional calculus]\label{corollary:continuous_holomorphic_functional_calculus}
Let $\mathcal{A}$ be a Fréchet algebra and $x \in \mathcal{A}_0$, assume that $D$ is a Cauchy domain of $ \sigma_{\mathcal{A}}(x)$ 
(\cref{definition:cauchy_domain_frechet_algebra}). If $\mathcal{A}$ has a unit set $e = 1_{\mathcal{A}}$, else, set $e = 1_{\mathcal{A}^+}$. Then, for every $\partial D' \subset D$ a contour of $\sigma_{\mathcal{A}}(a)$ (\cref{definition:contoru_compact_set}) such that $D' \subset D$, there is a continuous homomorphism of Fréchet algebras $\widehat{\Theta}_x: \text{Hol}(D) \to \mathcal{A}$, which is given by the following formulae,
$$ \widehat{\Theta}_x(f) = \frac{1}{2 \pi i} \int_{\partial D'} f(\lambda) (\lambda 1_{\mathcal{A}} - x)^{-1} d \lambda,  $$
where $\text{Hol}(D)$ is the Fréchet algebra of holmorphic functions over $D$ (\cref{example:holomorphic_functions_over_open_set}). Additionally, we have that, 
\begin{enumerate}
    \item If $\mathcal{C}$ is a maximal commutative subalgebra of $\mathcal{A}$, which contains $x$, then $f(x) \in \mathcal{A}_0 \cap \mathcal{C}$.    
    \item Denote by $u_0, \; u_1$ the complex functions $u_0(\lambda)  =1, \; u_1(\lambda) = \lambda$, then, $\widehat{\Theta}_x(u_0) = e$ and $ \widehat{\Theta}_x(u_1) = x$.
\end{enumerate}
\end{corollary}
\begin{proof}
Given that Fréchet algebras are pseudo complete (\cref{lemma:frechet_algebras_are_pseudo_complete}), \cref{remark:cauchy_domain_and_bounded_elements_psedo_complete_alg} tells us that $\sigma_{\mathcal{A}}(x)$ is a compact subset of $\mathbb{C}$, which implies that makes sense to talk about a contour of $\sigma_{\mathcal{A}}(x)$ (\cref{definition:contoru_compact_set}).

\cref{theorme:holomorphic_funcitonal_calculus_Frechet_algebras} tells us that $\widehat{\Theta}_x$ is an algebra homomorphism, so, we need to show that it is continuous. Recall that the topology of $\text{Hol}(D')$ is the topology of uniform convergence in compact subsets of $D'$ (\cref{example:holomorphic_functions_over_open_set}), thus, we need to prove that, if for all $K$ compact subset of $D'$ we have that $f_n \to f$ uniformly over $K$, then $\widehat{\Theta}_x(f_n) \to \widehat{\Theta}_x(f)$ inside $\mathcal{A}$.

By the definition of a contour (\cref{definition:contoru_compact_set}) we have that $\partial D'$ is a compact subset of $D$ and $\partial D \cap \sigma_{\mathcal{A}}(x) = \emptyset$, thus, \citep[Lemma 2.3.10]{fragoulopoulou_generalized_2022} tells us that there is $B\in  \mathfrak{B}_0$ such that $(\lambda 1_{\mathcal{A}} - x)^{-1} \in \mathcal{A}[\tilde{B}]$ for all $\lambda \in \partial D'$. The holomorphic functional calculus over $\mathcal{A}[\tilde{B}]$ is a continuous map (\cref{theorem:holomorphic_functional_calculus_banach_algebras}), therefore, we have that
$$ \frac{1}{2 \pi i} \int_{\partial D'} f_n(\lambda) (\lambda 1_{\mathcal{A}} - x)^{-1} d \lambda \to \frac{1}{2 \pi i} \int_{\partial D'} f(\lambda) (\lambda 1_{\mathcal{A}} - x)^{-1} d \lambda $$
inside $\mathcal{A}[\tilde{B}]$. Since the topology of $\mathcal{A}[\tilde{B}]$ is stronger than the topology of $\mathcal{A}$ (\citep[page 12]{fragoulopoulou_generalized_2022}), we have that 
$$ \frac{1}{2 \pi i} \int_{\partial D'} f_n(\lambda) (\lambda 1_{\mathcal{A}} - x)^{-1} d \lambda \to \frac{1}{2 \pi i} \int_{\partial D'} f(\lambda) (\lambda 1_{\mathcal{A}} - x)^{-1} d \lambda $$
inside $\mathcal{A}$, hence, $\widehat{\Theta}_x$ is a continuous map from $\text{Hol}(D')$ into $\mathcal{A}$.
\end{proof}

\begin{remark}[Holomorphic functional calculus and Riemann sums\index{Riemann sum}]\label{remark:holomorphic_func_calculus_integral_frechet_algebras}
Let $\mathcal{A}$ be a Fréchet algebra and $x \in \mathcal{A}_0$, then, in the proof of \cref{corollary:continuous_holomorphic_functional_calculus} it is mentioned that the element
$$ \frac{1}{2 \pi i} \int_{\partial D'} f(\lambda) (\lambda 1_{\mathcal{A}} - x)^{-1} d \lambda $$
can be computed as an element of $\mathcal{A}[\tilde{B}]$ using the holomorphic functional calculus over Banach algebras. The holomorphic functional calculus over Banach algebras gives an element 
$$ \frac{1}{2 \pi i} \int_{\partial D'} f(\lambda) (\lambda 1_{\mathcal{A}} - x)^{-1} d \lambda $$
which can be computed as a Bochner integral and also is the limit of Riemann sums (\cref{remark:holomorphic_calculus_is_a_Bochner_integral}), thus, given that the topology of $\mathcal{A}[\tilde{B}]$ is stronger than the topology of $\mathcal{A}$, the Riemann sums that converge to the element
$$ \frac{1}{2 \pi i} \int_{\partial D'} f(\lambda) (\lambda 1_{\mathcal{A}} - x)^{-1} d \lambda $$
inside $\mathcal{A}[\tilde{B}]$ also converge inside $\mathcal{A}$.
\end{remark}

\begin{remark}[Rational functions and holomorphic calculus\index{Rational functions}]\label{remark:holomorphic_funtions_and_holomorphic_calculus}
Let $\mathcal{A}$ be a Fréchet algebra and $x \in \mathcal{A}_0$, assume that $D$ is a Cauchy domain of $ \sigma_{\mathcal{A}}(x)$ 
(\cref{definition:cauchy_domain_frechet_algebra}), denote by $ \widehat{\Theta}_x$ the continuous map from $\text{Hol}(D)$ into $\mathcal{A}$ described in \cref{corollary:continuous_holomorphic_functional_calculus}. Then, for any polynomial $p(\lambda) = \sum_{n \leq d} c_n \lambda^d$ we have that $ \widehat{\Theta}_x(p) = \sum_{n \leq d} c_n x^n$, thus, for any rational function $r(\lambda) = p(\lambda) / q(\lambda)$ holomorphic over $\sigma_{\mathcal{A}}(x)$ we have that 
$$ \widehat{\Theta}_x(r) = p(x) (q(x))^{-1}. $$
Since the algebra of rational functions is dense in $\text{Hol}(D)$ (\citep[Corollary 4.86]{allan_introduction_2011}) we have that the holomorphic functional calculus for $x$ is uniquely determined by its values in rational functions. 
\end{remark}

There is also a spectral mapping\index{spectral mapping} for the holomorphic functional calculus over Fréchet algebras,

\begin{proposition}[Proposition 2.5.7 \citep{fragoulopoulou_generalized_2022}]\label{proposition:spectral_mapping_holomorphic_calculus_frechet_algebras}
Let $\mathcal{A}$ be a Fréchet algebra and $x \in \mathcal{A}$, then, for any $f \in F_x$,
$$ \sigma_{\mathcal{A}}( \widehat{\Theta}_x(f)) = f(\sigma_{\mathcal{A}}(x)). $$
\end{proposition}

\begin{remark}[Other formulations for the holomorphic functional calculus]\label{remark:other_formulations_holom_functional_calculus}
The holomorphic functional calculus is a well-known fact in the community, as you can check in \citep[Proof of lemma 1.2]{schweitzer_short_1992}, \citep[Introduction page 3]{gramsch_relative_1984}, and \citep[Lemma 1.3]{phillips_k-theory_1991} where it is stated, however, most of recent literature does not go into the details of its formulation and instead it provides references to fairly old and hard to find papers with those details.

The classical holomorphic calculus on Banach algebras has been generalized to non-normed algebra by several authors, for more information on that we recommend the introduction to \citep{arikan_holomorphic_2003}. Among those non-normed algebras many authors have focused on locally convex algebras, for example, \citep[Chapter VI]{waelbroeck_topological_1971} exposes various results on holomorphic calculus for those algebras. Most of the papers that establish the holomorphic functional calculus on locally convex algebras use mathematical techniques that I was not able to follow, also, some of those papers use notation and techniques from fairly old documents which I was not able to find. Fortunately, the exposition given in \citep{allan_spectral_1965} was simpler and more self contained than the other expositions, this exposition is also presented in \citep[Chapter 2]{fragoulopoulou_generalized_2022}.

In Fréchet m-convex algebras it is possible to come up with a formulation of the holomorphic functional calculus that relays on the Arens-Michael representation of a m-convex Fréchet algebra as a projective limit of Banach algebras (\citep[Theorem 3.3.7]{goldmann_uniform_1990} \citep{michael_locally_1952}), and it was used in \citep[Section 6.1]{goldmann_uniform_1990} to provide a holomorphic functional calculus for commutative m-convex Fréchet algebras. 

The approach taken for the holomorphic functional calculus lies within the realm of topological spaces, nonetheless, this is not the only approach towards functional calculus on algebras, bornologies are also used to work with functional calculus and have proven to be a viable and useful alternative (\citep[Chapter 2]{cuntz_topological_2007}). When using bornologies the projective limits become inductive limits, which makes them easier to handle; a bornology, compared to topology, cares about bounded sets instead of open sets (\citep[Definition 2.1]{cuntz_topological_2007}).
\end{remark}

\subsubsection{Holomorphic functional calculus for smooth sub algebras}
\label{section:holomorphic_functional_calculus_for_smooth_sub_algebras}

If we focus on Fréchet algebras that are smooth sub algebra the holomorphic functional calculus over Fréchet algebras has some nice properties that will be useful to work with the K theory of smooth sub algebras (\cref{sec:K_0_and_K_1_for_smooth_sub_algebras}) and the cyclic cohomology of smooth sub algebras (\cref{sec:pairing with K theory}).

\begin{lemma}[Topological and algebraic spectrum of Fréchet sub algebras]\label{lemma:topological_and_algebraic_spectrum_of_dense_frechet_sub_algebras}
Let $A$ be a C* algebra and $\mathcal{A}$ a sub algebra of $A$, assume that $\mathcal{A}$ has the following properties,
\begin{itemize}
    \item $\mathcal{A}$ is a Fréchet algebra whose topology is stronger than the topology of $A$,
    \item $\mathcal{A}$ is spectrally invariant with respect to $A$ i.e.
    $$\forall x \in \mathcal{A}, \; Sp_{\mathcal{A}}(x) = Sp_{A}(x), $$
    where, $\text{Sp}_{A}(x)$\index{$\text{Sp}_A(a)$} the algebraic spectrum\index{algebraic spectrum} of $x$ as an element of $A$ i.e.
    $$ Sp_{A}(a) = \{ \lambda \in \mathbb{C} | \nexists b \in A \text{ s.t. } b (\lambda 1_{A} - a) = (\lambda 1_{A} - a) b = 1_{A} \}.$$
\end{itemize}
Then, we have that, 
$$ \forall x \in \mathcal{A}, \; \text{Sp}_{\mathcal{A}}(x) = \sigma_{\mathcal{A}}(x), $$
and $\mathcal{A}= \mathcal{A}_0$. Additionally, 
$$ \text{Sp}_{\mathcal{A}}(x) = \bigcap \{ \text{Sp}_{\mathcal{A}[B]}(x): \; B \in \mathfrak{B}_0, x \in \mathcal{A}[B] \},  $$
and,
$$ \rho_{\mathcal{A}}(x) = \beta(x) = \inf \{ \rho_{\mathcal{A}[B]}(x): \; B \in \mathfrak{B}_0, x \in \mathcal{A}[B] \},  $$
where $\rho_{\mathcal{A}}(x)$ be the spectral radius of $x$ i.e. $\rho_{\mathcal{A}}(x) = \sup_{z \in \text{Sp}_{\mathcal{A}}(x)} |z|$.
\end{lemma}
\begin{proof}
We assume that $\mathcal{A}$ has a unit,  if $\mathcal{A}$ has no unit then we use $\mathcal{A}^+$ instead since both the algebraic and the topological spectrums of elements of $\mathcal{A}$ are defined as their spectrums on $\mathcal{A}^+$. Since The topology of $\mathcal{A}$ is stronger than the topology of $A$ we get that the map $i : \mathcal{A} \to A, \; i(x) = x$ is continuous, therefore, $G(\mathcal{A}) = i^{-1}(G(A))$ is open since $G(A)$ is open (\cref{lemma:some_properties_of_smooth_sub_algebras}). Since $G(\mathcal{A})$ is an open set, we have that the inversion over $\mathcal{A}$ is a continuous function by \citep[Corollary page 115]{waelbroeck_topological_1971}, also, from \cref{lemma:frechet_algebras_are_pseudo_complete} we know that $\mathcal{A}$ is pseudo-complete, thus, \citep[Theorem 2.3.13]{fragoulopoulou_generalized_2022} tell us that $\sigma_{\mathcal{A}}(x) = \overline{Sp_{\mathcal{A}(x)}}$. Since $Sp_A(x)$ is close (\cref{sec:banach_alg_top_of_spec}) and we have assumed that $Sp_A(x) = Sp_{\mathcal{A}}(x)$, then, $\sigma_{\mathcal{A}}(x) = Sp_{\mathcal{A}(x)}$. Additionally, since $Sp_A(x)$ is bounded (\cref{sec:banach_alg_top_of_spec}) we have that $\sigma_{\mathcal{A}}(x)$ is also bounded, thus, \citep[Corollary 2.3.14]{fragoulopoulou_generalized_2022} implies that $\mathcal{A} = \mathcal{A}_0$.

From \citep[Proposition 2.3.12]{fragoulopoulou_generalized_2022} we know that
$$ \text{Sp}_{\mathcal{A}}(x) = \bigcap \{ \sigma_{\mathcal{A}[B]}(x): \; B \in \mathfrak{B}_0, x \in \mathcal{A}[B] \}, $$
since $\mathcal{A}[B]$ are Banach algebras we have that all the elements are bounded, which implies that $\sigma_{\mathcal{A}[B]}(x) = Sp_{\mathcal{A}[B]}(x)$, which gives us the desired result.

Recall that $\rho_{\mathcal{A}}(x)$ is the spectral radius of $Sp_{\mathcal{A}}(x)$ (\cref{sec:banach_alg_top_of_spec}), therefore, from \citep[Theorem 2.3.11]{fragoulopoulou_generalized_2022} we get that $\beta(x) = \rho_{\mathcal{A}}(x)$, and from \citep[Proposition 2.3.12]{fragoulopoulou_generalized_2022} we get that
$$ \rho_{\mathcal{A}}(x) = \beta(x) = \inf \{ \rho_{\mathcal{A}[B]}(x): \; B \in \mathfrak{B}_0, x \in \mathcal{A}[B] \}.  $$
\end{proof}

\begin{corollary}[Topological and algebraic spectrum of smooth sub algebras]\label{corollary:topological_and_algebraic_spectrum_of_smooth_sub_algebras}
Let $A$ be a C* algebra and $\mathcal{A}$ a smooth sub algebra of $A$, then,
$$ \forall x \in \mathcal{A}, \; \text{Sp}_{\mathcal{A}}(x) = \sigma_{\mathcal{A}}(x), $$
and $\mathcal{A}= \mathcal{A}_0$. Moreover, 
$$ \text{Sp}_{\mathcal{A}}(x) = \bigcap \{ \text{Sp}_{\mathcal{A}[B]}(x): \; B \in \mathfrak{B}_0, x \in \mathcal{A}[B] \},  $$
and,
$$ \rho_{\mathcal{A}}(x) = \beta(x) = \inf \{ \rho_{\mathcal{A}[B]}(x): \; B \in \mathfrak{B}_0, x \in \mathcal{A}[B] \}.  $$
\end{corollary}
\begin{proof}
From \cref{lemma:some_properties_of_smooth_sub_algebras} we know that smooth sub algebras satisfy the conditions of \cref{lemma:topological_and_algebraic_spectrum_of_dense_frechet_sub_algebras}, therefore, this is a consequence of \cref{lemma:topological_and_algebraic_spectrum_of_dense_frechet_sub_algebras}.
\end{proof}

\begin{proposition}\label{proposition:holomorphic_calculus_coincide_in_smooth_sub_algebras_bounded_elements}
Let $A$ be a C* algebra and $\mathcal{A}$ a smooth sub algebra of $A$. Take $x \in \mathcal{A}$, assume that $D$ is a Cauchy domain of $ \sigma_{\mathcal{A}}(x)$ (\cref{definition:cauchy_domain_frechet_algebra}). If $\mathcal{A}$ has a unit set $e = 1_{\mathcal{A}}$, else, set $e = 1_{\mathcal{A}^+}$. Denote by $\widehat{\Theta}_x$ the continuous homomorphism of Fréchet algebras from $\text{Hol}(D)$ into $\mathcal{A}$ (\cref{corollary:continuous_holomorphic_functional_calculus}), and denote by $\Theta_x$ the continuous homomorphism of topological algebras from $\text{Hol}(D)$ into $A$ (\cref{theorem:holomorphic_functional_calculus_banach_algebras}), then, for any $f \in \text{Hol}(D)$ we have that
$$ \widehat{\Theta}_x(f) = \Theta_x(f). $$
\end{proposition}
\begin{proof}
We assume that $A$ has a unit, otherwise, work with $(A^+, \mathcal{A}^+)$.  By \cref{corollary:topological_and_algebraic_spectrum_of_smooth_sub_algebras} we know that $\sigma_{\mathcal{A}}(x) = Sp_A(x)$ for all $x \in \mathcal{A}$, thus, the domain of the holomorphic functional calculus over $\mathcal{A}$ is the same as the domain of the holomorphic functional calculus over $A$, that is, the algebra of holomorphic functions over the Cauchy domain $D$.

By \cref{remark:holomorphic_funtions_and_holomorphic_calculus} we know that for any rational function $r(\lambda) = p(\lambda) / q(\lambda)$ holomorphic on $Sp_A(x)$ we have that
$$ \widehat{\Theta}_x(r) = p(x) (q(x))^{-1}, $$
thus, the properties of the holomorphic funcitonal calculus on $A$ (\cref{theorem:holomorphic_functional_calculus_banach_algebras}) tell us that $\widehat{\Theta}_x(r) = \Theta_x(r)$ for any rational function holomorphic on $Sp_A (x)$. Since, the rational functions are dense in $\text{Hol}(D)$ (\citep[Corollary 4.86]{allan_introduction_2011}) and both both maps $\widehat{\Theta}_x, \Theta_x$ are continuous (\cref{corollary:continuous_holomorphic_functional_calculus}, \cref{theorem:holomorphic_functional_calculus_banach_algebras}), they coincide, that is, for all $f \in \text{Hol}(D)$ we have that
$$ \widehat{\Theta}_x(f) = \Theta_x(f). $$
\end{proof}

The setup we have given allows us to provide an alternative proof for the characterization of smooth sub algebras provided in \citep[Lemma 1.2]{schweitzer_short_1992}, notice that this characterization is key to prove that $M_n(\mathcal{A})$ is a smooth sub algebra of $M_n(A)$ is $\mathcal{A}$ is a smooth sub algebra of $A$ (\citep[Corollary 2.3]{schweitzer_short_1992}). 

\begin{proposition}\label{prop:invariace_holom_calculus_equal_spectral_invariance}
Let $A$ be a C* algebra and $\mathcal{A}$ a sub algebra with the following properties,
\begin{itemize}
    \item $\mathcal{A}$ is close under involution,
    \item $\mathcal{A}$ is a Fréchet algebra with a topology stronger than $A$,
\end{itemize}
then, $\mathcal{A}$ is invariant under the holomorphic functional calculus of $A$ iff $\mathcal{A}$ is spectrally invariant with respect to $A$.
\end{proposition}
\begin{proof}
From \cref{lemma:some_properties_of_smooth_sub_algebras} we know that, if $\mathcal{A}$ is invariant under the holomorphic functional calculus of $A$, then $\mathcal{A}$ is spectrally invariant with respect to $A$. So, let's look into the converse statement.

Denote by $\widehat{\Theta}_x$ the continuous homomorphism of Fréchet algebras from $\text{Hol}(D)$ into $\mathcal{A}$ (\cref{theorme:holomorphic_funcitonal_calculus_Frechet_algebras}), and denote by $\Theta_x$ the continuous homomorphism of topological algebras from $\text{Hol}(D)$ into $A$ (\cref{theorem:holomorphic_functional_calculus_banach_algebras}). Take $\mathcal{A}$ spectrally invariant with respect to $A$, then, from \cref{lemma:topological_and_algebraic_spectrum_of_dense_frechet_sub_algebras} we know that for all $x \in \mathcal{A}$, $\sigma_{\mathcal{A}}(x) = Sp_{\mathcal{A}}(x) = Sp_A(x)$, and $\mathcal{A} = \mathcal{A}_0$, therefore, by \cref{corollary:continuous_holomorphic_functional_calculus} we know that the map $\widehat{\Theta}_x$ takes the form of an integral. Notice that the domain of $\widehat{\Theta}_x$ and $\Theta_x$ is the same for all $x \in \mathcal{A}$ i.e. the functions that are holomorphic over a Cauchy domain of $Sp_A (x)$. 

Take $r(\lambda) = p(\lambda) / q(\lambda)$ a rational function holomorphic over $Sp_A (x)$, then, from \cref{remark:holomorphic_funtions_and_holomorphic_calculus} we know that
$$ \widehat{\Theta}_x (r) = p(x) (q(x))^{-1}, $$
and from \cref{theorem:holomorphic_functional_calculus_banach_algebras} we have that
$$ \Theta_x (r) = p(x) (q(x))^{-1}, $$
therefore, $\widehat{\Theta}_x (r) = \Theta_x (r)$ for all rational functions holomorphic over $Sp_A (x)$. Since the rational functions are dense in the algebra of holomorphic functions over $Sp_A (x)$ (\citep[Corollary 4.86]{allan_introduction_2011}), we have that 
$\widehat{\Theta}_x (f) = \Theta_x (f) $
for all holomorphic functions over $Sp_A (x)$. Since the range of $\widehat{\Theta}$ is a sub algebra of $\mathcal{A}$,for any $x \in \mathcal{A}$ and $f$ holomorphic over $Sp_A (x)$ we get that $\Theta_x (f) \in \mathcal{A}$ .
\end{proof}

Now we can look into the precise form that the holomorphic functional calculus (\cref{theorme:holomorphic_funcitonal_calculus_Frechet_algebras}) takes on smooth sub algebras,

\begin{theorem}[Holomorphic functional calculus on smooth sub algebras\index{smooth sub algebra!holomorphic functional calculus}]\label{theorem:holomorphic_funct_calculus_smooth_sub_algebras}
Let $A$ be a C* algebra and $\mathcal{A}$ a smooth sub algebra of $A$. Take $x \in \mathcal{A}$, assume that $D$ is a Cauchy domain of $ \sigma_{\mathcal{A}}(x)$ (\cref{definition:cauchy_domain_frechet_algebra}). If $\mathcal{A}$ has a unit set $e = 1_{\mathcal{A}}$, else, set $e = 1_{\mathcal{A}^+}$. Then, for every $\partial D' \subset D$ a contour of $\sigma_{\mathcal{A}}(a)$ (\cref{definition:contoru_compact_set}) such that $D' \subset D$, there is a continuous homomorphism of Fréchet algebras $\widehat{\Theta}_x: \text{Hol}(D) \to \mathcal{A}$, which is given by the following formulae,
$$ \widehat{\Theta}_x(f) = \frac{1}{2 \pi i} \int_{\partial D'} f(\lambda) (\lambda 1_{\mathcal{A}} - x)^{-1} d \lambda,  $$
where $\text{Hol}(D)$ is the Fréchet algebra of holmorphic functions over $D$ (\cref{example:holomorphic_functions_over_open_set}). Additionally, we have that, 
\begin{enumerate}
    \item If $\mathcal{C}$ is a maximal commutative subalgebra of $\mathcal{A}$, which contains $x$, then $f(x) \in \mathcal{C}$.
    \item Denote by $u_0, \; u_1$ the complex functions $u_0(\lambda)  =1, \; u_1(\lambda) = \lambda$, then, $\widehat{\Theta}_x(u_0) = e$ and $ \widehat{\Theta}_x(u_1) = x$.
    \item $Sp_A( \widehat{\Theta}_x(f) ) = f(Sp_A (x))$.
    \item Denote by $\Theta_x$ the continuous homormorphism from $\text{Hol}(D)$ into $A$ described in (\cref{theorem:holomorphic_functional_calculus_banach_algebras}), then, for all $f \in \text{Hol}(D)$ we have that,
    $$ \widehat{\Theta}_x (f) = \Theta_x(f). $$
    \item For any $f \in \text{Hol}(D)$ and $g$ holomorphic over $f(Sp_A(x))$, we have that $ \widehat{\Theta}_{f(x)}(g) = \widehat{\Theta}_x (g \circ f )$. 
\end{enumerate}
The element $\widehat{\Theta}_x(f)$ can be computed as the limit of Riemann sums taking values in $\mathcal{A}$.
\end{theorem}
\begin{proof}
From \cref{corollary:topological_and_algebraic_spectrum_of_smooth_sub_algebras} we know that all the elements of $\mathcal{A}$ are bounded i.e. $\mathcal{A} = \mathcal{A}_0$ and $Sp_{\mathcal{A}}(x) = \sigma_{\mathcal{A}}(x)$. Since $\mathcal{A}$ is spectrally invariant with respect to $A$ (\cref{lemma:some_properties_of_smooth_sub_algebras}), we get that $Sp_{A}(x) = \sigma_{\mathcal{A}}(x)$. Under this setting, the first two items are a consequence of \cref{corollary:continuous_holomorphic_functional_calculus} and the third item is a consequence of \cref{proposition:spectral_mapping_holomorphic_calculus_frechet_algebras}, also, the fourth item is a consequence of \cref{proposition:holomorphic_calculus_coincide_in_smooth_sub_algebras_bounded_elements}.

From \cref{proposition:properties_of_the_holomoprhic_functional_calculus}) we know that,
$$ \Theta_{f(x)}(g) = \Theta_x (g \circ f ),$$
so, given that $\widehat{\Theta}_x (f) = \Theta_x(f)$ (\cref{proposition:holomorphic_calculus_coincide_in_smooth_sub_algebras_bounded_elements}), then, we must have that $ \widehat{\Theta}_{f(x)}(g) = \widehat{\Theta}_x (g \circ f )$. 

From \cref{remark:holomorphic_func_calculus_integral_frechet_algebras} we know that $\widehat{\Theta}_x(f)$ can be computed as the limit of Riemann sums taking values in $\mathcal{A}$.
\end{proof}

\begin{remark}\label{remark:notation_uniqueness_holomorphic_funct_calculus}
Let $A$ be a C* algebra and $\mathcal{A}$ a smooth sub algebra of $A$, \cref{theorem:holomorphic_funct_calculus_smooth_sub_algebras} tells us that
$$  \widehat{\Theta}_x(f) = \Theta_x(f), $$
therefore, if $x \in \mathcal{A}$ and $f$ is an holomorphic function on a Cauchy domain of $Sp_A(x)$, we use the notation
$$ f(x) :=  \widehat{\Theta}_x(f),$$
as is done in \cref{theorem:holomorphic_functional_calculus_banach_algebras}.
\end{remark}

The following is a key fact for smooth sub algebras

\begin{proposition}\label{prop:matrix_alegbras_of_pre_C_star_algebras}
If $\mathcal{A}$ is a smooth sub algebra of $A$, then $M_n(\mathcal{A})$\index{$M_n(A)$} is a smooth sub algebra of $M_n(A)$ for any $n \in \mathbb{N}$.
\end{proposition}
\begin{proof}
From \cref{example:Frechet_algebras_matrix_algebras} we know that $M_n(\mathcal{A})$ is a Fréchet algebra, moreover, it has a topology stronger that the topology of $M_n(A)$ because convergence in $M_n(\mathcal{A})$ implies convergence of all the entries of the matrices, which in turn implies the convergence in $M_n(A)$ (\cref{proposition:C_star_norm_on_matrix_algebras}). Therefore, $M_n(\mathcal{A})$ is a sub *-algebra of $M_n(A)$ with a topology stronger that the topology of $M_n(A)$.

From \citep[Corollary 2.3]{schweitzer_short_1992} we get that if the set of invertible elements of $A$ is open in $A$, we have that $\mathcal{A}$ is spectrally invariant with respect to $A$ iff $M_n(\mathcal{A})$ is spectrally invariant with respect to $M_n(A)$, therefore, given that $G(A)$ is open in $A$ (\cref{proposition:GA_topological_group_and_open_in_banach_algebra}), \cref{prop:invariace_holom_calculus_equal_spectral_invariance} tell us that $M_n(\mathcal{A})$ is invariant under the holomorphic functional calculus of $M_n(A)$. The previous claims imply that $M_n(\mathcal{A})$ is a smooth sub algebra of $M_n(A)$ (\cref{def:smooth_sub_algebra}).
\end{proof}

Since the holomorphic functional calculus on smooth sub algebras coincides with the holomorphic functional calculus on C* algebras (\cref{theorem:holomorphic_funct_calculus_smooth_sub_algebras}), many results from C* algebras can be translated into the realm of smooth sub algebras,

\begin{proposition}[Polar decomposition of invertible elements\index{polar decomposition}]\label{proposition:polar_decomposition_of_invertible_elements_smooth_sub_algebras}
Let $A$ be a unital C* algebra and $\mathcal{A}$ a smooth sub algebra of $A$, denote by $\mathcal{A}_{\text{sa}}$ the set of self adjoint elements of $\mathcal{A}$, then,
\begin{itemize}
    \item If $a \geq 0$ with $a \in G(\mathcal{A})$, then there exists $b \in \mathcal{A}_{\text{sa}}$ such that $\exp(b)=a$. 
    \item If $a \in G(\mathcal{A})$, then there is a unique decomposition $a=b u$ with $b \in \mathcal{A}_{\text{pos}}$, and $u \in U(\mathcal{A})$. Moreover, $b=\left(a a^*\right)^{1 / 2}$.
    \item If $a \in U(\mathcal{A})$ then $b = 1_A$ and $u=a$, that is, the unitaries are a fixed point of the map $a \to u$.
\end{itemize}
We call $b$ the absolute value of $a$ and we denote by $|a|$, also, since $u$ is unique we denote it by $\omega(a)$.
\end{proposition}
\begin{proof}
Recall that the smooth sub algebras are closed under involution, so, if $a \in \mathcal{A}$ then $a^* \in \mathcal{A}$ (\cref{def:smooth_sub_algebra}). Given $a \in G(\mathcal{A})$, we have that $a a^*$ is invertible, also, $a a^*$ is a positive element of $\mathcal{A}$ we have that $Sp(a a^*) \subset (0,\inf)$ (\cref{proposition:cahracterization_of_positive_elements}). Since the function $z \mapsto z^{-1/2}$ is holomorphic over the spectrum of $a a^*$, hence, we can compute the element $|a|^{-1} = (a a^*)^{-1/2}$ using the holomorphic functional calculus over $A$ (\cref{corolllary:computing_minus_n_root_holomorphic_funcitonal_calculus}), and we get that $|a|^{-1} \in \mathcal{A}$ because $\mathcal{A}$ is closed under the holomorphic functional calculus of $A$ (\cref{def:smooth_sub_algebra}), thus, $\omega(a) =  a |a|^{-1}$ is an element of $\mathcal{A}$.
To check that the other claims hold you can follow the arguments in \cref{proposition:polar_decomposition_invertible_elements} and check that all the constructions make use of the holomorphic funcitonal calculus over $A$ and the continuous functional calculus over $A$ is used only to deduce properties of the elements e.g. properties of their spectrum.
\end{proof}

Since the holomorphic functional calculus on smooth sub algebras is a continuous homomorphism of Fréchet algebras (\cref{theorem:holomorphic_funct_calculus_smooth_sub_algebras}) we can use it to define smooth paths as in \cref{proposition:smoomth_paths_from_entire_functions},

\begin{proposition}[Smooth paths from entire functions in smooth sub algebras\index{smooth path} ]\label{proposition:smoomth_paths_from_entire_functions_smooth_sub_algebras}
Let $\mathcal{A}$ be a unital smooth sub algebra of a C* algebra $A$, then for any $a \in \mathcal{A}$ and $f \in \text{Hol}(\mathbb{C})$\index{$\text{Hol}(\mathbb{C})$} the function
$$ \gamma_{f,a} : \mathbb{C} \to \mathcal{A}, \; z \mapsto f(za) $$
is infinitely continuously differentiable i.e $\gamma_{f,a} \in C^{\infty}(\mathbb{C},\mathcal{A})$. 
\end{proposition}
\begin{proof}
We will use the holomorphic functional calculus over $\mathcal{A}$ (\cref{theorem:holomorphic_funct_calculus_smooth_sub_algebras}) and the topology of $\text{Hol}(\mathbb{C})$ (\cref{example:holomorphic_functions_over_open_set}),

\begin{itemize}
    \item \textbf{Continuity:} We want to proof that for any $a \in \mathcal{A}$ and  $f \in \text{Hol}(\mathbb{C})$ the function
    $$ \gamma_{f,a} : \mathbb{C} \to \mathcal{A}, \; z \mapsto f(za) $$
    is continuous, this means that for any $z \in \mathbb{C}$, if $z' \to z$ then $f(z' a) \to f(za)$ in $\mathcal{A}$. Given that the holomorphic functional calculus is a continuous map from $\text{Hol}(\mathbb{C})$ into $\mathcal{A}$ (\cref{theorem:holomorphic_funct_calculus_smooth_sub_algebras}), we will divide this problem into two pieces. First, we will show that the map $z \mapsto f \circ g_z$ with $g_z (a) = za$ is a continuous map from $\mathbb{C}$ into $\text{Hol}(\mathbb{C})$, then, given that $(f \circ g_z)(a) = f(za)$ we can use the continuity of the holomorphic functional calculus over $\mathcal{A}$ (\cref{theorem:holomorphic_funct_calculus_smooth_sub_algebras}) to show that the map $z \mapsto f(za)$ is continuous because is the composition of two continuous maps.
    
    Let $\{ K_i \}_{i \in \mathbb{N}}$ be a compact exhaustion of $\mathbb{C}$ (\cref{def:compact_exhaustion}), then, the topology of $\text{Hol}(\mathbb{C})$ is given by the set of seminorms (\cref{example:holomorphic_functions_over_open_set})
    $$ p_{i}(f) = \sup_{\omega \in K_i} | f(\omega) |, \; i \in \mathbb{N},$$
    thus, by the content of \cref{proposition:convergence_metrizable_lcs} we need to check that, given $n < \infty$ and $ \epsilon > 0$, we can find $\delta > 0$ such that if $| z' - z | \leq \delta$, then 
    $$p_{n}((f \circ g_{z'}) - (f \circ g_{z})) \leq \epsilon.$$ 
    Since $K_n$ is compact, there is $d \geq 0$ such that $K_n \subset \overline{B(0,d)}$, that is, we find a compact ball that contains $K_n$. For convenience, we set $| z' -z | \leq 1$, because we have that $| \omega z'| \leq d \max \{ |z|, |z| + 1 \}$ when $| z' -z| \leq 1$. 
    
    Set $d' =  d \max \{ |z|, |z| +1 \}$, then $f$ is uniformly continuous on $\overline{B(0, d')}$, meaning that there is a $\delta'$ such that, if $|\omega - \omega'| \leq \delta'$ and $\omega, \omega' \in \overline{B(0, d')}$, then, $ |f(\omega) - f (\omega')| \leq \epsilon$. Take $\omega \in K_n$, then $|\omega z - \omega z' | = | \omega | |z -z'|$, and we now that $d' \geq | \omega|$, so if we set 
    $$ \delta = \frac{\min \{\delta' ,1\}}{\max \{ d' ,1 \}} $$
    and we take $| z - z' | \leq \delta$, then, when $\omega \in \overline{B(0,d)}$ we can guaranty that $\omega z, \omega z' \in  B(0, d')$ and also that $|\omega z - \omega z' | \leq \delta'$, which in turn tell us that if $\omega \in K_n$, then $|(f \circ g_{z'})(\omega) - (f \circ g_{z}))(\omega) | \leq \epsilon$, or equivalently, then $\omega \in K_n$ we have that 
    $$p_{i}((f \circ g_{z'}) - (f \circ g_{z})) \leq \epsilon.$$
    
    According to \cref{proposition:convergence_metrizable_lcs}, the previous statement implies that $z \mapsto f \circ g_z$ is a continuous function from $\mathbb{C}$ into $\text{Hol}(\mathbb{C})$, therefore, the continuity of the holomorphic calculus over $\mathcal{A}$ (\cref{theorem:holomorphic_funct_calculus_smooth_sub_algebras}) tells us that $z \mapsto f(za)=(f \circ g_z)(a)$ is a continuous map from $\mathbb{C}$ to $A$ for every $a \in A$ and $f \in \text{Hol}(\mathbb{C})$.

    \item \textbf{Differentiable paths:} This demonstration is almost the same demonstration given in (\cref{proposition:smoomth_paths_from_entire_functions}), the only change lies in the fact that here we use the holomorphic functional calculus over $\mathcal{A}$ (\cref{theorem:holomorphic_funct_calculus_smooth_sub_algebras}) instead of the holomorphic functional calculus over $A$ (\cref{theorem:holomorphic_functional_calculus_banach_algebras}).

    Given that the map $a \mapsto Sp(a)$ is upper semi-continuous (\cref{proposition:upper_semi_continuity_spectrum_map}), according to \cref{remark:spectrum_upper_semi_continuity_and_function_evaluation} if $D$ is a Cauchy domain of $Sp(za)$ and $\gamma$ is the boundary of $D$, there must be $\eta >0$ such that, if $|z- z'| \leq \eta$ then $D$ is a Cauchy domain of $Sp(z'a)$. 
    
    Set $g_z(a) = za$, since $f(za) = (f \circ g_z)(a)$, we have that
    $$ \frac{\left( f(za) - f(z'a) \right)}{z - z'} =  \frac{\left( (f \circ g_z)(a) - (f \circ g_{z'})(a) \right)}{z - z'}  $$
    $$ = \frac{1}{2 \pi \mathrm{i}} \left(  \int_\gamma (f \circ g_z)(\lambda)(\lambda 1-a)^{-1} \mathrm{~d} \lambda - \int_\gamma (f \circ g_{z'})(\lambda)(\lambda 1-a)^{-1} \mathrm{~d} \lambda  \right) \frac{1}{z - z'}$$
    $$ = \frac{1}{2 \pi \mathrm{i}} \int_\gamma \left( \frac{(f \circ g_z)(\lambda) - (f \circ g_{z'})(\lambda)}{z - z'}  \right) (\lambda 1-a)^{-1} \mathrm{~d} \lambda ,$$
    moreover, if we use the fundamental theorem of calculus for complex integrals we get,
    $$ \frac{\left( f(za) - f(z'a) \right)}{z - z'}  =  \frac{1}{2 \pi \mathrm{i}} \int_\gamma \lambda \left( \frac{\int_{\lambda z'}^{\lambda z} f^{(1)}(\omega) d \omega }{\lambda z - \lambda z'}  \right) (\lambda 1-a)^{-1} \mathrm{~d} \lambda ,$$
    with $f^{(1)}(\omega) = \frac{d f}{dz} (\omega)$.
    Since $f^{(1)}$ is a continuous function over $\mathbb{C}$, the fundamental theorem of calculus for complex valued integrals tells us that for any $\lambda \in \mathbb{C}$, if $z' \to z$ then
    $$ \frac{\int_{\lambda z'}^{\lambda z} f^{(1)}(\omega) d \omega }{\lambda z - \lambda z'} \to  f^{(1)}(\lambda z). $$
    Since $f$ is an entire function, we know that $f^{(1)}$ is an holomorphic function over $\mathbb{C}$, therefore, it is uniformly continuous over compact sets. 
    
    Here comes the key part of the demonstration, notice that  the function $\lambda \mapsto f_{z,z'}(\lambda)$ given by
    $$ f_{z,z'}(\lambda) = \left( \frac{(f \circ g_z)(\lambda) - (f \circ g_{z'})(\lambda)}{z - z'}  \right) =  \lambda  \frac{\int_{\lambda z'}^{\lambda z} f^{(1)}(\omega) d \omega }{\lambda z - \lambda z'}$$
    is an holomorphic function over $\mathbb{C}$ when $z \neq z'$, so, we will ask for the limit of $f_{z,z'}(\cdot)$ as elements of the topological algebra $\text{Hol}(\mathbb{C})$ when $z' \to z$, which will turn out to be the function $\lambda \mapsto \lambda f^{(1)}(\lambda z)$, in which case the continuity of the holomorphic functional calculus over $\mathcal{A}$ (\cref{theorem:holomorphic_funct_calculus_smooth_sub_algebras}) would imply that 
    $$ \frac{d f(z'a)}{ dz'} |_{z} = \lim_{z' \to z} \frac{1}{2 \pi \mathrm{i}} \int_\gamma  f_{z,z'}(\lambda) (\lambda 1-a)^{-1} \mathrm{~d} \lambda $$
    $$= \frac{1}{2 \pi \mathrm{i}} \int_\gamma \lambda f^{(1)}(\lambda z) (\lambda 1-a)^{-1} \mathrm{~d} \lambda.$$
    
    Now we will look into how to show that 
    $$ (\lambda \mapsto f_{z,z'}(\lambda)) \to  (\lambda \mapsto \lambda f^{(1)}(\lambda z))$$
    inside $\text{Hol}(\mathbb{C})$. For $i \in \mathbb{N}$ set $K_i = \overline{B(0, i)}$, then, $\{ K_i \}_{i \in \mathbb{N}}$ is a compact exhaustion of $\mathbb{C}$ (\cref{def:compact_exhaustion}), then, the topology of $\text{Hol}(\mathbb{C})$ is given by the set of seminorms (\cref{example:holomorphic_functions_over_open_set})
    $$ p_{i}(f) = \sup_{\omega \in K_i} | f(\omega) |, \; i \in \mathbb{N},$$
    thus, by the content of \cref{proposition:convergence_metrizable_lcs} we need to check that, given $n < \infty$ and $ \epsilon > 0$, we can find $\delta > 0$ such that if $| z' - z | \leq \delta$, then 
    $$p_{n}\left( (\lambda \mapsto f_{z,z'}(\lambda)) -  (\lambda \mapsto \lambda f^{(1)}(\lambda z)) \right) \leq \epsilon.$$ 
    
    For convenience set $|z - z'| \leq 1$, under this setting, we have that, if $\lambda \in K_n$, then $z' \lambda \in B(0,n(|z| + 1))$. Chose $\delta' > 0$ such that, if $ \omega, \omega' \in \overline{B(0,n(|z| + 1))}$ and $|\omega - \omega'| \leq \delta '$,  then $|f^{(1)}(\omega) - f^{(1)}(\omega ') | \leq \epsilon/n$, under this setting, if we set
    $$ |z - z' | \leq \frac{\delta'}{n}, $$
    we have that
    $$ | \lambda z - \lambda z' | \leq |\lambda| |z - z'| \leq \delta ',  $$
    which guaranties that, for all $\lambda \in K_n$ the following holds 
    $$|f^{(1)}(z \lambda) - f^{(1)}(z' \lambda)| \leq \epsilon.$$ 
    Given that $\overline{B(0,n(|z| + 1))}$ is a convex set of $\mathbb{C}$, the line connecting $\lambda z$ with $\lambda z'$ lies inside $\overline{B(0,n(|z| + 1))}$, therefore, if $|z - z'| \leq \delta / n '$ we have that for any $0 \leq c \leq 1$ the following holds,
    $$|f^{(1)}(c z \lambda + (1-c) z' \lambda) - f^{(1)}(z \lambda)| \leq \epsilon/n.$$
    The fundamental theorem of calculus for complex valued functions tells us that,
    $$  \lambda f^{(1)}(\lambda z) = \lambda  \int_{\lambda z'}^{\lambda z} \frac{f^{(1)}(\lambda z)}{\lambda z - \lambda z'} d \omega, $$
    therefore,
    $$ \left| \lambda \frac{\int_{\lambda z'}^{\lambda z} f^{(1)}(\omega) d \omega }{\lambda z - \lambda z'} - \lambda f^{(1)}(\lambda z) \right| \leq \left( \sup_{\lambda \in K_n} |\lambda| \right)  \frac{\int_{\lambda z'}^{\lambda z} | f^{(1)}(\omega) - f^{(1)}(\lambda z) |d \omega }{ |\lambda z - \lambda z'|} .$$
    Since,
    $$ \int_{\lambda z'}^{\lambda z} | f^{(1)}(\omega) - f^{(1)}(\lambda z) |d \omega \leq \left( \sup_{0 \leq c \leq 1} |f^{(1)}(c z \lambda + (1-c) z' \lambda) - f^{(1)}(z \lambda)| \right) |\lambda z - \lambda z'| .$$
    if we use the bound on the values of $| f^{(1)}(\omega) - f^{(1)}(\lambda z) |$ when $\omega$ lies between $\lambda z'$ and $\lambda z$ we get that, if $\lambda \in K_n$, then
    $$ \left| \lambda \frac{\int_{\lambda z'}^{\lambda z} f^{(1)}(\omega) d \omega }{\lambda z - \lambda z'} - \lambda f^{(1)}(\lambda z) \right| \leq \left( \sup_{\lambda \in K_n} |\lambda| \right) \frac{\epsilon |\lambda z - \lambda z'|}{n} \frac{1}{|\lambda z - \lambda z'|} \leq \epsilon .$$
    The previous statement implies that if $z' \to z$, then $f_{z,z'}(\lambda) \to \lambda f^{(1)}(\lambda z)$ uniformly on $K_n$, under this setting, \cref{proposition:convergence_metrizable_lcs} tells us that $f_{z,z'}(\lambda) \to \lambda f^{(1)}(\lambda z)$ as elements of  $\text{Hol}(\mathbb{C})$, hence,
    $$ \frac{d f(z'a)}{ dz'} |_{z} = \frac{1}{2 \pi \mathrm{i}} \int_\gamma \lambda f^{(1)}(\lambda z) (\lambda 1-a)^{-1} \mathrm{~d} \lambda.$$
    We know that $f^{(1)} \circ g_z$ is holomorphic over $\mathbb{C}$, also, the function $\text{id}(\lambda) = \lambda$ is holomorphic over $\mathbb{C}$, from \cref{theorem:holomorphic_funct_calculus_smooth_sub_algebras} we know that $\text{id}(a) = a$, so, given that the holomorphic functional calculus over $\mathcal{A}$ is an homomorphism of algebras (\cref{theorem:holomorphic_funct_calculus_smooth_sub_algebras}), we get that $ (\text{id} f^{(1)} \circ g_z)(a) = a f^{(1)}(a), $ which implies that 
    $$ \frac{d f(z'a)}{ dz'} |_{z} = a f^{(1)}(za). $$
    Notice that the map $z \mapsto a f^{(1)}(za)$ is continuous because $f^{(1)}$ is a continuous function over $\mathbb{C}$. Since the function $\lambda \mapsto \lambda f^{(1)}(\lambda z)$ is holomorphic on $\mathbb{C}$, we can iterate the previous argument to show that
    $$ \frac{d f(z'a)}{ dz'} |_{z} = \frac{1}{2 \pi \mathrm{i}} \int_\gamma \lambda^n f^{(n)}(\lambda z) (\lambda 1-a)^{-1} \mathrm{~d} \lambda = a^{n} f^{(n)}(za),$$
    which implies that the function $z \mapsto f(za)$ is smooth.
\end{itemize}
\end{proof}

\begin{remark}[Homotopy from exponential function over smooth sub algebras]\label{remark:homotopy_from_exponential_functions_smooth_sub_algebra}
Let $A$ be a C* algebra with unit and $\mathcal{A}$ a smooth sub algebra of $A$, denote by $G_0(\mathcal{A})$ the set of invertibles in $\mathcal{A}$ that are homotopic to $1_{\mathcal{A}}$ inside $G(\mathcal{A})$. Using \cref{proposition:smoomth_paths_from_entire_functions_smooth_sub_algebras} if we take $f(x) = \exp(x)$ we have smooth homotopies of invertible elements,
$$ t \to \exp(ta), \; t \in [0,1], \; a \in \mathcal{A} ,$$
henceforth, $\exp{\mathcal{A}} \subset G_0(\mathcal{A})$, where $G_0(\mathcal{A})$ is the connected component of the identity of the topological group $G(\mathcal{A})$.\\
As in \cref{remark:homotopy_from_exponential_functions_smooth_sub_algebra} if we were more carefull with the conditions on \cref{proposition:smoomth_paths_from_entire_functions_smooth_sub_algebras} we could also have smooth maps like
$$ \gamma_{f,a} : (0,\infty) \to \mathcal{A}, \; z \mapsto f(za), \; \text{ when } \text{Sp}(a) \subset \Pi. $$
This map is used as a key step in the study of the semi group of homotopic projections of a smooth sub algebra in \citep[Lemma 3.43]{gracia-bondia_elements_2001} with $f(z) = z^{1/2}$ (the principal branch of the complex square root).
\end{remark}

You may guess the next step, since $G(\mathcal{A})$ is open for unital smooth sub algebras (\cref{lemma:some_properties_of_smooth_sub_algebras}), we could look into copying the proof from \cref{theorem: description_of_GA} to provide a characterization $G(\mathcal{A})$,

\begin{theorem}[Description of $G(\mathcal{A})$ ( Proposition 3.4.3 \citep{blackadar_k-theory_2012})]\label{theorem:description_of_GA_smooth_sub_algebras}
Let $\mathcal{A}$ be a smooth sub algebra of a unital C* algebra $A$, then:
\begin{itemize}
    \item $G_0(\mathcal{A})$ is an open-and-closed, normal subgroup of $G(\mathcal{A})$, and the components of $G(\mathcal{A})$ are precisely the cosets of $G_0(\mathcal{A})$ in $G(\mathcal{A})$
    \item $G_0(\mathcal{A})$ consists of all finite products of exponentials, so that
$$
G_0(\mathcal{A})=\left\{\exp \left(a_1\right) \exp \left(a_2\right) \cdots \exp \left(a_k\right): a_1, \ldots, a_k \in \mathcal{A}, k \in \mathbb{N}\right\} ;
$$
\item in the case where $A$ is commutative, $G_0(\mathcal{A})=\exp{\mathcal{A}}$
\end{itemize}
\end{theorem}
\begin{proof}
We denote by $G_0(\mathcal{A})$ the set of all invertibles of $\mathcal{A}$ that are homotopic to $1_{\mathcal{A}}$ through a path on invertibles, and we set $G_0 = G_0(\mathcal{A})$, also $G = G(\mathcal{A})$. Since the proof of \cref{theorem: description_of_GA} only uses the holomorphic calculus to come up with elements in $A$, you can translate it into $\mathcal{A}$ with the aid of \cref{remark:homotopy_from_exponential_functions_smooth_sub_algebra}. Also, recall that $G(\mathcal{A})$ is open and is a dense subset of $G(A)$ by \cref{lemma:some_properties_of_smooth_sub_algebras}.
\end{proof}

\begin{remark}[Canonical homotopies between invertibles in smooth sub algebras]\label{remark:canonical_homotopies_between_invertibles_smooth_sub_algebras}
Following an argument similar to \cref{remark:canonical_homotopies_between_invertibles_Banach_algebras} you can show that if two invertibles $x,y \in G(\mathcal{A})$ are homotopic we can construct a canonical path between them as 
$$ \gamma(t) = x (\exp \left(t a_1\right) \exp \left(t a_2\right) \cdots \exp \left(t a_k\right)), \; \gamma(0) = x , \; \gamma(1) = y.$$
The path $\gamma$ is a multiplication of continuous paths by \cref{remark:homotopy_from_exponential_functions_smooth_sub_algebra}, therefore is continuous i.e. an homotopy. Moreover, it is the multiplication of smooth paths, thus it is a smooth path. Notice that the product rule for the derivation still holds in this scenario.
\end{remark}

\begin{remark}[Smooth algebras as non-commutative smooth manifolds]\label{sec:smooth_algebras_non_commutative_smooth_manifolds}
The content of \citep{moerdijk_models_1991}, \citep{kainz_c-algebras_1987}, \citep{michor_characterizing_1994} it is mentioned that $C^{\infty}$ rings provide a nice setup for an algebraic approach towards differential geometry. Recall that C* algebras provide a nice setup for an algebraic approach towards the topology of locally compact Hausdorff spaces (\cref{section:non_commutative_geometry_dictionary}). The $C^{\infty}$ rings come up as rings where is possible to define mappings $f : A^{n} \to A^{m} $ with $f$ a smooth function $f : \mathbb{R}^n \to \mathbb{R}^m $ (\citep[Chapter 1, Section 1]{moerdijk_models_1991}), instead of just having mappings $p: A^{n} \to A^{m}$ with $p = (p_1, \cdots, p_m)$ and $p_l$ a polynomial in $n$ variables. 

The most common definition of a $C^{\infty}$ ring comes in the form of a functor $A : C^{\infty} \to \text{Sets}$ where, $C^{\infty}$ is the category of all $\mathbb{R}^n$ spaces with homomorphisms given by smooth functions, and $\text{Sets}$ is the category of sets with homomorphisms given by functions (\citep[Chapter 1, Section 1]{moerdijk_models_1991}), under this setting, the ring is understood as $A(\mathbb{R})$. A $C^{\infty}$ homomorphism is a natural transformation between the categories, so, for example, if $f,g : \mathbb{R} \to \mathbb{R}$ are smooth functions, they should generate maps $A(f), A(g) : A(\mathbb{R}) \to A(\mathbb{R})$ such that $A(f)\circ A(g) = A(f \circ g)$.

$C^{\infty}$ rings have a natural locally convex topology given by m-convex seminorms (\citep[Definition 2.3]{kainz_c-algebras_1987}), such that, $C^{\infty}$ homomorphisms are automatically continuous (\citep[Theorem 2.4]{kainz_c-algebras_1987}), much like C* homomorphisms are automatically continuous (\cref{proposition:automatic_continuity_C_star_algebras}). $C^{\infty}$ rings that correspond to smooth functions over a paracompact second countable manifold are Fréchet m-convex algebras (\citep[Theorem 2]{michor_characterizing_1994}), thus, it is sensible to look into non-commutative Fréchet algebras for a setup to generalise the algebras of smooth functions over smooth compact manifolds.
\end{remark}

\subsection{Fréchet $D^{*}_{\infty}$-subalgebras} \index{Fréchet $D^{*}_{\infty}$-subalgebras}
\label{sec:Frechet_d_infinity_subalgebras}

The Fréchet $D^{*}_{\infty}$-subalgebras arise are generalization of the example 
$$C^{\infty}([a,b]) \subset C([a,b]),$$
where $C^{\infty}([a,b])$ is a smooth sub algebra over $C([a,b])$. Fréchet $D^{*}_{\infty}$-subalgebras are smooth sub algebras (\cref{proposition:invariance_under_holomorphic_func_calculus}) and their structure resembles that of $C^{\infty}([a,b])$ in the sense that it has seminorms that satisfy a Leibniz like law for the seminorms (\cref{def:Frechet_d_infinity_subalgebras}). In \citep{bhatt_class_2013} and \citep{bhatt_smooth_2016} there are various examples of Fréchet $D^{*}_{\infty}$-subalgebras which makes the case for their usefulness, moreover, the smooth sub algebras of twisted crossed products take the form of Fréchet $D^{*}_{\infty}$-subalgebras (\cref{proposition:smooth_elements_as_D_infinity_Freche_subalgebra}). 

\begin{definition}[Fréchet $\left(D_{\infty}^*\right)$-subalgebra (Definition 1.2 \citep{bhatt_class_2013}]\label{def:Frechet_d_infinity_subalgebras}
Let $\left(A,\|\cdot\|_0\right)$ be a $C^*$-algebra. Let $\mathcal{A}$ be a dense ${ }^*$-subalgebra of $A$. Then $\mathcal{A}$ is called a Fréchet $\left(D_{\infty}^*\right)$-subalgebra of $A$ if there exists a sequence of seminorms $\left\{\|\cdot\|_i: 0 \leq i<\right.$ $\infty\}$ such that the following hold:
\begin{itemize}
    \item \textbf{m-convex * algebra:} For all $i, 1 \leq i<\infty$, for all $x, y$ in $\mathcal{B},\|x y\|_i \leq\|x\|_i\|y\|_i,\left\|x^*\right\|_i=\|x\|_i$.
    \item \textbf{Leibniz like law for seminorms:} For each $i, 1 \leq i<\infty$, there exists $D_i>0$ such that $\|x y\|_i \leq D_i\left(\|x\|_i\|y\|_{i-1}+\right.$ $\left.\|x\|_{i-1}\|y\|_i\right)$ holds for all $x, y$ in $\mathcal{B}$. $\left\{\|\cdot\|_i: 0 \leq i<\infty\right\}$. 
    \item \textbf{Fréchet *-algebra: }x $\mathcal{B}$ is a Hausdorff Fréchet ${ }*$-algebra with the topology $\tau$ defined by the seminorms $\left\{\|\cdot\|_i: 0 \leq i<\infty\right\}$.
\end{itemize}
\end{definition}

As it happens with smooth sub algebras, the matrix algebras of Fréchet $D^{*}_{\infty}$-subalgebras are still Fréchet $D^{*}_{\infty}$-subalgebras,

\begin{proposition}[PROPOSITION 2.6 \citep{bhatt_class_2013}]\label{prop:matrix_algebras_of_frechet_d_infinity_sub_algebras}
Let $\left(\mathcal{A},\left\{\|\cdot\|_i: 0 \leq i<\infty\right\}\right)$ be a Fréchet $\left(D_{\infty}^*\right)$-subalgebra of $A$. Then $M_n(\mathcal{A})$ is a dense ${ }^*$-subalgebra of $M_n(A)$; and it is also a Fréchet ${ }^*$-algebra with the projective tensorial topology defined by the family of seminorms $\|\cdot\|_{\gamma, i}$, which is the projective cross norm given by $\|\cdot\|_i$ on $\mathcal{B}$ and the $C^*$-norm $\|\cdot\|$ on $M_n(\mathbb{C})$. Moreover, under this assumptions the algebra $M_n(\mathcal{A})$ is a Fréchet $\left(D_{\infty}^*\right)$-subalgebra of $M_n(A)$.
\end{proposition}

Fréchet $D^{*}_{\infty}$-subalgebras are useful because they are closed under the holomorphic functional calculus of their ambient C* algebra, and are spectrally invariant\index{spectrally invariant},

\begin{proposition}[COROLLARY 3.2 \citep{bhatt_class_2013}]\label{proposition:invariance_under_holomorphic_func_calculus}
Let $\mathcal{A}$ be a Fréchet $\left(D_{\infty}^*\right)$-subalgebra of a $C^*$-algebra $A$
\begin{itemize}
    \item If $A$ has identity 1 , then $1 \in \mathcal{A}$ and $\left(\mathcal{A},\|\cdot\|_0\right)$ is a $Q$-normed algebra.
    \item The algebra $\mathcal{A}$ is hermitian, spectrally invariant in $A$, is closed under the holomorphic functional calculus of $A$ and is an algebra with a $C^*$-enveloping algebra having $C^*(\mathcal{A})=A$.
\end{itemize}
\end{proposition}

Notice that \cref{proposition:invariance_under_holomorphic_func_calculus} implies that Fréchet $D^{*}_{\infty}$-subalgebras are smooth sub algebras. As an additional interesting fact, Fréchet $D^{*}_{\infty}$-subalgebras are also closed under $C^{\infty}$ calculus\index{$C^{\infty}$ calculus} for the self-adjoint elements, in this context, $C^{\infty}$ calculus of self adjoint elements refers to computing the element $f(a)$ for any $a$ self adjoint and $f$ smooth over $\mathbb{R}$,

\begin{proposition}[COROLLARY 3.3 \citep{bhatt_class_2013}]\label{proposition:Frechet_d_infinity_algebras_closed_under_C_infinity_calculus}
Let $\mathcal{A}$ be a Fréchet $\left(D_{\infty}^*\right)$-subalgebra of a $C^*$-algebra $A$. Then $\mathcal{A}$ is closed under the $C^{\infty}$-functional calculus for self-adjoint elements of $A$.
\end{proposition}

\subsection{Examples}
\label{sec:example_smooth_sub_algebras}

Take $\mathcal{A}$ a sub *-algebra of a C* algebra $A$, then, how do we prove that $\mathcal{A}$ is a smooth sub algebra of A? to answer this question we would need to prove that $\mathcal{A}$ has the properties of a smooth sub algebra (\cref{def:smooth_sub_algebra}), which may prove to be highly non trivial, thus, we will piggyback on the work of mathematicians who have already given conditions for $\mathcal{A}$ to be a smooth sub algebra of $A$. Specifically, we will use both results from Bhatt et.al \citep{bhatt_class_2013} and results from Rennie \citep{rennie_smoothness_2003}. Our first example of smooth sub algebra is quite simple, but is the example that motivated the development of many techniques on smooth sub algebras, like Banach $D_{p}^{*}$-algebras and Fréchet $D^{*}_{\infty}$-subalgebras (\cref{sec:Frechet_d_infinity_subalgebras}), which will play an important role in our examples.

\begin{example}[Example 1.3 \citep{rennie_smoothness_2003}]\label{example:smooth_functions_on_interval}

Let $C[a, b]$ be the $C^*$-algebra of continuous functions on $[a, b]$. Recall that $C^{\infty}[a,b]$ is dense in $C[a,b]$ and is unital because $[a,b]$ is compact, moreover, $C^{\infty}[a, b]$ is a Fréchet algebra whose topology is defined by the sequence of norms $\left\{p_k\right\}$, where $p_k(f)=\sum_{i=0}^k \frac{\left\|f^{(i)}\right\|_{\infty}}{i !}$. This is the canonical example of Fréchet $D^{*}_{\infty}$-subalgebras, which in fact are invariant under holomorphic functional calculus (\cref{proposition:invariance_under_holomorphic_func_calculus}).

Also, if $\mathcal{A}$ is a Fréchet $\left(D_{\infty}^*\right)$-subalgebra of a $C^*$-algebra $A$. Then the following algebras can be given a sequence of norms that turn them into Fréchet $\left(D_{\infty}^*\right)$-subalgebra
\begin{itemize}
    \item $C([a, b], \mathcal{A}) \simeq C([a, b])\widehat{\otimes}_{\pi} \mathcal{A}$
    \item $C^{\infty}([a, b], \mathcal{A}) \simeq C^{\infty}([a, b]) \widehat{\otimes}_{\pi} \mathcal{A}$
    \item $C^{\infty}([a, b], A) \simeq C^{\infty}([a, b]) \widehat{\otimes}_{\pi} A$
\end{itemize}
were $\widehat{\otimes}_{\pi}$ is the projective tensor product (\cref{definition:projective_tensor_product_of_frechet_algebras}).
\end{example}

\begin{example}[Functions decaying at infinity]
The Schwarts algebra $\mathcal{S}(\mathbb{R}^d)$\index{$\mathcal{S}(\mathbb{R}^d)$} is dense in the C* algebra of continuous function decaying at infinity $C_0(\mathbb{R}^{d})$ \citep[Proposition 8.17]{folland_real_1999}. Moreover, it is generated by the action of $\mathbb{R}^d$ as a Lie group over $C_0(\mathbb{R}^d)$, where the action is given by the translation, therefore, by \citep[example 1.6]{bhatt_class_2013} $\mathcal{S}(\mathbb{R}^d)\subset C_0(\mathbb{R}^d)$ is a Fréchet $D^{*}_{\infty}$-subalgebra. 
\end{example}

\begin{example}[Algebra of infinite matrices with rapid decay (first example page 131 \citep{rennie_smoothness_2003})]\label{example:infinte_matrices_with_rapid_decay}
Let $\mathcal{H}$ be a separable Hilbert space and $F$ the algebra of finite rank operators on $\mathcal{H}$, then, we call $R D(H)$\index{$R D(H)$} the algebra of matrices with rapid decay \index{matrices with rapid decay} and is defined as
$$
R D(\mathcal{H})=\left\{\left(a_{i j}\right)_{i, j \in \mathbf{N}}: \sup _{i, j} i^k j^l\left|a_{i j}\right|<\infty, \; \forall k, l\right\} \in \mathbb{N}.
$$
$R D(H)$ can be provided with a locally convex topology given be the seminorms
$$
\begin{aligned}
& q_{k, l}\left(a_{i j}\right)=\sup _{i, j} i^k j^l\left|a_{i j}\right|, \quad k, l>0, \\
& q_{0,0}=\left\|\left(a_{i j}\right)\right\|=\text { operator norm, }
\end{aligned}
$$
and becomes a Fréchet *algebra under this topology, moreover, $R D(H)$ is the completion of $F$ in the topology determined by the seminorms above. 

Notice that $q_{0,0}$ is a $C^*$-norm, and the completion of $F$ in this norm is simply $\mathcal{K}$ (\cref{section:algebra_of_compact_operators}), the compact operators on $\mathcal{H}$. The explicit decay on the elements of $R D(H)$ guaranties that $Tr((a_{i j})_{i, j \in \mathbf{N}} = \sum_{n \in \mathbb{N}}a_{i,j}$ converges for any element of $R D(H)$, moreover, the explicit form of the topology over $R D(H)$ implies that $Tr$ is continuous linear functional on $R D(H)$, and also has the tracial property i.e. $Tr(ab) = Tr(ba)$.

Since $R D(H)$ is a smooth sub algebra of $\mathcal{K}$ were $Tr$ is a continuous linear functional, we may be inclined to believe that in any smooth sub algebra $Tr$ becomes a continuous linear functional. That is far from true, for example, \citep[example 1.7]{bhatt_class_2013} provides a smooth sub algebra (Fréchet $\left(D_{\infty}^*\right)$-subalgebra) of $\mathcal{K}$ that is bigger than that set of trace class operators, that is, it contains elements whose trace is infinite.
\end{example}

\begin{example}[Tensor product of smooth sub algebras (second example page 131 \citep{rennie_smoothness_2003})]\label{example:tensor_product_of_smooth_sub_algebras}

Let $A$ be a unital C* algebra and $\mathcal{A}$ a smooth sub algebra of $A$, also, let $\mathcal{B}$ be a local smooth sub algebra of a C* algebra $B$ (\citep[Definition 3]{rennie_smoothness_2003}), then, according to \citep[second example on page 131]{rennie_smoothness_2003} the projective tensor product $\mathcal{A} \widehat{\otimes}_{\pi} \mathcal{B}$ is a smooth sub algebra of $A \otimes B$. 

This example is quite relevant for us because $\mathcal{K}$ is a C* algebra with a trace that is not a continuous (\cref{section:algebra_of_compact_operators}), and $R D(\mathcal{H}) \subset \mathcal{K}$ is a smooth sub algebra with local units (\citep[defintions 2 and 3]{rennie_smoothness_2003}), were the trace becomes a continuous linear functional that is tracial i.e. $Tr(ab)=Tr(ba)$. This allow us to take a tracial state $\phi$ over $\mathcal{A}$ and define tracial state over $\mathcal{A} \widehat{\otimes}_{\pi} R D(\mathcal{H})$ as $\phi \widehat{\otimes}_{\pi} Tr$, which is relevant to the definition of cyclic cocycles on $\mathcal{A} \widehat{\otimes}_{\pi} R D(\mathcal{H})$ (\citep[Chapter 5]{prodan_bulk_2016}).

The elements of $\mathcal{A} \widehat{\otimes}_{\pi} R D(\mathcal{H})$ can be considered as infinite matrices with coefficients in $\mathcal{A}$ much like it happens in $A \otimes \mathcal{K}$ (\cref{section:stabilization_C_stal_algebra}), also, if $\{ p_n \}_{n \in \mathbb{N}}$ is the set of seminorms that provide the topology of $\mathcal{A}$ then according to \citep[Proposition 3.3.3]{prodan_bulk_2016} $\mathcal{A} \widehat{\otimes}_{\pi} R D(\mathcal{H})$ is characterized by

$$
\mathcal{A} \widehat{\otimes}_{\pi} R D(\mathcal{H})=\left\{\left(a_{i j}\right)_{i, j \in \mathbf{N}}: \sup _{i, j} i^k j^l p_n(a_{i j})<\infty, \; a_{ij} \in \mathcal{A}, \; \forall k, l, n \in \mathbb{N}\right\}.
$$

Additionally, if $\phi$ is a continuous linear functional on $\mathcal{A}$ and $(a_{i,j})_{i, j \in \mathbf{N}} \in \mathcal{A} \widehat{\otimes}_{\pi} R D(\mathcal{H})$, then 
$$(\phi \widehat{\otimes}_{\pi} Tr)\left( (a_{i,j})_{i, j \in \mathbf{N}} \right) = \sum_{i \in \mathbb{N}} \phi(a_{i,i}).$$

\end{example}

\begin{disclaimer}\label{disclaimer:tensor_product_smooth_sub_algebras}

The realm of smooth sub algebras is highly technical, so, there are some results that as you may noticed, are presented with little to none details on their proofs, only references. Among these results we would like to said a few words

\begin{itemize}
    \item We have not checked that $R D(H)$ is a smooth sub algebra of $\mathcal{K}$, this fact were taken from  \citep[page 131]{rennie_smoothness_2003}. The similarity of its semi norms compared to the semi norms of the smooth sub algebra inside $A \rtimes_{\alpha,\Theta} \mathbb{Z^{2}}$ (\cref{sec:derivations_twistted_crossed_product}) suggest that $R D(H)$ is also a Fréchet $D^{*}_{\infty}$-subalgebra, but we have not checked this.
    \item We have not found a proof for the fact that $\mathcal{A} \widehat{\otimes}_{\pi} \mathcal{B}$ is a smooth sub algebra of $A \otimes B$ in \cref{example:tensor_product_of_smooth_sub_algebras}, and as we mentioned, is a fact that is mentioned in \citep{rennie_smoothness_2003}. By the properties of the semi norms on a Fréchet $D^{*}_{\infty}$-subalgebra it looks plausible that the projective tensor product of two Fréchet $D^{*}_{\infty}$-subalgebras is again a Fréchet $D^{*}_{\infty}$-subalgebra, so if $R D(H)$ were a Fréchet $D^{*}_{\infty}$-subalgebra then we could try to check the aforementioned fact about projective tensor products to provide a demonstration for this fact, but we have not gone through the details of these claims. 
    \item We have not done the complete calculations to verify the characterization of $\mathcal{A} \widehat{\otimes}_{\pi} R D(\mathcal{H})$ as infinite matrices with rapid decay, however it seems to be a established fact in the literature \citep[Proposition 3.3.3]{prodan_bulk_2016}.
\end{itemize}

These are some of the black boxes of this chapter, and we had to settle with just providing references due to time constrains. If you happen to come across proofs for any of those facts we would love to hear about them, and we apologize for the lack of mathematical strictness in the presentation of these results. 

\end{disclaimer}

\chapter{Review of Fourier analysis}
\label{chapter:review_of_Fourier_analysis}

In the \cref{chap:fourier_analysis} we will look into translation of results of Fourier analysis into generalization of the C* algebras $C_0(G;A)$ for $G$ a locally compact abeliean gorups and $A$ a C* algebra, these results will come as generalizations of \cref{section:Fourier_tranform_on_banach_valued_functions}, \cref{section:Fourier_transform_and_Frechet_algebras} and \cref{section:Fourier_tranform_C_star_valued_functions_over_the_torus}.

\section{Main results}
\label{sec:fourier_analysis_main_results}

Let $G$ be a locally compact Hausdorff topological group, that is, topological group with a locally compact topology, recall that a topological group\index{topological group} is a group with a topology under which the multiplication ($G \times G \to G, \; (a,b) \mapsto ab$) and inversion ($G \to G, \;  a \mapsto a^{-1}$) are continuous. This types of groups are quite important for harmonic analysis because we can provide them with the structure of a measure space, here we will use the Borel $\sigma$-algebra ($\mathscr{B}(G)$)\index{$\mathscr{B}(G)$} and the measure $\mu$ will be left translation invariant i.e $\mu(A) = \mu(xA)$ for all $A \in \mathscr{B}(G)$ and $x \in G$. The measure $\mu$ is guaranteed to exists by the Riesz representation theorem\index{Riesz representation theorem} for positive continuous functionals on $C_c(G)$, also known as the Riesz-Markov theorem\index{Riesz-Markov theorem} (\citep[Theorem 3.18]{kantorovitz_introduction_2003}), because is possible to define a continuous positive functional on $C_c(G)$ that is left translation invariant and unique up to a multiplicative constant (\citep[Theorems 4.11, 4.12]{kantorovitz_introduction_2003}, \citep[Chapter 5]{de_chiffre_haar_2011}).

\begin{definition}[Haar measure\index{Haar measure}]\label{definition:Haar_measure}
Let $G$ be a locally compact Hausdorff topological group, then there is a left translation invariant measure $\mu$ over $\mathscr{B}(G)$ with the following properties:
\begin{itemize}
    \item \textbf{Radon measure:} $\mu$ is a regular Borel measure, and finite on compact sets (\citep[Definition 4.13]{kantorovitz_introduction_2003}, \cref{definition:Radon_measure})
    \item \textbf{Left translation invariant:}\index{left translation invariant measure} for any $S \in \mathscr{B}(G)$ and $g \in G$ we have that $\mu(S) = \mu(gS)$.
    \item $\mu$ is unique up to a multiplicative constant i.e. if $\mu_0$ and $\mu_1$ are Haar measures then there is $\lambda \in \mathbb{R}^+$ such that $\mu_0 = \lambda \mu_1$ (\citep[Definition 4.13]{kantorovitz_introduction_2003})
    \item By \citep[Proposition 2.1]{vinroot_haar_2008} $\mu$ has full support (\cref{definition:measure_with_full_support}) 
    \item $\mu(G) < \infty$ iff $G$ is compact (\citep[Proposition 2.1]{vinroot_haar_2008})
    \item $\hat{\mu}(A) := \mu(A^{-1})$ is a right translation invariant measure\index{right translation invariant measure} (\citep[Proposition 2.1]{vinroot_haar_2008})
    \item If $\phi$ is a continuous automorphism of $G$ with a continuous inverse then $(\mu \circ \phi)(A) := \mu(\phi(A))$ for $A \in \mathscr{B}(G)$ is a left Haar measure. 
\end{itemize}
\end{definition}

For every $x \in G$ the map $\phi(x): g \mapsto x^{-1} g x$ is a continuous automorphism of $G$ with continuous inverse, then $\mu \circ \phi_x$ is a Haar measure of $G$, which implies that there is some positive scalar $\delta_G (x)$ such that $(\mu \circ \phi_x)(A) = \delta_G (x) \mu(A)$ for all $A \in \mathscr{B}(G)$. The mapping $\delta_G : G \to \mathbb{R}^+$ is called the modular function and is a continuous homomorphism of groups (\citep[Proposition 2.2]{vinroot_haar_2008}), such that the measure obtained from the positive linear functional $f \to \int_G f(g) \delta_G (g) d \mu(g)$ is a right Haar measure on $G$ (\citep[Proposition 2.3]{vinroot_haar_2008}).

The groups where $\delta_G (x) = 1$ for all $x \in G$ are called unimodular\index{unimodular}, and are quite special because on those groups we have that $\mu(A) = \mu(A^{-1})$ for all $A \in \mathscr{B}(A)$, or in similar terms (\citep[page 120]{kantorovitz_introduction_2003}) 
$$\int_G f\left(x^{-1}\right) d \mu(x)=\int_G f(x) d \mu(x) \quad\left(f \in C_c(G)\right).$$

\begin{proposition}[Examples of unimodular groups]\label{proposition:examples_unimodular_groups}
Let $G$ be a locally compact Hausdorff topological group, then 
\begin{itemize}
    \item If $G$ is compact then $G$ is unimodular (\citep[Proposition 2.4]{vinroot_haar_2008})
    \item If $G$ is abelian then $G$ is unimodular (\citep[page 6]{vinroot_haar_2008})
\end{itemize}
For any unimodular group we have that $\mu(A) = \mu(A^{-1})$ for all $A \in \mathscr{B}(G)$.
\end{proposition}

\begin{remark}[Convolution and abelian group\index{convolution}]\label{remark:convolution_abelian_groups}
From \cref{example:banach_algebras} we know that $L^1(G)$ becomes a Banach algebra under the convolution, moreover, the convolution is commutative iff $G$ is abelian (\citep[Theorem 1.6.4]{deitmar_principles_2009}). Also, from \cref{example:banach_star_algebras} we know that $L^1(G)$ becomes a Banach *-algebra if $G$ is unimodular. Notice that if $A$ is a Banach *-algebra then \cref{example:L1_Bohner_spaces_and_convolution} tell us that $L^1(G;A)$ is a Banach *-algebra.
\end{remark}

In \cref{chap:fourier_analysis} we provide an special attention to the groups $\mathbb{Z}^d$ and $\mathbb{T}^d$ because these are the appropriate framework for the tight binding models (\cref{sec:motivation_from_physics}). The direct generalization of those models would make use of the group $\mathbb{R}^d$, that is, changing discrete models for continuous ones. Since all the groups we are interested in are abelian we will focus our attention into abelian locally compact groups\index{abelian locally compact groups}, which we refer to as LCA groups\index{LCA group}. This groups are very special because we can define homomorphism into $\mathbb{T}$ and those homomorphisms will have a nice relation with the Gelfand transform.

\begin{definition}[Dual group\index{dual group}]\label{definition:dual_group}

Let $G$ be a LCA then $\gamma: G \to \mathbb{T}$ is called a character\index{character of an abelian group} if 
$$ \gamma(x+y) = \gamma(x) \gamma(y), \; x,y \in G. $$
The set of all continuous characters of $G$ is denote by $\Gamma$ and is an abelian group under the operation
$$ (\gamma_1 + \gamma_2)(x) = \gamma_1 (x) \gamma_2(x), \; x \in G, \; \gamma_1, \gamma_2 \in \Gamma, $$
moreover, $\Gamma$ is a LCA under the topology of uniform convergence in compact sets of $G$ (\citep[Section 1.2.6]{rudin_fourier_1967}). $\Gamma$ has the following properties:
\begin{itemize}
    \item $\gamma(-x) = (- \gamma)(x) = (\gamma(x))^{-1} = (\gamma(x))^* $ 
    \item \textbf{Pontryagin duality:} Denote by $\hat{\Gamma}$ the dual group $\Gamma$, then every $x \in G$ may be regarded as a continuous character on $\Gamma$, there is a natural map $\alpha$ of $G$ into $\hat{\Gamma}$, defined by
$$
\alpha(x)(\gamma) = \gamma(x) \quad(x \in G, \gamma \in \Gamma) .
$$
The aforementioned map is an isomorphism of topological groups, or in informal terms, every LCA is the dual of its dual group. (\citep[Theorem 1.7.2]{rudin_fourier_1967}).

    \item Every compact LCA is the dual of a discrete LCA and every discrete LCA is the dual of a compact LCA (\citep[Section 1.7.3]{rudin_fourier_1967})
    \item If $G$ is compact then $\mu(G)=1$ iff the Haar measure on $\Gamma$ is the counting measure (\citep[Proposition 4.25]{folland_course_2016}).
    \item If $G$ is second countable then $\hat{G}$ is second countable (\citep[Theorem 7.6]{folland_course_2016}, \citep{1915124}).
\end{itemize}
\end{definition}

\textbf{From now on if $G$ is a compact abelian group then we set $\mu(G) =1$ such that the measure on $\Gamma$ is the counting measure.}

The Fourier transform is defined using the dual group, in that construction the convolution of functions over $L^1(G)$ plays a key role, so, let $f,g$ functions over $G$ then denote by $f \ast g$  the convolution of $f$ and $g$, that is, the function given by 
$$(f \ast g)(x) = \int_G f(x)g(y-x) dy,$$ 
when that sum converges. There are some cases where $f \ast g$ exists, for example, if $g \in L^{\infty}(G)$ and $f \in L^1(G)$ then $(f \ast g)$ is bounded and uniformly continuous (\citep[Section 1.1.6]{rudin_fourier_1967}), also, from previous examples we know that $\|f \ast g\| \leq \|f \| \|g \|$ if $f,g \in L^1(G)$.

\begin{definition}[Fourier transform\index{Fouriere transform}]\label{definition:Fourier_transform}

Let $G$ be a LCA and $\Gamma$ its dual gruop, $f \in L^1(G)$ and $\gamma \in \Gamma$, then the function defined by 
$$ \hat{f}(\gamma) = \int_{G} f(x) \gamma(-x) dx $$
is called the Fourier transform of $f$. Denote by 
$$A(\Gamma) = \{ \hat{f} | f \in L^1(G) \}.$$
\end{definition}

Notice that the Fourier transform is well defined because $(x \mapsto \gamma(-x)) \in L^{\infty}(G)$.

\begin{proposition}[Properties of the Fourier transform]\label{proposition:Fourier_transform_properties}
The Fourier transform has the following properties:
\begin{itemize}

    \item \textbf{Generalized Riemann-Lebesgue lemma:}\index{Riemann-Lebesgue lemma} $A(\Gamma) \subset C_0(\Gamma)$ (\citep[Theorem 1.2.4]{rudin_fourier_1967}).
    \item $A(\Gamma)$ is a self-adjoint sub-algebra of $C_0(\Gamma)$ that separates points of $\Gamma$ and vanishes nowhere (\citep[Theorem 1.2.4]{rudin_fourier_1967}), therefore the Stone-Weirstrass theorem tell us that $A(\Gamma)$ is dense in $C_0(\Gamma)$ (\citep[Stone-Weierstrass notes, Corollary 7]{quigg_real_2005}).
    \item \textbf{Continuity:} The Fourier transform is norm decreasing i.e $\| \hat{f} \|_{C_0(\Gamma)} \leq \| f \|_{L^1(G)}$ (\citep[Theorem 1.2.4]{rudin_fourier_1967})
    \item $\widehat{f \ast g} = \hat{f} \hat{g}$ (\citep[Theorem 1.2.4]{rudin_fourier_1967})
    \item $(f \ast \gamma)(x) = \gamma(x) \hat{f}(\gamma)$ (\citep[Theorem 1.2.4]{rudin_fourier_1967})
    \item \textbf{Injectivity:} if $f \in L^1(G)$ and $\hat{f} = 0$ then $f =0$ (\citep[Theorem 1.2.2]{rudin_fourier_1967})
    \item \textbf{Inversion formula:} If $f \in L^1(G)$ and $\hat{f} \in L^1(\Gamma)$ then $f(x) = \widehat{\hat{f}}(-x)$ almost everywhere, that is,
    $$ f(x) = \int_{\Gamma} \gamma(x) \hat{f}(\gamma) d \gamma, \text{ a.e. } x.$$
    If $f$ is continuous, these relations hold for every $x \in G$ (\citep[Theorem 4.33]{folland_course_2016}).
    \item \textbf{Gelfand transform:}\index{Gelfand transform} every non-zero homomorphism (involution preserving) $\phi :L^1(G) \to \mathbb{C}$ is obtained as $f \to \hat{f}(\gamma)$ for $\gamma \in \Gamma$, and we denote it by $\phi_{\gamma}$, moreover, if $\gamma_1 \neq \gamma_2$ then $\phi_{\gamma_1} \neq \phi_{\gamma_2}$ (\citep[Theorem 1.2.2]{rudin_fourier_1967}). 
\end{itemize}
\end{proposition}

\begin{remark}[Gelfand transform]\label{remark:Gelfand_transform_and_Fourier_transform}

Since $\Gamma$ is in one to one correspondence with the set of non-zero complex homomorphisms (involution preserving) of $L^1(G)$ (\cref{proposition:Fourier_transform_properties}) we can provide it with the final topology given by the homomorphisms $\overline{x}: \Gamma \to \mathbb{C}, \; \overline{x}(\gamma) = \hat{x}(\gamma)$ similar to what we did with the set of characters on the Gelfand transform (\cref{section:Gelfand_transform}). This topology is equivalent to the topology given by uniform convergence of characters on compact subsets of $G$ (\citep[Section 1.2.6]{rudin_fourier_1967}). Given this, the Fourier transform on $L^1(G)$ coincides with the Gelfand transform on $L^1(G)$, thus, you can compare \cref{theorem:gelfand_representation_theorem} and \cref{proposition:Fourier_transform_properties} for the parallel in those results.

In \citep[Theorem 4.69]{allan_introduction_2011} you can look for an explicit computation checking that the Fourier transform on $L^1(\mathbb{T})$ coincides with the Gelfand transform on $L^1(\mathbb{T})$, and in \citep[Theorem 4.75]{allan_introduction_2011} there is a similar computation for the case of $L^1(\mathbb{R})$.
\end{remark}

\begin{example}[Examples of Fourier transform]\label{example:Fourier_transform_and_dual_groups}
 
 If $G_1$ and $G_2$ are LCA then the Haar measure on $G_1 \oplus G_2$ (direct sum) is the product measure $\mu_1 \times \mu_2$, also, denote by $\hat{G_1}, \hat{G_2}$ the dual groups of $G_1, G_2$ respectively, then the dual group of $G_1 \oplus G_2$ is $\hat{G_1} \oplus \hat{G_2}$ with the identification $(\xi_1, \xi_2) (x_1, x_2) = \xi_1 (x_1) \xi_2 (x_2)$ for $\xi_i \in \hat{G_i}$ and $x_i \in G_i$ (\citep[Chapter 6]{katznelson_introduction_2004}, \citep[Proposition 4.7]{folland_course_2016}). 
 
The following are the most common examples of Fourier examples:

\begin{itemize}
    \item \textbf{Torus:} From \citep[Examples 1.2.7]{rudin_fourier_1967} we have that $\hat{\mathbb{T}} = \mathbb{Z}$ and the characters of $\mathbb{T}$ take the following form $\gamma_n(x) = x^n$ for $n \in \mathbb{Z}$, so, if $f \in L^1(\mathbb{T})$ then
    $$  \hat{f}(n) =\int_{\mathbb{T}}f(x) x^{-n} dx = \frac{1}{2 \pi} \int_{0}^{2 \pi}f(\exp(i\alpha)) \exp(-in\alpha) d \alpha, $$
    with $\; \alpha \in [0, 2 \pi)  \text{ and } x = \exp(i \alpha)$. Similarly, have that $\hat{\mathbb{Z}} = \mathbb{T}$  and $\gamma_x(n) = x^n$ for $x \in \mathbb{T}$, so, if $f \in L^1(\mathbb{Z})$ then
    $$  \hat{f}(x) = \int_{\mathbb{Z}}f(n) x^{-n} dn = \sum_{n = 1}^{\infty} f(n) \exp(-i n \alpha), \; \alpha \in [0, 2 \pi)  \text{ and } x = \exp(i \alpha).  $$
    Take $f \in C(\mathbb{T})$ then $f \in L^1(\mathbb{T})$, also, if $ f \in A(\mathbb{Z}) := \{ \hat{f}: f \in  L^1(\mathbb{Z})\} $, that is, if $f$ has an absolutely convergent Fourier series\index{Fourier series} then the inversion of the Fourier transformation (\cref{proposition:Fourier_transform_properties}) tell us that $f(\alpha) = \sum_{-\infty}^{\infty} e^{i \alpha n} \hat{f}(n)$. The algebra $ A(\mathbb{Z})$ is called the Wiener algebra\index{Wiener algebra} and denoted by $W(\mathbb{T})$ (\citep[Example 4.62]{allan_introduction_2011}), and is closed under inversion, that is, if $f(z) \neq 0$ for all $z \in \mathbb{T}$ then $1/f \in W(\mathbb{T})$ (\citep[Theorem 4.63]{allan_introduction_2011}).
    
    If we are dealing with d dimensional groups we have that $\widehat{\mathbb{T}^d} = \mathbb{Z}^d$ and $\widehat{\mathbb{Z}^d} = \mathbb{T}^d$. Additionally, we have
    $$  \hat{f}((n_1, \dots, n_d)) =  \int_{\mathbb{T}^d}f((x_1, \cdots, x_d)) (x_1^{-n_1})\cdots (x_d^{-n_d})  dx_1 \cdots d x_d $$
    $$= \frac{1}{(2 \pi)^d} \int_{0}^{2 \pi} \cdots \int_{0}^{2 \pi}  f((\exp(i\alpha_1), \cdots, \exp(i\alpha_d))) \exp(-in_1\alpha_1) \cdots \exp(-in_d\alpha_d) d \alpha_1 \cdots d \alpha_d, $$
    with $\; \alpha_i \in [0, 2 \pi)  \text{ and } x_i = \exp(i \alpha_i)$ and $n_i \in \mathbb{Z}$. Similarly, 
    $$ \hat{f}((x_1, \cdots, x_d)) = \sum_{- \infty}^{\infty} \cdots \sum_{-\infty}^{\infty} f((n_1, \cdots, n_d)) \exp(-i n_1 \alpha_1) \cdots \exp(-i n_d \alpha_d),$$
    with $\; \alpha_i \in [0, 2 \pi)$ , $x_i = \exp(i \alpha_i)$ and $n_i \in \mathbb{Z}$. Notice that the characters of $\mathbb{T}^d$\index{characters of $\mathbb{T}^d$} take the form
    $$ \gamma_{s}(\lambda) =  \left( \prod_{1 \leq j \leq d} \lambda_j^{s_j}  \right), \; s:= (s_1, \dots, s_d) \in \mathbb{Z}^d, \; \lambda := (\lambda_1, \cdots, \lambda_d) \in \mathbb{T}^d.$$
    
    \item \textbf{Reals:} From \citep[Examples 1.2.7]{rudin_fourier_1967} we have that $\hat{\mathbb{R}} = \mathbb{R}$ and $\gamma_y(x) = \exp(i y x)$ for $y \in \mathbb{R}$, therefore,
    $$ \hat{f}(y)= \int_{-\infty}^{\infty} f(x) \exp(-iyx) dx, \; f \in L^1(\mathbb{R}), \; y \in \mathbb{R} .$$
    Also for any $f \in A(\mathbb{R})$ i.e. $f \in C_0(\mathbb{R})$ and $\hat{f} \in L^1(\mathbb{R})$ we get the Fourier inversion formula
    $$ f(x) = \int_{\mathbb{R}} \exp(ixy) \hat{f}(y) dy .$$
    
\end{itemize}
\end{example}

One of the most important results from the Fourier analysis on LCA is the Plancherel theorem, which gives us an isomorphism of Hilbert spaces and will be key to our study of twisted crossed products

\begin{theorem}[Plancherel theorem\index{Plancherel theorem} (Section 1.6 \citep{rudin_fourier_1967})]\label{theorem:Plancherel_theorem}

Let $G$ be a LCA and $\Gamma$ its dual group, then the Fourier transform, when restricted to $L^1(G) \cap L^2(G)$ is an isometry with respect to the $L^2$-norms onto a dense linear sub-space of $L^2(\Gamma)$, hence it can be extended uniquely to an isometry of $L^2(G)$ onto $L^2(\Gamma)$. This unique extension will be denoted by
$$ \mathcal{F}: L^2(G) \to L^2(\Gamma), \; \mathcal{F}(f) = \hat{f} \text{ for }  f \in (L^1(G) \cap L^2(G)),$$
and we will use the notation $\mathcal{F}(g):= \hat{g}$.\index{$\mathcal{F}$ (Fourier transform)}

The aforementioned isomorphism is at the level of Hilbert space, thus is also respects the inner product on $L^2(G)$ and $L^2(\Gamma)$, a result known as the Parseval formula,
$$\langle f,g \rangle_{L^2(G)} = \int_{G} f(g) g(x)^* dx = \int_{\Gamma} \hat{f}(\gamma) (\hat{g}(\gamma))^* d \gamma = \langle \hat{f}, \hat{g} \rangle_{L^2(\Gamma)}.$$

Also, there is a formula for convolution in this case, if $f,g \in L^2(G)$ then $\widehat{fg} = \hat{f} \ast \hat{g}$ (\citep[Theorem 4.37]{folland_course_2016}). Notice that $fg$ might not be in $L^2(G)$ and $\hat{f} \ast \hat{g}$ might not be in $L^2(G)$, however the previous statement makes sense because $fg \in L^1(G)$ by Holder's inequality, and $\hat{f} \ast \hat{g} \in L^1(G)$ by Young's inequality (\cref{proposition:useful_inequalities_of_Bochner_L_p_spaces}).
\end{theorem}

The Parseval theorem can be used to show that $A(\Gamma)$, which we defined in \cref{definition:Fourier_transform}, consists of all the convolutions $f_1 \ast f_2$ for $f_1, f_2  \in L^2(\Gamma)$ (\citep[Theorem 1.6.3]{rudin_fourier_1967}).

\section{Fourier analysis and orthonormal basis}\label{section:Fourier_analysis_and_ortonormal_basis}

Assume that $G$ is discrete, then $\hat{G}$ is compact and we have a Hilbert space isomorphism (\cref{theorem:Plancherel_theorem})
$$ \mathcal{F}: l^2(G) \to L^2(\hat{G}), $$
moreover, the Kronecker delta function $\delta_g : G \to \mathbb{C}, \; \delta_g(x) =  \delta_{g,x}$ is an ortonormal basis of $l^2(G)$, therefore, $\{ \widehat{ \delta_g} \}_{g \in G}$ is an ortonormal basis of $L^2(\hat{G})$. Under this basis $L^2(G)$ can be interpreted as a direct sum of Hilbert spaces by \cref{lemma:hilbert_space_as_direct_sums}. Also, if either $G$ of $\hat{G}$ are second countable then \cref{section:algebra_continuous_functions_locally_comp_space} tell us that $L^2(G)$ and $L^2(\hat{G})$ are separable, so if $\{ \xi_i \}_{i \in \mathbb{N}}$ is a complete orthonormal base of $L^2(G)$ then $\{ \widehat{\xi_i} \}_{i\in \mathbb{N}}$ is a complete ortonormal base of $L^2(\hat{G})$.

 Since the norm of $f \in L^1(G)$ is given by $\| f \| = \sum_{g \in G} |f(g)|$ and the sum of an uncountable set of positive numbers can have maximum a countable amount of non-zero terms (\citep{rspuzio_uncountable_2013}), we have that finite combinations of $\delta_g$ are dense in $L^1(G)$. From the properties of the Fourier transform on $L^1(G)$ (\cref{proposition:Fourier_transform_properties}) we know that $A(G)$ is dense in $C_0(\hat{G})$ and the Fourier transform is a norm decreasing map into $C(\hat{G})$, thus, the set of finite linear combinations of functions $\widehat{\delta_g}, \; g \in G$ are dense in $C(\hat{G})$. 
 
 Let us look at a couple of examples:
 \begin{itemize}
     \item \textbf{Torus:} The Plancherel theorem (\cref{theorem:Plancherel_theorem}) tell us that 
     $$\{\overrightarrow{\alpha}  \mapsto \exp(i \overrightarrow{n} \cdot \overrightarrow{\alpha}) \}_{\overrightarrow{n} \in \mathbb{Z}^d} $$
     is a complete ortonormal base of $L^2(\mathbb{T}^d)$, because $ \overrightarrow{\alpha} \mapsto  \exp(-i \overrightarrow{n} \cdot \overrightarrow{\alpha})$ is the Fourier transform of $\delta_{\overrightarrow{n}}$.
     
     For $f$ a function over $\mathbb{T}^d$ whose Fourier transform exists, for example, if $f \in L^2(\mathbb{T}^d) \cup L^1(\mathbb{T}^d)$, define
     $$ S_N(f)(\overrightarrow{\alpha}) = \sum_{n \in V_N} \exp\left(i \overrightarrow{\alpha} \cdot \overrightarrow{n} \right) \hat{f}(\overrightarrow{n}) $$
     with $V_N = [-N,\dotsc,N]^d$. The sequence $\{ S_N(f)\}_{N \in \mathbb{N}}$ is often called the Fourier series\index{Fourier series} of $f$. 
     
     If $f \in C(\mathbb{T}^d)$ then $f \in L^2(\mathbb{T}^d)$, thus the Fourier transform gives a function $\hat{f} \in L^2(\mathbb{Z}^d)$ such that $\mathcal{F}^{-1} (\hat{f}) = f$. Also, the isomorphism of Hilbert spaces from the Plancherel theorem tell us that 
     $$f(\overrightarrow{\alpha}) = \sum_{\overrightarrow{n} \in \mathbb{Z}^d} \exp \left(i \overrightarrow{\alpha} \cdot \overrightarrow{n} \right) \hat{f}(\overrightarrow{n})  = \lim_{N \to \infty} S_N (\overrightarrow{\alpha})$$
     as elements in $L^2(\mathbb{T}^d)$, and the ordering of the elements in the sum is non important (\cref{lemma:properties_orthonormal_basis_hilbert_spaces}). From the proof of completeness of $L^2(G)$ (\citep[Theorem 1.23]{driver_funcitonal_2020}) we know that is possible to recover $f$ almost everywhere from a subsequence of $\{ S_N\}_{N \in \mathbb{N}}$, thus, we can recover a function in $L^2(\mathbb{T}^d)$ from its Fourier series. In \cref{section:convergence_of_the_Fourier_series} we will look how we can recover functions from $C(\mathbb{T}^d)$ using its Fourier coefficients.
     
     Also, from the previous discussion we get that finite linear combinations of elements of the form $\exp(i \overrightarrow{n} \cdot \overrightarrow{\alpha})$, which are called \textbf{trigonometric polynomials}, are dense in $C(\mathbb{T}^d)$, this result is known as the Weierstrass theorem (\citep[Theorem 3.6]{zygmund_trigonometric_2003}) and can also be obtained as an application of the Stone-Weierstrass theorem (\citep[Corollary 2.36]{allan_introduction_2011}.
     
     Additionally, since the involution commutes with integration for $\mathbb{C}$ valued functions, we have that
     $$ \widehat{f^*}(s) = (\widehat{f}(-s))^*, \; f \in L^1(\mathbb{T}^d), \; s \in \mathbb{Z}^d$$
     thus, the convolution formula for the Fourier transform along with the isomorphism from the Plancherel theorem (\cref{theorem:Plancherel_theorem}) tell us that
     $$ \widehat{f f^*}(0) = \sum_{s \in \mathbb{Z}^d} \widehat{f}(s)(\widehat{f}(s))^* = \| f \|_{L^2(\mathbb{T}^d)}. $$
     Notice that $\widehat{f f^*}(0) = 0$ iff $f = 0$, because the integral of positive function ($|f|^2$) is zero iff it is zero almost everywhere (\citep{2974242}), which translates in this context in $f = 0$. 
     
     \item \textbf{Reals:} $\mathbb{R}$ is a second countable LCA, therefore $H=L^2(\mathbb{R})$ has an ortonormal basis and is separable. There is not a unique orthonormal basis of $H$, because any linearly independent set of $H$ generates an orthonormal basis of $H$ as explained in \cref{chapter:hilbert_spaces_section}, however, we are interested in a basis the related well with the Fourier transform. This basis is the set of Hermite polynomials, which are defined through the formula $F_n(x)=\left(x-\frac{\mathrm{d}}{\mathrm{d} x}\right)^n \exp(-x^2 / 2)$ as $H_n=F_n /\left\|F_n\right\|_2$, and form an ortonormal set of functions on $L^2(\mathbb{R})$ (\citep[Proposition 2.88]{allan_introduction_2011}), moreover, $H_n$ are eigen-vectors of the Fourier transform i.e $\mathcal{F}(H_n) = (-i)^n H_n$.
     
     Let $\mathcal{H} = \{ p(x) \exp(-x^2 / 2) : \; p \text{ a polynomial } \}$, then $\mathcal{H}$ is geenrated by the Hermite polynomials. The set $\mathcal{H}$ is called the Hermite functions and is dense in $C_0(\mathbb{R})$ (\citep[Proposition 2.88]{allan_introduction_2011}) and in $L^p(\mathbb{R})$ for $1 \leq p < \infty$ (\citep[Theorem 2.92]{allan_introduction_2011}). The Fourier transform over $\mathbb{R}$ can be defined over $\mathcal{H}$ and extended to $L^2(\mathbb{R})$ by continuity (\citep[Section 2.19]{allan_introduction_2011}).
 \end{itemize}

\section{Convergence of the Fourier series}\label{section:convergence_of_the_Fourier_series}

It is a consequence of Carleson's theorem\index{Carleson's theorem} that 
$$\{ \sum_{n \in V_n} \exp{i \alpha \cdot n} \hat{f}(n) \}_{n \in \mathbb{Z}^d}$$
with $V_n = [-n, \cdots, n]^d$ converges to $f(\alpha)$ almost everywhere with $\alpha \in [0, 2\pi) ^d$ and $f \in L^2(\mathbb{T}^d)$ (\citep{fefferman_convergence_1971}), however, 
$$\{ \sum_{|n|\leq N, |m| \leq M} \exp{i (\alpha_1 n + \alpha_2 m)} \hat{f}(n,m) \}_{N,M \in \mathbb{N}}$$
does not in general converges to $f(\alpha_1, \alpha_2)$, actually, there is a function on $C(\mathbb{T}^2)$ such that the aforementioned sequence does not converge to $f(\alpha_1, \alpha_2)$ nowhere (\citep[Theorem 1]{fefferman_divergence_1971}). Notice that in the second sequence the rate of growth of $N$ and $M$ are not related, while on the first example they are the same, so, this change caused the sequence to diverge. This is one of many examples of the varied types of convergence that can be arise with trigonometric polynomials, since we are interested with C* algebras, that is, generalizations of algebras $C_0(X)$ we must look for a way to recover an element of $C(\mathbb{T}^d)$ using its Fourier coefficients.

First, we should mention that Carleson's theorem does not gives us uniform convergence of 
$$\{ \sum_{n \in V_n} \exp{i \overrightarrow{\alpha} \cdot \overrightarrow{n}} \hat{f}(\overrightarrow{n}) \}_{n \in \mathbb{Z}^d}$$
to $f$, just convergence almost everywhere, actually, there are continuous functions whose Fourier series diverges at a point (\citep[Chapter 2, theorem 2.1]{katznelson_introduction_2004}, \citep[chapter 3]{garcia_f_convergence_2018}). This undesired behaviour of the Fourier series comes from the properties of the Dirichlet kernel, which in one dimension takes the form
$$ D_N (t) = \sum_{k = -N } ^{n} \exp(ikt) = \frac{\sin((N + 1/2)t)}{\sin(t/2)}$$
and provide us with the Fourier series of $f$ by means of a convolution (\citep[Lemma 2.81]{allan_introduction_2011})
$$ S_N (f)(t) = \sum_{k = -N}^{N} \hat{f}(k)\exp(ikt) = (D_N \ast f)(t).$$
The Dirichlet kernel is a sequence of functions with unbounded $L^1$ norm and take both positive and negative values (\citep[page 92]{allan_introduction_2011}), and the  these traits makes them hard to study from a uniform convergence point of view (\citep[Chapter 2, section 1]{katznelson_introduction_2004}). Fortunately, there are many methods of summation for series, in particular, the Ces\`aro summation\index{Ces\`aro summation} of the Fourier series gives us a way of approximating continuous functions under the uniform norm, and it takes the form of a convolution with the Fejér kernel\index{Fejér kernel}, this is called the Fej\'er summation\index{Fej\'er summation}. The Fejér kernel in one dimension takes the form (\citep[Lemma 2.83]{allan_introduction_2011})
$$K_N (t) = \frac{1}{N+1}\sum_{k = 0}^{N} D_k(t) = \sum_{k = -N}^{N} \left(1 - \frac{|k|}{N+1}\right) \exp(ikt),$$
and the convolution of a function with the Fejér kernel gives us a sequence of functions that converge uniformly to continuous functions, a result known as the Fejér theorem (\citep[Theorem 2.84]{allan_introduction_2011})
$$ \lim_{n \to \infty} \sup_{x \in \mathbb{T}} \{ |f(x) - (K_n \ast f)(x) | \} =, 0$$
that is,
$$  \lim_{n \to \infty} \sup_{\alpha \in [0 ,2 \pi]} \{ |f(\exp(i \alpha)) - \sum_{k = -N}^{N} \left(1 - \frac{|k|}{N+1}\right) \hat{f}(k) \exp(ik\alpha) | \} = 0.  $$

The Fejér kernel is special because it is a positive normalized kernel, that is, it has the following properties ((\citep[page 12]{katznelson_introduction_2004})
\begin{itemize}
    \item \textbf{Positive:}  $K_N(\theta) \geq 0$ for all $\theta \in \mathbb{T}$.
    \item \textbf{Normalized:} $\int_{\mathbb{T}} K_N(\theta) d \theta = 1$.
    \item \textbf{kernel:} for all $0 < \delta < \pi $,
    $$ \lim_{n \to \infty} \int_{\delta}^{2 \pi - \delta} | K_n(t) | dt = 0  $$
\end{itemize}
Sequences of functions having those properties are also called summability kernels and also provide uniform convergence on continuous functions (\citep[Chapter 1, theorem 2.11]{katznelson_introduction_2004}). Furthermore, the Fejér kernel is useful to study the convergence of the Fourier series in other settings (\citep[Chapter 3, section 3]{zygmund_trigonometric_2003}).

As the Haar measure in the direct sum of groups takes the form of the product measure, the Dirichlet kernel and the Fourier Kernel in multiple dimensions take the form of multiplication of the kernels in one dimension (\citep{travaglini_fejer_1994}),
$$ D_N (t_1, t_2, \dotsc, t_d) = D_N(t_1) \cdot D_N(t_2 ) \cdot \dotsc \cdot D_N(t_d), \; t_i \in \mathbb{T}, \; (t_1, \dotsc, t_d \in [0,2 \pi)), $$
$$K_N (t_1, t_2, \dotsc, t_d) = K_N(t_1) \cdot K_N(t_2 ) \cdot \dotsc \cdot K_N(t_d), \; t_i \in \mathbb{T}, \; (t_1, \dotsc, t_d \in [0,2 \pi)),  $$
or explicitly, if $V_N = \{(s_1, \dotsc, s_d) \in \mathbb{Z}^d : \; |s_j| \leq N \}$ then
$$ D_N(t_1, \dotsc, t_d) = \sum_{s \in V_N} \exp\left(i \sum_{1 \leq j \leq d}s_j t_j \right) $$
$$ K_N(t_1, \dotsc, t_d) = \sum_{s \in V_N} \left( \prod_{1 \leq j \leq d} \left( 1 - \frac{|s_j|}{N+1} \right) \right)  \exp\left(i \sum_{1 \leq j \leq d}s_j t_j \right).$$
The Fejér kernel in $d$ dimensions is still a positive normalized kernel, and the convolution $(K_N \ast f)(t_1, t_2, \dotsc, t_d)$ converges uniformly to $f$ for $f \in C(\mathbb{T}^d)$ (\citep[Theorem 1]{baiborodov_approximation_1984}), notice that the convolution $(K_N \ast f)(t_1, t_2, \dotsc, t_d)$ corresponds to applying the Fej\'er summation to the function $f$.

\section{Fourier transform and Fréchet algebras}\label{section:Fourier_transform_and_Frechet_algebras}

The Fourier transform on $\mathbb{R}^d$ becomes a continuous automorphism over the Schwartz-algebra (\citep[Theorem 1.19]{bahouri_fourier_2011}), that is, the algebra of infinitely differentiable functions with decay greater that any polynomial, so, denote by $\mathcal{S}(\mathbb{R}^d)$ the Schwartz-algebra, then
$$ \mathcal{S}(\mathbb{R}^d) = \{ f \in C^{\infty}(\mathbb{R}^d ; \mathbb{C}) \;: \sup _{\substack{|\alpha| \leq k \\ x \in \mathbb{R}^d}}(1+|x|)^k\left|\partial^\alpha u(x)\right|<\infty \; \}$$ \index{$\mathcal{S}(\mathbb{R}^d)$}
with $|\alpha| \stackrel{\text { def }}{=} \alpha_1+\cdots+\alpha_d$, $\partial^\alpha f \stackrel{\text { def }}{=} \partial_1^{\alpha_1} \cdots \partial_d^{\alpha_d} f$ and $x^\alpha \stackrel{\text { def }}{=}$ $x^{\alpha_1} \cdots x^{\alpha_d}$.
$\mathcal{S}(\mathbb{R}^d)$ is a Fréchet *-algebra under the family of semi-norms 
$$
\|u\|_{k, \mathcal{S}} \stackrel{\text { def }}{=} \sup _{\substack{|\alpha| \leq k \\ x \in \mathbb{R}^d}}(1+|x|)^k\left|\partial^\alpha u(x)\right|<\infty ,
$$
moreover, the topology of $\mathcal{S}(\mathbb{R}^d)$ can be provided by various sets of semi-norms (\citep[Application II, section 3]{kantorovitz_introduction_2003}). In \cref{section:Fourier_analysis_and_ortonormal_basis} we provided an orthonormal basis for $L^2(\mathbb{R})$ that takes the form of the Hermite polynomials, each member of that basis takes the form $p(x) \exp{- x^2 /2}$, since the exponential function $x \mapsto \exp(-x^2 /2)$ decays faster than any polynomial can grow, the Hermite functions also belong to $\mathcal{S}(\mathbb{R})$. Also, $\mathcal{S}(\mathbb{R}^d)$ is a sub-algebra of $L^p(\mathbb{R})$ for $p \in [1, \infty]$, and the inclusion is a continuous map (\citep[Section 2.4.c]{hytonen_analysis_2016}).

$\mathbb{R}^d$ is locally compact, thus the semi-norms of $\mathcal{S}(\mathbb{R}^d)$ are devised to ensure that the functions decay fast enough such that any function of $\mathcal{S}(\mathbb{R}^d)$ also belong to any $L^p(\mathbb{R}^d)$, however, if we deal with $\mathbb{T}^d$ we do not need to set that requirement because if $f \in C(\mathbb{T})$ then $f \in L^p(\mathbb{T}^d)$ for $p \in [1, \infty]$. Therefore, the equivalent of the Schwartz algebra over $\mathbb{T}$ is the algebra of smooth functions $C^{\infty}(\mathbb{T})$\index{$C^{\infty}(\mathbb{T})$} provided with the set of semi-norms
$$ p_k(f) = \sup_{x \in \mathbb{T}} \{ | f^{(k)}(x) \}, $$ 
which turns $C^{\infty}(\mathbb{T})$ into a Fréchet algebra (\citep[Section I.5, Definition 2]{sugiura_unitary_1990}) with multiplication given by point-wise multiplications. The image of $C^{\infty}(\mathbb{T})$ under the Fourier transform is the algebra of rapidly decreasing sequences over $\mathbb{Z}$, which is denoted by 
$$\mathscr{S}(\mathbb{Z})=\left\{\varphi: \mathbb{Z} \rightarrow \mathbb{C} \mid \forall k \in \mathbb{N}, \; \exists K > 0 \text{ s.t. }  |n|^k |\varphi(n)| \leq K \quad\right\},$$\index{$\mathscr{S}(\mathbb{Z})$} 
and becomes a Fréchet algebra under the set of semi-norms (\citep[Section I.5, Definition 3]{sugiura_unitary_1990})
$$ q_k( \varphi) = \sup_{n \in \mathbb{Z}} |n^k \varphi(n)|,$$
and multiplication given by convolution. The Fourier transform becomes an isomorphism of Fréchet algebras between $C^{\infty}(\mathbb{T})$ and $\mathscr{S}(\mathbb{Z})$ (\citep[Section I.5,Theorem 5.3]{sugiura_unitary_1990}), and this isomorphism will be the motivation for the definition of dense Fréchet algebras inside twisted crossed products with $\mathbb{Z}^d$ in \cref{sec:derivations_twistted_crossed_product}. Notice that $C^{\infty}(\mathbb{T})$ is dense inside $C(\mathbb{T})$, because it contains all the trigonometric polynomials, which we know to be dense in $C(\mathbb{T})$. The algebra $C^{\infty}(\mathbb{T})$ is usually referred to as the \textit{Schwartz algebra of $\mathbb{T}$}, which is why it is denoted by an $\mathcal{S}$. There are various equivalent definitions of $\mathscr{S}(\mathbb{Z})$, for example, in \cref{proposition:characterization_smooth_sub_algebra_twsited_crossed_product} and \cref{remark:alternative_descriptions_of_smooth_sub_algebra} you can look a two of them. 

Also, if $f' \in C^{k}(\mathbb{T})$, then $\widehat{f^{(i)}} \in L^1(\mathbb{Z})$ for $0 \leq i \leq k-1$ and $\widehat{f^{(i)}}(n) = (in)^k \hat{f}(n)$ for all $n \in \mathbb{Z}$, moreover, $S_N(f^{(i)}) \to f^{(i)}$ for $0 \leq i \leq k-1$ in $C(\mathbb{T)}$ (\citep[Section I.5, Corollary of Theorem 5.1]{sugiura_unitary_1990}), we will come back to this fact in \cref{sec:derivations_twistted_crossed_product}.

The aforementioned results can be generalized into any connected compact Lie group (\citep[SectionII.8]{sugiura_unitary_1990}), via the study of its unitary representations and its Lie algebra, so, under this setting, we are working with a fairly simple group $\mathbb{T}$ with a Lie algebra $\mathbb{R}$ that has a trivial Lie bracket. 

The previous results can be generalized into $C(\mathbb{T}^d)$, such that,
\begin{itemize}
    \item $C^{\infty}(\mathbb{T}^d)$\index{$C^{\infty}(\mathbb{T}^d)$} is a Fréchet algebra under the set of seminorms
    $$ p_{k}(f) = \sup_{x \in \mathbb{T}^d} \left| \frac{\partial f}{\partial^{k_1} x_1, \cdots \partial^{k_d} x_d } \right|, \; k \in \mathbb{N}^d.$$
    \item the Fourier transform gives an isomorphism between the Fréchet algebra $C^{\infty}(\mathbb{T}^d)$ and the Fréchet algebra $\mathscr{S}(\mathbb{Z}^d)$, where $\mathscr{S}(\mathbb{Z}^d)$ the algebra of rapidly decreasing sequences over $\mathbb{Z}^d$ i.e.
    $$ \mathscr{S}(\mathbb{Z}^d) = \{ \phi: \mathbb{Z}^d \to \mathbb{C} | \forall k \in \mathbb{N}^d, \; \exists K >0 \textbf{ s.t. } |n|^k |\phi(k)| \leq K \}, \; |n|^k := |n_1|^{k_1} \cdots |n_d|^{k_d}, $$
    and its topology is given by the set of seminorms 
    $$ q_k(\phi) = \sup_{n \in \mathbb{Z}^d} |n|^k |\phi(n)|, \; k \in \mathbb{Z}^d, \; k \in \mathbb{N}^d.$$
    \item Over $C^{\infty}(\mathbb{T}^d)$ the maps 
    $$\partial_j : C^{\infty}(\mathbb{T}^d) \to C^{\infty}(\mathbb{T}^d), \; \partial_j (f) := \frac{\partial f}{ \partial x_j} $$
    are continuous, and
    $$ \widehat{\partial_j f}(n) = in_j \widehat{f}(n).  $$
    \item The topology of $C^{\infty}(\mathbb{T}^d)$ is generated by the set of seminorms
    $$
    p_n(f):= \sup_{x \in \mathbb{T}^d} |f(x)| +\sum_{k=1}^n \frac{1}{k!} \sum_{i_1, i_2, \ldots, i_k=1}^d \sup_{x \in \mathbb{T}^d} | \partial_{i_1} \partial_{i_2} \cdots \partial_{i_k} f |, \; n \geq 1, 
    $$ 
    where $\left\{\partial_{i_1}, \partial_{i_2}, \ldots, \partial_{i_n}\right\}$ be an ordered $n$-tuple from $\left\{\partial_1, \partial_2, \ldots, \partial_d\right\}$ and the sum 
    $$ \sum_{i_1, i_2, \ldots, i_k=1}^d \left\|\partial_{i_1} \partial_{i_2} \cdots \partial_{i_k} f\right\| $$
    runs over all the ordered $k$-tuple $\left\{\partial_{i_1}, \partial_{i_2}, \ldots, \partial_{i_k}\right\}$. In this case, $C^{\infty}(\mathbb{T}^d)$ is a smooth sub algebra of $C(\mathbb{T}^d)$ (\citep[Example 1.6]{bhatt_class_2013}).

    \item Notice that $ \alpha \mapsto \exp(i \overrightarrow{n} \cdot \overrightarrow{\alpha})$ belongs to $C^{\infty}(\mathbb{T}^d)$ for any $n \in \mathbb{Z}^d$, therefore, the algebra of trigonometric polynomials is a sub algebra of $C^{\infty}(\mathbb{T}^d)$.
\end{itemize}

\section{Fourier transform on $L^2(G;H)$ and $L^1(G;A)$}\label{section:Fourier_tranform_on_banach_valued_functions}

Let $G$ be a LCA and $H$ be a Hilbert space, then from \cref{proposition:tensor_product_and_L_2_functs} we know that $L^2(G) \otimes H \simeq L^2(G;H)$, so, since $\mathcal{F}: L^2(G) \to L^2(\hat{G})$ is a Hilbert space isomosphism (\cref{theorem:Plancherel_theorem}) then from \cref{proposition:opeators_tensor_product_hilbert_spaces} we can extend it to the following Hilbert space isomorphism
$$ \mathcal{F}\otimes Id_H : L^2(G;H) \to L^2(\hat{G};H) ,$$
which takes the following form on simple tensors 
$$ \mathcal{F}\otimes Id_H (f \otimes h) = \mathcal{F}(f) \otimes h. $$
Notice that $\mathcal{F}\otimes Id_H$ preserves the inner product, because $\mathcal{F}$ preserves the inner product, which is the reason for $ \mathcal{F}\otimes Id_H$ to be an isomorphism. 

Let $A$ be a C* algebra with $\pi_A:A \to B(H_A)$ a faithful representation and $G$ a second countable LCA, then from \cref{section:algebra_continuous_functions_locally_comp_space} we know that $C_0(G;A)$ has a faithful representation over $L^2(G;H)$ given by the multiplication operator i.e.
$$ \pi: C_0(G;A) \to B(L^2(G;H_A)), \; \pi(f)(r)(g) = \pi_A(f(g))(r(g)),$$
with $\; r \in L^2(G;H), \; g \in G, \; f \in C_0(G;A)$.
We can use the isomorphism $\mathcal{F}\otimes Id_{H_A}$ to interpret $C_0(G;A)$ also as an algebra of operators on $L^2(\hat{G};A)$ as follows
$$\hat{\pi} : C_0(G;A) \to B(L^2(\hat{G}; H_A)), \;  \hat{\pi}(f):= ((\mathcal{F}\otimes Id_{H_A}) \circ \pi(f) \circ (\mathcal{F}^{-1}\otimes Id_{H_A})).$$

The aforementioned generalization of the Fourier transform is special, that is, we need $H$ to be a Hilbert space, because is not the case that $\mathcal{F}\otimes Id_A$ can be extended into a bounded operator from $L^2(G;A)$ to $L^2(\hat{G};A)$ for $A$ an arbitrary Banach space, for example, $\mathscr{F} \otimes I_{L^1(\mathbb{T})}$ does not extend to a bounded operator from $L^2\left(\mathbb{T} ; L^1(\mathbb{T})\right)$ to $\ell^2\left(\mathbb{Z} ; L^1(\mathbb{T})\right)$ (\citep[Example 2.7.4]{hytonen_analysis_2016}), there are similar counter examples with $G = \mathbb{R}$ (\citep[Example 2.1.15]{hytonen_analysis_2016}). 

Let's look into an extension of the operator 
$$\mathcal{F} \otimes Id_A : L^1(G)\odot A \to C_0(\hat{G}) \odot A, \; (\mathcal{F} \otimes Id_A)(f \otimes a) = \mathcal{F}(f) \otimes a,$$
when $A$ is a Banach space, and we will later check that it coincides with $\mathcal{F} \otimes Id_H$ when $H$ is a Hilbert space. Take $A$ be a Banach space and $\gamma \in \hat{G}$, we know that $\gamma$ is a continuous function over $G$, thus if $f \in L^1(G;A)$ then from \cref{lemma:L_p_bochner_spaces_for_locally_compact_hausdroff_spaces} we know that $f \gamma$ is $\mu$-strongly measurable. Define
$$ \widetilde{f}(\gamma) = \int_{G} f(x) \gamma(-x) dx, $$
since
$$\| \widetilde{f}(\gamma) \| \leq \int_{G} \|f(x)\| |\gamma(-x)| dx \leq \| f \|_1,$$
then $\widetilde{f}(\gamma)$ exists and belongs to $A$ (\cref{prop:bochn_integra_condition}), moreover $\widetilde{f}: \hat{G} \to A$ is a bounded function. From \cref{theprem:Lp_approximation_simple_functions} we know that $L^1(G)\odot A$ is dense in $L^1(G;A)$, moreover, $C_c(G) \odot A$ is dense in $L^p(G;A)$ for $1 \leq p < \infty$ (\cref{lemma:L_p_bochner_spaces_for_locally_compact_hausdroff_spaces}). For $n \in \mathbb{N}$ take $f_n = \sum_{i\leq m(n)} f_{i,n} \otimes a_{i,n}$ with $f_{i,n} \in C_c(G)$ and $a_{i,n} \in A$ such that $\| f_n - f \|_1 \leq 1/n$, then $\{ f_n \}_{n \in \mathbb{N}}$ is a sequence such that
\begin{itemize}
    \item $\widetilde{f_n} = \sum_{i\leq m(n)}  \mathcal{F}(f_{i,n}) \otimes a_{i,n}$, thus if $A$ is a Hilbert space then $\widetilde{f_n} = (\mathcal{F}\otimes Id_A)(f_n)$.
    \item Since $\mathcal{F}(f_{i,n}) \in C_0(\hat{G})$ (\cref{proposition:Fourier_transform_properties}) then $\sum_{i\leq m(n)}  \mathcal{F}f_{i,n} \otimes a_{i,n} \in C_0(\hat{G})\odot A$, thus $\widetilde{f_n} \in C_0(\hat{G};A)$.
    \item $\|\widetilde{f_n}(g) - \widetilde{f}(g) \| \leq \| \widetilde{f_n -f} \|_1 \leq 1/n$, thus $\widetilde{f_n} \to \widetilde{f}$ under the supremum norm over the vector space of functions over $\hat{G}$ taking values in $A$.
\end{itemize}

By the previous computations $\{ \widetilde{f_n} \}_{n \in \mathbb{N}}$ is a Cauchy sequence under the supremum norm, and $\{ \widetilde{f_n} \}_{n \in\mathbb{N}} \subset C_{0}(\hat{G}) \odot A$, so, since $C_0(\hat{G}) \odot A$ is dense in $C_0(\hat{G};A)$ and the second is a Banach space under the supremum norm (\cref{sec:Continuous_functions_with_values_on_C_star_algebra}), then $\lim_{n \to \infty} \widetilde{f_n} \in C_0(\hat{G};A)$, that means $\widetilde{f} \in C_0(\hat{G};A)$. So, $f \mapsto \widetilde{f}$ is the unique norm decreasing extension of $\mathcal{F}$ to $L^1(G;A)$, thus, we will refer to it as $\mathcal{F}\otimes Id_A$, which by the way is defined through a Bochner integral. Consequently, $(\mathcal{F} \otimes Id_A)(L^1(G;A)) \subset C_0(\hat{G};A)$ and $(\mathcal{F} \otimes Id_A)(L^1(G;A))$ is dense in $C_0(\hat{G};A)$ as it happened for the Fourier transform taking values in $\mathbb{C}$ (\cref{proposition:Fourier_transform_properties}). If $G$ is $\mathbb{R}^d$ of $\mathbb{T}^d$ then other of results on Fourier analysis can be translated into $L^1(G;A)$, like the inversion the Fourier transform and the relation between convolution and the Fourier transform \citep[Section 2.4]{hytonen_analysis_2016}.

We have seen how $(\mathcal{F} \otimes Id_A)$ looks over $L^1(G;A)$, now, for completeness let's check that it coincides with $(\mathcal{F} \otimes Id_H)$ when $H$ is a Hilbert space. Recall that the definition of $(\mathcal{F} \otimes Id_H)$ when $H$ is a Hilbert space was given as the tensor product of Hilbert space bounded maps, thus, even though we know that it exists we do not know what particular forms it takes on arbitrary vectors, instead, we know that form is takes in simple tensors i.e. $g \otimes h$. On the other side, we know what is the particular form that $(\mathcal{F} \otimes Id_A)$ takes over $L^1(G;A)$ when $A$ is a Banach space, which is the form of a Bochner integral, therefore, it is worth checking that both definitions actually coincide, such that we can use them interchangeably.

Let $f \in L^1(G;H) \cap L^2(G;H)$, take $\{ f_n \}_{n \in \mathbb{N}}$ be a sequence of $\mu$-simple functions converging to $f$ almost everywhere and $\| f_n(g) \| \leq \| f(s) \|$ almost everywhere, as in \cref{remark:approximating_L_p_functions_by_step_functions}, then $f_n \to f$ in both $L^1(G;H), \; L^2(G;H)$. For $\mu$-simple functions we have that $\widetilde{f_n} = (\mathcal{F} \otimes Id_H)$, also, by definition of $(\mathcal{F} \otimes Id_H)$ we have that 
$$(\mathcal{F} \otimes Id_H)(f_n) \to (\mathcal{F} \otimes Id_H)(f)$$
in $L^2(\hat{G};H)$. Take $ \{(\mathcal{F} \otimes Id_H)(f_{n(j)}) \}_{j \in \mathbb{N}}$ converging to $(\mathcal{F} \otimes Id_H)(f)$ almost everywhere (\cref{proposition:useful_inequalities_of_Bochner_L_p_spaces}), then, $\widetilde{f_{n(j)}} \to \widetilde{f}$ over $C_0(\hat{G};H)$, meaning that $(\mathcal{F} \otimes Id_H)(f)$ is equal to $\widetilde{f}$ almost everywhere, consequently, we denote by $\mathcal{F} \otimes Id_H$ the unique extension of $\mathcal{F}$ to $L^1(G;H) \cap L^2(G;H)$ and we know that it takes the form of a Bochner integral.

Notice that translating results from scalar valued functions into vector valued functions is not straight forward, even for the simplest cases, for example, to best of our knowledge there is not a positive answer on whether the Fourier series converges almost everywhere for $f \in L^p(\mathbb{T};X)$ and $p > 1$ \citep[Open problem 4 in page 496]{hytonen_analysis_2016} (Carleson theorem).

Is also possible to translate resutls on Fréchet algebras into this context, for example,  \citep[Section 2.4.c]{hytonen_analysis_2016} has a great exposition for the case $G = \mathbb{R}^d$.

\section{Fourier transform on $C(\mathbb{T}^d,A)$ }\label{section:Fourier_tranform_C_star_valued_functions_over_the_torus}

Now we will look into the case $C(\mathbb{T}^d,A)$ for $A$ a C* algebra, as an application of \cref{section:Fourier_tranform_on_banach_valued_functions}, for simplicity, we go into the case $d=1$. Let $h \in L^2(\mathbb{T}) \otimes H_A$, then we know that it can be expressed as $h = \sum_{n \in \mathbb{Z}} e_n \otimes h_n$ with $h_n \in H_A$, under this representation we have
$$ (\mathcal{F} \otimes Id_{H_A})(h)(k) = h_k, $$
because $\mathcal{F}(e_k)(l) = \delta_{k,l}$. Therefore, 
$$S_N(h) = \sum_{|k| \leq N} e_i \otimes ((\mathcal{F} \otimes Id_{H_A})(h)(k)) \to h$$
in $L^2(\mathbb{T},H_A)$, as it happens in the case of scalar valued functions (\cref{section:faithful_representation_of_torus}).

Now that we know how $\mathcal{F} \otimes Id_H$ looks like let's dive into how this generalizes the results from \cref{section:faithful_representation_of_torus}. Let $A$ be a C* algebra and $\pi_A : A \to B(H_A)$ a faithful representation of $A$, fake $f \in C(\mathbb{T},A)$, then from \cref{section:algebra_continuous_functions_locally_comp_space} we know that there is a 
$$ \pi \otimes \pi_A : C(\mathbb{T};A) \to B(L^2(\mathbb{T}) \otimes H_A),$$
such that
$$ (\pi \otimes \pi_A)(f \otimes a)(g \otimes h) = (\pi(f)(g)) \otimes (\pi_A(a)(h)), \; f \in L^2(\mathbb{T}), h \in H_A, \; a \in A$$
and takes the equivalent form
$$ (\pi \otimes \pi_A)(f)(h)(x) = \pi_A(f(x))(h(x)), \; x \in \mathbb{T}, h \in L^2(\mathbb{T};H_A), \; f \in C(\mathbb{T},A). $$
From \cref{sec:tensor_product_of_bounded_opeartors_os_infinite_matrices} we know that $(\pi \otimes \pi_A)(f)$ has a matrix representation over $L^2(\mathbb{T}) \otimes H_A$ given the complete orthonormal basis $\{ \alpha \mapsto \exp(i n \alpha) \}_{n \in \mathbb{Z}}$ (\cref{section:Fourier_analysis_and_ortonormal_basis}). Let $\mathcal{T}$ be the *-algebra of trigonometric polynomials over $\mathbb{T}$, which we know that are dense in $C(\mathbb{T})$ from \cref{section:Fourier_analysis_and_ortonormal_basis}, thus, from \cref{sec:Continuous_functions_with_values_on_C_star_algebra} we can conclude that $\mathcal{T} \odot A$ is dense in $C(\mathbb{T};A)$, and \cref{lemma:tensor_products_and_generated_C_star_algebras} tell us that $C^*(\mathcal{T} \odot A) \simeq C(\mathbb{T}) \otimes A$. By the form that takes simple tensors, we know that 
$$\left( (\sum_{|k| \leq n} c_k (\alpha \mapsto \exp(i k \alpha))) \odot a \right) = \sum_{|k| \leq n} (\alpha \mapsto \exp(i k \alpha)) \odot (c_k a),$$
thus, the *-algebra generated by $\{ (\alpha \mapsto \exp(i n \alpha)) \odot a \}_{a \in A, n \in \mathbb{Z}}$ is dense in $C(\mathbb{T};A)$. Therefore, any element of $(\pi \otimes \pi_A)(C(\mathbb{T};A))$ can be approximated by elements of the form 
$$(\pi \otimes \pi_A) \left( (\sum_{|k| \leq n} (\alpha \mapsto \exp(i k \alpha)) \odot a\right) =   \sum_{|k| \leq n} \pi((\alpha \mapsto \exp(i k \alpha)) ) \otimes \pi_A(a) .$$

Denote by $e_k = (\alpha \mapsto \exp(i k \alpha))$, 
from \cref{proposition:properties_tensor_product_Hilbert_spaces} we know that $L^2(\mathbb{T}) \otimes H_A \simeq \sum_{n \in \mathbb{Z}} H_A$, also, from \cref{section:faithful_representation_of_torus} we have that $\pi(e_j)(e_k) = e_{k+j}$, thus

$$ \pi(e_j) \otimes \pi_A(a) (e_k\otimes h) = e_{k+j} \otimes \pi_A(a)(h), \; j,k \in \mathbb{Z}, \; h \in H_A,$$
which implies that, 
$$(\pi \otimes \pi_A)( e_j \otimes a) = [ \pi_A(a) \delta_{l, m + j} ]_{l,m \in \mathbb{Z}},$$
a diagonal matrix. Since convergence in the norm operator topology on $(\pi \otimes \pi_A)(C(\mathbb{T};A))$ implies uniform convergence of all the entries of the matrix, we get that the matrix representation of $(\pi \otimes \pi_A)(f)$ also consists of a matrix that is constant on diagonal i.e
$$(\pi \otimes \pi_A)(f) = [ \pi_A(a_{l-m}) ]_{l,m \in \mathbb{Z}},$$
for $\{ a_k\}_{k \in \mathbb{Z}} \subset A^{\mathbb{Z}}$, moreover, the elements $a_k$ come from the Fourier transform as in \cref{section:faithful_representation_of_torus}, which we now check.

Since $\mu$ is a left-invariant measure, then
$$ \int_{\mathbb{T}} K_n(y) d\mu(y)  = \int_{\mathbb{T}} K_n(x-y) d\mu(y),$$
for any $x \in \mathbb{T}$, thus
$$f(x) = \int_{\mathbb{T}} f(x) K_n(x-y) d \mu(y), $$
also,
$$ \left( \sum_{|k| \leq n}  \left(1 - \frac{|k|}{n+1}\right) e_k \otimes ((\mathcal{F} \otimes Id_{H_A})(f)(k) \right)(x) = (f \ast K_n)(x) = \int_{\mathbb{T}} f(y) K_n(x-y) d \mu(y).$$

Every continuous function between a compact metric space and a normed space is uniformly continuous (\citep{nlab:continuous_metric_space_valued_function_on_compact_metric_space_is_uniformly_continuous}), therefore, for $\epsilon > 0$ there is $\delta > 0$ such that $d(x,y) < \delta$ implies that $\| f(x) - f(y) \| \leq \epsilon/3$. Take $n$ such that 
$$\int_{\delta}^{2 \pi - \delta} |K_n (x)| d \mu (x) \leq \frac{\epsilon}{ 6 \max \{ 1, \| f\|_{C(\mathbb{T};A)} \} }, $$
let $I_{\delta} := [\delta, 2 \pi - \delta]$ and $J_{\delta} := \mathbb{T} / I_{\delta}$, then
$$  f(x) - (f \ast K_n)(x) =  \int_{\mathbb{T}} (f(x) - f(y))K_n(x-y) d \mu(y)$$
therefore,
$$ \| f(x) - (f \ast K_n)(x)\|$$
$$ \leq \left(\int_{x-I_{\delta}} \|f(x)-f(y)\| |K_n(x-y)| d \mu (y)\right) + \|\int_{x-J_{\delta}} (f(x) - f(y)) K_n(x-y) d \mu (y) \|, $$
$$ \leq 2 \|f\|_{C(\mathbb{T},A)}\left(\int_{x-I_{\delta}} |K_n(x-y)| d \mu (y)\right) + \int_{x-J_{\delta}} \|f(x) - f(y)\| |K_n(x-y) |d \mu (y) , $$
$$ \leq \frac{\epsilon}{3} + \int_{x-J_{\delta}} \frac{\epsilon}{3} |K_n(x-y) |d \mu (y) , $$
$$ \leq \frac{\epsilon}{3} +  \frac{\epsilon}{3} + \int_{\delta}^{2 \pi - \delta} |K_n (y)| d \mu (y) \leq \epsilon. $$

Since $(f \ast K_n) \to f$ in $C(\mathbb{T};A)$ then $(\pi \otimes \pi_A)(f \ast K_n) \to (\pi \otimes \pi_A)(f)$ in $B(L^2(\mathbb{T}) \otimes H_A)$, which tell us that
$$ (\pi \otimes \pi_A)(f) = [(\mathcal{F} \otimes Id_H)(f)(l-m) ]_{l,m \in \mathbb{Z}}. $$

The previous computations can be translated for $d > 1$, giving a generalization of the results exposed in \cref{section:n_torus}, that is, if $\{ K_n\}_{n \in \mathbb{N}}$ is the Fejér kernel in $d$ dimensions (\cref{section:convergence_of_the_Fourier_series}), we have that $(f \ast K_n) \to f$ in $C(\mathbb{T}^d;A)$ then $(\pi \otimes \pi_A)(f \ast K_n) \to (\pi \otimes \pi_A)(f)$ in $B(L^2(\mathbb{T}^d) \otimes H_A)$, which tell us that
$$ (\pi \otimes \pi_A)(f) = [(\mathcal{F} \otimes Id_H)(f)(l-m) ]_{l,m \in \mathbb{Z}^d}. $$

In \cref{sec:Fourier_analysis_twistted_crossed_product} we will use the Fejér kernel to provide us with a characterization of the elements of twisted crossed products with $\mathbb{Z}^d$, which will involve an argument similar to the one we have presented. Also, in \cref{sec:derivations_twistted_crossed_product} we will look into constructions involving Fréchet algebras similar to the ones in \cref{section:Fourier_transform_and_Frechet_algebras}, which provides an algebraic generalization of the derivations and Fréchet algebra inside $C(\mathbb{T})$, and allow to define a smooth sub algebra of $C(\mathbb{T}^d,A)$.

\chapter{More on K theory}
\label{chap:more_on_k_theory}

\section{K theory C* algebras}
\label{sec:more_on_k_theory_c_star_algebras}

\begin{definition}[c.f. Definition 4.2.1. \citep{blackadar_k-theory_2012} (Equivalente relations on idempotents)]\label{definition:equivalence_relations_idempotents}
Let $A$ be a C* algebra and $e$ and $f$ be idempotents in $A$, then, we introduce the following notation      
\begin{itemize}
    \item $e \sim f$ if there are $x, y \in A$ with $x y=e, y x=f$. (algebraic equivalence)\index{$\sim$ (algebraic equivalence)}
    \item $e \sim_s f$ if there is an invertible $z$ in $A^+$ with $z e z^{-1}=f$. (similarity)
    \item $e \sim_h f$ if there is a norm-continuous path of idempotents in $A$ from $e$ to $f$. (homotopy)
\end{itemize}
\end{definition}

In \citep[Corollary 4.2.3]{blackadar_k-theory_2012} it is proven that $\sim$ is an equivalence relation between idempotents of $A$. 

When we work with C* algebras we can establish the following equivalence relations,

\begin{definition}[c.f. Section 2.2 \citep{rordam_introduction_2000} (Equivalente relations on projections)]\label{definition:equivalence_relations_projections}
Let $A$ be a C* algebra and $p$ and $q$ projections in $A$, then, we introduce the following notation      
\begin{itemize}
    \item $p \sim q$ if there are $x \in A$ with $x x^*=p, \; x^* x=p$. (Murray-von Neumann equivalence)\index{$\sim$ (Murray-von Neumann equivalence)}
    \item $p \sim_u v$ if there is an unitary $u$ in $A^+$ with $u p u^{*}=q$. (unitary equivalence)
    \item $p \sim_h q$ if there is a norm-continuous path of projections in $A$ from $p$ to $q$. (homotopy)
\end{itemize}
\end{definition}

In the introduction of \citep[Section 2.2]{rordam_introduction_2000} it is proven that $\sim$ is an equivalence relation between projections of $A$. We will use the notation $\sim$ and $\sim_h$ for both idempotents and projections and their meaning will be described in the context of their usage.

The following result gives an important relation of invertibles and unitaries when we work with matrix algebras,

\begin{lemma}[Whitehead lemma (Proposition 3.4.1 \citep{blackadar_k-theory_2012}, Lemma 2.1.5 \citep{rordam_introduction_2000})]\label{lemma:Whitehead_lemma_c_star_alg}
Let $A$ be a unital Banach algebra, then, for $x,y \in G(A)$ we have that
$$
\left(\begin{array}{ll}
xy & 0 \\
 0 & 1
\end{array}\right) \sim_h \left(\begin{array}{ll}
x & 0 \\
0 & y
\end{array}\right) \;  \in G(M_2 (A))
$$
if $A$ is a unital C* algebra then this result translates into an homotopy inside $U(M_2 (A))$ if $x,y \in U(A)$.
\end{lemma}

According to \cref{lemma:Whitehead_lemma_c_star_alg} if $y = x^{-1}$ we are guarantying that the diagonal matrix $x \oplus y$ can be connected to the identity of $M_2(A)$, hence, due to \cref{theorem:description_of_UA} and  \cref{theorem: description_of_GA} it can be expressed as product of exponentials in $M_2(A)$. The Whitehead lemma plays an important role to establish the equivalence between the three equivalence relations of idempotents when one uses matrix algebras (\cref{proposition:equivalence_relations_on_idempotents}). 

\begin{proposition}[Equivalence relations on idempotents and projections]\label{proposition:equivalence_relations_on_idempotents}
Let $e,f \in Q(A)$ with $A$ a C* algebra, then:
\begin{enumerate}
    \item $e \sim_s f$ iff $e \sim f$ and $1_{A^+} - e \sim 1_{A^+} - f$ \citep[Proposition 4.2.5]{blackadar_k-theory_2012}.
    \item If $e \sim_h f$ in $Q(A)$, then $e \sim_s f$ by  \cref{proposition:canonical_paths_between_idempotents}.
    \item  If $e \sim f$, then $\left[\begin{array}{ll}e & 0 \\ 0 & 0\end{array}\right] \sim_s\left[\begin{array}{ll}f & 0 \\ 0 & 0\end{array}\right]$ in $M_2(A)$ \citep[Proposition 4.3.1]{blackadar_k-theory_2012}.
    \item If $e \sim_u f$, then $\left[\begin{array}{ll}e & 0 \\ 0 & 0\end{array}\right] \sim_h \left[\begin{array}{ll}f & 0 \\ 0 & 0\end{array}\right]$ in $M_2(A)$ \citep[Proposition 4.4.1]{blackadar_k-theory_2012}.
\end{enumerate}

Let $p,q \in P(A)$ with $A$ a C* algebra, then:
\begin{enumerate}
    \item $p \sim_u q$ iff $p \sim q$ and $1_{A^+} - e \sim 1_{A^+} - f$ \citep[Proposition 2.2.2]{rordam_introduction_2000}.
    \item If $p \sim_h q$, in $P(A)$ then $p \sim_u q$ by \cref{proposition:canonical_paths_between_projections}.
    \item  If $p \sim q$, then $\left[\begin{array}{ll}p & 0 \\ 0 & 0\end{array}\right] \sim_u\left[\begin{array}{ll}q & 0 \\ 0 & 0\end{array}\right]$ in $M_2(A)$ \citep[Proposition 2.2.t]{rordam_introduction_2000}.
    \item If $p \sim_u q$, then $\left[\begin{array}{ll}p & 0 \\ 0 & 0\end{array}\right] \sim_h \left[\begin{array}{ll}q & 0 \\ 0 & 0\end{array}\right]$ in $M_2(A)$ \citep[Proposition 2.2.t]{rordam_introduction_2000}.
\end{enumerate}
\end{proposition}

From \cref{proposition:equivalence_relations_on_idempotents} we know that the equivalence relations on idempotents
(\cref{definition:equivalence_relations_idempotents}) and the equivalence relations on projections (\cref{definition:equivalence_relations_projections}) become equivalent when we move into matrix algebras, that is why the group $K_0(A)$ can be defined using the equivalence relation $\sim$ (\citep[Chapter 6]{wegge-olsen_k-theory_1993}).

\subsection{Alternative proofs for the isomorphisms $K_j(A) \simeq K_j(\mathcal{A}), \; j=0,1$}
\label{sec:alternative_proofs_for_the_isomorphism_of_K_groups}

In the following, we present a little discussion on the proofs we found in the literature for the isomorphisms $K_0(\mathcal{A}) \simeq K_0(A)$, $K_1(\mathcal{A}) \simeq K_1(A)$ (\cref{theorem:isomorphism_K_0_and_k_1_for_smooth_sub_algebras}), where $A$ is an unital C* algebra and $\mathcal{A}$ is a smooth sub algebra of $A$.

\subsubsection{Path proposed by Connes}
\label{sec:path_proposed_by_Connes}

Most of the literature we reviewed take this result from \citep[Appendix C. Stability under Holomorphic Functional Calculus]{connes_noncommutative_2014}, more specifically from Proposition 3 of that appendix. Connes states that this is a consequence of the density theorem, for which he gives two references:
\begin{itemize}
    \item The book of K theory by Micheal Atiyah \citep{atiyah_k-theory_2018}, where the only reference to the density theorem we could find is in Lemma A9, which states:\\
    
    \begin{mdframed}
    LEMMA A9. Let $\pi: L \rightarrow M$ be a continuous linear map of Banach spaces with $\pi(L)$ dense in $M$ and let $U$ be an open set in $M$. Then, for any compact $X$
    $$
    \left[\mathrm{X}, \pi^{-1}(\mathrm{U})\right] \rightarrow[\mathrm{X}, \mathrm{U}]
    $$
    is bijective.\\
    \end{mdframed}

     So, if $\mathcal{A}$ were to be a Banach algebra we could apply this to establish an isomorphism between then the group of connected components of $G(M_n(\mathcal{A}))$ and the group of connected components of $G(M_n(A))$, because $G(M_n(A))$ is open in $M_n(A)$, and this provides us with the isomorphism $K_1(\mathcal{A}) \simeq K_1(A)$, unfortunately, $\mathcal{A}$ is a Fréchet algebra and not a Banach algebra. The proof of that Lemma relies on the properties of the Banach spaces $L^X$ and $M^X$ and the existence of partitions of unitities subordinate to finite open covers of $X$, so, we do not discard that the proof of Lemma A.9 can be generalized into the case where $L$ is a Fréchet space, but we currently do not know how to tackle that problem.
     
      Moreover, if \citep[Lemma A.9]{atiyah_k-theory_2018} were to hold for $\mathcal{A}$ a Fréchet algebra we could also use it to proof the isomorphism $K_0(\mathcal{A}) \simeq K_0(A)$ as follows: from \citep[Lemma 3.43]{gracia-bondia_elements_2001} is possible to show that there is an open set $Q_{U}(\mathcal{A})$ ($Q_U(A)$) of $\mathcal{A}$ ($A$), with the property $Q_{U}(\mathcal{A}) = i^{-1}(Q_{U}(A))$, and also it is a deformation retract of the set of idempotents of $\mathcal{A}$ ($A$). So, the semigroups $Q_{\infty}(\mathcal{A})$ and $Q_{\infty}(A)$ would be isomorphic, which leads to $K_0(\mathcal{A}) \simeq K_0(A)$.
     
     \item The second reference provided by Connes comes from the book of K theory by Max Karoubi \citep{karoubi_k-theory_1978}, where the only mention to the density theorem we could find is in problem 6.15, which states a similar result that \citep[Lemma A.9]{atiyah_k-theory_2018} and takes it one step further into inductive limits of Banach algebras. This results would be helpful if instead of inductive limits of Banach algebras it was stated for projective limits of Banach algebras, since any Fréchet m-convex algebra is isomorphic to a projective limit of Banach algebras (\cref{remark:other_formulations_holom_functional_calculus}), and Fréchet m-convex algebras encompass a lot of really interesting smooth sub algebras (\cref{sec:Frechet_d_infinity_subalgebras}). 
\end{itemize}

\subsubsection{Path proposed by Gracia-Bondia et. al.}
\label{sec:path_proposed_by_Gracia_Bondia}

The second reference we could find to the isomorphism $K_0(\mathcal{A}) \simeq K_0(A)$ is given in \citep[Theorem 3.44]{gracia-bondia_elements_2001}. As a preamble to the proof of such isomorphism, the holomorphic functional calculus of a smooth sub algebra is used to provide homotopies between elements of $\mathcal{A}$ (\citep[Lemma 3.43]{gracia-bondia_elements_2001}), however, as far as we know, no mention of such fact is stated in the book, neither the fact that the holomorphic calculus on $\mathcal{A}$ shares similar properties to the functional calculus on $A$. In \cref{sec:smooth_sub_algebras_functional_calculus} there is a discussion on those facts and we believe them to be a key fact if smooth sub algebras come into play.
    
On the other side, the proof of  \citep[Theorem 3.44]{gracia-bondia_elements_2001} relays of the homotopy type of open subsets of Fréchet spaces, which are stated to have the same homotopy type of CW-complexes according to results from \citep{milnor_spaces_1959}. In our opinion that is a very clever insight, unfortunately, we were not able to understand how the results from \citep{milnor_spaces_1959} give us a characterization of the homotopy of an open subset of Fréchet spaces, thus, we refrain from continuing towards that path.

\subsubsection{Path proposed by Bhatt et. al.}
\label{sec:path_proposed_by_phatt_at_al}

The third and last reference to those isomorphisms we found it in \citep[Theorem 3.1]{sj_bhatt_spectral_nodate}, which provides a proof for the isomorphisms $K_j(\mathcal{A}) \simeq K_j(A), \; j=0,1$ that takes into account the topology of both $\mathcal{A}$ and $A$ quite similarly to the approach of this document, more specifically, it implicitly uses of the functional calculus inside $\mathcal{A}$ and $A$ as follows, to proof the isomorphism $K_1(\mathcal{A}) \simeq K_1 (A)$ it builds on the isomorphism $U(M_n(\mathcal{A}))/ U_0 (M_n(\mathcal{A})) \simeq U(M_n(A))/ U_0(M_n(A))$, which happens to be a consequence of \cref{proposition:invertibles_and_untairies_of_smooth_algebras_are_dense} and \cref{proposition:cannonical_paths_invertibles_unitaries_smooth_sub_algebras}. To tackle the isomorphism $K_0(\mathcal{A}) \simeq K_0(A)$ it uses the Bott periodicity at the level of $\mathcal{A}$ and $A$, which is a great example on how various of the properties of the K theory of C* algebras translate into the K theory of smooth sub algebras, we didn't follow this path because one of our objectives is to explicitly provide the smooth paths between projections, such that, those can be used in the proof of the pairing between the K theory of $A$ and the continuous cyclic cohomology of $\mathcal{A}$ (\cref{proposition:pairing_topological_k_0_even_ciclyc_cohomology}).

\section{K theory for Banach algebras and generalizations}
\label{sec:k_theory_for_algebras_similar_to_Banach_algebras}

From \cref{remark:K_1_equivalence_of_invertibles} we know that, if $A$ is a unital C* algebra then 
$$K_1(A) = G_{\infty} (A) / \sim_{h}.$$
In arbitrary Banach algebras over $\mathbb{C}$ there is no involution, but there are invertible elements, thus $K_1(A) = G_{\infty} (A) / \sim_{h}$ is a good assignment for arbitrary Banach algebras. This is an abelian group and has many of the properties of $K_1$ in C* algebras \citep[Chapter 8]{blackadar_k-theory_2012}.

For $K_0(A)$ is necessary to turn into the set of idempotent elements, thus, for an arbitrary Banach algebra we define
$$
K_0(A) = Q_{\infty}(A) / \sim_{h}.
$$
This definition coincides with the definition we gave for C* algebras because $P_0(A)$ is a deformation retract of $Q_0(A)$ for any C* algebra (\cref{proposition:P_A_retraction_of_Q_A}). Moreover, the definition of $K_0(A)$ in term of homotopies of idempotents is equivalent to an algebraic analog of $K_0$ given in \cref{def:algebraic_k_0}, which is non trivial fact that is explored and generalized in \citep{cuntz_topological_2007}.

\begin{remark}[K theory for a bigger class of algebras]\label{remark:k_theory_for_a_bigger_class_of_algebras}
In \cref{theorem:isomorphism_K_0_and_k_1_for_smooth_sub_algebras} we saw that the groups $K_0$ and $K_1$ of a C* algebra can be recovered from smooth sub algebras, and most of the work was done by the invariance of smooth sub algebras under holomorphic calculus. So, we do not need the whole C* algebra to define its K groups, only a well behaved "skeleton" of the algebra, this is the main theme of the book \citep{blackadar_k-theory_2012} where these algebras are referred to as local C* algebras (or pre C* algebras), and all the results on exact sequences and liftings is generalized into this context.

Following a similar of \citep{blackadar_k-theory_2012} we could ask if the tools of K theory of C* algebras can be translated into other types of algebras with a functional calculus similar to the functional calculus over Banach algebras. This question has a positive answer in the form of local Banach algebras, which are algebras whose topology is defined in terms of bornologies rather than in terms of topologies \citep[Chapter 2]{cuntz_topological_2007}, and have similar properties to Banach algebras. Many of the results from K theory of C* algebras can be extended to local Banach algebras over $\mathbb{C}$, for example, Bott periodicity extends into this context \citep[Chapter 4]{cuntz_topological_2007} along with the six term exact sequence associated to crossed products \citep[Chapter 5]{cuntz_topological_2007}. Additionally, For a proposal of K theory of Fréchet m-convex algebras you can take a look at \citep{phillips_k-theory_1991} and \citep{perrot_secondary_2008}. 
\end{remark}

\section{Notes of algebraic K theory}
\label{sec:notes_on_algebraic_k_theory}

In the present section, if $\mathcal{A}$ is a C* algebra or a smooth sub algebra, we will denote by $K^{\text{top}}_j(\mathcal{A}), \; j=0,1$\index{$K^{\text{top}}_0(\cdot)$} \index{$K^{\text{top}}_1(\cdot)$} the groups defined in through homotopy equivalence relations over $\mathcal{A}$ as described in \cref{chap:K_theory}.

\begin{definition}[$K_0^{\text {alg}}(\mathcal{A})$ \index{$K_0^{\text{alg}}(\cdot)$} Definition 3.7 \citep{gracia-bondia_elements_2001}]\label{def:algebraic_k_0}
Let $\mathcal{A}$ be a unital algebra, we say that two idempotents $e, f \in Q_m(\mathcal{A})$ are equivalent, $e \sim f$, if and only if they are conjugate via some $v \in G L_{\infty}(\mathcal{A})$; that is, for some $n \in \mathbb{N}$ there is a $v \in G L_{m+n}(\mathcal{A})$ such that
$$
v\left(\begin{array}{cc}
e & 0 \\
0 & 0_n
\end{array}\right) v^{-1}=\left(\begin{array}{cc}
f & 0 \\
0 & 0_n
\end{array}\right) \text {. }
$$
The addition on the quotient $Q_{\infty}(\mathcal{A}) / \sim$ is defined by the rule
$$
[e]+[f]:=\left[\left(\begin{array}{ll}
e & 0 \\
0 & f
\end{array}\right)\right]=\left[\left(\begin{array}{ll}
f & 0 \\
0 & e
\end{array}\right)\right] \text {, }
$$
which is well defined since $e \oplus f \sim f \oplus e$. Therefore $V^{\text {alg }}(A):=Q_{\infty}(\mathcal{A}) / \sim$ is a commutative semigroup, and we denote by $K_0^{\text {alg }}(\mathcal{A})$ the Grothendieck group\index{Grothendieck group} of $V^{\text {alg }}(A)$.

If $\mathcal{A}$ is not unital then $K_0^{\text{alg}}(A)$ is defined as $K_0^{\text{alg}}$ of the algebraic unitization of $\mathcal{A}$.  
\end{definition}

\begin{remark}[Equivalent definitions for C* algebras]\label{remark:isomorphisms_of_finitely_generated_proejctiove_modules}
Let $\mathcal{A}$ be a C* algebra, then, the group $K_0^{\text {alg }}(\mathcal{A})$ is usually defined as the Grothendieck group of the semigroup of isomorphism classes of finitely generated projective right modules over $\mathcal{A}$. This definition coincides with the definition in terms of equivalence classes of idempotents, which is by far no trivial fact, for a discussion on this you can look into \citep[Sections 3.1 and 3.2]{gracia-bondia_elements_2001}. The equivalence of both definitions rests mainly on the results of \citep[lemma 3.13]{gracia-bondia_elements_2001}.

Also, if $A$ is a unital C* algebra, $K_0^{\text {alg }}(A) \simeq K_0^{\text {top }}(A)$ by \citep[Theorem 3.14]{gracia-bondia_elements_2001}, which is why the notation $K_0(A)$ is used interchangeably for both definitions when dealing with C* algebras.    
\end{remark}

The algebraic definition of $K_1$ takes a different route, first, notice that $G L_{\infty}(\mathcal{A})$ is a group, because, if $v,u \in G L_{n}(\mathcal{A})$ then $vu \in G L_{n}(\mathcal{A})$, also, if $m>n$, $v \in G L_{n}(\mathcal{A})$, $u \in G L_{m}(\mathcal{A})$ then 

$$uv =  u \left(\begin{array}{ll}
v & 0 \\
0 & 1_{m-n}
\end{array}\right) \in G L_{m}(\mathcal{A}),$$

is well defined for all elements of $G L_{\infty}(\mathcal{A})$, and

$$v^{-1}v =  
\left(\begin{array}{ll}
v^{-1} & 0 \\
0 & 1_{m-n}
\end{array}\right) \left(\begin{array}{ll}
v & 0 \\
0 & 1_{m-n}
\end{array}\right) = \left(\begin{array}{ll}
1_n & 0 \\
0 & 1_{m-n}
\end{array}\right) \in G L_{m}(\mathcal{A}).$$

Recall that we identify $v \in G L_{n}(\mathcal{A})$ with $\left(\begin{array}{ll}
v & 0 \\
0 & 1_{m-n}
\end{array}\right) \in G L_{m}(\mathcal{A})$ for any $m > n$. So, to get an commutative group out of $G L_{\infty}(\mathcal{A})$ we use the commutator of $G L_{\infty}(\mathcal{A})$,
$$ [G L_{\infty}(\mathcal{A}), G L_{\infty}(\mathcal{A})] = \{ [g_1,h_1] \cdots [g_n,h_n] \; |\;  [g_i,h_i] = g_i^{-1}h_i^{-1} g_i h_i,\; g_i,h_i \in G L_{\infty}(\mathcal{A}), \; n \in \mathbb{N} \} $$
which led to the following definition,

\begin{definition}[$K_1^{\text{alg}}(\mathcal{A})$ \index{$K_1^{\text {alg }}(\cdot)$} Page 131 \citep{gracia-bondia_elements_2001}]\label{def:algebraic_k_1}
For $\mathcal{A}$ a unital algebra, define
$$K_1^{\text {alg }}(\mathcal{A}) = G L_{\infty}(A)_{\text{ab}} = G L_{\infty}(\mathcal{A}) /  [G L_{\infty}(\mathcal{A}), G L_{\infty}(\mathcal{A})].$$
Under this definition, $v \sim u$ if $v = u (\Pi_{i\leq n} [g_i,h_i])$. The subscript $\text{ab}$ comes from the fact that $K_1^{\text {alg }}(\mathcal{A})$ is the abelienization of $G L_{\infty}(\mathcal{A})$.

\end{definition}

The relation between $K_1^{\text{alg}}(\mathcal{A})$ and $K_1^{\text{top}}(\mathcal{A})$ for C* algebras is not as simple as in the case of $K_0$, because there is not always an isomorphism $K_1^{\text{alg}}(\mathcal{A}) \simeq K_1^{\text{top}}(\mathcal{A})$, for example, $K_1^{\text{alg}}(\mathbb{C}) = G(\mathbb{C})$ with $G(\mathbb{C})$ the group of invertible elements in $\mathbb{C}$ (\citep[page 131]{gracia-bondia_elements_2001}), while $K_1^{\text{top}}(\mathbb{C}) = 0$ (\citep[example 8.1.8]{rordam_introduction_2000}). Turns out that there is a map 
$$\gamma : K_1^{\text{alg}}(\mathcal{A}) \to K_1^{\text{top}}(\mathcal{A}),$$ 
which is called the comparison map, and it becomes an isomorphism when $\mathcal{A}$ is a stable C* algebra (\citep{rosenberg1997algebraic}). A C* algebra $A$ is said to be stable if $A$ is isomorphic to $A \otimes \mathcal{K}$ (\citep{review_on_stable_c_star_algebras}), where $\mathcal{K}$ is the C* algebra of compact operators over a separable Hilbert space.

\chapter{Cyclic cohomology}
\label{chapter:cyclic_cohomology}

This short appendix pretends to give a rough idea of what is cyclic cohomology in order to make \cref{chap:non_commutative_geometry_and_topological_invariants_for_hamiltonians} more understandable. We focus on stating some results on cyclic cohomology and we do not delve into the technical details of its construction. Cyclic cohomology was developed as an approach to extend the notion of de Rham currents into the non commutative setting ( \citep[Chapter 3]{connes_noncommutative_2014}), it is deviced to work well with locally convex topological algebras (\citep[Chapter 3, Appendix B]{connes_noncommutative_2014}) and falls back into the de Rham homology when the algebras take the form of $C^{\infty}(M)$ with $M$ a compact manifold (\citep[Chapter 3, Section 2, Theorem 2]{connes_noncommutative_2014}). 


\section{Definition of Cyclic cohomology}
\label{sec:cyclic cohomology definition}

Let's look into the definition of cyclic cohomology,

\begin{definition}[c.f. Chapter 3 Section 1 \citep{connes_noncommutative_2014} (Cyclic cohomology) \index{cyclic cohomology}]\label{definition:cyclic_cohomology}
let $\mathcal{A}$ be a possibly non-commutative algebra over $\mathbb{C}$, then, the cyclic cohomology of $\mathcal{A}$ which we denote by $HC^*(\mathcal{A})$ \index{$HC^*(\mathcal{A})$} is the cohomology of the complex $(C^n_{\lambda},b)$, where
\begin{itemize}
    \item $C^n_{\lambda}$ is the space of $(n+1)$-linear functionals $\phi$ over $\mathcal{A}$ that satisfy the cyclicity condition,
    $$ \forall a^i \in \mathcal{A}, \;  \phi(a^1, \cdots, a^n, a^0) = (-1)^n \phi (a^0, a^1, \cdots, a^n).$$
    The elements of $C^n_{\lambda}$ are called cyclic $n$-cochains\index{cyclic cohomology!$n$-cochains} over $\mathcal{A}$. 
    \item $b_n$ is the Hochschild coboundary map given by
    $$  b_n (\phi) (a^0, \cdots, a^{n+1}) = \sum_{j=0}^{n} (-1)^{j} \phi (a^0, \cdots , a^j a^{j+1}, \cdots, a^{n+1}) $$
    $$   +  (-1)^{n+1} \phi (a^{n+1} a^0 , \cdots, a^n). $$
\end{itemize}

If $\mathcal{A}$ is a topological algebra and we ask for $\phi$ to a be a continuous linear functional over $\mathcal{A}$, we use the notation $HC_{\text{con}}^* (\mathcal{A})$\index{$HC_{\text{con}}^* (\mathcal{A})$} and call it the continuous cyclic cohomology\index{continuous cyclic cohomology} of $\mathcal{A}$. Notice that $b_n (\phi)$ is continuous $(n+1)-$linear functional over $\mathcal{A}$ because the multiplication is continuous in a topological algebra.
\end{definition}

For the complex
$$
\cdots \rightarrow C^{i-1}_{\lambda} \stackrel{b_{i-1}}{\rightarrow} C^i_{\lambda} \stackrel{b_i}{\rightarrow} C^{i+1}_{\lambda} \rightarrow \cdots
$$
we denote $Z^n_{\lambda}(\mathcal{A}):= \text{ker} (b_i)$\index{$Z^n_{\lambda}(\mathcal{A})$} and call it the \textbf{cyclic $n$-coccyles}\index{cyclic $n$-coccyles} over $\mathcal{A}$ or cyclic cocycles if it is clear by the context which is the value that $n$ takes, also, we denote $B^n_{\lambda}(\mathcal{A}) := \text{Im}(b_{i-1}) $\index{$B^n_{\lambda}(\mathcal{A})$} and call it the \textbf{$n$-coboundaries}\index{cyclic cohomology!$n$-coboundaries} over $\mathcal{A}$ (\citep[Definition 10.2]{gracia-bondia_elements_2001}). if we are working with continuous linear functionals then we will use the notation $Z^n_{\lambda, \text{con}}(\mathcal{A})$ and $B^n_{\lambda, \text{con}}(\mathcal{A})$. By definition of the cyclic cohomology we have that 
$$HC^n(\mathcal{A}) = Z^n_{\lambda} (\mathcal{A}) / B^n_{\lambda} (\mathcal{A})$$\index{$HC^n(\mathcal{A}) $}
and 
$$ HC_{\text{con}}^n(\mathcal{A}) = Z^n_{\lambda, \text{con}} (\mathcal{A}) / B^n_{\lambda, \text{con}} (\mathcal{A}).$$\index{$HC_{\text{con}}^n(\mathcal{A})$}


\begin{definition}[Cup product \index{cyclic cohomology!cup product} (Chapter 3, Section1, Theorem 12 \citep{connes_noncommutative_2014})]\label{definition:cup_product}
There is an homomorphism
$$ HC^{n}(\mathcal{A}) \otimes HC^{n}(\mathcal{B}) \to HC^{n+m}(\mathcal{A} \otimes \mathcal{B}), \; \phi \otimes \psi \mapsto \phi \# \psi,$$
we call $\phi \# \psi$\index{$\phi \# \psi$ (cup product of cyclic cohomology)} the cup product of $\phi$ and $\psi$.
\end{definition}

The cup product turns $HC^*(\mathbb{C})$ into polynomial ring with one generator $\sigma$ of degree $2$ (\citep[Chapter 3, Section1, Corollary 13]{connes_noncommutative_2014}), we have that if $\phi \in Z^n_{\lambda} (\mathcal{A})$ then (\citep[Chapter 3, Section1, Corollary 13]{connes_noncommutative_2014})
$$ \sigma \# \phi = \phi \# \sigma \in Z^{n+2}_{\lambda}(\mathcal{A}). $$
Denote by $S: HC^{n}(\mathcal{A}) \to HC^{n+2}(\mathcal{A})$ the map given by $S(\phi) = \phi \# \sigma$, then, $HC^*(\mathcal{A})$ becomes a $HC^*(\mathbb{C})$-bimodule. We won't dive into the technicalities of the cup product, you can take a look at \citep[Chapter 3, Section1]{connes_noncommutative_2014} for more information on the cup product and the properties of cyclic cohomology. The previous statement also implies that $HC^*(\mathcal{A})$ is divided into two $HC^*(\mathbb{C})$-bimodules invariant under multiplication by $\sigma$, those are
$$ HC^{\text{ev}}(\mathcal{A}):= \bigoplus_{k \in \mathbb{N}} HC^{2k}(\mathcal{A}), \; HC^{\text{odd}}(\mathcal{A}):= \bigoplus_{k \in \mathbb{N}} HC^{2k+1}(\mathcal{A}),$$\index{$HC^{\text{ev}}(\mathcal{A})$} \index{$HC^{\text{odd}}(\mathcal{A})$}
similarly, we have that
$$ HC^{\text{ev}}_{\text{con}}(\mathcal{A}):= \bigoplus_{k \in \mathbb{N}} HC^{2k}_{\text{con}}(\mathcal{A}), \; HC^{\text{odd}}_{\text{con}}(\mathcal{A}):= \bigoplus_{k \in \mathbb{N}} HC^{2k+1}_{\text{con}}(\mathcal{A}).$$\index{$HC^{\text{ev}}_{\text{con}}(\mathcal{A})$} \index{$HC^{\text{odd}}_{\text{con}}(\mathcal{A})$}

The following is an example of the cup product that will be useful for the pairing between K theory and cyclic cohomology (\cref{sec:how_to_construct_cocycles_on_M_n_A})

\begin{example}[Generalized trace]\label{example:generalized_trace}
Let $\psi$ be a trace on $B$, then, from \citep[Example 3.6.8 ]{khalkhali_basic_2013} we have that $\varphi \mapsto \varphi \# \psi$ defines a map
$$
H C^m( \mathcal{A}) \rightarrow H C^m( \mathcal{A} \otimes B),
$$
which takes the following form on simple tensors
$$
(\varphi \# \psi)\left(a^0 \otimes b^0, \ldots, a^m \otimes b^m\right)=\varphi\left(a^0, \ldots, a^m\right) \psi\left(b^0 b^1 \ldots b^m\right) .
$$
This construction is important for the pairing between cyclic cohomology and K theory (\cref{sec:how_to_construct_cocycles_on_M_n_A}), so, let $\psi=\text{tr}: M_n(\mathbb{C}) \rightarrow \mathbb{C}$ be the standard trace 
$$\text{tr} \left( (a_{i,j})_{i,j \leq n}  \right) = \sum_{i \leq n} a_{i,i} ,$$ 
then, the cup product allows us to define a map
$$
H C^m(\mathcal{A}) \rightarrow H C^m\left(M_k(A)\right)
$$
such that
$$
(\varphi \# \text{tr})\left(a^0 \otimes \delta_{i_0, j_0} , \ldots, a^m \otimes \delta_{i_m, j_m} \right)=\varphi\left(a^0, \ldots, a^m\right) \text{tr}\left(\delta_{i_0, j_0} \delta_{i_1, j_1} \ldots \delta_{i_m, j_m} \right),
$$
with $\delta_{i_k, j_k}$ the matrix with 1 in the entry $(i_k, j_k)$ and zero elsewhere.
\end{example}

The cyclic cohomology is a contravariant functor from the category of algebras into the category of modules (\citep[Chapter 3]{connes_noncommutative_2014}), which makes it a covariant functor at the level of topological spaces if we work with algebras of functions over topological spaces.

\begin{definition}[Induced morphism of modules (page 194 \citep{connes_noncommutative_2014})]\label{definition:induced_morphism_of_modules}
Let $\rho: \mathcal{A} \to \mathcal{B}$ be an algebra morphism, then, there is a morphism of complexes $\rho^*: C^n_{\lambda}(\mathcal{B}) \to C^n_{\lambda}(\mathcal{A})$ defined by
$$ \rho^*(\phi)(a^0, \cdots, a^1) = \phi(\rho(a^0), \cdots, \rho(a^n)) $$
and an induced map
$$ \rho^* : HC^*(\mathcal{B}) \to HC^{*}(\mathcal{A}), $$
which is a morphism of $HC^*(\mathbb{C})$-bimodules.
\end{definition}

The properties of cyclic cohomology still hold when working the continuous cyclic cohomology of locally convex algebras (\citep[Chapter 3, Appendix B]{connes_noncommutative_2014}), such that, $HC^*_{\text{con}} (\mathcal{A})$ is a module over the ring $HC_{\text{con}}^*(\mathbb{C})$ and $HC_{\text{con}}^*(\mathbb{C}) = HC^*(\mathbb{C})$. Additionally, to compute the continuous cyclic cohomology of a locally convex algebra there are various technicalities that need to be taken care of, like, using the description of projective tensor products as the set of continuous linear functionals on the product space (\cref{proposition:properties_projective_tensor_product}), and coming up with projective resolutions for the locally convex algebra at hand (\citep[Chapter 3, Appendix B Definition 1]{connes_noncommutative_2014}). If you want to go more into the details of computing the cyclic cohomology of algebras you can refer to \citep[Chapter 3, Appendix B]{connes_noncommutative_2014}, \citep{connes_non-commutative_1985}.

\section{Cohomology interesting and nuclear C* algebras}
\label{sec:when_is_cyclic_cohomology_interesting}


As an interesting fact on continuous cyclic cohomology, for any nuclear C* algebra $A$\index{C* algebra!nuclear}, the continuous continuous cyclic cohomology of $A$ takes the form (\citep[Remark 7 (page 113)]{khalkhali_basic_2013}), i.e.
$$
H C_{\text{con}}^{2 n}(A)=A^* \quad \text { and } \quad H C_{\text{con}}^{2 n+1}(A)=0,
$$
therefore, the cyclic cohomology of nuclear C* algebras has little to no information. Notice that the previous result is consistent with the mental image of cyclic cohomology being defined by derivations on algebras, since, cocycles defined over smooth sub algebras often to do not extend continuously to their C* algebras (\citep[page 446]{gracia-bondia_elements_2001}), for example, if we take $A = C(\mathbb{T})$, where the derivation 
$$ d : C^{\infty}(\mathbb{T}) \to C^{\infty}(\mathbb{T}) , \; d(f) := \frac{d f }{d \lambda}$$ 
is a densely defined unbounded operator over $C(X) \subset L^{2}(X)$ and the two cocycle over $X$ given by 
$$\phi(f,g) = \int_{\mathbb{T}} f d(g),$$ 
can not be extended continuously to $C(\mathbb{T})$ because $d$ cannot be extended continuously to $C(\mathbb{T})$; \citep[page 497]{spivak_calculus_1994}) provides an example of a sequence of functions over $\mathbb{R}$ which converge uniformly to the function $f(x) =0$ but whose derivatives are a sequence of functions that do not converge uniformly to a continuous function over $\mathbb{R}$, thus, the derivative as a map defined over $\mathcal{S}(\mathbb{R})$ (\cref{example:schwartz_space}) cannot be extended to a continuous map from $C_0(\mathbb{R})$ into $C_0(\mathbb{R})$.

\begin{remark}\label{remark:amenability_nuclearity_and_cicly_cohomology}
Inner derivations are those that take the forms $d(a) = ax -xa$ for some $x \in A$, a Banach $A$ is referred to be amenable\index{Banach algebra!amenable} if every bounded derivation from $A$ to a Banach-bimodule is inner, thus, amenable Banach algebras are those where bounded derivations take the simple form $d(a) = ax -xa$. The definition of amenability for C* algebras is motivated by the fact that for a locally compact group $G$, $G$ is amenable as a group (\citep[Proposition 11.2.5]{dales_introduction_2003})  iff $L^1(G)$ is amenable as a Banach algebra (\citep[Theorem IV.3.3.1]{blackadar_operator_2006}). Turns out that a C* algebra is amenable iff it is nuclear (\citep[Corollary IV.3.3.15]{blackadar_operator_2006}), therefore, an amenable C* algebra has the following continuous cyclic cohomology
$$
H C_{\text{con}}^{2 n}(A)=A^* \quad \text { and } \quad H C_{\text{con}}^{2 n+1}(A)=0.
$$
\end{remark}


\section{Pairing with K theory}
\label{sec:pairing with K theory}

Following the notation introduced in \cref{sec:notes_on_algebraic_k_theory}, if $A$ is a C* algebra or a smooth sub algebra, we will denote by $K^{\text{top}}_j(A), \; j=0,1$ the groups defined in through homotopy equivalence relations over $A$ as described in \cref{chap:K_theory}. If $A$ is an algebra, we will denote by $K^{\text{alg}}_j(A), \; j=0,1$ the groups defined through algebraic equivalence relations as described in \cref{sec:notes_on_algebraic_k_theory}.



Given that the C* algebra that is part of the Non-Commutative Brilluoin Torus is a nuclear C* algebras (\cref{remark:non_commu_brilluouin_torus_nuclear_C_star_algebra}), and the continuous cyclic cohomology of nuclear C* algebras has trivial information (\cref{sec:when_is_cyclic_cohomology_interesting}), we turn our attention towards smooth sub algebras, motivated by the fact that
$$ K_{i}(\mathcal{A}) \simeq K_{i} (A), \; i=0,1 $$
when $A$ is a unital C* algebra and $\mathcal{A}$ is a smooth sub algebra of $A$ (\cref{theorem:isomorphism_K_0_and_k_1_for_smooth_sub_algebras}). Notice that we can use the pairing at the level of unital C* algebras to assign a pairing between $HC^{*}_{\text{con}}(\mathcal{A}^{+})$ and $K_{0,1}(A)$ for each C* algebra $A$, because $K_1(A) = K_1(A^+)$ (\cref{definition:group_K_1}) and $K_0(A) \subset K_0(A^+)$ as a subgroup (\cref{definition:group_K_0}).

\subsection{Cocycles on $M_n(\mathcal{A})$}
\label{sec:how_to_construct_cocycles_on_M_n_A}

Let $\mathcal{A}$ be a C* algebra or a smooth sub algebra, since the elements of $K_{j}^{top}(\mathcal{A}), \; j = 0,1$ are equivalence classes of elements in $M_{\infty}(\mathcal{A})$, we need to find a way to extend a cyclic cocycle in $\mathcal{A}$ into a cyclic cocycle in $M_{\infty}(\mathcal{A})$, recall that $M_{\infty}(\mathcal{A})$ is the algebra whose elements are square matrices over $\mathcal{A}$ i.e. $M_{\infty}(\mathcal{A}) = \cup_{n \in \infty} M_n(\mathcal{A})$ (\cref{sec:groups_K_0_and_K_1} and \cref{sec:def_K_0_and_K_1_smooth_subalgebras}). Also, for any $n \in \mathbb{N}$ and $m>n$ we use the notation, 
$$
i_{n,m}: M_n(A) \rightarrow M_{m}(A): m \mapsto \left(\begin{array}{cc}
m & 0 \\
0 & 0_{m-n}
\end{array}\right),
$$
and $i_{n,m}$ is a canonical injective *-homomorphims for both C* algebras and Fr\'echet algebras. If we take $b \in M_{\infty}(\mathcal{A})$, then, there exists $n<\infty$ and $a \in M_n(\mathcal{A})$ such that $b = [a]_{0,1}$, hence, for $\phi \in Z^{n}_{\lambda}(\mathcal{A})$ we can try look for a way to define $\phi_{m} \in Z^{n}_{\lambda}(M_m(\mathcal{A}))$ and $\phi_{l} \in Z^{n}_{\lambda}(M_l(\mathcal{A}))$ such that if $a_0, \cdots, a_k \in M_m(\mathcal{A})$ and $l > m$ we have
$$ \phi_m(a_0, \cdots, a_k ) = \phi_l (i_{m,l}(a_0), \cdots, i_{m,l}(a_k)). $$

Recall that $M_n(\mathbb{C}) \otimes \mathcal{A} = M_n(\mathcal{A})$ for both C* algebras and smooth sub algebras (\cref{sec:C_star_alg_matrix_alg}, \cref{example:Frechet_algebras_matrix_algebras}), hence, we are motivated to use the cup product to define $\phi_n$. In this case, for each $\rho \in HC^{0}(M_n(\mathbb{C}))$ and $\phi \in HC^{m}(\mathcal{A})$ we can define an element $\rho \# \phi \in HC^{m}(M_n(\mathcal{A}))$, which is very fortunate because we can use the standard trace over $M_n(\mathbb{C})$ as exposed in \cref{example:generalized_trace},
$$ \text{tr}: M_n(\mathbb{C}) \to \mathbb{C}, \; (a_{i,j})_{i,j \leq n} \to \sum_{i \leq n} a_{i,i}.$$

Let $a_g \in M_m(\mathcal{A})$ for $g \leq n+1$ and $\phi \in HC^{n}(\mathcal{A})$, then $\text{tr}_{m} \# \phi \in HC^{n}(M_m(\mathcal{A}))$, also, denote by $\delta_{m,i,j} \in M_m(\mathcal{A})$ the matrix such that $\delta_{m,i,j}(l,k) = 1$ iff $i=l,\; j=k $, then
$$ a_g = \sum_{i,j \leq m} \delta_{m,i,j} a_g(i,j). $$
So, if $m' > m$ we have that $ i_{m,m'}(a_g) = \sum_{i,j \leq m} \delta_{m',i,j} a_g(i,j) + \sum_{i>m \text{ or } j>m} \delta_{m',i,j} 0$, and we end with
$$ \text{tr}_{m'} \# \phi (i_{m,m'}(a_1), \dots, i_{m,m'}(a_{n+1}))$$
$$ = \sum_{\substack{  i_1,j_1\leq m'\\ \dots \\i_{n+1},j_{n+1}\leq m'}  } \text{tr}_{m'}(\delta_{m',i_{1},j_{1}} \cdot \ldots \cdot \delta_{m',i_{n+1},j_{n+1}}) \phi(a_1(i_1,j_1), \dots, a_g(i_{n+1},j_{n+1}))  $$
$$ = \sum_{\substack{  i_1,j_1\leq m\\ \dots \\i_{n+1},j_{n+1}\leq m}  } \text{tr}_{m}(\delta_{m,i_{1},j_{1}} \cdot \ldots \cdot \delta_{m,i_{n+1},j_{n+1}}) \phi(a_1(i_1,j_1), \dots, a_g(i_{n+1},j_{n+1})),$$
$$ = \text{tr}_{m} \# \phi (a_1, \dots, a_{n+1}) $$
because $\phi(b_1,\cdots,b_{n+1}) = 0$ if $b_i = 0$ for some $i$. Therefore, $\phi \to \text{tr}_{m} \# \phi$ generates a cocycle on $M_m(\mathcal{A})$ which is consistent with the inclusion of $M_m(\mathcal{A})$ on $M_{m'}(\mathcal{A})$ when $m' > m$. 

\begin{lemma}[Generalized trace and continuous cyclic cohomology]\label{lemma:continuous_cyclic_cohomolgy_and_generalized_trace}
Let $\mathcal{A}$ be Fr\'echet algebra and $m \in \mathbb{N}$, then,
\begin{enumerate}
    \item If $\varphi \in C^{n}_{\lambda, \text{con}} (\mathcal{A})$, then, $\varphi \# \text{tr} \in C^{n}_{\lambda, \text{con}}(M_m(\mathcal{A}))$.
    \item If $\varphi \in B^{n}_{\lambda, \text{con}} (\mathcal{A})$, then, $\varphi \# \text{tr} \in B^{n}_{\lambda, \text{con}}(M_m(\mathcal{A}))$.
\end{enumerate}
\end{lemma}
\begin{proof}
\begin{enumerate}
    \item From \cref{example:generalized_trace} we know that 
    $$
    (\varphi \# \text{tr})\left(a^0 \otimes \delta_{i_0, j_0} , \ldots, a^m \otimes \delta_{i_m, j_m} \right)=\varphi\left(a^0, \ldots, a^m\right) \text{tr}\left(\delta_{i_0, j_0} \delta_{i_1, j_1} \ldots \delta_{i_m, j_m} \right),
    $$
    with $\delta_{i_k, j_k}$ the matrix with 1 in the entry $(i_k, j_k)$ and zero elsewhere. Since the projection $p_{i,j}: M_m(\mathcal{A}) \to \mathcal{A}, \; p_{i,j}(\sum_{k,l \leq m} a_{k,l} \otimes \delta_{k,l}) = a_{i,j}$ is a continuous map (\cref{example:Frechet_algebras_matrix_algebras}), and $\varphi$ is a continuous multilinear functional over $\mathcal{A}$, we have that $\varphi \# \text{tr}$ is a continuous multilinear functional over $\mathcal{A} \otimes M_m(\mathbb{C}) = M_n(\mathcal{A})$. Additionally, from \citep[Example 3.6.8 ]{khalkhali_basic_2013} we get that $\varphi \# \text{tr}$ is a cyclic cocycle over $M_m(\mathcal{A})$, therefore, we have that $\varphi \# \text{tr} \in C^{n}_{\lambda, \text{con}}(M_m(\mathcal{A}))$.
    \item Given that $(a \otimes \delta_{i,j})(b \otimes \delta_{k,l}) = (ab) \otimes (\delta_{i,j} \delta_{k,l} )$, if $\varphi = b \phi$ then we must that $\varphi \# \text{tr} = b (\phi \# \text{tr}) $, which implies that $\varphi \# \text{tr} \in B^{n}_{\lambda, \text{con}}(M_m(\mathcal{A})).$ 
\end{enumerate}
\end{proof}

\subsection{Pairing with $K_0$}
\label{sec:pairing_with_K0}

For a unital algerbas $\mathcal{A}$ there is a bilinear pairing\index{bilinear pairing} between $K_0^{\text{alg}}(\mathcal{A})$ and $HC^{\text{ev}}(\mathcal{A})$, and is constructed with $ \text{tr}_{m} \# \phi$ for $\phi$ in $HC^{2n}(\mathcal{A})$. The bilinear pairing takes the following form,

\begin{proposition}[c.f. Chapter 3, Section 3, Proposition 2   \citep{connes_noncommutative_2014}]\label{proposition:pairing_algebraic_k_0_even_ciclyc_cohomology}
Let $\mathcal{A}$ be an unital algebra, then, the following equality defines a bilinear pairing between $K_0^{\text{alg}}(\mathcal{A})$ and $HC^{\text{ev}}(\mathcal{A})$:
$$\langle[p],[\varphi]\rangle :=(m !)^{-1}(\text{tr}_{k} \# \varphi)(p, \ldots, p),$$
with $p \in P_k(\mathcal{A})$ and $\varphi \in Z_\lambda^{2 m}(\mathcal{A})$.
\end{proposition}

The pairing in \cref{proposition:pairing_algebraic_k_0_even_ciclyc_cohomology} can be extended to $K_0^{\text{top}}(\mathcal{A})$ and $HC^{\text{ev}}_{\text{con}}(\mathcal{A})$ for unital C* algebras and smooth sub algebras of unital C* algebras,

\begin{proposition}\label{proposition:pairing_topological_k_0_even_ciclyc_cohomology}
Let $\mathcal{A}$ be a unital C* algebra or a smooth sub algebra of a unital C* algebra, then, the following defines a bilinear pairing between $K_0^{\text{top}}(\mathcal{A})$ and $HC_{\text{con}}^{\text{ev}}(\mathcal{A})$:
$$\langle[p],[\varphi]\rangle :=(m !)^{-1}(\text{tr}_{k} \# \varphi)(p, \ldots, p)$$
with $p \in P_k(\mathcal{A})$ and $\varphi \in Z_{\lambda, \text{con}}^{2 m}(\mathcal{A})$.
\end{proposition}
\begin{proof}
There are two ways to prove that exists a pairing between $K_0^{\text{top}}(\mathcal{A})$ and $HC^{\text{ev}}_{\text{con}}(\mathcal{A})$,
\begin{itemize}
    \item We can take advantage of the connection between the equivalence relations for projections on $M_n(\mathcal{A})$ as stated in \cref{proposition:cannonical_paths_idempotents_projections_smooth_sub_algebras} and \cref{proposition:canonical_paths_between_idempotents}, which tell us that if $p \sim_{h} q$ in $P_n(\mathcal{A})$, then, we can find $v \in G_n(\mathcal{A})$ with $q = v^{-1} p v$, when $\mathcal{A}$ is either a C* algebra or a smooth sub algebra. From \cref{lemma:continuous_cyclic_cohomolgy_and_generalized_trace} we know that $\varphi \in B^{n}_{\lambda, \text{con}} (\mathcal{A})$, then, $\varphi \# \text{tr} \in B^{n}_{\lambda, \text{con}}(M_m(\mathcal{A}))$, therefore, $\text{tr}_{k} \# \varphi(p, \cdots, p) = 0$ because $\text{tr}_{k} \# \varphi(p, \cdots, p) = - \text{tr}_{k} \# \varphi(p, \cdots, p)$ by the properties of cyclic cocycles (\cref{definition:cyclic_cohomology}). From \citep[Chapter 3, Section 3, Lemma 1]{connes_noncommutative_2014} it is known that 
    $$ \text{tr}_{k} \# \varphi(p, \cdots, p) = \text{tr}_{k} \# \varphi(v^{-1} p v, \cdots, v^{-1} p v), $$
    additionally, by the proof of \citep[Chapter 3, Section 3, Proposition 2]{connes_noncommutative_2014} we knoe that $\langle[p],[\varphi]\rangle = \langle[p],[S\varphi]\rangle$, therefore, $\langle[p],[\varphi]\rangle$ is well defined.
    \item Recall that in unital C* algebras and unital smooth sub algebras if there is a homotopy of projections $p,q$ then there is also a smooth homotopy between $p$ and $q$ (\cref{proposition:cannonical_paths_idempotents_projections_smooth_sub_algebras} and \cref{proposition:canonical_paths_between_idempotents}). \citep[Lemma 4.1.4]{khalkhali_basic_2013} states that if $\phi$ is a $2m$ cyclic cocycle then $\phi (p_t, \ldots, p_t)$ is constant in $t$. By the proof of \citep[Chapter 3, Section 3, Proposition 2]{connes_noncommutative_2014} we know that $\langle[p],[\varphi]\rangle = \langle[p],[S\varphi]\rangle$, therefore, the previous discussion implies that $\langle[p],[\varphi]\rangle$ is well defined.
    
    which implies that $\langle[p],[\varphi]\rangle$ is well defined.
\end{itemize}
\end{proof}

Let $A$ be a C* algebra with unit and $\mathcal{A}$ a smooth sub algebra of $A$, then, from \cref{theorem:isomorphism_K_0_and_k_1_for_smooth_sub_algebras} we know that $K_0^{\text{top}}(\mathcal{A}) \simeq K_0^{\text{top}}(A)$, thus, if we use the aforementioned isomorphism along with \cref{proposition:pairing_topological_k_0_even_ciclyc_cohomology} we can establish a pairing between $HC^{\text{ev}}_{\text{con}}(\mathcal{A})$ and $K_0^{\text{top}}(A)$ for any C* algebra,

\begin{proposition}\label{proposition:pairing_topological_k_0_even_ciclyc_cohomology_C_star_algebras}
Let $A$ be a unital C* algebra and $\mathcal{A} \subset A$ a smooth sub algebra of $A$, then, the following equality defines a bilinear pairing between $K_0^{\text{top}}(A)$ and $HC^{\text{ev}}_{\text{con}}(\mathcal{A})$:
$$\langle[p],[\varphi]\rangle=(m !)^{-1}(\text{tr}_{k} \# \varphi)(p_s, \ldots, p_s)$$
where $p_s \in P_k(\mathcal{A})$, $[p_s] = [p]$ inside $K_0^{\text{top}}(A)$ and $\varphi \in Z_{\lambda, \text{con}}^{2 m}(\mathcal{A})$. 
\end{proposition}
\begin{proof}
From \cref{proposition:cannonical_paths_idempotents_projections_smooth_sub_algebras} we know that for any $p \in  P_k(A)$ there is a $p_s \in P_k(\mathcal{A})$ such that $[p] = [p_s]$ inside $K_0^{\text{top}}(A)$, therefore, we can use the isomorphism $K_0^{\text{top}}(\mathcal{A}) \simeq K_0^{\text{top}}(A)$ (\cref{theorem:isomorphism_K_0_and_k_1_for_smooth_sub_algebras}) along with \cref{proposition:pairing_topological_k_0_even_ciclyc_cohomology} to get the desired pairing.
\end{proof}

\subsection{Pairing with $K_1$}
\label{sec:pairing_with_K1}

Let $\mathcal{A}$ be a unital algebra, then, there is a bilinear pairing\index{bilinear pairing} between $K_1^{\text{alg}}(\mathcal{A})$ and $HC^{\text{odd}}(\mathcal{A})$, and is constructed with $ \text{tr}_{m} \# \phi$ for $\phi$ in $HC^{2n+1}(\mathcal{A})$. The bilinear pairing takes the following form,

\begin{proposition}[c.f. Chapter 3, Section 3, Proposition 3   \citep{connes_noncommutative_2014}]\label{proposition:pairing_algebraic_k_1_odd_ciclyc_cohomology}
Let $\mathcal{A}$ be an unital algebra, then the following defines a bilinear pairing between $K_1^{\text{alg}}(\mathcal{A})$ and $HC^{\text{odd}}(\mathcal{A})$:
$$\langle[u],[\varphi]\rangle := \frac{\Gamma \left( \frac{2n+1}{2} + 1 \right)^{-1}}{ 2^{2n+1} \sqrt{2 i} }  (\text{tr}_{k} \# \varphi)( u^{-1} -1 , u -1, u^{-1} -1, \ldots, u -1)$$
or $u \in U_k(\mathcal{A})$ and $\varphi \in Z_\lambda^{2 n +1}(\mathcal{A})$.
\end{proposition}

Notice that the cyclic property of the cocycles in $Z_\lambda^{2 n +1}(\mathcal{A})$ implies that 
$$  (\text{tr}_{k} \# \varphi)( u^{-1} -1 ,\ldots, u -1) = - (\text{tr}_{k} \# \varphi)( u -1 , \ldots, u^{-1} -1). $$
An important step in the proof of \cref{proposition:pairing_algebraic_k_1_odd_ciclyc_cohomology} is to show that
$$  (\text{tr}_{k} \# \varphi)( (uv)^{-1} -1 , \ldots, (uv) -1)  = $$
$$ (\text{tr}_{k} \# \varphi)( (u)^{-1} -1 , \ldots, u -1) + (\text{tr}_{k} \# \varphi)( (v)^{-1} -1 , \ldots, v -1),$$
which you can check in \citep[Chapter 3, Section 3, Proposition 3]{connes_noncommutative_2014}. Since any element of $K_1^{\text{alg}}(A)$ is expressed as $v = u (\Pi_{i\leq n} [g_i,h_i])$ with $[g_i,h_i] = g_i^{-1}h_i^{-1} g_i h_i$, the aforementioned property of the cocycles gives implies that
$$ (\text{tr}_{k} \# \varphi)( (u (\Pi_{i\leq n} [g_i,h_i]))^{-1} -1 , \ldots, (u (\Pi_{i\leq n} [g_i,h_i])) -1) =  (\text{tr}_{k} \# \varphi)( (u)^{-1} -1 , \ldots, u -1).$$ The aforementioned property of the evaluation
$$ (\text{tr}_{k} \# \varphi)( u^{-1} -1 ,\ldots, u -1) $$
will be useful to show that the pairing in \cref{proposition:pairing_algebraic_k_1_odd_ciclyc_cohomology} can be extended into the continuous cyclic cohomology.

\begin{proposition}[c.f. Remark 30.c on page 211 \citep{connes_noncommutative_2014} (Invariance of the cyclic cohomology under smooth paths)]\label{proposition:invariance_cyclic_cohomolgy_under_smooth_paths}
Let $\mathcal{A}$ be an algebra with unit and $\mathcal{B}$ a locally convex algebra. Take $\phi \in Z^n_{\lambda, \text{con}}(\mathcal{B})$, let $\rho_t, \; t \in [0,1]$ a continuous family of homomrphisms $\rho_t : \mathcal{A} \to \mathcal{B}$ such that for all $a \in \mathcal{A}$ the map $t \mapsto \rho_t (a) \in \mathcal{B}$ is of class $C^1$, then, 
$$ \rho_0^* = \rho_1^* : HC_{\text{con}}^*(\mathcal{B}) \to HC^*(\mathcal{A}). $$
\end{proposition}

\begin{proposition}\label{proposition:pairing_topological_k_1_odd_ciclyc_cohomology}
Let $\mathcal{A}$ be an unital C* algebra or a smooth sub algebra of a unital C* algebra, then, the following defines a bilinear pairing between $K_1^{\text{top}}(\mathcal{A})$ and $HC_{\text{con}}^{\text{odd}}(\mathcal{A})$:
$$\langle[u],[\varphi]\rangle := \frac{\Gamma \left( \frac{2n+1}{2} + 1 \right)^{-1}}{ 2^{2n+1} \sqrt{2 i} }  (\text{tr}_{k} \# \varphi)( u^{-1} -1 , u -1, u^{-1} -1, \ldots, u -1)$$
or $u \in U_k(\mathcal{A})$ and $\varphi \in Z_{\lambda, \text{con}}^{2 n +1}(\mathcal{A})$.
\end{proposition}
\begin{proof}
Recall that both C* algebras and smooth sub algebras are locally convex algebras. From the definition of $K_1^{\text{top}}(\mathcal{A})$ (\cref{definition:group_K_1}) we know that if $[u] = [v]$ then there must be $n \in \mathbb{N}$ such that $u \sim_h v$ inside $U_n(\mathcal{A})$, so, from \cref{proposition:cannonical_paths_idempotents_projections_smooth_sub_algebras} and \cref{proposition:cannonical_smooth_paths_unitaries_and_invetibles_C_star_algebras} we have that there are $h_j \in M_n(\mathcal{A})_{\text{sa}}$ such that
$$ u = v \prod_{j \leq k} \exp(i h_j).  $$
Denote $\xi (u) = (\text{tr}_{k} \# \varphi)( u^{-1} -1 , u -1, u^{-1} -1, \ldots, u -1)$, then, from the proof of \citep[Chapter 3, Section 3, Proposition 3]{connes_noncommutative_2014} we have that for any $\varphi \in Z_\lambda^{2 n +1}(\mathcal{A})$ the following equality holds
$$ \xi(uv) = \xi(u) + \xi(v),$$
so, if $u = v \prod_{j \leq k} \exp(i h_j)$, then
$$ \xi(u) = \xi(u) +  \sum_{j \leq k} \xi( \exp(i h_j))  .$$

Given $a \in \mathcal{A}$, from the holomorphic functional calculus we have the following family of algebra homomorphisms (\cref{theorem:holomorphic_funct_calculus_smooth_sub_algebras})
$$ \rho_t : \text{Hol}(\mathbb{C}) \to M_k(\mathcal{A}), \; \rho_t(f):= f(ta), $$
so, given that the map 
$$ t \mapsto f(ta), \; a \in M_k(\mathcal{A})_{\text{sa}}  $$
is smooth (\cref{proposition:smoomth_paths_from_entire_functions_smooth_sub_algebras}), we can use \cref{proposition:invariance_cyclic_cohomolgy_under_smooth_paths} to get that
$$ \rho_0 = \rho_1 : HC_{\text{con}}^*(M_n(\mathcal{A})) \to HC^*(\text{Hol}(\mathbb{C})). $$
Hence, $\rho_0 ([\text{tr}_{k} \# \varphi]) = \rho_1 ([\text{tr}_{k} \# \varphi])$, therefore, if we denote by $u_{\text{exp}}$ the function $x \mapsto \exp(i x)$ which belongs to $\text{Hol}(\mathbb{C})$ we have that $u_{\text{exp}}$ is an unitary in $\text{Hol}(\mathbb{C})$ and the pairing between $HC^{\text{odd}}(\text{Hol}(\mathbb{C}))$ and $K_1^{\text{alg}}(\text{Hol}(\mathbb{C}))$ tell us that
$$ \rho_0 ([\text{tr}_{k} \# \varphi])(u_{\text{exp}}^{-1} -1 , \cdots, u_{\text{exp}} -1 ) = \rho_1 ([\text{tr}_{k} \# \varphi])(u_{\text{exp}}^{-1} -1 , \cdots, u_{\text{exp}} -1 ). $$

By definition of $ \rho_0 ([\text{tr}_{k} \# \varphi]), \;  \rho_1 ([\text{tr}_{k} \# \varphi])$ (\cref{definition:induced_morphism_of_modules}) we have that
$$  \rho_0 (\text{tr}_{k} \# \varphi)(u_{\text{exp}}^{-1} -1 , \cdots, u_{\text{exp}} -1 ) = \xi(\exp(i t a)|_{t=0} = 0, $$
and
$$  \rho_1 (\text{tr}_{k} \# \varphi)(u_{\text{exp}}^{-1} -1 , \cdots, u_{\text{exp}} -1 ) =  \xi(\exp(i t a))|_{t=1} = \xi( \exp(i a)), $$
thus, we must have that $\xi( \exp(i a)) = 0$. The previous discussion implies that
$$ \xi( u) = \xi(v) +  \sum_{j \leq k} \xi( \exp(i h_j)) = \xi (v), $$
so, given that $\langle[p],[\varphi]\rangle = \langle[p],[S\varphi]\rangle$ by the proof of \citep[pChapter 3, Section 3, Proposition 3]{connes_noncommutative_2014}, we have that the bilinear pairing between $K_1^{\text{top}}(\mathcal{A})$ and $HC_{\text{con}}^{\text{odd}}(\mathcal{A})$ is well defined.
\end{proof}

Let $A$ be a C* algebra with unit and $\mathcal{A}$ a smooth sub algebra of $A$, then, from \cref{theorem:isomorphism_K_0_and_k_1_for_smooth_sub_algebras} we know that $K_1^{\text{top}}(\mathcal{A}) \simeq K_1^{\text{top}}(A)$, thus, if we use the aforementioned isomorphism along with \cref{proposition:pairing_topological_k_1_odd_ciclyc_cohomology} we can establish a pairing between $HC^{\text{odd}}_{\text{con}}(\mathcal{A})$ and $K_1^{\text{top}}(A)$ for any C* algebra,

\begin{proposition}\label{proposition:pairing_topological_k_1_odd_ciclyc_cohomology_C_star_algebras}
Let $A$ be a unital C* algebra and $\mathcal{A} \subset A$ a smooth sub algebra of $A$, then the 
following defines a bilinear pairing between $K_1^{\text{top}}(A)$ and $HC_{\text{con}}^{\text{odd}}(\mathcal{A})$:
$$\langle[u],[\varphi]\rangle := \frac{\Gamma \left( \frac{2n+1}{2} + 1 \right)^{-1}}{ 2^{2n+1} \sqrt{2 i} }  (\text{tr}_{k} \# \varphi)( u_s^{-1} -1 , u_s -1, u^{-1} -1, \ldots, u_s -1)$$
where $u_s \in U_k(\mathcal{A})$, $[u_s] = [u]$ inside $K_1^{\text{top}}(A)$ and $\varphi \in Z_{\lambda, \text{con}}^{2 n +1}(\mathcal{A})$.
\end{proposition}
\begin{proof}
From \cref{proposition:cannonical_paths_invertibles_unitaries_smooth_sub_algebras} we know that for any $u \in  U_k(A)$ there is a $u_s \in U_k(\mathcal{A})$ such that $[u] = [u_s]$ inside $K_1^{\text{top}}(A)$, therefore, we can use the isomorphism $K_1^{\text{top}}(\mathcal{A}) \simeq K_1^{\text{top}}(A)$ (\cref{theorem:isomorphism_K_0_and_k_1_for_smooth_sub_algebras}) along with \cref{proposition:pairing_topological_k_1_odd_ciclyc_cohomology} to get the desired pairing.
\end{proof}

\chapter{More on the Non-Commutative Brillouin Torus}
\label{chap:more_on_twisted_crossed_products}

\section{Twisted crossed products}
\label{sec:twisted_crossed_products_apendix}

\subsection{Group C* algebra and amenable groups}
\label{sec:group_C_algebra_and_amenable_group}

If $A = \mathbb{C}$, and both $\alpha, \zeta$ are trivial, then, the twisted crossed product simplifies to the enveloping C* algebra of $L^1(G)$ (\cref{sec:trivial_twisting_actions}), in this case, the enveloping C* algebra of $L^1(G)$ is called the \textbf{group C* algebra of $G$}\index{group C* algebra} and is denoted by $C^*(G)$\index{$C^*(G)$}; recall that the enveloping C* algebra of $L^1(G)$ is the C* algebra generated by $L^1(G)$ (\cref{sec:Universal_C_star_algebras}). In \cref{remark:untary_representations_and_star_representations} it is mentioned that there is a one-to-one correspondence between unitary representations of $G$ and *-representations of $L^1(G)$, which can be deduced also from the corresponding statement for twisted crossed products (\cref{theorem:characterization_of_twisted_crossed_products}) when we take $\alpha, \zeta$ to be trivial.

Take $G$ an abelian locally compact group, we know that $L^1(G)$ is commutative (\cref{remark:convolution_abelian_groups}), therefore, $C^*(G)$ is commutative. The Gelfand-representation theorem (\cref{theorem:commutative_gelfand_naimark_theorem}) tells us that there is a locally compact Hausdorff space $Q$ such that $C^*(G) \simeq C_0(Q)$, with $Q = \Phi_{C^*(G)}$ (the space of characters of $C^*(G)$). Under this setting, the Fourier transform over $G$ can be used to show that $Q = \hat{G}$, with $\hat{G}$ the dual group of $G$, the reason behind this result if the following: The Fourier transform provides *-representations of $L^1(G)$ given by the formula $\widehat{f \ast g} = \widehat{f} \widehat{g}$, and $\widehat{f} \in C_0(\hat{G})$ (\cref{definition:Fourier_transform}) (those are *-representations because $C_0(\hat{G})$ is a C* algebra), moreover, the *-representations of $L^1(G)$ are in one to one correspondence with the elements of $\hat{G}$ (\cref{remark:Gelfand_transform_and_Fourier_transform}), since $\mathcal{F}(L^1(G))$ is dense in $C_0(\hat{G})$ and the norm of the image of $L^1(G)$ under the *-representations is the supremum norm $C_0(G)$, we have that $C^*(G) \simeq C_0(\hat{G})$, that is, $\Phi_{C^*(G)} = \hat{G}$ (\citep[Theorem 5.18]{sundar_notes_202}).

By the previous argument, we can consider the the twisted crossed products to be the generalization of the group C* algebras, since they fall back to the group C* algebras when the action and the normalized 2-cocycle are trivial, and they generalize the notion of being the enveloping C* algebra of a $L^1$ Banach *algebra. Interestingly, $C^*(G)$ is nuclear iff $G$ is amenable (\citep[Theorem 4.2]{lance_nuclear_1973}).

Given that all locally compact abelian groups are amenable (\citep[Proposition 11.1.3]{dales_introduction_2003}) and $G$ is assumed to be abelian (\cref{def:separable_twisted_dynamical_system}), the content of  \citep[Corollary 3.9]{packer_twisted_1989} provides the following result,

\begin{theorem}[c.f. Corollary 3.9 \citep{packer_twisted_1989} (Twisted crossed products and nuclearity)]\label{theorem:twsited_crossed_product_amenable_nuclear}
Let $(A, G, \alpha, \zeta)$ be a separable twisted dynamical, then, if $A$ is nuclear we have that $A \rtimes_{\alpha, \zeta} G$ is nuclear.
\end{theorem}

\subsection{Discrete groups}
\label{sec:appendix_twisted_crossed_produt_discrete_groups}

There is one additional result gives a more complete description of the algebraic characterization of twisted crossed products with discrete separable groups,

\begin{lemma}\label{lemma:twisted_crossed_products_of_Generated_C_star_algebras}

Let $\mathcal{A}$ be a *-algebra and $A = C^*(\mathcal{A})$ and $G$ be a discrete, countable, and commutative group, denote by $L^1(G,\mathcal{A};\alpha,\zeta)_c$ the *-algebra of functions from $G$ into $\mathcal{A}$ that satisfy the same algebraic relations of $L_1(G,A;\alpha,\zeta)_c$ (\cref{def:algebra_generalized_trigonometric_polynomials}), additionally, let $(G,A,\alpha,\zeta)$ be a separable twisted dynamical system. Then, we have the following isomorphism,
$$ A \rtimes_{\alpha,\zeta} G \simeq C^*(L^1(G,\mathcal{A};\alpha,\zeta)_c), $$
or equivalently, $A \rtimes_{\alpha,\zeta} G$ is the C* algebra generated by the set
$$\mathcal{R} =  \{ a u_s :\; a \in \mathcal{A} , s \in G\},$$
subject to the following relations
\begin{itemize}
    \item \textbf{addition:} $a_0 u_s + a_1 u_s = (a_0 + a_1)u_s $ where $a_0 + a_1$ is an element of $A$.
    \item \textbf{multiplication:} $(a_0 u_s)( a_1 u_t) = (\zeta(s,t) a_0 \alpha(s)(a_1)) u_{s+t} $ where $\zeta(s,t) a_0 \alpha(s)(a_1)$ is an element of $A$.
    \item \textbf{Involution:} $(au_s)^* = \zeta(-s,s)^*\alpha(-s)(a^*) u_{-s}$ where $\zeta(-s,s)^*\alpha(-s)(a^*)$ is an element of $A$.
\end{itemize}
Where $a_i \in \mathcal{A}$ and $s,t \in G$.
\end{lemma}
\begin{proof}
First, we need to show that $C^*(L^1(G,\mathcal{A};\alpha,\zeta)_c)$ exists, then, we can proceed to proof that it is isomorphic to $A \rtimes_{\alpha,\zeta} G$. Following an argument similar to the one in the proof of \cref{proposition:twisted_crossed_product_restricts_to_discrete_conunatlbe_groups} we can check that for any 
representation $\pi$ of $L^1(G, \mathcal{A}; \alpha, \zeta)$ there is a representation of $\mathcal{A}$ given by $a \mapsto \pi(a u_0)$, thus, we must have that $\| a u_0 \| \leq \|a\|_A$ for all $a \in \mathcal{A}$ since $A = C^*(\mathcal{A})$; recall that the norm of $a \in \mathcal{A}$ as an element of $A$ is the supremum over the norm of all the representations of $\mathcal{A}$ (\cref{sec:Universal_C_star_algebras}). Similarly, is possible to check that $\| a u_s \| \leq \|a \|_A$ for all $a \in \mathcal{A}$, hence, any representation $\pi$ of $L^1(G,\mathcal{A};\alpha,\zeta)_c$ is continuous and
$$ \| \pi( \sum_{s \in G} f(s) u_s ) \| \leq \sum_{s \in G} \| f(s) \| = \| f\|_1.$$
Given that all the norm of $\pi(f)$ is bounded for any $\pi$ a representation of $L^1(G,\mathcal{A};\alpha,\zeta)_c$ the C* algebra generated by $L^1(G,\mathcal{A};\alpha,\zeta)_c$ exists. 

Since $L^1(G,\mathcal{A};\alpha,\zeta)_c$ is a sub *-algebra of $L^1(G,A;\alpha,\zeta)_c$, then, it is a sub *-algebra of $A \rtimes_{\alpha,\zeta} G$ due to \cref{proposition:twisted_crossed_product_restricts_to_discrete_conunatlbe_groups}, then, \cref{proposition:factoring_representations_with_enveloping_C_star_algebras} tells us that there must be a C* homomorphism
$$ \phi : C^*(L^1(G,\mathcal{A};\alpha,\zeta)_c) \to A \rtimes_{\alpha,\zeta} G $$
such that $\phi|_{L^1(G,\mathcal{A};\alpha,\zeta)_c} = i|_{L^1(G,\mathcal{A};\alpha,\zeta)_c}$ with $i$ the canonical inclusion of $L^1(G,A;\alpha,\zeta)_c$ on $A \rtimes_{\alpha,\zeta} G$. Additionally, $\phi$ is surjective because $\phi(L^1(G,\mathcal{A};\alpha,\zeta)_c)$ is dense in $A \rtimes_{\alpha,\zeta} G$, which comes from the density of $L^1(G,\mathcal{A};\alpha,\zeta)_c$ on $L^1(G,A;\alpha,\zeta)_c$. 

Since $\mathcal{A}$ is dense in $A$, $\| \pi(f) \| \leq \| f\|_1$ for every $f \in L^1(G,\mathcal{A};\alpha,\zeta)_c$ and $\pi$ a representation of $L^1(G,\mathcal{A};\alpha,\zeta)_c$, it is possible to extend $\pi$ into a representation of $L^1(G,A;\alpha,\zeta)_c$, thus, by \cref{proposition:factoring_representations_with_enveloping_C_star_algebras} there must be a C* homomorphism 
$$ \tilde{\phi}: A \rtimes_{\alpha,\zeta} G \to C^*(L^1(G,\mathcal{A};\alpha,\zeta)_c) $$
such that $\tilde{\phi}|_{L^1(G,\mathcal{A};\alpha,\zeta)_c} = \tilde{i}|_{L^1(G,\mathcal{A};\alpha,\zeta)_c}$ with $\tilde{i}$ the canonical inclusion of $L^1(G,\mathcal{A};\alpha,\zeta)_c$ in $C^*(L^1(G,\mathcal{A};\alpha,\zeta)_c)$. Additionally, $\tilde{\phi}$ is surjective because $\tilde{\phi}(L^1(G,A;\alpha,\zeta)_c)$ is dense in $C^*(L^1(G,\mathcal{A};\alpha,\zeta)_c)$.

Given that both $\phi$ and $\tilde{\phi}$ are surjective and $\phi|_{L^1(G,\mathcal{A};\alpha,\zeta)_c} = \tilde{\phi}^{-1}|_{L^1(G,\mathcal{A};\alpha,\zeta)_c}$, the automatic continuity of C* homomorphisms (\cref{proposition:automatic_continuity_C_star_algebras}) tell us that
$$ \forall f \in L^1(G,\mathcal{A};\alpha,\zeta)_c, \;  \| \phi(f) \| = \| i(f) \|. $$
Since $i(L^1(G,\mathcal{A};\alpha,\zeta)_c)$ is dense in $A \rtimes_{\alpha,\zeta} G$ we get that $\phi$ is a C* isomorphism and $\tilde{\phi}$ is its inverse.
\end{proof}

\subsection{Twisted crossed products with $\mathbb{Z}^{d}$}
\label{section:twisted_crossed_products_with_Z_apendix}

The characterization of normalized 2-cocycles on $\mathbb{Z}^d$ (\cref{proposition:characterization_of_+cocycles_over_integers}) tell us that there can not be a non-trivial normalized 2-cocycle over $\mathbb{Z}$, thus, we can only define crossed products with twisting actions of $\mathbb{Z}$ over $A$ with trivial normalized 2-cocycles. Recall that the twisted crossed product is a generalization of the crossed product because the crossed product\index{crossed product} corresponds to a twisted crossed product with a trivial normalized 2-cocycle (\citep[Definition II.10.3.7]{blackadar_operator_2006}). 
 
Let $(\alpha, \zeta)$ be a twisting action of $\mathbb{Z}^d$ on $A$ and $e_j = (0, \cdots, i, \cdots, 0)$ wiht a $1$ in the $j$ position, then, $\alpha_1: \mathbb{Z} \to \text{Aut}(A)$ defined by $\alpha_1(s) = \alpha(se_1)$ is an action of $\mathbb{Z}$ over $A$, and to this action there is an associated crossed product $A \rtimes_{\alpha_1} \mathbb{Z}$, which by \cref{lemma:twisted_crossed_products_with_Z} is the enveloping C* algebra of $\mathcal{P}(\mathbb{Z},A, \alpha_1)$.

Take $a \in \mathcal{P}(\mathbb{Z}^d,A, \alpha,\Theta)$ given by $a = \sum_{s_1 \in J_1} a(s_1) u_1^{s_1}$ with $J_1$ a finite set of $\mathbb{Z}$, then
$$ u_2^{s_2} \dotsc u_d^{s_d} (a) = \left(\sum_{s_1 \in J_1} \alpha\left(\sum_{2 \leq j \leq d}s_j e_j \right) (a(s_1)) \exp\left(i (s_1 e_1)^t \Theta (\sum_{2 \leq j \leq d} s_j e_j) \right) u_1^{s_1}\right) u_2^{s_2} \dotsc u_d^{s_d}. $$
Under this setup we have an action of $\mathbb{Z}^{d-1} = \mathbb{Z}^{d} / \mathbb{Z}$ over $\mathcal{P}(\mathbb{Z},A, \alpha_1)$ that takes the following form
$$ \hat{\alpha}_{d-1}(s_2, \dotsc, s_{d})(\sum_{s_1 \in J_1} a(s_1) u_1^{s_1}) = \sum_{s_1 \in J_1} \alpha\left(\sum_{2 \leq j \leq d}s_j e_j \right) (a(s_1)) \exp\left(i (s_1 e_1)^t \Theta (\sum_{2 \leq j \leq d} s_j e_j) \right) u_1^{s_1}. $$
By \citep[Theorem 4.1]{packer_twisted_1989} we get $ \hat{\alpha}_{d-1}$ can be extended to an action over $A \rtimes_{\alpha_1} \mathbb{Z} $, such that if we set $\Theta_{d-1} \in M_{d-1}(\mathbb{C})$ with 
$$\Theta_{d-1}(i,j) = \Theta(i+1,j+1),$$
then, $\zeta_{d-1}(x,y) = \exp(ix^t \Theta_{d-1}y)$ defines a normalized 2-cocycle over $\mathbb{Z}^{d-1}$ such that
$$ \left( A \rtimes_{\alpha_1} \mathbb{Z} \right) \rtimes_{\Theta_{d-1}, \hat{\alpha}_{d-1}} \mathbb{Z}^{d-1} \simeq A \rtimes_{\alpha,\Theta} \mathbb{Z}^{d},$$
with the isomorphism taking the expected form on finite polynomials,
$$ (\sum_{s_1 \in J_1} a(s_1) u_1^{s_1}) u_2^{s_2} \dotsc u_d^{s_d} \mapsto \sum_{s_1 \in J_1} a(s_1) u_1^{s_1} u_2^{s_2} \dotsc u_d^{s_d}. $$

Given that $\mathcal{P}(\mathbb{Z},A, \alpha_1)$ is dense in $A \rtimes_{\alpha_1} \mathbb{Z}$, then the algebra generated by elements of the form $$ a u_j \text{ with } a \in \mathcal{P}(\mathbb{Z},A, \alpha_1) , \;2 \leq j \leq d ,$$ is a dense *-algebra of $A \rtimes_{\alpha,\Theta} \mathbb{Z}^d$, moreover, algebraic manipulations show us that it is isomorphic to $\mathcal{P}(\mathbb{Z}^d,A, \alpha,\Theta)$ under the aforementioned map. The following proposition is a consequence of the previous discussion, 

\begin{proposition}
Let $(\mathbb{Z}^d,A,\alpha,\zeta)$ be a separable twisted dynamical system with $\zeta(x,y) = \exp(i x^t \Theta y)$, where $\Theta$ is a lower triangular matrix with zeros in the diagonal and entries in $[0,2\pi)$, then,
\begin{enumerate}
    \item Let $e_j = (0, \cdots, 1, \cdots, 0)$ with a $1$ in the $j$ position, set, $\alpha_1: \mathbb{Z} \to \text{Aut}(A)$ defined by $\alpha_1(s) = \alpha(se_1)$ be an action of $\mathbb{Z}$ over $A$, and denote by $A \rtimes_{\alpha_1} \mathbb{Z}$ the enveloping C* algebra of $\mathcal{P}(\mathbb{Z},A, \alpha_1)$. Then, there is an action of $\mathbb{Z}^{d-1} = \mathbb{Z}^{d} / \mathbb{Z}$ over $\mathcal{P}(\mathbb{Z},A, \alpha_1)$ that takes the following form
    $$ \hat{\alpha}_{d-1}(s_2, \dotsc, s_{d})(\sum_{s_1 \in J_1} a(s_1) u_1^{s_1}) = \sum_{s_1 \in J_1} \alpha\left(\sum_{2 \leq j \leq d}s_j e_j \right) (a(s_1)) \exp\left(i (s_1 e_1)^t \Theta (\sum_{2 \leq j \leq d} s_j e_j) \right) u_1^{s_1}, $$
    where $\sum_{s_1 \in J_1} a(s_1) u_1^{s_1} \in \mathcal{P}(\mathbb{Z},A, \alpha_1)$, and $\hat{\alpha}_{d-1}$ can be extended to an action of $\mathbb{Z}^{d-1}$ over $A \rtimes_{\alpha_1} \mathbb{Z}$.
    \item Denote $\Theta_{d-1} \in M_{d-1}(\mathbb{C})$ with 
    $$\Theta_{d-1}(i,j) = \Theta(i+1,j+1),$$
    then, and there is a C* algebra isomorphism
    $$ \left( A \rtimes_{\alpha_1} \mathbb{Z} \right) \rtimes_{\Theta_{d-1}, \hat{\alpha}_{d-1}} \mathbb{Z}^{d-1} \simeq A \rtimes_{\alpha,\Theta} \mathbb{Z}^{d},$$
    with the isomorphism taking the following form on finite polynomials,
    $$ (\sum_{s_1 \in J_1} a(s_1) u_1^{s_1}) u_2^{s_2} \dotsc u_d^{s_d} \mapsto \sum_{s_1 \in J_1} a(s_1) u_1^{s_1} u_2^{s_2} \dotsc u_d^{s_d}. $$
\end{enumerate}
\end{proposition}

Since $\Theta_{d-1}$ is a lower triangular matrix with zeros in the diagonal, we can iterate the definition of the aforementioned isomorphism to get a description of $A \rtimes_{\alpha,\Theta} \mathbb{Z}^{d}$ as an iterated crossed product
$$ A \rtimes_{\alpha_1} \mathbb{Z}  \rtimes_{\alpha_2} \mathbb{Z} \dotsc \rtimes_{\alpha_d}  \mathbb{Z}, $$
with $\alpha_k : \mathbb{Z} \to \text{Aut}(A \rtimes_{\alpha_1} \mathbb{Z}  \rtimes_{\alpha_2} \mathbb{Z} \dotsc \rtimes_{\alpha_{k-1}}  \mathbb{Z})$ given by
$$ \alpha_k(s_k)\left( \sum_{(s_1, \dotsc, s_{k-1}) \in J_{k-1}} a(s_1, \dotsc, s_{k-1}) u_1^{s_1} \dotsc u_{k-1}^{s_{k-1}} \right)  = $$
$$ \sum_{(s_1, \dotsc, s_{k-1}) \in J_{k-1}} \alpha\left(s_d e_d \right) (a(s_1, \dotsc, s_{k-1})) \exp\left(i \left(\sum_{1 \leq j \leq k-1} s_j e_j\right)^t \Theta (s_k e_k) \right) u_k^{s_k}, $$
and $J_{k-1}$ a finite subset subset of $\mathbb{Z}^{k-1}$. Notice that we have given an isomorphism $\alpha_k(s_k)$ on the dense sub algebra $\mathcal{P}(\mathbb{Z}^{k-1},A, \alpha,\Theta)$ and \citep[Theorem 4.1]{packer_twisted_1989} assure us that is can be extended to the whole C* algebra $A \rtimes_{\alpha_1} \mathbb{Z}  \rtimes_{\alpha_2} \mathbb{Z} \dotsc \rtimes_{\alpha_{d-1}}  \mathbb{Z}$.

Additionally, since $A$ is a sub C* algebra of $A \rtimes_{\alpha,\Theta} \mathbb{Z}^d$ (\cref{remark:canonical_inclusion_of_twisted_crossed_products}), we have that $A \rtimes_{\alpha_1} \mathbb{Z}  \rtimes_{\alpha_2} \mathbb{Z} \dotsc \rtimes_{\alpha_{k-1}}  \mathbb{Z}$ is a sub C* algebra of $A \rtimes_{\alpha,\Theta} \mathbb{Z}^d$ for very $k < d$.

\begin{proposition}\label{lemma:iterated_crossed_product}
Let $(\mathbb{Z}^d,A,\alpha,\zeta)$ be a separable twisted dynamical system with $\zeta(x,y) = \exp(i x^t \Theta y)$, where $\Theta$ is a lower triangular matrix with zeros in the diagonal and entries in $[0,2\pi)$, let $e_j = (0, \cdots, 1, \cdots, 0)$ with a $1$ in the $j$ position, then, $A \rtimes_{\alpha,\Theta} \mathbb{Z}^{d}$ is isomorphic to the iterated crossed product,
$$ A \rtimes_{\alpha_1} \mathbb{Z}  \rtimes_{\alpha_2} \mathbb{Z} \dotsc \rtimes_{\alpha_d}  \mathbb{Z}, $$
where, $\alpha_k : \mathbb{Z} \to \text{Aut}(A \rtimes_{\alpha_1} \mathbb{Z}  \rtimes_{\alpha_2} \mathbb{Z} \dotsc \rtimes_{\alpha_{k-1}}  \mathbb{Z})$ given by
$$ \alpha_k(s_k)\left( \sum_{(s_1, \dotsc, s_{k-1}) \in J_{k-1}} a(s_1, \dotsc, s_{k-1}) u_1^{s_1} \dotsc u_{k-1}^{s_{k-1}} \right)  = $$
$$ \sum_{(s_1, \dotsc, s_{k-1}) \in J_{k-1}} \alpha\left(s_d e_d \right) (a(s_1, \dotsc, s_{k-1})) \exp\left(i \left(\sum_{1 \leq j \leq k-1} s_j e_j\right)^t \Theta (s_k e_k) \right) u_k^{s_k}, $$
and $J_{k-1}$ a finite subset subset of $\mathbb{Z}^{k-1}$.
\end{proposition}

\subsection{Fourier analysis}
\label{sec:Fourier_analysis_twistted_crossed_product_apendix}

In this section, we will expand on the properties of the Fourier coefficients. By definition of the twisted crossed product we know that $L^1(\mathbb{Z}^d,A;\alpha, \Theta)$ is a sub *-algebra of $A \rtimes_{\alpha,\Theta}\mathbb{Z}^d$ (\cref{definition:twsited_crossed_product}), which lead us to point out how the twisted crossed products are a generalization of the group C* algebras in \cref{sec:group_C_algebra_and_amenable_group}. Under this setting, the Banach *-algebra $L^1(\mathbb{Z}^d,A;\alpha,\Theta)$ is a generalization of the Wiener algebra\index{Wiener algebra}, which is mentioned in \cref{example:Fourier_transform_and_dual_groups}, and the elements of $L^1(\mathbb{Z}^d,A;\alpha,\Theta)$ take the form of an infinite series, with this we mean that
$$  \sum_{s \in \mathbb{Z}^d} \Phi_s(p) u^s $$
converges inside $A \rtimes_{\alpha,\Theta}\mathbb{Z}^d$. Not all the elements of $A \rtimes_{\alpha,\Theta}\mathbb{Z}^d$ can be described as an infinite series of its Fourier coefficients, as we discuss in \cref{section:convergence_of_the_Fourier_Series_twisted_crossed_products}.

\subsubsection{Weights, traces and states}
\label{section:fourier_analysis_weight_traces_and_states_apendix}

In our efforts to generalize the C* algebra $C(\mathbb{T}^d)$, we have provided non-trivial commutation relations for the generators $\{a u_j\}_{a \in A, 1 \leq j \leq d}$, and we got in return the C* algebra $A \rtimes_{\alpha,\Theta}\mathbb{Z}^d$, which has a faithful representation that resembles the one of $C(\mathbb{T}^d)$, and its elements have Fourier coefficients that resembles those of elements of $C(\mathbb{T}^d)$. Along this path, we have replaced $\mathbb{C}$ with a C* algebra $A$, now, we will look at how we can use the zeroth Fourier coefficient ($\Phi_0(p)$) to provide a generalization of the $L^2$ norm of an element of $C(\mathbb{T}^d)$, using the fact that the set of positive elements of a C* algebra ($A_{\text{pos}}$) share various properties with the set of positive elements of $\mathbb{C}$ (\cref{proposition:cahracterization_of_positive_elements}).

\begin{lemma}[Generalized norm of an element in $ A \rtimes_{\alpha,\Theta}\mathbb{Z}^d$]\label{lemma:generalized_norm_of_twisted_crossed_product}
The zeroth Fourier coefficient
$$ \Phi_0 : A \rtimes_{\alpha,\Theta}\mathbb{Z}^d \to A , $$
has the following properties,
\begin{itemize}
    \item $\Phi_0((A \rtimes_{\alpha,\Theta}\mathbb{Z}^d)_{\text{pos}}) = A_{\text{pos}}$.
    \item If $p \in (A \rtimes_{\alpha,\Theta}\mathbb{Z}^d)_{\text{pos}}$ then $\Phi_0(p) = 0$ iff $p =0$.
\end{itemize}
\end{lemma}
\begin{proof}
The following assertions come from the characterization of positive elements of a C* algebra given in  \cref{proposition:cahracterization_of_positive_elements}. Since, any positive element of $A \rtimes_{\alpha,\Theta}\mathbb{Z}^d$ looks like $p p^*$, the computations done in \cref{lemma:Fourier_coefficients_multiplication_and_involution} tell us that
\begin{itemize}
    \item Given the sequence of generalized trigonometric polynomials $\{ p^{(n)} \}_{n \in \mathbb{N}}$ that corresponds to the generalized Fejér summation of $p$ (\cref{definition:generalized_fourier_sequence_and_fejer_summation}), we have that
    $$ \Phi_0(p p^*) =  \Phi_0\left(\lim_{n \to \infty} p^{(n)} (p^{(n)})^*\right) $$
    $$= \lim_{n \to \infty} \Phi_0\left(p^{(n)} (p^{(n)})^*\right) = \lim_{n \to \infty} \sum_{x \in V_n} \prod_{j=1}^d\left(1-\frac{\left|x_j\right|}{n+1}\right)^{2} \Phi_x(p) \Phi_{x}(p)^*,$$
    additionally, $\Phi_x(p) \Phi_{x}(p)^* \in A_{\text{pos}}$. Since the set of positive elements of $A$ is closed and the sum of positive elements is positive we have that $\Phi_0(p p^*) \in A_{\text{pos}}$. The mapping is surjective because $\Phi_0(\sqrt{a}u^0) = a$ for any $a \in A_{\text{pos}}$.
    
    \item Since the norm of a sum of positive elements is greater or equal to the norm of each of the positive elements being added, we have that
    $$ \| \sum_{x \in V_n} \prod_{j=1}^d\left(1-\frac{\left|x_j\right|}{n+1}\right)^{2} \Phi_x(p) \Phi_{x}(p)^* \| \geq  \| \prod_{j=1}^d\left(1-\frac{\left|x_j\right|}{n+1}\right)^{2} \Phi_x(p) \Phi_{x}(p)^* \| .$$
    Also, from the continuity of the map $\Phi_0$ we get that 
    $$\Phi_0(p p^*) = \lim_{n \to \infty} \Phi_0\left(p^{(n)} (p^{(n)})^*\right),$$
    therefore,
    $$\| \Phi_0(p p^*) \| \geq \| \prod_{j=1}^d\left(1-\frac{\left|x_j\right|}{n+1}\right)^{2} \Phi_x(p) \Phi_{x}(p)^* \| \; \forall x \in \mathbb{Z}^d \text{ and } n \in \mathbb{N},$$
and we end up with $\| \Phi_0(p p^*) \| \geq \| \Phi_x(p) \|^2$. Consequently, $\| \Phi_0(p p^*) \| = 0$ iff $\| \Phi_x (p) \| = 0$ for all $x \in \mathbb{Z}^d$, or equivalently, if $p = 0$.
\end{itemize}
\end{proof}

The map $p \mapsto \Phi_0(p p^*)$ falls back to 
$$ f \mapsto \sum_{s \in \mathbb{Z}^d} |\mathcal{F}(f)(s) |^2 $$
when we work with the algebra $C(\mathbb{T}^d)$, and the isomorphism given by the Fourier transform (\cref{theorem:Plancherel_theorem}) allow us to say that the aforementioned map takes the form
$$ f \mapsto \|f \|^2. $$
When we move into the realm of $\mathbb{C}$ valued functions, we have that $|\mathcal{F}(f)(s) |^2 \in \mathbb{R}^+$, and we can get rid of the weights associated to the Fejér sum because the series has only positive numbers. Interestingly enough, the set of positive elements in a C* algebras behave much like positive real numbers, such that
$$ \Phi_0 (p p^*) = \sum_{s \in \mathbb{Z}^d} \Phi_s (p) \Phi_s (p)^*, $$
makes sense because the sum can take any ordering, so, let's check why this happens.

Recall that $\{ p^{(n)} \}_{n \in \mathbb{N}}$ is the generalized Fejér summation of $p$ and $\{ S^{(n)}(p) \}_{n \in \mathbb{N}}$ is the generalized Fourier series of $p$ \cref{definition:generalized_fourier_sequence_and_fejer_summation}). First, $\{ \Phi_0(p^{(n)} (p^{(n)})^*) \}_{n \in \mathbb{N}}$ is a converging sequence of positive elements, and 
$$ \Phi_0(p^{(n)} (p^{(n)})^*) \leq \Phi_0(p^{(m)} (p^{(m)})^*)$$
for $n \leq m$, since 
$$ \prod_{j=1}^d\left(1-\frac{\left|x_j\right|}{n+1}\right)^{2} \Phi_x(p) \Phi_{x}(p)^* \leq \prod_{j=1}^d\left(1-\frac{\left|x_j\right|}{m+1}\right)^{2} \Phi_x(p) \Phi_{x}(p)^*   $$
when $n \leq m$ and $x \in V_n$. This implies that the aforementioned sequence is a Cauchy sequence, thus, for every $\epsilon > 0$ there is $N > 0$ such that if $n,m \geq N$ we have that 
$$ \|  \Phi_0(p^{(m)} (p^{(m)})^*) - \Phi_0(p^{(n)} (p^{(n)})^*) \| = $$
$$\|  \sum_{x \in V_m} \prod_{j=1}^d\left(1-\frac{\left|x_j\right|}{m+1}\right)^{2} \Phi_x(p) \Phi_{x}(p)^* - \sum_{x \in V_n} \prod_{j=1}^d\left(1-\frac{\left|x_j\right|}{n+1}\right)^{2} \Phi_x(p) \Phi_{x}(p)^* \| \leq \epsilon,$$
so, assume that $m \geq n$, then, $ \Phi_0(p^{(m)} (p^{(m)})^*) - \Phi_0(p^{(n)} (p^{(n)})^*)$ consists of the sum of two positive elements, each one of norm less than or equal to $\epsilon$ due to \cref{proposition:cahracterization_of_positive_elements},
$$ \Phi_0(p^{(m)} (p^{(m)})^*) - \Phi_0(p^{(n)} (p^{(n)})^*) = $$
$$\sum_{x \in V_n} \left[ \prod_{j=1}^d  \left(1-\frac{\left|x_j\right|}{m+1}\right)^{2} - \prod_{j=1}^d  \left(1-\frac{\left|x_j\right|}{n+1}\right)^{2} \right] \Phi_x(p) \Phi_{x}(p)^* + $$
$$\sum_{\substack{ x \in V_m \\  x \notin V_n}} \prod_{j=1}^d\left(1-\frac{\left|x_j\right|}{m+1}\right)^{2} \Phi_x(p) \Phi_{x}(p)^*. $$

Notice that
$$ \Phi_0(S^{(n)}(p) (S^{(n)}(p) )^* ) = \sum_{x \in V_n} \Phi_x(p) \Phi_{x}(p)^*,$$
so, 
$$ \| \Phi_0 (p p^*) - \Phi_0(S^{(n)}(p) (S^{(n)}(p) )^* ) \| \leq $$
$$\| \Phi_0 (p p^*) - \Phi_0(p^{(m)} (p^{(m)})^*) \|  + \| \Phi_0(p^{(m)} (p^{(m)})^*) - \Phi_0(S^{(n)}(p) (S^{(n)}(p) )^* ) \|, $$
the first term can be made as small as desired because the sequence converges, hence, let's focus on the second one. Assume that $m \geq n$, then
$$  \Phi_0(p^{(m)} (p^{(m)})^*) - \Phi_0(S^{(n)}(p) (S^{(n)}(p) )^* ) =   $$
$$   \sum_{\substack{ x \in V_m \\  x \notin V_n}} \prod_{j=1}^d\left(1-\frac{\left|x_j\right|}{m+1}\right)^{2} \Phi_x(p) \Phi_{x}(p)^* -  \left( \sum_{x \in V_n} \left[ 1 - \prod_{j=1}^d  \left(1-\frac{\left|x_j\right|}{m+1}\right)^{2} \right] \Phi_x(p) \Phi_{x}(p)^*  \right) ,$$
we have mentioned that the first term can be small as desired as a consequence of the properties of positive elements in a C* algebra, also, if $n$ is fixed and $m$ can be taken arbitrarily large the second term can be made arbitrarily small. Consequently, if $N \in \mathbb{N}$ such that
\begin{itemize}
    \item $ \| \Phi_0 (p p^*) - \Phi_0(p^{(m)} (p^{(m)})^*) \| \leq \epsilon/3 $ when $n \geq N$,
    \item $ \| \Phi_0(p^{(m)} (p^{(m)})^*) - \Phi_0(p^{(n)} (p^{(n)})^*) \| \leq \epsilon/3 $ when $n,m \geq N$,
\end{itemize}
we have that,
\begin{itemize}
    \item for any $l \geq N$ there is $M_l \geq N, l$ such that, when $k \geq M_l$
    $$ \| \sum_{x \in V_l} \left[ 1 - \prod_{j=1}^d  \left(1-\frac{\left|x_j\right|}{k+1}\right)^{2} \right] \Phi_x(p) \Phi_{x}(p)^* \| \leq \epsilon /3 $$
\end{itemize}
and
$$ \| \Phi_0 (p p^*) - \sum_{x \in V_l} \Phi_x(p) \Phi_{x}(p)^* \| \leq    $$
$$ \| \Phi_0 (p p^*) - \Phi_0(p^{(k)} (p^{(k)})^*) \| + \| \sum_{\substack{ x \in V_k \\  x \notin V_l}} \prod_{j=1}^d\left(1-\frac{\left|x_j\right|}{m+1}\right)^{2} \Phi_x(p) \Phi_{x}(p)^* \| $$
$$+ \|  \left( \sum_{x \in V_l} \left[ 1 - \prod_{j=1}^d  \left(1-\frac{\left|x_j\right|}{k+1}\right)^{2} \right] \Phi_x(p) \Phi_{x}(p)^*  \right) \|  $$
$$ \leq \epsilon.  $$

The previous calculations tell us that
$$ \sum_{x \in V_l} \Phi_x(p) \Phi_{x}(p)^* \xrightarrow[l \to \infty]{} \Phi(p p^*)_0, $$
additionally, if $P \subset \mathbb{Z}^d$ and $V_l \subset P$ then
$$ \| \Phi(p p^*) -  \sum_{x \in P} \Phi_x(p) \Phi_{x}(p)^* \| \leq \| \Phi(p p^*) - \sum_{x \in V_l} \Phi_x(p) \Phi_{x}(p)^* \| $$
$$+ \| \sum_{x \in P} \Phi_x(p) \Phi_{x}(p)^* - \sum_{x \in V_l} \Phi_x(p) \Phi_{x}(p)^* \|. $$
Since $\sum_{x \in P} \Phi_x(p) \Phi_{x}(p)^* \geq \sum_{x \in V_l} \Phi_x(p) \Phi_{x}(p)^*$ and $\Phi(p p^*) \geq \sum_{x \in P} \Phi_x(p) \Phi_{x}(p)^*$ we have that
$$ \Phi(p p^*) -  \sum_{x \in V_l} \Phi_x(p) \Phi_{x}(p)^* \geq \sum_{x \in P} \Phi_x(p) \Phi_{x}(p)^* - \sum_{x \in V_l} \Phi_x(p) \Phi_{x}(p)^*,$$
thus, for $\sum_{x \in P} \Phi_x(p) \Phi_{x}(p)^*$ to be close to $ \Phi(p p^*)$ we only require $P$ to contain the appropriate $V_l$, ie,
$$ \| \Phi(p p^*) -  \sum_{x \in V_l} \Phi_x(p) \Phi_{x}(p)^* \| \leq \epsilon/2.$$

We can condensate the previous discussion on the following lemma,

\begin{lemma}[Absolute convergence of $\Phi_0 (p p^*)$]\label{lemma:absolute_convergence_of_zero_fourier_coefficient}
Let $p \in A \rtimes_{\alpha,\Theta}\mathbb{Z}^d$, then, 
$$ \Phi_0 (p p^*) = \sum_{s \in \mathbb{Z}^d} \Phi_s(p) \Phi_s (p)^*,  $$
where the order in the sum does not matter.
\end{lemma}

\begin{corollary}\label{lemma:absolute_convergence_of_zero_fourier_coefficient_inverse_order}
Using the identities for the involution on $A \rtimes_{\alpha,\Theta}\mathbb{Z}^d$ (\cref{lemma:Fourier_coefficients_multiplication_and_involution}), we get that, if $p \in A \rtimes_{\alpha,\Theta}\mathbb{Z}^d$, then, 
$$ \Phi_0 (p^* p) = \sum_{s \in \mathbb{Z}^d} \alpha(s) \left(\Phi_{-s} (p)^* \Phi_{-s}(p)  \right),  $$
where the order in the sum does not matter.
\end{corollary}

The map $p \mapsto \Phi_0(p)$ is a generalization of a faithful continuous weight over $A \rtimes_{\alpha,\Theta} \mathbb{Z}^d$ (\cref{remark:weights_and_continuity}), since it is a continuous linear map over $( A \rtimes_{\alpha,\Theta} \mathbb{Z}^d)_{\text{pos}}$ taking values over $A$ and for $p \in ( A \rtimes_{\alpha,\Theta} \mathbb{Z}^d)_{\text{pos}}$ we have that $\Phi_0(p) = 0$ iff $p =0$ (\cref{lemma:generalized_norm_of_twisted_crossed_product}). The continuous linear map $\Phi_0$ allows us to promote continuous linear functionals over $A$ into continuous linear functionals over $A \rtimes_{\alpha,\Theta}\mathbb{Z}^d$, such that, those functionals become traces if they are traces over $A$ and are invariant under the maps $\alpha(x)$,

\begin{remark}[Continuous linear functionals on  $A \rtimes_{\alpha,\Theta}\mathbb{Z}^d$]\label{remark:continuous_linear_functionals_twsited_crossed_product}
Let $\eta$ be a continuous linear functional on $A$, then, we can create a continuous linear functional on $A \rtimes_{\alpha,\Theta}\mathbb{Z}^d$ by setting 
$$\hat{ \eta} : A \rtimes_{\alpha,\Theta}\mathbb{Z}^d \to \mathbb{C}, \; \hat{ \eta}(p) := \eta(\Phi_0(p)).$$
Now let's look at what happens if $\eta$ is tracial and $\eta$ is invariant under $\alpha(x)$, in this case, the zeroth Fourier coefficient of the Fej\'er summation of $p$ takes the following form,
\begin{align*}
\hat{\eta} ( p^{(n)} (p^{(n)})^*) = \sum_{x \in V_n} \prod_{j \leq d} \left(1 - \frac{|x_j|}{n+1} \right)^2 \eta( \Phi_{x}(p) \Phi_{x}(p)^*) &\\
\hat{\eta} ( (p^{(n)})^* p^{(n)}) = \sum_{x \in V_n} \prod_{j \leq d} \left(1 - \frac{|x_j|}{n+1} \right)^2 \eta(  \alpha(x)( \Phi_{-x}(p)^* ) \alpha(x)(\Phi_{-x}(p))) &
\end{align*}
Recall that $\alpha(x)$ is a C* homomorphism for any $x \in \mathbb{Z}^d$, thus, 
$$ \alpha(x)( \Phi_{-x}(p)^* ) \alpha(x)(\Phi_{-x}(p)) = \alpha(x)( \Phi_{-x}(p)^* \Phi_{-x}(p)) ,$$
so, if $\eta$ is $\alpha$ invariant, that is, if $\eta \circ \alpha(x) (a) = \eta (a)$ for all $a \in A, \; x \in \mathbb{Z}^d$, we get that
$$ \hat{\eta} ( (p^{(n)})^* p^{(n)}) = \sum_{x \in V_n} \prod_{j \leq d} \left(1 - \frac{|x_j|}{n+1} \right)^2 \eta( \Phi_{-x}(p)^* \Phi_{-x}(p)) .$$
Since $\eta$ is tracial we have that $\eta( \Phi_{-x}(p)^* \Phi_{-x}(p)) = \eta( \Phi_{-x}(p) \Phi_{-x}(p)^*)$, which implies that
$$ \hat{\eta} ( (p^{(n)})^* p^{(n)}) = \hat{\eta} ( p^{(n)} (p^{(n)})^*). $$
Finally, we get $\hat{\eta}(p p^*) = \hat{\eta}(p^* p)$ from the fact that 
$$\hat{\eta} (p^{(n)} (p^{(n)})^*) \to \hat{\eta} (p p^*),$$ 
$$\hat{\eta} ( (p^{(n)})^* p^{(n)}) \to \hat{\eta} (p^* p),$$
as $n \to \infty$. Additionally, from \cref{lemma:generalized_norm_of_twisted_crossed_product} we know that for $p \in (A \rtimes_{\alpha,\Theta}\mathbb{Z}^d)_{\text{pos}}$, $\Phi_0(p) = 0$ iff $p=0$, thus, if $\eta$ is faithful then $\hat{\eta}$ is also faithful.
\end{remark}

\begin{remark}[Weight and traces on  $A \rtimes_{\alpha,\Theta}\mathbb{Z}^d$]\label{remark:weights_and_traces_twsited_crossed_product}
In a similar fashion, if $\eta$ is a weight over $A$, then, we can define a weight over A $\rtimes_{\alpha,\Theta} \mathbb{Z}^{d}$ by setting $$\hat{ \eta} : A \rtimes_{\alpha,\Theta}\mathbb{Z}^d \to \mathbb{C}, \; \hat{ \eta}(p) := \eta(\Phi_0(p)).$$
We have that
$$ \Phi_0(p^{(n)} (p^{(n)})^*) \leq \Phi_0(p^{(m)} (p^{(m)})^*) $$
when $n \leq m$. Under this setting \cref{remark:weights_and_continuity} tell us that if $\phi$ is a lower semicontinuous weight we have that
$$ \lim_{n \to \infty} \eta \left(\Phi_0 (p^{(n)} (p^{(n)})^*) \right) = \eta(\Phi_0 (p p^*)) $$
over the extended positive reals, that is, over $[0, \infty]$. Therefore, is possible to approximate the value of $\hat{\eta}$ using the Fejér sums.

Following a similar argument as in \cref{remark:continuous_linear_functionals_twsited_crossed_product}, if $\eta$ is a trace and is $\alpha$ invariant ($\eta = \eta \circ \alpha(x), x \in \mathbb{Z}^d$) then,
$$\hat{\eta} (p^{(n)} (p^{(n)})^*) \to \hat{\eta} (p p^*),$$ 
$$\hat{\eta} ( (p^{(n)})^* p^{(n)}) \to \hat{\eta} (p^* p),$$
and
$$ \hat{\eta} ( (p^{(n)})^* p^{(n)}) = \hat{\eta} ( p^{(n)} (p^{(n)})^*), $$
which implies $\hat{\eta}(p p^*) = \hat{\eta}(p^* p)$, that is, $\hat{\eta}$ is a trace. Additionally, if $\eta$ is faithful then $\hat{\eta}$ is also faithful.
\end{remark}




    



\subsubsection{Convergence of the Fourier series}
\label{section:convergence_of_the_Fourier_Series_twisted_crossed_products}

We have shown that $p^{(n)} \to p$ in the topology given by the C* norm on $A \rtimes_{\alpha,\Theta}\mathbb{Z}^d$, so, we can ask if 
$$\sum_{x \in \mathbb{Z}^{d}} \Phi_x (p)u^x \to p,$$
that is, if the generalized Fourier series converges in $A \rtimes_{\alpha,\Theta}\mathbb{Z}^d$. As you may guess, this is not the case in general, for example, if $A =  \mathbb{C}$, and both $\alpha$, $\Theta$ are trivial, then 
$$A \rtimes_{\alpha,\Theta}\mathbb{Z}^d \simeq C^{*}(u_1, ..., u_d, \; u_i u_j = u_j u_i, \; u_i u_i^{*} = u_i^{*} u_i  = 1) \simeq C(\mathbb{T}^{d}),$$ 
and is known that there are continuous functions over $\mathbb{T}$ whose Fourier series diverges in a point (\cref{section:convergence_of_the_Fourier_series}) i.e. the Fourier series does not converge to the function in the supremum norm (norm of $C(\mathbb{T}) \simeq C^{*}(u, \; u u^{*} = u^{*} u  = 1)$). On the other hand, the Fej\'er summation of $f$ provides a sequence of trigonometric polynomials that converge uniformly to $f$ if $f \in C(\mathbb{T}^d)$ (\cref{section:convergence_of_the_Fourier_series}), which is why we used the generalized Fej\'er summation to provide a sequence of non-commutative polynomials convergent to $a$ for every $a \in A \rtimes_{\alpha,\Theta}\mathbb{Z}^d$. 

Also, in the context of $A \rtimes_{\alpha,\Theta}\mathbb{Z}^d$ we have replaced the converge of 
$$ \sum_{x \in \mathbb{Z}^d} | \mathcal{F}(f)(x) |^2 = \sum_{x \in \mathbb{Z}^{d}} | \Phi_x (f) |^{2} \in \mathbb{R}^+$$
for $f \in C(\mathbb{T}^{d})$ by the fact (\cref{lemma:absolute_convergence_of_zero_fourier_coefficient})
$$\Phi_0(p p^*) = \sum_{x \in \mathbb{Z}^d} \Phi_x(p) \Phi_{x}(p)^* \in A_{\text{pos}}$$
for every $p \in A \rtimes_{\alpha,\Theta}\mathbb{Z}^d$. Following this reasoning, we can use $\Phi_0$ to provide $A \rtimes_{\alpha,\Theta}\mathbb{Z}^d$ with a positive definite inner product over $A$, let 
$$ \langle p, q \rangle = \left(\Phi_0(p q^*)\right)^* , \; p,q \in A \rtimes_{\alpha,\Theta}\mathbb{Z}^d,$$
then from \cref{lemma:generalized_norm_of_twisted_crossed_product} we know that $\langle p,p \rangle \geq 0$ and $\langle p, p \rangle = 0$ iff $p = 0$, also, using the relation between involution and Fourier coefficients (\cref{lemma:Fourier_coefficients_multiplication_and_involution}) we get that $(\langle a, b \rangle)^* = \langle b, a \rangle$. Following \cref{lemma:multiplicative_identity_of_twisted_crossed_product}, set the scalar multiplication on $A \rtimes_{\alpha,\Theta}\mathbb{Z}^d$ as
$$r_m : (A \rtimes_{\alpha,\Theta}\mathbb{Z}^d) \times A \to A \rtimes_{\alpha,\Theta}\mathbb{Z}^d, \; r_m\left( p,a \right) = p (a u^0),$$
then $\langle p, q a \rangle = \langle p,q \rangle a$ for $p,q \in A \rtimes_{\alpha,\Theta}\mathbb{Z}^d$ and $a \in A$. If $A$ is a unital C* algebra then we also have that $\langle u^x, u^y \rangle = 1_A\delta_{x,y}$ for $x,y \in \mathbb{Z}^d$, in the same way as $z \to \frac{1}{(2 \pi)^{n/2}} \exp{ i \langle z, x \rangle}$ are an ortonormal functions over $[0, 2 \pi]^d$. This description of $ A \rtimes_{\alpha,\Theta}\mathbb{Z}^d$ coincides with the first condition of a Hilbert C* module (\citep{nlab:hilbert_module}) that we discussed on \cref{section:non_commutative_geometry_dictionary}, however, at the moment we do not know if $A \rtimes_{\alpha,\Theta}\mathbb{Z}^d$ is complete with respect to the norm $\| p \| = \sqrt{\| \langle p, p \rangle \|}$, which is the second condition for $A \rtimes_{\alpha,\Theta}\mathbb{Z}^d$ to be a Hilbert C* module\index{Hilbert C* module}. Our best guess would be that it is not the case, instead, $A \rtimes_{\alpha,\Theta}\mathbb{Z}^d$ should be a pre-Hilbert module (\citep[Definition 15.1.1]{wegge-olsen_k-theory_1993}), because we are dealing with a generalization of the C* algebra $C(\mathbb{T}^d)$ which is not complete with respect to the $L^2$ norm on $L^2(\mathbb{T}^d)$.

Summation techniques for the Fourier series can be generalized into the context of twisted crossed products, such that one can ask for conditions on the convergence of 
$$\sum_{x \in \mathbb{Z}^{d}} \Lambda(x) \Phi_x (p)u^{x},$$
where $\Lambda: \mathbb{Z}^d \to \mathbb{C}$. For example, we have already looked into a summation technique that allowed us to approximate any element of $A \rtimes_{\alpha,\Theta}\mathbb{Z}^d$, which we called the generalized Fej\'er summation. For a review on summation techniques over twisted crossed products, you can take a look at \citep[section 3]{bedos_fourier_2016}.

In most references an arbitrary element of $A \rtimes_{\alpha,\Theta}\mathbb{Z}^d$ is represented as $\sum_{x \in \mathbb{Z}^{d}} \Phi_x (p)u^{x}$, for example \citep[section 5]{carey_index_2014} or \citep[chapter 3]{prodan_bulk_2016}. The notation $\sum_{x \in \mathbb{Z}^{d}} \Phi_x (p)u^{x}$ can be misleading because it seems to suggest that the series $\sum_{x \in \mathbb{Z}^{d}} \Phi_x (p)u^{x}$ converges in the norm topology inside $A \rtimes_{\alpha,\Theta}\mathbb{Z}^d$, but that is not the case in general as we said before. This notation must be taken with a grain of salt, cause it refers to the facts that
\begin{itemize}
    \item The non-commutative polynomial are dense in $A \rtimes_{\alpha,\Theta}\mathbb{Z}^d$,
    \item An element of $A \rtimes_{\alpha,\Theta}\mathbb{Z}^d$ is uniquely determined by its Fourier coefficients,
\end{itemize}
but it does not refer to norm convergence on $A \rtimes_{\alpha,\Theta}\mathbb{Z}^d$ unless otherwise stated.


\subsection{Fourier analysis on tensor products}
\label{sec:Fourier_analysis_on_tensor_products}

  The Commutative Gelfand-Naimark theorem (\cref{theorem:commutative_gelfand_naimark_theorem}) tells us that any commutative C* algebra takes the form $C_0(\Omega)$ with $\Omega$ a locally compact Hausdorff space, under this setting, given an arbitrary C* algebra $A$, we can uniquely define a C* algebra $C_0(\Omega) \otimes A$ where the algebraic tensor product $C_0(\Omega) \odot A$ is dense, this happens because $C_0(\Omega)$ is a nuclear C* algebra (\cref{example:nuclear_C_star_algebras}). From \cref{sec:Continuous_functions_with_values_on_C_star_algebra} we know that 
$$ C_0(\Omega) \otimes A \simeq C_0(\Omega,A), $$
thus, taking tensor products in this case amounts to work with continuous functions over $\Omega$ decaying at infinity and taking values in $A$.

This setting is important to us because it is also related to $A \rtimes_{\alpha,\Theta}\mathbb{Z}^d$, in the sense that $A \rtimes_{\alpha,\Theta}\mathbb{Z}^d$ arises as a generalization of the C* algebra $C(\mathbb{T}^d,A)$ (\cref{sec:trivial_twisting_actions}). So, in the case of a trivial twisted crossed product we have that $A \rtimes_{\alpha,\Theta}\mathbb{Z}^d \simeq C(\mathbb{T}^d,A)$, and, if $A$ is a nuclear C* algebra we have that $C(\mathbb{T}^d,A)$ is also a nuclear C* algebra (\cref{label:nuclearity_of_tensor_product_and_associativity}), which implies that
$$ C(\mathbb{T}^d,A) \otimes B \simeq (C(\mathbb{T}) \otimes A) \otimes B \simeq C(\mathbb{T}) \otimes (A \otimes B) \simeq C(\mathbb{T}^d, A \otimes B) .$$

This result can be generalized into the context of $A \rtimes_{\alpha,\Theta}\mathbb{Z}^d$, and is important for the study of topological insulators in \citep{prodan_bulk_2016} because:

\begin{itemize}
    \item \citep[Chapter 5]{prodan_bulk_2016}: Tensor products with the C* algebra of matrices $M_n(\mathbb{C})$ are used to deal with elements in matrix algebras over $A \rtimes_{\alpha,\Theta}\mathbb{Z}^d$ (\cref{sec:top_inva_over_the_non_commutative_brillouim_torus}). These are used to establish a pairing between the K theory of $A \rtimes_{\alpha,\Theta}\mathbb{Z}^d$ and the cyclic cohomology of $\mathcal{A}_{\alpha, \Theta}$ (\cref{sec:pairing with K theory}), and to define the groups $K_j, \; j=0,1$ of the C* algebra $A \rtimes_{\alpha,\Theta}\mathbb{Z}^d$ (\cref{sec:k_theory_for_C_star_algebras}).
    \item \citep[Proposition 3.2.4]{prodan_bulk_2016}: Tensor products with the C* algebra of compact operators (\cref{section:stabilization_C_stal_algebra}) are used to deal with boundary Hamiltonians on homogeneous materials (\cref{sec:applications_into_solid_state_physics}). The C* algebra $(A \rtimes_{\alpha,\Theta}\mathbb{Z}^d)\otimes \mathcal{K}$ is part of a three-term exact sequence that also contains the C* algebras $A \rtimes_{\alpha,\Theta}\mathbb{Z}^d$, and the Toeplitz extension of $A \rtimes_{\alpha,\Theta}\mathbb{Z}^d$ (\cref{sec:a_sneak_and_peak_into_the_toeplitz_extension}).
\end{itemize}


Suppose that we have an amenable discrete group $G$, a nuclear C* algebra $A$ and an action of $G$ on $A$ i.e.
$$ \alpha : G \to \text{Aut}(A), $$
then, we can define an action of $G$ on $A \otimes B$ for any nuclear C* algebra $B$ by setting
$$ \hat{\alpha} : G \to \text{Aut}(A \otimes B), \; \hat{\alpha}(i) = \alpha(i) \otimes \text{id}_{B} .$$
The action of $G$ on $A \otimes B$ defines a new twisted crossed product $(A\otimes B) \rtimes_{\hat{\alpha}, \zeta} G$ which will turn out to be isomorphic to $(A \rtimes_{\alpha, \zeta} G) \otimes B$ when $A$ and $B$ are separable, and one of them is nuclear. This implies that we can carry a Fourier analysis over $(A \rtimes_{\alpha, \Theta} \mathbb{Z}^d) \otimes B$. Since $A$ and $B$ are separable, by \cref{remark:tensor_product_separable_C_star_algebras} we have that $A \otimes B$ is separable, thus by \cref{theorem:twsited_crossed_product_amenable_nuclear} there is a faithful representation of $(A\otimes B) \rtimes_{\hat{\alpha}, \zeta} G$ on $L^2(G,H_A \otimes H_B)$.

\begin{proposition}\label{proposition:isomorphism_tensor_product_twisted_crossed_product}
Let $G$ be countable, discrete and commutative group, let $A,B$ separable C* algebras and $B$ a nuclear C* algebra, then  
$$ (A \otimes B) \rtimes_{\hat{\alpha}, \zeta} G \simeq  (A \rtimes_{\alpha, \zeta} G) \otimes B,$$
and the isomorphism is the extension of the map given by
$$ (a \otimes b)u_s  \mapsto (a u_s) \otimes b .$$
\end{proposition}
\begin{proof}
This comes as an application of the algebraic characterization of twisted crossed products when $G$ is a countable, discrete and commutative group, which is exposed in \cref{lemma:twisted_crossed_products_of_Generated_C_star_algebras}, \cref{proposition:twisted_crossed_product_restricts_to_discrete_conunatlbe_groups}.

From \cref{proposition:twisted_crossed_product_restricts_to_discrete_conunatlbe_groups} we know that $A \rtimes_{\alpha, \zeta} G$ is the enveloping C* algebra of $L^1(G,A;\alpha,\zeta)_c$, therefore \cref{lemma:tensor_products_and_generated_C_star_algebras} tell us that $(A \rtimes_{\alpha, \zeta} G) \otimes B$ is the enveloping C* algebra of 
$$ L^1(G,A;\alpha,\zeta)_c \odot B .$$ 
On the other side, given that $B$ is a nuclear C* algebra, from \cref{remark:nuclear_C_star_algebra_enveloping_C_star_algebra} we know that $A \otimes B$ is the enveloping C* algebra of $A \odot B$, and \cref{lemma:twisted_crossed_products_of_Generated_C_star_algebras} tell us that $(A \otimes B) \rtimes_{\hat{\alpha}, \zeta} G$ is the enveloping C* algebra of 
$$L^1(G, A \odot B; \hat{\alpha},\zeta).$$ 

The *-algebras $L^1(G, A \odot B; \hat{\alpha},\zeta)$ and $L^1(G,A;\alpha,\zeta)_c \odot B$ are canonically isomorphic by the map 
$$ (a \otimes b) u_s \mapsto (a u_s) \otimes b, $$
therefore their enveloping C* algebras must also be isomorphic since they correspond to the enveloping C* algebra of the same *-algebra. Consequently, we have the isomorphism
$$ (A \otimes B) \rtimes_{\hat{\alpha}, \zeta} G \simeq  (A \rtimes_{\alpha, \zeta} G) \otimes B,$$
and the isomorphism is the extension of the map given by
$$ (a \otimes b)u_s  \mapsto (a u_s) \otimes b .$$

\end{proof}

\begin{remark}[Tensor product and twisted crossed products with $\mathbb{Z}^d$]\label{remark:tensor_product_twisted_crossed_product_with_interegers}
If $G = \mathbb{Z}^d$ then the isomorphism between $(A \rtimes_{\alpha, \Theta} \mathbb{Z}^d) \otimes B \simeq (A \otimes B) \rtimes_{\hat{\alpha}, \Theta} \mathbb{Z}^d$
is the extension of the map
$$ \phi((a u^x) \otimes b) =  (a \otimes b)u^x , \; a \in A, \; b \in B, \; x \in \mathbb{Z}^d.$$
\end{remark}

\subsubsection{Traces}
\label{sec:tensor_product_twisted_crossed_product_traces}
 
If the algebra $B$ has no identity, then $(A \rtimes_{\alpha, \Theta} G) \otimes B$ has no identity (\cref{lemma:multiplicative_identity_of_twisted_crossed_product}). Also, if $\tau_A$ is a tracial state on $A$ and $\tau_B$ is a tracial state over $B$ then we can extend it to a state on $A \otimes B$ by \cref{proposition:extending_C_star_homomorphisms_into_tensor_products}, which additionally is a trace (as it can be seen by checking its behavior over finite sums of simple tensors on $A \otimes B$). In this case, we can define a tracial state over $ (A\otimes B) \rtimes_{\hat{\alpha}, \Theta} G \simeq  (A \rtimes_{\alpha, \Theta} G) \otimes B$ by
$$ \hat{\tau}(a) = (\tau_A \otimes \tau_B )(\Phi_0(a)), $$
with $\Phi_0$ the generalized zeroth Fourier coefficient (\cref{lemma:computation_of_Fourier_coefficients}), moreover, $ \hat{\tau}$ is faithful if $\tau_A \otimes \tau_B$ is faithful (\cref{remark:continuous_linear_functionals_twsited_crossed_product}).

\begin{example}[Tensor product with $M_n (\mathbb{C})$]\label{example:tensor_product_with_M_n_C}
Let $A$ be a C* algebra with a faithful tracial state $\eta$ which commutes with $\alpha$, then $A \rtimes_{\alpha,\Theta}\mathbb{Z}^d$ has faithful tracial state, which is also continuous and is defined as $\hat{\eta}(a) = \eta(\Phi_0(a))$ (\cref{remark:continuous_linear_functionals_twsited_crossed_product}). From the discussion on \cref{sec:C_star_alg_matrix_alg} we know that we can define a tracial state on $(A \rtimes_{\alpha,\Theta}\mathbb{Z}^d) \otimes M_n(\mathbb{C})$ as follows
$$ (\alpha \otimes \text{tr}) :  (A \rtimes_{\alpha,\Theta}\mathbb{Z}^d) \otimes M_n(\mathbb{C}) \to \mathbb{C}, \; (\alpha \otimes \text{tr})\left(\sum_{1 \leq i,j \leq n} a_{i,j} \otimes e_{i,j}\right) = \sum_{1 \leq i \leq n} \hat{\alpha}(a_{i,i})$$
with $a_{i,j} \in A \rtimes_{\alpha,\Theta}\mathbb{Z}^d$, moreover, since $\hat{\alpha}(a) = \Phi_0(a)$ then 
$$ (\alpha \otimes \text{tr})\left(\sum_{1 \leq i,j \leq n} a_{i,j} \otimes e_{i,j}\right) = \sum_{1 \leq i \leq n} \alpha(\Phi_0( a_{i,i})) .$$

By \cref{remark:tensor_product_twisted_crossed_product_with_interegers} an alternative way of defining the same tracial state is through $M_n(A) \rtimes_{\hat{\alpha}, \Theta} \mathbb{Z}^d$ as follows
$$\hat{\alpha}: M_n(A) \rtimes_{\hat{\alpha}, \Theta} \mathbb{Z}^d \to \mathbb{C}, \;  \hat{\alpha}(a) = (\alpha \otimes \text{tr})(\Phi_0 (a)) = \sum_{1 \leq i \leq n} \alpha \left( (\Phi_0(a))_{i,i} \right), $$
with $a \in M_n(A) \rtimes_{\hat{\alpha}, \Theta} \mathbb{Z}^d$.
\end{example}

\begin{example}[Tensor product with $\mathcal{K}$]\label{example:tensor_product_with_K}
From \cref{section:algebra_of_compact_operators} we know that $\mathcal{K}$ has a lower semicontinuous trace, which is given by $\text{Tr}(a) = \sum_{i \in \mathbb{N}}a_{i,i}$. So, if $A$ is a C* algebra with a lower semicontinuous weight (trace) $\eta$, then $\eta \otimes \text{Tr}$ is a lower semicontinuous weight (trace) (\cref{section:stabilization_C_stal_algebra}), therfore, $p \mapsto \eta(\Phi_0 (p))$ defines a lower semicontinuous weight (trace) over $(A \rtimes_{\alpha,\Theta}\mathbb{Z}^d)\otimes \mathcal{K} \simeq (A \otimes  \mathcal{K})\rtimes_{\tau,\sigma} \mathbb{Z}^{d} $. The trace cannot be made into a continuous one because is unbounded. Also, any element of $(A \rtimes_{\alpha,\Theta}\mathbb{Z}^d)\otimes \mathcal{K}$ can be seen as an infinite algebra $b = \{ b_{i,j} \}_{i,j \in \mathbb{N}}, \; b_{i,j} \in A \rtimes_{\alpha,\Theta}\mathbb{Z}^d$ with $b_{i,j} \to 0$ as $i+j \to \infty$ (\cref{section:stabilization_C_stal_algebra}).
\end{example}

\subsubsection{Derivations and smooth algebras}
\label{sec:tensor_products_smooth_algebras}

Derivations for twisted crossed products can be defined using the action of $\mathbb{T}^{d}$ over $(A \rtimes_{\alpha,\Theta} \mathbb{Z}^d) \otimes B$, and take the familiar form over non-commutative polynomials, 
$$ \partial_j p = \sum_{s \in \mathbb{Z}^{d}} i s_j \Phi_s (p) u^{s} , \; p \in (\mathcal{A \otimes B})_{\alpha, \Theta}, \; \Phi_s (p) \in A \otimes B ,$$
where $(\mathcal{A \otimes B})_{\alpha, \Theta}$ is the sub *algebra of $(A \otimes B) \rtimes_{\Theta, \hat{\alpha}} \mathbb{Z}^d$ consisting of elements whose Fourier coefficients decay faster than any polynomial, as described in \cref{proposition:characterization_smooth_sub_algebra_twsited_crossed_product}. Recall that $(\mathcal{A \otimes B})_{\alpha, \Theta}$ is a Fréchet $D^{*}_{\infty}$-subalgebra of $(A \rtimes_{\alpha, \Theta} \mathbb{Z}^d) \otimes B$ (\cref{proposition:smooth_elements_as_D_infinity_Freche_subalgebra}). Also, if $\eta_a, \eta_b$ are continuous traces over $A,B$ respectively, then, $\eta_a \otimes \eta_b$ is a continuous trace over $A \otimes B$ and $p \mapsto \eta_a \otimes \eta_b (\Phi_0(p) )$ is continuous linear functional over $(\mathcal{A \otimes B})_{\alpha, \Theta}$ (\cref{remark:properties_of_derivations}).

\begin{example}[Matrix algebras and twisted crossed products]\label{example:matrix_algebras_and_twisted_crossed_products}
From \cref{proposition:C_star_norm_on_matrix_algebras} we know that $M_n(A \rtimes_{\alpha,\Theta} \mathbb{Z}^d)$ is isomorphic to $(A \rtimes_{\alpha,\Theta} \mathbb{Z}^d) \otimes M_n(\mathbb{C})$, and \cref{proposition:isomorphism_tensor_product_twisted_crossed_product} tell us that $(A \rtimes_{\alpha,\Theta} \mathbb{Z}^d) \otimes M_n(\mathbb{C})$ is isomorphic to $M_n(A) \rtimes_{\alpha,\Theta} \mathbb{Z}^d$, hence, each element of $p \in M_n(A \rtimes_{\alpha,\Theta} \mathbb{Z}^d)$ is uniquely determined by the set $\{ \Phi_s(p) \}_{s \in \mathbb{Z}^d}$, where $\Phi_s(p)$ is a matrix over $A$. Given that the norm of a matrix with entries in a C* algebra is bigger or equal to the norm of each one of its elements (\cref{sec:C_star_alg_matrix_alg}), for any $i,j \leq n$ the set $\{ \Phi_s(p)_{i,j} \}_{s \in \mathbb{Z}^d}$ correspond to the Fourier coefficients of an element of $A \rtimes_{\alpha,\Theta} \mathbb{Z}^d$, where $\Phi_s(p)_{i,j}$ denotes the element $(i,j)$ of the $n \times n$ matrix $\Phi_s(p)$. 

Since the norm of a matrix with entries in a C* algebra is lower than the sum of the norm of all the entries (\cref{sec:C_star_alg_matrix_alg}), any set of $n^2$ elements from $\mathcal{A}_{\alpha, \Theta}$ define an element of $M_n(A \rtimes_{\alpha,\Theta} \mathbb{Z}^d)$ whose Fourier coefficients decay faster than any power, thus, it is a part of the smooth sub algebra of $M_n(A \rtimes_{\alpha,\Theta} \mathbb{Z}^d)$ (\cref{proposition:characterization_smooth_sub_algebra_twsited_crossed_product}). Hence, the smooth sub algebra of $M_n(A \rtimes_{\alpha,\Theta} \mathbb{Z}^d)$ corresponds to the Fr\'echet algebra $M_n(\mathcal{A}_{\alpha,\Theta})$, which coincides with the properties of smooth sub algebras (\cref{prop:matrix_alegbras_of_pre_C_star_algebras}), additionally, if $\{ (\Phi_s(p)_{i,j})_{i,j \leq n} \}_{s \in \mathbb{Z}^d}$ are the Fourier coefficients of an element of $M_n(\mathcal{A}_{\alpha,\Theta})$, we have that $  \Phi_s(\partial_l (p))_{i,j} = \partial_l( \Phi_s(p)_{i,j})$.
\end{example}

\begin{example}[Smooth sub algebras of $(A \rtimes_{\alpha,\Theta}\mathbb{Z}^d)\otimes \mathcal{K}$]\label{example:smooth_sub_algebra_of_tensor_product_wiht_non_continuous_trace}
Let $A$ be a C* algebra with a faithful continuous trace $\eta$, then, from the previous discussion we know that we can define a smooth sub algebra $\mathcal{A \otimes K}_{\alpha, \Theta}$ of $(A \rtimes_{\alpha,\Theta}\mathbb{Z}^d)\otimes \mathcal{K}$, whose the Fréchet topology is given by
$$
\|x\|_n:=\|x\|_0+\sum_{k=1}^n \frac{1}{k!} \sum_{i_1, i_2, \ldots, i_k=1}^d\left\|\partial_{i_1} \partial_{i_2} \cdots \partial_{i_k} x\right\|,
$$
 where $\left\{\partial_{i_1}, \partial_{i_2}, \ldots, \partial_{i_k}\right\}$ is an ordered $k$-tuple from $\left\{\partial_1, \partial_2, \ldots, \partial_d\right\}$, and $\left\{\partial_1, \partial_2, \ldots, \partial_d\right\}$ are the commutative derivations densely defined over $(A \otimes  \mathcal{K})\rtimes_{\Theta,\hat{\alpha}} \mathbb{Z}^{d}$. Under this setting, $\mathcal{A \otimes K}_{\alpha, \Theta}$ becomes a Fréchet $D^{*}_{\infty}$-subalgebra of $(A \rtimes_{\alpha,\Theta}\mathbb{Z}^d)\otimes \mathcal{K}$, with a faithful lower semicontinuous trace given by $\eta \otimes \text{Tr}$, nonetheless, $\eta \otimes \text{Tr}$ is not continuous, because there are elements of $A$ with infinite trace. For example, take $a \in A_{\text{pos}}$ and $t \in \mathcal{K}_{\text{pos}}$ with $\text{Tr}(k) =\infty, \; \eta(a) \neq 0$, then, 
 $$\eta \otimes \text{Tr} (au^{0} \otimes k) = \infty.$$
 
 To obtain a smooth sub algebra of $(A \rtimes_{\alpha,\Theta}\mathbb{Z}^d)\otimes \mathcal{K}$ with a continuous linear trace is necessary to resort to the projective tensor product of Fréchet algebras, in particular, let $\mathcal{RD}(l^2(\mathbb{N}))$ be the algebra of infinite matrices with rapid decay (\cref{example:infinte_matrices_with_rapid_decay}), then, from \citep[second example page 131]{rennie_smoothness_2003} we know that $\mathcal{A}_{\alpha, \Theta} \otimes_{\pi} \mathcal{RD}(l^2(\mathbb{N}))$ is a smooth sub algebra of $(A \rtimes_{\alpha,\Theta}\mathbb{Z}^d) \otimes \mathcal{K}$. In $\mathcal{A}_{\alpha, \Theta} \otimes_{\pi} \mathcal{RD}(l^2(\mathbb{N}))$ we can define both the derivations and traces as follows, let $\text{tr}$ be a continuous trace over $A \rtimes_{\alpha,\Theta}\mathbb{Z}^d$, then, define $\hat{\partial_j} = \partial_i \otimes_{\pi} \text{id}$ and $\hat{\text{tr}} = \text{tr} \otimes_{\pi} \text{Tr}$, under this setting those are well defined and continuous over the whole algebra $\mathcal{A} \otimes_{\pi} \mathcal{RD}(l^2(\mathbb{N}))$ because $\text{tr}$ is continuous on $\mathcal{A}_{\alpha,\Theta}$ and $\text{Tr}$ is continuous on $\mathcal{RD}(l^2(\mathbb{N}))$ (\cref{proposition:properties_projective_tensor_product}). You can refer to \citep[Proposition 3.3.3]{prodan_bulk_2016} for a description of a countable family of seminorms that define the topology of $\mathcal{A}_{\alpha, \Theta} \otimes_{\pi} \mathcal{RD}(l^2(\mathbb{N}))$, notice that the set of seminorms provided in \citep[Proposition 3.3.3]{prodan_bulk_2016} comes as a generalization of the seminorms that define the topology of $\mathcal{RD}(l^2(\mathbb{N}))$ (\cref{example:infinte_matrices_with_rapid_decay}).
\end{example}

\begin{remark}[Properties of the derivations]\label{remark:properties_of_derivations}
Expanding on the non-commutative geometry dictionary (\cref{section:non_commutative_geometry_dictionary}), given a continuous linear functional over $A$, which we refer to as $\eta$, is possible to define a continuous linear functional over $A \rtimes_{\alpha,\Theta}\mathbb{Z}^d$ (\cref{remark:continuous_linear_functionals_twsited_crossed_product}), which we called $\hat{\eta}$. The continuous linear functional $\hat{\eta}$ behaves like a generalization of the integration of a function over the closed manifold $\mathbb{T}^d$, because
$$ \hat{\eta}(\partial_j p) = \hat{\eta}(0) = 0, \; \hat{\eta}( (\partial_j p)q) = - \hat{\eta}( p(\partial_j q)), $$
for $p,q \in \mathcal{A}_{\alpha, \Theta}$ and $1 \leq j \leq d$. 

We can take this analogy one step further and assume that $\eta$ is a lower semicontinuous trace (weight) over $A$, in which case we have that $\hat{\eta}$ becomes a trace (weight) over $A \rtimes_{\alpha,\Theta}\mathbb{Z}^d$ (\cref{remark:weights_and_traces_twsited_crossed_product}). In this setting, $\hat{\eta}$ is a generalization of integration functions over $C(\mathbb{T})$ with respect to a unbounded measure over $\mathbb{T}$ i.e. a measure $\mu$ such that $\mu(\mathbb{T}) =  \infty$ (\cref{remark:weights_and_continuity}).
\end{remark}

\section{Twisted transformation group C* algebras}
\label{sec:twsited_transformation_group_C_star_algebras}

\begin{lemma}\label{lemma:twisted_transformation_group_C_star-alg_and_tensor_products}
Let $C(\Omega) \rtimes_{\alpha, \Theta} \mathbb{Z}^d$ be a twisted transformation group C* algebra such that $\Omega$ is compact, assume that there is a measure $\mu$ over $\Omega$ and $\Omega$ has finite measure i.e. $\mu(\Omega) < \infty$, additionally, assume that $\alpha$ is an isometric action (\cref{definition:isometric_action_twisted_transf_group_C_star_algebras}). Let $D$ be a separable and nuclear C* algebra with a faithful representation $\pi_D : C \to B(H_D)$, then, $C(\Omega) \otimes D)\rtimes_{\alpha, \Theta} \mathbb{Z}^d$ is a sub C* algebra of $C(\Omega, B(L^2(\mathbb{Z}^d))\otimes D)$ and each element $q$ of $C(\Omega) \otimes D)\rtimes_{\alpha, \Theta} \mathbb{Z}^d$ takes the form
$$ \omega \mapsto \pi_{\omega} \otimes \pi_{D} (q).$$
\end{lemma}
\begin{proof}
First, we need to check that the the map $\omega \mapsto \pi_{\omega} \otimes \pi_{D} (q)$ is continuous, if this is the case, then it will be an element of the C* algebra $C(\Omega, B(L^2(\mathbb{Z}^d))\otimes D)$. Recall that $\pi_{\omega} \otimes \pi_D$ is a representation of  $(C(\Omega) \rtimes_{\alpha, \Theta} \mathbb{Z}^d) \otimes D$ over the Hilbert space $L^2(\mathbb{Z}^d) \otimes H_D$, and takes the following form on simple tensors (\cref{proposition:opeators_tensor_product_hilbert_spaces}) 
$$\pi_{\omega} \otimes \pi_D (p \otimes d ) = \pi_{\omega}(p) \otimes \pi_D(d).$$

Since $(C(\Omega) \rtimes_{\alpha, \Theta} \mathbb{Z}^d) \otimes D$ is the closure of $(C(\Omega) \rtimes_{\alpha, \Theta} \mathbb{Z}^d) \odot D$ (\cref{definition:Nuclear_C_star_algebra}), for every $q \in (C(\Omega) \rtimes_{\alpha, \Theta} \mathbb{Z}^d) \odot D$ and any $\epsilon > 0$, there is an element of the form $ \sum_{i \leq m} p_i \otimes d_i$ with $ \| q - \sum_{i \leq m} p_i \otimes d_i \| \leq \epsilon/3 $. Since the C* homomorphisms  are norm decreasing (\cref{proposition:automatic_continuity_C_star_algebras}), we have that
$$ \forall \omega \in \Omega, \; \| \pi_{\omega} \otimes \pi_D(\sum_{i \leq m} p_i \otimes d_i) - \pi_{\omega}\otimes \pi_D(q) \| \leq \epsilon/3.  $$
As an application of the triangle inequality we get
$$ \| \pi_{\omega_0} \otimes \pi_D (q) - \pi_{\omega_1}\otimes \pi_D(q) \| \leq \| \pi_{\omega_0}\otimes \pi_D(q) - \pi_{\omega_0}\otimes \pi_D(\sum_{i \leq m} p_i \otimes d_i) \| + $$
$$\| \pi_{\omega_0}\otimes \pi_D(\sum_{i \leq m} p_i \otimes d_i) - \pi_{\omega_1}\otimes \pi_D(\sum_{i \leq m} p_i \otimes d_i) \| + \| \pi_{\omega_1}\otimes \pi_D(\sum_{i \leq m} p_i \otimes d_i) - \pi_{\omega_1}\otimes \pi_D(q) \|, $$
so, the first and third terms are bounded above by $\epsilon/3$, thus, to show that the map $ \omega \mapsto \pi_{\omega}\otimes \pi_D (q) $ is continuous we need to show find $\delta$ such that, if $d(\omega_0, \omega_1) \leq \delta$ then 
$$\| \pi_{\omega_0} \otimes \pi_D (\sum_{i \leq m} p_i \otimes d_i) - \pi_{\omega_1} \otimes \pi_D(\sum_{i \leq m} p_i \otimes d_i) \|.$$
Since $\pi_{\omega_0} \otimes \pi_D(\sum_{i \leq m} p_i \otimes d_i) = \sum_{i \leq m} \pi_{\omega_0}(p_i) \otimes \pi_D(d_i)$, we have that,
$$ \| \pi_{\omega_0}\otimes \pi_D(\sum_{i \leq m} p_i \otimes d_i) - \pi_{\omega_1}\otimes \pi_D(\sum_{i \leq m} p_i \otimes d_i) \| \leq$$
$$ \sum_{i \leq m} \| (\pi_{\omega_0}(p_i) - \pi_{\omega_1}(p_i))\otimes \pi_D( d_i) \|. $$
From \cref{proposition:opeators_tensor_product_hilbert_spaces} we get the following relation,
$$\| (\pi_{\omega_0}(p_i) - \pi_{\omega_1}(p_i))\otimes \pi_D( d_i) \| = \| (\pi_{\omega_0}(p_i) - \pi_{\omega_1}(p_i)) \| \| \pi_D( d_i) \|,$$
thus, we can use the fact that the map $\omega \mapsto \pi_{\omega}(p)$ is continuous to show that the map $\omega \mapsto \pi_{\omega} \otimes \pi_D(p_i \otimes d_i)$ is also continuous, which in turn implies that the map $\omega \mapsto \pi_{\omega} \otimes \pi_D (\sum_{i \leq m} p_i \otimes d_i)$ is also continuous. Given that the map $\omega \mapsto \pi_{\omega} \otimes \pi_D (\sum_{i \leq m} p_i \otimes d_i)$ is continuous, the inequalities previously displayed implies that the map $\omega \mapsto \pi_{\omega} \otimes \pi_D(q)$ is continuous for any $q \in (C(\Omega) \rtimes_{\alpha, \Theta} \mathbb{Z}^d) \otimes D$.

We know that $C(\Omega) \rtimes_{\alpha, \Theta} \mathbb{Z}^d$ is a sub C* algebra of $C(\Omega, B(L^2(\mathbb{Z}^d)))$, thus, a faithful representation $\pi_f$ of $C(\Omega, B(L^2(\mathbb{Z}^d)))$ falls back into a faithful representation of $C(\Omega) \rtimes_{\alpha, \Theta} \mathbb{Z}^d$. Given that $D$ is a nuclear C* algebra (\cref{definition:Nuclear_C_star_algebra}), the C* algebra $C(\Omega, B(L^2(\mathbb{Z}^d))) \otimes D$ is the closure of $C(\Omega, B(L^2(\mathbb{Z}^d))) \odot D$ under the norm coming from the representation $\pi_f \otimes \pi_D$, similarly, the C* algebra $(C(\Omega) \rtimes_{\alpha, \Theta} \mathbb{Z}^d) \otimes D$ is the closure of $((C(\Omega) \rtimes_{\alpha, \Theta} \mathbb{Z}^d) \odot D$ under the norm coming from the representation $\pi_f \otimes \pi_D$, hence, $((C(\Omega) \rtimes_{\alpha, \Theta} \mathbb{Z}^d)\otimes D$ is a sub C* algebra of $C(\Omega, B(L^2(\mathbb{Z}^d))) \odot D$. There is an inclusion map $i : C(\Omega) \rtimes_{\alpha, \Theta} \mathbb{Z}^d \to C(\Omega, B(L^2(\mathbb{Z}^d)))$ which takes the following form (\cref{remark:faithful_representation_of_C_0_Omega_rtimes_Z}),
$$ i(p) \mapsto (\omega \mapsto \pi_{\omega}(p)), $$
and this inclusion map extends into an inclusion $i \otimes \text{id}_D : (C(\Omega) \rtimes_{\alpha, \Theta} \mathbb{Z}^d) \otimes D \to C(\Omega, B(L^2(\mathbb{Z}^d))) \otimes D$, which takes the form $\omega \mapsto \pi_{\omega} \otimes \pi_D (q)$ for any $q \in (C(\Omega) \rtimes_{\alpha, \Theta} \mathbb{Z}^d) \odot D$. Since the map $\omega \mapsto \pi_{\omega} \otimes \pi_D(q)$ is continuous for any $q \in (C(\Omega) \rtimes_{\alpha, \Theta} \mathbb{Z}^d) \odot D$, and $(C(\Omega) \rtimes_{\alpha, \Theta} \mathbb{Z}^d) \odot D$ is dense inside $(C(\Omega) \rtimes_{\alpha, \Theta} \mathbb{Z}^d) \otimes D$, the map $i \otimes \text{id}_D$ takes the form 
$$ i \otimes \text{id}_D (q) = (\omega \mapsto \pi_{\omega} \otimes \pi_D (q)). $$
\end{proof}

\begin{remark}[$C(\Omega) \rtimes_{\alpha, \Theta} \mathbb{Z}^d$ as a continuous field of C* algebras \index{continuous field of C* algebras}]\label{remark:twsited_crossed_product_as_continuous_field_of_C_star_alg}
Let $C(\Omega) \rtimes_{\alpha, \Theta} \mathbb{Z}^d$ be a twisted transformation group C* algebra such that $\Omega$ is compact, assume that there is a measure $\mu$ over $\Omega$ and $\Omega$ has finite measure i.e. $\mu(\Omega) < \infty$, additionally, assume that $\alpha$ is an isometric action (\cref{definition:isometric_action_twisted_transf_group_C_star_algebras}), then, \cref{remark:faithful_representation_of_C_0_Omega_rtimes_Z} tells us that for $p \in C(\Omega) \rtimes_{\alpha,\zeta} \mathbb{Z}^d$
$$ \| p \|_{C(\Omega) \rtimes_{\alpha, \Theta} \mathbb{Z}^d} = \| p \|_{C(\Omega, B(L^2(\mathbb{Z}^d)))} = \sup_{\omega \in \Omega} \| \pi_{\omega}(p) \|, $$
therefore, $C(\Omega) \rtimes_{\alpha, \Theta} \mathbb{Z}^d$ becomes a continuous field of C* algebras ( see \citep[Section IV.1.6]{blackadar_operator_2006}), with, 
$$C(\Omega) \rtimes_{\alpha, \Theta} \mathbb{Z}^d \simeq (\Gamma, \coprod_{\omega \in \Omega} A_\omega),$$
where,
$$A_\omega = \pi_{\omega}(C(\Omega) \rtimes_{\alpha, \Theta} \mathbb{Z}^d), $$
$$\Gamma = \{ f : \Omega \to  \coprod_{\omega \in \Omega} A_\omega | f(\omega) = \pi_{\omega}(p), \; p \in C(\Omega) \rtimes_{\alpha, \Theta} \mathbb{Z}^d \}.$$
\end{remark}

\begin{remark}[$(C(\Omega) \rtimes_{\alpha, \Theta} \mathbb{Z}^d) \otimes D$ as a continuous field of C* algebras]\label{remark:tensor_product_with_twsited_crossed_product_as_continuous_field_of_C_star_alg}
Let $C(\Omega) \rtimes_{\alpha, \Theta} \mathbb{Z}^d$ be a twisted transformation group C* algebra such that $\Omega$ is compact, assume that there is a measure $\mu$ over $\Omega$ and $\Omega$ has finite measure i.e. $\mu(\Omega) < \infty$, additionally, assume that $\alpha$ is an isometric action (\cref{definition:isometric_action_twisted_transf_group_C_star_algebras}). Let $D$ be a separable and nuclear C* algebra with a faithful representation $\pi_D : C \to B(H_D)$, then, from \cref{proposition:isomorphism_tensor_product_twisted_crossed_product} we know that 
$$(C(\Omega) \otimes D)\rtimes_{\alpha, \Theta} \mathbb{Z}^d \simeq (C(\Omega) \rtimes_{\alpha, \Theta} \mathbb{Z}^d) \otimes D.$$ 

From \cref{lemma:twisted_transformation_group_C_star-alg_and_tensor_products} we know that $(C(\Omega) \otimes D)\rtimes_{\alpha, \Theta} \mathbb{Z}^d \subset C(\Omega, B(L^2(\mathbb{Z}^d))) \otimes D$, hence, $(C(\Omega) \otimes D)\rtimes_{\alpha, \Theta} \mathbb{Z}^d$ is also a continuous field of C* algebra with  
$$D_\omega = \pi_{\omega}\otimes \pi_D ((C(\Omega) \otimes D) \rtimes_{\alpha, \Theta} \mathbb{Z}^d), $$
$$\Gamma = \{ f : \Omega \to  \coprod_{\omega \in \Omega} D_\omega | f(\omega) = \pi_{\omega} \otimes \pi_D(p), \; p \in (C(\Omega) \otimes D) \rtimes_{\alpha, \Theta} \mathbb{Z}^d \}.$$
\end{remark}

We can use the structure of $C(\Omega) \rtimes_{\alpha, \Theta} \mathbb{Z}^d$ as a sub C* algebra of $C(\Omega, B(L^2(\mathbb{Z}^d)))$ to describe the spectrum of the elements in $C(\Omega) \rtimes_{\alpha, \Theta} \mathbb{Z}^d$

\begin{lemma}[Spectrum of elements in $C(\Omega) \rtimes_{\alpha, \Theta} \mathbb{Z}^d$]\label{elmma:spectrum_of_elements_in_twisted_crossed_product}
Following the notation in \cref{remark:tensor_product_with_twsited_crossed_product_as_continuous_field_of_C_star_alg}, let $D$ be a separable and nuclear C* algebra with a unit and having a faithful representation $\pi_D: D \to H_D$, if $p \in (C(\Omega) \otimes D) \rtimes_{\alpha, \Theta} \mathbb{Z}^d$, then,
$$ Sp(p) = \bigcup_{\omega \in \Omega} Sp(\pi_\omega \otimes \pi_D (p)). $$
\end{lemma}
\begin{proof}
Given that both $C(\Omega)$ and $D$ have a unit, then, $C(\Omega) \times D$ has a unit, so, from \cref{lemma:multiplicative_identity_of_twisted_crossed_product} we know that the identity of $(C(\Omega) \otimes D) \rtimes_{\alpha,\Theta} \mathbb{Z}^d$ is $1_{C(\Omega)} \otimes 1_D u^0$ with $1_{C(\Omega)}$ the function over $\Omega$ taking the constant value $1$. Since $(C(\Omega) \otimes D) \rtimes_{\alpha,\Theta} \mathbb{Z}^d$ is a sub C* algebra of $C(\Omega, B(L^2(\mathbb{Z}^d)) \otimes D)$ (\cref{lemma:twisted_transformation_group_C_star-alg_and_tensor_products}), we can use
\cref{lemma:representation_on_L2_Z} to get,
$$ \pi_{\omega}\otimes \pi_D(1_{C(\Omega)} \otimes 1_D u^0) = 1_{B(L^2(\mathbb{Z}^d)\otimes H_D)}, $$
hence, the identity of $C(\Omega, B(L^2(\mathbb{Z}^d)) )\otimes D $ is the function 
$$\omega \to \pi_{\omega} \otimes \pi_D(1_{C(\Omega)} \otimes 1_D u^0),$$
that is, the identity of $C(\Omega, B(L^2(\mathbb{Z}^d)) )\otimes D $ belongs to $(C(\Omega) \otimes D) \rtimes_{\alpha,\Theta} \mathbb{Z}^d$.
Given the present setting, the content of
\cref{corollary:spectrum_and_ambient_C_star_algebra} implies that for $p \in (C(\Omega) \otimes D) \rtimes_{\alpha,\Theta} \mathbb{Z}^d$ 
$$Sp_{(C(\Omega) \otimes D )\rtimes_{\alpha,\Theta} \mathbb{Z}^d}(p) = Sp_{C(\Omega,B(L^2(\mathbb{Z}^d))\otimes D)  }(p).$$
Since $p \in  (C(\Omega) \otimes D) \rtimes_{\alpha,\Theta} \mathbb{Z}^d$ takes the form of the map $\omega \mapsto \pi_{\omega} \otimes \pi_D (p)$ when viewed as an element of $C(\Omega, B(L^2(\mathbb{Z}^d)) \otimes D)$ (\cref{lemma:twisted_transformation_group_C_star-alg_and_tensor_products}), the content of \cref{lemma:spectrum_elements_C_b_X_A} implies that
$$ Sp(p) = \bigcup_{\omega \in \Omega} Sp(\pi_\omega \otimes \pi_D (p)). $$
\end{proof}

We can use the structure of $C(\Omega) \rtimes_{\alpha, \Theta} \mathbb{Z}^d$ as a sub-algebra of $C(\Omega, B(L^2(\mathbb{Z}^d)))$ to provide a description of the continuous functional calculus over $C(\Omega) \rtimes_{\alpha, \Theta} \mathbb{Z}^d$,

\begin{lemma}[Continuous calculus and self adjoint elements]\label{lemma:continuous_calculus_and_self_adjoint_elements}
Let $p$ be a self adjoint element of $C(\Omega) \rtimes_{\alpha,\Theta} \mathbb{Z}^d$, set $p = \{ p_{\omega} \}_{\omega \in \Omega}$ with $p_{\omega} :=  \pi_{\omega}(p)$, then, for any $f$ continuous in the spectrum of $p$ we have that
$$ f\left( \{ p_{\omega} \}_{\omega \in \Omega} \right) = \{ f(p_{\omega}) \}_{\omega \in \Omega}. $$
\end{lemma}
\begin{proof}
The content of \cref{remark:faithful_representation_of_C_0_Omega_rtimes_Z} implies that any $p \in C_0(\Omega) \rtimes_{\alpha,\Theta} \mathbb{Z}^d$ is uniquely determined by the values it takes under the family of representations $\{ \pi_{\omega} \}_{\omega}$, we have that $f(p)$ is uniquely determined by the values $\{ \pi_{\omega} (f(p)) \}_{\omega \in \Omega}$. The map 
$$ \pi_{\omega}: C(\Omega) \rtimes_{\alpha,\Theta} \mathbb{Z}^d \to B(L^2(\mathbb{Z}^d)) , \;  p_{\omega}:=  \pi_{\omega}(p)$$
is a C* algebra homomorphism (\cref{lemma:representation_on_L2_Z}) for any $\omega \in \Omega$, also, from \cref{proposition:properties_continuous_functional_calculus} we know that the continuous functional calculus commutes with C* algebra homomorphisms, thus
$$ f \circ \pi_{\omega} = \pi_{\omega} \circ f, $$
therefore,
$$ f\left( \{ p_{\omega} \}_{\omega \in \Omega} \right) = \{ f(p_{\omega}) \}_{\omega \in \Omega}. $$
\end{proof}

The following is a weaker version of the estimate on the decay of Fourier coefficients of projections computed with continuous calculus, an estimate that is done by Prodan et.al. using the Combes-Thomas estimate (\citep[Proposition 2.4.2]{prodan_bulk_2016}) and provides a stronger bound on the decay of the Fourier coefficients, nonetheless, for the ultimate goal using continuous cyclic cocycles our approach is enough, as exposed in \cref{sec:top_inva_over_the_non_commutative_brillouim_torus}.

\begin{lemma}[Decay on the Fourier coefficients on projections from gaped self-adjoint elements]\label{lemma:decay_fourier_coefficients_projections_from_gapped_self_adjoint_elemetns}
Let $\mathcal{C}(\Omega)_{\Theta,\alpha}$ be the smooth sub algebra of $C(\Omega) \rtimes_{\alpha,\Theta} \mathbb{Z}^d$ defined using the action of $\mathbb{T}^d$ on $C(\Omega) \rtimes_{\alpha,\Theta} \mathbb{Z}^d$ (\cref{definition:smooth_sub_algebra_of_twisted_crossed_product_with_Z}), and take $p \in \mathcal{C}(\Omega)_{\Theta,\alpha}$ a self adjoint element, then
\begin{itemize}
    \item Assume that $g$ is a gap in the spectrum of $\pi_{\omega}(p)$ for every $\omega \in \Omega$ (\cref{definition:spectral_gap}). Let, $\chi_g: \mathbb{R} \to \mathbb{R}$ be any decreasing function such that $\chi_g(x) = 1$ if $x \leq g$ and $\chi_g(x) = 0$ if $x \geq g$, then, $\chi_g(p)$ is a projection and
    $$ \chi_g(p) \in \mathcal{C}(\Omega)_{\Theta,\alpha}, \text{ with } \chi_g(p) = \{ \chi_g(p_{\omega}) \}_{\omega \in \Omega}. $$
    Also, for every $x \in \mathbb{N}$ there is $K_x < \infty$ such that 
    $$ |s|^{x} \sup_{\omega \in \Omega} | \Phi_s ( \chi_g (p)) (\omega) | \leq K_x. $$
    \item If $\mu \notin Sp(\pi_{\omega}(p))$ for all $\omega \in \Omega$, then $\mu \notin Sp(p)$ and 
    $$ \chi(p \leq \mu) \in  \mathcal{C}(\Omega)_{\Theta,\alpha}, \text{ with } \chi_{\mu}(p) = \{ \chi_{\mu}(p_{\omega}) \}_{\omega \in \Omega}. $$
    Also, for every $x \in \mathbb{N}$ there is $K_x < \infty$ such that 
    $$ |s|^{x} \sup_{\omega \in \Omega} | \Phi_s ( \chi_{\mu} (p)) (\omega) | \leq K_x. $$
\end{itemize}

\end{lemma}
\begin{proof}
\begin{itemize}
    \item From \cref{elmma:spectrum_of_elements_in_twisted_crossed_product} we have that
    $$ Sp(p) = \bigcup_{\omega \in \Omega} Sp(\pi_{\omega}(p)), $$
    thus, 
    $$ g \cap Sp(p) = \emptyset. $$
    The $C^{\infty}$ Urysohn lemma (\citep[Lemma 8.18]{folland_real_1999}) tells us that, given $S \subset \mathbb{R}$ with $S$ compact such that, $$S \cap Sp(p) = \{ \lambda \in Sp(p) | \lambda \leq g \},$$
    if $U$ is open and $S \subset U$, then, there is $f \in C^{\infty}(\mathbb{R})$ such that $f|S = 1$ and $\text{supp}(f) \subset U$. Under this setting, since $g \notin Sp(p)$, we can take $U$ such that
    $$ U \cap \{ \lambda \in Sp(p) | \lambda > g\} = \emptyset,$$
    therefore, $f|_{Sp(p)} = \chi_g |_{Sp(p)}$. From the continuous functional calculus (\cref{theorem:continuous_functional_calculus}) we have that
    $$ f(p) = \chi_g (p) $$
    and $f(p)$ is a projection. From the invariance of $\mathcal{C}(\Omega)_{\Theta,\alpha}$ under $C^{\infty}$ calculus (\cref{corollary:smooth_elements_are_invariant_under_holomoprhic_and_c_infty_calculus}), we get that
    $$ \chi_g(p) = f(g) \in \mathcal{C}(\Omega)_{\Theta,\alpha}.$$
    From the characterization of $\mathcal{C}(\Omega)_{\Theta,\alpha}$ using the decay of Fourier coefficients, we get that for every $x \in \mathbb{N}$ there is $K_x < \infty$ such that 
    $$ |s|^{x} \sup_{\omega \in \Omega} | \Phi_s (p) (\omega) | \leq K_x. $$ 
    Additionally, form \cref{lemma:continuous_calculus_and_self_adjoint_elements} we know that
    $$ \chi_g(p) = \{ \chi_g(\pi_{\omega}(p) ) \}_{\omega \in \Omega}.$$
    \item From \cref{elmma:spectrum_of_elements_in_twisted_crossed_product} we have that
    $$ Sp(p) = \bigcup_{\omega \in \Omega} Sp(\pi_{\omega}(p)), $$
    hence, $ \mu \notin Sp(p) $, thus, there must be $g \subset \mathbb{R}$ open and convex such that $\mu \in g$ and $g \cap Sp(p) = \emptyset$, then, the previous item gives us the desired properties.
\end{itemize}
\end{proof}


\begin{proposition}\label{proposition:trace_over_non_commutative_brillouin_torus}
Let $C(\Omega) \rtimes_{\alpha, \Theta} \mathbb{Z}^d$ be the C* algebra part of the Non-Commutative Brillouin Torus (\cref{definition:non_commutative_brillouin_torus}), denote by $\mu$ the normalized Radon measure over $\Omega$ with full support which is described in \cref{sec:topological_algebras_for_disordered_crystals}, then, the following is a trace over $C(\Omega) \rtimes_{\alpha, \Theta} \mathbb{Z}^d$
$$ \mathscr{T}: \mathcal{C}(\Omega)_{\alpha, \Theta} \to \mathbb{C}, \; \mathscr{T}(p) := \int_{\Omega} \Phi_{0}(p)(\omega) d \mu(\omega). $$
\end{proposition}
\begin{proof}
Also, given that $\mu$ is a Radon measure (\cref{definition:Radon_measure}) with full support (\cref{definition:measure_with_full_support}) over $\Omega$, from (\cref{section:non_commutative_geometry_dictionary}) we know that the following defines a faithful continuous trace over $C(\Omega)$,
$$\eta: C(\Omega) \to \mathbb{C}, \; \eta(f):= \int_{\Omega} f(\omega) d \mathbb{P}(\omega).$$
Additionally, since $\Omega$ is defined as the product of $\mathbb{Z}^d$ identical spaces, if, $U \in \Omega$ is open, then 
$$ \mu(U) = \mu(\varrho(s)(U)), \; s \in \mathbb{Z}^d, $$
therefore, 
$$\eta(\alpha(s)(f)) = \int_{\Omega} \alpha(s)(f)(\omega) d \mu (\omega) = \int_{\Omega} f(\omega) d \mu (\omega) = \eta(f), \; s \in \mathbb{Z}^d.$$
Given the present setting, the C* algebra
$$ C(\Omega)\rtimes_{\alpha, \Theta} \mathbb{Z}^d $$
has a faithful continuous trace given by (\cref{remark:continuous_linear_functionals_twsited_crossed_product})
$$ \mathscr{T}:C(\Omega)\rtimes_{\alpha, \Theta} \mathbb{Z}^d \to \mathbb{C}, \;  \mathscr{T}(p) := \eta(\Phi_0(p)),$$
that is, 
$$ \mathscr{T}: \mathcal{C}(\Omega)_{\alpha, \Theta} \to \mathbb{C}, \; \mathscr{T}(p) := \int_{\Omega} \Phi_{0}(p)(\omega) d \mu(\omega). $$
\end{proof}


\section{Translating concepts from $\mathbb{T}^n$ into the Non-Commutative Brillouin Torus}
\label{sec:translating_concepts_from_T_n_to_non_comm_brillo_torus}

In this section we present analogies between $A \rtimes_{\alpha, \zeta} \mathbb{Z}^d$ and $C(\mathbb{T}^d)$ that have been explored in \cref{chap:fourier_analysis} and \cref{chap:more_on_twisted_crossed_products}, those analogies come as an algebraic and analytic generalization of the techniques used in the Fourier analysis for $\mathbb{T}^d$. In the following $A$ will denote a second countable C* algebra, and we will follow the notations used for the definition of the Twisted crossed product (\cref{definition:twsited_crossed_product}),

\begin{remark}\label{remark:generated_C_atra_algebras_translating_concepts}
Twisted crossed products as generated C* algebras,
\begin{itemize}
    \item Commutative (\cref{section:n_torus}): $C(\mathbb{T}^d)$ is the C* algebra generated by elements of the form $u_j$ with $1 \leq j \leq d$ which follow the commutation relations
    $$ u_j u_i = u_i u_j.  $$   
    \item Non-Commutative (\cref{lemma:twisted_crossed_products_with_Z}): $A \rtimes_{\alpha, \zeta} \mathbb{Z}^d$ is the C* algebra generated by elements of the form $a u_j$ with $a \in A, \; 1 \leq j \leq d$ which follow the commutation relations
    $$ a_0 u_i a_1 u_j = a_0 \alpha(e_i)(a_1) u_i u_j $$ 
    $$  a_0 \alpha(e_i)(a_1) u_i u_j = \zeta(i,j) a_0 \alpha(e_i)(a_1) u_j u_i. $$
\end{itemize}
\end{remark}

\begin{remark}\label{remark:enveloping_c_star_algebra_translating_concepts}
Twisted crossed products as enveloping C* algebras,
\begin{itemize}
    \item Commutative (\cref{sec:group_C_algebra_and_amenable_group}): For every representation of $L^1(\mathbb{Z}^d)$
    $$\pi : L^1(\mathbb{Z}^d) \to B(H)$$
    there is a unique representation 
    $$ \tilde{\pi}: C(\mathbb{T}^d) \to B(H).$$
    Moreover, $C(\mathbb{T}^d)$ is the enveloping C* algebra of $L^1(\mathbb{Z}^d)$.
    \item Non-Commutative (\cref{theorem:characterization_of_twisted_crossed_products}): There is a Banach *-algebra $L^1(\mathbb{Z}^d,A;\alpha, \zeta)$ such that, any representation
    $$ \pi: L^1(\mathbb{Z}^d,A;\alpha, \zeta) \to B(H) $$
    extends uniquely to a representation
    $$ \tilde{\pi}: A \rtimes_{\alpha, \zeta} \mathbb{Z}^d \to B(H).$$
    Moreover, $A \rtimes_{\alpha, \zeta} \mathbb{Z}^d$ is the enveloping C* algebra of $L^1(\mathbb{Z}^d,A;\alpha, \zeta)$.
\end{itemize}
\end{remark}

\begin{remark}\label{remark:nuclearity_translating_concepts}
Twisted crossed product and nuclear C* algebras,
\begin{itemize}
    \item Commutative (\cref{example:nuclear_C_star_algebras}): The C* algebra $C(\mathbb{T}^d)$ is nuclear.
    \item Non-Commutative (\cref{theorem:twsited_crossed_product_amenable_nuclear}): If $A$ is a nuclear C* algebra then $A \rtimes_{\alpha, \zeta} \mathbb{Z}^d$ is a nuclear C* algebra.
\end{itemize}
\end{remark}

\begin{remark}\label{remark:fourier_coefficients_translating_concepts}
Twisted crossed products and fourier coefficients,
\begin{itemize}
    \item Commutative (\cref{section:n_torus}): Every $f \in C(\mathbb{T}^d)$ is uniquely determined by a set 
    $$\{ \mathcal{F}(f)(s) \}_{s \in \mathbb{Z}^d}, \; \mathcal{F}(f)(s) \in \mathbb{C},$$
    where
    $$ \mathcal{F}(f)(s) =  \int_{\mathbb{T}^d} \gamma_{-s}(\lambda) f(\lambda) d \mu(\lambda). $$
    In this expression, $ \gamma_{-s}(\lambda) $ is the character over $\mathbb{T}^d$ given by $\lambda^{-s}$.
    \item Non-Commutative (\cref{lemma:computation_of_Fourier_coefficients}): Every $p \in A \rtimes_{\alpha, \zeta} \mathbb{Z}^d$ is uniquely determined by a set 
    $$\{ \Phi_s (p) \}_{s \in \mathbb{Z}^d}, \; \Phi_s(p) \in A,$$
    where 
    $$ \Phi_s(p) u^s = \int_{\mathbb{T}^d} \gamma_{-s}(\lambda) \tau(\lambda)(p) d \mu(\lambda).$$
    In this expression, $\tau$ is a strongly continuous action of $\mathbb{Z}^d$ over $A \rtimes_{\alpha, \zeta} \mathbb{Z}^d$ and $ \gamma_{-s}(\lambda) $ is the character over $\mathbb{T}^d$ given by $\lambda^{-s}$.
\end{itemize}
\end{remark}

\begin{remark}\label{remark:faithful_representation_translating_concepts}
Twisted crossed products and faithful representations,
\begin{itemize}
    \item Commutative (\cref{section:n_torus}): $C(\mathbb{T}^d)$ has a faithful representation over
    $$ \sum_{\mathbb{Z}^d} \mathbb{C} \simeq L^2(\mathbb{Z}^d) $$
    given by infinite matrices $\pi(f) \in B(\sum_{\mathbb{Z}^d} \mathbb{C})$, whose $(l,m)$ component is
    $$ \mathcal{F}(f)(l-m). $$ 
    \item Non-Commutative (\cref{lemma:Fourier_coefficients_and_right_regular_representation}): Let $\pi: A \to B(H)$ be a faithful representation, then, $A \rtimes_{\alpha, \zeta} \mathbb{Z}^d$ has a faithful representation over 
    $$\sum_{\mathbb{Z}^d} H  \simeq L^2(\mathbb{Z}^d,H) $$
    given by infinite matrices $\tilde{\Pi}(p) \in B(\sum_{\mathbb{Z}^d} H)$, whose $(l,m)$ component is
    $$[\zeta(l,m-l) \pi(\alpha(l)( \Phi_{m-l}(p)))]_{l,m \in \mathbb{Z}^d}.$$\end{itemize}
\end{remark}

\begin{remark}\label{remark:l_2_norm_translating_concepts}
Twisted crossed product and the $L^2$ norm,
\begin{itemize}
    \item Commutative (\cref{section:Fourier_analysis_and_ortonormal_basis}): For every $f \in C(\mathbb{T}^d)$,
    $$\mathcal{F}(ff^*)(0) = \sum_{z \in \mathbb{Z}^d} |\mathcal{F}(f)(s)|^2 $$
    such that
    $$ \mathcal{F}(ff^*)(0) = 0 $$
    iff $f = 0$.
    \item Non-Commutative (\cref{lemma:absolute_convergence_of_zero_fourier_coefficient}): For every $p \in A \rtimes_{\alpha, \zeta} \mathbb{Z}^d$,
    $$\Phi_0(p p^*) = \sum_{z \in \mathbb{Z}^d} \Phi_s(a) \Phi_s(a)^*$$
    such that
    $$ \Phi_0(p p^*) = 0 $$
    iff $p = 0$.
\end{itemize}
\end{remark}

\begin{remark}\label{remark:tensor_product_translating_concepts}
Twisted crossed products and tensor products,
\begin{itemize}
    \item Commutative (\cref{proposition:isomorphism_tensor_product_of_commutative_C_star_algebras}): Let $B$ be any C* algebra, then
    $$C(\mathbb{T}^d) \otimes B \simeq C(\mathbb{T}^d,B).$$
    \item Non-Commutative (\cref{proposition:isomorphism_tensor_product_twisted_crossed_product}): Let $B$ be any C* algebra, then
    $$ ( A \rtimes_{\alpha, \zeta} \mathbb{Z}^d )\otimes B \simeq (A \otimes B) \rtimes_{\hat{\alpha}, \zeta} \mathbb{Z}^d, $$
    with 
    $$ \hat{\alpha}(s)(a \otimes b) = \alpha(s)(a) \otimes b. $$
\end{itemize}
\end{remark}

\begin{remark}\label{remark:iterated_tensor_product_translating_concepts}
Twisted crossed products and iterated tensor products,
\begin{itemize}
    \item Commutative (\cref{section:n_torus}): We have that
    $$C(\mathbb{T}^d) \simeq \underbrace{C(\mathbb{T}) \otimes \cdots \otimes C(\mathbb{T})}_{d \text{ times }},$$
    and for any $1 \leq j \leq d-1$, $C(\mathbb{T}^j)$ is a sub C* algebra of $C(\mathbb{T}^d)$.
    \item Non-Commutative (\cref{lemma:iterated_crossed_product}): $A \rtimes_{\alpha, \zeta} \mathbb{Z}^d$ can be written as an iterated crossed product with $\mathbb{Z}$, that is, there are $d$ canonical C* algebras $\{ A_i \}_{0 \leq i \leq d-1}$ and group actions $\alpha_i : \mathbb{Z} \to \text{Aut}(A_{i-1})$ such that
    $ A_{i+1} \simeq A_i \rtimes_{\alpha_i} \mathbb{Z} $ and
    $$ A \rtimes_{\alpha, \zeta} \mathbb{Z}^d \simeq  A_{d-1} \rtimes_{\alpha_d} \mathbb{Z}, $$
    with $A_0 = A$. For $1 \leq i \leq d-1$, $A_i$ is the C* algebra generated by elements of the form
    $$ a u_j, \; 1 \leq j \leq i, \; a \in A $$
    following the commutation relations
    $$ a_0 u_k a_1 u_l = a_0 \alpha(e_k)(a_1) u_k u_l $$ 
    $$  a_0 \alpha(e_k)(a_1) u_k u_l = \zeta(k,l) a_0 \alpha(e_k)(a_1) u_l u_k. $$
    Additionally, for $0 \leq j \leq d-1$, $A_{j}$ is a sub C* algebra of $A_{d-1} \rtimes_{\alpha_d} \mathbb{Z}$.
\end{itemize}
\end{remark}

\begin{remark}\label{remark:smooth_sub_algebras_translating_concepts}
Twisted crossed products and smooth sub algebras,
\begin{itemize}
    \item Commutative (\cref{section:Fourier_transform_and_Frechet_algebras}): $C^{\infty}(\mathbb{T}^d)$ is Fréchet m-convex algebra with 
    $$ C^{\infty}(\mathbb{T}^d) \subset C(\mathbb{T}^d), $$
    such that, $f \in C^{\infty}(\mathbb{T}^d)$ iff for every $n \in \mathbb{N}^d$ there is $K_n < \infty$ with
    $$ |s|^n | \mathcal{F}(f)(s) | < K_n. $$
    Additionally, $C^{\infty}(\mathbb{T}^d) $ is invariant under the holomorphic functional calculus of $C(\mathbb{T}^d)$, and the set of self adjoint elements of $C^{\infty}(\mathbb{T}^d)$ is invariant under the smooth functional calculus on $\mathbb{R}$ defined over $C(\mathbb{T}^d)$.
    \item Non-Commutative (\cref{corollary:smooth_elements_are_invariant_under_holomoprhic_and_c_infty_calculus}): There is Fréchet m-convex algebra $\mathcal{A}_{\alpha, \zeta}$ with
    $$\mathcal{A}_{\alpha, \zeta} \subset  A \rtimes_{\alpha, \zeta} \mathbb{Z}^d,$$
    such that, $p \in \mathcal{A}_{\alpha, \zeta}$ iff for every $n \in \mathbb{N}^d$ there is $K_n < \infty$ with
    $$ |s|^n \| \Phi_s(p) \| < K_n. $$
    Additionally, $\mathcal{A}_{\alpha, \zeta}$ is invariant under the holomorphic functional calculus of $A \rtimes_{\alpha, \zeta} \mathbb{Z}^d$, and the set of self adjoint elements of $\mathcal{A}_{\alpha, \zeta}$ is invariant under the smooth functional calculus on $\mathbb{R}$ defined over $A \rtimes_{\alpha, \zeta} \mathbb{Z}^d$.
\end{itemize}
\end{remark}

\begin{remark}\label{trigonometric_polunomials_translating_concepts}
Twisted crossed products and trigonometric polynomials,
\begin{itemize}
    \item Commutative (\cref{section:Fourier_transform_and_Frechet_algebras}): Define
    $$ f_{\mathcal{R}}: \mathbb{T}^d \to \mathbb{C}, \; f_s(\lambda) = \sum_{s \in \mathcal{R}, \; \mathcal{R} \subset \mathbb{Z}^d} \exp(i \lambda \cdot s ),$$
    let
    $$ \text{Tri}(\mathbb{T}^d) = \{ f_{\mathcal{R}} | \; |\mathcal{R}| < \infty \}, $$
    then $\text{Tri}(\mathbb{T}^d)$ is the set of trigonometric polynomials over $\mathbb{T}^d$ and
    $$ \text{Tri}(\mathbb{T}^d) \subset C^{\infty}(\mathbb{T}^d). $$
    \item Non-Commutative (\cref{lemma:smooth_elements_are_a_sub_algebra}): Let 
    $$ \mathcal{P}(A, \mathbb{Z}^d, \alpha, \zeta) = \{ \sum_{s \in \mathcal{R} , \; \mathcal{R} \subset \mathbb{Z}^d } a_s u^s| \; |\mathcal{R}| < \infty \}, $$
    then, $\mathcal{P}(A, \mathbb{Z}^d, \alpha, \zeta)$ is called the set of non-commutative trigonometric polynomials and
    $$ \mathcal{P}(A, \mathbb{Z}^d, \alpha, \zeta) \subset  \mathcal{A}_{\alpha, \zeta}.$$
\end{itemize}
\end{remark}

\begin{remark}\label{remark:derivartions_translating_concepts}
Twisted crossed products and derivations,
\begin{itemize}
    \item Commutative (\cref{section:Fourier_transform_and_Frechet_algebras}): The derivations
    $$\partial_j: C^{\infty} (\mathbb{T}^d) \to  C^{\infty} (\mathbb{T}^d),$$
    given by,
    $$ \partial_j f := \frac{\partial f}{ \partial \lambda_j} ,$$
    are continuous and commute.
    
    \item Non-Commutative (\cref{lemma:algebra_of_smooth_elements_is_Frechet_space}): The maps given by
    $$\partial_j : \mathcal{A}_{\alpha, \zeta} \to \mathcal{A}_{\alpha, \zeta},$$
    with,
    $$ \Phi_s(\partial_j a) = i s_j \Phi_s(a),$$
    are continuous commuting derivations over $\mathcal{A}_{\alpha, \zeta}$.
\end{itemize}
\end{remark}

\begin{remark}\label{remark:trace_translating_concepts}
Twisted crossed products and traces,
\begin{itemize}
    \item Commutative (\cref{section:n_torus}): The map $ f \mapsto \mathcal{F}(f)(0)$ with
    $$\mathcal{F}(f)(0) = \int_{\mathbb{T}^d} f(\lambda) \frac{d \mu(\lambda)}{ (2 \pi)^{d} } $$
    is a faithful continuous trace over $C(\mathbb{T}^d)$.
    \item Non-Commutative (\cref{section:fourier_analysis_weight_traces_and_states_apendix}): We have that
    $$ \Phi_0( (A \rtimes_{\alpha, \zeta} \mathbb{Z}^d)_{\text{pos}} ) = A_{\text{pos}}, $$
    also, if $\gamma$ is a trace over $A$ such that 
    $$\gamma \circ \alpha(s) = \gamma $$
    for all $s \in \mathbb{Z}^d$, then
    $$ \gamma \circ \Phi_0  $$
    is a trace over $A \rtimes_{\alpha, \zeta} \mathbb{Z}^d$, which is continuous if 
    $\gamma$ is continuous and is faithful if $\gamma$ is faithful.
\end{itemize}
\end{remark}

\begin{remark}[Continuous field of C* algebras]\label{remark:contirnuous_field_c_stra_algebras_translating_concepts}
Twisted crossed products ans continuous fields of C* algebras,
\begin{itemize}
    \item Commutative (\cref{sec:Continuous_functions_with_values_on_C_star_algebra}): We have that 
    $$C(\mathbb{T}^d)\otimes C_0(\Omega) \simeq C_0(\Omega, C(\mathbb{T}^d)),  $$
    and
    $$ C_0(\Omega, C(\mathbb{T}^d)) \subseteq C_b(\Omega, C(\mathbb{T}^d)),   $$
    such that, for $f \in C_0(\Omega, \mathbb{T}^d)$,  
    $$\|f\| = \sup_{\omega \in \Omega} \| f(\omega)\| $$
    \item Non-Commutative (\cref{remark:faithful_representation_of_C_0_Omega_rtimes_Z}): Let $C(\Omega) \rtimes_{\alpha, \Theta} \mathbb{Z}^d$ be a twisted transformation group C* algebra such that $\Omega$ is compact, assume that there is a Radon measure with full support $\mu$ over $\Omega$ and $\Omega$ has finite measure i.e. $\mu(\Omega) < \infty$, additionally, assume that $\alpha$ is an isometric action (\cref{definition:isometric_action_twisted_transf_group_C_star_algebras}). Then,
    $$ C(\Omega) \rtimes_{\alpha, \zeta} \mathbb{Z}^d \subseteq C(\Omega, B(L^2(\mathbb{Z}^d)))  $$
    and $p \in C(\Omega) \rtimes_{\alpha, \zeta} \mathbb{Z}^d$ can be seen as $p = \{ p_{\omega}\}_{\omega \in \Omega} \subset C(\Omega, B(L^2(\mathbb{Z}^d)))$, with
    $$\|p \|_{C(\Omega) \rtimes_{\alpha, \zeta} \mathbb{Z}^d} = \sup_{\omega \in \Omega} \| p_{\omega} \|_{B(L^2(\mathbb{Z}^d))}$$
\end{itemize}
\end{remark}

\section{A sneak and peak into the Toeplitz extension}
\label{sec:a_sneak_and_peak_into_the_toeplitz_extension}

Let $A$ be any C* algebra, then, $A \rtimes_{\alpha, \zeta} \mathbb{Z}^d$ is part of a three-term exact sequence, which, following the notation on \cref{remark:iterated_tensor_product_translating_concepts} looks like
$$
\boldsymbol{0 \longrightarrow A_{d-1} \otimes \mathcal{K} \stackrel{\psi}{\longrightarrow}  T\left(A_{d-1} \rtimes_{\alpha_{d}} \mathbb{Z}\right) \stackrel{\pi}{\longrightarrow} A_{d-1} \rtimes_{\alpha_{d}} \mathbb{Z} \longrightarrow 0}, 
$$
where (\citep[Section 2]{pimsner_exact_1980}):
\begin{itemize}
    \item $\boldsymbol{A_{d-1} \rtimes_{\alpha_{d}} \mathbb{Z} \simeq A \rtimes_{\alpha, \zeta} \mathbb{Z}^d}$.
    \item $\boldsymbol{T\left(A_{d-1} \rtimes_{\alpha_{d}} \mathbb{Z}\right)}$ is the sub C* algebra of $(A_{d-1} \rtimes_{\alpha_{d}} \mathbb{Z}) \otimes \mathcal{T}$ generated by elements of the form $ \boldsymbol{u_d \otimes v} $ and $ \boldsymbol{\tilde{a} \otimes 1} $, with 
    \begin{itemize}
        \item $\tilde{a} \in A_{d-1}$,
        \item $\mathcal{T}$ is the \textbf{Toeplitz C* algebra} (\cref{section:Toeplitz_algebra}), that is, the C* algebra generated by a partial isometry $v$,
        $$v^* v =1, \; 1- v v^* = P, \; P = P^*, \; P^2 = P.  $$ 
    \end{itemize} 
    The C* algebra $\boldsymbol{T\left(A_{d-1}\right)}$ is called the \textbf{Toeplitz extension}\index{Toeplitz extension} of the crossed product $A_{d-1} \rtimes_{\alpha_{d}} \mathbb{Z}$.
\end{itemize}

Set $\Omega$ and $\alpha$ as in the Non-Commutative Brillouin Torus (\cref{definition:non_commutative_brillouin_torus}), denote $A = C(\Omega)$, then, according to \citep[Proposition 4.2.1]{prodan_bulk_2016} $\boldsymbol{\alpha_j(s)}$ is homotopic to the identity map over $\boldsymbol{A_{j-1}}$ for any $\boldsymbol{1 \leq j \leq d, \; s \in \mathbb{Z}}$, this is a key fact to compute the K groups of the C* algebras in the short exact sequence (\citep[Proposition 4.2.4]{prodan_bulk_2016}). The K groups of the C* algebras in the short exact sequence take the following form (\citep[Proposition 4.2.4]{prodan_bulk_2016})
$$ \boldsymbol{ K_j(A_{l-1} \rtimes_{\alpha_{l}} \mathbb{Z}) =  \mathbb{Z}^{2^{l-1}}, \; j = 0,1, \; 1 \leq l \leq d} $$
$$ \boldsymbol{K_j(A_{l-1} \otimes \mathcal{K}) = K_j(T\left(A_{l-1} \rtimes_{\alpha_{l}} \mathbb{Z}\right)) = \mathbb{Z}^{2^{l-2}}, \; j = 0,1, \; 1 \leq l \leq d.}$$

\subsection{Applications into solid state systems}
\label{sec:applications_into_solid_state_physics}

If we set $A = M_n(C(\Omega))$, then, according to \citep[Section 3.2]{prodan_bulk_2016} the short exact sequence
$$
\boldsymbol{0 \longrightarrow A_{d-1} \otimes \mathcal{K} \stackrel{\psi}{\longrightarrow}  T\left(A_{d-1} \rtimes_{\alpha_{d}} \mathbb{Z}\right) \stackrel{\pi}{\longrightarrow} A_{d-1} \rtimes_{\alpha_{d}} \mathbb{Z} \longrightarrow 0}, 
$$
describes the relation between the following two objects,
\begin{itemize}
    \item A homogeneous system in $d$ dimensions with no boundary, in which case its Hamiltonian belongs to $A_{d-1} \rtimes_{\alpha_{d}} \mathbb{Z} \simeq M_n(C(\Omega)) \rtimes_{B_+,\alpha} \mathbb{Z}^d$.
    \item A homogeneous system in $d$ dimensions with a boundary in the $d$ dimension, in which case its hamiltonian belongs to $T\left(A_{d-1} \rtimes_{\alpha_{d}} \mathbb{Z}\right)$.
\end{itemize}

Under this setting the C* algebra $A_{d-1} \otimes \mathcal{K}$ contains the Hamiltonians whose action can be approximated arbitrarily by operators in a finite neighborhood of the boundary, therefore, it is referred to as the \textbf{algebra of boundary observables}. Similarly, $A \rtimes_{\alpha, B_+} \mathbb{Z}^d$ is referred to as the \textbf{algebra of bulk observables}, and $T\left(A_{d-1} \rtimes_{\alpha_{d}} \mathbb{Z}\right)$ is referred to as the \textbf{algebra of half space obsevables}.

Using the smooth sub algebra of $A_{d-1}$ and the smooth sub algebra of $\mathcal{K}$ given by the infinite matrices with rapid decay (\cref{example:infinte_matrices_with_rapid_decay}), it is possible to define a smooth sub algebra of $A_{d-1} \otimes \mathcal{K}$, which we call $\mathcal{E}_d$ (\citep[Proposition 3.3.3]{prodan_bulk_2016}). Since the canonical trace over $\mathcal{K}$ becomes a continuous linear functional over the algebra of infinite matrices with rapid decay, it is possible to use the canonical trace over $\mathcal{K}$ along with the derivations densely defined in $A_{d-1}$ to provide continuous cyclic cocycles over $\mathcal{E}_d$ (\citep[Section 5.3]{prodan_bulk_2016}). Interestingly enough, these cocycles along with the maps $\text{Ind}$ and $\text{Exp}$ associated with the aforementioned short exact sequence provide the right tool to prove that
\begin{itemize}
    \item if $d=2,3$ and the Bulk Gap Hypothesis holds, then, the current along the $j$ direction of the boundary, with $1 \leq j \leq d $, is proportional to the entry $(j,d)$ of the conductivity tensor, therefore, it is quantized (\citep[Corollaries 7.2.1 and 7.2.2]{prodan_bulk_2016}),
    \item if $d=1,2$, the Bulk Gap Hypothesis holds and the Hamiltonian has chiral symmetry, then, the chiral polarization on the boundary can be computed by an odd Chern number and is quantized (\citep[Corollaries 7.3.2 and 7.3.3]{prodan_bulk_2016}).
\end{itemize}
The previously mentioned properties of homogeneous materials are known as the \textbf{bulk boundary correspondence for homogeneous materials} because the previous statements set a relation between physical observables on the boundary of homogeneous materials and the value of physical observables on the bulk of homogeneous materials (infinite model with no boundaries). 

From the results mentioned above, if $\{H_{\omega}\}_{\omega \in \Omega}$ is a covariant family of Hamiltonians in dimension $2$ or $3$ with Fermi projection given by $P_F$ and $\text{Ch}_{\{j,d\}}(P_F) \neq 0$, then, the current on the $j$ direction of the boundary is not zero and is equal to the transverse conductivity of the bulk model. Additionally, since  $\text{Ch}_{\{j,d\}}(P_F)$ is invariant under deformations of $H$ inside the weak disorder regime, the current on the boundary is also resilient under such deformations, so, given that such models require a boundary to become conductive and their conductivity is resilient under small deformations they are called \textbf{topological insulators}. 

We refrain from diving into the \textbf{bulk boundary correspondence for homogeneous materials} due to time constraints, nonetheless, we would like to emphasize both the importance of such results in the study of topological insulators and its importance in the study of continuous cyclic cocycles over $C(\Omega, M_n(\mathbb{C}))\rtimes_{\alpha, B_+} \mathbb{Z}^d$, since it is key to the computation of the range of $\text{Ch}_I$ (\citep[Theorem 5.7.1]{prodan_bulk_2016}).

\chapter{Personal comments}
\label{chapter:personal_comments}

\section{Personal motivation}
\label{sec:personal_motivation}

Our main reference is Prodan and Baldes book about topological insulators and K theory \citep{prodan_bulk_2016}. This document is devised as a guide for newcomers mathematicians and physicists that would like to understand some of Prodan and Baldes's developments in \citep{prodan_bulk_2016} while keeping a good grasp of the mathematical tools behind those constructions. You may wonder why we are taking such an endeavor when there are tons of books about K theory, non commutative geometry and topological insulators, well, that is a fair question, and to be honest the answer lays in the challenges I faced while trying to follow the book of Prodan and Baldes. This document is the result of my personal journey through that book and many other references about non commutative geometry, and I would like to give a version of the story that is really helpful for students like myself. So, I devised this document as a guide for newcomers, like a document that my past version would have valued a lot, and I hope that it is also of some utility to you.

\subsection{Disclaimer}
\label{sec:disclaimer}

The literature on non-commutative geometry is bast and in practice never ending, many new papers are appearing every day and the learning curve is really step, such that it takes many hours of work just to get grasp of some concepts. Therefore, the concepts presented in this document are far from novel, even more, I would call them standard in the community. Most of the results used in this document are not given a proof inside this document, for example, the chapter on cyclic cohomology is a recount on results and explanations of concepts, whilst only a few actual computations are done. That is common theme through the document, and is both a statement of practicality and a reflection of my limitations as a mathematician. I have tried my best to provide a clear path through my reasoning while not spending an absurd amount of time going through the details of all the tools necessary to traverse that path, therefore, I have focused on the details I though were missing the in literature that I followed and I do my best to provide good reference for the other results. 

Non-commutative geometry is a highly interdisciplinary field and there are many perspectives to tackle the same problem. The approach in this document is highly biased by my background and my objective, which is to analyse topological insulators, that is why this document has a strong focus on the topological side of the algebras along with the analytical tools inside those algebras. We pay little attention to other faces of non commutative geometry, for example, we do not go into the details of the algebraic structure of neither the algebras we use, nor we dive into the geometry of the spaces under studied. Non-commutative geometry also has a rich category description and most of its constructions have a categorical stand point of view that is intellectually attractive, however we do not dive into that analysis. If you were hooping to get an algebraic, geometric or categorical analysis out of this document that is not the case, nonetheless, is highly probable that some of the tools that are presented in this document will be useful in your future incursions into non-commutative geometry, so it may not be such a waste of time to join me in my journey.

If you are looking for another points of view on non-commutative geometry you will find them in the bibliography, however, as the bibliography is by no means a small, I recommend you to star with the classics \citep{connes_noncommutative_2014}, \citep{gracia-bondia_elements_2001}, \citep{khalkhali_basic_2013}. Representation theory is another subject that shines up due to its absence, if you want to dive into the representation theory of C* algebras and Von Neumann algebras we recommend \citep{kadison_fundamentals_1983_V1} and \citep{kadison_fundamentals_1983_V2}. For the representation theory of groups and harmonic analysis we recommend the book \citep{dales_introduction_2003}.

If you happen to have questions or comments about this document you can contact me through email \href{mailto:jsflorez@gmail.com}{jsflorez@gmail.com}, in StackExchange as \href{https://math.stackexchange.com/users/1143665/trucupey}{trucupey} and in Mathoverflow as \href{https://mathoverflow.net/users/507127/trucupey}{trucupey}.

\printindex

\cleardoublepage
\phantomsection
\addcontentsline{toc}{chapter}{Bibliography}
\bibliographystyle{apalike}
\bibliography{thesis}


\includepdf{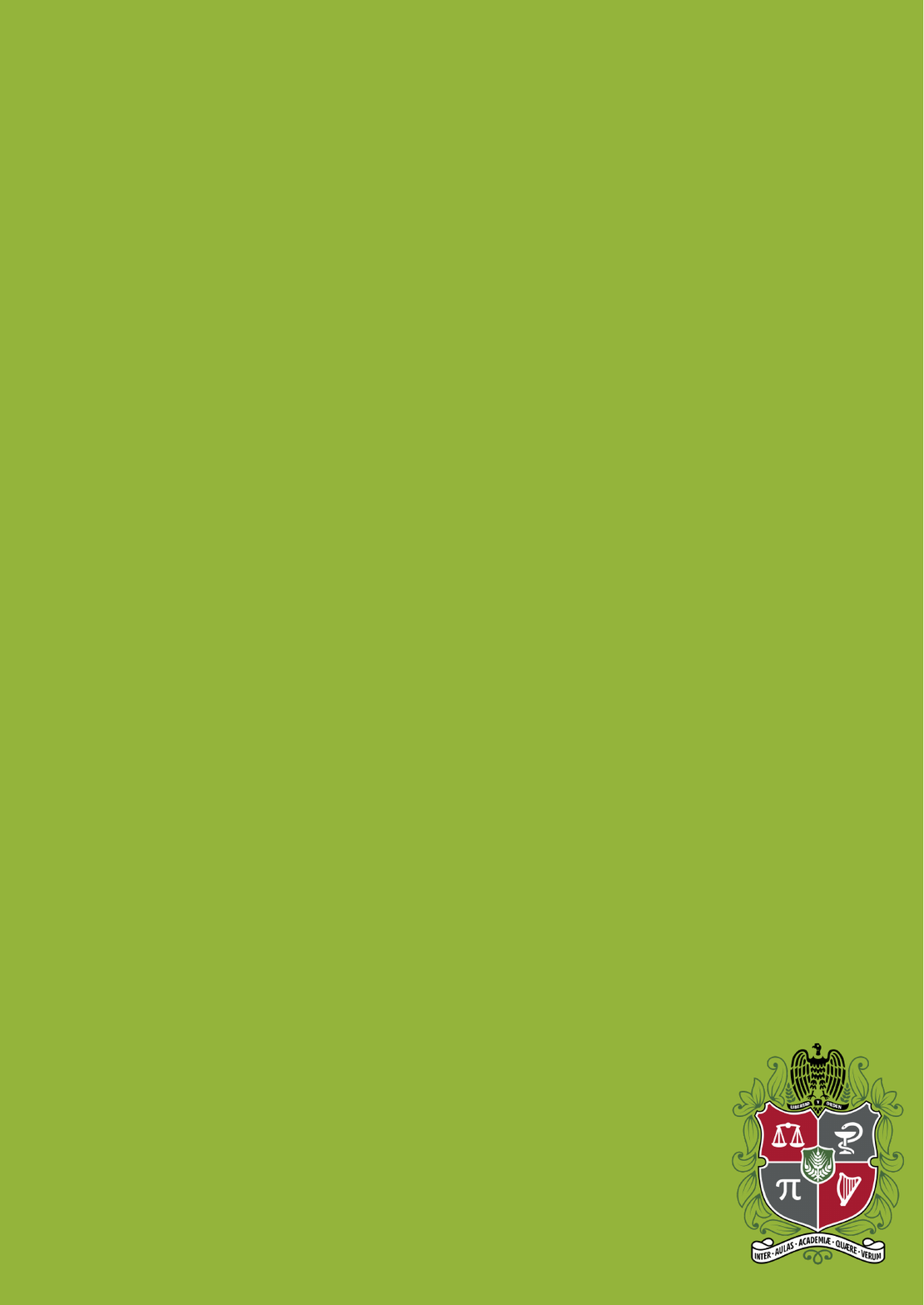}

\end{document}